\documentclass[11pt,a4paper,oneside,openright]{book}
\usepackage[a4paper,lmargin=30mm, textwidth=160mm, tmargin=20mm, textheight=257mm]{geometry}

\usepackage[onehalfspacing]{setspace}

\usepackage[defaultlines=4,all]{nowidow}

\usepackage[utf8]{inputenc}
\usepackage[T1]{fontenc}
\usepackage{lmodern}
\usepackage[english]{babel} 

\usepackage[bottom]{footmisc}

\usepackage[useregional]{datetime2}
\DTMlangsetup[en-GB]{ord=raise,monthyearsep={,\space}}

\usepackage{fancyhdr}

\fancypagestyle{plain}{
	\fancyhf{}
	\fancyfoot[R]{\thepage}
	
}

\usepackage[titletoc,toc,title]{appendix}

\usepackage{etoolbox}
\makeatletter
\newcommand{\thechaptername}{}
\newcounter{chapter@last@figure}
\newcounter{chapter@last@table}

\pretocmd{\caption}{
	\ifnum\pdfstrcmp{\@captype}{figure}=0
		\ifnum\value{chapter}=\value{chapter@last@figure}
		\else
			\addtocontents{lof}
			{\protect\numberline{\bfseries\thechapter\quad\thechaptername}}
		\fi
	\fi
	
	\ifnum\pdfstrcmp{\@captype}{table}=0
		\ifnum\value{chapter}=\value{chapter@last@table}
		\else
			\addtocontents{lot}
			{\protect\numberline{\bfseries\thechapter\quad\thechaptername}}
		\fi
	\fi
	
	\ifnum\pdfstrcmp{\@captype}{algorithm}=0
		\ifnum\value{chapter}=\value{chapter@last@algorithm}
		\else
			\addtocontents{loa}
			{\protect\numberline{\bfseries\thechapter\quad\thechaptername}}
		\fi
	\fi
\expandafter\setcounter\expandafter{chapter@last@\@captype}{\value{chapter}}
}{}{failure}

\makeatother

\usepackage{bibentry} 
\nobibliography*

\usepackage{amsmath,amsfonts,amsthm,bm,amssymb,mathtools,amssymb} 

\usepackage{braket}

\usepackage{bbold} 
\usepackage{bigints} 
\usepackage{relsize} 
\usepackage{mathabx} 
\DeclareMathOperator{\sech}{sech} 

\usepackage{graphicx}
\usepackage{subcaption}
\usepackage[export]{adjustbox} 
\usepackage{float}

\usepackage{xcolor} 
\usepackage{comment} 

\usepackage[pagebackref]{hyperref} 

\usepackage{color}   
\hypersetup{
    colorlinks=true, 
    linktoc=all,     
    linkcolor=cyan,  
    citecolor=blue
}

\begin{document}

\frontmatter

\begin{titlepage}
\begin{singlespace}
\begin{center}

\centering
\includegraphics[width=0.2\textwidth]{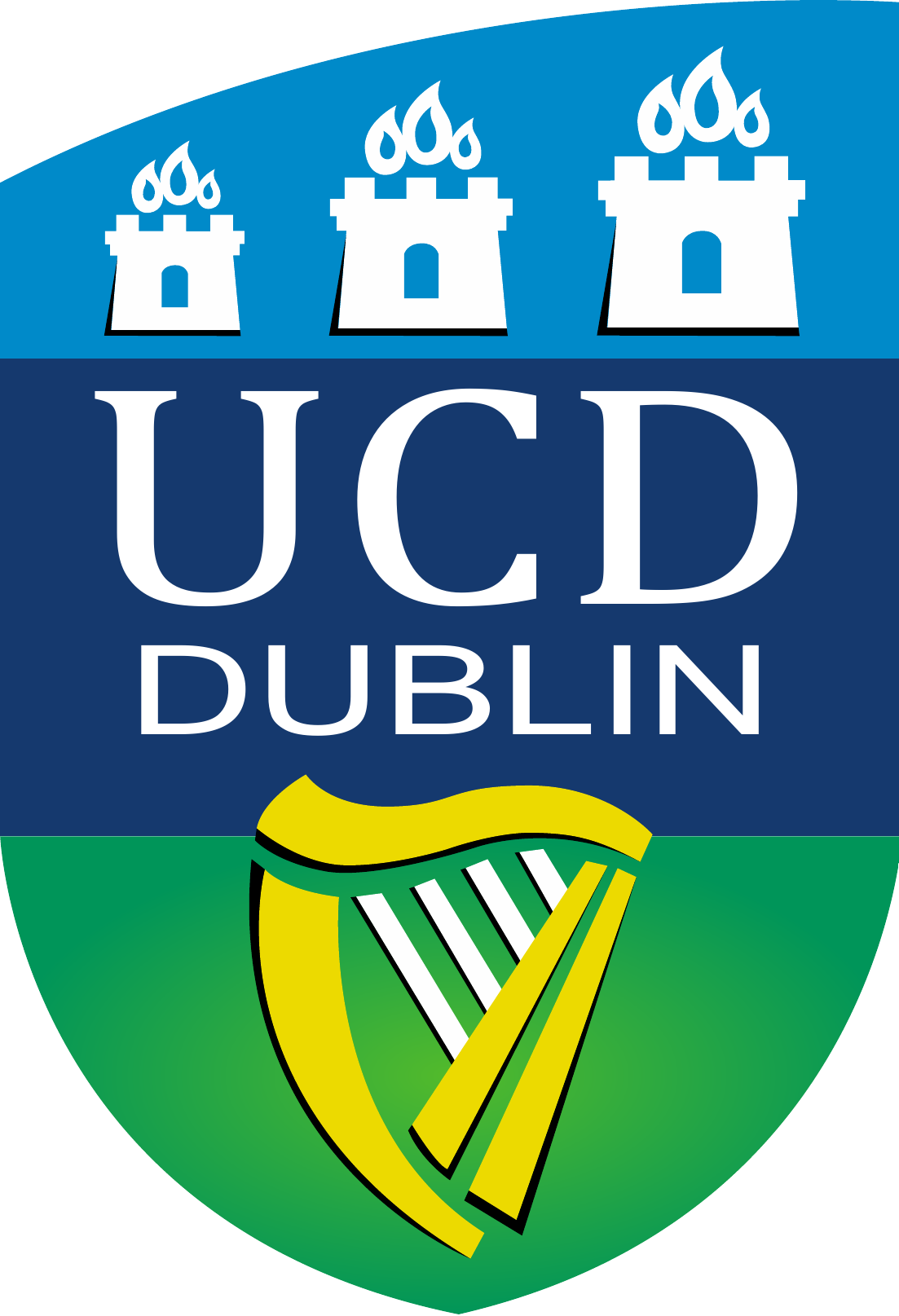} 
\par\vspace*{4\baselineskip}

\Huge
\textbf{Quantum transport \\ in interacting nanodevices}\\
\huge
From quantum dots to single-molecule transistors
\vfill

\Large
by

\vfill

\textbf{Emma L. Minarelli}

\small
17209152

\vfill

\Large
The thesis is submitted to University College Dublin in fulfilment of the requirements for the degree of Doctoral Philosophy.

\vfill

\textbf{UCD School of Physics}

\vfill

\large

\begin{tabular}{rl}
\textbf{Head of School}: Prof.\ Emma Sokell 
\end{tabular}
\vfill
\begin{tabular}{rl}
\textbf{Principal Supervisor}: Ass. Prof. Andrew K. Mitchell 
\end{tabular}
\vfill
\begin{tabular}{rl}
\textbf{Doctoral Studies Panel}: & Prof. Ronan McNulty \\
& Prof. Vladimir Lobaskin
\end{tabular}
\vfill

\Large
September 5, 2022

\end{center}
\end{singlespace}
\end{titlepage}	

\begin{singlespace}	
	
\tableofcontents
	
\begin{doublespace}
\chapter*{Abstract}
The enormous interest in industrial application of semiconductor components has led to the development of unprecedented control over the manufacture of electronic devices on the nanometer scale. This allows to perform highly controllable and fine-tuned experiments in the quantum regime where exotic effects can nowadays be measured. Among those, breakthrough measurements of electrical conductance experimentally confirmed the Kondo effect - a many-body quantum effect involving macroscopic entanglement. In quantum dot devices, enhanced conductance below a characteristic energy scale is the signature of Kondo singlet formation. Precise predictions of quantum transport properties in similar nanoelectronics devices are therefore desired to design optimal functionality and control.\\
Standard mesoscopic transport methods suffer from limitations in nanostructure specifics, set-up design, energy, temperature and voltage regime of applicability. To overcome these issues, such that we obtain modelling flexibility and accurate conductance predictions, in this thesis we analytically derive alternative and improved quantum transport formulations having as their starting point scattering theory in the Landauer-B{\"u}ttiker formula, linear response theory in the Kubo formula, nonequilibrium Keldysh theory in the Meir-Wingreen formula and Fermi liquid theory in the Oguri formula. We perform a systematic benchmark of our exact expressions, comparing with the standard approaches using a state-of-the-art numerical renormalization group techniques (NRG). The new formulations not only reproduce literature results, but also show higher accuracy and computational efficiency, as well as a wider applicability under regimes and conditions out of reach by existing methods. We also derive generalized effective models for multi-orbital two-lead interacting nanostructures in both Coulomb blockade and mixed-valence regime, which yield reusable conductance predictions directly in terms of the effective model parameters.\\
We conclude by applying our novel formulations to complex nanoelectronics systems, including a single-molecule benzene transistor, a charge-Kondo quantum dot made from graphene and a semiconductor triple quantum dot.
\end{doublespace}

\newpage
\chapter*{List of publications}
This thesis is based on the following articles:
\begin{enumerate}
	\item \underline{Emma L. Minarelli}, Jonas B. Rigo, Andrew K. Mitchell, \textit{Linear response quantum transport through interacting multi-orbital nanostructures}, preprint $2209.01208$ \cite{transport} ~;
	\item \underline{Emma L. Minarelli}, Jonas B. Rigo, Andrew K. Mitchell, \textit{Two-Channel Charge-Kondo physics in graphene quantum dots}, Nanomaterials 2022, 12(9), 1513 \cite{minarelli2022two} ~;
	\item \underline{Emma L. Minarelli}, and Andrew K. Mitchell, \textit{Resonant spin-flip conductance in charge-Kondo circuits}, in preparation \cite{CK}~;
	\item Sudeshna Sen, \underline{Emma L. Minarelli}, Jonas B. Rigo  and Andrew K. Mitchell,\textit{Theory of a benzene transistor: symmetry and strong correlations and quantum interference},  in preparation \cite{benzene}~.\\
	
Other publication to which the author has contributed:
 	\item \underline{Emma L. Minarelli}, Kim P{\"o}yh{\"o}nen, Gerwin A.R. Van Dalum, Teemu Ojanen, Lars Fritz, \textit{Engineering of Chern insulators and circuits of topological edge states}, Phys. Rev. B 99, 165413 \cite{minarelli2019engineering} ~.
\end{enumerate}


\newpage
\chapter*{Statement of Original Authorship}
\vspace{4cm}
I hereby certify that the submitted work is my own work, was completed while registered as a candidate for the degree stated on the Title Page, and I have not obtained a degree elsewhere on the basis of the research presented in this submitted work.

\newpage
\chapter*{Acknowledgements}
The last page of the thesis is a good moment to think back about the intense years of work that brought me to this point. It is really true that knowing the path is not the same as walking it. In these last moments before the submission, I would like to spend words on the people who have walked with me during these years.\\

\noindent{I} thank the PI in our research group Andrew Mitchell for the work completed together. Thanks for showing me what scientific research is about, for teaching me the optimal way to make scientific communication, for presenting me the crucial importance of collaboration through networking that is sense of scientific community and for your support at the initial stages of my PhD.\\
I thank the former PostDoc in our research group  Prof. Sudeshna Sen for the continuous support and discussion. I do value your knowledge across various topics in condensed matter from which I have learnt a lot. Especially, I enjoyed our collaboration on the benzene molecule project - leading to Chapter 7.\\ 
I thank my group mate and close friend Jonas Rigo for teaching me the know-how of the NRG code used in the group. Thank you for producing in record time all the plots in Chapter 6 (as the numerical background required for those was beyond me), for proofreading introduction, conclusion and the subsection on NRG in the thesis. Across the years, thank you for sharing with me your viewpoints and life convictions: as those are just the opposite of mine, I quite enjoyed chatting with you and learnt a lot from you. Lastly, I am grateful for your enthusiastic support throughout the last 12 days prior to the thesis submission: it is really true that we believed we could, so we did. I'm sure you picked where this quote comes from.\\
I thank Prof. Pietro Ballone for being such a reliable and experienced academic person I could always ask for feedback and advice across the years.\\
I thank the School manager Mrs. Bairbre Fox for being, from the very first day I ever came to Rep. of Ireland (and you were the very person I met at the School) to a couple of days prior to the thesis submission, an absolute anchor of help. You shared with me the know-how of the School with genuine and smart comments. Your advices, ranging from strategical moves to let me be introduced in the Dubliner physics community to information about living in Dublin, have been an invaluable help.\\

\noindent{On} a personal note, I want to thank the people who, either for the whole experience or for significant moments of it, made me laugh, filled me with joy, supported me and above all, shared their life experience with me. These beautiful people, who are mentioned according to the country where we met, are: Alice Bertoldo, Anna Capriati, Folkert Kuipers, Domingo Gallegos and Danna Oassis, Charlotte Beneke, Katerina Bandsalat, Luka Jibuti, Sara Previdi, Ilias Samathrakis, Jenny Lazenbnik, Anastasia Azareva, Zoheb Khan, Aarif Amod, Pallavi Singh, Maurizio Maffei, Verena Backhaus, Laura Zubi, Lauren Pompey, Federica Bucci, Samantha Calcagnile, Eleonora Sarti, Sara Morselli, Rebecca Travaglini, Floriana Amicone, Katerina Verteletsky, Susanna Bracco, Caitlin DeLatte and Gerwin van Dalum who shares with me several crucial aspects of my PhD experience and whose PhD thesis also inspired me in writing some sections of Chapter 2. \\
A special thanks goes to Chiara Latini, who is the older sister I always wanted, rather than because of blood but because of choice.\\

\noindent{Last} but not least, I am grateful to my parents Antonella Balboni and Enzo Minarelli who followed my journey as a student walking from Bologna to Berlin, from Berlin to Utrecht and coming now to an end in Dublin. Thanks for your endless support and encouragements, thanks for your sincere love and affection. Thanks for being there for me both in the harsh and in the joyful moments. Many things happened during these years. However, I have always been faithful and truly passionate to my principles, interests and goals. This is definitely something I have learnt from you. Obtaining the Doctor title because of this PhD degree is the highest result (and satisfaction) one can reach as a student. For this reason (and many more), I would like to dedicate this whole work to you.

\begin{flushright}
	Emma L. Minarelli\\
	September 5, 2022\\
	Cento, Ferrara - Italy
\end{flushright}

\newpage
\vspace*{3cm}

\begin{flushright}
To my parents Antonella and Enzo, \\
who valued education above anything else.
\end{flushright}
\vfill 

\listoffigures
\listoftables

\newpage
\chapter*{List of Symbols}
\begin{tabular} {cp{0.8\textwidth}}
$\mathcal{A}(\omega),\mathcal{A}_{0}(\omega)$ & spectral function of interacting, noninteracting system\\
$\epsilon_{F}$ & Fermi level \\
$f^{Eq}$ & thermal equilibrium Fermi-Dirac distribution \\
$f^{\mu_{\alpha}}$ & nonequilibrium Fermi-Dirac distribution at chemical potential $\mu_{\alpha}$ with respect to $\alpha$-lead \\
$G^{x}_{0}(\omega, t, T,V_{bias})$ & free-particle propagator or bare Green's function for $x=R,A,<,>$ retarded, advanced, lesser, greater respectively\\
$G^{x}(\omega, t, T,V_{bias})$ & particle propagator or fully interacting Green's function for $x=R,A,<,>$ retarded, advanced, lesser, greater respectively\\
$\mathcal{G}^{C}(\omega,T, V_{bias})$ &  electrical conductance due to $\Delta V_{bias}$\\
$\mathcal{G}^{E}(\omega,T, V_{bias})$ & heat conductance due to $\Delta T/T$ \\
$\mathcal{G}_{0} =e^2 /h$ & quantised electrical conductance \\
$\hslash \equiv 1$ & Planck constant \\
$\mathbb{1}$ & identity operator or matrix, according to the context \\
$\hat{I}^{Q}(t)$ & current operator for given source operator $\hat{Q}$ \\
$k_{B} \equiv 1$ & Boltzmann constant\\
$K_{Q,Q^{\prime}}(t,t^{\prime})$ & current-current correlator  \\
$\chi(t,t^{\prime})$ & dynamical susceptibility or linear response function \\
$\rho(\omega),\rho_{0}(\omega)$ & local density of states of interacting, noninteracting system \\
$\left\lbrace \cdot ~,~ \cdot\right\rbrace ~~,~~ \left[ \cdot~,~\cdot\right]$ & fermionic anti-commutator and bosonic commutator\\
\end{tabular}
	
\end{singlespace}


\mainmatter
\chapter{Introduction}
\subsection*{Condensed matter theory and emergent phenomena}
The standard model of particle physics offers to the best of our knowledge a perfect description of the electron. It tells us that the electron is a point-like, Fermionic particle that carries a mass and couples to gravitational, weak and electromagnetic fields. This description, albeit sufficient to understand a single specimen, is not telling anything about the rich and vast physics that can be observed when one considers not just one electron, but many \cite{anderson1972more}.\\
At low enough temperatures, atoms can bond together in structures of sometimes higher and sometimes lower regularity and form what we call \textit{condensed matter}. A consequence of the bonding of atoms in condensed matter is the creation of a region that is densely populated by electrons. The electrons in such a system cannot be treated separately because they strongly interact via the Coulomb force and hence condensed matter systems exhibit phenomena that emerge from the interplay of many electrons. These phenomena can all be traced back to the basic properties of a single electron, however they cannot be predicted by considering only one isolated electron. In this regard, the most important properties of the single electron are its Fermionic nature and electromagnetic interaction. The Pauli-exclusion principle states that two or more Fermions can never occupy the same quantum state and thus leads to a form of interaction. This elementary interaction is what causes the emergence of the Fermi surface in solids, which is a key concept to understand and distinguish between conductors and insulators. Interestingly the Fermi surface is in many cases still a valid description, even when considering interactions through the formation of effective, Fermionic quasiparticles dressed by interactions \cite{baym2008landau,Altland}. The rich range of physical properties and materials originates in the inherent complexity of quantum many-body systems, where new physics can emerge, especially at low temperatures.

\subsection*{Strongly interacting models: the Kondo effect}
There are cases when the noninteracting quasiparticle description is no longer sufficient, and interactions must be considered. Arguably, the most basic interacting system is the Anderson impurity model, that lets a strongly localized and interacting orbital hybridize with a noninteracting electronic bath \cite{Anderson1961}. Generalized impurity models are composed of a genuinely interacting central region - the so-called ``impurity`` - characterized by few degrees of freedom, coupled to a noninteracting bath which has a macroscopically large number of degrees of freedom. Regardless of this apparent simplicity, perturbative approaches to this problem do not converge and a deep understanding comes only from non-perturbative methods such as the Renormalization Group theory \cite{Wilson1975}. The quest for a solution to the Anderson model has led, among many developments, to the discovery of the Kondo physics – which describes first local moment formation and then at lower temperatures the formation of an entangled many-body singlet of all itinerant electrons of the free bath with extremely localized electrons in the impurity orbital. This is a true many-body phenomenon, that emerges from electronic properties in different atomic orbitals \cite{Krishna1980renormalization}. However, in quantum mechanics we often encounter the term ``curse of dimensionality`` which refers to the exponential growth of complexity in problems with respect to the number of particles in the system. Since many-body physics essentially considers an infinite number of particles, it is immediately clear that to solve problems like the Kondo problem a lot of ingenuity is required to circumvent this infinite complexity of many-body physics.

\subsection*{Nanostructures and nanoelectronics}
However, this field does not only offer challenges, but also opportunities. The rise of semiconductor-based electronics and their ongoing miniaturization has led to the development of unprecedented control over quantum matter and materials in the nanometer scale. The semiconductor industry perceives the ever more influential quantum effects in ever smaller electronics as an annoyance that needs correction. But many physicists have noticed the potential to harness these phenomena in devices like the semiconductor quantum dot \cite{cronenwett1998tunable} or single-molecule electronics. Such nanostructures have unique electronic properties, like a discrete excitation spectrum and are interesting in their own right as \textit{artificial atoms} \cite{Kastner_SET1992}.  However, nanostructures like quantum dots or molecules can be incorporated in an electronic circuit, by attaching them to leads. In the lead-coupled configuration, the electrons tunnelling from one lead to the others, require transition into a state of the nanostructure. Hence, through fine tuning of plunger voltages applied to the connected lead-nanostructure system, tunnelling can either be enhanced or suppressed through the nanostructure. This sort of control allows to manipulate even more intricate many-body effects, like the aforementioned Kondo effect in interacting nanostructures. One can harness the power of these effects to build electronic devices of unprecedented efficiency and size \cite{song2011single}. In order to realize such devices a deep understanding of the underlying mechanisms beyond the equilibrium electronic state and quantum transport properties are required.

\subsection*{Quantum transport}
In order to reach high control and precise tuning in any electronic device, it is essential to obtain an understanding of how a finite electronic gradient is maintained over time (potential difference of voltage bias). A finite electronic flux (current) develops between two leads with different chemical potentials. In such a situation, where one lead can be identified as the \textit{source} and the other lead as the \textit{drain}, we obtain an unbalanced electronic distribution in the coupled system. In this configuration, electrons have to traverse the nanostructure in order for a current to flow.  In this scenario, systems can exhibit many-body effects like the Kondo effect and thus, it is crucial to take strong correlations into account when attempting to predict the electrical conductance in nanoelectronic devices. The study of these systems and their properties go under the umbrella term of \textit{quantum transport} \cite{Jauho_QuantumKineticsTransport,Nazarov}. This can also be generalized to thermoelectronic transport arising when both a voltage bias and a temperature gradient exists between leads.

\subsection*{Theory versus experiments}
Advanced fabrication techniques have allowed to measure conductance properties directly in nanodevices exhibiting exotic phases. It is one of the aims of the theory  to predict, or to explain where possible, these observations to gain a proper understanding of the mechanisms behind quantum transport. Only through such knowledge is it possible to systematically improve or design devices for applications in future electronics. Ultimately, the test of the theory is comparison with the experimental quantum transport data.

\subsection*{Standard mesoscopic transport techniques}
Despite the continuous development and extensions of quantum transport methodologies, we find in the literature several restrictions to their application to general nanostructures. In many cases, this limits calculations or crude approximations are required. Here we list the common constraints required in standard transport formulae: $(i)$ noninteracting model, $(ii)$ single-orbital nanostructure, $(iii)$ set-up configuration satisfying proportionate coupling condition, $(iv)$ either \textit{ac} or \textit{dc} conductance regime and $(v)$ zero temperature i.e. Fermi liquid system. These conditions irremediably  limit the investigation in arbitrary nanostructures and challenge the engineering of devices suitable for experimental measurement of observables. It is a goal of this thesis to overcome some of these restrictions.

\subsection*{A step forward}
This thesis aims to go beyond these standard requirements such that: we can overcome physical limitation in composition and design of set-ups, we present formulations involving quantities whose experimental observation is feasible and we increase the accuracy of numerical results. \\
In this work: $(a)$ we analytically derive alternative transport formulations that are less affected by the faults described in the literature and so are more useful in practical implementation; $(b)$ we analytically derive improved transport formulations whose applicability is in regimes and/or conditions usually out of reach by known results and so exceed the standard methods validity. We corroborate the analytical calculations and predictions by numerical results using a state-of-the-art numerical method, the numerical renormalization group technique. \\
With the compelled combination of analytical and numerical verifications of our new transport formulations, we apply those to $(i)$ interacting models, $(ii)$ multi-orbital nanostructures, $(iii)$ genuinely non-proportionate coupling set-ups, $(iv)$ both \textit{ac} and \textit{dc} conductance regimes and $(v)$ finite temperature scales.

\subsection*{Outline of the thesis}
This thesis comprises Chapters $2,3$ as theoretical background followed by Chapters $4,5$ on developing new methodologies and theory, and finally Chapters $6,7,8$ on applications.\\
Chapter $2$, \textit{Fundamentals}, provides the essentials of the field theory formalism, the models and the physics subject of this work. These are respectively, Green's functions as propagators in equilibrium and nonequilibrium theory, quantum impurity models with the primary example of the Anderson and Kondo models, the key concept underlying Kondo physics that is the perturbative renormalization method, and the complete description of the Kondo effect via numerical renormalization group technique (NRG). We conclude with generalities on the set-up employed in our applications - the generalized two-channel Kondo model.\\
Chapter $3$, \textit{Theory of mesoscopic quantum transport}, contains both theory and analytical derivation of the most important transport techniques used in the literature. Following the historical development in the field, we start with the transport in noninteracting models at finite temperature using scattering theory in the Landauer-B{\"u}ttiker formula, the transport under linear response theory in the Kubo formula, the transport for interacting model under nonequilibrium condition at finite temperature in the Meir-Wingreen formula and the transport for Fermi liquid models in the Oguri formula. We conclude with a discussion on artificial atoms where the Kondo effect can be implemented, including also the electrical conductance properties as function of applied voltages, temperature scale and a brief comment on the pioneering experimental verifications of the Kondo effect.  \\
The following five chapters are our new contributions to the field of mesoscopic transport.\\
Chapter $4$, \textit{Improved calculations for quantum transport in interacting multi-orbital nanostructures}, includes the analytical derivations of alternative and improved quantum transport formulations. \\
In Sec.\ref{sec:AlternativeLandauer}, we introduce the concept of structured leads and differentiate it for equivalent and inequivalent lead components. We derive for these generalised impurity models the analytical expressions for Kubo, Meir-Wingreen and Oguri conductance formulae with the aim of finding Landauer-type formulae. 
In Sec.\ref{sec:AlternativeMW}, we have a two fold approach to derive alternative formulation of the Meir-Wingreen current formula. In the first method, we attempt to calculate an expression for lesser impurity Green's function exhibiting an explicit voltage bias dependence. In the second method, we derive a corrected Fermi distribution out of equilibrium condition under perturbation expansion. Taking inspiration from the Ng ansatz, we use this new distribution to derive a nonequilibrium lesser impurity self-energy such that we have an expression for Meir-Wingreen formula exact up to the order of perturbation expansion used in the calculation. 
In Sec.\ref{sec:ImprovedKubo}, we derive an improved version of the Kubo conductance formula and substantiate our analytical findings with numerical results in NRG showing better numerical accuracy of our new formulation for electrical transport compared to the literature one. We mention also the breakdown of the NRG implementation for the heat Kubo formula - both standard and improved versions.
In Sec.\ref{sec:AlternativeOguri}, we introduce the concept of extended impurity model. For noninteracting system, we calculate an alternative formulation of Fisher-Lee conductance using $\mathrm{T}$-matrix. For interacting cases, we find an alternative Oguri formula within Fermi liquid regime characterised by a renormalized $\mathrm{T}$-matrix.\\
Chapter $5$, \textit{Effective models: emergent proportionate coupling}, presents quantum transport results for systems with microscopic set-ups lacking the proportionate coupling (PC) condition. We analyse two important cases. 
In Sec.\ref{sec:CBPC}, we consider a non-PC two-channel Kondo model in Coulomb blockade regime for both even and odd charge ground-states. We derive effective low energy models and study the RG-flow of the $\mathrm{T}$-matrix spectrum in its effective basis via NRG to find the channel decoupling temperature and the renormalized couplings. Using the alternative Oguri and improved Kubo formulae, we compute electrical conductance by means of the renormalized parameters.  Except for the case of the even electron ground-state with integer spin $S=0$ state where the effective model shows already the PC property, systems with even electron ground-states and higher integer spins or systems with odd electron ground-states with half-integer spins require also the RG analysis in order to find the decoupling condition. In the odd charge case, we study spin $S=1/2$ ground-state in $(i)$ absence of potential scattering, $(ii)$ \textit{sd}-symmetry and $(iii)$ broken \textit{sd}-symmetry.
In Sec.\ref{sec:MVPC}, we consider a non-PC two-channel Kondo model in mixed-valence regime. We derive effective low energy models and analyse the $\mathrm{T}$-matrix spectrum in its effective single-channel form via NRG. We use the Meir-Wingreen formula in PC version under linear response at zero temperature to evaluate electrical conductance by means of the effective parameters. We study three different cases, namely spinless system, spinful system at singlet-doublet and at doublet-triplet mixed-valence points. Their corresponding effective models are mapped to a noninteracting resonant level model, an infinitely strongly interacting Anderson impurity model and a two Fermionic orbital system, respectively. These are the minimal effective models in terms of which to understand quantum transport properties.\\
Chapter $6$, \textit{Application and comparison of quantum transport techniques}, gives a quantitative demonstration of the alternative and the improved quantum transport formulations from Chapter $4,5$ using NRG. We carry out the numerical benchmark using the triple quantum dot system in both proportional and non-proportional coupling because of the rich physics in the model and its versatility in different regimes and configuration, see Secs.\ref{sec:TQDpc},\ref{sec:TQDnopc}. Hence, the triple quantum dot allows us to test very challenging transport scenarios with the aid of a unique model and to render the benchmark comparable and at the same time expressive.\\
Chapter $7$, \textit{Transport in molecular junction: benzene transistor}, presents an extensive investigation of the transport properties of the benzene single-molecule coupled to two equivalent metallic leads. We study \textit{meta} and \textit{para} coupling configurations (both in non-PC): the different internal symmetries lead to vastly different phenomenology in the model. We find effective models at mixed-valence singlet-doublet and doublet-triplet crossovers that exhibit emergent proportionate coupling condition contrary to the bare model. Further, we carry out exact calculation in NRG to compute the molecular entropy and the electrical conductance using improved Kubo formula to characterise the low energy physics of the benzene junction.\\
Chapter $8$, \textit{Two-channel charge-Kondo physics in graphene quantum dots}, takes inspiration from the charge-Kondo paradigm to design a new device that exhibits a Kondo effect while employing non-metallic leads. In particular, we choose both impurity and lead composed of graphene, which shows a pseudogap density of states at the Fermi level. In this setting, we do not expect any Kondo effect unless particle-hole symmetry is broken. We use NRG in the full density matrix formulation to compute dynamical quantities such as $\mathrm{T}$-matrix, with electrical conductance calculated via improved Kubo formula and static quantities such as the impurity entropy and the magnetic susceptibility to investigate the various non-trivial regimes of the $3D$ version of the phase diagram as a function of system parameters and channel asymmetry.\\
Chapter $9$, \textit{Discussion and conclusion}, brings to an end this thesis. We summarise the most important analytical expressions and numerical results we derive in this work. We conclude with comments on a broad context of applicability of our findings and suggest interesting future directions.


\chapter{Fundamentals}
This chapter contains the groundwork for the thesis:  we present essential mathematical tools and concepts which we will extensively use and develop in the later chapters.\\
The outline of the chapter starts with section \ref{sec:Gtheo} to introduce the many-body Green's functions as a technique to study important system properties both in equilibrium and nonequilibrium conditions. In section \ref{sec:QImpModel} we introduce the general models of interest in this work: impurity systems, exemplified by the Anderson and Kondo models. This section offers also the opportunity to explain several important analytical techniques, their limiting cases and significant physical concepts we will refer throughout the thesis. Then, in section \ref{sec:RGtheo} we present a key concept behind impurity models: the renormalization of energy scales. We also present its numerical version, the numerical renormalization group technique: all the numerical results in this thesis are calculated using it. In the last section \ref{sec:2CK} we illustrate a special case of the Kondo model, that is its two-channel version whose rich renormalization flow of couplings leads to highly nontrivial physics.

\section{The Green's functions}\label{sec:Gtheo}
The Green's function is an important technique to study systems both in the classical and quantum theories. This thesis utilises field theory methods in operator formalism through Green's functions and this initial section sets the generalities of the technique.\\
In classical field theory, the Green's function is a powerful method for solving inhomogeneous differential equations. In the context of quantum field theory, the Green's function represents the \textit{propagator} that is the physical quantity giving access to the most important many-body system properties \cite{Coleman,Mahan}. Considering the time domain, the Green's functions are defined either in imaginary-time - where its analyticity and poles location is studied - or in real-time. Within the purposes of this thesis, we work only with the latter and we classify the functions according their contour ordering: the time-ordered Green's function for equilibrium states and the contour-ordered Green's function for out-of-equilibrium states \cite{Economou}.\\
In this thesis we adopt the \textit{double-time temperature-dependent} Green's function definition as introduced by Zubarev \cite{Zubarev1960}:
\begin{equation}\label{eq:defG}
G(t,t^{\prime}) = \langle\langle  \Psi_{H}(t); \Psi_{H}^{\dagger}(t^{\prime}) \rangle\rangle ~,
\end{equation}
where the notation $\langle\langle ;  \rangle\rangle$ indicates the average is taken over grand-canonical ensemble. The definition in Eq.\ref{eq:defG} is given in \textit{Heisenberg picture}: time dependent operators $ \Psi(t)$ evolve according to $\Psi(t) = e^{iHt}\Psi(t=0)e^{-iHt}$ and time-independent states $ \psi_{0}$ evolve as $ \psi_{0} = e^{iHt} \psi(t)$, having no time dependence in the full $\hat{H}$. We define also the total time derivative of the operator $ \Psi(t)$ as $\dot{\Psi}(t) = i [H,\Psi(t)],~\hbar=1$ where we have taken the original operator in Schr{\"o}dinger picture being time independent so $(\partial_{t}\Psi_{S})(t)=0$ and the time derivative given in the notation as $\dot{\Psi}(t)\doteq\partial_{t}\Psi(t)$ is computed by the commutator with $\hat{H}$ only.\\
In this work we will always refer to Fermions, such that the operators satisfy the anti-commutation relations at the \textit{same} time $t$, namely:
\begin{equation}
	\begin{aligned}
		& \lbrace \Psi_{a}(t), \Psi_{b}(t)\rbrace = \lbrace \Psi^{\dagger}_{a}(t), \Psi_{b}^{\dagger}(t) \rbrace =0 ~,\\
		&\lbrace \Psi_{a}(t), \Psi_{b}^{\dagger}(t) \rbrace = \delta_{a,b} ~.
	\end{aligned}
\end{equation}
In the next sections we illustrate first the Green's function formalism in equilibrium and nonequilibrium theory, then the relation among different Green's functions by means of the fluctuation-dissipation theorem in equilibrium.

\subsection{Time-Ordered Green's function: equilibrium theory}
In \textbf{equilibrium theory} Green's functions are defined on a single real-time axis, the so-called \textit{time-ordered contour} $\mathcal{C}_{t}$  \cite{Jauho_QuantumKineticsTransport}, that is running from $t\rightarrow -\infty$ to $t\rightarrow +\infty$.\\
On $\mathcal{C}_{t}$, we introduce the \textbf{casual or time-ordered equilibrium} Green's function $G^{c}(t,t^{\prime})$
\begin{equation}\label{eq:defGcasual}
	\begin{aligned}
		G^{c}(t,t^{\prime}) &= -i \langle  T_{\mathcal{C}_{t}} \big[ \Psi(t) \Psi^{\dagger}(t^{\prime}) \big] \rangle \\
		&= -i \theta(t-t^{\prime}) \langle \Psi(t) \Psi^{\dagger}(t^{\prime})\rangle +i \theta(t^{\prime}-t) \langle \Psi^{\dagger}(t^{\prime}) \Psi(t)\rangle ~,
	\end{aligned} 
\end{equation}
where $T_{C_{t}} [\dots]$ is the time-ordering operator for Fermions \cite{Lancaster},
\begin{equation} 
	T_{\mathcal{C}_{t}} \big[ \Psi(t) \Psi^{\dagger}(t^{\prime}) \big] = 
	\begin{cases}
		\Psi(t) \Psi^{\dagger}(t^{\prime}) & \text{if}\ t>t^{\prime} \\
		-\Psi^{\dagger}(t^{\prime})\Psi(t)  & \text{if}\ t<t^{\prime}.
	\end{cases}
\end{equation}
For completeness, we introduce also the \textbf{anticasual or antitime-ordered equilibrium} Green's function $G^{\widetilde{c}}(t,t^{\prime})$
\begin{equation}
	\begin{aligned}
		G^{\widetilde{c}}(t,t^{\prime}) &= -i \langle  T_{\widetilde{\mathcal{C}}_{t}} \big[ \Psi(t) \Psi^{\dagger}(t^{\prime}) \big] \rangle \\
		&= -i \theta(t^{\prime}-t) \langle \Psi(t) \Psi^{\dagger}(t^{\prime})\rangle +i \theta(t-t^{\prime}) \langle \Psi^{\dagger}(t^{\prime}) \Psi(t)\rangle ~,
	\end{aligned} 
\end{equation}
where $T_{\widetilde{\mathcal{C}}_{t}} [\dots]$ is the antitime-ordering operator for Fermions.\\
Then, on $\mathcal{C}_{t}$ contour we introduce four other types of equilibrium Green's functions:
\begin{equation}\label{eq:defGeq}
	\begin{aligned}
		&G^{R}(t,t^{\prime}) = -i \theta(t-t^{\prime}) \langle \lbrace \Psi(t), \Psi^{\dagger}(t^{\prime}) \rbrace \rangle ~,\\
		&G^{A}(t,t^{\prime}) = i \theta(t^{\prime}-t) \langle \lbrace \Psi(t), \Psi^{\dagger}(t^{\prime}) \rbrace \rangle  ~,\\
		& G^{<}(t,t^{\prime}) = i \langle\Psi^{\dagger}(t^{\prime})\Psi(t) \rangle ~,\\
		&G^{>}(t,t^{\prime}) = -i \langle\Psi(t)\Psi^{\dagger}(t^{\prime})\rangle ~.
	\end{aligned}
\end{equation}
The reason for such variety of function as listed in Eq.\ref{eq:defGeq} it is understood in terms of their different physical interpretation. The \textbf{retarded} $G^{R}$ is non-vanishing events happening at time interval $t\geq t^{\prime}$ whereas the \textbf{advanced} $G^{A}$ indicates non-vanishing ones in the opposite interval $t\leq t^{\prime}$. The functions $G^{R/A}$ are used in Linear Response theory, they determine the spectral properties and density of states of the system - as we will see in the following. The \textbf{lesser} $G^{<}$ definition is related to the single-particle density, thus it is referred as particle propagator. On the contrary, the \textbf{greater} $G^{>}$ shows inverted operators order, hence it is related to the hole density and it is called hole propagator.  The functions $G^{> / <}$ are used to calculate density and currents - as we will see later.\\
By means of this physical interpretation, we can rewrite the retarded and advanced Green's functions in Eq.\ref{eq:defGeq} as
\begin{equation}
	\begin{aligned}
		& G^{R}(t,t^{\prime}) = \theta(t-t^{\prime}) \big( G^{>}(t,t^{\prime}) -  G^{<}(t,t^{\prime}) \big) ~,\\
		&G^{A}(t,t^{\prime}) = \theta(t^{\prime}-t) \big( G^{<}(t,t^{\prime}) -  G^{>}(t,t^{\prime}) \big) ~.
	\end{aligned}
\end{equation}
From the equilibrium Green's functions defined on real-time $\mathcal{C}_{t}$ contour in Eq.\ref{eq:defGeq}, we can apply two type of transformations. The first one converts the propagators in real frequency $\omega$ using the \textbf{Fourier transform}, that is:
\begin{equation}
G^(\omega)= \int_{-\infty}^{+\infty} dt e^{i\omega t} G(t) ~.
\end{equation}
The second one is used to transform the propagator in complex domain $z=\omega \pm i \eta$ for $\eta \rightarrow 0^{+}$. The sign $+(-)$ indicates the integral converges only the upper (lower) half-imaginary plane. This is the \textbf{Laplace transform}, namely:
\begin{equation}
	G(z)= \int_{0}^{\infty} dt e^{i\omega z} G(t) ~.
\end{equation}
We complete the section by introducing the \textbf{Dyson equation}. We consider the interacting self-energy $\Sigma(\omega,T)$ as function of energy $\omega$, temperature $T$. In its general term, the self-energy consists of the sum
\begin{equation}\label{eq:defSelEngeneral}
\Sigma(\omega,T) = \Sigma^{U}(\omega,T) +  \Delta(\omega,T) ~,
\end{equation}
that is the system interacting part $\Sigma^{U}$ - here it is modelled as Hubbard-type interactions - plus the so-called hybridization $\Delta$ or noninteracting part.\\
The general Dyson equation structure reads
\begin{equation}
	G(\omega,T) = G_{0}(\omega,T) + G_{0}(\omega,T)\Sigma(\omega,T) G(\omega,T) ~,
\end{equation}
where $G(\omega,T)$ and $G_{0}(\omega,T)$ indicate the full interacting and bare Green's functions, respectively.\\
We can write a specific version of Dyson equation according to the Green's functions listed in Eq.\ref{eq:defGeq} i.e. for $G^{R/A}$ it reads
\begin{equation}\label{eq:defDysonRA}
	\begin{aligned}
		G^{R/A}(\omega,T) &= G_{0}^{R/A}(\omega,T) + G_{0}^{R/A}(\omega,T)\Sigma^{R/A}(\omega,T)G^{R/A}(\omega,T)\\
		&= \big[ [G_{0}^{R/A}(\omega,T)]^{-1} - \Sigma^{R/A}(\omega,T) \big]^{-1}
	\end{aligned}
\end{equation}
and for $G^{< / >}$ we have the Keldysh-Dyson equation
\begin{equation}\label{eq:defDysonKeldysh}
	G^{</>}(\omega,T) = G^{R}(\omega,T) \Sigma^{</>}(\omega,T) G^{A}(\omega,T) ~.
\end{equation}
The noninteracting counterparts of these equations can be found in Eq.\ref{A:eq:g_Def}.\\
For general interacting many-body systems, an exact analytical solution of $\Sigma$ might not exist and approximate or alternative solutions are required.

\subsection{Contour-Ordered Green's function: nonequilibrium theory}
\begin{figure}[H]
\centering
\includegraphics[width=0.75\linewidth]{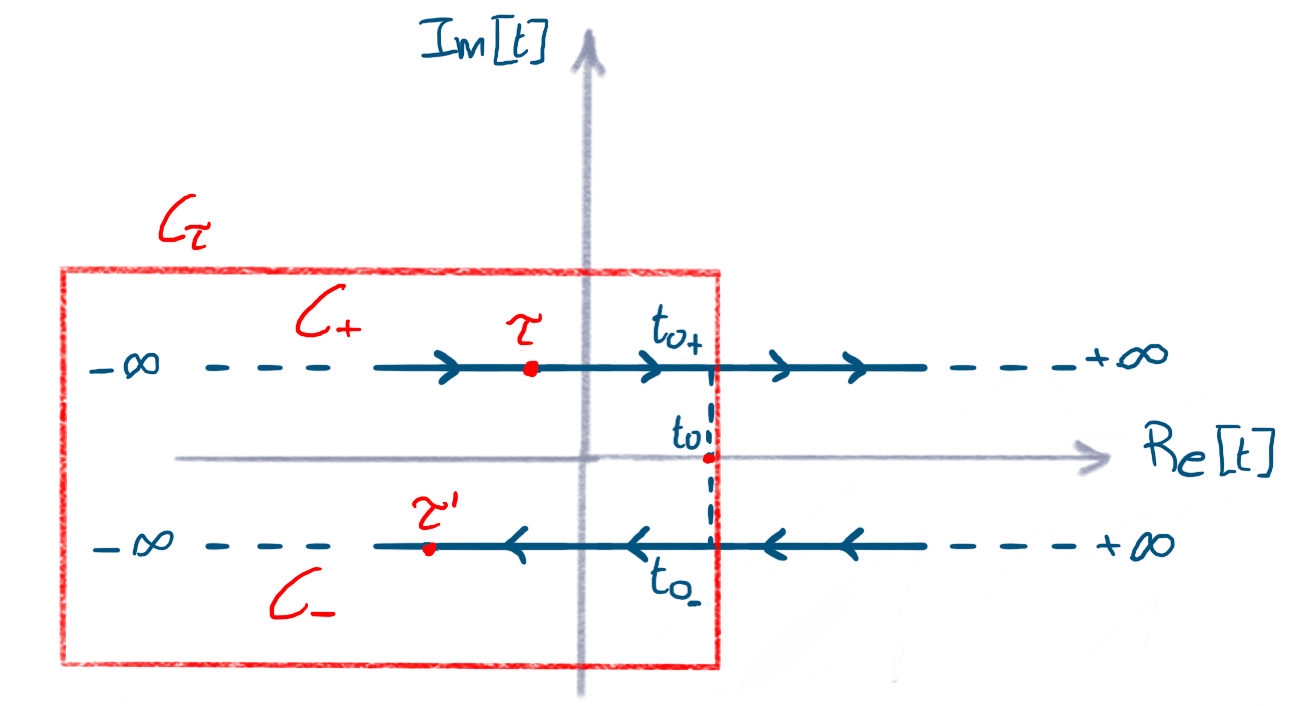}
\caption[The Schwinger-Keldysh contour $\mathcal{C}_{\tau}$]{The Schwinger-Keldysh contour $\mathcal{C}_{\tau}$ resulting from the union of the two counter propagating real-time axes $\mathcal{C}_{+} = (-\infty, t_{0+}]$ and $\mathcal{C}_{-} = [t_{0-},-\infty)$ for $t_{0\pm}=t_{0}\pm i\eta$. On the contour $\mathcal{C}_{\tau}$  the nonequilibrium Green's function in Eq.\ref{eq:defGneq} are defined.}\label{F2:KeldyshCont}
\end{figure}

In general terms, quantum transport is a proper out-of-equilibrium process because of applied bias voltage or temperature gradient. In this thesis we often discuss models in regimes away form equilibrium condition and so in this section we intend to present the necessary formalism for the out-of-equilibrium condition. The technical mathematical implication on nonequilibrium theory are beyond our scope in this introduction and we refer to further details in advanced literature as in the \cite{Jauho_QuantumKineticsTransport,Ryndyk,Spicka_NGF,Maciejko_NonEquilibrium}.\\ 
In equilibrium many-body systems, the \textit{adiabatic theorem} holds in the sense that interactions are switched on/off adiabatically i.e. infinitesimally slowly such that the occupation numbers in the system are not changed. This ensures both initial and final quantum states are connected and differ at most only  by a phase factor. On the contrary, in \textbf{nonequilibrium theory} for general time-dependent Hamiltonian, the adiabatic theorem doesn't apply: initial states at $t \rightarrow -\infty$ and far in future states at $t \rightarrow +\infty$ are not related any more. Irreversible symmetry breaking effects happen between states at $t \rightarrow -\infty$ and $t \rightarrow +\infty$.  \\
We conclude the fundamental difference between equilibrium and nonequilibrium approach is that in the latter case far in the future time $t \rightarrow +\infty$, zero-temperature systems will not return to their \textit{initial} ground state and finite temperature systems will not return to their \textit{initial} thermal equilibrium distribution. In order to establish a consistent nonequilibrium formalism, we need to modify the real time-ordered contour such that every observable is described solely by the asymptotic quantum states in the remote past and no reference to future states is required.\\ 
Prior to any path deformation allowing to solve nonequilibrium transport problems, as a standard literature assumption  \cite{Jauho_QuantumKineticsTransport}, we assume the system is in a noninteracting state in the infinite past. This assumption is also know as \textit{finite-size nonequilibrium system} \cite{Maciejko_NonEquilibrium}. It indicates that the continuous arguments in the Green's functions are replaced by their discretized version and the continuous energy spectrum by discrete levels. The Green's functions become finite-sized matrices and several simplifications in the Keldysh equations are, as a consequence, possible. By means of the finite-size system assumption, there is \textit{no} initial correlation calculated at $t \rightarrow -\infty$. Furthermore, in this thesis we are interested only in \textit{steady-state phenomena} happening at the completion of all the transient states. With this respect, correlations decay in time such that at any positively large interval of time there is no signature left of any initial correlations calculated at distant past.\\
Having in mind these considerations, we continue with the contour deformation that eliminates the asymptotic $t\rightarrow + \infty$ curve. It is achieved by evolving the system from $t \rightarrow -\infty$ to an instant of interest $t=t_{0}$ - when the steady state is reached - and from it by evolving the system back to $t \rightarrow -\infty$. In presence of interactions, the contour consists also of a third branch that extends into the imaginary axis $(t,t-i\beta)$. In view of the finite-size system assumption for steady-state processes, disregarding initial correlations is equivalent to neglect the imaginary strip such that we are left only with the two real axes \cite{Maciejko_NonEquilibrium,Stefanucci2013}. These two counter propagating real-time axes, namely $\mathcal{C}_{+} = (-\infty, t_{0+}]$ and $\mathcal{C}_{-} = [t_{0-},-\infty)$, are then combined by joining them at instant $t_{0}$ that is the real-time projection of $t_{0\pm}=t_{0}\pm i\eta$, see sketch in Fig.\ref{F2:KeldyshCont}. The result is a semi-closed contour: that identifies the \textit{Schwinger-Keldysh contour} $\mathcal{C}_{\tau} =\mathcal{C}_{+}\bigcup\mathcal{C}_{-}$, where contour-ordered operators are defined \cite{Keldysh1965, Stefanucci2013}. As indicated by the different time notation, the composition of branches in the Keldysh contour results in complex-time variables.\\
In the context of transport problems, the absence of initial correlations indicates that in the remote past the nanostructure is fully \textit{decoupled} from the reservoirs. Each region is found in its noninteracting, thermal equilibrium state with respective chemical potentials. It is common practice to describe the system modelling by eigenbasis and matrix elements defined in this disconnected, noninteracting model.\\
\noindent{On} $\mathcal{C}_{\tau}$ we introduce the \textbf{contour-ordered nonequilibrium} Green's functions $G(\tau, \tau^{\prime})$: 
\begin{equation}
	G(\tau, \tau^{\prime}) = -i \langle  T_{\mathcal{C}_{\tau}} \big[ \Psi(\tau) \Psi^{\dagger}(\tau^{\prime}) \big] \rangle ~,
\end{equation}
where $T_{\mathcal{C}_{\tau}}[\dots]$ is the contour-ordering operator for Fermions. \\
In order to define Green's functions on $\mathcal{C}_{\tau}$, we need to find the correspondence between both real-time ordered and contour-ordered functions as defined on $\mathcal{C}_{+},\mathcal{C}_{-}$ and on $\mathcal{C}_{\tau}$, respectively. According to which branch the contour arguments $\tau,\tau^{\prime}$ belong to, we can identify different real-time Green's functions. The correspondence between $G(\tau, \tau^{\prime})$ and the type of $G(t,t^{\prime})$ is obtained by projecting the complex-time arguments from the Keldysh-contour onto the real-time axis \cite{LangrethTheorem,LangrethLinearRespNEq1976}. This correlation is indeed guaranteed by the \textit{equal formal structure} of the two theories at zero temperature \cite{JauhoWingreeMeir1994} - as we discuss below. In conclusion, on the Schwinger-Keldysh contour, Green's functions have both time and branch dependence. According to the complex-time arguments location on either one or both counter propagating branches $\mathcal{C}_{+},\mathcal{C}_{-}$, we obtain four different expressions corresponding to the real-time Green's functions:
\begin{equation}\label{eq:defGneq}
	G(\tau,\tau^{\prime})=
	\begin{cases}
		G^{C}(t,t^{\prime}) =
		-i \langle T_{C_{t}} \big[ \Psi(t) \Psi^{\dagger}(t^{\prime}) \big] \rangle ~, \quad \tau,\tau^{\prime} \in \mathcal{C}_{+}\\
		G^{<}(t,t^{\prime})=i \langle \Psi^{\dagger}(t^{\prime})   \Psi(t) \rangle ~, \quad \tau \in \mathcal{C}_{+},\tau^{\prime} \in \mathcal{C}_{-}\\
		G^{>}(t,t^{\prime})= 
		-i \langle  \Psi^{\dagger}(t)   \Psi^{\dagger}(t^{\prime}) \rangle ~, \quad \tau \in \mathcal{C}_{-},\tau^{\prime} \in \mathcal{C}_{+}\\
		G^{\widetilde{C}}(t,t^{\prime})= 
		-i \langle \widetilde{T}_{t}  \big[ \Psi(t) \Psi^{\dagger}(t^{\prime})\big] \rangle ~, \quad \tau,\tau^{\prime} \in \mathcal{C}_{-} ~.
	\end{cases}
\end{equation}
In Eq.\ref{eq:defGneq}, only three out of these four expressions are linearly independent due to two \textit{Keldysh identities}:
\begin{equation}\label{eq:identityKeldysh}
	\begin{aligned}
		& \text{I identity:} \quad G^{C}+G^{\widetilde{C}}=G^{<} +G^{>} ~, \\
		& \Rightarrow G^{R} = G^{C} - G^{<} \quad , \quad G^{A} = G^{C} - G^{>} \\
		& \text{II identity:} \quad G^{R} - G^{A} = G^{>} - G^{<} \equiv +2i \mathit{Im} G^{R} ~,
	\end{aligned}
\end{equation}
where the second equality in $\text{II}$ identity is calculated using Eq.\ref{eq:defSpectral}. We remark that the identities in Eq.\ref{eq:identityKeldysh} are valid for $(i)$ \textit{both interacting and noninteracting} systems, $(ii)$ in time and frequency domain. Furthermore, analogous identities hold for the self-energy - see for instance Eq.\ref{A:eq:delta_Def}.\\
We can turn now our discussion on the equivalence occurring between equilibrium and nonequilibrium theories. Formally, the perturbation expansion of nonequilibrium Green's function has the same structure of the equilibrium expansion at zero-temperature \cite{JauhoWingreeMeir1994}. This crucial aspect of nonequilibrium theory has important consequences in many-body physics, whose analysis goes beyond the scope of our discussion. For our purposes, this formal analogy allows us to borrow the familiar equilibrium field theory techniques at zero temperature (such as perturbation expansion, Wick's theorem, Dyson equation, equation of motion technique etc.) and to apply those to the contour-ordered Green's functions. In practice, the connection between real-time and complex-time variables Green's function is achieved by the \textit{Langreth theorem}, more details are presented on Appendix \ref{app:MW}.\\ 
We conclude this section by commenting that the contour-ordered Green's functions obey the same equilibrium Dyson equation structure if calculated at strictly zero temperature 
\begin{equation}
	G(\omega, T=0) = G_{0}(\omega, T=0) + G_{0}(\omega, T=0)\Sigma(\omega, T=0) G(\omega, T=0) ~.
\end{equation}
By means of this formal analogy at $T=0$, we can write the nonequilibrium Dyson equations for retarded, advanced propagators using Eq.\ref{eq:defDysonRA} and for lesser, greater propagators using Eq.\ref{eq:defDysonKeldysh}. \\ 
Having completed the presentation of equilibrium and nonequilibrium Green's function formalism, in the next section we discuss the correspondence between different Green's functions.

\subsection{Fluctuation-dissipation theorem}
In this section we show the intrinsic relation among the equilibrium Green's functions defined in Eq.\ref{eq:defGeq}. This discussion gives us the opportunity to introduce other important quantities we will extensively use throughout this thesis. The expressions are given in general form. We return to their version in two-channel models applied to quantum transport processes in Chapter 3.\\
We define the \textbf{spectral function} as summation of particle and hole energy distribution:
\begin{equation}
\mathcal{A}_{\sigma}(\omega) = \frac{i}{2 \pi} [G^{>}_{\sigma}(\omega)-G^{<}_{x\sigma}(\omega)] =\frac{i}{2 \pi} [G^{R}_{\sigma}(\omega)-G^{A}_{x\sigma}(\omega)] = -\frac{1}{\pi} \mathit{Im} [G^{R}_{\sigma}(\omega)]~,
\end{equation}
for $\sigma$ spin and we make use of the Keldysh identities in Eq.\ref{eq:identityKeldysh}.\\
We define the Fermionic distribution or \textbf{Fermi-Dirac distribution} in thermal equilibrium
\begin{equation}\label{eq:defFermieq}
f^{Eq}(\omega)= \dfrac{1}{1 + e^{\beta(\omega - \mu)}} ~,
\end{equation}
where $\beta=(k_{B}T)^{-1}$ for Boltzmann constant $k_{B}$ and chemical potential $\mu$ for equilibrium models. \\
Using these definitions, we introduce now the \textbf{fluctuation-dissipation theorem} \cite{KuboFDT1966, KuboStatisticalPhysics1985}. It states that taking a system almost at equilibrium condition, its \textit{response or dissipation} caused by a weak external perturbation (slowly switched-on) it is equivalent to its \textit{spontaneous fluctuations} which tend to move the system away from the equilibrium condition. This significant statistical relation has enormous physical implication. Within this work, we encounter this relation in the context of linear response theory, as we see in the next section. Here, we aim now to present the theorem in the context of equilibrium Green's function formalism. Its version holding for time-independent Fermionic system in equilibrium - either noninteracting or interacting:
\begin{equation}
\begin{aligned}
& G^{<}_{\sigma}(\omega)= i f^{Eq}(\omega) 2\pi\mathcal{A}_{\sigma}(\omega)  \\
&G^{>}_{\sigma}(\omega) = -i (1-f^{Eq}(\omega)) 2\pi \mathcal{A}_{\sigma}(\omega) ~.
\end{aligned}
\end{equation}
The physical insight behind these fluctuation-dissipation expressions - see for completeness Eqs. \ref{eq:defFDtheo} - is that the particle $G^{<}$ and hole $G^{>}$ fluctuations are related to the dissipation $\mathit{Im} [G^{R}]$ by a proportionality factor that is the equilibrium Fermion distribution \cite{Coleman, DiVentra}. We emphasize that these expressions are applicable only to thermal equilibrium models. Out-of-equilibrium phenomena - which are exactly the kind of events happening in quantum transport  - systems do not obey this relation any longer but require generalized nonequilibrium distribution functions \cite{ness2014nonequilibrium}. Then, approximate or alternative solutions are required, as we will see later in this work.\\
The concept of propagator defined by the Green's function is the theoretical cornerstone to study our systems and many of the results in this thesis will be based on it. In the next section, we discuss the general physical systems main subject of this work.

\section{The quantum impurity models}\label{sec:QImpModel}
In this section we introduce the general model we are interested in: that is the \textbf{quantum impurity system}, see the set-up sketch in Fig.\ref{F3:LandauerScheme}. This is a general class of models, defined by the localised interacting quantum system or \textit{impurity}, and the quantum baths or \textit{leads}, and their hybridization - to form a nontrivial strongly-correlated system. By construction, the impurity has a few degrees of freedom with arbitrary interactions and it couples to large noninteracting systems, the leads. The latter are macroscopic electrons reservoirs at thermal equilibrium. Among these elements, local impurity-leads coupling can be tuned accordingly. \\
We note that in general terms, the leads can be either Bosonic or Fermionic modes. In this thesis, we focus on the latter only: the electrons populating the leads form a conduction band whose distribution is regulated by the external application of a voltage bias $V_{bias}$ and hold at temperature $T$. \\
The impurity can designed into several forms. It can be a real physical impurity that dopes a host system, for instance Fe magnetic impurity in Au - which indeed initially motivated the study of impurity models \cite{Berg_1934ExpResistivity}. We comment again below on the historical root of impurity physics since it sheds light on crucial aspects for our discussions. The impurity can be the so called \textit{artificial atoms} \cite{Kastner_Artificialatoms1994} or \textit{semiconductor quantum dots} coupled to leads: single or multiple configuration of quantum dots, with reciprocal interactions governed by direct coupling or indirect exchange. \\
In this thesis we study generalised quantum impurity models and then go on to examine specific example systems comprising: a large quantum dot; a triple quantum dot configuration; and a benzene as single-electron molecule, all coupled to regular metallic leads. We also consider a set-up where the quantum dot and the leads are built from graphene components. These complex impurity models are characterised by a highly nontrivial set-up configuration, with many-body interactions occurring on the impurity.

\subsection*{Theory of magnetic local moments: a brief historical regression}
\begin{figure}[H]
\centering
\includegraphics[width=0.75\linewidth]{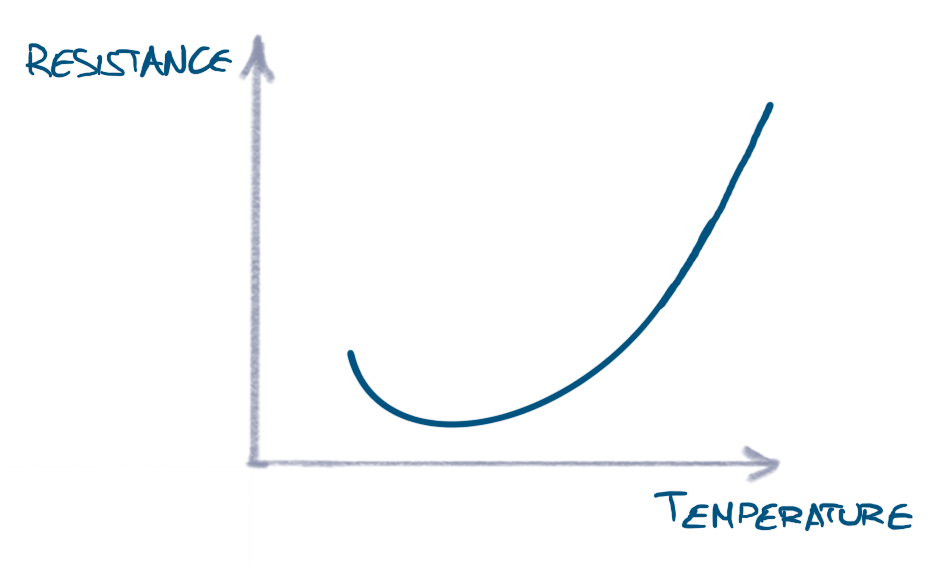}
\caption[Sketch of the experimental data for resistance versus temperature in Au]{Sketch of the resistance versus temperature for Au: the experimental data \cite{Berg_1934ExpResistivity} show a minimum at certain low temperature, attributed to magnetic impurities.}\label{F2:ResistanceOrig}
\end{figure}

Impurity models have originally developed in the context of the \textit{magnetic moment of impurities} (such as Fe, Mn) embedded in non-magnetic host metals (such as Au, Ag). Phenomenologically, the resistivity formula at low temperature $T$ is given by:
\begin{equation}
R(T) = R_{0} + aT^{2} + bT^{5}  ~,
\end{equation} 
where $R_{0}$ is the temperature independent contribution for scattering from static impurities, the quadratic term indicates the electron-electron scattering and the last term is due to the phonon scattering. The resistivity is expected to monotonically decrease as function of temperature. Furthermore, it is expected to reach a finite temperature-independent value as $T \rightarrow 0$ that is proportional to the number of defects, since no phonons or electronic scattering is left at zero temperature. However, the experimental observations showed an unexpected resistivity minimum \cite{Berg_1934ExpResistivity,Longinotti_1964Resistivity}, see sketch in Fig.\ref{F2:ResistanceOrig}. The measured profile was proportional to the impurity concentration $R_{0}$ and it occurred only in presence of magnetic impurities. Notably, on the contrary of non-magnetic impurities or potential scattering site, the magnetic impurities have internal dynamical degrees of freedom such that inelastic scattering events of conduction electrons can happen. \\
\noindent{These} bizarre observations were resolved in the pioneer work by Jun Kondo \cite{Kondo1964}. He considered the existence of a resistivity minimum as result of an effective single-impurity effect and so he discarded interactions due to correlation among localised spin of different impurities. His insight brought him to perturbatively calculate the resistivity due to scattering by magnetic impurities \cite{Hewson}. To leading order in the impurity-conduction electrons exchange coupling $J$, the first order correction to the resistivity is derived as
\begin{equation}
R(T) = R_{0} [1 - J\rho_{\alpha} \log (k_{B}T/D)] + \mathcal{O}(J)^{2} ~,
\end{equation}
where $D$ is the \textit{energy cut-off} typically the conduction electron \textit{bandwidth} and $\rho_{\alpha}$ is the density of states of the $\alpha$-conduction electrons at the Fermi level. Adding this correction to the phenomenological equation, the final expression correctly reproduces the experimental measurements of a minimum in resistivity at low temperature that is a proper logarithmic increase in resistivity.\\
Although Kondo resolved the unexpected resistance minimum observation, this perturbative result opened another issue. Being based on perturbation theory, his derived equation is supposed to remain small with decreasing temperature. However, it actually diverges as $T \to 0$ and the perturbation result is no longer applicable. It is worthwhile notice that the divergence at zero temperature was not experimentally measured. That implies there is an earlier threshold temperature at which the pertubative result breaks down. The temperature that identifies the regime below which the perturbative calculation by Kondo is not applicable any more is 
\begin{equation}\label{eq:Tk}
k_{B}T_{K} \backsim D e^{-1/J\rho_{\alpha}} ~,
\end{equation}
where $T_{K}$ is the \textbf{Kondo temperature}.\\
From the original measurement, the non-trivial low-energy physics of magnetic impurity models was already evident. The physical understanding of the system behaviour at $T\lesssim T_{K}$ came to be known as the \textbf{Kondo problem}. In particular, by advancing in the theory, the role of the Kondo temperature became predominant: it was understood that it contains all the microscopic details of the low-energy model. Any result expressed in terms of the rescaled temperature i.e. $T/T_{K}$ is \textit{universal} in the sense that the low-energy physics details are identical despite of the differences in the microscopic model.\\
In the next section, after introducing the concepts of energy scaling, we will discuss in quantitative terms the physical limits in the Kondo problem: for given Kondo temperature, we encounter $T \gg T_{K},~ T \sim  T_{K},~ T \ll T_{K}$ regimes. The perturbative result for the resistance is only a superficial understanding of the highly non-trivial many-body physics revealed in the Kondo problem.\\
The core material of the section is the detailed discussion on the two fundamental impurity models  whose importance goes beyond the actual impurity physics field: the Anderson model and the Kondo model. These sections not only allow us to present them but also to introduce several techniques, system properties and concepts we will extensively use in later chapters.

\subsection{The Anderson single-impurity model}
The first derived impurity paradigm to capture the magnetic local moments formation in metallic hosts and their mutual interactions is the Anderson model (AM) \cite{Anderson1961,Anderson_LocalisedMoments1978,Coleman2002LocalMoment}. The importance of the AM roots in its parameter defined over several energy scales and emergent physics once appropriate low energy scale is reached. Later, we will introduce the concept of energy scaling: using a progressive reduction of energy cut-off we will understand the AM as an umbrella covering different other impurity models - only occurring at smaller energies.\\
The Anderson model consists of an isolated interacting impurity whose localised \textit{d-} or \textit{f-} orbital energies are within the conduction band of the host metal, to which it hybridises and it causes a modification of the impurity wavefunction.\\
The \textbf{Anderson Hamiltonian model} reads: 
\begin{equation}\label{eq:AM}
\hat{H}^{AM} = \underbrace{\sum_{\mathbf{k}\sigma} \epsilon_{\mathbf{k}} \hat{n}_{\mathbf{k}\sigma}}_{\hat{H}_{bath}} + \underbrace{\sum_{\mathbf{k}\sigma} \Big [ V_{\mathbf{k}} c^{\dagger}_{\mathbf{k}\sigma}d_{\sigma} + V^{\star}_{\mathbf{k}}d^{\dagger}_{\sigma}c_{\mathbf{k}\sigma}\Big]}_{\hat{H}_{hyb}} + \underbrace{\epsilon_{d}\sum_{\sigma}\hat{n}_{d\sigma} + U\hat{n}_{d\uparrow}\hat{n}_{d\downarrow}}_{\hat{H}_{atomic}} ~,
\end{equation}
this equation represents the \textit{single-channel model} and it is written in \textit{mixed-representation}, where we use momentum-space $\mathbf{k}$ for $\hat{H}_{bath}$ operators and real-space representation for impurity operators. In the expression we define the number operator for the bath $\hat{n}_{\mathbf{k}\sigma}=c^{\dagger}_{\mathbf{k}\sigma}c_{\mathbf{k}\sigma}$ for $c^{\dagger}_{\mathbf{k}\sigma}(c_{\mathbf{k}\sigma})$ creation (annihilation) operator in the bath and the number operator for the impurity $\hat{n}_{d\sigma}=d^{\dagger}_{\sigma}d_{\sigma}$ for $d^{\dagger}_{\sigma}(d_{\sigma})$ creation (annihilation) operator for the impurity. In Eq.\ref{eq:AM}, $\hat{H}_{atomic}$ describes a \textit{single-impurity} in (non)degenerate \textit{d-}state such that its electronic occupancy can be at most two spin-$1/2$ Fermions. Hence, the impurity occupancy is either empty $\ket{0}$, singly $\ket{\sigma},\sigma=\lbrace \uparrow, \downarrow \rbrace$ or doubly $\ket{\updownarrows}$ occupied states. Note that $\ket{\sigma}$ corresponds to the \textit{local moment} state whose origin stems from specific energy scales requirements - as we discuss below. \\
Considering that a sufficiently strong Coulomb repulsion among electrons prevents any finite electron flux, a Coulomb repulsive potential term is introduced for state occupied by two electrons with opposite spins. This energy penality is then required only for doubly occupied state, which is then the least energetically favoured state in the model considering the three type of states described within $\hat{H}^{AM}$ . The Coulomb interaction term for the doubly occupied state in the impurity \textit{d-}orbital is modelled as
\begin{equation}
U =  \int d\mathbf{r} \int d\mathbf{r}^{\prime} \frac{e^{2}}{|\mathbf{r}-\mathbf{r}^{\prime}|}|\phi_{d}(\mathbf{r})|^{2}|\phi_{d}(\mathbf{r}^{\prime})|^{2} ~,
\end{equation} 
where $|\phi_{d}|^{2}$ is the electron density in a single atomic orbital $\phi_{d}$ \cite{Hewson}. This term is the actual source of the magnetism in the model. In $\hat{H}_{atomic}$, we define the single-impurity energy $\epsilon_{d}$ and $d^{\dagger}_{\sigma}$ the operator creating a \textit{d-}electron with $\sigma$ spin in the atomic \textit{d-}orbital. We note that the impurity is local in space and its operator does not carry any momentum index.\\
The single-impurity is embedded in a \textit{host metal} that is the noninteracting electronic bath $\hat{H}_{bath}$. The bath continuum energy spectrum $\epsilon_{\mathbf{k}}$ takes a finite energy range due to the finite bandwidth $\epsilon_{\mathbf{k}} \in [-D,+D]$ of the single metallic conduction band. In $\hat{H}_{bath}$, we define $c^{\dagger}_{\mathbf{k}\sigma}$ the operator creating an itinerant conduction electron with $\mathbf{k}$ momentum and $\sigma$ spin. \\
The localised \textit{d-}orbital impurity electrons scatter off the metallic bath and form a \textit{virtual bound state} \cite{Friedel1958}. This corresponds to a resonance at $\epsilon_{d}+\mathit{Re}\Sigma_{dd}(\omega)$ level. In case the resonance originates at the Fermi level in the conduction band density of states, the virtual bound state forms the so-called Kondo resonance as we discuss more later. Moreover, we can interpret the bound state formation as tunnelling process between the \textit{d-}orbital and the itinerant conduction band electron of the metal. We refer to this process as the \textit{hybridization} between the impurity wavefunction bound states  $\phi_{d}$ and the itinerant conduction electron Bloch waves $\psi_{\mathbf{k}}$. The resulting hybridization matrix included in $\hat{H}_{hyb}$ reads $\mathbb{V}_{\mathbf{k}} = \braket{\phi_{d}|\hat{H}_{hyb}|\psi_{\mathbf{k}}}$.\\ 
This system is also called the single-impurity single-channel Anderson model, as we have a 1-channel system with one impurity coupled to one electronic reservoir. Despite its simple structure, the AM effectively describes many different many-body systems and is often used as limiting test for more complicated systems.\\
In the remainder of this section we introduce two limiting cases of the Anderson impurity model: the system without and with hybridization \cite{Coleman2002LocalMoment}. In the former condition, the system reduces to an isolated interacting impurity. In the latter one, the coupled impurity-bath system can be further analysed either in presence or in absence of Coulomb potential interaction.
  
\subsection*{The atomic limit of the Anderson model}
As the first limiting case of the general model, we consider an isolated impurity with associated Coulomb repulsive potential. This is a decoupled system where no tunnelling events occur with the bath. The model Hamiltonian does not include any hybridization type of term like $\hat{H}_{hyb}$ and so it equal to $\hat{H}_{atomic}$ only - see its definition in Eq.\ref{eq:AM}. This system has a total of four atomic quantum states and we can distinguish those according to their three energy configurations for the \textit{d-}orbital electron: $(i)$ singly occupied state $E_{1d}=\epsilon_{d}$ for each $\sigma=\lbrace \uparrow, \downarrow \rbrace$, $(ii)$ empty state $E_{0d}=0$ and $(iii)$ doubly occupied state $E_{2d}=U+2\epsilon_{d}$. The configuration $(i)$ is two-fold spin degenerate and so corresponds to an associated magnetic moment of an effective localised spin $S=1/2$. The other two configurations have no spin degeneracy and so no magnetic moment is formed. \\
The atomic limit of the Anderson model can be analysed by means of its phase diagram which presents different behaviour according to energy level.\\
In order to generate \textbf{local moments} from the atomic limit of the Anderson model, we look at the repulsive Coulomb potential phase at $U>0$. Here, we can make energetically favoured the singly occupied state by imposing $\epsilon_{d} < \epsilon_{F}$ and $U + \epsilon_{d} > \epsilon_{F}$ with $\epsilon_{F}$ the Fermi level, here after set to $\epsilon_{F}=0$. This energy scale implies that the Coulomb repulsive potential is then larger than the on-site impurity energy such that it is favoured to add only one electron to the \textit{d-}orbital. As consequence, the singly occupied state is located at the lowest-energy level state: this is the local moment state, an effective spin-$1/2$ degree of freedom. Thus, the singly occupied state is a \textit{two-fold degenerate magnetic spin doublet ground-state}.

\subsection*{The noninteracting coupled Anderson model}
A second limiting case of the model considers the single-impurity without Coulomb potential term in $\hat{H}_{atomic}$ coupled to the conduction Fermi sea bath: this is the noninteracting version of the Anderson model in Eq.\ref{eq:AM}, called also \textbf{resonant level model}. In this system, the \textit{d-}orbital impurity electrons scatter off the itinerant conduction electrons bath which are fully delocalised states except at the vicinity of the impurity region. Thus, the conduction electron wavefunction forms a Bloch state as it is exponentially decaying with distance from the impurity region. As consequence, the impurity electrons scattering off the conduction band states develop \textit{virtual bound states} - since the actual bound state exists only in vicinity of the impurity where the conduction electron wavefunction is fully localised. Furthermore, these scattered electrons are immersed in the continuum electrons bath and so impurity bound states can hybridise with the conduction electron states. Such a hybridization causes the broadening of the \textit{d-}level and the formation of \textit{noninteracting resonance} in conduction band density of states.\\
A quantitative discussion on the noninteracting Anderson model requires to introduce some important techniques  we will extensively apply in thesis.\\

\noindent{We} start with a concrete analysis on the virtual bound state resonance and its attributes. To achieve this comprehension, we need to calculate the impurity Green's function - here it is a local, noninteracting many-body object.\\
In its general terms,  we can analytically the derive Green's functions using \textbf{equation of motion theory} (EoM) \cite{Hewson, Flensberg, Lacroix1981_UInfAMGreenFunc,Dias_ImpSolver2019}. In general, the EoM method to evaluate Green's functions allows to include information about the on-site correlations - which are not considered in this section - and to leave out those from the leads. This method is based on consecutive differentiating the Green's function in order to obtain a series of coupled differential equations. For noninteracting models - as the $U=0$ Anderson model - or for general simple one dimensional models, the equations form a closed set and the problem is exactly solvable, leading to coupled equation sets written in form of continued fraction equation \cite{ContinuedFraction_Mori1965}. On the contrary, in presence of interactions the set of equation cannot be solved exactly. Approximate solutions are obtained by truncating the recursive equation for instance by neglecting some correlations or by approximating the interaction contribution causing the  the particle-hole resonances \cite{MeirWingreenLee_StronglyInteractingElec_PRL1991}. Of importance in our studies, the EoM technique is applicable to both nonequilibrium and equilibrium Green's functions.  \\
We can now introduce the set of equations derived using equation of motion approach to describe the virtual bound state resonance.\\
Initially we write the retarded Green's function in complex domain denoted $G^{R}_{a,b}(z) =\langle\langle \Psi_{a}; \Psi^{\dagger}_{b} \rangle\rangle_{z}$, where the complex variable is defined as $z=\omega + i \eta, \eta >0$ to make convergence in the upper half-plane. We can convert the complex frequency results into to the corresponding real frequency domain by applying analytic continuation of $z$ to real axis for $\eta \rightarrow 0^{+}$. By differentiating $G^{R}_{a,b}(z)$ we obtain the two equivalent equations:
\begin{equation}\label{eq:EoM}
\begin{aligned}
\langle\langle \Psi_{a}; \Psi^{\dagger}_{b} \rangle\rangle_{z} &= (1/z) \Big[ \langle \lbrace \Psi_{a},\Psi^{\dagger}_{b}\rbrace\rangle + \langle\langle\Psi_{a}; [\hat{H}_{full}, \Psi^{\dagger}_{b}] \rangle\rangle_{z} \Big] ~,\\
&= (1/z) \Big[ \langle \lbrace\Psi_{a},\Psi^{\dagger}_{b}\rbrace\rangle - \langle\langle [\hat{H}_{full},\Psi_{a}];\Psi^{\dagger}_{b} \rangle\rangle_{z}  \Big] ~,
\end{aligned}
\end{equation}
that is derived using the definitions for time derivatives of operators in the Heisenberg picture. This is the type of equation of motion structure we will apply throughout this thesis.\\
We use the equation of motion in Eq.\ref{eq:EoM} to calculate the retarded noninteracting impurity Green's function $G^{R}_{0,dd\sigma}(z)$:
\begin{equation}\label{eq:G_0dd}
\langle\langle d_{\sigma};d^{\dagger}_{\sigma}  \rangle\rangle_{z}= G^{R}_{0,dd\sigma}(z)  = \frac{1}{z-\epsilon_{d} -\Delta(z)} ~,
\end{equation}
where $\Delta(z)$ is the total \textbf{hybridization function} describing the impurity bath coupling, and follows as
\begin{equation}
\Delta(z)=\sum_{\mathbf{k}\sigma} |V_{\mathbf{k}}|^{2} G^{0,R}_{bath,\mathbf{k}\sigma}(z) ~,
\end{equation}
where $V_{\mathbf{k}}$ is the hybridization coupling strength as in Eq.\ref{eq:AM} and $G^{0,R}_{bath,\mathbf{k}\sigma}$ with superscript zero indicates the \textit{isolated} noninteracting, thermal-equilibrium bath. It is straightforward to derive $G^{0,R}_{bath,\mathbf{k}\sigma}$ applying EoM 
\begin{equation}
\langle\langle c_{\mathbf{k}\sigma};c^{\dagger}_{\mathbf{k}\sigma}  \rangle\rangle_{z} = G^{0,R}_{bath,\mathbf{k}\sigma}(z) =\sum_{\mathbf{k}} \frac{1}{z-\epsilon_{\mathbf{k}}} ~,
\end{equation} 
such that 
\begin{equation}
	\Delta(z) =\sum_{\mathbf{k}} \frac{|V_{\mathbf{k}}|^{2}}{z-\epsilon_{\mathbf{k}}}~,
\end{equation}
and we obtain \textit{hybridization function for the noninteracting Anderson's model}. The analytic continuation to the real frequency follows as $G^{0,R}_{bath,\mathbf{k}\sigma}(\omega)\rightarrow G^{0,R}_{bath,\mathbf{k}\sigma}(\omega+i0^{+})$ with $\Delta(\omega)\rightarrow\Delta(\omega+i0^{+})$.
The hybridization function $\Delta(z)$ is clearly complex: its real and imaginary parts are derived using the Dirac identity and they take different form according to the conduction band features \cite{Altland}. The calculation of the complex $\Delta(z) = \mathit{Re}\Delta + i\mathit{Im}\Delta$ implies to calculate the real and imaginary part of $G^{0,R}_{bath,\mathbf{k}\sigma}$. It is standard to take the so-called electronic density of states \textbf{wide-band limit} (WBL), which we now discuss for finite and infinite band \cite{JauhoWingreeMeir1994}.\\
In case of finite WBL i.e. the conduction is constant within a finite range $\epsilon_{\mathbf{k}} \in [-D,+D]$, the real part of the hybridization function 
is the function principal value and it is given by 
\begin{equation}
\mathit{Re}\Delta(\omega) = \frac{V^{2}}{2D} \log \bigg[ \frac{D+\omega}{D-\omega} \bigg]
\end{equation}
and the imaginary part reads
\begin{equation}
\mathit{Im}\Delta(\omega) = \frac{- \pi V^{2}}{2D}  \theta(D-|\omega|) ~,
\end{equation}
that is constant within the bandwidth. In the above equations we define $V^2=\sum_{\mathbf{k}} |V_{\mathbf{k}}|^2$.\\
In case of infinite WBL i.e. $D\rightarrow +\infty$, the real part vanishes. 
We note $\mathit{Im}\Delta(\omega)$ is related the noninteracting uncoupled bath spectral function, and hence free density of states. We use these definitions to rewrite the imaginary component of hybridization function 
\begin{equation}
\mathrm{\Gamma}(\omega) \equiv -\mathit{Im}\Delta(\omega) = -\Big( -\pi\sum_{\mathbf{k}\sigma} |V_{\mathbf{k}}|^{2} \rho_{0}(\omega) \Big)  = - \sum_{\mathbf{k}\sigma} |V_{\mathbf{k}}|^{2} \mathit{Im}G_{bath,\mathbf{k}\sigma}^{0,R}(\omega)  ~,
\end{equation}
from now on we refer to this result for $\mathrm{\Gamma}$ the \textit{gamma or levelwidth function} as corresponding to the imaginary part of the noninteracting self-energy in Eq.\ref{eq:defSelEngeneral}. This equation acquires general definition in \textit{matrix} form as given in Eq.\ref{eq:LevelWidth}. \\
By means of these equations, we can finally characterise the virtual bound state resonance. The resonance in the isolated impurity spectral function
is $\delta$-function pinned at \textit{d-}orbital energy level as
$\propto ~ \delta(\omega-\epsilon_{d})$. Then, by hybridizing with the Fermi sea states, the Dirac delta smears into a \textit{Lorentzian function} with width at half-max equal to $\mathrm{\Gamma}$ and pinned at $\epsilon_{d}$, namely
\begin{equation}\label{eq:DoS_0dd}
\rho_{0,dd\sigma}(\omega) = 
\mathcal{A}_{0,dd\sigma}=-\sum_{\sigma}\frac{1}{\pi}\mathit{Im}[ G^{R}_{0,dd\sigma}(\omega)]
= \frac{\Gamma/\pi}{(\omega-\epsilon_{d})^{2}+\Gamma^{2}} ~,
\end{equation}
where we have used the imaginary part of the expression for noninteracting impurity Green's function in Eq.\ref{eq:G_0dd}. The Eq.\ref{eq:DoS_0dd} defines the density of states in single-particle picture: we revisit in the next section how the spectral function is modified in presence of the Coulomb interaction.\\

\noindent{We} take a step further in our discussion and we focus on the scattering process among impurity electrons and surrounding Fermi sea. In particular, the successive scattering on the conduction electrons is successfully described using \textbf{scattering theory} \cite{Mahan, Economou}.\\
By recalling the general resolvent Green's function equation $(\omega \pm i \eta -H)G^{R/A}(\omega)= 1$ where $1$ is the identity function, we manipulate it to get the Dyson equation for $G^{R}_{bath,\mathbf{k}\mathbf{k}^{\prime}\sigma}$ in the noninteracting Anderson model:
\begin{equation}
G^{R}_{bath,\mathbf{k}\mathbf{k}^{\prime}\sigma}(\omega) = \delta_{\mathbf{k}\mathbf{k}^{\prime}} G^{0,R}_{bath,\mathbf{k}\mathbf{k}^{\prime}\sigma}(\omega) + \big( \delta_{\mathbf{k}\mathbf{k}^{\prime}} G^{0,R}_{bath,\mathbf{k}\mathbf{k}^{\prime}\sigma}(\omega) \big) VG^{R}_{bath,\mathbf{k}\mathbf{k}^{\prime}\sigma}(\omega) ~,
\end{equation}
this is noninteracting version in momentum representation of Eq.\ref{eq:defDysonRA} but now calculated for $G^{R}_{bath,\mathbf{k}\mathbf{k}^{\prime}\sigma}$ the propagator for the conduction electrons under successive scattering with impurity  \textit{d-}orbital electrons with strength $V$ and $\delta_{\mathbf{k}\mathbf{k}^{\prime}} G^{0,R}_{bath,\mathbf{k}\mathbf{k}^{\prime}\sigma} \equiv G^{0,R}_{bath,\mathbf{k}\mathbf{k}\sigma}$ the propagator for the isolated noninteracting bath term. From this expression, the system Hamiltonian $\hat{H}$ is composed by the noninteracting part $\hat{H}_0$ plus $V$ that is the additional interaction due to the impurity. We can treat the latter perturbatively and that means by iteratively solving the Dyson equation in terms of power series in $V$, we derive the \textbf{$\mathrm{T}$-matrix equation} in momentum space representation. We introduce the bath matrix $\mathbb{G}^{R}_{bath,\mathbf{k}\mathbf{k}^{\prime}\sigma}$ with $dim(N\times N)$ and $\mathrm{T}$-matrix equation reads:
\begin{equation}\label{eq:TmatrixEq}
\mathbb{G}^{R}_{bath,\mathbf{k}\mathbf{k}^{\prime}\sigma}(\omega) = \delta_{\mathbf{k}\mathbf{k}^{\prime}} \mathbb{G}^{0,R}_{bath,\mathbf{k}\mathbf{k}^{\prime}\sigma}(\omega) + \mathbb{G}^{0,R}_{bath,\mathbf{k}\mathbf{k}\sigma}(\omega) \cdot \mathbb{T}_{\mathbf{k}\mathbf{k}^{\prime}}(\omega) \cdot \mathbb{G}^{0,R}_{bath,\mathbf{k}^{\prime}\mathbf{k}^{\prime}\sigma}(\omega) ~,
\end{equation}
where the $\mathrm{T}$-matrix $\mathbb{T}_{\mathbf{k}\mathbf{k}^{\prime}}$ is a \textit{local} object that encapsulates all information about the physical scattering processes happening at the impurity region and its general matrix form reads
\begin{equation}
\mathbb{T}_{\mathbf{k}\mathbf{k}^{\prime}}(\omega)=\mathbb{V} + \mathbb{V}\cdot\mathbb{G}^{0,R}_{\mathbf{k}\mathbf{k}^{\prime}\sigma}(\omega)\cdot\mathbb{V} + \mathbb{V}\cdot\mathbb{G}^{0,R}_{\mathbf{k}\mathbf{k}^{\prime}\sigma}(\omega)\cdot\mathbb{V}\cdot\mathbb{G}^{0,R}_{\mathbf{k}\mathbf{k}^{\prime}\sigma}(\omega)\cdot\mathbb{\mathbb{V}} + \dots  =
\mathbb{V}\cdot(\mathbb{1}- \mathbb{G}^{0,R}_{\mathbf{k}\mathbf{k}^{\prime}\sigma}(\omega)\cdot\mathbb{V})^{-1} ~,
\end{equation}
for $\mathbb{V}$ with $dim(N\times N)$ the matrix form of the interaction $V$.\\
In particular, the scattering $\mathrm{T}$-matrix equation for the noninteracting $U=0$ Anderson's model is equal to  
\begin{equation}
\mathbb{T}_{dd}(\omega)=\mathbb{V}\cdot \left( \sum_{\sigma} \mathbb{G}^{R}_{0,dd\sigma}(\omega) \right)\cdot\mathbb{V} ~,
\end{equation}
where $\mathbb{G}^{R}_{0,dd\sigma}$ is the \textit{full} impurity propagator matrix version of the definition in Eq.\ref{eq:G_0dd} - full in the sense of broadened due to bound states hybridizing with Bloch conduction states. \\
This nomenclature e.g. full, bare propagators, it is used to adhere to the standard \textbf{Feynman diagrams technique} \cite{Coleman}. The full propagator usually refer to the system in presence of interaction, here it indicates the bath coupled to the impurity in presence of scattering events. The full propagator is then calculated using the bare propagator identified by the isolated system and its successive interactions with the impurity. These events are connected through vertices according to diagrammatic rules.\\

\noindent{We} continue the discussion by recalling that the $\mathrm{T}$-matrix is related to the $\mathrm{S}$-matrix defined as $S(\omega) = exp(2i \delta(\omega))$, where $\delta(\omega)$ is the scattering phase shift.\\ 
At zero temperature, it can be shown that the phase shift evaluated at the Fermi surface sets the charge occupation $\hat{n}_{d}$ in the ground-state. We can see this by integrating the noninteracting impurity version of the density of states Eq.\ref{eq:DoS_0dd} up to the Fermi level  $\epsilon_{F}$, namely
\begin{equation}\label{eq:FSR}
\hat{n}_{d} = 
 \sum_{\sigma} \int^{\epsilon_{F}}_{-\infty} d\omega \rho_{0,dd\sigma}(\omega) = \frac{2}{\pi} \delta(\epsilon_{F})
\end{equation}
to get the ground-state impurity occupation. Note that here we have assumed $\delta(-\infty) =0$ and the prefactor two takes into account the spin degeneracy. This result is known as \textbf{Friedel sum rule} (FSR) \cite{Friedel1958}. This result holds only for genuine Fermi liquid systems, also in the interacting case.\\
We apply now the FRS to find the phase shift value at \textit{d-}level to set the charge occupation $\hat{n}_{d}$ at the bound state resonance. We obtain the phase shift is equal to $\delta(\omega =\epsilon_{F})=\pi/2$: at the resonant level we have \textit{unitary scattering}. As consequence, $n_{d}\equiv1$ that is the impurity ground-state is singly occupied. This is exactly valid at the particle-hole symmetry regime, otherwise $n_{d}\neq 1,\delta(\omega =\epsilon_{F})\neq\pi/2$ in general. \\
Another usual version of the FRS is obtained by considering that the tangent of $\delta(\omega)$ can be calculated from the ratio of the imaginary and real part of $G^{R}_{0,dd\sigma}$ in Eq.\ref{eq:G_0dd}. With few more steps, we can write the spectral function in terms of scattering phase shift, namely
\begin{equation}\label{eq:SpecLangKRes}
\mathcal{A}_{dd\sigma}(\omega=\epsilon_{F}) \equiv  \mathcal{A}_{0,dd\sigma}(\omega=\epsilon_{F}) = \frac{1}{\pi \Gamma}\frac{1}{1 + [\tan^{-1}\delta(\epsilon_{F})]^{2}} =  \frac{1}{\pi \Gamma} \sin^{2}\delta(\epsilon_{F})  ~,
\end{equation}
where we have used the definition of the noninteracting spectral function Eq.\ref{eq:defSpectral} with the Friedel sum rule evaluation at the Fermi level \cite{Langreth1966}. This expression is actually adiabatically invariant: the resonance at the Fermi level appears also in presence of finite Coulomb interaction - as we discuss in the next section.\\
 
\noindent{We} conclude the analysis on $U=0$ Anderson's model considering that in its essence, the noninteracting AM basically describes a unique Fermionic level whose electrons tunnel to an noninteracting, thermal-equilibrium electronic reservoir. This is indeed equivalent to the so called \textbf{noninteracting resonant level model} (NI-RLM) \cite{Hewson}: a single-level impurity - it can be at most occupied by two spin-$1/2$ Fermions to obey the Pauli's exclusion principle - couples to a bath formed by identical noninteracting Fermions. Despite their simple description, these models are often used to understand more elaborate systems. \\
Because the NI-RLM is used in Sec.\ref{sec:MVPC}, we write here its real-space representation for completeness. We can calculate that by applying the Fourier transformation 
\begin{equation}
	c^{\dagger}_{k \sigma} = \sum_{n=-\infty}^{+\infty} c^{\dagger}_{n \sigma}e^{i k n} ~,
\end{equation} 
to the $U=0$ version of the Anderson model in Eq.\ref{eq:AM}. The continuous model Hamiltonian in momentum space is mapped to real-space representation where $n^{th}$-label indicates the site number on the bath chain - and physically each site corresponds to an orbital. The resulting NI-RLM reads as:
\begin{equation}\label{eq:NI-RLM}
	\begin{aligned}
	\hat{H}^{NI-RLM} &\equiv \hat{H}^{AM}(U=0) \\
	&= \sum_{\sigma} \left(
	\sum_{n=0}^{+\infty} ( \epsilon_{n\sigma} c^{\dagger}_{n \sigma}c_{n, \sigma} + t_{n}c^{\dagger}_{n \sigma}c_{n+1 \sigma}  + t^{\star}_{n}c^{\dagger}_{n+1 \sigma}c_{n \sigma} ) + \epsilon_{d} d^{\dagger}_{\sigma}d_{\sigma} 
	+ V\sum_{\sigma} ( d^{\dagger}_{\sigma}c_{0\sigma}+ c^{\dagger}_{0\sigma}d_{\sigma} ) \right) ~,
	\end{aligned}
\end{equation}
where the Fermionic bath becomes one dimensional semi-infinite chain with tight-binding hopping term $t_{n}$. Its $0^{th}$-site - identified in the operators $c^{\dagger}_{0\sigma},c_{0\sigma}$ - couples to the single-level $\epsilon_{d}$. We note that the locality of the impurity implies its invariance under change of representation. Furthermore, the real-space representation form of $\hat{H}^{NI-RLM}$ highlights its intrinsic feature of satisfying proportionate coupling condition, as discuss in Sec.\ref{sec:PC}. As important conclusion: any model mapped to $\hat{H}^{NI-RLM}$ exhibits the PC property. This crucial property will be used in later chapters.\\

\subsection*{The finite- and infinite- $U$ Anderson model}
We now consider the full model with an interacting single-impurity coupled to the Fermi sea as modelled in Eq.\ref{eq:AM}.\\
We start again by calculating the retarded interacting impurity Green's function $G^{R}_{dd\sigma}(z)$ and we avail again of EoM technique in Eq.\ref{eq:EoM} to calculate it:
\begin{equation}\label{eq:G_dd}
\langle\langle d_{\sigma};d^{\dagger}_{\sigma}  \rangle\rangle_{z}= G^{R}_{dd\sigma}(z) =\frac{1}{z-\epsilon_{d}-\Delta(z)-\Sigma(z)} ~,
\end{equation}
and $G^{R}_{dd\sigma}(\omega),\Sigma(\omega)$ are obtained by letting the complex variable $z=\omega+i\eta$ for $\eta>0$ and then taking the limit to $\eta\rightarrow 0^+$ as before.\\
In presence of a finite $U\hat{n}_{d\uparrow}\hat{n}_{d\downarrow}$ term in the model Hamiltonian, the impurity propagator includes also the self-energy term $\Sigma(z)$ such that the equation cannot be analytically solved exactly and it can be computed by NRG. Rather than discussing how to calculate an expression for $G^{R}_{dd}$, we focus on the physical implication of the interaction in the coupled system. \\
Under adiabatic switching-on interactions, the single \textit{d-}level impurity resonance splits in three parts. Slowly increasing the interactions in the model allows us to keep the \textit{d-}level occupancy constant by progressively shifting the level itself. Even though the spectral function is reshaping, $\mathcal{A}_{dd\sigma}$ has to be normalised to one and thus its weight has to be conserved while the level is shifted. In order to keep the normalization of $\mathcal{A}_{dd\sigma}$, we need to introduce a renormalization wavefunction weight, the so-called \textit{quasiparticle weight} $Z$, that corresponds to the modified area under the spectral resonance due to finite $U$.\\
It can be shown that by continuously slow switching-on interactions, there is a unique correspondence between excitations in the $U$ finite model and those in the $U=0$ model. Under such conditions, the system locally behaves as \textbf{Fermi liquid} (FL) \cite{Nozieres1974,Hewson_RenormalPT-FL1993,Hewson}.\\
We combine now the results for the noninteracting AM and new features emerging from the interacting AM to derive an expression for the full impurity propagator $G^{R}_{dd\sigma}(z)$. We take the conduction band in the standard infinite WBL approximation such that the only hybridization contribution is its imaginary part $\Delta(z)=-i\mathrm{\Gamma}$. At $T=0$, the self-energy function takes a specific form for a Fermi liquid model whose $\mathit{Im}\Sigma^{U}$ vanishes for decreasing energies. Here, we specify the vanishing component is the \textit{interacting} part of the self-energy, see its general definition in Eq.\ref{eq:defSelEngeneral}. In conclusion, the renormalised impurity Green's function reads
\begin{equation} \label{eq:G_ddFL}
\lim_{\omega,\eta\rightarrow 0^+} G^{R}_{dd\sigma}(\omega + i \eta) = \lim_{\omega,\eta\rightarrow 0^+} \frac{1}{(\omega - \epsilon^{\star}) + i (\eta + \Gamma^{\star})} ~,
\end{equation}
this Green's function indicates the \textit{d-}level resonance weight is renormalized by $Z \ll 1$ and it is located at position $\epsilon^{\star}= Z(\epsilon_{d} + \mathit{Re}\Sigma(\omega))$ that is the so called dressed particle-energy due to interactions and it has renormalised half-width at half-max given by $\mathrm{\Gamma}^{\star} = Z \mathrm{\Gamma}$. This renormalised resonance is the \textbf{Abrokosov-Suhl resonance} according to its original discovery but it is generally named as \textbf{Kondo resonance} \cite{Abrikosov1965,Suhl1965,Georges2016}. This is because within the energy window $\epsilon_{d} < \omega < \epsilon_{d}+U$, the renormalised resonance fits the local moment region of the $\lbrace U\rbrace$ phase-diagram and so the corresponding ground-state is the spin degenerate singlet state. Notably, the spectral function of the renormalized \textit{d-}level resonance in finite $U$ system is equivalent to the spectral function of the noninteracting AM as derived using the FSR in Eq.\ref{eq:SpecLangKRes}. This indicates that the peak position starts at $\omega=\epsilon_{d}$ in the $U=0$ model. Then, it shifts to $\omega=\epsilon_{F}$ in the interacting model having $\mathcal{A}_{dd\sigma}(\omega=\epsilon_{F})$ is an adiabatic invariant determined solely by the scattering phase: hence, the Kondo resonance at the Fermi level is a universal property of impurity models regardless interactions. As last comment, we also note that the vanishing $\mathit{Im}\Sigma$ as Fermi liquid property is also applied in the derivation of the Friedel sum rule \cite{Langreth1966}.\\
For completeness, we also give the useful result of the imaginary part of the impurity Green's function in Eq.\ref{eq:G_dd} under the Fermi liquid condition defined in the limits as in Eq.\ref{eq:G_ddFL}, namely
\begin{equation}
\lim_{\omega,\eta\rightarrow 0^+} \mathit{Im} G^{R}_{dd\sigma}(\omega) = \lim_{\omega,\eta\rightarrow 0^+} \frac{-(\eta+\Gamma^{\star})}{(\omega-\epsilon^{\star})^{2} + (\eta +\Gamma^{\star})^{2}} ~,
\end{equation}
as written using the above definitions of renormalized both single-particle energy and hybridization \cite{HewsonOguri_RenormalizedImp_2004}.\\
Away from the lowest energy states around the Fermi level, we encounter the other two peaks emerging in the interacting model. At $\omega= \epsilon_{d}$ is the \textit{hole-like peak} for charge fluctuations at \textit{valence} transition to empty states i.e. $\ket{\sigma} \rightarrow \ket{0}$ and at $\omega= \epsilon_{d} +U$ is the \textit{electron-like peak} for charge fluctuation at \textit{valence} transition to doubly occupied states i.e. $\ket{\sigma} \rightarrow \ket{\updownarrows}$. The two peaks are separated in energy by $U$ and each has half-width at half-max related to $\Gamma$.\\ 
Recalling that the charge fluctuation lifetime $\tau$ is proportional to the inverse of $U$, we can comment that fluctuations to empty and doubly occupied states happens within finite lifetime $\tau$ in an interacting AM with finite but relatively small $U$. For finite time interval, the system can be found in $\ket{0},\ket{\updownarrows}$ states in addition to the singly occupied one. \\

\noindent{We} complete this section presenting the limiting case of the interacting model: that is by setting the Coulomb repulsive potential to be large i.e. $U\rightarrow +\infty$ and the corresponding model is known as \textbf{infinite-U Anderson model} \cite{Lacroix1981_UInfAMGreenFunc}. In practice, this limit is generally approximated when a Coulomb repulsive potential exceeding in large amount the hybridization and this condition is typically obtained for high temperature superconductors, gigantic magnetoresistance and heavy electron materials \cite{Coleman_FLInfU_2007}. The hole-like and electron-like valence fluctuation bands are then infinitely separated in energy $U$ from the low-energy states such that no valence transitions from singly occupied states to either empty or doubly occupied states occur any more.  This is because transitions to the lowest excited high-energy states are inaccessible due to very large energy gap among those and the singlet ground-state local moment.\\ 
Focusing on the low-energy scales of the infinite-$U$ model, the effective model presents a  \textit{single energy configuration state} that is the singly occupied one \cite{Coleman_FLInfU_2007}. On the contrary of the interacting AM for relatively small $U$, the charge fluctuation lifetime $\tau \rightarrow 0$ for $U \rightarrow \infty$ and so the system never occupies $\ket{0},\ket{\updownarrows}$ states in finite time interval.\\
The very large $U$ limit ensures negligible double occupied states and this assumption has been used to facilitate electrical conductance calculation in quantum dots set-up - see Sec.\ref{sec:Kexp} for more detail about this system - where generally $U \sim 1~meV, V^{2} \sim 1-10 ~\mu eV$ and condition $U \gg V$ is approximately correct \cite{MeirWingreen_AMoutEquilib_PRL1993}.\\
On the level of the many-body Hilbert space, the infinite $U$ limit implies some subspaces of total Hilbert space are projected out by means of perturbation expansion in the hybridization. The resulting system at low energy and temperature scales has suppressed any multi-occupancy states. This motivates derivation of effective spin models as now described.

\subsection{The impurity Kondo model}
The second fundamental impurity model we aim to discuss is the \textbf{Kondo model} whose physics is framed in concept of \textit{energy scale renormalization}.\\
As general fact in many-body physics, we often encounter processes developing for several energy scales. In cases where bare and emergent energy scales are so widely spread, the technique able to describe system properties at any energy scales is the \textbf{renormalization group technique} (RG) \cite{Wilson_RGEpsilonExp1974,Hewson}. In this present context we aim to introduce general renormalization concepts to aid our discussion on the Kondo model. We will return to it a in quantitative fashion in the next section Sec.\ref{sec:RGtheo}.\\
The renormalization group consists of including the high-energy physics of the microscopic model into a few effective parameters controlling the corresponding low-energy physics. Each step of the renormalization approach progressively reduces the energy scale cut-off and that is achieved by \textit{integrating out} the high-energy degrees of freedom \cite{Wilson1975}. The RG procedure is in essence an \textit{energy scaling} towards lower ground-state. And so the renormalized model Hamiltonian reflects a lower-energy model with smaller energy spectrum than the original one. The elimination of higher-energy states implies a reduction of degrees of freedom. Crucially, such a reduction - rather than neglecting essential model properties - allows us to study the otherwise untreatable bare microscopic model through only its lower-energy part. This is because the scaling approach indicates that the low-energy physics properties do not dramatically depend of the high-energy physics details. Furthermore, it can be shown that systems sharing the same low-energy excitation spectrum as result of scaling from different microscopic models form a \textit{universality class}.\\
Renormalization technique is indeed a very powerful strategy often encountered in various physics fields when the real system phenomena develop within a very large energy spectrum interval. Examples include the physics of quark confinement. As consequence, the system can be effectively modelled only through a renormalized model Hamiltonian.\\
We continue now by applying the RG technique to impurity models: this allows us to understand, from energy scaling perspective, the physics captured by the various models. We start considering a general impurity microscopic model whose conduction bandwidth $D_{0}$ spans over arbitrarily large energy scales. The local moments formation happens at the order of the Coulomb energy - that is at the energy scale of few $\sim eV$. On the other hand, the Kondo effect develops at a scale of the order of few $\sim meV$ \cite{Coleman}. For the energy range, these crucial events characterising the Kondo physics are separated by a large energy interval. The energy scale required to fully capture the model spans over many orders. In this scenario, the renormalization group is the ideal technique to express the high-energy physics effects into few, controllable set of low-energy parameters \cite{Wilson1975}. \\
Using the RG terminology, we can classify the models Hamiltonian we previously encountered and go beyond them by further reducing the energy cut-off. Starting with the initial microscopic model at high energy scale, the AM Hamiltonian in Eq.\ref{eq:AM} identifies the first scaling step in energy renormalization with a reduced energy cut-off $D^{\prime}<D_{0}$. By eliminating the high-energy excitations bands from the original microscopic model, the system energy spectrum spans now a range within $D^{\prime} \in (|\epsilon_{d}| ; \epsilon_{d}+U)$ with hole and charge bands separated by an amount $U$ energy.\\
Then, we can reduce further the energy cut-off $D^{\prime}$ by sending $U \rightarrow \infty$: the resulting low-energy model is the infinite-$U$ AM as presented already. This is the second scaling step with reduced energy cut-off $D^{\prime\prime} < D^{\prime}$. The renormalised Hamiltonian with low-energy spectrum $D^{\prime\prime} < \epsilon_{d}+U$ does not include the doubly occupied states. The low-energy system behaves as an effective orbital with spin degenerate ground-state that couples with the conduction electron band through virtual states. \\
Carrying on with the scaling procedure, we can reduce even further the energy cut-off to configure the low-energy physics to be determined solely by the ground-state Kondo singlet - that is formed by the complete screening of the impurity spin by the surrounding conduction electron. When the doublet $S=1/2$ hybridizes with the surrounding conduction electron sea, the hybridization induces virtual charge fluctuation that brings spin exchange between doublet and electrons sea. The virtual processes reach the excited states i.e. doubly $\ket{\updownarrows}$ and empty $\ket{0}$ which are separated in energy from the singlet state  
\begin{equation}
\begin{aligned}
&E_{\ket{\updownarrows}}- E_{\ket{\sigma}} +  \approx U + \epsilon_{d} > \epsilon_{F} ~,\\
& E_{\ket{\sigma}} - E_{\ket{0}} \approx - \epsilon_{d} > \epsilon_{F} ~.
\end{aligned}
\end{equation}
As consequence, the reduced energy cut-off $D^{\prime\prime\prime}$ now lies between $\epsilon_{F} < U + \epsilon_{d}$ with $\epsilon_{d}<\epsilon_{F}$.\\
This further energy scaling step from the infinite-U AM requires a new type of impurity model: that is the Kondo model. By introducing its features,  we can benefit from the discussion on the energy renormalization process.\\ 
The core process in the rescaling is the spin exchange happening during virtual charge processes at energy cut-off $D^{\prime\prime\prime}$. It  is determined by the coupling $J$ between singlet state on the impurity - that is effectively a single level occupied by a spin-$1/2$ -  with conduction band electrons. The exchange is perturbatively calculated and at second-order perturbation in the hybridization term $V$, the exchange coupling is given by 
\begin{equation}
J ~ \propto ~ \frac{1}{ E_{\ket{0}}- E_{\ket{\sigma}} } + \frac{1}{ E_{\ket{\updownarrows}} - E_{\ket{\sigma}}}  ~,
\end{equation}
where the expression gives emphasis to the energy gap separating each spin state. According to the energy regime in $D^{\prime\prime\prime}$ cut-off, the exchange coupling $J$ results as \textit{antiferromagnetic} (AFM).\\
This renormalised Hamiltonian defined at energy cut-off $D^{\prime\prime\prime}$ i.e. the Kondo model comprises antiferromagnetic coupling between the Fermi sea electrons in the bath - itinerant, noninteracting electrons - with the single localised ground-state singlet of the impurity. We conclude such  interactions are correctly modelled by the so-called \textbf{\textit{sd-}model} \cite{ZenerI_1951,ZenerII_1951}. The terminology refers originally the localised \textit{d-} or \textit{f-} atomic levels of magnetic impurities that couple with the free, itinerant electrons \textit{s-} or \textit{p-} atomic levels of the non-magnetic metallic host where the impurities are inserted in. \\
We have all the elements to define the model Hamiltonian of the renormalised system at energy regime lower than $D^{\prime\prime\prime}$. First, we have AFM exchange coupling between the impurity singlet and the Fermi electron sea
\begin{equation}
\hat{H}_{sd} = J \hat{\mathbf{s}}_{cond}(0) \cdot \hat{\mathbf{S}}_{d}
= \sum_{\mathbf{k}\mathbf{k}^{\prime}} \sum_{\sigma\sigma^{\prime}}  J_{\mathbf{k}\mathbf{k}^{\prime}}\Big( c^{\dagger}_{\mathbf{k}\sigma}(0)  \boldsymbol{\sigma}_{\sigma\sigma^{\prime}} c_{\mathbf{k}^{\prime}\sigma^{\prime}}(0) \Big) \cdot \hat{\mathbf{S}}_{d} ~,
\end{equation}
where $\hat{\mathbf{s}}_{cond}(0)$ is the spin operator of the conduction electrons in the bath calculated at the impurity site defined as at the origin and  $\boldsymbol{\sigma}_{\sigma\sigma^{\prime}}$ is the Pauli matrix vector, $\hat{\mathbf{S}}_{d}$ is the impurity spin operator at \textit{d-}level and $J$ coupling constant is the Heisenberg spin exchange interaction. Recalling the spin operator components $\hat{S}^{z}$, $\hat{S}^{\pm}= \hat{S}^{x} \pm i \hat{S}^{y}$ that in $\hat{H}_{sd}$ correspond to spin-flip events, the explicit form of the \textit{exchange coupling Hamiltonian} is given by
\begin{equation}
\begin{aligned}
\hat{H}_{sd} &= \sum_{\mathbf{k}\mathbf{k}^{\prime}} \frac{J_{\mathbf{k}\mathbf{k}^{\prime}}}{2} \Big( c^{\dagger}_{\mathbf{k}\downarrow}c_{\mathbf{k}^{\prime}\uparrow}\hat{S}_{d}^{+} +c^{\dagger}_{\mathbf{k}\uparrow}c_{\mathbf{k}^{\prime}\downarrow}\hat{S}_{d}^{-} + (c^{\dagger}_{\mathbf{k}\uparrow}c_{\mathbf{k}^{\prime}\uparrow} - c^{\dagger}_{\mathbf{k}\downarrow}c_{\mathbf{k}^{\prime}\downarrow})\hat{S}_{d}^{z} \Big) \quad \text{isotropic} \\
& = \frac{1}{2}\sum_{\mathbf{k}\mathbf{k}^{\prime}}  \Big( J^{\bot}_{\mathbf{k}\mathbf{k}^{\prime}} \big( c^{\dagger}_{\mathbf{k}\downarrow}c_{\mathbf{k}^{\prime}\uparrow}\hat{S}_{d}^{+} +c^{\dagger}_{\mathbf{k}\uparrow}c_{\mathbf{k}^{\prime}\downarrow}\hat{S}_{d}^{-} \big) +
J^{z}_{\mathbf{k}\mathbf{k}^{\prime}}(c^{\dagger}_{\mathbf{k}\uparrow}c_{\mathbf{k}^{\prime}\uparrow} - c^{\dagger}_{\mathbf{k}\downarrow}c_{\mathbf{k}^{\prime}\downarrow})\hat{S}_{d}^{z}   \Big) \quad \text{anisotropic} ~.
\end{aligned}
\end{equation}
Secondly, in the \textit{sd-}model we have the \textit{potential scattering Hamiltonian} 
\begin{equation}
\hat{H}_{W} = \sum_{\mathbf{k}\mathbf{k}^{\prime}\sigma} W_{\mathbf{k}\mathbf{k}^{\prime}\sigma} c^{\dagger}_{\mathbf{k}\sigma}c_{\mathbf{k}^{\prime}\sigma} ~.
\end{equation}
The $\hat{H}_{W}$ term is spin conserving, hence it does not contribute to impurity singlet spin  state and it is usually neglected since it does not modify the general local moment state. However, it is an indicator of the particle-hole asymmetry regime in the model. The reasoning behind the scattering potential will become clear in the next section where we illustrate the relation  between the Anderson model and the \textit{sd-}model. \\
In conclusion, the \textbf{isotropic Kondo model} \cite{Kondo1964} is defined by 
\begin{equation}\label{eq:K}
\hat{H}^{K} =  \underbrace{\sum_{\mathbf{k}\sigma} \epsilon_{\mathbf{k}} c^{\dagger}_{\mathbf{k}\sigma}c_{\mathbf{k}\sigma}}_{\hat{H}_{bath}} +
 \underbrace{\sum_{\mathbf{k}\mathbf{k}^{\prime}\sigma} W_{\mathbf{k}\mathbf{k}^{\prime}\sigma} c^{\dagger}_{\mathbf{k}\sigma}c_{\mathbf{k}^{\prime}\sigma}}_{\hat{H}_{W}} +
\underbrace{J \hat{\mathbf{s}}_{cond}(0) \cdot \hat{\mathbf{S}}_{d}}_{\hat{H}_{sd}} ~.
\end{equation}
The Hamiltonian in Eq.\ref{eq:K} is the renormalised Hamiltonian at lower-energy scale of the infinite-U Anderson model. By means of the energy scaling approach, rather than understanding the Kondo model as alternative impurity model from the other previously introduced i.e. AM and infinite-U AM, we identify $\hat{H}^{K}$ as the lower-energy scale version of those impurity models that are simply occurring at higher-energy scales. All those energy rescaled impurity models correspond to the very same original microscopic model. At each step in the renormalisation procedure, we learn a different physics aspect of the impurity system. However, when we reach the last step in the RG procedure by deriving the Kondo model, we understand that the impurity system can be modelled, in its essence, by an exchange coupling term $\hat{H}_{sd}$ between impurity and bath, a potential scattering term $\hat{H}_{W}$ and the bath itself. The RG machinery is applicable in the same format at any energy scale: the RG allows us to access the system physics at very different energies and to understand the minimal parameters governing its low-energy dynamic.\\

\noindent{Considering} that every rescaled Hamiltonian is born from the previous higher-energy model, we may ask if there is any relation between these rescaled models. The affirmative answer to this question is the subject of the next section.  

\subsection{The Schrieffer-Wolff transformation: from Anderson to Kondo impurity model}
We previously presented the Kondo impurity model as energy rescaled version of original microscopic one. We aim in the present section to relate $\hat{H}^{K}$ to its counterpart  Anderson model $\hat{H}^{AM}$. In order to derive this, we select a specific parameter range in phase diagram of the AM and we avail of perturbation theory to calculate its parameters.\\
We consider the AM in the local moment regime where its Coulomb potential is repulsive i.e. $U>0$. The favoured energetic configuration places the singly occupied state as ground-state and so the Fermi level lies between the levels $\epsilon_{d} <\epsilon_{F}< \epsilon_{d}+ U$. Under this energy level configuration, the lower-energy version of the AM Hamiltonian is equivalent to the Kondo model and this relation is calculated using the \textbf{Schrieffer-Wolff transformation} (SW) \cite{SchriefferWolff1966,SchriefferCoqblin1969}. This transformation corresponds to a one-step in the energy scaling procedure where the valence fluctuations to empty and doubly occupied states are eliminated from the original Anderson impurity model. At lower-energy shell, the resulting model is effectively a single level occupied at most by a spin-$1/2$. \\
Operatively, the SW transformation is a generalization of \textit{degenerate perturbation calculation} at second order in the hybridization term \cite{Lowdin_DPT_1951}. Hence, we derive an effective model by projecting the AM Hamiltonian onto the singly occupied states. The fundamental ingredient of the SW transformation is the \textit{projector operator} $\hat{P}_{i}$ for $i$ state. They are defined from the isolated impurity wavefunction $\ket{\Psi}$ for $i=0,1,2$ electrons components on the impurity as follows:
\begin{equation}
\begin{aligned}
&\hat{P}_{0}= (1- \hat{n}_{d\uparrow})(1- \hat{n}_{d\downarrow}) = \ket{0}\bra{0}~, \\
&\hat{P}_{1}= \hat{n}_{d\uparrow} +\hat{n}_{d\downarrow} - 2\hat{n}_{d\uparrow}\hat{n}_{d\downarrow}= \ket{\uparrow}\bra{\uparrow}+ \ket{\downarrow}\bra{\downarrow} ~,\\
&\hat{P}_{2}= \hat{n}_{d\uparrow}\hat{n}_{d\downarrow} =\ket{\uparrow\downarrow}\bra{\uparrow\downarrow}~,
\end{aligned}
\end{equation}
respectively on the empty, singly and doubly occupied states. Each electronic wavefunction per $i$ state is then found from  $\hat{P}_{i}\ket{\Psi}=\ket{\Psi_{i}}$. We act the projectors $\hat{P}_{i}$ on the Schr{\"o}dinger equation 
\begin{equation}
\sum_{j}\hat{H}_{ij}\ket{\Psi_{i}} =E\ket{\Psi_{i}} ~,
\end{equation}
where the equation is evaluated for all $i,j=0,1,2$ electron impurity states. In particular, $\hat{H}_{ij}= \hat{P}_{i}\hat{H}\hat{P}_{j}$ and its components are Hermitian satisfying $\hat{H}_{ji}= \hat{H}_{ij}^{\dagger}$. Furthermore, in absence of magnetic field as it is calculated in the literature, the eigenenergy $E_{i}\equiv E$. We note already that the final model will not contain the components $\hat{H}_{02}=\hat{H}_{20}$: there is no term in the AM Hamiltonian that adds or removes two electrons on the impurity simultaneously.\\
We now apply the projectors to the AM Hamiltonian to calculate the effective Hamiltonian $\hat{H}_{eff}$ corresponding to lower-energy model with singly occupied states only. This is obtained by eliminating $\ket{\Psi_{0}},\ket{\Psi_{2}}$ from the expression $\hat{P}_{1}  \hat{H} \hat{P}_{1} \ket{\Psi_{1}}$ to get: 
\begin{equation}
\underbrace{\big( \hat{H}_{11} + \hat{H}_{12}(E\mathbb{1}-\hat{H}_{22})^{-1}\hat{H}_{21} + \hat{H}_{10}(E\mathbb{1}-\hat{H}_{00})^{-1}\hat{H}_{01} \big )}_{\hat{P}_{1}  \hat{H} \hat{P}_{1}=\hat{H}_{eff}}\ket{\Psi_{1}} = E \ket{\Psi_{1}} ~.
\end{equation} 
The $\hat{H}_{eff}$ describes three single-electron exchange processes: $(i)$ events due to $\hat{H}_{ii}$ elements with conservation of electronic occupancy at the impurity site, $(ii)$  events due to $\hat{H}_{10}$ element with an initial electron hopping off the impurity and a final electron hopping on the impurity and $(iii)$ events due to $\hat{H}_{12}$ element with an initial electron hopping on the impurity and a final electron hopping off the impurity. We remark that $(ii),(iii)$ events happen through virtual intermediate processes with conservation of the electronic occupation at the impurity site. These are the only remaining processes surviving at lower-energy scales: the model is effectively describing only the singly occupied states. \\
Having an expression $\hat{H}_{eff}$, we need to identify its term correctly. By applying the proper anti-commutation relations and by recognising the spin operators, we determine a term that is equivalent to the exchange coupling in the \textit{sd-}model
\begin{equation}
J_{eff} = \sum_{\mathbf{k}\mathbf{k}^{\prime}} J_{\mathbf{k}\mathbf{k}^{\prime}} = 2\sum_{\mathbf{k}\mathbf{k}^{\prime}} V_{\mathbf{k}}V^{\star}_{\mathbf{k}^{\prime}} \bigg( \frac{1}{\epsilon_{\mathbf{k}} -\epsilon_{d}} + \frac{1}{U + \epsilon_{d} -\epsilon_{\mathbf{k}^{\prime}}} \bigg) ~,
\end{equation} 
such that $J=8V^2/U$ at particle-hole symmetry regime with $\epsilon_{d}=-U/2$. Next, the term that corresponds to the potential scattering in the \textit{sd-}model 
\begin{equation}
W_{eff} = \sum_{\mathbf{k}\mathbf{k}^{\prime}} W_{\mathbf{k}\mathbf{k}^{\prime}} =\sum_{\mathbf{k}\mathbf{k}^{\prime}} \frac{V_{\mathbf{k}}V^{\star}_{\mathbf{k}^{\prime}}}{2} \bigg( \frac{1}{\epsilon_{\mathbf{k}} -\epsilon_{d}} - \frac{1}{U + \epsilon_{d} -\epsilon_{\mathbf{k}^{\prime}}} \bigg) ~.
\end{equation}
The expressions $J_{eff},W_{eff}$ are valid at $|\epsilon_{\mathbf{k}}| \ll |\epsilon_{d} - \epsilon_{F}|$ and $|\epsilon_{\mathbf{k}}| \ll |U +\epsilon_{d} - \epsilon_{F}|$. We note that these energy level restrictions are indeed the same we already saw in the energy rescaling of the RG procedure but now in the context of second order perturbation result. Furthermore, within the energy level configuration adopted, we quantitatively verify that the exchange coupling is antiferromagnetic i.e. $J_{eff}>0$. Regarding the potential scattering, $W_{eff}$ vanishes at particle-hole symmetry i.e. $\epsilon_{d}= U/2$ and so a nonzero $W_{eff}$ potential indicates how close to the particle-hole symmetric regime the energy spectrum of the impurity model. However, since the Hamiltonian $\hat{H}_{W}$ is spin conserving, hence no spin-flip of the impurity spin is due to $W_{eff}$ and for the scope of studying the impurity spin dynamics we often neglect the potentials scattering term. \\ 
We conclude the SW transformation recovers the Kondo impurity model in Eq.\ref{eq:K} up to $\mathcal{O}(V^{2}/U)^{2}$: the effective model perturbatively derived is constrained to $n_{d}=1$ subspace only.  By means of the perturbation result, the Kondo model is the effective lower-energy model Hamiltonian of the Anderson model. The equivalence at low-energy shell between the two models is a powerful relation we will make use of often in the thesis.\\

\noindent{With} this section we conclude our discussion of the impurity models. Each of them covers a different energy range of phase diagram; consequently, different physics aspect characterises each energy shell for the system with an impurity coupled to an electronic bath. By progressively reducing the energy cut-off, the RG method discloses the various models at different energy scales. Each model represents the lower-energy version of the previous one up to the original real microscopic model. \\
In the next section, to complete the basics for this thesis we present in a consistent form the energy renormalization procedure required to fully study the physics underlined in the Kondo problem, that is the physics happening at energy scales smaller than the Kondo temperature. Although the Schrieffer-Wolff transformation allows the exact mapping between $\hat{H}^{AM}$ and $\hat{H}^{K}$, the low temperature physics of the Kondo model is still uncovered. Different approaches are required as the renormalization group theory, as we discuss in the next section. 

\section{The renormalization concept}\label{sec:RGtheo}
\begin{figure}[h]
	\centering
	\includegraphics[width=0.75\linewidth]{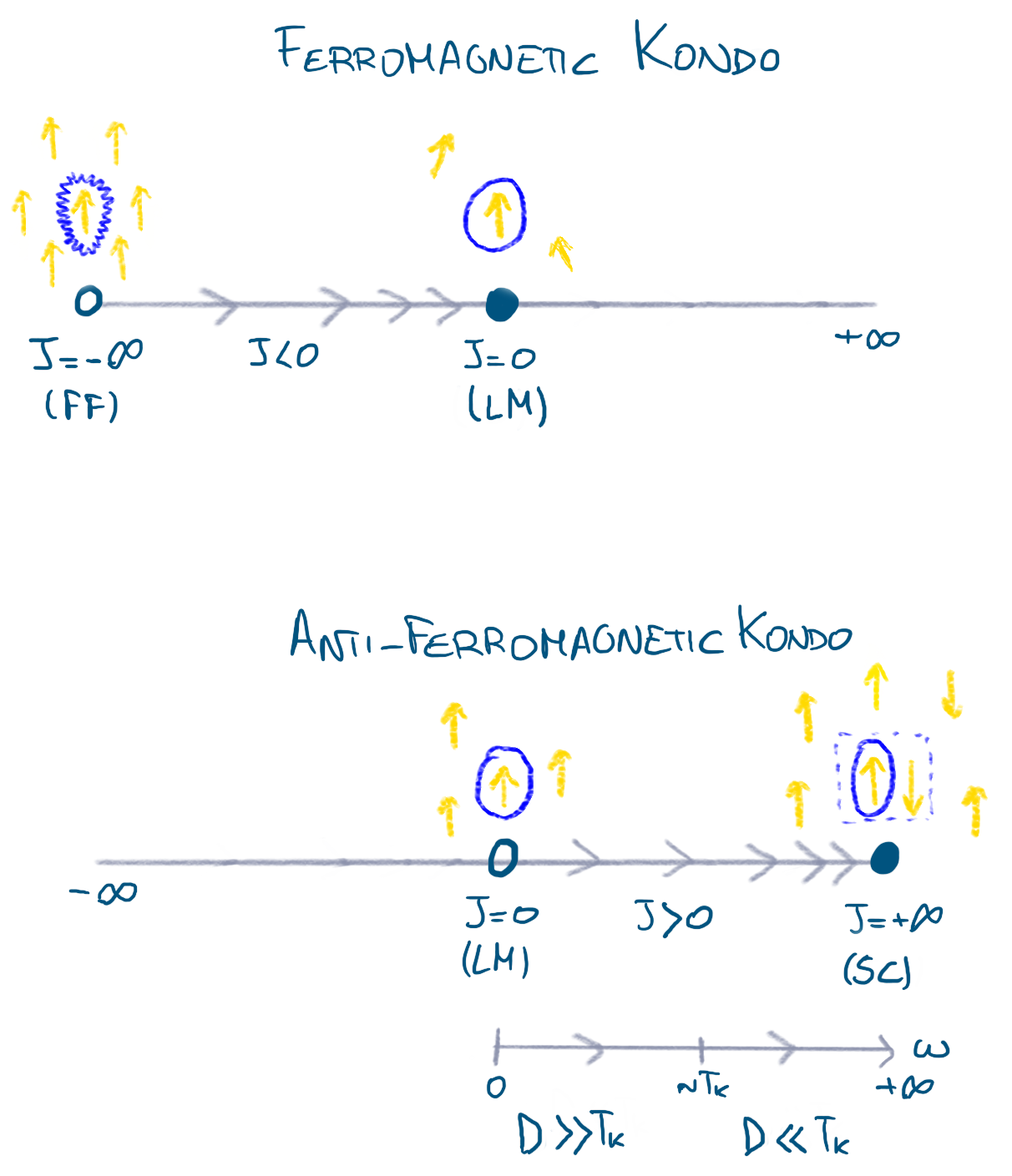}
	\caption[Schematic of the RG-flow in the one-channel Kondo model]{Schematic of the RG-flow in the one-channel Kondo model \cite{Coleman}. The fixed-points are classified as attractive (full circle) or repulsive (empty circle). In the ferromagnetic coupling regime $J<0$ (above) the coupling constant runs from ferromagnetically frozen (FF) to free spin or local moment (LM) fixed-points. The antiferromagnetic coupling regime $J>0$ (below) from local moment (LM) to strong-coupling (SC) fixed-points, here the corresponding Kondo scale is also given.}\label{F2:RG1CK}
\end{figure}

In the previous section on impurity models in Sec.\ref{sec:QImpModel} we have introduced some energy renormalization concepts when they were necessary. This section aims to collect this knowledge into a more comprehensive framework such that the Kondo problem is completely explained. \\
As we saw already, theoretical studies on the formation of magnetic local moments from impurities located in non-magnetic host metals were initially oriented in understanding the unexpected minimum in the $R(T)$ curve resistivity versus temperature, see sketch of the measured profile in Fig.\ref{F2:ResistanceOrig}. After some decades from the original experimental measurements, the explanation to the resistivity minimum was found by Kondo \cite{Kondo1964}. He performed a perturbative expansion of the resistivity in the coupling $J$ between impurity and conduction electrons: the first order correction is proportional to $-J\rho\log(k_{B}T/D)$, where $D$ is the conduction bandwidth. Previously we argued that, as result from the perturbation expansion, this correction should stay small by lowering the temperature. Although this was not experimentally measured, the resistivity equation as derived by Kondo diverges at $T\rightarrow0$ indicating the breakdown of the perturbation theory approach. This last statement is an oversimplification: we should actually take into account also the exchange coupling sign and we should distinguish $R(T\rightarrow0)$ accordingly \cite{Abrikosov1965}. For ferromagnetic (FM) coupling $J<0$, the series of the correction contribution to the resistivity at $T\rightarrow 0$ leads to a convergent series, hence no infra-red divergences in the resistivity equation i.e. $R(T \approx 0)\rightarrow 0$ as we have from the log term in the resistivity expression. On the contrary, in case of antiferromagnetic (AFM) coupling $J>0$, the resistivity correction leads to a divergent series such that the equation $R(T\approx  0)\rightarrow \infty$. As we saw already, the energy scale breaking down the convergence is at Kondo temperature $T_{K}$ in Eq.\ref{eq:Tk}.\\
From this analysis we learn that a perturbative treatment of the Kondo problem is insufficient to study the impurity model at $T\leq T_{K}$. An early version of the concept of energy scale renormalization is developed by P.W. Anderson and G. Yuval  in terms of \textit{running coupling constant} \cite{AndersonYuval_RunningCoupling1970}. This terminology means that the couplings depend on the energy cut-off $J(D)$. We say that the couplings continuously \textit{flow} as function of the cut-off by progressively decreasing the temperature. During the energy rescaling, the system can undergo two different events according to where the energy cut-off is placed compared to the relevant energy scale in the model. One case is the \textit{crossover}: it occurs when the cut-off exceeds the characteristic high-energy scale of the system, such that any lower-energy state exists only as virtual process. When interactions mediated by these virtual fluctuations are incorporated into the model Hamiltonian, the system intrinsically changes its form. The other case is the \textit{fixed-point} (FP): it occurs when the cut-off falls below the characteristic lowest-energy scale of the system, such that any further energy bandwidth reduction does not modify the model Hamiltonian and the system remains invariant under further scaling steps. To sum up: by reducing the energy cut-off, we can think of the couplings as progressively flowing through different crossovers until the fixed-point is reached. The corresponding fixed-point Hamiltonian encloses the core of the low-energy physics.\\
Formally, this initial idea of renormalization by Anderson and Yuval is operatively constructed in the so-called \textbf{poor's man scaling} (PMS) \cite{AndersonPoor1970}. In its essence, an effective Hamiltonian living in lower energy \textit{shell} is calculated by systematically eliminating the outer energy shells where excited states are found. This energy scaling approach consists of two steps rather than three as it is performed in general renormalization schemes - as we discuss at the end of this section and we continue with presenting the PMS approach. The first step consists of scaling the original energy cut-off as $D \rightarrow D/b$, provided $b>1$ to determine the energy shell width $D(1-1/b)$ to be eliminated. This energy rescaling allows to divide the conduction electron band energy into two regions: the external energy shells at $ D/b < |\epsilon_{\mathbf{k}}| \leq D$ and the internal one at  $|\epsilon_{\mathbf{k}}| \leq D/b$. The former ones are those we systematically eliminate and the latter one is the state we want to keep. We proceed with the second step in eliminating the outermost shells using the $\mathrm{T}$-matrix formulation in Eq.\ref{eq:TmatrixEq} to describe the induced interactions due to the formation of virtual bound states when atomic orbital of the impurity hybridises with the conduction band state \cite{AndersonPoor1970}.\\ 
The two-step scaling approach in the PMS results into the following \textit{renormalised effective Kondo Hamiltonian} 
\begin{equation}
	\begin{aligned}
\hat{H}^{K}(D) = &\sum_{|\epsilon_{\mathbf{k}}|<D,\sigma} \epsilon_{\mathbf{k}} c^{\dagger}_{\mathbf{k}\sigma}c_{\mathbf{k}\sigma} + 
 \frac{J^{\bot}(D)}{2} \sum_{\mathbf{k}\mathbf{k}^{\prime}} \big( c^{\dagger}_{\mathbf{k}\downarrow}c_{\mathbf{k}^{\prime}\uparrow}\hat{S}_{d}^{+} +c^{\dagger}_{\mathbf{k}\uparrow}c_{\mathbf{k}^{\prime}\downarrow}\hat{S}_{d}^{-} \big) +
\frac{J^{z}(D)}{2} \sum_{\mathbf{k}\mathbf{k}^{\prime}} (c^{\dagger}_{\mathbf{k}\uparrow}c_{\mathbf{k}^{\prime}\uparrow} - c^{\dagger}_{\mathbf{k}\downarrow}c_{\mathbf{k}^{\prime}\downarrow})\hat{S}_{d}^{z} +\\ 
&+\sum_{\mathbf{k}\mathbf{k}^{\prime}\sigma} W_{\mathbf{k}\mathbf{k}^{\prime}\sigma}(D) c^{\dagger}_{\mathbf{k}\sigma}c_{\mathbf{k}^{\prime}\sigma} ~.
\end{aligned}
\end{equation}
This expression exactly recovers the original Kondo model when $D=D_{0}$ is the bare bandwidth, 
otherwise it has the very same form of $\hat{H}^{K}$ in Eq.\ref{eq:K} but now with rescaled couplings. These terms indicate the couplings under energy rescaling are redefined to incorporate both bare couplings and renormalised couplings emerging from the PMS. This is because the renormalization procedure finds an effective version of the couplings  but leaves invariant the bare couplings $J^{\bot},J^{z}$. By noticing the extra terms have the same form of the bare coupling $J$, we can directly absorb them by just redefining the couplings into $J^{\bot}(D), J^{z}(D)$. As we know from previous discussion, the exchange couplings include impurity spin-flip events and so their renormalised terms are kept in addition to the bare ones to redefine the coupling in $\hat{H}^{K}(D)$. On the contrary, the term derived from renormalising the potential scattering does \textit{not} scale on reducing $D$. Thus, it is generally omitted in the final $\hat{H}^{K}(D)$ equation: it does not depend on the impurity spin and so it simply contributes as an energy shift.\\ 
The PMS is an iterative method: the energy bandwidth is reduced until the couplings flow towards a FP of the theory. We then can write the couplings running along the renormalization flow in form of differential equations
\begin{equation}
\begin{aligned}
&\frac{d J^{z}}{d \log D} = -2 \rho {(J^{\bot})}^{2}, \quad\frac{d J^{\bot}}{d \log D} = -2 \rho J^{z} J^{\bot} \quad\text{anisotropic} ~,\\
&\frac{d J}{d \log D} = -2 \rho J^{2} \quad\text{isotropic} ~,
\end{aligned} 
\end{equation}
where these equations are calculated using the $\mathrm{T}$-matrix approach again. For the sake of conciseness, we now discuss the isotropic renormalization flow equation only - noticing that here we are discussing only the one-channel version of the Kondo model. The most important property of the one-channel physics is that its RG-flow can reach a Fermi liquid FP - on the contrary of the multichannel Kondo physics where also non-Fermi liquid FP may arise \cite{Affleck_2Ch1991}. In the next section we will discuss the case of two-channel Kondo physics, as it is the main configuration treated in this thesis. \\
We return now to present the RG-flow. We say that $J=0$ is a fixed-point in the system and its features will depend upon the coupling being larger or lesser than zero i.e. being antiferromagnetic or ferromagnetic.\\ 
At ferromagnetic coupling $J<0$, the scaling reduces the coupling strength such that by lowering the energies the coupling goes to zero logarithmically. The $J=0$ is an \textbf{attractive free spin FP} meaning that as $T\rightarrow 0$ the impurity local moment is fully decoupled from the conduction electrons band. At the lowest energy scale, the coupling decreases reaching $J=-\infty$ that corresponds to a \textbf{repulsive ferromagnetically frozen FP}. Generally speaking, the FM regime is the trivial case of the Kondo model.\\
The opposite regime is antiferromagnetic coupling $J>0$ where the scaling increases the coupling strength. This regime is indeed the nontrivial case of the Kondo model since the system experiences three different stages according to the energy scale. At energy cut-off much larger than the Kondo temperature $D\gg T_{K}$, the coupling $J\ll D$ is now near the \textbf{repulsive local moment FP}. Then, the coupling $J$ increases continuously while the cut-off progressively reduces up to the energy scale coinciding with $T_{K}$ where crossover to an \textbf{attractive strong-coupling FP} occurs. At this energy scale $D \approx T_{K}$, the impurity spin starts being screened by the surrounding conduction electrons. Around the scale dictated by the Kondo temperature, the system is a proper many-body object. However, by lowering even further the energy scale at $D \ll T_{K}$, the local moment is completely screened and the impurity spin is fully locked in the cloud of conduction electrons. In the limit $D\rightarrow 0$, we observe the formation of a nonmagnetic spin-singlet impurity ground-state. Under this limit, the system forms a \textbf{Fermi liquid} - we comment more on this aspect below - and the strong-coupling fixed-point is identified by an infinite coupling i.e. $J(D\rightarrow 0) = \infty$.\\
\noindent{By} means of the energy rescaling, we have disclosed the highly nontrivial physics of AFM Kondo model and we can now connect these results to a comprehensive understanding of the resistivity curve. As we just saw, the Kondo model RG-flow comprises of two FPs, the local moment and the strong-coupling at $J=0,~J=\infty$ respectively. At  high-energy scale much larger than $T_{K}$, the coupling among impurity-conduction electrons is almost negligible. That is indeed defining the local moment phase. According to the RG-flow of couplings, from this high-energy scale the system flows continuously away. The fact that the strong-coupling phase emerges only at low energy and temperature scales as opposed to the local moment phase occurring at higher energy scale resembles the \textit{asymptotic freedom} states typical of particle high-energy physics \cite{tHooft_AsymptoticFree1999}. The consequence of the temperature lowering is a decrement in the resistivity curve. At $T=T_{K}$, the system switches to strong-coupling phase with a progressive increment of the coupling flowing to the new regime. This is the reason for identifying $T_{K}$ the crossover energy scale between phases. In this energy scale, two crucial aspects take place. The first, the conduction electrons wavefunction is fully localised at the impurity spin region. This indicates the start of the initial stage of the impurity screening process. The second, the switching to strong-coupling phase introduces a deep conductance enhancement that is accompanied by very low resistivity. This is the reason why the resistance curve stops decreasing and it registers a minimum at $T=T_{K}$. By lowering even further the energy scale, the additional increment of the coupling strength corresponds to an even stronger impurity screening. As consequence, the conductance begins to reduce its value and the resistance grows again despite of the lower energy regime. However, in the limit $D \ll T_{K} \rightarrow 0$, we reach the lowest-energy scale of the model. The singlet formation is completed and the impurity spin degree is frozen - as opposed to its freedom at high-energy scale. The system couplings flow to the final strong-coupling FP being $J=\infty$. At this energy scale, excitations are properly described by Fermi liquid theory. The surrounding conduction electrons become inert to the newly formed nonmagnetic impurity state. The final impurity screened state at the lowest energy scales is important to ensure the resistance curve saturation. Otherwise, the curve would indefinitely diverge due to the continuous uncompleted singlet state formation at each rescaled energy. A schematic of the RG-flow of the Kondo model in both coupling regimes is given in Fig.\ref{F2:RG1CK}.\\

\noindent{We} return now to focus on the scale $D \ll T_{K}$. This regime is characterised by an attractive strong-coupling FP where the running coupling flows towards it. Ultimately, the coupling growth takes infinite value. It has to be remarked that this result $J=\infty$ derived in the PMS method is clearly an artefact of the perturbative calculation. We already argued this method breaks down at any energy scales comparable to the Kondo temperature. We conclude that the PMS calculation obtains correct results that have been lately verified by exact methods \cite{Krishna1980renormalization,Tarasevych_FRGexact_2018}. For instance, the PMS properly predicts the infinite growth of $J$ and the scaling invariance of $T_{K}$. However, the PMS is not able to explain fully the low-energy physics of magnetic impurities. Moreover, it does not have access to \textit{unstable fixed-point}. In the Kondo model RG-flow these are located at $J=-\infty$ in the FM coupling regime and at $J=0$ in the AFM coupling regime. The $D \ll T_{K}$ demands alternative nonperturbative approaches which were indeed discovered after some years.\\ 
The theoretical advancements have been made with exact analytical solutions of the Kondo model derived by means of the \textit{Bethe ansatz} developed independently by N. Andrei \cite{Andrei_BetheAnsatz1980} and P. Wiegmann \cite{Wiegmann_BetheAnsatz1981}; by means of \textit{Fermi liquid theory} by P. Nozi\`{e}res \cite{Nozieres1974}. Within the scope of our discussion, the work by Nozi\`{e}res is particularly relevant. He showed that the ground-state of the Kondo model at energy scale smaller than the Kondo temperature is actually a spin singlet. That is formed by two possible configuration of the impurity spin-$1/2$ coupled to a conduction electron such that the resulting singlet state reads: $1/\sqrt{2} (\ket{\uparrow_{imp} \downarrow_{cond}} - \ket{\downarrow_{imp} \uparrow_{cond}})$. We remark that $\ket{\uparrow_{cond}},\ket{\downarrow_{cond}}$ are complex many-body objects. Moreover, Nozi\`{e}res showed that the low-energy excitations in the Kondo model can be understood in the framework of Fermi liquid theory because the magnetic susceptibility and the heat capacitance of the nonmagnetic singlet at $D\rightarrow 0$ are the same of a Fermi liquid.\\
Next to these seminal analytical results, a numerical algorithm to exactly solve the Kondo problem has been showed by K. Wilson - as we will return below to comment more on this \textit{numerical energy scaling technique}. Crucially, these simulations numerically confirmed that the exchange coupling takes infinite value at the strong-coupling FP \cite{Krishna1980renormalization}. Moreover, the numerics could unambiguously calculate the Kondo ground-state as spin singlet - hence, giving numerical evidence of the previous analytical previsions.\\ 
As conclusion to this discussion, we comment in general terms the \textbf{renormalization group method} (RG) \cite{Wilson_RGEpsilonExp1974}. Its name requires some comments: the energy renormalization procedure as we presented so far does not have any inverse operation, so it would be more correct to refer to it as semi-group method. However, for historical reasons, we will adopt the common nomenclature. \\
As we already know, the Poor's man scaling approach presented before is based on a two-step scaling, although generic renormalization approaches require a three-step procedure. The first two are identical to the PMS approach: rescaling the energy cut-off and eliminating the higher-energy momentum shells. Often, this second step is calculated by considering that each degree of freedom is a variable that can be integrated out in path-integral formalism - and so it is standard to phrase this step as integrating-out higher-energy degree of freedom \cite{Wilson_RGEpsilonExp1974}. The last step requires to rescale back the conduction bandwidth to its original values after each renormalization cycle.\\
We can quantify the previous observation that the Schrieffer-Wolff transformation is analogous of the step-two in RG method, by eliminating the excited empty- and doubly- occupied energy states. The Poor's man scaling captures the essential feature of the RG operation: the scaling of the energy band and the elimination of excited states. Although the RG method is born within a purely many-body low-energy scale problem - we also recall the early work by L. P. Kadanoff on renormalization in real-space \cite{Kadanoff2000} - its applicability goes beyond the condensed matter field and extends to general problems in physics.

\subsection{The numerical renormalization group technique}
\begin{figure}[h!]
	\centering
	\includegraphics[width=0.75\linewidth]{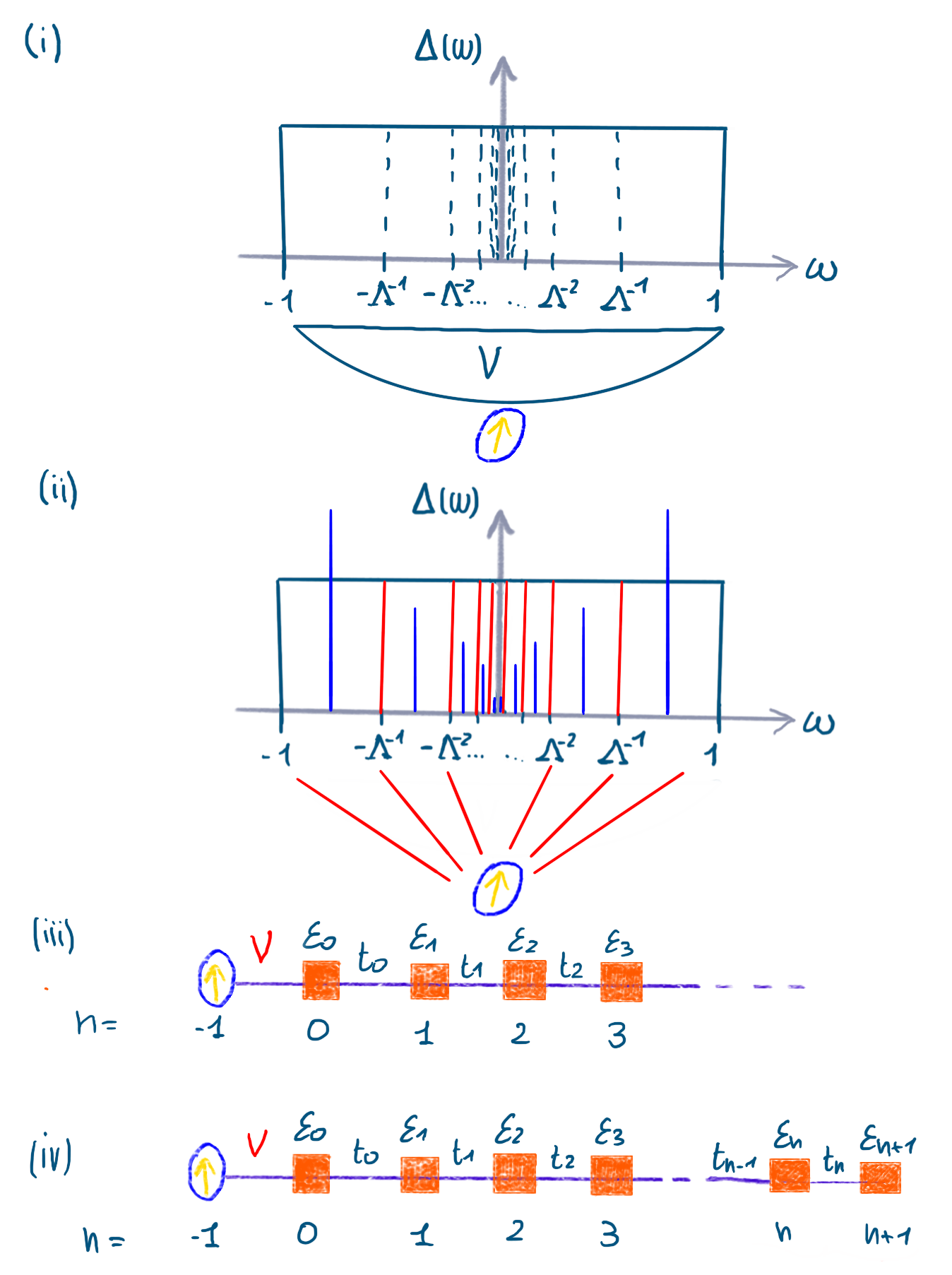}
	\caption[Schematic of the NRG protocol]{Schematic of the NRG protocol \cite{Bulla2008}: $(i)$ the conduction band is logarithmically discretized through the parameter $\Lambda$, $(ii)$ the continuum energy spectrum in each interval is approximated to a single state, $(iii)$ the discretized model is mapped to the Wilson chain representation, where the impurity coupled to the first bath site is the only physical part of the chain, $(iv)$ adding iteratively a site to construct a finite chain length such that after truncation the renormalized Hamiltonian is found and $(v)$ calculate dynamical and thermodynamic quantities (which is not illustrated in the schematic).}\label{F2:NRG}
\end{figure}

As mentioned before, the Kondo model at energy scale smaller than $T_{K}$ requires a nonperturbative approach. In this chapter we introduced only analytical solutions so far. However, a numerical version of the energy scaling has been invented by Wilson, the so-called \textbf{numerical renormalization group technique} (NRG) \cite{Wilson1975,Bulla2008}. This approach is able to find nonperturbative solutions to the Kondo problem. The exact solution is recovered in NRG upon restoring the continuum limit of the bath.\\
The NRG method is concisely detailed by a five-step protocol adapted to general impurity models. Here below we present this protocol, highlighting its more physical aspects and we leave its technicalities to the numerical paper by Wilson cited above. A schematic of this protocol is drawn in Fig.\ref{F2:NRG}.\\
\!\!Following standard literature \cite{Bulla2008}, we consider as initial model the familiar single-impurity Anderson Hamiltonian in Eq.\ref{eq:AM} - whose conduction band forms a continuum energy spectrum.\\
The first step consists of \textit{logarithmically discretizing} the conduction band by introducing the parameter $\Lambda$. The band is split in an infinite set of finite intervals at $x_{n} = \pm \Lambda^{-n},~n=0,1,2,\dots$, provided $\Lambda>1$. Within each interval $d_{n}=x_{n+1}-x_{n}$, the band shows a continuous energy spectrum which is directly coupled to the impurity.
Within each energy interval $d_{n}$, we can find a discrete basis set of orthonormal wave functions. This basis ascribes completely the bath modes in $d_{n}$ and is known as \textit{energy shell}. The continuum limit is restored by taking the limit $\Lambda \to 1$. \\
In the second step, we retain only the \textit{zero-mode} Fourier basis function from every energy shell. In the limiting case of the infinitely wide conduction band, see Eq.\ref{eq:DoSinfWBL}, the zero-mode is the only non-vanishing one. By means of this simplification, the conduction band is transformed into an infinite set of discrete states, each corresponding to a different energy shell.\\
In the third step, the discretized Hamiltonian is mapped to a semi-infinite, one dimensional chain modelled as tight-binding model. This is the so-called \textit{Wilson chain}  representation. The isolated impurity is coupled through the original hybridization $V$ to the first site of the chain - sometimes also referred to as the $0^{th}$ Wilson chain site. It is important to stress that this chain site is the only physical one. The rest of the bath chain $n=1,2,3\dots$ is attached to the $0^{th}$-site. The chain is characterized by a recursive relation for the single-particle energy $\epsilon_{n}$ and hopping $t_{n}$ as it calculated in Eq.31 in reference \cite{Bulla2008}. We remark that this relation always leads to  exponentially decaying $t_{n}\sim \Lambda^{-n/2}$ for a metallic density of states.  This specific feature of the Wilson chain  implies that sites further away from the impurity contribute lesser and lesser to the overall system. Hence, truncation of the chain at finite length $N$ is an excellent approximation of whole infinite chain - provided that the number of kept sites is large enough. More discussion on the choice of discretizing the conduction band in logarithmic intervals and the reliability of the NRG method applied to the Anderson model can be found in reference \cite{KristianMurthyWWI1980}.\\
The fourth step comprises the construction of a finite chain length starting from the isolated impurity $\hat{H}_{-1}\equiv \hat{H}_{atomic}$. Then, the chain length is increased by adding one site after the other to $\hat{H}_{-1}$. Upon consequent energy rescaling, we obtain the iterative construction prescription:
\begin{equation}\label{eq:2H_WC}
\begin{aligned}
&\hat{H}_{0} = \Lambda^{-1/2} \left( \hat{H}_{atomic} + \sum_{\sigma} \epsilon_{0} c^{\dagger}_{0\sigma} c_{0\sigma}+ V  \sum_{\sigma}(d^{\dagger}_{\sigma}c_{0\sigma} + c^{\dagger}_{0\sigma}d_{\sigma}) \right) ~, \\
&\hat{H}_{n+1} = \sqrt{\Lambda}\hat{H}_{n} + \Lambda^{n/2}  \sum_{\sigma}
\left(  \epsilon_{n+1} c^{\dagger}_{n+1\sigma} c_{n+1\sigma} +t_{n} (c^{\dagger}_{n\sigma} c_{n+1\sigma} + c^{\dagger}_{n+1\sigma} c_{n\sigma})\right) ~.
\end{aligned}
\end{equation}
In order to make this procedure tractable for numerical implementation, a further simplification must be taken. Upon diagonalising $\hat{H}_{n}$, only a constant amount of the lowest lying energy levels are retained. This is the so-called \textit{truncation process}, it prevents the Hamiltonian from exponentially growing in dimension. This is justified because the influence of high-energy states on low-energy states can be related to the hopping between Wilson chain sites which amounts to a small perturbation of the order of $\mathcal{O}(1/\sqrt{\Lambda})$. Carrying out this iterative procedure yields the renormalised Hamiltonian $\hat{H}_{N}$ and its desired eigenspectrum and eigenstates. With the completion of this step, the full solution of the Kondo problem is obtained. This is a \textit{solution} in the sense that the time-independent Schr{\"o}dinger equation requires the diagonalization of the Hamiltonian - as it is achieved in the above discussed fourth step of the NRG protocol. In order to extract physically useful observables, it is necessary to include an additional step.\\
In the fifth and last step of the protocol, we are dealing with calculating the quantities of interest. Those could be dynamical quantities e.g. Green's functions or thermodynamic quantities e.g. impurity entropy. This can be achieved using the Lehmann representation of dynamical quantities - which is computationally very expensive \cite{Bulla2008}.\\
With this we have concluded the discussion on the NRG protocol. We also have gained a complete quantitative understanding of the Kondo physics thanks to the numerical results obtained from the  NRG algorithm. In this thesis all the results are calculated with full-density-matrix (FDM) NRG using the complete Anders-Schiller basis (AS-basis) \cite{AndersSchiller2005,AndersSchiller2006,Weichselbaum2007}. As discussed, the NRG process requires the discarding of majority of states per iteration. As a result, the remaining states from the final iteration span only a fraction of the actual Fock space of the entire initial system. However it was shown that, by combining the higher-energy discarded states with the whole set of states from the last iteration, a complete basis of the Fock space is obtained. This is what is known as AS-basis. By means of this set of states, a more accurate computation of dynamical quantities can be performed. For instance, it was shown that using this strategy, the spectral function correctly normalise to the unit.\\

\section{The two-Channel Kondo model}\label{sec:2CK}
\begin{figure}[H]
	\centering
	\includegraphics[width=0.75\linewidth]{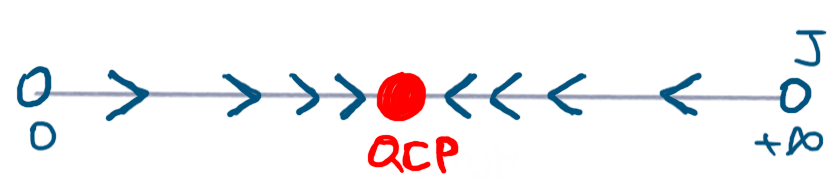}
	\caption[Schematic of renormalization flow of the overscreened Kondo effect]{Schematic of renormalization flow of the overscreened Kondo effect in the two-channel isotropic Kondo model. $(i)$ Symmetric channels $J_L =J_R \equiv J$: both $J=0,J=+\infty $ indicate unstable, repulsive fixed-points (empty circle) and those flow towards an intermediate new stable fixed-point (full red circle) with non-Fermi liquid feature, that is the so-called quantum critical fixed-point (QCP). $(ii)$ Asymmetric channels $J_L \neq J_R$: the RG-flow similarly as described in $(i)$, with the addition that all the FPs are now unstable. }\label{F2:QCP}
\end{figure}

\begin{figure}[t]
	\centering
	\includegraphics[width=0.75\linewidth]{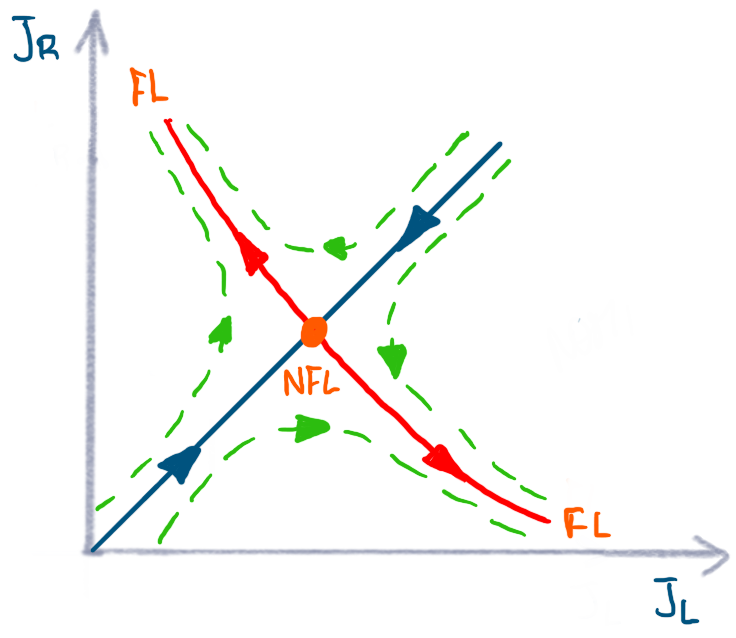}
	\caption[The RG-flow of overscreened 2C-Kondo model]{The RG-flow of overscreened 2C-Kondo model. Channel symmetric case (blue): couplings flow along bisector towards the quantum critical non-Fermi liquid, stable FP (NFL). In presence of small detuning: (red) at $T^{\star} \ll T_{K}$, couplings flow outwards from the unstable non-Fermi liquid to FL fixed-point and this is the Fermi liquid crossover line, also called separatrix; (green) channel asymmetric case at $T <T^{\star}$, couplings run initially towards unstable non-Fermi liquid FP (the smaller asymmetry, the smaller $T^{\star}$, the closer the curve is to red, blue paths) then they flow away from it eventually reaching the Fermi liquid region.}\label{F2:2CRG}
\end{figure}

The model Hamiltonian discussed in the previous two sections, namely on quantum impurity models in Sec.\ref{sec:QImpModel} and on renormalization theory in Sec.\ref{sec:RGtheo}, is given in Eq.\ref{eq:K} and it applies strictly to a \textit{single-channel} model with an impurity spin-$1/2$. In this section, we want to extend our understanding of the Kondo model. There are two possible variants \cite{Coleman}: $(i)$ the multi-channel Kondo model where $k$ different channels contribute to the impurity screening, $(ii)$ the higher impurity $S$-spin with $S>1/2$. The resulting $k$-channel $S$-spin Kondo model Hamiltonian reads
\begin{equation}
\hat{H}^{K}_{k,S} = \sum_{l=1}^{k}\left( \sum_{\mathbf{k}\sigma}
\epsilon_{\mathbf{k}} c^{\dagger}_{\mathbf{k}l\sigma} c_{\mathbf{k}l\sigma} + J_{l} \hat{\mathbf{s}}_{l}(0) \cdot \hat{\mathbf{S}} \right) ~,
\end{equation}
where $\hat{\mathbf{S}}$ is the impurity spin-$S$ operator and $\hat{\mathbf{s}}_{l}(0) =\sum_{\mathbf{k}\mathbf{k}^{\prime}} c^{\dagger}_{\mathbf{k}l\sigma}(0)\boldsymbol{\sigma}_{\sigma\sigma^{\prime}} c_{\mathbf{k}^{\prime}l\sigma^{\prime}}(0)$ is the spin operator of the conduction band for $l^{th}$-channel evaluated at the impurity site that is placed at the origin.\\
We consider the Kondo model in its antiferromagnetic coupling regime where we have a characteristic Kondo scale $T_{K}$. In the $\hat{H}^{K}_{k,S}$ system, the local moment FP is repulsive in analogy to the standard one-channel model. However, the strong-coupling FP behaviour highly depends on both number of channels and impurity spin size: in general terms, three distinct cases may occur. At $k=2S$ there is \textbf{perfect screening}: this case has the similar strong coupling physics of the standard one-channel Kondo model as previously discussed.\\ 
In case $J_{l}$ is the same for all $k$-channels, then there is only one Kondo scale. Whereas, if $J_{l}$ is different in every channel, we find as many Kondo scales as the number of channels that can quench the impurity spin. These multiple Kondo scales is due to the residual interaction of the impurity spin-$S$ with the conduction band in $l^{th}$-channel. This process implies that initially there is either excess or insufficient screening of the impurity degree of freedom depending on the number of available channels. Hence, one of the two possibilities is the \textbf{underscreening} case at $k<2S$: $J_{eff}<0$, hence its RG-flow runs as ferromagnetic coupling in the one-channel i.e. it flows to zero by reducing the energy scales since $J_{eff}=0$ is a stable, attractive FP. \\
\noindent{The} other case is the \textbf{overscreening} at $k>2S$: $J_{eff}>0$ implies that both bare and renormalized couplings are AFM and so they flow towards the strong-coupling FP. This makes both the local moment FP at $J_{eff}=0$ and the strong-coupling FP at $J_{eff}=+\infty$ unstable and repulsive fixed-points. It has been conjectured by Nozi\`{e}res \cite{Nozieres_KondoRealMetal1980}, lately verified also through other analytical methods, that there must be an \textit{intermediate} fixed-point $J^{\star}$ towards which $J_{eff}$ is flowing to by reducing the energy scales. This point is a new type of fixed-point dictating the low-energy physics of the overscreened Kondo model: it is \textit{not} of Fermi liquid type, as it can be deduced by the temperature dependence in the thermodynamic quantities. Furthermore, the presence of this intermediate point indicates that there is no signature of a Kondo scale namely a particular energy scale where universal $T/T_{K}$ behaviour can be defined. We assist at a scale-invariant condition where the correlation length, at which the impurity spin is affected by the conduction electron, diverges \cite{Affleck_2Ch1991}. This situation is indeed typical of a critical point. Hence, this new intermediate fixed-point is referred as \textbf{quantum critical non-Fermi liquid FP}, a sketch of this new RG-flow is given in Fig.\ref{F2:QCP}. As we discuss more below, in case of symmetric channels this critical point is the unique stable FP of the RG-flow; otherwise, it is always unstable \cite{Andrei_MultiChannelCritical1984}.\\
\noindent{In} this thesis we will primarily encounter the \textit{overscreened} condition, hence in the remaining of this section we discuss it the context of $k=2$ channel and impurity spin-$1/2$ system \cite{van2020electric}.\\
Initially, we consider a system in absence of external magnetic field with symmetric impurity-bath couplings $J_{L}=J_{R}$ and with independent baths i.e. no direct coupling between left, right bath $J_{LR}=J_{RL}=0$. We can now study the usual three regimes with respect to the Kondo temperature as we did in the one-channel case such that we highlight the non-trivial low-energy flow diagram \cite{Nozieres_KondoRealMetal1980} and its transport regimes \cite{Pustilnik_2CTransport2004} of the \textbf{symmetric two-channel Kondo model} (2C-Kondo). Considering that the couplings RG-flow end always in the non-Fermi liquid FP, see Fig.\ref{F2:QCP}, rather than characterising the system through Kondo temperature which is only defined for Fermi liquid systems, we use $T^{NFL}$ for the discussion below.\\
At $T \gg T^{NFL}$, the running coupling constant flows away from the local moment FP, resembling the one-channel system regime. Due to the weak coupling regime, there is zero charge transport between the two baths passing through the impurity. By lowering the temperature, the coupling strength increases progressively in the RG-flow sense. When temperatures approximate $T^{NFL}$, both baths attempt to screen the impurity. In the regime $T \ll T^{NFL}$, channels competition continues, although the channel symmetry makes none of them succeeding in a complete impurity screening to form the Kondo singlet. This condition  leads to the so-called \textit{spin frustration}. Although the summation of left-right and right-left currents across the system is conserved, considering the electronic transport measured between baths the conductance increases as consequence of their equally enhanced coupling with the impurity.\\ 
As we discussed before, this low-energy regime is characterised by a spin overscreened condition and the effective coupling flows inwards the intermediate quantum critical, Fermi liquid and stable FP. The signature of its non-Fermi liquid physics is found in the finite residual ground-state impurity entropy $S_{imp}(T \rightarrow 0)= 1/2 \log 2$ \cite{coleman1995simple2CK}: this is a further evidence of the absence of Kondo singlet formation in the overscreened state. This result is opposed to the one in models with asymmetric channel coupling: the low-energy system effectively recovers the one-channel model with Fermi liquid FP and the singlet formation implies a vanishing $S_{imp}(T\rightarrow 0)=0$. Other signatures of unconventional behaviour of the intermediate FP is the magnetic spin susceptibility $\chi_{imp}(T)$  proportional to $\log T$ \cite{Affleck_2Ch1991} and the fractional charges leading to non-integer entropy value - as opposed to the standard temperature scaling of $\chi_{imp}(T)$ \cite{coleman1995simple2CK} and to integer charge with vanishing entropy in SC phase characteristic of the Fermi liquid physics \cite{Affleck_CriticalOverscreened_1991}. The exotic features of the intermediate QCP can be properly associated with Majorana degree of freedom. The Majorana is a fermionic particle which is identical to its own anti-particle. Two Majorana fermions can be combined to form one Dirac fermion. This is reason why Majorana are sometimes referred as \textit{half} of Dirac fermion.\\
\noindent{As} next in our discussion, we note that the critical feature of the intermediate FP in the two-channel set-up makes the model very sensitive to small external perturbation. We refer to this configuration as general \textbf{asymmetric 2C-Kondo model} $J_{L}\neq J_{R}$ and generally finite magnetic field $B$, where the small detuning could be for instance the  application of a weak magnetic field or the creation of coupling asymmetry.\\ 
The role of the detuning is to introduce an additional energy scale  $T^{\star}$ lower than the Kondo scale such that the spin frustrated state survives only within the interval $T^{\star} < T < T_{K}$. At $T \ll T^{\star}$, the frustration is completely lifted: only one channel is effectively coupled to impurity spin and the system is effectively modelled as one-channel in low-energy scale. The completed singlet ground-state formation means the effective coupling flows towards the standard stable strong-coupling Fermi liquid FP. Having $J_{LR}=J_{RL}=0$, the system is always in a zero conductive state. In case we ease this condition, in this set-up, the lack of transport through the impurity can be understood in terms either of magnetic field freezing the spin state or of one bath decoupled due to the channel asymmetry.\\
We note that $T^{\star}$ identifies the \textit{Fermi liquid crossover} between non-Fermi liquid to Fermi liquid physics. The transition is possible through a finite detuning parameter that becomes non negligible only at $T<T^{\star}$. Below the crossover, functions are invariant once expressed in terms of $T/T^{\star}$, provided $T^{\star} \ll T_{K}$. This shows the universal behaviour of the  Fermi liquid crossover \cite{Fritz_Crossover2CK_2011,Mitchell_UniversalLowT2CK2012}. A general overview on the RG-flow of the overscreened Kondo effect in two-channel model is sketched in Fig.\ref{F2:2CRG}.\\
\noindent{For} an experimental point of view,  the two-channel model observation challenges the set-up design. It is required independent leads such that each corresponding quantum point contact is a distinct and tunable channel to connect leads to the nanostructure \cite{GoldhaberGordon-2CK-2007Exp} - as we discuss more in Sec.\ref{sec:Kexp}.\\ 
We comment that the two-channel Kondo physics - even three-channel Kondo - also realised in charge-Kondo system \cite{2CK-RenormalisationFlow_IftikharPierre2015Exp,iftikhar2018tunable} as we discuss further in Chapter \ref{ch:graphene}.\\
With this knowledge we have presented the essential theoretical and analytical background as foundation of this thesis. We are now in the position to dive into the core material: the quantum transport phenomena and its application to dedicated impurity models.

\chapter{Theory of mesoscopic quantum transport}
\begin{figure}[H]
\centering
\includegraphics[width=0.75\linewidth]{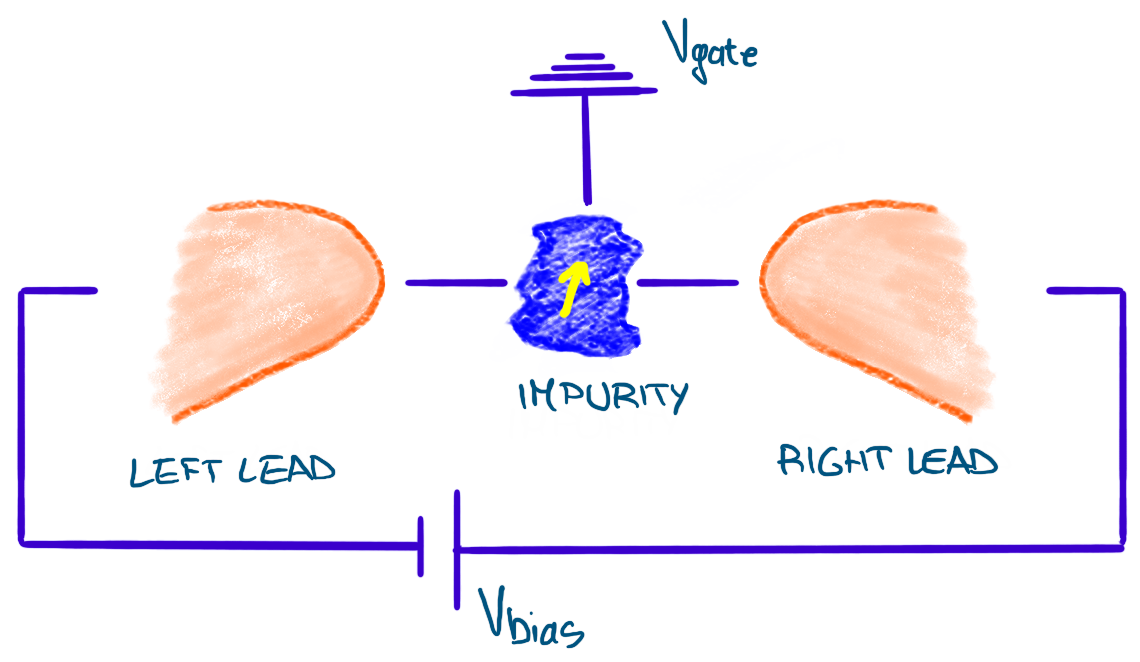}
\caption[Cartoon of the quantum impurity model]{Cartoon of the quantum impurity model composed of two leads hybridising with the impurity. The system is embedded in regular electronic circuit, meaning with voltage bias and voltage gate to tune the electronic distribution on the leads and on the impurity, respectively.}\label{F3:ImpScheme}
\end{figure}

As a vast subject in condensed matter physics, quantum transport processes bridge from macroscopic measurable effects to microscopic quantum behaviour - where the system is treated as a many-body object. One of its main observables is the \textbf{electrical conductance} which is directly probed in experimental quantum devices, designed with a nanostructure coupled between two leads.\\ 
In general terms, transport phenomena are characterised according to the \textit{length scales} involved in the system namely the sample size $L$, the elastic free mean-path $l_{el}$ and the quantum-mechanical coherence length $l_\phi$. In this thesis, we primarily refer to \textbf{mesoscopic quantum transport} that is characterised by: \\
$(i)$ $L<l_\phi$ indicating the transport is coherent and fully described by quantum effects (the opposite regime would indicate classical transport);\\
$(ii)$ $L<l_{el}$ saying that transport is ballistic (opposed to diffusive regime) where electrons do not inelastically scatter while they move through the sample such that only elastic scattering events occur. \\
We are interested in studying quantum transport in generic \textbf{quantum impurity models}. In Sec.\ref{sec:QImpModel}, we have detailed the physics of the two fundamental systems namely the Anderson and the Kondo impurity models and their limiting cases. Here, we address the discussion on the technical calculation used to determine conductance in general impurity models. We classify the methodologies according to their regime of applicability and present their analytical derivations.\\
In its essence, the \textit{model Hamiltonian} of an impurity system consists of three elements
\begin{equation}\label{eq:defImpModelGen}
\hat{H} = \hat{H}_{leads} + \hat{H}_{imp} + \hat{H}_{hyb} ~,
\end{equation} 
with $\hat{H}_{leads}$ describing the electronic \textit{reservoirs} placed adjacent of the central nanostructure region where electrons are kept in thermal equilibrium - once the reservoirs are wired with a voltage bias we refer to them as \textit{leads}. $\hat{H}_{imp}$ describes the nanostructure as generalized \textit{quantum impurity}, where electron-electron interactions are strong. The coupling between leads and impurity is given by $\hat{H}_{hyb}$, see Fig.\ref{F3:ImpScheme}. 
The system is taken to be in an initial equilibrium state among the three uncoupled regions e.g. the leads on the two sides and to the central area. The leads show a continuum energy spectrum and electron-electron interactions are assumed to be negligible due to the Coulomb screening effect. It is assumed the coupling among regions is established adiabatically such that all initial transient phenomena are discarded. As soon as a finite electronic tunnelling between leads and impurity is activated at finite bias voltage, the system enters in a nonequilibrium state. The impurity usually presents local interactions and exhibits a few discrete energy levels.\\ 
\noindent{Having} in mind this impurity system modelling, this chapter offers a roadmap through the standard literature \textit{operative schemes} to characterise quantum transport. Primarily, we are interested in calculating dynamical quantities such as the electrical conductance $\mathcal{G}^{C}$ and the heat conductance $\mathcal{G}^{E}$. 
The transport \textit{between the leads} (different from transport process \textit{within the leads}) due to a finite bias voltage gives a finite conductance: this is the \textbf{differential conductance} defined as 
\begin{equation}
\mathcal{G}^{ac}_{\alpha}(\omega,T,V_{bias}) = \frac{\partial I_{\alpha}(\omega,T,V_{bias})}{\partial V_{bias}} ~,
\end{equation}
for $\alpha=L,R$ leads, with $eV_{bias}=\mu_{L}-\mu_{R}$ related to the leads chemical potential and $\omega$ the frequency of the \textit{ac}-frequency of $V_{bias}$. This is a general definition \cite{DiVentra} applicable for electrical transport under $\Delta  V_{bias}$ and for heat transport under temperature variation. We can use this definition to calculate the electronic gradient between the leads at vanishing voltage bias
\begin{equation}\label{eq:DiffCondLRac}
\mathcal{G}^{ac}_{\alpha}(\omega,T) = \frac{\partial I_{\alpha}(\omega,T,V_{bias})}{\partial V_{bias}} \biggr\rvert_{V_{bias}\rightarrow 0}~,
\end{equation}
that is the \textbf{linear response differential conductance} subject to the periodic external field \\
$ V_{bias}(\omega,t)  \propto \allowbreak |\Delta V_{bias}| \cos(\omega t)$ where $\omega$ is indeed the \textit{driving frequency}. 
Furthermore, once we take the limit of vanishing $\omega$, we obtain the \textbf{dc differential conductance} under linear response,
\begin{equation}\label{eq:DiffCondLRdc}
\mathcal{G}^{dc}_{\alpha}(T) = lim_{\omega\rightarrow0}\mathcal{G}^{ac}_{\alpha}(\omega,T)~.
\end{equation}
Eqs.\ref{eq:DiffCondLRac},\ref{eq:DiffCondLRdc} are the cornerstone conductance formulae we refer to in the thesis.\\

\noindent{In} this chapter, the discussion mostly follows the historical development of the analytical techniques and hence, as outline we start with section \ref{sec:Landauer} where we present the transport theory for noninteracting models by means of scattering theory in the Landauer formula, in section \ref{sec:Kubo} the transport theory under linear response due to small external perturbation in the Kubo formula for conductance, in section \ref{sec:MeirWingreen} the transport theory for interacting models defined on Keldysh contour by means of nonequilibrium theory in the Meir-Wingreen formula, in section \ref{sec:Oguri} the transport for interacting models within the Fermi liquid theory in the Oguri argument. At the end of the chapter, in section \ref{sec:Kexp} we present the typical device amenable for Kondo effect verification through electrical conductance measurements: the quantum dot system.\\
Before we start with the core of the chapter, we introduce the two-channel generalization of the main physical quantities needed in the study of quantum transport.

\subsection*{Physical quantities for two-channel quantum transport}
In Sec.\ref{sec:Gtheo} we introduce the general form of the main physical quantities used in this thesis. In that section, the discussion is set without any model Hamiltonian for the sake of generality. In Sec.\ref{sec:QImpModel} we introduce the most important impurity models studied in this work. Furthermore, in the beginning of this chapter we have defined the model Hamiltonian for quantum transport systems in Eq.\ref{eq:defImpModelGen}. We can now discuss the quantum impurity models in the two channel configuration and then introduce the physical quantities used in the study of quantum transport in these two-channel models.\\
In the single-channel Anderson model presented in Eq.\ref{eq:AM}, the two-channel version for asymmetric channels is straightforward to obtain by introducing a summation over $\alpha=L,R$ leads, namely 
\begin{equation}
\hat{H}^{2CAM} = \sum_{\alpha=L,R}\sum_{\mathbf{k}\sigma}\left(  \epsilon_{\mathbf{k}} c^{\dagger}_{\alpha\mathbf{k}\sigma}c_{\alpha\mathbf{k}\sigma} + ( V_{\alpha\mathbf{k}} c^{\dagger}_{\alpha\mathbf{k}\sigma}d_{\sigma} + V^{\star}_{\alpha\mathbf{k}}d^{\dagger}_{\sigma}c_{\alpha\mathbf{k}\sigma}) \right) + \epsilon_{d}\sum_{\sigma}\hat{n}_{d\sigma} + U\hat{n}_{d\uparrow}\hat{n}_{d\downarrow}  ~.
\end{equation}
Instead for symmetric channels model, the hybridizations reduce to $V_{L\mathbf{k}}=V_{R\mathbf{k}}\equiv V_{\mathbf{k}}$. By using a symmetric linear combination of these couplings, the system satisfies the proportionate coupling property and it is cast back to the single-channel model in Eq.\ref{eq:AM} - as we will discuss in detail in Sec.\ref{sec:PC}.\\
In the isotropic Kondo model in Eq.\ref{eq:K}, the two-channel version is achieved by introducing the summation over $\alpha\neq \beta = L,R$ leads, namely
\begin{equation}
	\hat{H}^{2CK} =  \sum_{\alpha\neq \beta =L,R}\sum_{\mathbf{k}\sigma}\left(  \epsilon_{\mathbf{k}} c^{\dagger}_{\alpha\mathbf{k}\sigma}c_{\beta\mathbf{k}\sigma} +
\sum_{\mathbf{k}^{\prime}} W_{\mathbf{k}\mathbf{k}^{\prime}\sigma} c^{\dagger}_{\alpha\mathbf{k}\sigma}c_{\beta\mathbf{k}^{\prime}\sigma} \right) +
\sum_{\alpha\neq \beta =L,R}J_{\alpha\beta} \hat{\mathbf{s}}_{cond}(0) \cdot \hat{\mathbf{S}}_{d} ~.
\end{equation}
From $\hat{H}^{2CK}$ we have that both exchange coupling $J$  and potential scattering $W$ have non-diagonal matrix structure. Hence, the $\mathbb{J},\mathbb{W}$ matrices of  $dim(2\times2)$ in the channel space for $nm$ matrix elements follow as:
\begin{equation}
	\begin{aligned}
		[\mathbb{J}]_{nm}=
		 \begin{bmatrix}
			J_{LL} & J_{LR} \\  J_{LR} & J_{RR}
		\end{bmatrix}_{nm}  \quad ~, \quad
	   [\mathbb{W}]_{nm}= 
		 \begin{bmatrix}
			W_{LL} & W_{LR} \\  W_{LR} & W_{RR}
		\end{bmatrix}_{nm}   ~,
	\end{aligned}
\end{equation}
where the $J_{\alpha\alpha},~W_{\alpha\alpha}$ describe processes between $\alpha$-bath and the impurity whereas $J_{\alpha\beta},~W_{\alpha\beta}$ refer to inter-channel processes with spin-flip events. In case $J_{\alpha\alpha}= J_{\beta\beta}$ and $W_{\alpha\alpha}= W_{\beta\beta}$, the system presents $\alpha\beta$-symmetry. By performing suitable transformation, it is possible to make diagonal these matrices - and this mathematical operator allows channel decoupling at low-energy scale as we apply in Sec.\ref{sec:CBPC}.\\
We start with the definition of the \textbf{spectral function}, namely:
\begin{equation}\label{eq:defSpectral}
	\begin{aligned} 
		&(i)~ \text{general:}~ \mathcal{A}_{x\sigma}(\omega) = \frac{i}{2 \pi} [G^{>}_{x\sigma}(\omega)-G^{<}_{x\sigma}(\omega)] =\frac{i}{2 \pi} [G^{R}_{x\sigma}(\omega)-G^{A}_{x\sigma}(\omega)] = -\frac{1}{\pi} \mathit{Im} [G^{R}_{x\sigma}(\omega)] ~, \\
		&(ii)~ \text{non interacting systems:} \begin{cases}
			\mathcal{A}_{0,\alpha\mathbf{k}\sigma}(\omega) = \delta(\omega-\epsilon_{\mathbf{k}}) ~, \\
			\mathcal{A}_{0,dd\sigma}(\omega) = \delta(\omega-\epsilon_{d})  ~.
		\end{cases}
	\end{aligned} 
\end{equation}
where in $(i)$ we use the Keldysh identities in Eq.\ref{eq:identityKeldysh}, in  $(ii)$ we use the Dirac identity and the subscript $0$ identifies noninteracting quantities. The subscript $x=\alpha\mathbf{k}$ indicates the Green's function for $\alpha=L,R$ leads and momentum $\mathbf{k}$ derived from $\hat{H}_{bath}$, $x=dd$  indicates the impurity Green's function derived from $\hat{H}_{imp}$.\\ 
From the spectral function definition in Eq.\ref{eq:defSpectral}, we define the \textbf{local density of states}:
\begin{equation}\label{eq:defDos}
	\begin{aligned}
	&(i) ~ \hat{H}_{bath}:~	\rho_{\alpha\sigma}(\omega) = \int \frac{d^{n}\mathbf{k}}{{(2 \pi)}^{n}} \mathcal{A}_{\alpha\mathbf{k}\sigma}(\omega) ~, \\
	&(ii)~\hat{H}_{imp}:~ \rho_{dd\sigma}(\omega) \equiv  \mathcal{A}_{dd\sigma}(\omega) = -\frac{\mathit{Im}G^{R}_{dd\sigma}(\omega)}{\pi} ~,
	\end{aligned} 
\end{equation}
where corresponding noninteracting density of states $\rho_{0,\alpha\sigma},\rho_{0,dd\sigma}$ are calculated by means of $\mathcal{A}_{0,\alpha\mathbf{k}\sigma},\mathcal{A}_{0,dd\sigma}$ functions respectively. The association between local density of states and spectral function for $\hat{H}_{bath}$ in expression $(i)$ finds its meaning also within the Wilson chain representation we introduce in Sec.\ref{sec:RGtheo}. Only the $0^{th}$ site is attached to the impurity and it is the only physical site in the whole semi-infinite chain. In the logarithmic discretization step in the NRG protocol, the $c^{\dagger}_{0\sigma},c_{0\sigma}$ operators in Eq.\ref{eq:2H_WC} are defined by integrating over all momenta $\mathbf{k}$ \cite{Bulla2008}. Hence, we can safely associate the local density of states on $n=0$ site with the spectral function as defined in $(i)$ in Eq.\ref{eq:defDos}. About expression $(ii)$, the association is straightforward since the impurity site does not carry any momentum. \\
In Eq.\ref{eq:defFermieq}, we give the definition for the equilibrium Fermi distribution. In transport processes, the presence of an external perturbation is required to shift the electronic population and so to preserve an electronic gradient across the model. Such perturbation can be implemented physically by applying a voltage bias to the system reservoirs. Then, the electronic distribution refers to the constant (in time) difference in chemical potential in its reservoirs i.e. $\Delta\mu= e V_{bias} \equiv \mu_{L}-\mu_{R}$, where the left lead is the \textit{source} and the right lead is the \textit{drain} in the system. Due to the unbalanced chemical potential in the two leads, there is a net electronic flux in the system and finite electronic conductance is measured. In this situation, the \textbf{Fermi-Dirac distribution} reads
\begin{equation}\label{eq:defFermiNeq}
	f^{\mu_{\alpha}}(\omega)= \dfrac{1}{1 + e^{\beta( \omega - \mu_{\alpha})}} ~,
\end{equation}
for $\mu_{\alpha}$ and $\beta_{\alpha}=(k_{B}T_{\alpha})^{-1}$ the chemical potential and the temperature for $\alpha$ corresponding to the bath of reference. We remark that we cannot write an analytical expression for out-of-equilibrium systems, hence Eq.\ref{eq:defFermiNeq} is still an equilibrium electronic distribution but now for a two-channel model where the electronic gradient is preserved by shifting the chemical potential in one lead with respect to the other.\\
In analogy to the general definition for the \textbf{fluctuation-dissipation theorem} in Sec.\ref{sec:Gtheo}, we introduce now the version for Fermionic uncoupled system in the generic two-channel system, namely:
	\begin{equation}\label{eq:defFDtheo}
		\begin{aligned}
			& G^{<}_{x\sigma}(\omega)= i f^{\mu_{\alpha}}(\omega) 2\pi\mathcal{A}_{x\sigma}(\omega)  \\
			&G^{>}_{x\sigma}(\omega) = -i (1-f^{\mu_{\alpha}}(\omega)) 2\pi \mathcal{A}_{x\sigma}(\omega) ~,
		\end{aligned}
	\end{equation}
where the expressions hold both interacting and noninteracting models using the corresponding spectral function in Eq.\ref{eq:defSpectral} and $x=\alpha\mathbf{k},dd $ for bath and impurity Green's function. \\
We now discuss the noninteracting impurity model since it contains important quantities. Recalling the retarded Green's function $G^{R}_{0,dd\sigma}(\omega)$ defined in Eq.\ref{eq:G_0dd}, the total \textbf{hybridization function} $\Delta(\omega)$ for the two-channel model follows as:
\begin{equation}\label{eq:HybFunc}
	\Delta(\omega)
	= \sum_{\alpha=L,R}\Delta_{\alpha}(\omega)=
	\sum_{\alpha\mathbf{k}\sigma} |V_{\alpha\mathbf{k}}|^{2} G^{0,R}_{bath,\mathbf{k}\sigma}(\omega) ~,
\end{equation}
where $V_{\alpha\mathbf{k}}$ is the hybridization coupling strength with respect to the $\alpha$-lead as in Eq.\ref{eq:AM} and $G^{0,R}_{bath,\mathbf{k}\sigma}$ where the superscript zero indicates the \textit{isolated} noninteracting, thermal-equilibrium bath.\\
In the wide-band limit approximation, we have $\mathit{Re}\Delta=0$ and $\mathit{Im}\Delta(\omega) =- \pi \sum_{\alpha\mathbf{k}}|V_{\alpha\mathbf{k}}|^{2}\delta(\omega -\epsilon_{\mathbf{k}})$. Hence, the imaginary part of the hybridization is indeed the noninteracting spectral function $\mathcal{A}_{0,\alpha\mathbf{k}\sigma}(\omega)$, see Eq.\ref{eq:defSpectral} and  we can use it to get the corresponding noninteracting bath density of states $\rho_{0,\alpha\sigma}(\omega)$ with respect to $\alpha$-lead, see in Eq.\ref{eq:defDos}, that is a $\delta$-function peaked at the single-particle energy $\epsilon_{\mathbf{k}}$. We use these definitions to rewrite the imaginary component of hybridization function for the infinite band as
\begin{equation}\label{eq:DoSinfWBL}
	\begin{aligned}
		&\text{total levelwidth function:} ~ \mathrm{\Gamma} \equiv \sum_{\alpha=L,R}\mathrm{\Gamma}^{\alpha}\\
		&(i) \quad \mathrm{\Gamma}^{\alpha}(\omega) \equiv -\mathit{Im}\Delta_{\alpha}(\omega) = -\Big( -\pi\sum_{\mathbf{k}\sigma} |V_{\alpha\mathbf{k}}|^{2} \rho_{0,\alpha\sigma}(\omega) \Big)  = - \sum_{\mathbf{k}\sigma} |V_{\alpha\mathbf{k}}|^{2} \mathit{Im}G_{bath,\mathbf{k}\sigma}^{0,R}(\omega)  ~,\\
		&(ii) \!\quad \mathrm{\Gamma}^{\alpha} \equiv -\mathit{Im}\Delta_{\alpha} = + \pi |V_{\alpha}|^{2} \rho_{0,\alpha\sigma} ~,
	\end{aligned}
\end{equation}
from now we refer to this result as $\mathrm{\Gamma}^{\alpha}$ \textit{gamma or levelwidth function} as corresponding to the imaginary part of the noninteracting self-energy in Eq.\ref{eq:defSelEngeneral}. This equation acquires general definition in \textit{matrix} form as given in Eq.\ref{eq:LevelWidth}. Eq.\ref{eq:DoSinfWBL} is detailed for  $(i)$ the most general and $(ii)$ the most common in literature configurations. In $(i)$ case, see also Eq.\ref{eq:4defEqIneq}, we refer to \textbf{structured leads} meaning that the density of states is function of energy $\omega$  and the leads can be either \textbf{equivalent} i.e. $\rho_{0,\alpha\sigma}(\omega)=\rho_{0,\alpha^{\prime}\sigma}(\omega) \equiv \rho_{0,\sigma}(\omega)$ or \textbf{inequivalent} i.e.  $\rho_{0,\alpha\sigma}(\omega) \neq \rho_{0,\alpha^{\prime}\sigma}(\omega)$. In $(ii)$ case, we give the standard result used in the literature that is equivalent  \textit{metallic} leads with \textit{constant} density of states inside the wide-flat band i.e. $\rho_{0,\alpha}=\rho_{0,\alpha^{\prime}}\equiv\rho_{0}$. Unless explicitly stated, we adopt the standard condition as defined in $(ii)$ and we return later in Sec.\ref{sec:AlternativeLandauer} on the generalization we introduce in $(i)$.\\
With these equations we conclude the generalization of the fundamental definitions given in Sec.\ref{sec:Gtheo} and we are now ready to discuss the theories of quantum transport.

\section{Theory of transport in noninteracting models: the Landauer formula and scattering theory} \label{sec:Landauer}
The very first formulation of transport through a mesoscopic nanostructure is developed in the context of noninteracting systems. This approach is depicted in a general physical intuitive picture of electrons moving, bunching back and forth through an intermediate structure. In this section, the transport process is framed by single-particle picture formulation using scattering theory results. \\
In a sense, we consider this section as preliminary for the succeeding transport theories presented in this chapter, and used later in the thesis, which require a more sophisticated treatment. Furthermore, the formulae derived in this section have to be regarded as benchmark formulae scheme: in absence or almost negligible interactions, any complex model must recover those as limiting cases. We now illustrate the noninteracting transport equations and their corresponding generalization following their historical derivation \cite{StoneSzafer_LandauerRevisited}.
 
\subsection*{The Landauer formula}
The key idea underlying the physics of transport in noninteracting structures is to interpret it as a \textit{scattering process}. The transport understood in terms of spacial variation of currents in metallic conduction was initially developed by R. Landauer \cite{Landauer1957} and below we aim to present this simple but very effective approach.\\ 
\begin{figure}[H]
\centering
\includegraphics[width=0.75\linewidth]{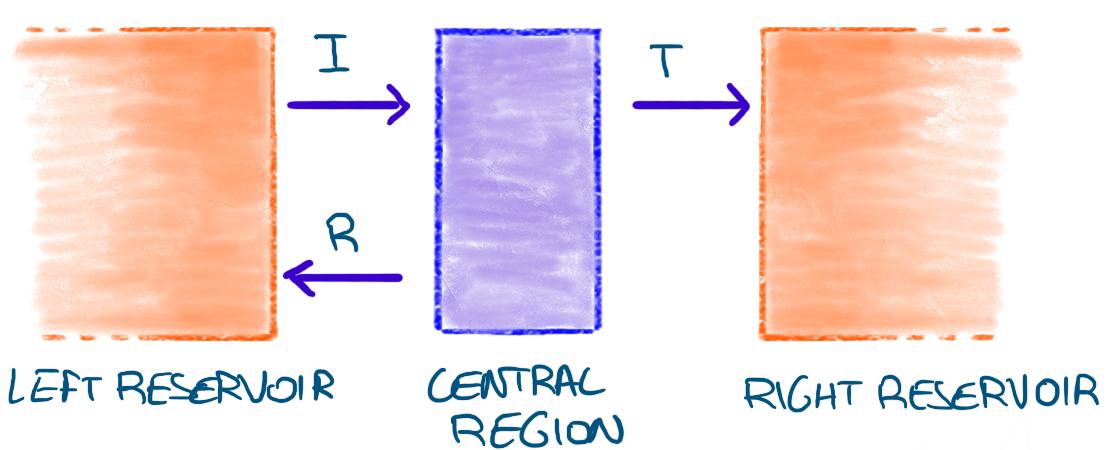}
\caption[Cartoon of the Landauer transport formulation]{Cartoon of the Landauer transport formulation: the electron motion through the sample is described as scattering events characterised by incident $\mathcal{I}$, reflection $\mathcal{R}$ and transmission $\mathcal{T}$ probabilities.}\label{F3:LandauerScheme}
\end{figure}
\noindent{In} the Landauer formulation, the conductance $\mathcal{G}$ is determined by the electron motion through one dimensional conductor between two phase-randomizing reservoirs. In practice, this is the calculation of an incident electrical flow - resulting into a finite current $I$ - scattering onto an array of obstacles. We remark that the obstacles length $L$ should suit the mesoscopic transport conditions \cite{Landauer1970}, as discussed in the introduction in this chapter. As we know already, this specific length scale ensures the electron motion is quantum mechanical coherent in the whole sample. A further requirement in the Landauer approach is that the contact between reservoirs and sample are \textit{reflectionless}. This allows for the complete electron thermalization occurs at electronic temperature and chemical potentials of the reservoir where they are absorbed and emitted from. This property is applicable to each electron, before it is emitted every time into the central area. In practice, the reflectionless property can be engineered for instance by contact having a "horn" shape.\\
The scattering processes were originally formalised by means of the total reflection probability of the array $\mathcal{R}$. Without loss of generality, we assume the current is flowing from left to right reservoir. As consequence, the system has filled right-moving states in left reservoir and unfilled left-moving states in right reservoir such that the entire system manifests an asymmetrical electrochemical potential. The intuitive electron counting occurring during the scattering event is expressed in terms of an unequal accumulation of density carriers $1+\mathcal{R}$ and $1-\mathcal{R}$ on left and right side of the sample, respectively. Applying \textit{diffusion equation} to associate the current with the density gradient $-2\mathcal{R}/L$ \cite{Nazarov}, the diffusion coefficient $D$ is derived and it is used in the Einstein Relation \cite{Kubo1957} to determine the resistivity of the sample. Hence, the \textbf{original Landauer formula} to express electrical conductance reads
\begin{equation}
\mathcal{G}^{C} = \dfrac{e^{2}}{h}\dfrac{\mathcal{T}}{\mathcal{R}} ~,
\end{equation}
where $\mathcal{T},\mathcal{R}$ are \textit{transmission, reflection probabilities} associated to the current moving back and forth while flowing through the conductor, see in Fig.\ref{F3:LandauerScheme} a sketch of the system. Here, in the electrical conductance equation we have used the superscript $C$ indicating \textit{charge} current as source of conductance to distinguish it from the heat conductance. To pin down the discussion, we present now the Landauer formula derivation to some detail \cite{Nazarov, DiVentra}. \\
We start with introducing the physical components in the set-up, namely the leads and the central region. Those will be in contact through tunnelling junctions. The \textit{central area} is a  nanostructure - considering its typical size in mesoscopic conductors - and it is the only region in the system where scattering events occur. The \textit{leads} are the source of the electrons moving in ballistic-type motion - since in this area no scattering events happen. The model accounts for two semi-infinite reservoirs and they are kept in unbalanced electrochemical state by applying at those an external voltage bias $V_{bias}$. Under these nonequilibrium conditions identified by an unequal chemical potential on the two leads $\mu_{L} \neq \mu_{R}$, a finite current $I$ can be measured across the sample. For simplicity, $V_{bias}$ is taken as a time-independent function: hence, we need to maintain constant the electronic gradient across the sample in order to preserve a net current flow. The electrons are kept in this thermal condition, that is constant in time i.e. steady state but unequal on the two leads, by means of reflectionless contacts as set-up requirement. Last, the \textit{tunnelling junctions} made by general Ohmic contacts allow transfer of electron from and to the lead and the central region.\\
The usefulness of the physical configuration  we just described is to treat the two adjacent leads as the \textit{boundary condition} of scattering events. We can formulate then a lead Hamiltonian element comprehensive of the actual single-particle lead Hamiltonian plus some one dimensional confining potential $V_{conf}(r)$. The electronic wave functions inside the sample assumes an asymptotic behaviour with longitudinal and transverse components. From each transverse quantum mode $n=1,2,\dots,N$ we can derive the transmission $\mathcal{T}$ and the reflection $\mathcal{R}$ probabilities as ratio of incoming and outgoing current waves. The probability conservation $\mathcal{T}+\mathcal{R}=1$ directly entails current conservation through the sample i.e. $I_{L}=-I_{R}$. Hence, this calculation defines the scattering events by means of the probability associated to an electron either to be reflected back to the lead where it was emitted or to be transmitted through the nanostructure. Then, these probabilities are used to calculate the total current flowing in the sample. The necessary condition to get $I\neq 0$ is obtained by removing the equilibrium condition in the sample applying a constant voltage bias between leads. The electronic population at each lead is then regulated by the equilibrium Fermi-Dirac distribution in Eq.\ref{eq:defFermiNeq}, such that $f^{\mu_{L}},f^{\mu_{R}}$ are the weights of the corresponding electronic wave functions.\\
By means of these elements, at finite temperature and in presence of an external voltage bias the expression for the charge current is given by
\begin{equation}
I(T,V_{bias}) = \frac{2e}{h} \int_{-\infty}^{+\infty} dE \left( f^{\mu_{L}}(E-\mu_{L}) - f^{\mu_{R}}(E-\mu_{R}) \right) \sum_{n=1}^{N}\mathcal{T}_{n}(E,T,V_{bias}) ~,
\end{equation} 
where $\mathcal{T}(E,T,V_{bias})$ indicates the \textit{total energy dependent transmission probability} per $n$-mode through the sample and it is a function of temperature and voltage bias applied in the system. As a consequence, the total current inherits the same dependence. The probability ranges  between $0 \leq \mathcal{T}(E) \leq 1$, where perfect reflection happens at $\mathcal{T}=0$ and perfect transmission at $\mathcal{T}=1$. The transmission numbers are deduced in unit of quantised conductance. However, we note those values are probed only in the set-up with a special configuration, namely the split-gate geometry used at the \textit{quantum point contacts}. In the remainder of this presentation on the Landauer approach we discuss more on this aspect and we calculate the current equation at particular temperatures and voltages values to show the effects of those limitations on the system.\\
A first case regards the system at zero temperature $T \rightarrow 0$. The noninteracting many-body electron cloud forms a degenerate Fermi gas when the thermal energy in unit of $k_{B}T$ is smaller than the Fermi energy $\epsilon_{F}$ \cite{FermiSeaformation_ExpHeidelberg_2013}. Then, the chemical potential of each lead $\mu_{\alpha}$ equates $\epsilon_{F}$, where we not restricting $\epsilon_{F}$ value. Thus, only states occupying this level - or in its proximity - are involved in the hopping processes. The corresponding Fermi-Dirac distribution takes the equilibrium form in Eq.\ref{eq:defFermieq} computed at the Fermi level i.e. $f^{\mu}(E-\epsilon_{F})= \theta(E-\epsilon_{F})$ and its form is a step function. The transmission $\mathcal{T}$ at $T=0$ is also evaluated at $\epsilon_{F}$.\\
A second case considers the system in zero bias condition $e V_{bias} =\mu_{L}-\mu_{R} \rightarrow 0$ such that the equilibrium state is restored, now with electronic population distributed according to the equilibrium chemical potential $\mu$ in both leads. This condition allows a perturbative expansion of the Fermi-Dirac function in order of $\mathcal{O}(\mu_{L}-\mu_{R})$, infinitesimal difference approximating the equilibrium value. At strictly zero bias $\mu_{L}-\mu_{R}=0$, we observe vanishing total current $I(T,V_{bias}=0)$. Otherwise at small but finite $V_{bias}$, we get from the expansion $I(T,V_{bias}\simeq0) \gtrsim 0$ such that we can calculate a finite differential conductance under linear response as in Eq.\ref{eq:DiffCondLRac}. At $\mathcal{G}^{dc}_{\alpha}(\omega\to 0,T,V_{bias}\rightarrow 0)$, we find that the transmission probability is finite and  may reach its max value.\\
We can now combine both limiting cases and calculate the total charge current at $T=0$ and $V_{bias} \rightarrow 0$ 
\begin{equation}
I(T=0,V_{bias}\to 0) = \frac{2e^{2}V_{bias}}{h} 
\sum_{n=1}^{N}\mathcal{T}_{n}(E=\epsilon_{F},T=0,V_{bias}\to 0) ~,
\end{equation}
where the prefactor stands for the spin degrees of freedom in absence of spin-orbit scattering. On the contrary, the factor of two disappears in case of both spin-orbit scattering and nonzero magnetic field \cite{Beenakker_RandomMatrix1997} but then it is restored by $\sum_{\sigma}\mathcal{T}_{n\sigma}$. As mentioned before, we can measure finite conductance only from transport between the leads. Using the linear response differential conductance definition in Eq.\ref{eq:DiffCondLRac}, we obtain
\begin{equation}
\begin{aligned}
\mathcal{G}^{C}(T=0) &= \frac{dI(T=0,V_{bias})}{dV_{bias}} \biggr\rvert_{V_{bias}\rightarrow0} ~,\\
&= \frac{2e^{2}}{h} \sum_{n=1}^{N}\mathcal{T}_{n}(\epsilon_{F},T=0) ~,
\end{aligned}
\end{equation}
that is the \textbf{electrical conductance Landauer formula} for \textit{two perfect channels} at zero temperature, meaning the derivation is computed on the typical two-probes set-up with one spinful channel per probe and corresponding $V_{bias}$ \cite{Beenakker_QuantumTransport1991}.\\
Among the many important consequences of the Landauer expression, it shows the intrinsic quantised nature of conductance. In condition of absolute no reflection but maximum transmission, in split-gate geometry set-ups we can probe the differential conductance can reach its max value namely
\begin{equation}
\mathcal{G}^{C}(T=0,V_{bias}\to 0) =  \frac{2e^{2}}{h}  \equiv\mathcal{G}_{0} ~.
\end{equation}  
At $T \simeq 0$, by varying the voltage gate - that is another voltage applied now on the nanostructure to tune its electrostatic population - the conductance versus temperature curve conforms to steps taking place at integer values of $\mathcal{G}_{0}$. This remarkable result demonstrates the Landauer formula is able to disclose the conductance quantization in units of $e^{2} /h$ per channel - $N=2$ in this case for the two spin degenerate channels. Moreover, as we will see later, in set-ups with $N$ perfect channels, the maximum conductance of the multichannel model is simply given by $\mathcal{G}^{C}(T=0,V_{bias}\to 0) = N \mathcal{G}_{0} $. The curve $\mathcal{G}(\mathcal{G}_{0})$ as function of $V_{gate}$ acquires an oscillatory behaviour, as described by the so-called Coulomb blockade model discuss later in Sec.\ref{sec:Kexp}. This typical conductance curve \textit{staircase} is observed in experiments, see on the right of Fig.\ref{F3:exp1}\textit{(A)}. \\
In conclusion, the Landauer formula embraces the transport physics through scattering events defined by transmission and reflection probabilities. Its disadvantages are the restricted applicability two-probes two-channel samples only in noninteracting impurity models by means of transmission probabilities. These limitations are overcame in the Fisher-Lee and Landauer-B{\"u}ttiker formulae.

\subsection{The Fisher-Lee formula}
As we know from the last discussion, the electronic transport through a nanostructure relies on an unbalanced electrochemical condition across the sample that is nonequilibrium state. Rather than in terms of temporal correlations, the original Landauer's insight conceives electron transport as a result of successive scattering events. \\
Further understanding on this approach applicable to noninteracting systems is the translation from probability coefficients into \textit{transmission and reflection amplitudes} - as it is incorporated in the \textbf{Fisher-Lee formula} \cite{FisherLeeRelation1981,StoneSzafer_LandauerRevisited}. These amplitudes are identified as static scattering properties of the central area, the target, with respect to the boundary conditions imposed at the two adjacent leads. Hence, these amplitudes characterise the plane-waves of electrons involved in scattering processes inside the sample. Those coefficients are arranged to compose the matrix elements of $\mathbb{S}(E)$, the $\mathrm{S}$-scattering matrix of the process. The relation between these elements and the electron wave functions is achieved by introducing the Green's functions of the scattering problem - as we show to some detail in this section.\\
The Fisher-Lee formula uses as a starting point the Kubo formula for electrical conductance in Eq.\ref{eq:defKuboLRel} as derived by means of Linear Response theory \cite{Kubo1956}. Within this formalism, the transport equation structure is not contingent on properties of central area, in particular with respect to the presence of local interactions as indeed this formulation considers only noninteracting impurity models. This derivation is built upon elements of scattering theory \cite{Mahan,Economou}: no restriction is made on the number of channels and the equations are implicitly calculated at zero temperature. \\
Assuming a \textit{dc}-regime, this constraint allows to evaluate the incident electronic waves travelling from $\alpha=L,R$ left and right leads towards the target as superposition of plane waves $\phi_{\alpha}$: those enter into the formalization of the  scattering event. In some sense, the leads provide electronic wave-guides from which we can determine suitable basis to be used in the scattering problem \cite{Beenakker_RandomMatrix1997}. \\
A stationary scattering state $\Psi_{\alpha}$ is defined in the \textit{Lippmann-Schwinger equation} as 
\begin{equation}
\Psi^{\pm}_{\alpha} = \phi^{\pm}_{\alpha} + G^{\pm} V \phi^{\pm}_{\alpha} ~,
\end{equation}
where the superscript $\pm$ indicates outgoing and incoming waves for the \textit{single-particle Green's functions} $G^{\pm}$, the plane waves $\phi^{\pm}_{\alpha}$ and $V$ is the local potential scattering. The transmission amplitude corresponding to outgoing waves is proportional to $ \bra{ \phi^{+}_{r} } V \ket{ \Psi^{+}_{L}}$. Using these equations, it is possible to associate the \textbf{transmission} and \textbf{reflection amplitudes} $t_{nn^{\prime}},r_{nn^{\prime}}$ to the asymptotic form of the Green's functions and to arrange those terms as matrix elements of $\mathbb{S}(E)$, namely
\begin{equation}\label{eq:FisherLee_Smatrix}
\mathbb{S} (E)= \begin{pmatrix}
\mathbb{r}_{LL}(E) & \mathbb{t}_{LR}(E) \\  \mathbb{t}_{RL}(E) & \mathbb{r}_{RR}(E) 
\end{pmatrix} ~,
\end{equation}
where the diagonal elements represent the reflection matrices to left and right channel with $dim(N \times N)$, the off-diagonal elements represent the transmission matrices from left to right and reverse with $dim(N \times N)$. As presented in Eq.\ref{eq:FisherLee_Smatrix}, we are highlighting the energy dependence included in the amplitude terms. Considering each electronic wave packet as $n^{th}$-channel transferring electrons from and to the target, we incorporate the $n=1,\dots,N$-incoming and $n=1,\dots,N$-outgoing amplitude coefficients from both leads into the vector $\mathbf{a}^{\pm}_{\alpha,n}$ to obtain the matrix relation
\begin{equation}
\left(a^{+}_{R,1} ~ a^{+}_{R,2} ~ \dots ~ a^{+}_{R,N}; a^{-}_{R,1} ~ a^{-}_{R,2} ~ \dots ~ a^{-}_{R,N}\right)^{T} = \mathbb{S} (E) \left(a^{-}_{L,1} ~ a^{-}_{L,2} ~ \dots ~ a^{-}_{L,N}; a^{+}_{L,1} ~ a^{+}_{L,2} ~ \dots ~ a^{+}_{L,N} \right)^{T} ~,
\end{equation}
and so we conclude the $\mathrm{S}$-matrix captures the essence of scattering process by linking the amplitudes of all possible events to electronic wave packets at a given energy $E$.\\
By means of these results, we study now the continuity equation for current defined in the context of transmission and reflection amplitudes \cite{Flensberg}. In absence of an external magnetic field, the system obeys to time-reversal symmetry namely $\mathbb{S}=\mathbb{S}^{T}$ and the off-diagonal terms in Eq.\ref{eq:FisherLee_Smatrix} are identical $\mathbb{t}_{LR} = \mathbb{t}_{RL}$. Clearly, the current must be conserved in all $n^{th}$-channel by equating incoming and outgoing electron fluxes. Hence, it has to be ensured the unitary property of $\mathrm{S}$-matrix i.e. $\mathbb{S}^{-1}=\mathbb{S}^{\dagger}$ and so we obtain the identity $\mathbb{S}^{\dagger}\mathbb{S} = \mathbb{S}\mathbb{S}^{\dagger} = \mathbb{1}$. The latter condition is indeed the direct translation of the probability conservation $\mathcal{T}+\mathcal{R} =1$ used in the Landauer formula - a connection made by the joint work of Fisher and Lee.\\
In order to complete the calculation, we derive the differential conductance equation using the transmission amplitudes as featuring in the $\mathrm{S}$-matrix in Eq.\ref{eq:FisherLee_Smatrix}. As we mentioned, their expression is defined by asymptotic Green's functions and so we derive the under linear response \textbf{electrical dc-conductance Fisher-Lee formula} \cite{FisherLeeRelation1981},
\begin{equation}\label{eq:FisherLee}
\begin{aligned}
\mathcal{G}^{dc}(T=0) &\stackrel{(i)}{=} \frac{2e^{2}}{h} \sum_{n,n^{\prime}=1}^{N} |t_{nn^{\prime}}(E,T=0,V_{bias})|^{2} \biggr\rvert_{V_{bias},E\rightarrow0 } ~,\\
& \stackrel{(ii)}{=}\frac{2e^{2}}{h} Tr [\mathbb{t}^{\dagger}(E,T=0,V_{bias})\mathbb{t}(E,T=0,V_{bias})]\biggr\rvert_{V_{bias},E\rightarrow0 } ~,\\
&\stackrel{(iii)}{=} \frac{2e^{2}}{h} Tr \Big[\widetilde{\mathbb{T}}(E,T=0,V_{bias}) \Big] \biggr\rvert_{V_{bias},E\rightarrow0 }~.
\end{aligned}
\end{equation}
These equivalent expressions are just based on different assumptions: $(i)$ is obtained considering the time-reversal symmetry in the model, $(ii)$ is calculated using the unitary property of $\mathrm{S}$-matrix and $(iii)$ is given in terms of the generalized transfer matrix defined in Eq.\ref{eq:GenTransfMat}. As mentioned initially, the conductance has been evaluated only in \textit{dc-}regime at zero temperature. \\
For completeness, we note that so far we have referred to $\mathcal{G}^{dc}$ in terms only of transmission amplitudes $[\mathbb{t}]_{nn^{\prime}}$. However, by means of the $\mathrm{S}$-matrix structure in Eq.\ref{eq:FisherLee_Smatrix}, the term $\mathcal{T}_{n}=Tr [\mathbb{t}^{\dagger}\mathbb{t}]$ indicates both the transmission probabilities - as it were in the Landauer formulation - and the eigenvalues of the matrix $\mathbb{t}^{\dagger}\mathbb{t}$. We deduce it is equivalent to work with either amplitudes or eigenvalues \cite{Beenakker_RandomMatrix1997}.\\
We close this section with some final observations. The Fisher-Lee formula generalizes the Landauer derivation since its validity extends to multi-channel systems, without restriction on the noninteracting nature of the mesoscopic sample. Moreover, by introducing Green's functions, the conductance derivation acquires a rigorous mathematical structure in terms of $\mathrm{S}$-matrix elements. Hence, in the Fisher-Lee formula the only restriction left is that calculations are performed at zero temperature - still within noninteracting two channel impurity models. The temperature and number of channels limitations are removed in the Landauer-B{\"u}ttiker formula, as we see in the next section.

\subsection{The Landauer-B{\"u}ttiker formula}
The development on appropriate methodology to study noninteracting mesoscopic transport reaches its completion by extending the validity of the Fisher-Lee relation in Eq.\ref{eq:FisherLee_Smatrix} to multiple channel set-ups, at finite temperature and in the \textit{ac-}regime - still in the context of scattering theory. As a reference, we still have the system sketched in Fig.\ref{F3:LandauerScheme}, but now with several arrows pointing inwards and outwards from the central region. This has been investigated in the derivation by M. B{\"u}ttiker and R. Landauer \cite{LandauerButtiker1985} which we concisely show now for a sample with two-probe $\alpha=L,R$-leads coupled to the central region by $N$ channels.\\ 
In general, we take an arbitrary number of channels per lead so we define $n^{i}_{L} \neq n^{i}_{R}$ for $i=1, \dots ,N$. As a main property, the channels are independent. At finite temperature, the longitudinal component of the electronic wavefunctions have different electron velocity per channel $v^{i}_{\alpha} \neq v^{i+1}_{\alpha}$ with $v_{\alpha}(\mathbf{k})=\partial \epsilon_{\mathbf{k}_{\alpha}}/\partial\mathbf{k}_{\alpha}$ defined in momentum space. In case the system is isotropic, as we consider in this work, then the velocity reads as $v_{\alpha}(E)$ where we solve the equation $\epsilon_{\mathbf{k}_{\alpha}}=E$ for $\mathbf{k}_{\alpha}$. These velocities play the role of correction according to the sample size, as we see in the end. Continuing with the Landauer derivation, it is framed by the $\mathrm{S}$-matrix evaluated for independent electrons i.e. noninteracting systems - as it has been derived in the Fisher-Lee formulation. However, to extend its viability to general systems, some assumptions are required as we show now below.\\
First, it is assumed all the incident carriers from the left reservoir are prepared at same chemical potential $\mu^{i}_{L} \equiv \mu_{L}$ and similarly for all carriers from the right reservoir $\mu^{i}_{R} \equiv \mu_{R}$. In general we take unequal chemical potential on each side $\mu_{L} \neq \mu_{R}$ to ensure finite current across the sample. These chemical potentials identify the unbalanced but constant thermal state in each reservoir population prior to any scattering processes. They also indicate condition of fast thermalization for incident electrons at the proximity of leads. Furthermore, by assuming the same chemical potential in each channel per sample side, all channels carry the same current $I$. This current is then flowing due to applied external voltage bias $V_{bias} = (\mu_{A}-\mu_{B})/e$, here defined with respect to perfectly conducting leads potential. These are not the reservoirs in contact with the central region such that we have in general $\mu_{A} \neq \mu_{L}$ and $\mu_{B} \neq \mu_{R}$. \\
Secondly, to account for finite temperature conditions, the Fermi distributions in the reservoirs are included in the current equation as a weight and they are also calculated over the whole continuum energy spectrum for positive values. This assumption is completely heuristic, since these distributions are introduced by hand in the derivation.\\
We implement now these requirements within the framework of scattering theory and we start by introducing the zero temperature current of a two-probe model, namely
\begin{equation}
I(T=0,V_{bias}) = ev \frac{\partial n(E)}{\partial E} (\mu_{L} - \mu_{R}) 
= ev \frac{1}{h/2 v(E)} \mathcal{T}(\mu_{L} - \mu_{R}) ~,
\end{equation}
where $v(E)$ is the velocity for a isotropic system, $\partial n(E) / \partial E$ is the density of states in a one dimensional sample and $\mathcal{T}$ the total transmission probability as we saw in the Landauer approach. The corresponding finite temperature current of two-probe model in linear response reads,
\begin{equation}
I(T,V_{bias}\to 0) = \frac{e}{h} \int_{0}^{+ \infty} dE \sum_{\alpha=L,R}\frac{-\partial f^{\mu_{\alpha}}(E-\mu_{\alpha})}{\partial E} \underbrace{(\mu_{L}-\mu_{R})}_{eV_{bias}} \mathcal{T}(E) ~.
\end{equation}
Recalling the current flux conservation implies $\mathcal{T}_{LR} =\mathcal{T}_{RL} \equiv \mathcal{T}$ and the linear variation of Fermi distribution is given by $f^{\mu_{L}}-f^{\mu_{R}} \simeq - (\partial f^{\mu_{R}}(E)/ \partial E) (\mu_{L}-\mu_{R})$ and analogously for $- (\partial f^{\mu_{L}}(E)/ \partial E)$, we can manipulate this the finite temperature current equation into
\begin{equation}\label{eq:LB_2probe}
I(T,V_{bias}\to 0) = \frac{2e}{h} \int_{0}^{+ \infty} dE \left( f^{\mu_{L}}(E-\mu_{L})\mathcal{T}_{LR}(E) - f^{\mu_{R}}(E-\mu_{R})\mathcal{T}_{RL}(E) \right) ~.
\end{equation}
The generalization of Eq.\ref{eq:LB_2probe} for multiple channels sample is straightforward resolved by introducing a summation upon $N$-channels for the transmission probability i.e. $\sum_{i,j}^{N} \mathcal{T}_{ij}$.\\ 
By means of the current definition in Eq.\ref{eq:LB_2probe}, the electrical conductance $\mathcal{G}$ is determined by tracking each scattering process as transmission and reflection probabilities, then by incorporating those as elements of the $\mathrm{S}$-matrix. We present now the conductance equation in two common situations.\\
In the case of \textit{homogeneous multiple channel}, the set-up is built with the same number of channels $n^{i}_{L} = n^{i}_{R} \equiv N $ and the same longitudinal velocity per channel $v^{i}_{L}(E) = v^{i}_{R}(E) \equiv v_{i}(E)$ in both sample sides. Generalizing now to the \textit{ac-}conductance case at finite temperature, one obtains
\begin{equation}
\mathcal{G}^{ac}(\omega,T,V_{bias}\to 0)= \frac{2e^{2}}{h/2} \sum_{i}^{N} \mathcal{T}_{i}(E) \dfrac{2 / v_{i}(E)}{(1+\mathcal{R}_{i}(E) - \mathcal{T}_{i}(E))/ v_{i}(E)} ~.
\end{equation}
In case these requirements are lifted, we obtain the general form that is the \textbf{electrical \textit{ac-}conductance Landauer-B{\"u}ttiker-formula}:
\begin{equation}
\mathcal{G}^{ac}(\omega,T,V_{bias} \to 0)= \frac{2e^{2}}{h/2} \dfrac{\sum^{n^{i}_{L}}_{i}\mathcal{T}_{i}(E)}
{1 +
\sum^{n^{i}_{L}}_{i} \frac{1}{v^{i}_{L}(E)} \left( \sum^{n^{i}_{L}}_{i} \dfrac{\mathcal{R}_{i}}{v^{i}_{L}}(E) \right)
-  \sum^{n^{i}_{R}}_{i} \dfrac{1}{v^{i}_{R}(E)} \left( \sum^{n^{i}_{R}}_{i} \dfrac{\mathcal{T}_{i}}{v^{i}_{R}} (E)\right) } 
\end{equation}
and this is the most general formulation for electrical conductance for noninteracting mesoscopic system as it is calculated at $T \neq 0$, in the \textit{ac-}regime, for an arbitrary number of channels per sample side with correction on $\mathcal{T},\mathcal{R}$ probabilities due to the different incoming/outgoing carrier velocities. Once again we see that the scattering approach is a well controlled tool to tackle transport problem for noninteracting models, since it relies on exactly solvable noninteracting Green's functions to construct the $\mathrm{S}$-matrix.\\ 
We observe that in the limit of zero temperature, where both reservoirs population are filled up to the Fermi energy, in the \textit{dc-}regime, the Landauer-B{\"u}ttiker formula reduces to the original Landauer formula for a single-channel per lead configuration i.e. $\mathcal{G}^{dc}(\omega\to 0,T=0,V_{bias}\to 0) = 2e^{2} / h \sum_{n} \mathcal{T}_{n}$. We deduce that the Landauer formulation is indeed the limiting case of the transport equation for multi-channel noninteracting systems in \textit{ac-}regime. Furthermore, we can confirm now the multiplicative property of the \textit{max conductance} $\mathcal{G}_{0}$ as the \textit{quantised unit} carried by each channel at max transmission condition, for small bias and zero temperature: $\mathcal{G}(\omega\to 0,T=0,V_{bias}\to 0) = N \mathcal{G}_{0}$. \\
As a last comment of the section, it is interesting to discuss the correction on the transmission and reflection probabilities due to finite sample size. For very small sample size, where ballistic mesoscopic quantum transport occurs, the electronic wavefunctions of incoming carriers reach the opposite side of the scattering barrier in a  plane-wane shape. Thus, each longitudinal velocity $v^{i}_{\alpha}(E)$ characterizes the wave composition and velocities are required to fully capture the scattering process in small systems. On the contrary, in larger sample size, the electronic wavefunctions evolve in an exponential decaying form within the scattering barrier size. Once they arrive at the opposite side, all incoming carriers have approximately reached the same longitudinal velocity. Hence, the conductance calculation does not require any channel renormalization due to different incoming and outgoing velocities but just a correction to the unique final longitudinal velocity. We infer that, by comparing the typical sample size, the mesoscopic scale meets the latter size requirement. At given conductance value, the increment of sample size and the consequent increment on the number of channels require transmission probability scaling as $\mathcal{T}_{i} \sim 1/N$. Thus, the Landauer-B{\"u}ttiker formula reduces to the Fisher-Lee type of conductance formula \cite{Beenakker_QuantumTransport1991} - but now still valid at $T \neq 0$, in \textit{ac-}regime for multiple channel set-up.\\
We conclude this section mentioning that the Landauer-B{\"u}ttiker formula offers a very general approach to study mesoscopic transport: both the original Landauer and the Fisher-Lee formulations can be studied as limit cases of it. In general, it is taken as benchmark since any conductance formulae applicable in interacting models should recover the Landauer-B{\"u}ttiker formula once interactions are removed.\\ 
We include now in our discuss the extension to multiple probes as derived in a proper format by B{\"u}ttiker.

\subsection*{The B{\"u}ttiker multiprobe device}
This section serves to complete the comprehensive survey on mesoscopic transport though noninteracting structures. We note that the presented Landauer-B{\"u}ttiker formula suffers from a dis-homogeneous treatment of the leads. Some are considered as source/sink to carry current, some other are used to measure voltage bias without affecting the actual transport process. Hence, each the incoming and outgoing carriers see a different part of the system according to the physical location of the channel. This unequal approach motivates an extension of the Landauer-B{\"u}ttiker formula  to overcome this ambiguity.\\
The first description of an actual multi-probe mesoscopic device, where the carriers see at equal foot all the terminals, is developed by M. B{\"u}ttiker \cite{Buttiker_ConductionSymmetryReview1988}. In this work, the mathematical essence of the original Landauer-B{\"u}ttiker formula is kept but now the system comprises terminals to measure voltages that also affect the current flow. Hence, both the voltage terminals and the ones to drive the current as probes are included in the conductance equation.\\
We shortly demonstrate this generalization by rewriting the current at finite temperature used in the Landauer-B{\"u}ttiker for a general $\alpha$-lead as
\begin{equation}
I(T,V_{bias}\to 0) = \frac{e}{h} \int_{0}^{+ \infty} dE \sum_{\alpha \neq \alpha^{\prime}} \frac{-\partial f^{\mu_{\alpha}}(E-\mu_{\alpha})}{\partial E} \underbrace{(\mu_{\alpha}-\mu_{\alpha^{\prime}})}_{e(V_{\alpha}-V_{\alpha^{\prime}})} Tr[\mathbb{t}^{\dagger}(E)\mathbb{t}(E)]   ~.
\end{equation}
Following similar manipulations to those already used before, it can be derived a general current equation that is the \textbf{multiprobe B{\"u}ttiker current formalism} \cite{Buttiker1986,DiVentra}:
\begin{equation}
\begin{aligned}
I(T,V_{bias}\to 0) & \stackrel{(i)}{=} \frac{e^{2}}{h} \int_{0}^{+ \infty}  dE \frac{-\partial f^{\mu_{\alpha}}(E-\mu_{\alpha})}{\partial E} \sum_{\alpha \neq \alpha^{\prime}} 
\left(\mathcal{T}_{\alpha \alpha^{\prime}}(E)V_{\alpha} - \mathcal{T}_{\alpha^{\prime} \alpha}(E)V_{\alpha^{\prime}}\right) ~,\\
& \stackrel{(ii)}{=}\frac{2e}{h} \int_{0}^{+ \infty} dE \sum_{\alpha \neq \alpha^{\prime}}  \left(  f^{\mu_{\alpha}}(E-\mu_{\alpha})\mathcal{T}_{\alpha \alpha^{\prime}}(E) - 
f^{\mu_{\alpha^{\prime}}}(E-\mu_{\alpha^{\prime}})\mathcal{T}_{\alpha^{\prime} \alpha}(E)  \right)~,
\end{aligned}
\end{equation}
where we have shown it in two different versions to highlight the physics behind the multiple channel set-up: $(i)$ demonstrates the linear relation between the current $I$ for applied voltage bias $V_{\alpha}$ through the total transmission probabilities $\mathcal{T}$, $(ii)$ is the direct extension of the two-probe case and it shows the current $I$ is given by the difference between the current flowing towards the $\alpha$-probe from all the other probes and the current flowing outwards from $\alpha$-probe towards all the other probes. This very intuitive result demonstrates the scattering theory accuracy to describe transport problem in absence of interactions. 

\subsection*{Pros and cons in the Landauer approach}
At the end of this section where we have discussed in detail the development of the transport theory in noninteracting structures, we want to make some final remarks.\\ 
The scattering theory sets the fundamental tools to approach the  phenomenology of transport in noninteracting systems at finite temperatures. Its exactly solvable mathematical formulation using  equilibrium, noninteracting Green's functions, see definitions  in Sec.\ref{sec:Gtheo}, and its straightforward applicability to systems in various condition make the machinery introduced by Landauer and B{\"u}ttiker extremely versatile to many physics problems. \\
Early attempts to derive an interacting version of Landauer equation were made for instance for zero temperature systems with electron-electron interactions modelled as scattering phase-shift \cite{Langreth1966}. Another one was obtained for infinite one-dimensional system at finite temperature, with interactions parametrized as zero or twice the momenta at Fermi level: at vanishing energy frequency, in this configuration it was derived a single-channel Landauer formula \cite{ApelRice1DInterLan1982}. However, these were not completely successful results. More recently, leads partitioning was showed to give Landauer-type transport \cite{Lopes2020landauerPartitioned}.\\
In the context of our work, we seek for methodology beyond the noninteracting conditions: this is indeed the crucial limitation of the Landauer formalism. This motivates both the succeeding standard approaches as presented in this chapter and, foremost, the alternative implementations for interacting system we derive in Sec.\ref{sec:AlternativeLandauer}. 

\section{Theory of transport in models under linear response: the Kubo formula}\label{sec:Kubo}
For both equilibrium and out-of-equilibrium regimes, quantum transport is generally a process modelled according to the dynamics between the thermal equilibrium electronic reservoirs and the nanostructure, whose features highly affect the transmission across the sample - see a cartoon of a quantum impurity model in Fig.\ref{F3:ImpScheme}.\\
In generic elaborate models, it is convenient to determine the transport process with no restriction on the temperature, local potential on the central region, or specific coupling at the quantum point contacts between leads and impurity. Furthermore, in general we wish to allow the nanostructure to be interacting.\\
An important subject in the theory of mesoscopic transport through general nanostructure is therefore the study of conductance under \textit{linear response theory} (LR) \cite{Coleman,Szunyogh2009}. This vast theoretical framework finds its application in quantum transport by investigating the effect of small but finite electronic gradient across the sample, which induces a current linearly dependent on the voltage bias. This gives access to the corresponding differential conductance under linear response, see Eqs.\ref{eq:DiffCondLRac},\ref{eq:DiffCondLRdc}, is evaluated in the limit of vanishing voltage bias. Evidences of the linear dependence in $V_{bias}$ is tracked in the expression for the current Eq.\ref{eq:defIeom} entering in the dynamical susceptibility and in the Fermi distribution Eq.\ref{eq:defFermiNeq} at unbalanced electronic distribution. Thus, under LR, instead of searching for the ideal temperature window and the most effective connection among the elements, we identify an appropriate external force or \textit{perturbation} that drives the system out of equilibrium, and estimate the resulting current to leading, in this case linear, order perturbatively.\\
In this section, we commence with presenting the linear response theory in its general terms and then we introduce the standard framework: the Kubo formula.\\

\noindent{We} need to introduce a different nomenclature form transport, distinct from the scattering theory in Sec.\ref{sec:Landauer} and the nonequilibrium theory in Sec.\ref{sec:MeirWingreen}. Instead of expressing the transport with respect to temperature, coupling between nanostructure-bath and interactions, we introduce the \textit{source operator} $\widehat{Q}_{\alpha}$, where $\alpha$ indicates which bath the source term applies to. This operator is chosen according to the transport process of interest - thus, we recognise already the versatility of this formulation.\\
In thermal equilibrium condition, the electrical population in the bath is regulated through the equilibrium Fermi distribution in Eq.\ref{eq:defFermieq}. At infinitely distant past, each bath mode is set at $T$ temperature and the difference in chemical potential $\Delta \mu$ between baths is identically zero, hence $V_{bias}=0$. Thus, in the system there is no current flow and no net transport occurs through the sample. By  contrast, the presence of an \textit{external perturbation} $\Delta\Omega(t)$, which can be time dependent, produces an electrical gradient across the systems such that a net flow results. We stress that by applying any external force to the model, the system is driven away from its initial equilibrium state - and so nonequilibrium theories are required for any transport phenomena, see more discussion in Sec.\ref{sec:MeirWingreen}. However, the advantage of the linear response condition is that the first order nonequilibrium correction to observables can be related to the equilibrium correlation functions.\\
In order to describe these two quantum states i.e. the initial equilibrium state with no transport and the final out-of-equilibrium state with transport process, we introduce the general model Hamiltonian
\begin{equation}
	\hat{H} = \hat{H}_{0} + \hat{H}^{\prime}(t) ~,
\end{equation}
where the full model Hamiltonian is given by the time-independent system initially at thermal equilibrium $\hat{H}_{0}$ plus the perturbed one $\hat{H}^{\prime}(t) =\widehat{Q}_{\alpha}(t)\Delta\Omega(t)$. As presented here, although these states seem to be disconnected, as we see later, by assuming a slow switch-on of a weak perturbation, then the two states \textit{are} smoothly connected.\\
We define the \textbf{current operator} $\widehat{I}$ given some source $\widehat{Q}_{\alpha}$ as:
\begin{equation}\label{eq:defIeom}
	\begin{aligned}
		\langle \widehat{I}^{Q}\rangle(t) &=  \langle \dfrac{d\widehat{Q}_{\alpha}(t)}{dt} \rangle \equiv \langle \dot{Q}_{\alpha}(t) \rangle ~,\\
		&= + i \langle[\hat{H},\widehat{Q}_{\alpha}(t)]\rangle ~,
	\end{aligned} 
\end{equation}
where in the second line we have used the Heisenberg equation of motion: the commutator $[.. , ..]$ is computed for the full Hamiltonian $\hat{H}$ with the current operator, and $\langle \dots \rangle$ indicates the thermal expectation value. Of relevance for this work, the current in Eq.\ref{eq:defIeom} is studied for electrical and heat transport. In the former case, the source is $\widehat{Q}_{\alpha}(t) = \widehat{N}_{\alpha}$ number operator for $\alpha$-bath such that the \textbf{charge current} is given by
\begin{equation}\label{eq:defChargeCurr}
	\widehat{I}^{C}(t) = \langle \dot{N}_{\alpha}(t) \rangle= -i e \langle[\hat{H},\widehat{N}_{\alpha}(t)]\rangle ~.
\end{equation}
In the latter case, the source is $\widehat{Q}_{\alpha}(t)= \widehat{H}_{\alpha}$ bath Hamiltonian so that the \textbf{energy current} reads
\begin{equation}\label{eq:defEnCurr}
	\widehat{I}^{E}(t)  = \langle \dot{H}_{\alpha}(t) \rangle= i  \langle[\hat{H},\widehat{H}_{\alpha}(t)]\rangle ~.
\end{equation}
Furthermore, by recalling the first law of thermodynamics $dE = dQ + \mu dN$ for $E$ energy, $Q$ heat and $N$ number of particle, we deduce the \textbf{heat current} as
\begin{equation}
\widehat{I}^{Q}(t)= \widehat{I}^{E}(t)- \mu \widehat{I}^{N}(t) ~.
\end{equation} 
We note that the particle current can be understood as an equivalent concept of the \textit{thermodynamic work} but now in the context of impurity physics models. However, this is beyond the scope of this thesis and we refer elsewhere for appropriate discussions.\\ 
The importance of treating transport in terms of general current operators and perturbation - as key elements in the linear response theory - it is that we can tackle both electrical and heat transport phenomena, as we see in the next section. 

\subsection{The Kubo formula for conductance}
In this section we show how the transport calculation under linear response can be addressed on both electric and heat processes, despite their unequal treatment at microscopic level. Before presenting this formulation, we briefly reflect on such an intrinsic difference.\\
In case of electronic flow through the sample, the charge variation probed at the baths attached to the central nanostructure is regulated by applying an external voltage bias $V_{bias}$ to the baths themselves - we refer to those as \textit{source} and \textit{drain} of an electrical circuit. This perturbation corresponds to the Hamiltonian, namely
\begin{equation}
\hat{H}^{\prime}(t) = -e |\Delta V_{bias}| \widehat{N}_{\alpha}\cos(\omega t) ~,
\end{equation}
where $\omega$ is the driving frequency encapsulated in a sinusoidal function. As side comment, we note the perturbation equation form is a time-independent operator times time-dependent function $\Delta\Omega(t)$: such combination is particularly amenable to be Fourier transformed into frequency domain \cite{WeberArfken}. In particular, the Fourier transformation of the perturbation gives 
\begin{equation}
\int^{+\infty}_{-\infty} dt \Delta\Omega(t) e^{i  \widetilde{\omega} t} ~ \propto ~ \delta(\omega + \widetilde{\omega}) + \delta(\omega - \widetilde{\omega}) ~,
\end{equation}
and that implies in Fourier space $\hat{H}^{\prime}(\omega \equiv \pm \widetilde{\omega})$.\\
On the contrary, in case of heat flow, due to a \textit{temperature gradient} there is no equivalent perturbation that can be added into the bare model Hamiltonian. The reason stems on the lack of an actual microscopic meaning of temperature, since temperature is a statistical property of the leads at thermal equilibrium. Instead we simply consider a model with leads held at different temperatures.\\
These considerations bring us insights on which necessary assumptions are required to described both electrical and heat transport using linear response formalism. There are \textit{four assumptions} inherent in the linear response Kubo approach to quantum transport \cite{DiVentra}, as we list now: \\
$(i)$ it is to consider closed but not isolated systems, such that they can undergo external forces;\\
$(ii)$ at infinite distant past $t\rightarrow -\infty$, the system state stays in thermal equilibrium with baths at $T$ temperature;\\
$(iii)$ at $t_{0}$, a small force perturbs the system and the model Hamiltonian at $t>t_{0}$ becomes $\hat{H} = \hat{H}_{0} + \hat{H}^{\prime}(t)$. The response of the system under such perturbation is linear in the perturbation coupling. We therefore assume the \textit{adiabatic approximation}: the time evolution of the system induced by the perturbation is faster than the energy relaxation time that brings the system to global equilibrium. The corresponding energy relaxation length $l_{rel}$ compared to the mesoscopic length scales discussed already sets as $L< l_{\phi} \lesssim l_{rel}$; \\
$(iv)$ There is no energy shifts in the unperturbed system in the presence of the bath. This implies: $(a)$ the correlations are neglected in the distant past $t\rightarrow -\infty$ and so also at $t\rightarrow +\infty$, such that at $t_{0}$ the system is in steady state and $(b)$ the adiabatic switching on and off of perturbation smoothly connects equilibrium and out-of-equilibrium states - ultimately this greatly simplifies practical calculations.\\
By means of these assumptions, the transport under linear response encompasses equilibrium quantities only - as the calculation is made only with respect to the system parameters at $t<t_{0}$. We conclude this formalism is indeed consistent also for heat transport events, since no reference to nonequilibrium parameters as $\Delta T$ is made, provided $\Delta T$ is infinitesimal.\\

\noindent{Having} motivated the concept of linear response theory physically, we turn now our discussion to a quantitative level and we introduce the key elements for calculating transport under linear response ultimately leading to the Kubo formulation.\\
As starting point, we define the \textit{current-current correlator} as the following two-point correlation function, namely
\begin{equation}\label{eq:defCurrCurrCor}
K_{Q,Q^{\prime}} (t, t^{\prime}) = \langle\langle\widehat{I}_{Q}(t);\widehat{I}_{Q^{\prime}}(t^{\prime}) \rangle\rangle_{0} ~,
\end{equation}
where we use the double-bracket notation evaluated in absence of any perturbation. By recalling that the initial model $\hat{H}_{0}$ is time independent, the correlation is a function only of differences in time coordinates and so we can write $K_{Q,Q^{\prime}} (t, t^{\prime}) = K_{Q,Q^{\prime}} (t- t^{\prime})$. Then, we define the \textit{dynamical susceptibility} or \textit{response function} as
\begin{equation}\label{eq:defSuscept}
	\begin{aligned}
		\chi (t, t^{\prime}) &\equiv \chi (t-t^{\prime}) ~,\\
		&= K^{R}_{Q,Q^{\prime}} (t- t^{\prime})  
		= - i \theta( t - t^{\prime}) \langle [\widehat{I}_{Q}(t),\widehat{I}_{Q^{\prime}}(t^{\prime})]\rangle_{0} ~,
	\end{aligned}
\end{equation}
where we have used the definition of retarded correlator, see Eq.\ref{eq:defGeq}. The full current expectation value is given by the integration over the susceptibility multiplied for the perturbation, that is to say
\begin{equation}
	\langle\widehat{I}_{Q}\rangle(t) = \int^{+\infty}_{-\infty} dt^{\prime}\chi (t-t^{\prime})\Delta\Omega(t^{\prime}) + \mathcal{O}( \hat{H}^{\prime})^{2} ~,
\end{equation}
where we have emphasised the result is \textit{linear order} in perturbation expansion. This integral is of the form of a convolution product such that it is straightforward to perform its Fourier transform to get the current expectation value in frequency domain, namely
\begin{equation}
	\langle\widehat{I}_{Q}\rangle(\omega) = \chi(\omega) \Delta\Omega(\omega) +\mathcal{O}( \hat{H}^{\prime})^{2}  ~,
\end{equation}
where the resulting frequency coincides with the driving one $\omega \equiv \widetilde{\omega}$ and the susceptibility is evaluated under adiabatic approximation according to
\begin{equation}
	\chi(\omega) = \dfrac{i}{\omega}\big( K^{R}_{Q,Q^{\prime}}(\omega) - K^{R}_{Q,Q^{\prime}}(0) \big) ~,
\end{equation}
with the corresponding current-current retarded correlation given by
\begin{equation}\label{eq:defKuboLR}
	K^{R}_{Q,Q^{\prime}}(\omega) = \int^{+\infty}_{-\infty} dt^{\prime} \Big(-i\theta( t- t^{\prime}) \langle [\widehat{I}_{Q}(t),\widehat{I}_{Q^{\prime}}(t^{\prime})]\rangle_{0} \Big) e^{i \omega ( t- t^{\prime})} ~.
\end{equation}
Of fundamental importance, the Eq.\ref{eq:defKuboLR} is the \textbf{Kubo formula for linear response conductance}: it calculates the linear susceptibility as response function of a general current in a system subjected to perturbation on its baths \cite{Kubo1956,Kubo1957}. Crucially, Eq.\ref{eq:defKuboLR} is computed only in terms of the unperturbed model.\\
We have now defined all the elements we need for introducing the charge and heat Kubo formulas. Notably, the Kubo approach finds its application in several domains, for instance in systems in presence of magnetic field, in the calculation of dielectric function and for conductivity. In this work we are interested in the Kubo formula Eq.\ref{eq:defKuboLR} for \textit{conductance} hence to evaluate the differential conductance in \textit{ac-} and \textit{dc-} regimes, see Eqs.\ref{eq:DiffCondLRac},\ref{eq:DiffCondLRdc}. We continue now by presenting the conductance formula for general source operator $\widehat{Q}_{\alpha}$ and then by describing its expression for electrical and heat transport.\\
The \textit{linear response function} in Eq.\ref{eq:defSuscept} for \textit{conductance} reads
\begin{equation}\label{eq:defSusceptCondAC}
\chi(\omega, T) =\Big( \dfrac{e^{2}}{h}\Big) 2\pi \dfrac{-\mathit{Im}K^{R}_{Q,Q^{\prime}}(\omega, T)}{\omega}~,
\end{equation}
where we indicate $\mathcal{G}_{0}= e^{2}/h$ the unitary quantised conductance for spinless Fermions. Of interest for this work, we often look at the low frequency states that is the \textit{dc-}regime. We can derive the \textit{dc-limit conductance} from Eq.\ref{eq:defSusceptCondAC} taking the limit for $\omega \rightarrow 0$, namely
\begin{equation}\label{eq:defSusceptCondDC}
\chi^{dc}(T) = \lim_{\omega \rightarrow 0} \dfrac{e^{2}}{h}2\pi  \dfrac{-\mathit{Im}K^{R}_{Q,Q^{\prime}}(\omega,T)}{\omega} ~,
\end{equation}
where we make use of the analyticity properties of the current-current correlator in Eq.\ref{eq:defCurrCurrCor} and those are $\mathit{Re}K^{R}(\omega) = \mathit{Re}K^{R}(-\omega)$ and $\mathit{Im}K^{R}(\omega) = -\mathit{Im}K^{R}(-\omega)$ \cite{IzumidaSakai1997}.\\
Finally, the \textbf{Kubo formula for electrical conductance} is obtained by inserting the definition of charge current in Eq.\ref{eq:defChargeCurr} into the response function in Eq.\ref{eq:defSusceptCondAC} and so we have
\begin{equation}\label{eq:defKuboLRel}
\mathcal{G}^{C}(\omega,T) = \dfrac{e^{2}}{h} 2\pi \dfrac{-\mathit{Im}K^{R}_{N_{\alpha},N_{\alpha^{\prime}}}(\omega, T)}{\omega} ~,
\end{equation}
this is the \textit{ac-}regime conductance, compare it to Eq.\ref{eq:DiffCondLRac}, with the current-current correlator for variation of charge number on the bath defined as
\begin{equation}
	K^{R}_{N_{\alpha},N_{\alpha^{\prime}}}(\omega, T) = 
	\int^{+\infty}_{-\infty} dt^{\prime} \Big(-i\theta(t^{\prime}) \langle [\dot{N}
	_{\alpha}(t^{\prime}),\dot{N}_{\alpha^{\prime}}]\rangle_{0} \Big) e^{i \omega t^{\prime}}
\end{equation}
and its corresponding \textit{dc-}limit follow similarly as in Eq.\ref{eq:defSusceptCondDC} by taking the zero-frequency limit of Eq.\ref{eq:defKuboLRel}, compare it to Eq.\ref{eq:DiffCondLRdc}. \\
The \textbf{Kubo formula for heat conductance} in \textit{ac-}regime is analogously obtained by using the definition of energy current in Eq.\ref{eq:defEnCurr} so we have
\begin{equation}\label{eq:defKuboLRhea}
	\mathcal{G}^{E}(\omega, T) = \frac{1}{hT} 2\pi \dfrac{-\mathit{Im}K^{R}_{H_{\alpha},H_{\alpha^{\prime}}}(\omega, T)}{\omega}  ~,
\end{equation}
where $1/hT$ is the unitary quantised heat conductance for spinless Fermions, with the current-current correlator for variation of energy on the bath defined as
\begin{equation}
	K^{R}_{H_{\alpha},H_{\alpha^{\prime}}}(\omega, T) = 
	\int^{+\infty}_{-\infty} dt^{\prime} \Big(-i\theta(t^{\prime}) \langle [\dot{H}_{\alpha}(t^{\prime}),\dot{H}_{\alpha^{\prime}}]\rangle_{0} \Big) e^{i \omega t^{\prime}}
\end{equation}
and its corresponding \textit{dc-}limit is obtained as mentioned before.\\
We can now summarise our findings to bring forward our discussion. Once we have worked out the susceptibility - either in \textit{ac} Eq.\ref{eq:defSusceptCondAC} or in \textit{dc} Eq.\ref{eq:defSusceptCondDC} regimes as derived from Kubo formula in Eq.\ref{eq:defKuboLR}, we can calculate the conductance characterising the transport events for electrical flux in Eq.\ref{eq:defKuboLRel} and for heat process in Eq.\ref{eq:defKuboLRhea}.\\
Furthermore, we can relate the transport coefficients for charge and heat currents to the current-current correlator in Eq.\ref{eq:defCurrCurrCor}. This result \cite{Luttinger1964} into the matrix product between the components for isolated charge $\chi^{CC}$, isolated heat $\chi^{QQ}$, thermopower $\chi^{CQ},\chi^{QC}$ and the variation in voltage bias and in temperature, namely
\begin{equation}
	\begin{pmatrix}
		\widehat{I}^{C} \\
		\widehat{I}^{Q}
	\end{pmatrix} =
	\begin{pmatrix}
		\chi^{CC}  & \chi^{CQ}\\
		\chi^{QC} & \chi^{QQ}
	\end{pmatrix}
	\begin{pmatrix}
		\Delta V_{bias}\\
		\Delta T / T
	\end{pmatrix} ~.
\end{equation}
The heat current reads $\widehat{I}^{Q}= \kappa \Delta T$ with $\kappa$ the heat conductance and we have two possible cases: $(i)$ at $\Delta V_{bias}=0$, $ \kappa = \chi^{QQ}/T$ and $(ii)$ at $\widehat{I}^{C}=0$, $ \kappa =(\chi^{QQ} - \chi^{QC}\chi^{CQ}/\chi^{CC})/T$. The electrical current at $\Delta T =0$ reads $\widehat{I}^{C}= \chi^{CC} \Delta V_{bias}$ with $\chi^{CC}$ indicating the electrical conductance.\\

\noindent{We} conclude this section with an emphasis on the manifestation of the \textit{Fluctuation-Dissipation theorem} in the Kubo formulation, as indeed presented originally by Kubo \cite{KuboFDT1966}.\\
We consider a general observable $\widehat{A}$ and we calculate its correlation function $K_{A,A^{\prime}}(\omega)$ using Eq.\ref{eq:defCurrCurrCor}. We study then its linear response to small perturbation $\Delta A(t)$ by calculating the response function or susceptibility $\chi_{A,A^{\prime}}(\omega)$ in Eq.\ref{eq:defSuscept}. It is a statistical mechanics proof to show that the imaginary part of the response function represents the energy dissipation of the system. Thus, we can derive the relation between spontaneous fluctuations and energy dissipation given by
\begin{equation}
	\underbrace{K_{A,A^{\prime}}(\omega)}_{fluctuations} = 2 \Big( 1 + \dfrac{1}{e^{\beta \omega}-1} \Big)\underbrace{\mathit{Im}\chi_{A,A^{\prime}}(\omega)}_{dissipation} ~, 
\end{equation}
where the term $(e^{\beta \omega}-1)^{-1}$ is the Bosonic distribution and so this equation is the Bosonic analogous of the fluctuation-dissipation equation presented for Fermions in Eq.\ref{eq:defFDtheo}.\\
From this final discussion, we evince that the Kubo approach to calculate linear response function under external perturbation underlies the intrinsic relation occurring between the \textit{spontaneous fluctuations} to bring the system at equilibrium - or to restore it in case external forces have already been applied - and the \textit{dissipations} to drive the system towards nonequilibrium states.

\subsection*{Pros and cons in the Kubo approach}
The Kubo formula for electrical and heat conductance in Eqs.\ref{eq:defKuboLRel},\ref{eq:defKuboLRhea} in their version for both \textit{ac-} and \textit{dc-} regimes, offer a solid formulation applicable to general models - with no constraints imposed to the system characterisation. Despite of the substantial theoretical framework of transport under linear response, the Kubo formulas hide two significant issues.\\
The first is related to the practical evaluation of the current-current correlator in Eq.\ref{eq:defCurrCurrCor} for the electrical conductance. Due to its structure, the evaluation of $K^{R}_{N_{\alpha},N_{\alpha^{\prime}}}$ using the equation of motion technique leads to cumbersome expressions that do not close in continued fraction form. At each cycle, the new expression is producing mixed term involving factors from each baths and the dot - this is a \textit{local} correlator. We conclude that in case the device is engineered with particularly articulated contacts to the central region, the linear response Kubo formula Eq.\ref{eq:defKuboLR} is model-specific. In involves $\dot{N}_{\alpha}$ operators which are obtained as $i\left[\hat{H}_{hyb},\hat{N}_{\alpha}\right]$ and hence depend on junction details. Hence, it is fully determined by the hybridizing nanostructure orbitals and by the local lead orbitals. Under this condition, which we are generally interested in for advanced modelling, the usage of the standard Kubo formula is again cumbersome.\\
The second issue has its numerical reasoning. In the \textit{dc-}limit regime, we have tacitly hidden the imminent \textit{indeterminate} result at strictly zero frequency \cite{IzumidaSakai1997}. A well-defined and finite linear response conductance in the \textit{dc-}regime implies  $\mathit{Im}K^{R}_{X,X^{\prime}}(\omega\rightarrow0,T)\sim \omega$ for $\hat{X}=\hat{N},\hat{H}$. Furthermore, the Kubo prescription involves dividing by frequency and taking the low-$\omega$ limit, which is numerically poorly controlled. In fairness, the interesting physics of lead-coupled interacting nanostructures is often controlled by properties emerging only at low-energy scales such as the Kondo temperature $T_{K}$ given in Eq.\ref{eq:Tk}. The renormalization group structure of the low-energy physics, see Sec.\ref{sec:RGtheo}, means that the system behaviour in the limit $\omega\rightarrow0$ can be extracted for finite $|\omega| \ll T,T_{K}$. However, in this regime, $\mathit{Im}K^{R}_{X,X^{\prime}}$ is itself \textit{very} small. Any numerical implementation of the correlator must therefore be very accurate over a sufficient range of $|\omega| \ll T,T_{K}$ such that the conductance can be deduced. \\
We mention that even with highly accurate methods such as NRG, this computation represents a challenge \cite{Wilson1975,Bulla2008}. A further issue with the NRG implementation is that, even with the full density matrix approach \cite{Weichselbaum2007,AndersPruschkeNRGGF2006}, dynamical information is not reliably obtained for $|\omega|\ll T$. This can make it difficult to extract the \textit{dc-}limit conductance numerically using the standard Kubo formula as presented in Eqs.\ref{eq:defKuboLRel},\ref{eq:defKuboLRhea}. \\
It is therefore desirable to derive a Kubo version that exempts from those limitations such that the formulation finds its effective applicability in any frequency range and for any impurity model configuration. We advance in these directions in Sec.\ref{sec:ImprovedKubo}.

\section{Theory of transport in interacting models: the Meir-Wingreen formula and Keldysh formalism}\label{sec:MeirWingreen}
As emphasized in Sec.\ref{sec:Landauer}, transport is intrinsically a nonequilibrium process. The Keldysh formalism developed in the context of  many-body nonequilibrium theory, as we introduced it in Sec.\ref{sec:Gtheo}, presents the appropriate methodology to study transport in out-of-equilibrium condition. Within such framework, the most general transport approach can be derived to investigate a completely unrestricted class of systems regardless the presence of interactions on the nanostructure, the geometric configuration of the set-up and the temperature regime. \\
This section illustrates the fundamentals on the methodology that is the current formula derived by Y. Meir and N.S. Wingreen: their formalism is intrinsically determined by the local properties of the central structure. While we present the enormous advantages of this formulation on a theoretical ground, we highlight already the crucial difficulties in determining any exact solutions in practical application on elaborate systems - except if some approximations are taken.\\
In this section we showcase the common literature strategies to solve the Meir-Wingreen formula in nonequilibrium mesoscopic transport processes. We corroborate our discussion with the full derivation of the expression in the Appendix \ref{app:MW} and we consider it as reference for the details on nonequilibrium technical calculations.

\subsection*{Transport out-of-equilibrium: the Meir-Wingreen formula for time-independent current}
We have introduced the nonequilibrium formulation in Sec.\ref{sec:Gtheo}: we present now how this theory comes into play in the context of mesoscopic interacting transport by means of the \textbf{Meir-Wingreen current formula} (MW) defined on the Schwinger-Keldysh contour $\mathcal{C}_{\tau}$ \cite{MeirWingreen_1992,JauhoWingreeMeir1994}, see also the Fig.\ref{F2:KeldyshCont}.\\
In a system in steady-state, the \textit{time-independent current Meir-Wingreen formula} reads 
\begin{equation}\label{eq:MW}
\begin{aligned}
I =  \frac{ie}{2h}\sum_{\sigma}\bigintsss d\omega Tr \bigg[ 
& \Big( f^{\mu_{L}} (\omega-\mu_{L}) \Gamma^{L} -  f^{\mu_{R}}(\omega-\mu_{R}) \Gamma^{R} \Big) \Big( G_{dd\sigma}^{R}(\omega) -G_{dd\sigma}^{A}(\omega)\Big)+\\
&+ \Big( \Gamma^{L} - \Gamma^{R} \Big) G_{dd\sigma}^{<}(\omega) \bigg] ~.
\end{aligned}
\end{equation}
Eq.\ref{eq:MW} includes highly nontrivial elements: its fully detailed analytical derivation and its extension to time-dependent form are discussed in the Appendix \ref{app:MW}.  \\
We comment now on which elements are underlying in the MW formula. The most evident is Fermi-Dirac distribution defined in Eq.\ref{eq:defFermiNeq}, where the leads are held at chemical potential $\mu_{\alpha}$ and at temperature $T$. The voltage bias enter through a finite, time-independent $eV_{bias}= (\mu_{L}-\mu_{R})$. Otherwise at $\mu_{R} \equiv \mu_{L}$, the distributions converge to the common equilibrium potential difference function $f^{Eq}(\omega)$ and $I \equiv0$ - see further comments on this aspects in the Appendix \ref{app:MW}. Then, the specific form of the MW formula is due to the current calculated in the presence of noninteracting leads. The system is coupled to the leads and its tunnelling processes are regulated by hybridization function enclosed in the gamma function $\mathrm{\Gamma}^{\alpha}$ with respect to $\alpha$-lead. We note a wide band flat approximation of the conduction band is implicitly assumed in the baths such that $\mathrm{\Gamma}^{\alpha}$ has the standard definition as in Eq.\ref{eq:DoSinfWBL} and it directly depends on local quantity of the central interacting region. Lastly, but definitely most crucially, the MW formula depends on many-body interacting Green's functions of central region $G_{dd\sigma}$ as defined in Eq.\ref{eq:defGeq}, \ref{eq:defGneq}. Physically, these correlation functions are local quantities, fully determined by the geometry and the energy levels of the central region. Mathematically, these are nonequilibrium functions defined on the Schwinger-Keldysh contour. The difficulty is indeed calculating the nonequilibrium Green's functions. To sum up: once all these quantities are calculated, the \textit{Meir-Wingreen current formula can be formally evaluated}.\\
After this general introduction on the MW formula, we present the advantages and disadvantages of it. The latter gives motivation for implementing constraints in the model such that the MW formula can be calculated in practice. However, as we will mention at the end of this section, these literature strategies are not yet enough to tackle more elaborate structures like those we are interested in this thesis. Hence, these comments orient us towards alternative methodologies as derived in this thesis.

\subsection*{$(I)$ Pros and cons in the Meir-Wingreen approach}
As presented in Eq.\ref{eq:MW}, the MW formula is the most general method to calculate mesoscopic quantum transport in interacting nanostructure with no limitation on the specific form of the tunnel couplings nor on the temperature range. However, evaluating the MW equation in its original formulation presents significant challenges. Below we discuss some of these issues.
\begin{enumerate}
\item \textit{Requirement to calculate nonequilibrium Green's functions explicitly:}\\
The differential conductance gives non-vanishing values only in the presence of finite $V_{bias}$. The MW formula requires an explicit differentiation of Green's functions with respect to the applied voltage bias, see Eq.\ref{eq:DiffCondLRac}. In the MW formula in Eq.\ref{eq:MW}, the first multiplication of term shows an explicit voltage dependence inside the Fermi-Dirac functions. By contrast, the second term proportional to $G^{<}$ has no term containing an explicit $V_{bias}$ parameter. In principle, an implicit voltage dependence is enclosed in the self-energy $\Sigma(\omega,T,V_{bias})$. In this scenario, we would need to derive an exact expression for the nonequilibrium, interacting self-energy in order to perform the derivative with respect to the voltage. However, it is not possible to find exact analytical expressions for the nonequilibrium, interacting self-energy in general.\\
Early work by Meir and Wingreen \cite{MeirWingreenLee_StronglyInteractingElec_PRL1991} showed already, the asymmetric explicit voltage dependence in the MW formula. In that study, the linear-response or differential conductance is calculated only with respect to the first term of Eq.\ref{eq:MW} i.e. $\int df^{\mu_\alpha}/dV \mathit{Im} G^{R}$. Even if we invoke some approximation to neglect the current information given by the term proportional to $G^{<}$, we are still left with the unsolved problem of deriving an analytical expression for the nonequilibrium, interacting self-energy to feed in $G^{R}$. This is equivalent to finding the nonequilibrium, interacting self-energy - an open problem.
\item \textit{Absence of a known nonequilibrium  distribution formula:}\\
As we have seen in the fluctuation-dissipation theorem, see Eq.\ref{eq:defFDtheo}, an expression for the equilibrium $G^{</>}$ can be found but in terms of the known Fermi-Dirac distribution and the equilibrium retarded Green's function. However, the equivalent nonequilibrium distribution is not in general know. This also the reason why in the literature the equations \ref{eq:defFDtheo} are often regarded as \textit{ansatz}.
\end{enumerate}
We present now in the following the constraints adopted in the literature yielding to solvable variants of the MW equation.

\subsection{Solution for noninteracting models: the pseudo-equilibrium case}
We start with a simple but very effective strategy under which the MW formula in Eq.\ref{eq:MW} yields exact analytical solution: the removal of the interacting term in $\hat{H}_{imp}$ such that we are left with a mesoscopic transport process in noninteracting model at finite temperature. And the self-energy expression reduces only to its hybridization component, see Eq.\ref{eq:defSelEngeneral}. \\
We introduce some basics of \textit{nonequilibrium, noninteracting Keldysh formalism} using the contour-ordered Green's functions presented in Sec.\ref{sec:Gtheo}.\\
In nonequilibrium and noninteracting models, the perturbative scheme provided in the Dyson equation is analytically exact \cite{Maciejko_NonEquilibrium}. The \textit{noninteracting Dyson equation} for the retarded correlator in nonequilibrium formalism reads
\begin{equation}\label{eq:Dyson-RA-NEQ-NI}
\sum_{\sigma}G^{R}_{\sigma}(\omega)= \sum_{\sigma} \Big( G^{R}_{0\sigma}(\omega) 
+G^{R}_{0\sigma}(\omega) \underbrace{|V|^{2}G^{R}_{0\sigma}(\omega)}_{\Delta^{R}} G^{R}_{\sigma}(\omega) \Big)~,
\end{equation}
compare it with its full form in Eq.\ref{eq:defDysonRA}. We note that the nonequilibrium, noninteracting retarded hybridization $\Delta^{R}$ depends purely on isolated, noninteracting leads. Hence, the Eq.\ref{eq:Dyson-RA-NEQ-NI} is exactly solvable and in practical calculation it is accustomed to define the following identity valid for nonequilibrium and noninteracting models
\begin{equation}\label{eq:Id-GgammaG-NEQ}
G^{R}G^{A} \equiv   \frac{i(G^{R}-G^{A})}{\Gamma^{L}+\Gamma^{R}} ~,
\end{equation}
where the numerator is equal to the spectral function given in Eq.\ref{eq:defSpectral} and the denominator $\sum_{\alpha}\mathrm{\Gamma}^{\alpha}= \mathrm{\Gamma}^{L}+\mathrm{\Gamma}^{R} \equiv \Gamma$ is the total gamma function contribution from both equivalent, metallic leads in wide-flat band approximation as given in Eq.\ref{eq:DoSinfWBL}.\\
The \textit{noninteracting Dyson-Keldysh equation} for the lesser correlator, see its full form in Eq.\ref{eq:defDysonKeldysh} is given by
\begin{equation}
\sum_{\sigma}G^{<}_{\sigma}(\omega)=\sum_{\sigma} \Big( G^{R}_{\sigma}(\omega) \underbrace{|V|^{2}G^{<}_{0\sigma}(\omega)}_{\Delta^{<}}G^{A}_{\sigma}(\omega) \Big)~,
\end{equation}
where the noninteracting self-energy part namely the lesser hybridization function is defined as
\begin{equation}
\Delta^{<}(\omega) = i f^{\mu_{L}}(\omega) \Gamma^{L} +i f^{\mu_{R}}(\omega) \Gamma^{R}  ~,
\end{equation}
see more details on the noninteracting self-energy expression in Eq.\ref{A:eq:delta_Def}.\\
On mathematical perspective, since these equations are formally derived for contour-ordered noninteracting Green's functions on Schwinger-Keldysh contour, the presence of interactions imply the one-body potential is diagonal on $\mathcal{C}_{\tau}$. That means the interacting lesser component in the self-energy Eq.\ref{eq:defSelEngeneral} vanishes $\Sigma^{U,</>}=0$, leaving nonzero only $\Delta^{</>}$ term.\\ 
At a first glimpse, the last two expressions seem to offer a final resolution to MW equation, at least under noninteracting conditions. However, we are still left with the problem of which form is taken by the nonequilibrium distributions required in $\Delta^{<}$ expression. A solution is found in the so-called \textbf{pseudo-equilibrium distribution} \cite{MeirWingreen_1992,JauhoWingreeMeir1994}, as we aim to present it here adhering to the literature. It allows us to relate lesser Green's functions to retarded ones out of equilibrium.\\
As general model representative of noninteracting systems we can think of the \textit{noninteracting resonant level model}, as the one we presented in the context of the noninteracting coupled Anderson impurity model in Sec.\ref{sec:QImpModel}, see Eq.\ref{eq:NI-RLM}. In nonequilibrium conditions, this single-level model has exact analytical solutions: all its relevant quantities i.e. Green's functions, self-energies, gamma functions are scalars. Under these conditions and at finite temperature, we can derive the \textit{noninteracting, pseudo-equilibrium Fermi distribution}
\begin{equation}\label{eq:defFermiPseudoEq}
\overline{f}^{Eq}(\omega) \equiv \frac{f^{\mu_{L}}(\omega)\Gamma^{L}+f^{\mu_{R}}(\omega)\Gamma^{R}}{\Gamma^{L}+ \Gamma^{R}}  ~,
\end{equation}
using the definition of gamma function in Eq.\ref{eq:DoSinfWBL}  and the equation trivially reduces to the standard equilibrium Fermi distribution in Eq.\ref{eq:defFermieq} $\overline{f}^{Eq} \rightarrow f^{Eq}$ if $\mu_{L}=\mu_{R}$ - that is, at equilibrium. In case of structured leads, the expression \ref{eq:defFermiPseudoEq} still holds and we use the gamma function as $\Gamma^{\alpha}(\omega)$, see definition in Eq.\ref{eq:4defEqIneq}.\\
By Fourier transforming in energy the standard definition of  uncoupled, noninteracting lesser functions given in Eq.\ref{A:eq:g_Def} and by inserting in it $\overline{f}^{Eq}$, we end up to calculate the \textbf{pseudo-equilibrium form of lesser Green's function}, namely
\begin{equation}\label{eq:G<PseudoEq}
\begin{aligned}
\overline{G}^{<}_{dd\sigma}(\omega) 
& \stackrel{(i)}{\equiv} i \overline{f}^{Eq}(\omega) 2\pi\mathcal{A}_{0,dd\sigma}(\omega)= i f^{\mu_{L}}(\omega)G_{0,dd\sigma}^{R}(\omega)\Gamma^{L}G_{0,dd\sigma}^{A}(\omega) +
i f^{\mu_{R}}(\omega)G_{0,dd\sigma}^{R}(\omega)\Gamma^{R}G_{0,dd\sigma}^{A}(\omega) ~,\\
& \stackrel{(ii)}{\equiv} 2\pi i \overline{f}^{Eq}(\omega) \rho_{0,dd\sigma}(\omega) ~,
\end{aligned}
\end{equation}
where the expressions $(i)$ is derived by means of the identity in Eq.\ref{eq:Id-GgammaG-NEQ} and $(ii)$ is derived using the definition of the noninteracting version of the local density of states in Eq.\ref{eq:defDos}. All the correlation functions appearing in $\overline{G}^{<}_{dd\sigma}$ - see equation \ref{eq:G<PseudoEq} - are nonequilibrium, noninteracting retarded/advanced Green's functions that are solved in exact analytical form using the Dyson equation \ref{eq:Dyson-RA-NEQ-NI}. \\
Hence, we infer that by introducing a constraint on absence of interactions, it is possible to implement in standard single-level systems a fully solvable solution for $G^{<}_{dd\sigma}$ as we have shown in Eq.\ref{eq:G<PseudoEq}.\\

\noindent{At} this stage, we have obtained a tractable form of local lesser function $G^{<}_{dd}$ and we can calculate the corresponding noninteracting version of the MW formula. A straightforward plug-in of the identity in Eq.\ref{eq:Id-GgammaG-NEQ} and of the pseudo-equilibrium $\overline{G}^{<}_{dd}$ function in Eq.\ref{eq:G<PseudoEq} into the original MW current formula in Eq.\ref{eq:MW} we find 
\begin{equation}
I =  \frac{e}{h} \sum_{\sigma}\int d\omega  
\Big( f^{\mu_{L}} (\omega)-  f^{\mu_{R}}(\omega) \Big) Tr \Big[4 \Gamma^{L}G^{R}_{0,dd\sigma}(\omega) \Gamma^{R}G^{A}_{0,dd\sigma}(\omega) \Big] ~,
\end{equation} 
as literature result this is the MW current calculated under pseudo-equilibrium condition \cite{MeirWingreen_1992,JauhoWingreeMeir1994}. We discuss now the terms appearing in this formula. The element $f^{\mu_{L}}-f^{\mu_{R}}$ is the difference of \textit{contact occupation} functions hence these are assumed to be at equilibrium with respect to the corresponding lead electronic occupation. In order to investigate in depth the physical implication of this expression, we write down the explicit form of factors inside the trace by using the definition of gamma function in Eq.\ref{eq:DoSinfWBL}, namely
\begin{equation}\label{eq:GenTransfMat}
Tr \Big[4 \Gamma^{L}G^{R}_{0,dd\sigma}(\omega) \Gamma^{R}G^{A}_{0,dd\sigma}(\omega) \Big] 
 =Tr \Big[4 \underbrace{(-\pi \rho_{0,L} |V_{L}|^{2} G^{R}_{0,dd\sigma}(\omega))}_{\mathbb{t}(\omega)}
\underbrace{ (-\pi \rho_{0,R} |V_{R}|^{2} G^{A}_{0,dd\sigma}(\omega)) }_{\mathbb{t}^{\dagger}(\omega)}\Big] \doteq Tr [\widetilde{\mathbb{T}}(\omega)] ~,
\end{equation}
where, in more general terms, the gamma function is actually a matrix as defined in Eq.\ref{eq:LevelWidth}. Hence, the underlined factors can be associated to the \textit{elastic transmission coefficients}, the left to right process $\mathbb{t}(\omega)$ and the inverse one $\mathbb{t}^{\dagger}(\omega)$ as matrix elements of the $\mathrm{T}$-matrix seen previously. In Eq.\ref{eq:GenTransfMat}, we also introduce the useful notation of \textit{generalized transfer matrix} $\widetilde{\mathbb{T}}(\omega)$: it indicates the matrix multiplication among the among the matrix form of gamma functions for left and right leads and the noninteracting impurity Green's functions for direct and inverse transmission processes. Thus, the term $\widetilde{\mathbb{T}}(\omega)$ encapsulates electrons transmission in noninteracting models.\\
In this context, these factors are formally derived using contour-ordered noninteracting Green's functions Schwinger-Keldysh contour. Hence, by means of this physical association, at finite temperature the \textbf{nonequilibrium noninteracting MW formula} reads as,
\begin{equation}\label{eq:MW->Landauer}
\begin{aligned}
I &=  \frac{e}{h} \sum_{\sigma} \int d\omega  
\big( f^{\mu_{L}} (\omega)-  f^{\mu_{R}}(\omega) \big) Tr \Big[ \mathbb{t}(\omega)\mathbb{t}^{\dagger}(\omega)\Big] ~,\\
&\equiv \frac{e}{h} \sum_{\sigma} \int d\omega  
\big( f^{\mu_{L}} (\omega)-  f^{\mu_{R}}(\omega) \big) Tr \Big[\widetilde{\mathbb{T}}(\omega) \Big] ~,
\end{aligned}
\end{equation}
where the expression is fully solvable within the pseudo-equilibrium state for noninteracting models in presence of coupling between leads and central region and in the second line we write it for the generalized transfer matrix as defined in Eq.\ref{eq:GenTransfMat}. This current equation is a remarkable output from the formal derivation on Schwinger-Keldysh contour at finite temperature leading to exact solutions. \\
Furthermore, the Eq.\ref{eq:MW->Landauer} coincides with the previous noninteracting formulation of mesoscopic transport i.e. the Landauer-B{\"u}ttiker formula in Eq.\ref{eq:LB_2probe} at finite temperature, although the transmission coefficients in the two approaches are derived by different means. On one side, the Landauer-B{\"u}ttiker formalism transforms the transmission/reflection probabilities of electrons moving inwards/outwards a barrier into scattering amplitudes of asymptotic noninteracting Green's functions. The methodology is developed within the framework of single-particle scattering theory such that the amplitudes are embedded in the matrix elements of the $\mathrm{S}$-matrix describing the process, see Sec.\ref{sec:Landauer}. On the other hand, within the context of nonequilibrium theory, the MW current formula under the constraint of noninteracting or pseudo-equilibrium nanostructure reduces to the version of Eq.\ref{eq:MW->Landauer} at finite temperature. In this form, elastic coefficients are identified by the multiplication of constant gamma function times nonequilibrium noninteracting Green's functions. In the original MW formula, it is not possible to identify such coefficients: the full nonequilibrium interacting many-body problem has to be solved and we know no exact expression can be found for these correlation functions. Hence, no identification of elastic transmission coefficients can be determined in the interacting nonequilibrium model in general.\\
We conclude that  at finite temperature and in absence of interactions in the mesoscopic nanostructure, the MW formula gives analytical solution and it is equivalent to the Landauer-B{\"u}ttiker formula in Eq.\ref{eq:LB_2probe}.\\

\noindent{For} completeness, we mention a precedent work by C. Caroli \cite{Caroli1970} where an analogue to the noninteracting MW formulation in Eq.\ref{eq:MW->Landauer} is derived using Keldysh formalism to calculate the noninteracting propagators. In this modelling, the effective tunnelling Hamiltonian consists of the lead contribution and the transfer term to track the probability of an electron to tunnel through a barrier. As we can see from the terminology, this work borrows the physical intuition of describing the transport as scattering problem and then it frames it into the more robust nonequilibrium theory - used to evaluate these electronic transmissions and the same total transfer matrix in Eq.\ref{eq:GenTransfMat} is derived in Caroli's work. However, no direct relation to the Landauer formula is made and no direct demonstration of how many-body interactions can be incorporated into the nonequilibrium picture. Hence, the work by Caroli has its importance as a precursor in the usage of nonequilibrium technique in the context of mesoscopic transport, though it presents some limitation which makes this approach less versatile.

\subsection{Solution for models under proportionate coupling}
In this section we tackle the problem of solving the time-independent MW formula by imposing some restrictions to its terms. As we observed already, the challenging part is due to $G_{dd}^{<}$ factor \cite{Dias_noPC2017}: there is no exact analytical expression neither for its interacting self-energy $\Sigma_{dd}^{<}$ nor for its nonequilibrium distribution. Indeed, reliable computational schemes only exists in certain cases in certain parameter regimes. A quite common strategy is to make the term proportional to $G_{dd}^{<}$ vanishing. This is achieved by imposing \textit{proportionality between the couplings} using a real scalar $\gamma$ such that
\begin{equation}
\begin{aligned}
&\mathbf{V}_{L\mathbf{k}n} = \gamma \mathbf{V}_{R\mathbf{k}n} ~,~ \forall ~\mathbf{k},n \\
&\text{and,}~ \epsilon_{L\mathbf{k}} = \epsilon_{R\mathbf{k}} ~,~ \forall ~\mathbf{k} 
\end{aligned}
\end{equation}
where the coupling is between the impurity orbital degree $n$ to the zero-orbital (local orbital) of $\alpha$-lead. Each impurity degree of freedom is defined per channel.\\
This proportionality $\gamma$ is inherited by the corresponding hybridization function, as defined in Eq.\ref{eq:DoSinfWBL}, such that the proportionate coupling condition is usually written as direct proportionality between gamma functions. We need to introduce the generalisation of these functions: the \textbf{gamma or levelwidth functions} properly have a \textit{matrix} structure, namely
\begin{equation}\label{eq:LevelWidth}
[ \mathbb{\Gamma}^{\alpha}(\omega)]_{nm} = \pi  \mathbf{V}^{\star}_{\alpha n} \mathbf{V}_{\alpha m} \rho_{0,\alpha\sigma}(\omega) ~,
\end{equation}
where $| \mathbf{V}_{\alpha n}|^2 = \sum_{\mathbf{k}} |\mathbf{V}_{\alpha\mathbf{k} n}|^2$ as before and the expression is defined on $n,m$ matrix element indicating the impurity degrees of freedom. We use a similar version of this equation in the Meir-Wingreen formula derivation in Eq.\ref{A:eq:LevelWidth}. In Eq.\ref{eq:LevelWidth}, we use the general definition of structured leads, see Eq.\ref{eq:4defEqIneq}. The complete discussion on the leads generalization is given in Sec.\ref{sec:AlternativeLandauer}.\\
Hence, the proportionality has to be fulfilled in \textit{each matrix element} of $[\mathbb{\Gamma}^{\alpha}(\omega)]_{nm}$. We define the time-independent \textbf{proportionate coupling} model (PC) a system whose gamma functions obey to the following matrix equalities, namely
\begin{equation} \label{eq:defPC} 
\begin{aligned}
&(i) \quad [\mathbb{\Gamma}^{L}(\omega)]_{nm}= \gamma[\mathbb{\Gamma}^{R}(\omega)]_{nm} ~,~ \forall~ n,m ~\text{and}~ \forall~ \omega \\
&(ii) \quad \![\mathbb{\Gamma}^{L}]_{nm}= \gamma[\mathbb{\Gamma}^{R}]_{nm} ~,~ \forall~ n,m \\
&\text{total levelwidth matrix:} ~ \mathbb{\Gamma}(\omega) \equiv \sum_{\alpha=L,R}\mathbb{\Gamma}^{\alpha}(\omega) ~,
\end{aligned}
\end{equation} 
where $\gamma$ is a real scalar. The fundamental requirement underlying the PC condition is the \textit{leads equivalence}, otherwise the different density of states per lead would break the equality in Eq.\ref{eq:defPC}. We remark that in the definition, we give in $(i)$ case the generalised version for structured, equivalent leads see Eq.\ref{eq:4defEqIneq} and in $(ii)$ case the standard form for equivalent, metallic leads in wide-band approximation - see also Eq.\ref{eq:DoSinfWBL}.\\
The PC constraint has been introduced originally in the context of the MW formula \cite{MeirWingreen_1992} but it reached its formal development in later works \cite{MeirWingreen_AMoutEquilib_PRL1993,JauhoWingreeMeir1994}. Here, we follow this literature to present the MW formula under proportionate coupling condition - hence we use the definition $(ii)$ from Eq.\ref{eq:defPC} in the section.\\
For general finite temperature models, we want to derive which conditions render the MW formula satisfying the PC property. We impose the PC condition in Eq.\ref{eq:defPC} to the current expression calculated with respect to the left lead in Eq.\ref{A:eq:J_L}. Furthermore, in steady state we use current conservation $I=I_{L}=-I_{R}$ and we can symmetrize the current to get $I= (I_{L}+I_{R})/2=(I_{L}-I_{R})/2$ \cite{JauhoWingreeMeir1994} to write:
\begin{equation}
I =x I_{L} - (1-x)I_{R} 
\end{equation}
and we insert in the original MW equation, Eq.\ref{eq:defPC}, 
\begin{equation}
\begin{aligned}
I = \frac{i e}{h} \sum_{\sigma} \int d\omega Tr \bigg[
&\Gamma^{L} [\gamma x - (1-x)]G^{<}_{dd\sigma}(\omega) + \\
& \hspace*{1.5cm}+ \Gamma^{L} \big( \gamma x f^{\mu_{L}} (\omega) - (1-x)f^{\mu_{R}}(\omega) \big) \big(  G^{R}_{dd\sigma}(\omega)  - G^{A}_{dd\sigma}(\omega) \big)  \bigg] ~.
\end{aligned}
\end{equation}
From this equation, the parameter $x$ is arbitrary and it can be fixed as $x= 1/(1+\gamma)$ such that the term proportional to $G^{<}_{dd}$ vanishes. At finite temperature, the resulting  \textbf{nonequilibrium interacting MW formula in PC} reads
\begin{equation}\label{eq:MW->PC}
\begin{aligned}
I &= \frac{ie}{h} \sum_{\sigma} \int d\omega \big( f^{\mu_{L}} (\omega)-  f^{\mu_{R}}(\omega \big)
Tr \bigg[ 4\frac{\Gamma^{L}\Gamma^{R}}{\Gamma^{L}+\Gamma^{R}} 
\big(  G^{R}_{dd\sigma}(\omega)- G^{A}_{dd\sigma}(\omega) \big) \bigg] ~,\\
&= \frac{2e}{h} \sum_{\sigma} \int d\omega \big( f^{\mu_{L}} (\omega)-  f^{\mu_{R}}(\omega \big) Tr \bigg[ \underbrace{ 4\frac{\Gamma^{L}\Gamma^{R}}{\Gamma^{L}+\Gamma^{R}}   \big(  - \mathit{Im} G^{R}_{dd\sigma}(\omega) \big) }_{\widetilde{\mathbb{T}}_{PC}(\omega)}\bigg] ~,
\end{aligned}
\end{equation}
where in the second line we have identified the difference of nonequilibrium retarded and advanced interacting Green's functions as spectral function in Eq.\ref{eq:defSpectral}. Furthermore, we recognise the terms inside the squared parenthesis as an equivalent expression for the generalized transfer matrix $\widetilde{\mathbb{T}}(\omega)$ defined in Eq.\ref{eq:GenTransfMat} but now under PC so we call it $\widetilde{\mathbb{T}}_{PC}(\omega)$. By multiplying top and bottom the expression in squared parenthesis in Eq.\ref{eq:MW->PC} for the total levelwidth function $\Gamma^{L}+\Gamma^{R}$, we find
\begin{equation}\label{eq:GenTransfMatPC}
	\begin{aligned}
	&\widetilde{\mathbb{T}}_{PC}(\omega) \stackrel{(i)}{=}4\frac{\Gamma^{L}\Gamma^{R}}{\Gamma^{L}+\Gamma^{R}}\Big(\frac{\Gamma^{L}+\Gamma^{R}}{\Gamma^{L}+\Gamma^{R}}\Big)(-\mathit{Im}G^{R}_{dd\sigma}(\omega)) \doteq
	4\frac{\Gamma^{L}\Gamma^{R}}{(\Gamma^{L}+\Gamma^{R})^{2}} t_{dd}(\omega)~,\\
	&\stackrel{(ii)}{=} 4\frac{|V_{L}|^{2}|V_{R}|^{2}}{(|V_{L}|^{2}+|V_{R}|^{2})^{2}}(-\pi\rho_{0})
	\mathit{Im}\underbrace{(V^{\star}_{L}G^{R}_{dd\sigma}(\omega)V_{L}+V^{\star}_{R}G^{R}_{dd\sigma}(\omega)V_{R})}_{T_{dd}(\omega)} \doteq
	4\frac{|V_{L}|^{2}|V_{R}|^{2}}{(|V_{L}|^{2}+|V_{R}|^{2})^{2}} t_{dd}(\omega)~,\\
	&\stackrel{(iii)}{\propto}
	4\frac{J_{L}J_{R}}{(J_{L}+J_{R})^{2}}t_{dd}(\omega) ~,
	\end{aligned}
\end{equation}
where in $(i)$ we just rearrange the terms, in $(ii)$ we use the definition of levelwidth function for equivalent metallic leads in Eq.\ref{eq:DoSinfWBL} and the association of $\mathrm{T}$-matrix in Eq.\ref{eq:TmatrixEq} with Green's function as given by the Fisher-Lee relation \cite{FisherLeeRelation1981}, in $(iii)$ we make use of the Schrieffer-Wolff transformation results $J_{\alpha}~ \propto~ |V_{\alpha}|^{2}$ as for the Anderson impurity model to write the dimensionless prefactor. In all the expressions in Eq.\ref{eq:GenTransfMatPC} we identify the definition of $\mathrm{T}$-\textit{matrix spectral function} $t_{dd}(\omega)$, namely
\begin{equation}\label{eq:TmatSpe}
	t_{dd}(\omega) = - \pi \rho_{0}\mathit{Im}T_{dd}(\omega) \equiv (\Gamma^{L}+\Gamma^{R})(-\mathit{Im}G^{R}_{dd\sigma}(\omega)) ~.
\end{equation}
We observe that, for structured leads - see definition in Eq.\ref{eq:4defEqIneq} - the density of states $\rho_{0,\alpha\sigma}(\omega)$ may vanish, for instance in the case of pseudogap spectrum as we study in Chapter \ref{ch:graphene}. In this scenario, an alternative definition of $t_{dd}(\omega)$ is adopted, see Eq.\ref{Eq:9Tmatspec}.\\
The spectrum of the $\mathrm{T}$-matrix can take values $0 \leq t_{dd}(\omega) \leq1$ where the upper and lower bound are understood in terms of spectral function $0 \leq \mathcal{A}_{dd\sigma}(\omega)\leq1$ defined in Eq.\ref{eq:defSpectral}. The low-energy behaviour of $t_{dd}(\omega=T=0)$ is obtained by means of FSR given in Eq.\ref{eq:FSR} as
\begin{equation}
	t_{dd}(\omega=0) = \sin^{2}\left(\frac{\pi}{2}n_{imp}\right) ~,
\end{equation}
where $n_{imp}$ represents the so-called thermodynamic \textit{excess charge} due to the charge accumulation on the impurity for single-channel model in Fermi liquid regime \cite{Logan2014CommonNFL}. Furthermore, $n_{imp} \equiv \hat{n}_{d}$ in the wide band limit approximation.\\
We define the \textbf{geometric factor} as the dimensionless ratio, namely
\begin{equation}\label{eq:defGeomFactor}
	\widetilde{\Gamma}=4\frac{\Gamma^{L}\Gamma^{R}}{(\Gamma^{L}+\Gamma^{R})^{2}} ~,
\end{equation}
that is completely determined by the coupling configuration between leads-nanostructure and it is a well-defined feature emerging from the PC condition in the model. Both expressions in Eq.\ref{eq:defGeomFactor} correctly return the quantised conductance if calculated for symmetric leads at $\omega=T=0$ regime. As before, Eq.\ref{eq:defGeomFactor} is given here for equivalent, metallic leads but it can be generalised to equivalent, structured leads as well by inserting $\mathrm{\Gamma}^{\alpha}(\omega)$. \\
Hence, the MW formula under PC in Eq.\ref{eq:MW->PC} can be equivalently written as one of the generalized transfer matrix under PC  expressions as showed in Eq.\ref{eq:GenTransfMatPC} as it is more convenient in the problem.\\
We also note that in Eq.\ref{eq:MW->PC} no manipulation has been made on the correlation functions: these are the original contour-ordered, fully interacting many-body Green's functions. As consequence, although $Tr[ \widetilde{\mathrm{\Gamma}} \mathcal{A}]$ exhibits similarity in structure with the Landauer-B{\"u}ttiker formula in Eq.\ref{eq:LB_2probe}, it cannot be cast to some sort of elastic transmission coefficients as we show now analytically.\\
After a few manipulations in the nonequilibrium interacting Dyson Eq.\ref{eq:defDysonRA}, we get
\begin{equation}
	\Sigma^{R/A} = (G_{0\sigma}^{R/A})^{-1}- (G^{R/A}_{\sigma})^{-1}
	\Rightarrow G^{R}_{\sigma}-G^{A}_{\sigma} = G^{R}_{\sigma} (\Sigma^{R} -\Sigma^{A}) G^{A}_{\sigma} \equiv  G^{R}_{\sigma} \Sigma G^{A}_{\sigma}
\end{equation}
and we insert it into the MW formula under PC as derived in Eq.\ref{eq:MW->PC} to obtain
\begin{equation}\label{eq:MW->PC-SelfEnergy}
	I = \frac{i e}{h} \sum_{\sigma} \int d\omega 
	Tr \Big[\frac{ \Gamma^{L}\Gamma^{R}} {-i \big( \Gamma^{L}+\Gamma^{R} \big)}  G_{dd\sigma}^{R}(\omega)  \Sigma_{dd}(\omega) G_{dd\sigma}^{A}(\omega)  \Big] ~,
\end{equation}
where in the denominator $-i(\mathrm{\Gamma}^{L}+\mathrm{\Gamma}^{R})= -i \mathrm{\Gamma} \equiv i\mathit{Im}\Delta$ coincides with the noninteracting part of the self-energy that is the hybridization function for equivalent, metallic leads as defined in Eq.\ref{eq:DoSinfWBL}.\\
In the expression in Eq.\ref{eq:MW->PC-SelfEnergy}, compare it with the noninteracting version of the MW formula in Eq.\ref{eq:MW->Landauer}, the signature of interactions through the presence of the self-energy prevents $G^{R} \Sigma G^{A}$ from any association to an elastic transmission coefficient. On a fundamental level, the self-energy breaks the correspondence between the scattering amplitudes as defined in the Landauer-B{\"u}ttiker formula in Eq.\ref{eq:LB_2probe} and the contour-ordered noninteracting Green's functions as shown in the nonequilibrium noninteracting Dyson Eq.\ref{eq:Dyson-RA-NEQ-NI}. By means of this crucial difference, we demonstrate the MW formula in PC in Eq.\ref{eq:MW->PC-SelfEnergy} deviates from the Landauer-B{\"u}ttiker expression due to interactions.\\
In conclusion, once a model satisfies the PC condition, the MW formula assumes a simplified version given in Eq.\ref{eq:MW->PC} and that is used as current input to calculate the differential conductance in Eq.\ref{eq:DiffCondLRac}. Usually, this current expression is obtained by applying numerical schemes to compute $\mathit{Im}G^{R}$. In this thesis we use the NRG technique for computing equilibrium Green's function, as presented in Sec.\ref{sec:RGtheo}. \\
We comment that the PC condition is a less restrictive constraint to be designed in set-ups and it brings concrete advantages in practical calculations. However, this useful property is not commonly found in general systems but only some special models obey to the PC property - as we present now.

\subsection{Proportionate coupling models}\label{sec:PC}
The definition in Eq.\ref{eq:defPC} explains the proportionate coupling condition at the level of matrix elements equivalence. Now, we shift the discussion on the physical implementation of this condition. At macroscopic level, a mesoscopic quantum system fulfils PC if all its nanostructure components are \textit{equally coupled} to each lead by the same hybridization constant. Looking at the physical set-up configuration, we can directly check this and eventually tune differently the impurities arrangement. The reason behind the simplified mathematical treatment of system under PC it is because certain linear combinations of the leads states can decouple, yielding an effective single-channel description. Electrons then tunnel from this effective single-bath to the central nanostructure through a unified hybridization term. On an operative level, the single-lead nanostructure configuration allows to factorize the tunnelling Hamiltonian $\hat{H}_{hyb}$ and this eases the equations. We can  conclude that the proportionate coupling as macroscopic property also affects the system at microscopic level.\\
We start with two special models in proportionate coupling and then we end with a list of some very simple but commonly used models in non-PC condition.\\
\begin{figure}[H]
\centering
\includegraphics[width=0.8\linewidth]{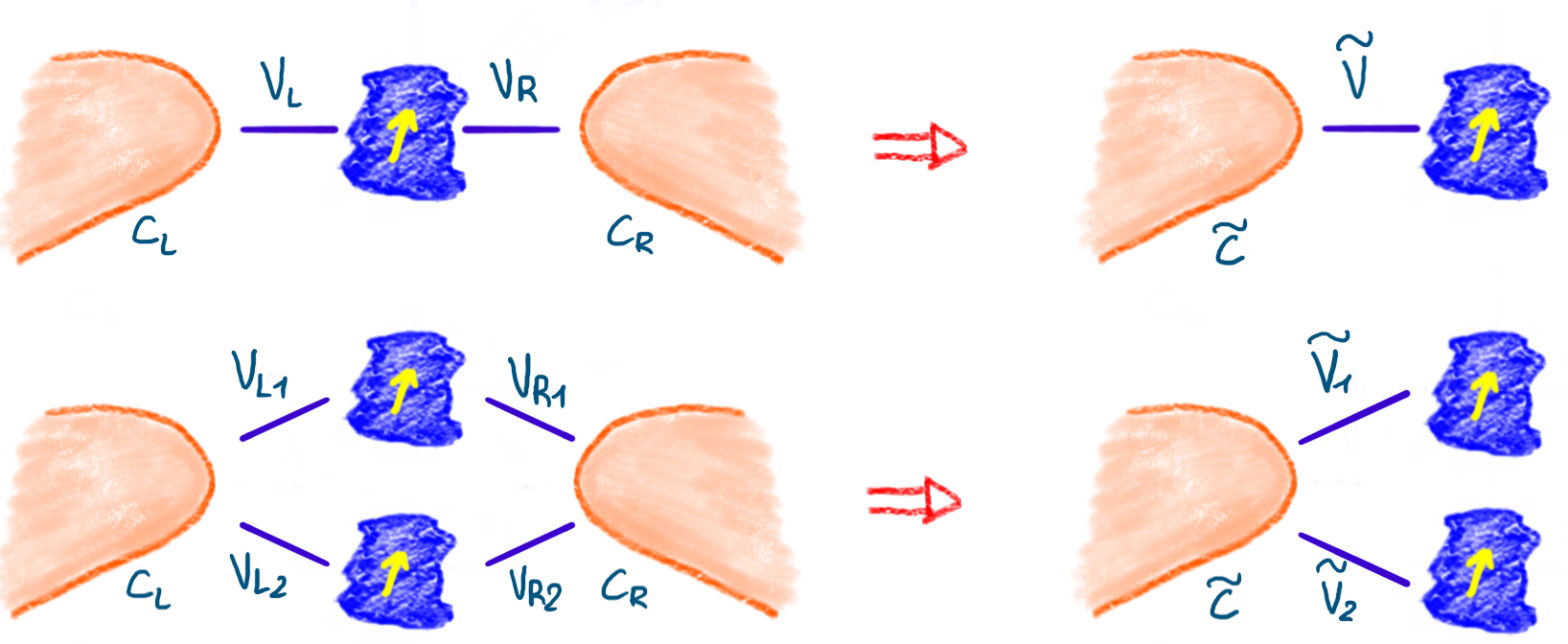}
\caption[Cartoon of standard proportionate coupling models]{Cartoon of two basic proportionate coupling models. \textit{(top)} the single-impurity model and \textit{(bottom)} two-impurity in parallel configuration model. See the main text for the corresponding Hamiltonian equation.}\label{F3:PC}
\end{figure}
\noindent{The} first model under PC is the \textit{single-impurity model with single-energy level}, which can be interacting as in the single-impurity Anderson model (SIAM) \cite{Anderson1961} in Eq.\ref{eq:AM} or not as in the noninteracting resonant level model in Eq.\ref{eq:NI-RLM}. In this system, all the quantities are scalar. Hence, the equality among matrix elements as required in the PC definition in Eq.\ref{eq:defPC} reduces to a scalar equation equivalence, namely 
\begin{equation}
 \Gamma^{L}= \gamma  \Gamma^{R} \Leftrightarrow V_{L\mathbf{k}} = \gamma V_{R\mathbf{k}} ~, \forall~\mathbf{k} ~,
\end{equation}
which is therefore automatically always in PC - for systems with two equivalent leads.\\
From this expression we can unify the two decoupled leads into a single one whose corresponding hybridization reads $V_{\mathbf{k}}$ such that $V_{\mathbf{k}} \equiv V_{L\mathbf{k}}$ and $  \gamma V_{\mathbf{k}} \equiv V_{R\mathbf{k}}$. We use those to define the hybridization Hamiltonian under PC as
\begin{equation}
\hat{H}_{hyb} = \sum_{\mathbf{k}\sigma} \Big( V_{\mathbf{k}} d_{\sigma}^{\dagger} \underbrace{(c_{L\mathbf{k}\sigma} + \gamma c_{R\mathbf{k}\sigma})}_{\sqrt{1+\gamma^{2}}\widetilde{c}_{\mathbf{k}\sigma}} + V^{\star}_{\mathbf{k}}\underbrace{(c_{L\mathbf{k}\sigma}^{\dagger} + \gamma c_{R\mathbf{k}\sigma}^{\dagger})}_{\sqrt{1+\gamma^{2}}\widetilde{c}^{\dagger}_{\mathbf{k}\sigma}} \Big)
=\sum_{\mathbf{k}\sigma} \overline{V}_{\mathbf{k}}\left(d_{\sigma}^{\dagger}\widetilde{c}_{\mathbf{k}\sigma} + \widetilde{c}_{\mathbf{k}\sigma}^{\dagger}d_{\sigma} \right) ~,
\end{equation}
where in the second equality we assume the couplings are real such that $V_{\mathbf{k}} = V^{\star}_{\mathbf{k}}$ and we define $\overline{V}_{\mathbf{k}} = V_{\mathbf{k}} \sqrt{1+\gamma^{2}}$ to satisfy the Fermionic anti-commutation relation for $\lbrace\widetilde{c}_{\mathbf{k}\sigma},\widetilde{c}^{\dagger}_{\mathbf{k}\sigma}\rbrace =1$. See a cartoon of the lead combination in the single-impurity model and its reduction to proportionate coupling configuration with a unique lead coupled via single-point contact to single-electronic impurity level in Fig.\ref{F3:PC}. \\
In particular, the MW formula in Eq.\ref{eq:MW} for the SIAM then takes the following compact form
\begin{equation}
I = \frac{2e}{h} \sum_{\sigma} \int d\omega \big( f^{\mu_{L}} (\omega)-  f^{\mu_{R}}(\omega) \big) \left( - \widetilde{\Gamma}\mathit{Im}G_{dd\sigma}^{R}(\omega)\right) ~,
\end{equation}
that is the Eq.\ref{eq:MW->PC} with the term in round parenthesis is the generalized transfer matrix under PC in Eq.\ref{eq:GenTransfMatPC} with the geometric factor $\widetilde{\Gamma}$ in Eq.\ref{eq:defGeomFactor}. Elaborate systems \cite{Dias_2QD-SIAM_2008} can be studied by means of their low energy effective modelling which final configuration with single effective lead coupled to a single effective site is precisely the SIAM we just discussed. \\
The second example of model in PC by construction is the \textit{parallel multi-impurity configuration} with the same proportionality coupling to each bath. To pin down the analysis, we take the two-impurity parallel configuration. The matrix equality in Eq.\ref{eq:defPC} is calculated now for matrices with $dim(2\times2)$ yielding the matrix relation 
\begin{equation}
\begin{aligned}
& [\mathbb{\Gamma}^{L}]_{nm} = \gamma  [\mathbb{\Gamma}^{R}]_{nm} ~,~\text{with}~ n,m=1,2 \\
& \hspace*{-1cm} \Leftrightarrow  \pi \rho_{0,L}  \begin{pmatrix} V_{L1}V^{\star}_{L1} & V_{L1}V^{\star}_{L2} \\ V_{L2}V^{\star}_{L1} & V_{L2}V^{\star}_{L2} \end{pmatrix} = 
\gamma\pi \rho_{0,R}  \begin{pmatrix} V_{R1}V^{\star}_{R1} & V_{R1}V^{\star}_{R2} \\ V_{R2}V^{\star}_{R1} & V_{R2}V^{\star}_{R2} \end{pmatrix}  ~.
\end{aligned}
\end{equation}
We can repeat now the same leads decoupling and then recombination of those but now for each impurity separately. This results in $V_{n\mathbf{k}} \equiv V_{Ln\mathbf{k}}$ and $ \gamma V_{n\mathbf{k}} \equiv V_{Rn\mathbf{k}}$ for $n=1,2$ impurity. We can now use these PC hybridization terms to factorize $\hat{H}_{hyb}$ and we get
\begin{equation}
\begin{aligned}
\hat{H}_{hyb} &= \sum_{\mathbf{k}\sigma} \Big( V_{1\mathbf{k}} d^{\dagger}_{1\sigma}\underbrace{(c_{L\mathbf{k}\sigma} + \gamma c_{R\mathbf{k}\sigma})}_{\sqrt{1+\gamma^{2}}\widetilde{c}_{\mathbf{k}\sigma}} + V_{2\mathbf{k}} d^{\dagger}_{2\sigma}\underbrace{(c_{L\mathbf{k}\sigma} + \gamma c_{R\mathbf{k}\sigma})}_{\sqrt{1+\gamma^{2}}\widetilde{c}_{\mathbf{k}\sigma}} +~ H.c. \Big) ~,\\
&= \sum_{\mathbf{k}\sigma} \left( \overline{V}_{1\mathbf{k}} ( d^{\dagger}_{1\sigma} \widetilde{c}_{\mathbf{k}\sigma} +\widetilde{c}^{\dagger}_{\mathbf{k}\sigma}d_{1\sigma} ) +
\overline{V}_{2\mathbf{k}} ( d^{\dagger}_{2\sigma} \widetilde{c}_{\mathbf{k}\sigma} +\widetilde{c}^{\dagger}_{\mathbf{k}\sigma}d_{2\sigma} )  \right) ~,
\end{aligned}
\end{equation}
that confirms the two-impurity in parallel set-up presents indeed a PC configuration, see the cartoon in Fig.\ref{F3:PC}. In the equation, in the second equality we assume real coupling such that $V_{n\mathbf{k}} = V^{\star}_{n\mathbf{k}}$ and we define $\overline{V}_{n\mathbf{k}} = V_{n\mathbf{k}}\sqrt{1+\gamma^{2}}$ for $n=1,2$ to satisfy again the Fermionic anti-commutation relation for $\widetilde{c},\widetilde{c}^{\dagger}$ operators.\\
We now end with a discussion of commonly encountered non-proportionate coupling models.\\ 
For example, we consider systems with non parallel impurity configurations e.g. \textit{serial configuration of multi-impurity}: in the model each impurity is directly coupled only to one lead and $\hat{H}_{hyb}$ cannot be factorized. We see this clearly for instance in the two-impurity serial configuration whose tunnelling Hamiltonian $\hat{H}_{Hyb} =\sum_{\mathbf{k}} \left( V_{L\mathbf{k}}d^{\dagger}_{1\sigma}c_{L\mathbf{k}\sigma} + V_{R\mathbf{k}}d^{\dagger}_{2\sigma}c_{R\mathbf{k}\sigma} + H.c. \right) $ cannot be decoupled since the impurity one tunnels only towards left lead and similarly the impurity 2 towards right lead only.\\
However, \textit{parallel configuration of multi-impurity} also does not satisfy in general the PC: each impurity has a different coupling per each lead. We show this in the two-impurity parallel set-up with  tunnelling Hamiltonian
\begin{equation}
\hat{H}_{hyb} = \sum_{\mathbf{k}\sigma} \Big( V_{1\mathbf{k}} d^{\dagger}_{1\sigma}\underbrace{(c_{L\mathbf{k}\sigma} + \gamma_{1} c_{R\mathbf{k}\sigma})}_{\sqrt{1+\gamma_{1}^{2}}\widetilde{c}_{1}} + V_{2\mathbf{k}} d^{\dagger}_{2\sigma}\underbrace{(c_{L\mathbf{k}\sigma} + \gamma_{2} c_{R\mathbf{k}\sigma})}_{\sqrt{1+\gamma_{2}^{2}}\widetilde{c}_{2}} +~H.c. \Big) ~,
\end{equation}
where $\gamma_{1} \neq \gamma_{2}$ determines no decoupling on $\hat{H}_{hyb}$ can be performed.\\
In most cases, any \textit{multi-orbital nanostructures} show macroscopically a lack of the PC property. An example we will encounter in this this is ionic model \cite{Hewson}: each orbital has its own coupling term and the hybridization cannot be factorized. In general, systems with \textit{dangling impurity} connected only to other impurities but detached from the leads present a non-PC configuration \cite{Dias_2QD_2013}.\\
From this short presentation, we conclude it is more common to work with mesoscopic models whose microscopic description does not fulfil the proportionate coupling constraint in Eq.\ref{eq:defPC}.

\subsection*{$(II)$ Pros and cons in the Meir-Wingreen approach}
In this section, we saw the standard strategies used in the literature to tackle the Meir-Wingreen current formula.  We have presented and derived results for systems at finite temperature. In particular, we have highlighted the correspondence between contour-ordered Green's functions and asymptotic Green's functions as element of $\mathrm{S}$-matrix only under absence of interactions, culminating in the MW formula recovering the Landauer equation as seen in Eq.\ref{eq:MW->Landauer}. And also in this case, our arguments have been developed in standard finite temperature theory. Hence, for any nonzero temperature, the equality between nonequilibrium correlation functions and scattering amplitudes holds only for noninteracting mesoscopic models. We conclude that the standard strategies employed for quantum transport calculation under MW formula are too restrict at noninteracting impurity structure, at proportionate coupling model and at strictly zero temperature regime. However, we can go beyond these standard approaches; this is the main focus of this thesis.\\
In the Sec.\ref{sec:AlternativeMW}, we present two derivations for calculating approximatively the nonequilibrium $G_{dd}^{<}$ aiming for an expression with an explicit voltage bias dependence. We successfully reach a final equation for the lesser propagator that takes inspiration from the well-known Ng ansatz. In this derivation, we also find the correction to the ansatz due to interacting impurity. \\
Another venue of improvements is for mesoscopic transport at zero temperature, which is framed within Fermi liquid theory as it is developed in the Oguri argument in Sec.\ref{sec:Oguri}. We note that both the nonequilibrium and Fermi liquid theories presents challenging technical steps.  As an outcome of our investigation in mesoscopic transport, we find that we can go beyond these known methods and we derive an alternative formulation of the Oguri formula that liaises with the $T=0$ MW result from the literature - see the relevant section in Sec.\ref{sec:AlternativeOguri}. \\
For models under proportionate coupling, we saw in Sec.\ref{sec:PC} its validity is confined to few macroscopic structures exhibiting this feature. Considering the usefulness of this constraint, we can beyond the lack of the macroscopic PC property. As result from this thesis, starting from systems whose microscopic models lack the PC property, we derive effective Hamiltonian at low energy and low temperature scales which presents an incipient PC condition. We discuss prominent cases of emergent PC in system within Coulomb blockade regime in Sec.\ref{sec:CBPC} and at mixed-valence transition in Sec.\ref{sec:MVPC}.\\
In conclusion, this section motivates the research in alternative quantum transport calculations implemented in versatile models.

\section{Theory of transport in Fermi liquid models: the Oguri argument}\label{sec:Oguri}
As a last topic on the standard theory of mesoscopic transport theory we present the argumentation introduced by A. Oguri \cite{Oguri_FermiLiquidTheoryConductance1997}. Following the historical development of methodologies,  many of the preceding theories and formulations converge in Oguri's derivation. Those are the noninteracting scattering theory introduced in Sec.\ref{sec:Landauer}, the linear response theory introduced in Sec.\ref{sec:Kubo}, the Kubo formula for conductance as given in Eq.\ref{eq:defKuboLR} and the microscopic Fermi liquid treatment of models as we present in Eq.\ref{eq:G_ddFL}.\\
The core of the Oguri argument is to relate the linear response Kubo formula for the electrical conductance in interacting models at thermal equilibrium with the transmission/reflection/incident coefficients defined through asymptotic form of interacting single-particle Green's functions. This connection is valid only at zero temperature and at zero frequency condition such that the system behaves as a Fermi liquid.\\
In this section we present the original Oguri argument for a general one dimensional model and then we detail its formulation for a one dimensional lattice.

\subsection*{Electrical conductance in Fermi liquid interacting models}
In the original work by A. Oguri \cite{Oguri_FermiLiquidTheoryConductance1997}, the mesoscopic transport process is studied in the standard impurity system, see Fig.\ref{F3:ImpScheme}, where the current flows through a finite interacting region and noninteracting leads sit adjacent to it. We consider as a model Hamiltonian the Anderson impurity model given in Eq.\ref{eq:AM}. The argumentation developed by Oguri is based on three fundamental assumptions \cite{Oguri_FermiLiquidTheoryConductance1997}: $(i)$ the interactions are limited to a small area such that scattering treatment of electronic flow is admitted, $(ii)$ the system size allows for a description of the low-energy states in Bloch waves and thus Fermi liquid theory holds at low energies and $(iii)$ the perturbation expansion in powers of the Coulomb interaction $U$ is well-defined. These assumptions are required in calculating the Kubo conductance formula. We comment more on those for a general one dimensional model while we introduce the analytical calculations.\\
The first assumption regards the set-up definition. The model is composed by two semi-infinite, noninteracting, thermal equilibrium leads coupled to a finite size central region where Coulomb interaction and local scattering potential are adiabatically switched on. 
Furthermore, it is consistent to study the asymptotic form of the single-particle Green's functions to define the \textit{scattering amplitudes}, namely the transmission $t_{nn^{\prime}}$ and the reflection $r_{nn^{\prime}}$ amplitudes when considering $T=0$ \cite{Mahan}. Those are used to describe the current flowing across the sample within a well-defined scattering theory treatment as we discuss previously in Sec.\ref{sec:Landauer}. By means of the Fisher-Lee matrix relation in Eq.\ref{eq:FisherLee_Smatrix}, the conservation of transmission and reflection electrical probabilities translate into unitary property of the scattering $\mathrm{S}$-matrix. Using the \textit{optical theorem} for the off-diagonal elements \cite{Flensberg, Mahan} of the $\mathrm{T}$-matrix or \textit{discontinuity equation} of the transmission amplitude, the current conservation across the sample can be expressed equivalently to 
\begin{equation}
\sum_{nn^{\prime}} \mathit{Im}t_{nn^{\prime}(\omega)} = \sum_{nn^{\prime}} \mathit{Im}G^{0,R}_{0,bath}(\omega)| t_{nn^{\prime}}(\omega)|^{2} \equiv - \pi \rho_{0,\alpha}(\omega)\sum_{nn^{\prime}}| t_{nn^{\prime}}(\omega)|^{2} ~,
\end{equation}
for the isolated, noninteracting leads Green's function and the last equality is obtained using Eq.\ref{eq:defDos}. This expression has different interpretations according to the impurity property. In absence of interactions, the amplitude $t_{nn^{\prime}}$ coincides with the well-defined transmission amplitudes properly defined in scattering theory. Under such condition, the discontinuity equation on the imaginary plane $\mathit{Im}t_{nn^{\prime}}$ holds and so does the unitary property of the scattering matrix. On the contrary, in presence of interactions, both the discontinuity equation and the unitarity conditions are satisfied \textit{only} at zero frequency and at zero temperature regime. This special condition is consistently chosen according to the system size - as we discuss more below.\\ 
Considering the leads region occupies a larger area than the central one, it results that the wavelength of low-energy states is much longer than the central region dimension. This system size makes the Fermi wavelength - for which the electrons possess a maximum energy - to be of the order of the Bloch wavefunction defined on the first Brillouin zone \cite{PustilnikGlazman_review2004}. The Bloch waves are typically used to described the low-energy states in a Fermi liquid system. Hence, the second assumption regards the system dimensionality as compatible to a proper microscopic \textit{Fermi liquid} description. In particular, we make use of Fermi liquid property  such as $\mathit{Im}\Sigma^{U}(\omega\rightarrow0,T=0) \to 0$ \cite{Nozieres1974,Hewson_RenormalPT-FL1993,Hewson} where $\Sigma^{U}$ is the interacting component of the self-energy - see Eq.\ref{eq:defSelEngeneral}. This vanishing part entails the renormalization of impurity energy at low-energy scales as we present in Eq.\ref{eq:G_ddFL}. Thus, to reconnect with the optical theorem result we introduce above, in case $\mathit{Im}\Sigma^{U}\equiv0$ is satisfied, the expression for $\mathit{Im}t_{nn^{\prime}}$ holds and we can successfully construct scattering amplitudes to describe the electrical flow.\\
The last assumption relies on the Fermi liquid treatment of the microscopic model such that the perturbative expansion in Coulomb potential $U$ is well-defined \cite{Yamada970AMPert}. The outcome from the expansion is that $\mathit{Im}\Sigma^{U}$ 
is proportional to $\omega^{2}$ at low $\omega$ at $T=0$. This is an exact result within expansion of the order $\mathcal{O}(U^{2})$ that is for our purposes here. However, in general it is also exact to all orders in $U$. We can derive this expression by means either of diagrammatic expansion of the vertex \cite{Yamada316ImSelfDiag} or of $\mathrm{T}$-matrix approach \cite{Yamada1286AMPertTMat}. Furthermore, these findings are in perfect agreements with the Fermi liquid description. Hence, at zero temperature we have $\mathit{Im}\Sigma^{U} ~ \propto ~ \omega^{2}\stackrel{\omega\rightarrow 0}{\longrightarrow}0$.\\
In conclusion, the crucial aspect in the Oguri derivation is that starting from a fully interacting impurity, under vanishing both energy and temperature regime, the system becomes a Fermi liquid. Thus, we can use microscopic Fermi liquid properties and we can construct scattering coefficients to describe transport activity.\\
Considering these analytical results, we use this understanding to calculate the correction due to interactions in the current-current correlator defined in the \textit{dc-}regime Kubo formula for electrical conductance in Eq.\ref{eq:defKuboLRel}. If the currents are measured with probes \textit{in the leads}, in the diagrammatic expansion there is \textit{no} contribution from the imaginary part of the interacting vertex i.e. $\mathit{Im}\Sigma^{U}(\omega=0,T=0)\to 0$. However, there are finite contribution from the real part of the interacting vertex i.e. $\mathit{Re}\Sigma^{U}(\omega=0,T=0) \neq 0$. The corresponding correlator can be written in terms of the generalized transfer matrix given in Eq.\ref{eq:GenTransfMat}, namely 
\begin{equation}
\mathcal{G}^{dc}(\omega\rightarrow0,T=0) = \frac{2e^{2}}{h}|\widetilde{\mathbb{T}}(\omega\rightarrow0)|^{2} ~,
\end{equation}
where $\widetilde{\mathbb{T}}$ is determined by the product of the gamma functions in matrix form given in Eq.\ref{eq:LevelWidth} and the renormalised impurity Green's function in Eq.\ref{eq:G_ddFL}, here given in matrix form as follows: 
\begin{equation}
	\begin{aligned}
	&\text{full system:}\quad [\mathbb{G}(\omega)]^{-1} = z \mathbb{1} - \epsilon_{d}\mathbb{1} -\mathbb{\Delta}(\omega)-\mathbb{\Sigma}^{U}(\omega) ~, \\
	&\text{Fermi liquid system:} \quad [\mathbb{G}(\omega=0)]^{-1}=z \mathbb{1} - \mathbb{H}^{\star} ~,\\
	&\hspace*{3.5cm} \text{with}~
	\mathbb{H}^{\star} = \epsilon_{d}\mathbb{1}+ \mathit{Re}\mathbb{\Delta}(\omega=0)+ \mathit{Im}\mathbb{\Delta}(\omega=0) + \mathit{Re}\mathbb{\Sigma}^{U}(\omega=0) ~,
	\end{aligned}
\end{equation}
where the hybridization matrix $\mathit{Re}\mathbb{\Delta}(\omega=0)$ vanishes in case of wide band limit. The equation for $\mathcal{G}^{dc}(\omega\rightarrow0,T=0)$ given above is then Landauer-type of formula but now in \textit{dc-}regime only and derived from interacting impurity models at $\omega=T=0$ regime: that is the outcome from the so-called \textbf{Oguri argument}.\\
This equation for $\mathcal{G}^{dc}$ displays a familiar form to those we presented in previous sections but now it is derived from different regimes and so it targets systems at zero temperature. In this section we are dealing with an interacting model that, under the constraints in energy and temperature ranges, its \textit{dc-}limit electrical conductance is determined solely by the generalized transfer matrix. On the contrary, in the scattering theory description used in the Landauer approach in Sec.\ref{sec:Landauer}, the starting point is a noninteracting model at finite temperature to calculate electrical conductance in \textit{ac-}regime. The theory to connect these two descriptions such that a Landauer-type of formula is obtained is the Fermi liquid. Within this theory, the excitations at low energy and low temperature are substituted by dressed particles or \textit{quasiparticles} that are free particles with infinitely long lifetime. Hence, the current-current correlator has the \textit{same} diagrammatic expansion if computed for the renormalised impurity propagators at $\omega\rightarrow0,T=0$ Fermi liquid condition and for the noninteracting propagators at finite $\omega,T$.\\

\begin{figure}[t]
\centering
\includegraphics[width=1.05\linewidth]{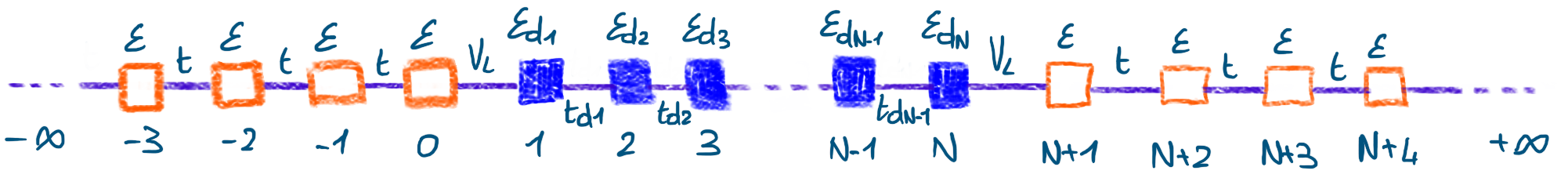}
\caption[Schematic of the set-up in the Oguri argument]{The schematic shows the one dimensional lattice version of an interacting non-proportionate coupling impurity model with two leads, seen in the main text the equation Anderson impurity model Hamiltonian. Full sites (blue) indicate the interacting sites of the finite-size impurity chain with single-particle energy and hopping detailed at each site. At the two sides, empty sites (orange) indicate the noninteracting sites of the semi-infinite lead chains with homogeneous parameters.}\label{F3:Oguri}
\end{figure}

\noindent{We} continue the analysis by considering impurity system on one dimensional lattice with tight-binding modelling  \cite{Oguri_QMCTransport1997}. Without loss of generality, in this discussion we consider two leads, but this derivation can be generalised to multiple leads set-up.\\
Considering the tight-binding chain structure \cite{Altland}, the real space representation is the most appropriate and so the model is given by an infinite chain with sites labelled according to the index $n$, see Fig.\ref{F3:Oguri}. The noninteracting sites are symmetrically spanning from $-\infty < n \leq 0$ and from $N+1 \leq n < +\infty$: these are characterised by same single-particle energy $\epsilon$ and uniform hopping $t$. The interacting sites are located in between the two semi-infinite chains, occupying from $1 \leq n \leq N$.  These sites are instead characterised by single-particle energy $\epsilon_{dn}$, hopping $t_{dn}$ and interaction $U$ such that local potential and disorder may be tuned at each site accordingly. At the two sides of the interacting region, the edge noninteracting sites - i.e. $n=0,n=N+1$ on the left and right lead respectively - are directly coupled though hybridization term with the first interacting chain site. And the choice of time translational invariant interactions ensure those are smoothly vanishing at the interface with the semi-infinite chains  \cite{Oguri_TransmissionProb_2001}.\\
\!\!\!\noindent{We} introduce now for the one dimensional lattice system in Fig.\ref{F3:Oguri} - that is a non-proportionate coupling model - the real space version of the single-impurity Anderson model as utilised by Oguri \cite{Oguri_QuasiParticlePRB2001}, namely
\begin{equation}
\begin{aligned}
&\hat{H}^{AM} = \hat{H}_{leads}+\hat{H}_{imp}+\hat{H}_{hyb} ~,\\
&= \sum_{\sigma} \Bigg(  \sum_{n=0}^{-\infty}
\big( tc^{\dagger}_{n\sigma}c_{n+1\sigma} + t^{\star}c^{\dagger}_{n+1\sigma}c_{n\sigma} + \epsilon\hat{n}_{n\sigma} \big) 
 +  \sum_{n=N+1}^{+\infty}
\big( tc^{\dagger}_{n\sigma}c_{n+1\sigma} + t^{\star}c^{\dagger}_{n+1\sigma}c_{n\sigma} + \epsilon\hat{n}_{n\sigma} \big) \\
&  \hspace*{2cm}+ \sum_{n=1}^{N} \big(  \epsilon_{dn}\hat{n}_{dn\sigma} 
+t_{dn}d^{\dagger}_{n\sigma}d_{n+1\sigma} + t^{\star}_{dn}d^{\dagger}_{n+1\sigma}d_{n\sigma} \big)
+ U \hat{n}_{dn \uparrow}\hat{n}_{dn \downarrow} \\
&\hspace*{2cm}+\big( V_{L} c^{\dagger}_{ 0\sigma}d_{1\sigma} +
V^{\star}_{L}d^{\dagger}_{1\sigma}c_{ 0\sigma} +
 V_{R} c^{\dagger}_{ N+1\sigma}d_{N\sigma} +V^{\star}_{R}d^{\dagger}_{N\sigma}c_{N+1\sigma} \big) \Bigg) ~.
\end{aligned}
\end{equation}
We proceed now with the general approach described at the beginning of the section to derive the \textit{dc-}regime Kubo formula for electrical conductance. We expand the Green's functions in order of $\mathcal{O}(U)^{2}$ in terms of the one-particle bare propagators corresponding to the irreducible, fully contracted components of the diagrammatic expansion. From the perturbative result, we write the Dyson equation Eq.\ref{eq:defDysonRA} for the real space $n$ impurity sites model, namely
\begin{equation}
G_{nn^{\prime}}(\omega,T) = G_{0,nn^{\prime}}(\omega,T)
+ \sum_{ll^{\prime}=1}^{N}  G_{0,nl}(\omega,T)\Sigma^{U}_{ll^{\prime}}(\omega,T)G_{l^{\prime}n^{\prime}}(\omega,T) ~,
\end{equation}
where the interacting component of the self-energy is evaluated solely on the central region sites through the summation in $ll^{\prime}$ indices.\\
Having a Fermi liquid impurity ground-state ensures that in the limit of vanishing energy and temperature there is no vertex correction due to interactions in the current-current correlator used in the Kubo formula. The resulting conductance expression on the real space system for $N$ impurity sites then reads
\begin{equation}
\mathcal{G}^{dc}_{N}(\omega\rightarrow0,T=0) = \frac{2e^{2}}{h} \Big( 2 \Gamma^{L}(\omega=0)G_{d1dN\sigma}^{R}(\omega=0,T=0)
2 \Gamma^{R}(\omega=0)G_{d1dN\sigma}^{A}(\omega=0,T=0)
\Big) ~,
\end{equation}
where the levelwidth function in Eqs.\ref{eq:LevelWidth},\ref{eq:DoSinfWBL} is now defined for equivalent and structured leads in real space representation as
\begin{equation}
\Gamma^{\alpha}(\omega=0) = -\sum_{\sigma}|V_{\alpha}|^{2} \mathit{Im}G^{0,R}_{0,\alpha n\sigma}(0) = +\pi \sum_{\sigma}|V_{\alpha}|^{2} \rho_{0,\alpha n \sigma} (0)  ~,
\end{equation}
with $G_{0,\alpha n}^{0}$ is the isolated, noninteracting lead Green's function at the $n^{th}$ site, namely $G_{0,L 0}^{0}$ for left lead and $G_{0,R N+1}^{0}$ for right lead. The fully interacting impurity Green's function Eq.\ref{eq:G_ddFL} is now defined in Fermi liquid regime, here give in matrix form for real space representation as follows
\begin{equation}
	\begin{aligned}
&[\mathbb{G}(\omega=0)]_{d1dN\sigma} = [z\mathbb{1}-\mathbb{H}^{\star}]^{-1}_{d1dN\sigma} ~,
&\text{with} ~ \mathbb{H}^{\star}= \mathit{Im}[\mathbb{\Delta}(\omega=0)]_{ll^{\prime}} + \mathit{Re}[\mathbb{\Sigma}^{U}(\omega=0)]_{ll^{\prime}}~,
\end{aligned}
\end{equation}
where we assume WBL approximation.\\
We conclude that by restricting the parameter range to zero energy and zero temperature as appropriate for the Fermi liquid regime, the fully interacting impurity model reduces to an effective noninteracting one. Hence, in the above expression for $\mathcal{G}^{dc}$ we recognise the quantised conductance term is multiplied by the generalised transfer matrix $\widetilde{\mathbb{T}}$ given in Eq.\ref{eq:GenTransfMat}. Using these identification, we obtain
\begin{equation}\label{eq:Oguri}
\begin{aligned}
\mathcal{G}^{dc}_{N}(\omega\rightarrow0,T=0) &=\\
&=\frac{2e^{2}}{h} \sum_{\sigma} 2\underbrace{ (- |V_{L}|^{2})\mathit{Im}G^{0,R}_{0,L0L0\sigma}(\omega=0) G_{0,d1dN\sigma}^{A}(\omega=0,T=0)}_{\mathbb{t}(\omega=0)} \times \\
& \hspace*{2.5cm} \times 2\underbrace{ (-|V_{R}|^{2})\mathit{Im}G^{0,R}_{0,RN+1RN+1\sigma}(\omega=0)G_{0,d1dN\sigma}^{A}(\omega=0,T=0)}_{\mathbb{t}^{\dagger}(\omega=0)} ~,\\
&= \frac{2e^{2}}{h}
 4 \Gamma^{L}(\omega=0)\Gamma^{R}(\omega=0)|G_{0,d1dN\sigma}^{R}(\omega=0,T=0)|^{2} ~,\\
&\equiv \mathcal{G}_{0} \widetilde{\mathbb{T}}(\omega\rightarrow0)  ~,
\end{aligned}
\end{equation}
where in the first equality we use the $\mathrm{S}$-matrix elements defined in Eq.\ref{eq:FisherLee_Smatrix} and evaluate the gamma function according to the chain side - see for the sake of clarity Fig.\ref{F3:Oguri}, 
in the second equality we use the definition of gamma function in Eq.\ref{eq:DoSinfWBL} and in the last equivalence the generalised transfer matrix for transmission processes in noninteracting models as introduced in Eq.\ref{eq:GenTransfMat}. The Eq.\ref{eq:Oguri} is the so-called \textbf{Oguri formula}.\\
This familiar expression is not a completely unexpected result. In this section we have presented the Oguri argument with the aim to calculate the \textit{dc-}regime electrical conductance in the Kubo formula starting with a interacting impurity model under Fermi liquid regime. At $\omega,T$ vanishing, we assist at the impurity energy renormalization and $\mathit{Im}\Sigma^{U}=0$: those characteristics allow to link interacting and noninteracting models \cite{HewsonOguri_RenormalizedImp_2004}. Thus, the final conductance expression in Eq.\ref{eq:Oguri} has lost any interacting features. It shows then:\\
$(i)$ the same form of the Fisher-Lee formula given in Eq.\ref{eq:FisherLee}. However, in that context, the initial impurity system is already noninteracting. The matrix association in Eq.\ref{eq:FisherLee_Smatrix} is the main ingredients to associate $\widetilde{\mathbb{T}}$ to the Landauer-B{\"u}ttiker formula in Eq.\ref{eq:LB_2probe}, resulting also in the finite temperature \textit{ac-}conductance formula, see discussion in Sec.\ref{sec:Landauer}; \\
$(ii)$ its current has the same form of Meir-Wingreen formula in Eq.\ref{eq:MW->Landauer}, where the expression is derived in general out-of-equilibrium and noninteracting models at finite temperature. This equation coincides also with the Caroli formula \cite{Caroli1970}, as noted by Oguri himself \cite{Oguri_QMCTransport1997}.\\
The latter point necessitates more comments where we find that the Oguri argument recovers the Meir-Wingreen formula. However, also the opposite is true. If we consider the current equation Eq.\ref{eq:MW->Landauer} at zero energy and zero temperature regime, we obtain
\begin{equation}
I = \frac{e}{h}\sum_{\sigma} \big( \Theta(\omega-\mu_{L}) - \Theta(\omega-\mu_{R}) \big) Tr \Big[4 
\pi|V_{L}|^{2}\rho_{0,L\sigma}(0)G^{R}_{dd\sigma}(0)\pi|V_{R}|^{2}\rho_{0,R\sigma}(0)G^{R}_{dd\sigma}(0) \Big] ~,
\end{equation}
where the Fermi distributions reduce to Heaviside functions and the impurity correlators included in the definition of the generalized transfer matrix are now renormalized as given in  see Eq.\ref{eq:G_ddFL} because the system enters in Fermi liquid regime. Then, taking vanishing voltage bias, we calculate the linear response conductance as defined in Eq.\ref{eq:DiffCondLRdc}. The final expression is \textit{equivalent} to the Oguri formula in Eq.\ref{eq:Oguri}.  We infer that on one side, we have the Meir-Wingreen formulation based on contoured-ordered Green's function on a Keldysh contour in models without interacting impurity; on the other side we have the Oguri argument based on Green's function defined within Fermi liquid regime for renormalized impurity model. Hence, we notice the two transport formulations converge to the same transport expression only in the low-energy regime. \\
As side comment we also highlight that under Fermi liquid regime $\omega=T=0$ and linear response $eV_{bias}=\mu_{L}-\mu_{R} \to 0$, the MW current under proportionate coupling for interacting systems in Eq.\ref{eq:MW->PC-SelfEnergy} reduces to the noninteracting MW formula in Eq.\ref{eq:MW->Landauer} - that is the Landauer-B{\"u}ttiker formula. Thus, as remarkable result, at zero temperature and in the limit of zero voltage bias, the MW formula is equivalent to the Landauer-B{\"u}ttiker.  Hence, we conclude that within the Fermi liquid description the the Meir-Wingreen formulation in Eq.\ref{eq:MW}, the Landauer-B{\"u}ttiker formula \ref{eq:LB_2probe},\ref{eq:MW->Landauer} and the Oguri argument in Eq.\ref{eq:Oguri} are equivalent expressions and can be derived one from another to study quantum transport processes.\\ 

\noindent{In} conclusion, the essential aspect on the Oguri argument is the validity of a microscopic Fermi liquid treatment of the impurity model at low energy and low temperature regime. This allows us to reformulate the problem in terms of the Landauer-type of formula but now with interacting Green's functions. Although the final expression shares similarities with the Fisher-Lee and Landauer formula and the Meir-Wingreen formula, the Oguri argument is the only method among those presented so far that is based in Fermi liquid theory. Hence, this transport method is particularly suitable for a specific parameter range.

\subsection*{Pros and cons in the Oguri approach}
In this section, we introduce the conductance formula derived for interacting models satisfying the Fermi liquid regime as originally presented by Oguri. This is a solid argumentation for models can be treated as Fermi liquid at zero temperature and frequency conditions. Hence, we obtain an exact expression to calculate the \textit{dc-}regime electrical conductance.\\ 
Although the Fermi liquid picture is usually the standard one, we note that non-Fermi liquid states can also be designed in prepared set-ups. Those states occur in overscreened Kondo case: they are characterised by thermodynamic quantities exhibiting temperature dependence and the term $\mathit{Im}\Sigma$ is no longer vanishing. It is then desirable than to find an alternative formulation based on the single-particle Green's function methods as used in the Oguri derivation but now valid also for non-Fermi liquid models such that we can avoid the complexity of the vertex correction on the conductance. Moreover, going beyond the Fermi liquid regime gives access to the \textit{ac-}regime conductance. We note that in case the nanostructure hosts free local moments as the temperature vanishes i.e. we are in presence of the so-called \textit{singular Fermi liquid} arising in underscreened Kondo regime \cite{Nozieres_KondoRealMetal1980} or from ferromagnetic couplings \cite{VarmaSingularNFL2002}, $\mathit{Im}\Sigma^{U}=0$ holds asymptotically at $T\rightarrow 0$ - hence, the Fermi liquid description is recovered.\\
The actual limitation in the Oguri argument is the zero temperature limit. This may be particularly problematic when comparing with results of experiments performed at base temperatures larger than emergent energy scales - such as the $T_{K}$ Kondo temperature. That is because experiments often appear to show non-Fermi liquid signatures due to the fact the low temperature limit is not reached in practice, and non negligible electron-electron interactions yield a finite $\mathit{Im}\Sigma$. It is then of interest to find expression for finite temperature systems.\\
In conclusion, we seek for alternative formulation of the Oguri argument such that finite frequency and finite temperature regimes can be addressed. This discussion is presented in Sec.\ref{sec:AlternativeOguri}.

\section{Kondo effect in artificial atoms}\label{sec:Kexp}
In this chapter we have extensively presented the fundamental quantum transport techniques for mesoscopic devices. In this section, we give insights on the practical realization of set-up capable to detect the Kondo effect. Across several decades of technical advancing, the experimental verification of the Kondo physics is closely related to the continuous technological development in nanoelectronic devices able to open new horizons in the study of quantum transport. A cornerstone in conductance measurements is the realization of semiconductors heterostructure characterised by electron gas confined in a small region that is connected to electronic reservoirs through tunable tunnel contacts. These devices made indeed possible the first experimental verification of the Kondo effect \cite{Goldhaber-Gordon-Kondo1998Exp} - as we discuss at the end of this section. \\
In this section we introduce one realization of artificial atom that is the quantum dot. Then, we describe the electrical conductance dependence and regimes in a quantum dot system for specific parameters such as temperature and applied voltages $V_{gate},V_{bias}$. We conclude our discussion with a short section on the pioneer experiments on semiconductor quantum dot nanodevice able to probe the Kondo effect. The increasing manufacture precision and fine control over macroscopic quantities developed over the years made possible realistic implementation where the main properties of the Kondo effect could be extensively verified. \\

\begin{figure}[H]
\centering
\includegraphics[width=1.05\linewidth]{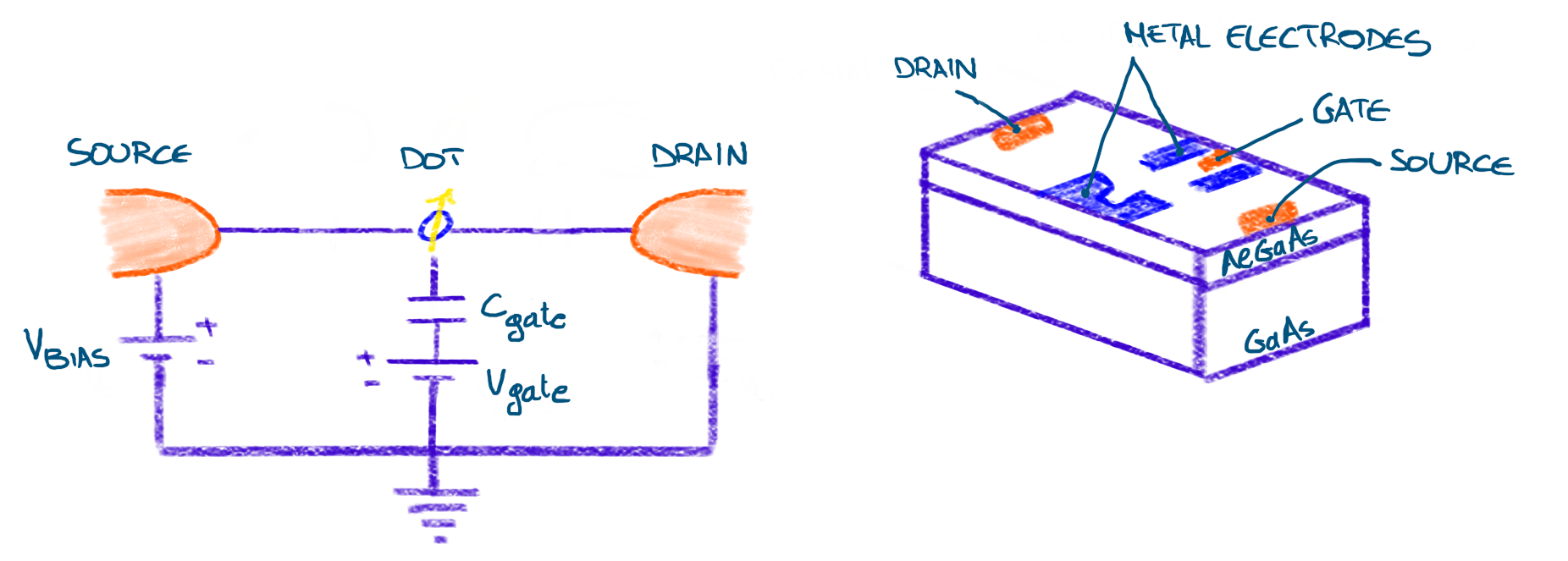}
\caption[Schematic of quantum dot]{The schematics show: $(left)$ the circuit for a quantum dot connected to source and drain by tunnelling junctions with corresponding $V_{bias} = (\mu_{s}-\mu_{d})/2$, $(right)$ the $GaAs/Al_{x}Ga_{1-x}As_{x}$ heterostructure for constructing zero dimensional electron gas as described in the main text.}\label{F3:QD}
\end{figure}

\noindent{In} the early days of the nanodevice development, it was verified that once the conductors fabrication reaches small size, the physical dimension of the device determines the electronic properties of the system. Quantitatively, we know already small sample extension indicates its size $L$ is only larger than the Fermi wavelength but it is smaller than $l_{el},l_{\phi},l_{rel}$ length scales. This paves the way for quantum mesoscopic transport measurements. The geometric properties of the set-up enforce a confinement in the electron gas and this determines quantization \textit{both} in energy and in charge. The confined electron gas takes then a discrete spectrum of energy levels and its electronic occupancy follows integer steps. Moreover, considering that these very small conductors display level spacing between consecutive levels of order of $meV$ - rather than $eV$ as in ordinary metal - these systems are usually called \textit{artificial atoms} \cite{Kastner_SET1992,Kastner_Artificialatoms1994}. An experimental realization of artificial atom is the \textbf{quantum dot} (QD): this is a zero dimension electron gas, meaning that the electrical confinement is applied to \textit{all} three dimensions \cite{Flensberg}. Given the important role covered by the QD system in this thesis, we discuss now in some detail its physical processes.\\
We first want to identify how to implement the confinement inside the semiconductor in experimental set-up. In order to do so, we analyse the typical realization of QD consisting of an heterostructure with a semiconducting layer of $GaAs$ at the bottom and insulating layer of $Ga_{1-x}Al_{x}As_{x}$ with $x$ doping on the top, see the schematic in Fig.\ref{F3:QD}. At zero temperature, there are no free charges in the conduction bands of a general undoped semiconductor. However, by doping the $Ga_{1-x}Al_{x}As_{x}$ substrate with $Si$ atoms, then free conduction electrons are provided. Considering that the bottom of conduction band of $Ga_{1-x}Al_{x}As_{x}$ is higher than the lowermost conduction band of $GaAs$, the electrons moving from the former to the latter substrate can gain energy. Hence, the free positive carriers from $Si$ accumulate only at the $GaAs$ interface: this creates confinement in the direction perpendicular to the interface and so we obtain the formation of a two dimensional electron gas. Said in other words, the positive charges on the $Si$ atoms develop a strong electric field oriented perpendicularly to the interface resulting into a potential well where electrons are confined in. The remaining ionized $Si$ donors left in $Ga_{1-x}Al_{x}As_{x}$ layer creates an electrostatic energy that grows by increasing the number of transmitted electrons. This increment in energy gained due to transferring electrons to $GaAs$ layer continues until it is in balance with the growth in electrostatic energy. Regarding the other two remaining directions, the spacial confinement is then achieved because of electric fields imposed by very small additional confinement electrodes - usually placed at the top of $Ga_{1-x}Al_{x}As_{x}$ substrate. It is generally assumed that the voltage on these spacial constrictions is fixed such that confinement potential is static as well. This allows to manipulate the electron density at will using top gate voltages.\\
Now that the electron gas is confined in all the three dimensions so we have a proper QD device, we can embed it into an electrical circuit as shown in the schematic in Fig.\ref{F3:QD}. Of relevance for our discussion, the various types of voltage applied on the QD cover different functions. We start with considering the primary operation in a QD, that is the electrons tunnelling. In practice, the electronic transport is achieved through dot and reservoirs junctions - those are the \textit{source, drain quantum point contacts} and are realized for instance with metallic Ohmic contacts. The coupling on these contacts is usually kept small in order to avoid charge fluctuations and dispersive effects. The corresponding potential applied on these contacts is the \textit{voltage bias} and it is understood in terms of electronic waveguides. Inside the nanostructure, the waveguide width is controlled by the \textit{voltage gate}. Such a voltage tunes the number of electronic modes supported by the waveguide i.e. $N ~\propto ~V_{gate}$, and each mode contributes with a unit of quantised conductance $e^2/h$. Hence, the voltage gate is varied to adjust the electrostatic energy of the electron gas confined in the potential well formed perpendicularly to the interface between the layers. \\
From this analysis, bias and gate voltages assume an essential role in the QD functioning. On one side, the $V_{bias}$ determines the electronic transport activation: a minimum bias is required for finite electronic flow. Once the last electronic propagating mode reaches its pinch-off, the mode extinguishes and there is no net electronic flow. This phenomenon is the so-called Coulomb blockade - as we return on it in more details later. Here we comment that in the blockade regime only one electronic mode is supported: thus, these devices work as \textbf{single-electron transmission} mode \cite{Kastner_SET1992}. On the other, the $V_{gate}$ governs the internal electrostatic population of the QD because it is applied directly on the confined electron gas inside the device. Thus, we have access to unitary variations of electrons inside the dot. We note also that each of these mentioned voltages is completed by its own capacitance element i.e. $C_{s}, C_{d}, C_{gate}$, as shown in the Fig.\ref{F3:QD}.\\
\noindent{We} continue our discussion analysing properties and regimes of the electrical conductance $\mathcal{G}$ as function of temperature $T$ and voltages $V_{bias},V_{gate}$. From the dependence in $T$, we acquire information of the relevant energy scales of the system and which conductance regimes contribute to the electronic transport. From the dependence in $V_{bias},V_{gate}$, we understand the electronic occupancy of the dot and the equilibrium or steady state condition of the system. We corroborate our presentation  with some qualitative conductance plots to display the various conductance curves. These sketches also anticipate some of the quantitative results we present in later chapters of the thesis. \\
In the spirit of energy renormalization and relevant energy scales, there are three energy scales dictating the conductance behaviour as function of temperature in general QD systems \cite{PustilnikGlazman_review2004}. The first regime is due to the energy $X_{i}$ required to add an electron with opposite spin to the one already occupying the orbital $i$ such the state becomes doubly occupied $\ket{\uparrow\downarrow}$. This term can be identified in the on-site Coulomb repulsion for Anderson impurity model, see Eq.\ref{eq:AM}, or in the \textbf{charging energy} $E_{C}$ governing the dot occupancy in the so-called \textit{constant-interaction model} often used to model experimental set-up. 
The second regime is given by the energy $X^{\prime}_{ij}$ indicating the capacitive interaction between different orbitals $i \neq j$. The \textbf{level spacing} $\delta E$ refers to the energy gap between single-particle dot levels. The last regime is identified by the Kondo temperature $T_{K}$ defined in Eq.\ref{eq:Tk} which controls the crossover from weak to strong asymptotic Kondo regime. We infer the typical energy scales generically run as $T_{K} \ll X^{\prime}_{ij} \ll  X_{i}$ and the corresponding model Hamiltonian for the isolated quantum dot with multiple orbitals $i,j$ reads as
\begin{equation}
\hat{H}_{dot}= \overbrace{\sum_{ij\sigma} t_{ij} d^{\dagger}_{i\sigma}d_{j\sigma}}^{\text{single-particle levels}} + \overbrace{\sum_{i} X_{i}\hat{n}_{i\uparrow}\hat{n}_{i\downarrow} + \sum_{i \neq j}  X^{\prime}_{ij} \sum_{\sigma}\hat{n}_{i\sigma}\hat{n}_{j\sigma}}^{\text{interactions}} ~,
\end{equation}
where the first term indicates the hopping between orbitals with $t_{ij}$ internal dot couplings, the last two terms describes the density-density interactions  within an extended Hubbard model approximation as we describe above. The leads also may have a finite level spacing $\delta E_{\mathbf{k}}$, but this is assumed to be negligible, $\delta E_{\mathbf{k}}\ll T,T_{K}$. Hence, leads describe faithfully a metal conduction band.\\
\begin{figure}[H]
\centering
\hspace*{-1.4cm}\includegraphics[width=1.2\linewidth]{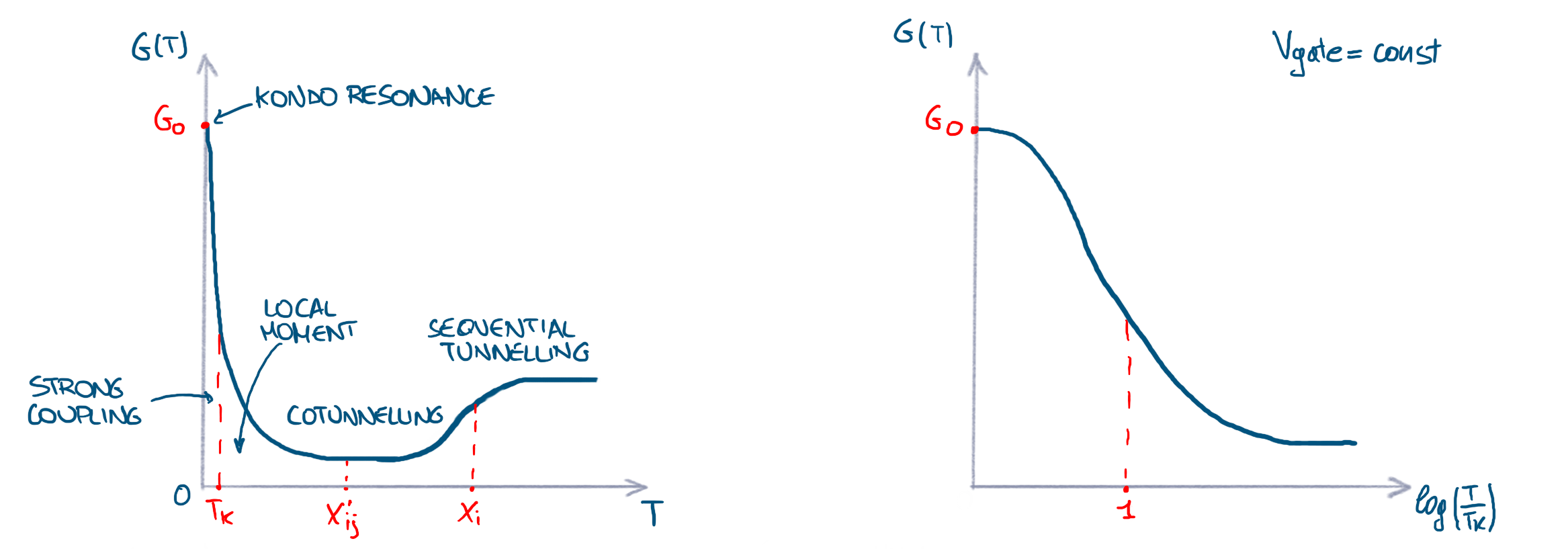}
\caption[Schematic of $\mathcal{G}$ as function temperature in quantum dot model]{The illustrations show: (left) the conductance curve as function of temperature with the different transport regimes, (right) the universal conductance curve at fixed $V_{gate}$. See the main text for more discussion. These curves were observed in the experiment \cite{Kouwenhoven2000Exp} of which we report a plot with experimental data in Fig.\ref{F3:exp2}\textit{(b)}. }\label{F3:QD_T}
\end{figure}
\noindent{We} present now the quantum transport regimes as temperature is progressively lowered, see the conductance curve sketched in Fig.\ref{F3:QD_T} on the left. This discussion is given in qualitative fashion since the specific transition temperatures from a regime to another are highly depending from the set-up parameters.\\
At very high temperature, the transport is in  \textbf{sequential tunnelling} regime: due to the high thermal energy of electrons, the conductance is insensitive to both the value of $X_{i}$ and the applied $V_{gate}$. In this regime, the time spent by an electron on the dot is the largest time scale of the process. Hence, this time interval is larger than the coherent transport time and we infer there is no coherent tunnelling between successive events.  \\
By lowering the temperature, the conductance assumes $V_{gate}$ dependence and its curve is described by the \textbf{Coulomb blockade model} \cite{AverinCBSET1986}. Only when the voltage is tuned at the charge degeneracy point between different states - i.e. mixed-valence regime on the dot characterised by high fluctuation in charge and spin - then the electron transmission is \textit{activationless}. This implies there is no minimum potential required to move one electron from/to the dot. Hence, the conductance is finite and it exhibits sharp peaks. Otherwise, in order to obtain finite conductance within integer charge state, it is needed a finite activation energy i.e.  $V_{bias} >X_{i}$.  If this threshold is not reached, the system shows very low conductance that is a  Coulomb blockade valley. By varying $V_{gate}$, we can explore the different charge states in the system. As consequence, the conductance assumes an \textit{oscillatory} behaviour typical of a sequence of Coulomb blockade valleys and peaks. This phenomenon becomes more pronounced by lowering and lowering the temperature. The peaks at the mixed-valence transitions get sharper feature; the Coulomb blockade valleys in the odd charge sector grow to become plateaux whereas the valleys corresponding to even charge sector persists up to very low $T$. Later we return on this discussion when we study the conductance dependence from the voltages in even and odd charge sectors when the system is in Kondo regime. \\
At medium temperature and increased dot-lead coupling, the transport enters in \textbf{cotunnelling} regime where the transfer of one electron from one lead to another is accompanied by an intermediate state that is \textit{virtual} -  because its energy level is larger than the initial one \cite{NazarovCoTun1990}. This is also a single quantum process, as opposed to the sequential tunnelling. In order to correctly capture the physics of these events we need higher-order tunnelling process which corresponds to the second order $\mathcal{O}(V_{s}V_{d}/U_{i})$ perturbation contribution. Considering the nature of the intermediate state, successive tunnelling events occur in shorter time interval than the coherent one - thus the naming of coherent transport. In the cotunnelling, there are three kinds of transport. The first two regimes characterised the dot ground-state with spin-$0$:
\begin{itemize}
\item the \textbf{inelastic cotunnelling}, where the energy possessed by the electron on the lead before and after the tunnelling is unequal. Furthermore, the virtual intermediate state exhibits a particle-hole pair. This regime does not contribute to the linear response conductance, hence for our purposes we can neglect it \cite{NazarovCoTun1990};
\item the \textbf{elastic cotunnelling}, where the tunnelled electron has the same energy on the lead before and after the transmission. This transport regime is linear in $V_{bias}$, hence it is relevant for linear response conductance study. Moreover, in this regime activationless transport contributes to finite conductance. That means one can measure finite current, the so-called \textit{leakage current}, also at $V_{bias}$ values lower than the threshold one \cite{Flensberg}. And so nonzero conductance is registered even in voltage regime where there should have been a Coulomb blockade valley.
\end{itemize}
The last transport regime is identified for the dot ground-state with a nonzero spin:
\begin{itemize}
\item the \textbf{spin-flip cotunnelling}. In case of an odd number of electrons on the dot ground-state, its corresponding spin is half-integer. Hence, the intermediate virtual state can be accompanied by a flip of the lead electron spin at the end of the transmission with a simultaneous flip of the spin on the dot.
\end{itemize}
Once the transport regime is set at the cotunnelling regime, the conductance curve reaches a minimum. In what follows, we adopt the nomenclature of the regimes defined in the energy renormalization procedure, see Sec.\ref{sec:RGtheo}. Although the sequence of oscillation from valleys to peaks in adherence of the Coulomb blockade model is preserved, by lowering even further the temperature we approach the $T \gg T_{K}$ interval and the system then enters in the \textit{local moment} regime where the cloud of conduction electrons is weakly coupled to the dot. Hence, there is weak electronic transmission and the conductance remains low. At $T\sim T_{K}$, it occurs the crossover between asymptotic transmission regimes. At $T \ll T_{K}$, the screening of the spin on dot with the surrounding conduction electrons is completed. It follows the formation of the many-body Kondo singlet spin state and the system undergoes to \textit{strong coupling} regime. As consequence of the locked dot spin, there is no net electronic flow \textit{through dot-lead} junctions. However, the electronic flow \textit{between leads} is enhanced in this regime and this establishes a sudden boost to conductance up to the quantised value $\mathcal{G}_{0}$ that can be reached at zero temperature. At temperature smaller than $T_{K}$, the transport is a consequence of a \textit{non} perturbative conductance regime where the logarithmic divergence is found in the perturbation theory. 
As we discussed in detail in Sec.\ref{sec:RGtheo}, the renormalization group technique is the correct method to describe the physics at $T \leq T_{K}$ temperature interval that is in general for all $T\ll U_{i}$.\\
As evidence from our presentation, the conductance starts from a medium value due to the electronic thermalization on the dot, then decreases gradually by decreasing $T$ and at the end, when the temperature approaches $T_{K}$, $\mathcal{G}$ grows up to the Kondo resonance at $T=0$ \cite{Georges2016}. Thus, the signature of Kondo effect in QDs is given by unexpected \textit{enhancement in electrical  conductance} at very low temperature, in analogy with the anomalous increment in resistivity at low $T$ measured in magnetic impurities incorporated in metals - see discuss at the beginning of Sec.\ref{sec:QImpModel}. This is because the scattering events of conduction electrons from the QD contribute to the source-drain current that would otherwise be zero without the dot-lead electrons scattering in presence of finite $V_{bias}$. By contrast, in metals with impurities the enhanced scattering hinders bulk transport and increases sample resistance.\\ 
We remark that at \textit{fixed} $V_{gate}$, i.e. constant electronic population on the dot, the QD system has a unique and well-defined $T_{K}$. As presented so far, the conductance as function of temperature $T$ follows an oscillatory behaviour consistent with the Coulomb blockade model. If we plot the conductance as function of the rescaled temperature $T/T_{K}$, it follows a \textit{universal curve}, see sketch in Fig.\ref{F3:QD_T} on the right. This universal behaviour is clearly not valid at very high temperature but it becomes more and more confirmed at low temperatures because of the energy rescaling.\\
\begin{figure}[H]
\hspace*{-1.0cm}
\hspace*{-0.4cm}\includegraphics[width=1.15\linewidth]{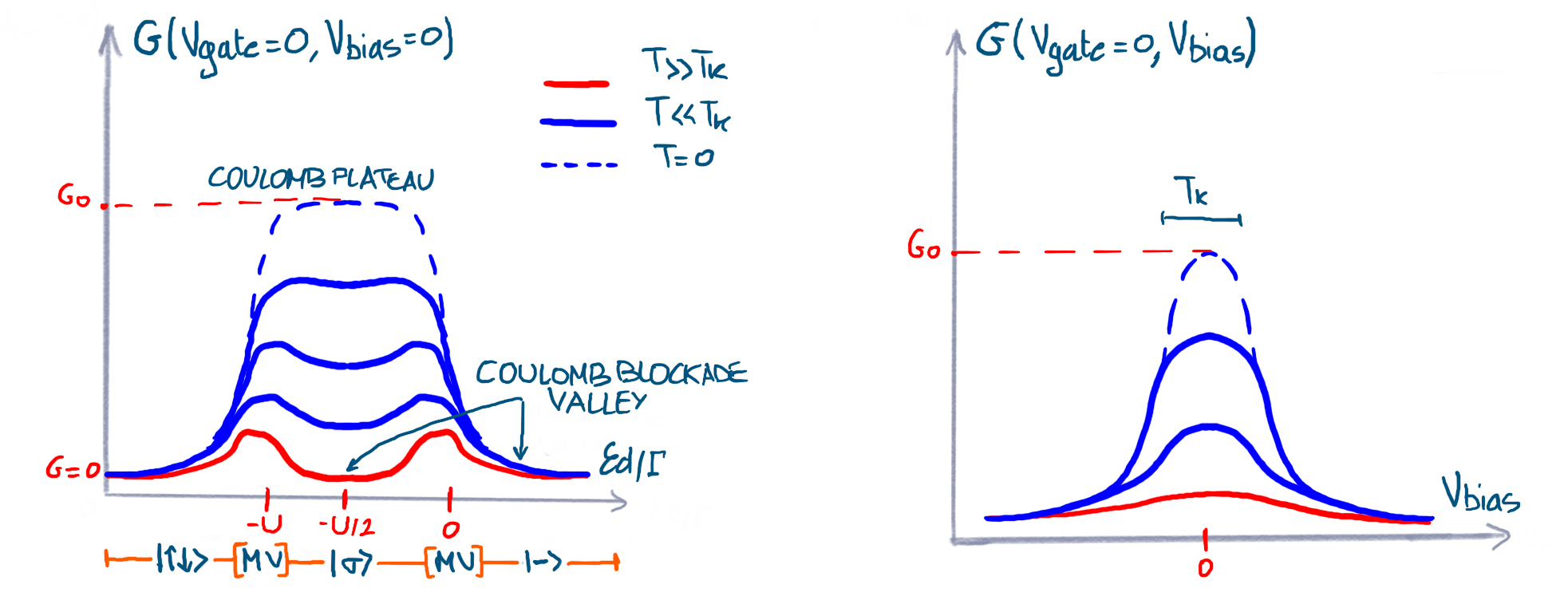}
\caption[Schematic of $\mathcal{G}$ in single-level quantum dot model]{The illustrations sketch the conductance for a single-level QD at $V_{gate}=0$ for high (red line), low (blue line) and zero temperature (blue dash line). $(left)$ the linear response conductance curve as function of dot single-particle energy $\epsilon_{d}$ over gamma $\Gamma$ at $V_{bias} \rightarrow 0$, doubly, singly and empty occupied states separated by mixed-valence regime transitions are also indicated. We report a plot with experimental data of this observation in Fig.\ref{F3:exp2}\textit{(b)} from experiment \cite{Kouwenhoven2000Exp}. $(right)$ the conductance curve as function of finite $V_{bias}$ with zero peak bias at $V_{gate}=V_{bias}=T=0$. We report a plot with experimental data of this observation in Fig.\ref{F3:exp1}\textit{(right,B)} from experiment	\cite{Kouwenhoven2000Exp}. See the main text for more discussion.}\label{F3:QD_V}
\end{figure}
\noindent{We} shift our discussion to the conductance as function of either voltage gate or voltage bias while keeping the other voltage fixed, at temperature regime comparable to $T_{K}$. In this respect, it is instructive to restrict the presentation to a \textbf{single-level QD}. This means the level can be either empty, singly or doubly occupied and the filling process occurs only once. Hence, this is identical to the single-impurity Anderson model. In order to do so, we set $V_{gate}=0$. 
We show now two scenarios: the conductance curve either at zero or at finite voltage bias.\\
We consider first the case of vanishing $V_{bias}$ such that the conductance formula to use is clearly the linear response conductance as defined in Eq.\ref{eq:DiffCondLRac}. The conductance curve, as shown in the sketch in Fig.\ref{F3:QD_V} on the left, differs accordingly to both charge sector and temperature interval. At $T \gg T_{K}$, the conductance displays Coulomb blockade valleys in both even (doubly occupied and empty states) and odd (singly occupied state) sectors with the exception of finite transmission at the mixed-valence regime where narrow peaks appear. By lowering the temperature at $T \ll T_{K}$, in the even sector persists zero conductance. However, in presence of an odd number of electron on the dot, it is possible to completely screen the unpaired spin such that the Kondo singlet is formed. As consequence, the conductance registers the formation of a Coulomb plateau determining finite electronic transmission - as opposed to the valleys at higher temperature. At $T=0$, the plateau sets reaches up to the quantised value $\mathcal{G}_{0}$ at the particle-hole symmetric point with $V_{gate}=0$. Hence, once the system is tuned deep inside the singly occupied state level, finite conductance is measured through the QD at low temperature. We observe also that the single-level QD reproduces one cycle of  oscillatory conductance. We deduce that in case of multiple levels, the conductance shows a repetition of this very same pattern.\\
We consider next the case with finite $V_{bias}$ .
As we can see in the sketch in Fig.\ref{F3:QD_V} on the right, the conductance curve differs according to the temperature regime. At $T \gg T_{K}$, the conductance is almost constant independent of the finite $V_{bias}$. Then, by lowering the temperature, the conductance curve becomes sharper and enhanced progressively around $V_{bias}=0$. At $T=0$ \textit{and} $V_{bias}=0$, we observe the so-called \textit{zero bias peak} meaning that the conductance reaches the maximum quantised value and has width of the order of $T_{K}$. Within this regime, we can use linear response theory to calculate the conductance and this is case of activationless transport. This anomalous peak is the signature of the nonzero electronic transmission between direct coupling of the leads determining finite conductance measured across the system. Under this parameter regime, the tunnelling process is independent from thermal energy and potentials.\\
\begin{figure}[H]
\centering
\includegraphics[width=0.8\linewidth]{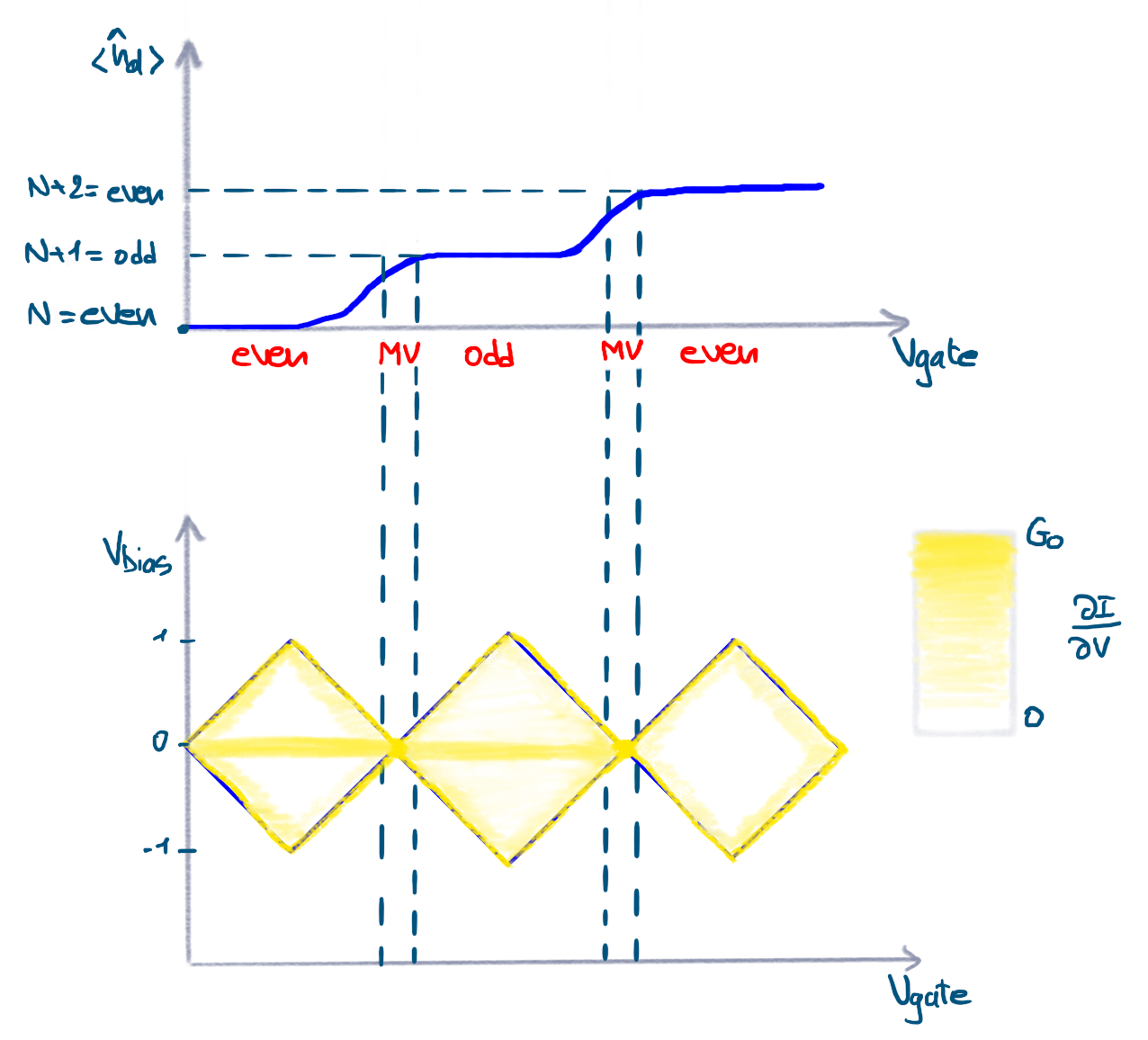}
\caption[Schematic of $\mathcal{G}$ in multiple-level quantum dot model]{The illustrations sketch the stability diagram conductance for a multi-level QD as function of $V_{gate}$ and $V_{bias}$: $(top)$ dot occupancy $\braket{\hat{n}_{d}}$ starting at even number $N$  of electrons, $(bottom)$ differential conductance with Coulomb diamond structure following $\braket{\hat{n}_{d}}$ parity. Experimental data of this observation is found in the experiment \cite{DeFranceschi_ElectronCotunneling2001Exp}. See the main text for more discussion.}\label{F3:QD_VV}
\end{figure}
\noindent{As} last part of our discussion, we present the conductance curve as function of both $V_{bias}$ and $V_{gate}$ at temperature relevant to the Kondo regime. As opposed to the previous case, at finite $V_{gate}$, the dot explores multiple energy levels. In this scenario we have a \textbf{multi-level QD}. Its signature is a sequence of conductance oscillations that is the repetition of the single-level pattern we previously discuss. In the sketch given in Fig.\ref{F3:QD_VV}, we show the dot occupancy given for an arbitrary number of electron $N$ spreads on different levels: this is the so-called \textit{stability diagram}. Even at strictly zero temperature, the curve never follows a sharp step path but it smears at the transition between two different charge sectors by tuning accordingly $V_{gate}$. In this smooth charge variation we identify the mixed-valence regime of the dot, clearly a degenerate crossing. The reason behind the soft transition is due to the presence of the leads coupled through junction to the dot. In the figure, we also present the corresponding differential conductance with the characteristic diamond structures found by interpolating lines departing from finite values of  $V_{bias},V_{gate}$ \cite{DeFranceschi_ElectronCotunneling2001Exp}. We comment more on this periodic pattern. Finite conductance is found only $(i)$ at every charge degenerate crossing i.e. at the mixed-valence regimes $(ii)$ at each odd charge sector. In particular, the differential conductance is enhanced at $V_{bias}\rightarrow 0$ that is observed in the inner region of the diamond of odd charge sectors. \\
Hence, in a multiple orbital QD we study the condition for finite electronic transmission as function of the dot occupancy. By tuning simultaneously the value of  $V_{bias},V_{gate}$ we calculate the differential conductance $\partial I/ \partial V$. By tuning only $V_{gate}$ and the following different temperature regime at vanishing $V_{bias}\rightarrow 0$ we calculate the linear response conductance $\partial I/ \partial V |_{V_{bias}\rightarrow 0}$. Combining this information, we have a complete survey of the conductance properties and physics regime according to temperature and voltage parameters. \\
The Kondo effect is a many-body problem based on first experimental observation in early $1930s$ \cite{Berg_1934ExpResistivity}, see also an illustration in Fig.\ref{F2:ResistanceOrig} and formalised only in mid $1960s$ \cite{Kondo1964}. Despite of the extensive theoretical and, lately also numerical \cite{Krishna1980renormalization}, investigations developed across decades, it remains a challenging task the practical realization of devices capable to measure the Kondo effect. Only at end of $1990s$ \cite{Goldhaber-Gordon-Kondo1998Exp}, appropriate technological advancing made possible the first experimental observation of the physics behind the Kondo singlet formation - as we discuss more in the next section.

\subsection{Kondo effect in quantum dots: experimental verification}
\begin{figure}[H]
	\centering
	\subfloat[]{\includegraphics[width=0.35\linewidth]{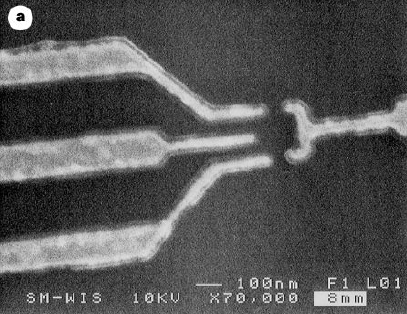}}
	\hfill
	\subfloat[]{\includegraphics[width=0.6\linewidth]{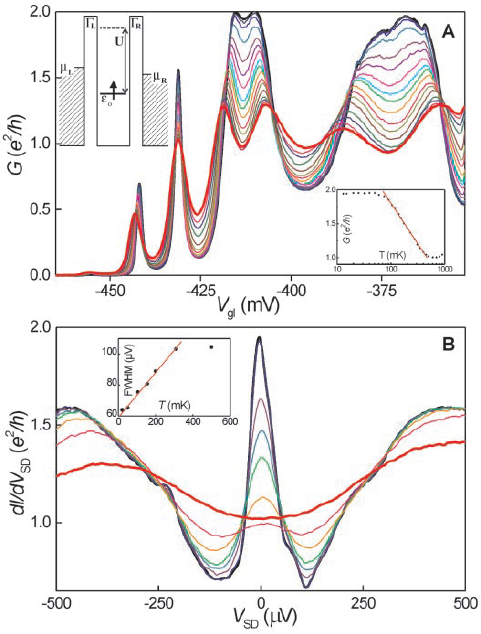}}
	\caption[Original Kondo effect data: set-up, Coulomb oscillation and differential conductance]{\textit{(left)} Scanning electron microscope image showing top view of sample. Three gate electrodes, the one on the right and the upper and lower ones on the left, control the tunnel barriers between reservoirs of two dimensional electron gas (at top and bottom) and the droplet of electrons. The middle electrode on the left is used as a gate to change the energy of the droplet relative to the two dimensional electron gas (imagine and caption from  \cite{Goldhaber-Gordon-Kondo1998Exp}). \textit{(right)} $(A)$ Coulomb oscillations of conductance versus voltage gate left $V_{gl}$ at $B=0.4~ T$ for different temperatures ranging from $15~ mK$ (thick black trace) up to $800~mK$ (thick red trace). The voltage gate right $V_{gr}$ is fixed at $2448~ mV$. $(B)$ Differential conductance $\partial I/\partial V_{SD}$ versus \textit{dc} $V_{bias}$ between source and drain contacts $V_{SD}$ for $T$ ranging from $15~ mK$ (thick black trace) up to $900~ mK$ (thick red trace), at voltage gate left $V_{gl}=-413~mV$ and $B = 0.4~ T$ (imagines and caption from \cite{Kouwenhoven2000Exp}). }\label{F3:exp1}
\end{figure}

\begin{figure}[H]
		\centering
		\subfloat[]{\includegraphics[width=0.5\linewidth]{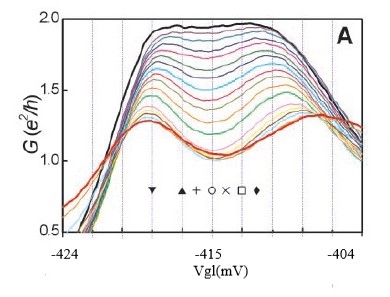}}
		\hfill
		\subfloat[]{\includegraphics[width=0.5\linewidth]{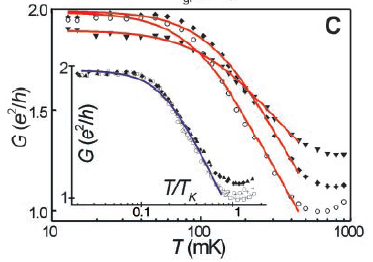}}
	\caption[Original Kondo effect data: unitary limit and universality]{\textit{(a)} Unitary limit of conductance as function of $V_{gl}$ for $T$ ranging from $15~ mK$ (thick black trace) where $\mathcal{G}/\mathcal{G}_{0}\simeq 2$ up to $900 ~mK$ (thick red trace) and $V_{gr}=-488~mV$ and $B= 0.4 ~T$. \textit{(b)} The red curves indicate $\mathcal{G}(T)$  varying from cotunnelling regime to Kondo resonance at fixed voltage gate as extracted from \textit{(a)} with $V_{gl}=-411,-414,-418~mV$ (various symbols). In the inset: over a temperature range of $2.6~K$, that is several times larger than $T_{K}$,  the conductance is universal function of the one-scaling parameter $T/T_{K}$. (images from \cite{Kouwenhoven2000Exp}). }\label{F3:exp2}
\end{figure}

In the previous section, we explain the physics of the most versatile set-up utilised in Kondo effect measurements that is the semiconductor quantum dot. Furthermore, with the aid of plot schematics we present the conductance curve dependence on temperature, energy scale and applied voltages. In this conclusion section, we aim to briefly mention the crucial experimental verifications of the Kondo effect. The theoretical knowledge we present in the Secs.\ref{sec:QImpModel},\ref{sec:RGtheo},\ref{sec:2CK} is now converted in realistic devices tuned by macroscopic voltage parameters.\\
The pioneering experimental realization of the Kondo effect in single-electron transmission devices for spin-$1/2$ model is achieved by M.A. Kastner's group in 1998 \cite{Goldhaber-Gordon-Kondo1998Exp}. They used quantum dots made of $GaAs$/$GaAlAs$ heterostructures, see the original image from scanning electron microscope on the left of Fig.\ref{F3:exp1}. They measured conductance at $V_{bias}=0$ through a dot in dependence of $V_{gate}$. Tuning the gate voltage results in adding or removing an electron to the dot, as the energy spectrum of the dot gets shifted. The different parity of electronic population in the set-up allows to measure the typical periodic pattern in the conductance curve. The Coulomb blockade regime is alternated by Coulomb peaks where enhanced electronic scattering occurs in correspondence of Kondo singlet formation. The experimental data measured indeed peaks which formed pairs separated by valleys. This confirms the sequence of odd and even number of electrons in the dot. The peaks became better resolved with decreasing temperature, suggesting that the range of increased conductivity between the pairs of peaks results from a Kondo resonance.\\
\noindent{Half} a year later, L.P. Kouwenhoven and colleagues \cite{cronenwett1998tunable} neatly showed in the linear response conductance the oscillation from Columb blockade valleys to Kondo peaks according to the number of electron on the dot. In the differential conductance, the odd valleys were in correspondence of pronounced peaks at $V_{bias}=0$. In particular, it was observed that at decreasing temperature values, the conductance peak increases in the middle of the Kondo valley with odd electron occupancy. See experimental data in coloured lines measured by these authors a couple of year later in panels right in Fig.\ref{F3:exp1}. M.A. Kastner's group at the end of 1998 could show also the crossover between the strong coupling Kondo phase and the mixed-valence regime \cite{Goldhaber_MVKondo1998}.\\
\noindent{The} original measurement of Kondo effect in integer spin quantum dot model is realised by L.P. Kouwenhoven and co-workers in mid 2000 \cite{DeFranceschi-Kouwenhoven_KondoSpin1-2000Exp}. They studied spin-flip events in an even charge sector characterised by spin $S=1$ multiplet that is degenerate with a singlet spin state. This yields to high spin fluctuation and strong Kondo effect. In the same year, the same group experimentally verified the Kondo resonance or the so-called unitary limit in the conductance, as signature of the Kondo effect arising at low-temperature \cite{Kouwenhoven2000Exp}. In the real data, see plot $(a)$ in Fig.\ref{F3:exp2}, the lowest temperature curve reaches $\mathcal{G}/\mathcal{G}_{0}\simeq2$. Furthermore, they verified the universal behaviour of the conductance as function of $T/T_{K}$, data on this observations are in plot $(b)$ in Fig.\ref{F3:exp2}.\\
\noindent{In} general, the electronic reservoir used in these experiments is a many-body system, that is the so-called \textit{Fermi sea}. The critical size and energy condition of the Fermi sea determine the collective behaviour of bath which hybridizes with a single-level. This was shown experimentally by S.Jochim's group in 2013 \cite{FermiSeaformation_ExpHeidelberg_2013}.\\
The experimental set-ups mentioned so far are actually single-channel Kondo models. The systems were usually composed by a semiconductor quantum dot with an odd number of electrons such that it has an excess spin $S = 1/2$ coupled to two electrical leads. Thus, this spin-$1/2$ local moment is coupled to two electrical leads called the infinite reservoirs. The two infinite reservoirs cooperate to form a single-channel Kondo effect to screen the dot spin.\\
From an experimental point of view, as it became evident over a decade of research, realistic device hosting two channel Kondo model presents a challenging set-up design. It is required independent leads such that each corresponding quantum point contact is a distinct and tunable channel connecting the leads to the nanostructure. The first concrete realization in this direction of physically separate electron baths coupled to a local artificial magnetic impurity was reached by D.Goldhaber-Gordon and colleagues in 2007 \cite{GoldhaberGordon-2CK-2007Exp}. In the experiment realizing the two channel spin Kondo model, the quantum dot with $S = 1/2$ is coupled to two electrical leads or infinite reservoirs and in addition to one finite reservoir. Introducing one finite reservoir has important consequences: the finite reservoir can be effectively represented as a large quantum dot with a quantized energy level spectrum and a charging energy. As we know, at low temperature, the charging energy can forbid charge transfer to or from the finite reservoir causing the Coulomb blockade regime. Thus, this finite reservoir also attempts to screen the excess dot spin-$1/2$, forming an independent single-channel Kondo effect. The two independent channels i.e. the finite reservoir and the infinite reservoir can be tuned to be equally coupled to the spin-$1/2$, forming the sought-after two channel Kondo effect. \\
This novel engineered set-up disclosed the study of the two channel system and its interesting physics. The renormalisation flow of two channel system was observed originally in the charge-Kondo model by F. Pierre's group in 2015 \cite{2CK-RenormalisationFlow_IftikharPierre2015Exp}. This experiment was also the pioneer realization of the charge-Kondo paradigm. Few years later in 2018, the same group experimentally measured quantum criticality in the three channel charge-Kondo model \cite{iftikhar2018tunable}.\\
We conclude mentioning that the Kondo effect has also been observed in carbon nanotubes \cite{nygaard_Kondonanotubes2000}, in molecular nanodevice \cite{Scott_KondoMol2010} as we study in Chapter \ref{ch:benz} and graphene sheets \cite{chen2011tunable} as we study in Chapter \ref{ch:graphene}.


\chapter{Improved calculations for quantum transport in interacting multi-orbital nanostructures}
In this chapter we address our discussion to methods and insight leading to improved calculations for physical quantities involved in the quantum transport calculation. This chapter offers a valuable analytical source of calculations for deriving important quantities under problematic regimes and/or for elaborate system configurations - as those are not fully addressed in the current literature.\\
The outline of this chapter is the following. In section \ref{sec:AlternativeLandauer}, we introduce the concept of structured leads and their different manipulation according to their equivalent and inequivalent composition. In section \ref{sec:AlternativeMW}, we study the derivation of the lesser impurity Green's function out of equilibrium and consider an alternative formulation of the Meir-Wingreen current formula. In section \ref{sec:ImprovedKubo}, we present an improved version of the Kubo formula for conductance under linear response. In  section \ref{sec:AlternativeOguri}, we show an alternative formulation of the Oguri formula by introducing the concept of extended impurity model.\\
Most of the results presented in this chapter are published in the reference \cite{transport}.

\section{Conductance for systems with structured leads}\label{sec:AlternativeLandauer} 
In this section our objective is the generalization on the leads composition to an arbitrary density of states. We used this already in Eqs.\ref{eq:DoSinfWBL},\ref{eq:LevelWidth},\ref{eq:defPC}. We consider a general interacting impurity model as given in Eq.\ref{eq:defImpModelGen} with interaction and we present the following generalization for the conduction electron bath.\\
We aim to introduce a general density of states with \textit{energy dependence}, namely $\rho_{0,\alpha\sigma}(\omega)$ and we define this general leads structure as \textbf{structured leads}. Here, $\alpha=L,R$ are the leads labels and the subscript $0$ denotes the free density of states of the isolated leads without being coupled with the dot. This condition opposes to the standard practice of assuming metallic conduction band in the leads such that the density of states is a constant in energy i.e. $\rho_{0,\alpha\sigma}$.\\
In general, the corresponding density of states might not be the same for all the leads in the set-up. Thus, for structured leads models, we define
\begin{equation} \label{eq:4defEqIneq} 
\boxed{
\begin{aligned}
&\textbf{equivalent leads:} \quad \rho_{0,\alpha\sigma}(\omega) = \rho_{0,\alpha^{\prime}\sigma}(\omega) \equiv \rho_{0,\sigma}(\omega) \quad \text{and} \quad  \epsilon_{\mathbf{k}\alpha}=\epsilon_{\mathbf{k}\alpha^{\prime}} \equiv \epsilon_{\mathbf{k}} ~,~\forall~ \alpha,\alpha^{\prime} ~\text{and}~ \forall~\mathbf{k}\\
&\textbf{inequivalent leads:} \quad \rho_{0,\alpha\sigma}(\omega) \neq \rho_{0,\alpha^{\prime}\sigma}(\omega) \quad \text{and} \quad \epsilon_{\mathbf{k}\alpha} \neq \epsilon_{\mathbf{k}\alpha^{\prime}} ~~ \forall ~ \mathbf{k} 
\end{aligned} }~.
\end{equation}
We remark the proportional coupling property is \textit{always} satisfied for equivalent leads set-up with the same gamma matrices $[\mathbb{\Gamma}^{\alpha}(\omega)]_{nm}$, see the definition given in Eq.\ref{eq:defPC}.\\
The two conditions in Eq.\ref{eq:4defEqIneq}, not only necessitate of a different implementation in the physical set-up but also have significant impact on the quantum transport for the model. And this understanding is indeed the crucial point in the present section.\\
In the following, we detail the derivation for the structured, equivalent leads and then for the structured, inequivalent ones where emphasis is placed on the differences between the two modellings. The systems with equivalent leads offer the opportunity to recover the Meir-Wingreen and the Landauer formulas. On the contrary, in the inequivalent, noninteracting model we derive a version of the Kubo formula, modified by the highly non-trivial electronic condition in the leads.

\subsection{Systems with equivalent leads}
We consider the standard Anderson impurity model in Eq.\ref{eq:AM} but now for structured source and drain leads. Instead of the nomenclature adopted so far in terms of left and right leads, here we present the calculations for $\alpha=s,d$ namely \textit{source} and \textit{drain} leads - this implies a finite $V_{bias}$ is applied and current flows from source to drain.\\
As first goal, we aim to calculate the Kubo formula for electrical conductance. Then, from this expression we can derive an analogue equation to the Landauer and Meir-Wingreen formula but now for structured, equivalent leads. We present now the series of manipulations required to obtain the Kubo formula in the generalized system \cite{PustilnikKondoMagnetic2001,Flensberg}.\\
As initial operation, we perform the rotation of the leads states in channel space: namely from the original source/drain charge basis into \textit{diagonal} even/odd basis allowing for independent charge conservation per sector. This transformation is defined via the unitary matrix $\mathbb{U}_{sd}$  with $dim(2\times2)$ and it reads
\begin{equation}
	\begin{aligned}
&\mathbf{C}_{eo,\mathbf{k}\sigma} = \mathbb{U}_{sd} \mathbf{C}_{sd,\mathbf{k}\sigma} ~, \\
&\hspace*{-3cm} \text{with the vectors:}~ \mathbf{C}_{sd,\mathbf{k}\sigma} = \begin{pmatrix}
	c_{s,\mathbf{k}\sigma} \\ c_{d,\mathbf{k}\sigma} \end{pmatrix} ~,~
 \mathbf{C}_{eo,\mathbf{k}\sigma} =\begin{pmatrix}
 	c_{e,\mathbf{k}\sigma} \\ c_{o,\mathbf{k}\sigma} \end{pmatrix} ~.
\end{aligned}
\end{equation} 
For proportionate coupling models, such as the AIM, proper choice of the matrix $\mathbb{U}_{sd}$ allows to decouple the odd lead combination from the impurity. Note that this decoupling is only possible under proportionate coupling condition.
This implies the rotated Hamiltonian is decomposed into $\hat{H}_{e,imp}$ representing the noninteracting even combination of lead electrons coupled to the single-impurity level and $\hat{H}_{o}$ representing the noninteracting odd combination of lead electrons. We proceed now with computing the Kubo formula for electrical conductance of the rotated Hamiltonian.\\
In this derivation we find convenient to define the \textit{local lead orbital} at the nanostructure position $c_{\alpha\sigma}=\sum_{\mathbf{k}}U^{\alpha}_{\mathbf{k}}c_{\alpha\mathbf{k}\sigma}$ with $\sum_{\mathbf{k}}|U^{\alpha}_{\mathbf{k}}|^2 =1$. The coupling then reads $V^2 = V^2_{s}+V^2_{s}$. Using local orbital basis we avoid the summation over momenta $\mathbf{k}$.\\
We define the charge current as in Eq.\ref{eq:defChargeCurr} but now in the original source/drain basis, namely
\begin{equation}
\widehat{I}^{C}_{\hat{N}_{d}}(t) = \langle \dot{N}_{d}(t) \rangle= -i e \langle[\hat{H},\widehat{N}_{d}(t)]\rangle ~,
\end{equation}
and we impose current conservation with respect to the drain current using a linear combination such that the final current expression is given only in terms of the \textit{odd} electron combination, namely 
\begin{equation}
\widehat{I}_{\hat{N}_{o}} =-i\sum_{\sigma} \langle \frac{1}{\sqrt{|V_{s}|^{2}+|V_{d}|^{2}}} \left( V_{s}V_{d} c^{\dagger}_{o\sigma} d_{\sigma} - V^{\star}_{s}V^{\star}_{d}d_{\sigma}^{\dagger}c_{o\sigma} \right) \rangle ~,
\end{equation}
where the independent charge flavour conservation per sector implies the operators belong to different Hilbert spaces. As consequence, a general string composed by the multiplication of odd lead electrons and impurity operators can be factorised.\\
We define the retarded current-current correlator for the original \textit{sd} basis as in Eq.\ref{eq:defCurrCurrCor}, namely
\begin{equation}
K_{\hat{N}_{s},\hat{N}_{d}} (t, t^{\prime}=0) = \langle\langle\widehat{I}_{\hat{N}_{s}}(t);\widehat{I}_{\hat{N}_{d}}(0) \rangle\rangle ~,
\end{equation}
through the dynamical retarded susceptibility as in Eq.\ref{eq:defSuscept}, 
\begin{equation}
\chi (t)= K^{R}_{\hat{N}_{s},\hat{N}_{d}} (t, T)  
		= -i \theta( t) \langle [\widehat{I}_{\hat{N}_{s}}(t),\widehat{I}_{\hat{N}_{d}}(0)]\rangle ~,
\end{equation}
which is evaluated at equilibrium, without the bias perturbation, $V_{bias}=0$. The corresponding current-current correlator for the odd electron combination sector is then given by $ \langle\langle\widehat{I}_{\hat{N}_{o}}(t);\widehat{I}_{\hat{N}_{o}}(0) \rangle\rangle$ and we refer to it still with notation $K_{\hat{N}_{s},\hat{N}_{d}} (t, t^{\prime}=0)$ for the sake of clarity.\\
By means of the independent charge sector conservation, we can decompose $K^{R}_{\hat{N}_{s},\hat{N}_{d}}$ into product of local impurity Green's function $G_{dd\sigma}^{</>}$ and noninteracting odd combination of lead electrons Green's function $G_{0,o\mathbf{k}\sigma}^{</>}$. Hence, after few manipulation, the current-current correlator reads
\begin{equation}
\begin{aligned}
K^{R}_{\hat{N}_{s},\hat{N}_{d}} (t, T) = -i \theta(t)\sum_{\sigma}
\frac{|V_{s}|^{2}|V_{d}|^{2}}{|V_{s}|^{2}+|V_{d}|^{2}} & \Big( 
G_{0,o\sigma}^{<}(-t)G_{dd\sigma}^{>}(t)
-G_{dd\sigma}^{<}(t)G_{0,o\sigma}^{>}(-t) +\\
& \hspace*{0.5cm} +G_{dd\sigma}^{<}(-t)G_{0,o\sigma}^{>}(t) -G_{0,o\sigma}^{<}(t)G_{dd\sigma}^{>}(-t)\Big)
\end{aligned}
\end{equation}
and using the Hermitian property of the contour-ordered Green's functions we can show that the above expression is imaginary i.e. $K^{R}_{\hat{N}_{s},\hat{N}_{d}}(t) \equiv \mathit{Im} K^{R}_{\hat{N}_{s},\hat{N}_{d}}(t)$.\\
The Kubo formula for linear response is obtained from the Fourier transform of $\mathit{Im} K^{R}_{\hat{N}_{s},\hat{N}_{d}}$ as in Eq.\ref{eq:defKuboLR}, namely
\begin{equation}
\begin{aligned}
&\mathit{Im}K^{R}_{\hat{N}_{s},\hat{N}_{d}} (\omega, T) = \int^{+\infty}_{-\infty} dt ~\mathit{Im} K^{R}_{\hat{N}_{s},\hat{N}_{d}}(t, T) e^{i \omega t}  ~,\\
&= \frac{-i}{2} \sum_{\sigma} \frac{|V_{s}|^{2}|V_{d}|^{2}}{|V_{s}|^{2}+|V_{d}|^{2}} \int^{+\infty}_{-\infty} \frac{d\omega_{1}}{2\pi}
 \Big( 
-G_{0,o\sigma}^{<}(\omega_{1})G_{dd\sigma}^{>}(\omega_{1}+\omega)
+G_{dd\sigma}^{<}(\omega_{1}+\omega)G_{0,o\sigma}^{>}(\omega_{1}) +\\
& \hspace*{6cm} -G_{dd\sigma}^{<}(\omega_{1}-\omega)G_{0,o\sigma}^{>}(\omega_{1}) +G_{0,o\sigma}^{<}(\omega_{1})G_{dd\sigma}^{>}(\omega_{1}-\omega)\Big) ~,
\end{aligned}
\end{equation}
that is a thermal equilibrium result and so we can apply the fluctuation-dissipation theorem in Eq.\ref{eq:defFDtheo}. In particular, for the odd basis Green's functions, we write the expression in terms of the noninteracting, structured density of states $\rho_{0,o\sigma}(\omega)$ and after few steps we get
\begin{equation}
\begin{aligned}
&\mathit{Im}K^{R}_{\hat{N}_{s},\hat{N}_{d}} (\omega, T) =  \frac{1}{2} \sum_{\sigma} \frac{|V_{s}|^{2}|V_{d}|^{2}}{|V_{s}|^{2}+|V_{d}|^{2}} 2\pi\int^{+\infty}_{-\infty} d\omega_{1}
  \rho_{0,o\sigma}(\omega_{1}) \times \\
& \hspace*{0.5cm}  \times \Big(
\mathcal{A}_{dd\sigma}(\omega_{1}-\omega) \big( f^{Eq}(\omega_{1}-\omega)  - f^{Eq}(\omega_{1})\big) - \mathcal{A}_{dd\sigma}(\omega_{1}+\omega) \big( f^{Eq}(\omega_{1}+\omega)  - f^{Eq}(\omega_{1})\big) \Big) ~.
\end{aligned}
\end{equation}
The integration over the whole energy spectrum allows to perform a shift of variables such that we end with the following expression
\begin{equation}
\begin{aligned}
\mathit{Im}K^{R}_{\hat{N}_{s},\hat{N}_{d}} (\omega, T) &=  \frac{1}{2} \sum_{\sigma} \frac{|V_{s}|^{2}|V_{d}|^{2}}{|V_{s}|^{2}+|V_{d}|^{2}}  2\pi\int^{+\infty}_{-\infty} d\widetilde{\omega}
 \mathcal{A}_{dd\sigma}(\widetilde{\omega}) \times\\
 & \hspace*{1cm}\times \Big( 
(\rho_{0,o\sigma}(\widetilde{\omega} - \omega) - \rho_{0,o\sigma}(\widetilde{\omega}+\omega) ) f^{Eq}(\widetilde{\omega}) +\\
&\hspace*{2cm} - (\rho_{0,o\sigma}(\widetilde{\omega} - \omega)f^{Eq}(\widetilde{\omega} - \omega)- \rho_{0,o\sigma}(\widetilde{\omega} + \omega)f^{Eq}(\widetilde{\omega} + \omega))\Big)~.
\end{aligned}
\end{equation}
We can now insert this resulting expression for $\mathit{Im}K^{R}_{\hat{N}_{s},\hat{N}_{d}} (\omega, T)$ into the definition of linear response function for conductance in Eq.\ref{eq:defSusceptCondAC} to obtain the Kubo formula for electrical conductance as in Eq.\ref{eq:defKuboLRel}, namely
\begin{equation}\label{eq:4GacKEq}
\begin{aligned}
&\mathcal{G}^{C,ac}(\omega,T) = \mathcal{G}_{0} 2\pi \dfrac{-\mathit{Im}K^{R}_{\hat{N}_{s},\hat{N}_{d}}(\omega, T)}{\omega} ~,\\
&= \mathcal{G}_{0} \frac{-4\pi^{2}}{2\omega} \sum_{\sigma} \frac{|V_{s}|^{2}|V_{d}|^{2}}{|V_{s}|^{2}+|V_{d}|^{2}} \int^{+\infty}_{-\infty} d\widetilde{\omega}
 \mathcal{A}_{dd\sigma}(\widetilde{\omega}) \times \Big( 
(\rho_{0,o\sigma}(\widetilde{\omega} - \omega) - \rho_{0,o\sigma}(\widetilde{\omega}+\omega) ) f^{Eq}(\widetilde{\omega}) + \\
&\hspace*{4cm} - (\rho_{0,o\sigma}(\widetilde{\omega} - \omega)f^{Eq}(\widetilde{\omega} - \omega)- \rho_{0,o\sigma}(\widetilde{\omega} + \omega)f^{Eq}(\widetilde{\omega} + \omega))\Big) ~,
\end{aligned}
\end{equation}
that is the \textbf{ac-Kubo formula for structured, equivalent  leads}.\\
In similar spirit to Eq.\ref{eq:defSusceptCondDC}, we take the limit $\omega\rightarrow0$ in $\mathcal{G}^{C,ac}$ to get the \textit{dc-}Kubo formula. Using the pre-factor $1/(-2\omega)$ and the infinitesimal difference quotient, we reconstruct the definition of derivative and we get 
\begin{equation}
\mathcal{G}^{C,dc}(T) = \mathcal{G}_{0} 4\pi^{2} \sum_{\sigma} \frac{|V_{s}|^{2}|V_{d}|^{2}}{|V_{s}|^{2}+|V_{d}|^{2}}
\int^{+\infty}_{-\infty} d\widetilde{\omega} \mathcal{A}_{dd\sigma}(\widetilde{\omega})\rho_{0,o\sigma}(\widetilde{\omega}) \left( \frac{-\partial f^{Eq}(\widetilde{\omega})}{\partial\widetilde{\omega}} \right) ~,
\end{equation}
where by multiplying top and bottom for $\pi \rho_{0,o
\sigma}(\widetilde{\omega})$ and we identify the definition of gamma function for structured leads as in Eq.\ref{eq:DoSinfWBL} to get  
\begin{equation} \boxed{
4\pi \sum_{\sigma} \frac{|V_{s}|^{2}|V_{d}|^{2}}{|V_{s}|^{2}+|V_{d}|^{2}} \rho_{0,\sigma}(\omega) \frac{\pi\rho_{0,\sigma}(\omega)}{\pi\rho_{0,\sigma}(\omega)} \equiv 4\frac{\Gamma^{L}(\omega)\Gamma^{R}(\omega)}{\Gamma^{L}(\omega)+\Gamma^{R}(\omega) }  \doteq \mathrm{\Gamma}(\omega)\widetilde{\mathrm{\Gamma}}} ~,
\end{equation} 
with $\rho_{0,\sigma}(\omega)$ the density of states for structured, equivalent leads and with the definition of the geometric factor in Eq.\ref{eq:defGeomFactor}.\\
Hence, the \textbf{dc-Kubo formula} is rewritten as
\begin{equation}\label{eq:4GdcStrucEquiv}
	\begin{aligned}
\mathcal{G}^{C,dc}(T) &= \mathcal{G}_{0}  \sum_{\sigma}  \int^{+\infty}_{-\infty} d\widetilde{\omega} \pi  \frac{4\Gamma^{L}(\widetilde{\omega} )\Gamma^{R}(\widetilde{\omega} )}{\Gamma^{L}(\widetilde{\omega} )+\Gamma^{R}(\widetilde{\omega} )} \mathcal{A}_{dd\sigma}(\widetilde{\omega}) \left( \frac{-\partial f^{Eq}(\widetilde{\omega})}{\partial\widetilde{\omega}} \right) ~, \\
&= \mathcal{G}_{0} \widetilde{\mathrm{\Gamma}}\sum_{\sigma}\int^{+\infty}_{-\infty} d\widetilde{\omega} \underbrace{\pi \mathrm{\Gamma}(\omega)\mathcal{A}_{dd\sigma}(\widetilde{\omega})}_{t_{ee\sigma}(\omega)} \left( \frac{-\partial f^{Eq}(\widetilde{\omega})}{\partial\widetilde{\omega}} \right) ~,
\end{aligned}
\end{equation} 
where $t_{ee\sigma}(\omega)$ is the $\mathrm{T}$-\textit{matrix spectral function} for the even lead combination, see general definition given in Eq.\ref{eq:TmatSpe}.\\
Furthermore, if we take the zero temperature limit we obtain
\begin{equation}
\mathcal{G}^{C,dc} (T=0)= \mathcal{G}_{0} \widetilde{\mathrm{\Gamma}}  \sum_{\sigma}\pi\mathrm{\Gamma}(\omega=0) \mathcal{A}_{dd\sigma}(\omega=0) ~.
\end{equation}  
In conclusion, we derive the expression for the Kubo formula for electrical conductance for structured and equivalent leads. Note that this derivation holds for any model under PC condition. Now, we proceed with the following two short sections to present how from this equation derived from interacting impurity models we can reconstruct the Meir-Wingreen and Landauer formulas.\\

\noindent{By} analogy to Eq.\ref{eq:4GdcStrucEquiv}, we write the equivalent Meir-Wingreen formula for the current in nonequilibrium condition, namely:
\begin{equation}\label{eq:4LEquiv} \boxed{
\begin{aligned}
\hat{I} &= \frac{2e}{h} \sum_{\sigma} \int^{+\infty}_{-\infty} d\omega 4\pi \frac{|V_{s}|^{2}|V_{d}|^{2}}{|V_{s}|^{2}+|V_{d}|^{2}} 
\rho_{0,o\sigma}(\omega) \pi \mathcal{A}_{dd\sigma}(\omega) \Big( f^{\mu_{L}}(\omega) -f^{\mu_{R}}(\omega) \Big) \\
&\equiv\frac{2e}{h}\widetilde{\mathrm{\Gamma}} \sum_{\sigma} \int^{+\infty}_{-\infty} d\omega 
\mathrm{\Gamma}(\omega) (-\mathit{Im}G_{dd\sigma}^{R}(\omega))\Big( f^{\mu_{L}}(\omega) -f^{\mu_{R}}(\omega) \Big)  
\end{aligned} } ~,
\end{equation}
that is the \textbf{Meir-Wingreen current for structured equivalent leads} valid for fully interacting nonequilibrium impurity model. We note that the appearance of the $\widetilde{\mathrm{\Gamma}}$ geometric factor states Eq.\ref{eq:4LEquiv} fulfils the \textit{proportionate coupling} condition, see Eq.\ref{eq:defPC}. This is a significant result because we have derived the very same expression as the original Meir-Wingreen formula under PC, see Eq.\ref{eq:MW->PC}, but from the Kubo formula as starting point. It is has to be noted that Eq.\ref{eq:4LEquiv} was already derived in  Eq.2 of reference \cite{MeirWingreen_AMoutEquilib_PRL1993}. However, in that case, the derivation is based on a genuine nonequilibrium framework as opposed to the general equilibrium and interacting formulation we present in this section. Furthermore, although the gamma function in \cite{MeirWingreen_AMoutEquilib_PRL1993} takes both frequency and spin dependence giving the same analytical form of our expression Eq.\ref{eq:4LEquiv}, the concept of structured leads and its implication on a set-up design are not yet unravelled. Considering this additional physical understanding, we introduce a novelty in the derived Eq.\ref{eq:4LEquiv}. \\
The crucial aspect of our derivation is the lead equivalence allowing to a direct decomposition of the Hilbert space that determines leads decoupling in the system. This aspect simplifies enormously the treatment of complicated strings of leads and impurity operators. Hence, as first important outcome from the structured equivalent leads in interacting equilibrium  impurity model we obtain the very same Meir-Wingreen current formula.

\subsection*{The Oguri formula for structured, equivalent leads}
In the context of structured, equivalent leads for interacting impurity models under PC, we can find an interesting avenue to recover a Landauer-type of formula. \\
We consider here again the Anderson impurity model defined in Eq\ref{eq:AM} fro simplicity. We start with the interacting impurity Green's function in  Eq.\ref{eq:G_dd} with the assumption of wide-flat conduction band such that the structured gamma function contributes only in its imaginary part. Then, at zero temperature and for decreasing energies, the system behaves as Fermi liquid model. Using the renormalized single-particle impurity energy $\epsilon_{d}^{\star}$ and the renormalized hybridization $\Gamma^{\star}$ as in Eq.\ref{eq:G_ddFL}, for structure leads system under Fermi liquid condition we can write the impurity Green's function as follows
\begin{equation}
\lim_{\omega,\eta \rightarrow 0} G_{dd\sigma}^{R}(\omega+i\eta) = \lim_{\omega,\eta \rightarrow 0} \frac{1}{\omega-\epsilon_{d}^{\star}(\omega) +i(\eta +\Gamma^{\star}(\omega))} ~, 
\end{equation}
where we define the renormalised impurity level $\epsilon_{d}^{\star}(\omega)=\epsilon_{d}+ \mathit{Re}\Sigma(\omega)$ and the renormalised gamma function $\Gamma^{\star}(\omega)=\Gamma+\mathit{Im}\Delta(\omega)$ in WBL approximation. The corresponding imaginary part at taken limits reads
\begin{equation}
\mathit{Im}G_{dd\sigma}^{R}(0) = \frac{\Gamma^{\star}(0)}{(\epsilon_{d}^{\star}(0))^{2} +(\Gamma^{\star}(0))^{2} } \equiv |G_{dd\sigma}^{R}(0)|^{2} (\mathit{Im}\Delta(0)) =
|G_{dd\sigma}^{R}(0)|^{2} (-\Gamma^{\star}(0)) ~,
\end{equation}
where in the last step we use Eq.\ref{eq:DoSinfWBL}. In the AIM under Fermi liquid regime the renormalised gamma function reduces to $\Gamma^{\star}\equiv\Gamma$.\\
In order to bring at completion the calculation, we use the above expression for $\mathit{Im}G_{dd\sigma}^{R}(\omega=0)$ to find the spectral function $\mathcal{A}_{dd\sigma}(\omega=0)$ via Eq.\ref{eq:defSpectral}. Then, we insert this result into the \textit{dc}-regime conductance in Eq.\ref{eq:4GdcStrucEquiv} taking also $T\to 0$ limit and we find:
\begin{equation}\boxed{
\mathcal{G}^{C,dc} (\omega=0,T\to 0)= \mathcal{G}_{0} \widetilde{\mathrm{\Gamma}} (\mathrm{\Gamma}^{L}(0)+\mathrm{\Gamma}^R(0))^2 \sum_{\sigma} |G^{R}_{dd\sigma}(0)|^2} ~,
\end{equation}
which correctly recover the Oguri result in Eq.\ref{eq:Oguri} at $\omega=T=0$  Fermi liquid regime in the presence of structured leads, for the Anderson impurity model.\\

\noindent{In} conclusion from this section, we find that the generalization to structured leads allows the treatment for quantum transport in set-up beyond the standard constant metallic conduction band. We find a generalization of the Kubo formula for electrical conductance for structured equivalent leads in Eq.\ref{eq:4GacKEq}, oppose to literature results where the energy dependence of the conduction band is neglected \cite{Sindel2005}. Moreover, under the requirement of leads equivalence as defined in Eq.\ref{eq:4defEqIneq}, we gain easier analytical calculations and, foremost, we find again expressions that recover the two fundamental quantum transport approaches i.e. the nonequilibrium under the Meir-Wingreen current formula and the Oguri formula. \\
We continue now our discussion with the case of inequivalent leads.

\subsection{Systems with inequivalent leads}
As introduction to this section, we consider the general set-up with structured leads as it is predominant feature of the current section and we frame the discussion into the classification given in Table \ref{table:strucEqIneq}.
\begin{table}[H]
	\centering
\begin{tabular}{ |c||c|c| } 
 \hline
  & $\rho_{0,\sigma}(\omega)$ & $\rho_{0,L\sigma}(\omega)\neq\rho_{0,R\sigma}(\omega)$ \\
 \hline\hline
 $U=0$ & Landauer formula Eq.\ref{eq:LB_2probe} & generalised Kubo Eq.\ref{eq:4GacKInequiv} \\   \hline\hline
 $U$& generalised Landauer Eq.\ref{eq:4LEquiv} & \textit{no Landauer-type}\\
  \hline
\end{tabular}  ~.
\caption[Summary for structured equivalent vs inequivalent leads]{Summary structured lead for equivalent vs inequivalent leads and interacting vs noninteracting impurity model.}\label{table:strucEqIneq}
\end{table}
\noindent{The} first column reports both the literature and the generalization we presented in the previous section for set-up with equivalent leads meaning the definition in the first line of Eq.\ref{eq:4defEqIneq}. The second column provides a new scenario to be explored: the case of set-up with inequivalent leads, see second line in Eq.\ref{eq:4defEqIneq}. We can already argue that it is not possible to cast any interacting impurity models without the same density of states in each lead into a Landauer-type of formula. The bottleneck is the operations one can legitimately enforce on the dynamical retarded susceptibility in Eq.\ref{eq:defSuscept} and the impossibility in this case of satisfying the PC condition.\\
In this section, we aim to derive an analogue of the Kubo formula for electrical conductance starting form the standard Anderson impurity model in Eq.\ref{eq:AM} but now for structured and inequivalent leads. At completion of our calculations, we find two types of term: one is reducible to the expression in Eq.\ref{eq:4GacKEq} as we obtain for equivalent leads set-up whereas the other one shows a different expression structure  representing the correction due to the intrinsic different nature of each lead. As outcome, we derive a version of the Kubo formula comprises different density of states in the leads.\\
Considering the similarity with the previous calculation for equivalent leads, the analytical steps leading to the same type of equations format not repeated here and we present only the final new expressions. We highlight the difference and difficulties of this machinery in inequivalent leads configuration. Ultimately, we also realize the incompatibility of this approach in case of inequivalent leads in interacting impurity models.\\

\noindent{We} start again with the standard Anderson impurity model.
The charge current is defined per each charge sector as in Eq.\ref{eq:defChargeCurr}. On the contrary of the case where the density of states is the same in each channel, there is no unitary transformation to decouple one effective lead combination because the transformed basis are now diagonal. \\
Hence, as given in Eq.\ref{eq:defSuscept}, the dynamical retarded susceptibility in the general charge sectors case reads,
\begin{equation}
\begin{aligned}
&\chi (t) = K^{R}_{\hat{N}_{\alpha},\hat{N}_{\alpha^{\prime}}} (t, T)  = -i \theta( t) \langle [\widehat{I}_{\hat{N}_{\alpha}}(t),\widehat{I}_{\hat{N}_{\alpha^{\prime}}}(0)]\rangle ~,\\
&\hspace*{-0.5cm}= \sum_{\mathbf{k}\mathbf{k}^{\prime}\sigma} \Big( -i\theta( t) iV_{\alpha^{\prime}\mathbf{k}^{\prime}}iV^{\star}_{\alpha\mathbf{k}} \langle [c^{\dagger}_{\alpha^{\prime}\mathbf{k}^{\prime}\sigma}(0)d_{\sigma}(0),d^{\dagger}_{\sigma}(t)c_{\alpha\mathbf{k}\sigma}(t)]\rangle
+i\theta( t) iV_{\alpha\mathbf{k}}iV^{\star}_{\alpha^{\prime}\mathbf{k}^{\prime}} \langle [c^{\dagger}_{\alpha\mathbf{k}\sigma}(t)d_{\sigma}(t),d^{\dagger}_{\sigma}(0)c_{\alpha^{\prime}\mathbf{k}^{\prime}\sigma}(0)]\rangle\\
&-i\theta( t)iV_{\alpha\mathbf{k}}iV_{\alpha^{\prime}\mathbf{k}^{\prime}}\langle [c^{\dagger}_{\alpha\mathbf{k}\sigma}(t)d_{\sigma}(t),c^{\dagger}_{\alpha^{\prime}\mathbf{k}^{\prime}\sigma}(0)d_{\sigma}(0)]\rangle
+i\theta( t) iV^{\star}_{\alpha^{\prime}\mathbf{k}^{\prime}}iV^{\star}_{\alpha\mathbf{k}} \langle [d^{\dagger}_{\sigma}(0)c_{\alpha^{\prime}\mathbf{k}^{\prime}\sigma}(0),d^{\dagger}_{\sigma}(t)c_{\alpha\mathbf{k}\sigma}(t)]\rangle \Big) ~,\\
& \doteq [K^{R}_{\hat{N}_{\alpha},\hat{N}_{\alpha^{\prime}}}]^{Equiv} + [K^{R}_{\hat{N}_{\alpha},\hat{N}_{\alpha^{\prime}}}]^{Inequiv} ~,
\end{aligned} 
\end{equation}
where in the summation we recognise the first two terms are the contribution for equivalent leads, with notation $[K^{R}_{\hat{N}_{\alpha},\hat{N}_{\alpha^{\prime}}}]^{Equiv}$ - note that previously we have $\alpha\equiv\alpha^{\prime} \doteq o$ for the odd combination of lead electron due to the full decoupling. Whereas the last two terms in the sum are the addition due to the intrinsic inequivalent electronic nature of the conduction band, with notation $[K^{R}_{\hat{N}_{\alpha},\hat{N}_{\alpha^{\prime}}}]^{Inequiv}$. While we carrying on the derivation, we will confirm this association to equivalent and inequivalent contributions. \\
We focus now on $U=0$ case. In $K^{R}_{\hat{N}_{\alpha},\hat{N}_{\alpha^{\prime}}}$, we deal with a noninteracting two-particle Green's function model. As textbook approach \cite{Lancaster,Jauho_QuantumKineticsTransport}, the correlator can be factorized into product of two one-particle Green's function by means of the Wick's theorem for Fermionic fields in time domain. The theorem applies only to time-ordered string of fields $G_{\alpha\alpha^{\prime}}^{c}(t,0)$ defined in Eq.\ref{eq:defGcasual} and transforms those into normal-ordered string of fully contracted fields. The reasoning for this is because either expressions partially contracted or without particle number conservation or equal-time time-ordered correlators vanish inside the normal ordering symbol. Once the nonzero one-particle Green's functions are calculated, we need to select the retarded component from the casual or time-ordered equilibrium Green's function with the aid of the Heaviside functions - see its Green's function structure in Eq.\ref{eq:defGcasual}.\\
At the end of lengthy manipulations, we obtain the two groups of lesser and greater Green's functions. \\
In the first set there are the local, noninteracting impurity Green's functions $G^{</>}_{0,dd\sigma}$. Because of the property of locality in the correlator, it is consistent to apply the fluctuation-dissipation theorem as defined in Eq.\ref{eq:defFDtheo} for simplifying this type of expression - in the same spirit as we proceed for the equivalent leads. \\
In the second set there are the nonlocal, mixed and noninteracting Green's functions namely: \\
$G^{</>}_{0,\alpha\alpha^{\prime}\sigma},G^{</>}_{0,\alpha^{\prime}\alpha\sigma},G^{</>}_{0,\alpha d\sigma},G^{</>}_{0,d\alpha\sigma}$. Those highly nontrivial correlations, rather than being related to the physical spectral function as described in the fluctuation-dissipation theorem \cite{MeirComplexStructures2005,Jauho_QuantumKineticsTransport,Ryndyk}, they show the following dependence
\begin{equation}
\begin{aligned}
& G^{<}_{x\sigma}(\omega)= i f^{Eq}(\omega) i\big(G^{R}_{x\sigma}(\omega) -G^{A}_{x\sigma}(\omega)\big)  \\
&G^{>}_{x\sigma}(\omega) = -i (1-f^{Eq}(\omega)) i\big(G^{R}_{x\sigma}(\omega) -G^{A}_{x\sigma}(\omega)\big) 
\end{aligned} ~,
\end{equation}
where $x=\alpha\alpha^{\prime},\alpha^{\prime}\alpha,\alpha d,d\alpha$ and we refer to these equations as the \textit{nonlocal relations} between particle/hole fluctuations and dissipations. Few comments on these expressions before proceeding with the derivation. The Green's function $G^{</>}_{0,\alpha\alpha^{\prime}\sigma}$ indicates the crossed correlation occurring between two different leads and $G^{</>}_{0,\alpha^{\prime}\alpha\sigma}$ just the reverse event. This factor generates finite electrical conductance due to the nonzero electronic flux moving from $\alpha$-lead through the impurity into $\alpha^{\prime}$-lead and viceversa. The mixed Green's functions $G^{</>}_{0,\alpha d\sigma},G^{</>}_{0,d\alpha\sigma}$ determine the electrical conductance manifesting directly between a lead and the adjacent impurity. By current conservation, these must ultimately be related.\\
We continue now with the calculation: as outcome of the Wick's theorem we decompose the current-current correlator $K^{R}_{\hat{N}_{\alpha},\hat{N}_{\alpha^{\prime}}} (t, T)$ into product of two one-particle lesser or greater Green's functions in real time, which present property either of locality or of nonlocality. The actual Kubo formula for linear response is given in the Fourier space as in Eq.\ref{eq:defKuboLR}. In real frequency domain, we simplify the Green's functions invoking their local and nonlocal relation to the dissipations as discussed above. In order to calculate those relations, we make use of the following retarded correlators:\\
$(i)$ the noninteracting impurity Green's function as defined in Eq.\ref{eq:G_0dd};\\
$(ii)$ the noninteracting crossed leads Green's function is determined by
\begin{equation}
\langle\langle c_{\alpha\mathbf{k}\sigma};c^{\dagger}_{\alpha^{\prime}\mathbf{k}^{\prime}\sigma} \rangle\rangle_{\omega} =\lim_{\eta\rightarrow 0}G^{R}_{0,\alpha\alpha^{\prime}\mathbf{k}\mathbf{k}^{\prime}\sigma}(\omega+i\eta ) 
\end{equation}
and we can use the $\mathrm{T}$-matrix equation as defined in Eq.\ref{eq:TmatrixEq} to determine its equation
\begin{equation}
\mathbb{G}^{R}_{\alpha\alpha^{\prime}\mathbf{k}\mathbf{k}^{\prime}\sigma}(\omega) =  \mathbb{G}^{0,R}_{\alpha\alpha^{\prime}\mathbf{k}\mathbf{k}^{\prime}\sigma}(\omega) + \mathbb{G}^{0,R}_{\alpha\alpha\mathbf{k}\mathbf{k}\sigma}(\omega) \cdot \mathbb{T}_{dd\mathbf{k}\mathbf{k}^{\prime}\sigma}(\omega) \cdot \mathbb{G}^{0,R}_{\alpha^{\prime}\alpha^{\prime}\mathbf{k}^{\prime}\mathbf{k}^{\prime}\sigma}(\omega) ~,
\end{equation}
where we use the notation of isolated $dim(2\times2)$ Green's function matrices $\mathbb{G}^{0}$ in the channel space, with element $\mathbb{G}^{0,R}_{\alpha\alpha^{\prime}}$. We define the $\mathrm{T}$-matrix expression as the local component determined at the impurity, namely
\begin{equation}
\mathbb{T}_{dd\mathbf{k}\mathbf{k}^{\prime}\sigma}(\omega) =  \Big( |\mathbf{V}_{\alpha\mathbf{k}}|^{2} + |\mathbf{V}^{\star}_{\alpha^{\prime}\mathbf{k}^{\prime}}|^{2} \Big) \mathbb{G}_{0,dd\sigma}^{R}(\omega) ~.
\end{equation}
$(iii)$ the mixed Green's functions for $x=\alpha,\alpha^{\prime}$ using the equation of motion technique brings to
\begin{equation}
\begin{aligned}
&\langle\langle c_{\alpha\mathbf{k}\sigma} ;d^{\dagger}_{\sigma}  \rangle\rangle_{\omega} = \lim_{\eta\rightarrow 0}\sum_{\mathbf{k}\sigma} G^{R}_{0,\alpha\mathbf{k} d\sigma}(\omega+i\eta) \quad\Rightarrow ~ G^{R}_{0,\alpha\mathbf{k} d\sigma}(\omega) = \sum_{\mathbf{k}\sigma}G^{0,R}_{0,dd\sigma}(\omega) V_{\alpha\mathbf{k}}G^{R}_{0,\alpha\alpha\mathbf{k}\sigma}(\omega) ~,\\
&\langle\langle d_{\sigma}; c^{\dagger}_{\alpha\mathbf{k}\sigma}\rangle\rangle_{\omega} = \lim_{\eta\rightarrow 0}\sum_{\mathbf{k}\sigma} G^{R}_{0,d\alpha\mathbf{k} \sigma}(\omega+i\eta) \quad\Rightarrow ~ G^{R}_{0,d \alpha\mathbf{k}\sigma}(\omega) = \sum_{\mathbf{k}\sigma}\frac{1}{\omega - \epsilon_{\alpha\mathbf{k}} - |V_{\alpha\mathbf{k}}|^{2}G^{0,R}_{0,\alpha\alpha\mathbf{k}\sigma}(\omega)} ~,
\end{aligned}
\end{equation}
where in the first line $G^{R}_{0,\alpha\alpha\mathbf{k}\sigma}$ can be written by means of $\mathrm{T}$-matrix equation as in Eq.\ref{eq:TmatrixEq} and in the second line the term $G^{0,R}_{0,\alpha\alpha\mathbf{k}\sigma}$ has continued fraction format. In conclusion, due to the lack of interaction, we are able to reduce all the different Green's functions to two kinds only, namely $G_{0,\alpha\alpha\mathbf{k}}$ and $G_{0,dd}$ such that all the correlators are transformed into those local objects. \\ 
We have now calculated all the elements required for the linear response function for conductance in Eq.\ref{eq:defSusceptCondAC} to obtain the Kubo formula for electrical conductance as in Eq.\ref{eq:defKuboLRel}. We split the result as follows:
\begin{equation}\label{eq:4GacKInequiv} \boxed{
\begin{aligned}
&\mathcal{G}^{C,ac}(\omega,T) = \mathcal{G}_{0} 2\pi \dfrac{-\left( \mathit{Im}[K^{R}_{\hat{N}_{\alpha},\hat{N}_{\alpha^{\prime}}}]^{Equiv}  +  \mathit{Im}[K^{R}_{\hat{N}_{\alpha},\hat{N}_{\alpha^{\prime}}}]^{Inequiv}\right)}{\omega} (\omega, T)  ~, \\
&\text{the equivalent contribution:}\\
&\mathcal{G}_{0} 2\pi \dfrac{-\mathit{Im}[K^{R}_{\hat{N}_{\alpha},\hat{N}_{\alpha^{\prime}}}]^{Equiv}}{\omega}(\omega, T)  \\
&= \mathcal{G}_{0} \frac{4\pi}{2\omega} \sum_{\mathbf{k}\sigma,\alpha\alpha^{\prime}} \int^{+\infty}_{-\infty} d\widetilde{\omega}
 \mathcal{A}_{0,dd\sigma}(\widetilde{\omega}) ~\Big( 
(V_{\alpha^{\prime}\mathbf{k}}V_{\alpha\mathbf{k}}^{\star}\mathit{Im}G^{R}_{0,\alpha\alpha^{\prime}\mathbf{k}\sigma}(\widetilde{\omega} + \omega) -V_{\alpha\mathbf{k}}V_{\alpha^{\prime}\mathbf{k}}^{\star}G^{R}_{0,\alpha\alpha^{\prime}\mathbf{k}\sigma}
(\widetilde{\omega}-\omega) ) f^{Eq}(\widetilde{\omega}) + \\
&\hspace*{2cm} - (V_{\alpha^{\prime}\mathbf{k}}V^{\star}_{\alpha\mathbf{k}}\mathit{Im}G^{R}_{0,\alpha\alpha^{\prime}\mathbf{k}\sigma}(\widetilde{\omega} + \omega)f^{Eq}(\widetilde{\omega} + \omega)- V_{\alpha\mathbf{k}}V_{\alpha^{\prime}\mathbf{k}}^{\star}\mathit{Im}G^{R}_{0,\alpha\alpha^{\prime}\mathbf{k}\sigma}(\widetilde{\omega} - \omega)f^{Eq}(\widetilde{\omega} - \omega))\Big) ~,\\
&\text{the inequivalent contribution:}\\
&\mathcal{G}_{0} 2\pi \dfrac{-\mathit{Im}[K^{R}_{\hat{N}_{\alpha},\hat{N}_{\alpha^{\prime}}}]^{Inequiv}}{\omega}(\omega, T)  \\
&=\mathcal{G}_{0} \frac{-8\pi}{2\omega} \sum_{\mathbf{k}\sigma,\alpha\alpha^{\prime}}\int^{+\infty}_{-\infty} d\widetilde{\omega} 
\mathit{Im}G^{R}_{0,\alpha\mathbf{k} d\sigma} (\widetilde{\omega}) \Big(
 V_{\alpha\mathbf{k}}V_{\alpha^{\prime}\mathbf{k}}\mathit{Im}G^{R}_{0,d\alpha^{\prime}\mathbf{k}\sigma} (\widetilde{\omega}+\omega)f^{Eq}(\widetilde{\omega})(1-f^{Eq}(\widetilde{\omega}+\omega)) +\\ 
 &  \hspace*{6cm}+ V^{\star}_{\alpha^{\prime}\mathbf{k}}V^{\star}_{\alpha\mathbf{k}}\mathit{Im}G^{R}_{0,d\alpha^{\prime}\mathbf{k}\sigma} (\widetilde{\omega}-\omega)f^{Eq}(\widetilde{\omega}-\omega)(1-f^{Eq}(\widetilde{\omega}))\Big) \\
&\hspace*{4cm}+\mathit{Im}G^{R}_{0,d\alpha\mathbf{k} \sigma} (\widetilde{\omega})
\Big(  V_{\alpha\mathbf{k}}V_{\alpha^{\prime}\mathbf{k}}\mathit{Im}G^{R}_{0,\alpha^{\prime}\mathbf{k}d}(\widetilde{\omega}+\omega)f^{Eq}(\widetilde{\omega}+\omega)(f^{Eq}(\widetilde{\omega})-1) +\\ 
&\hspace*{6cm}+V^{\star}_{\alpha^{\prime}\mathbf{k}}V^{\star}_{\alpha\mathbf{k}}\mathit{Im}G^{R}_{0,\alpha^{\prime}\mathbf{k}d\sigma} (\widetilde{\omega}-\omega)f^{Eq}(\widetilde{\omega})(f^{Eq}(\widetilde{\omega}-\omega)-1))\Big) ~,
\end{aligned}  }~,
\end{equation}
that is \textbf{ac-Kubo formula for structured, inequivalent leads} in generalised Landauer-type forms valid for noninteracting model. We observe that the expression for $\mathit{Im}[K^{R}_{\hat{N}_{\alpha},\hat{N}_{\alpha^{\prime}}}]^{Equiv}$ has the same form and argument dependence as in Eq.\ref{eq:4GacKEq}, but now each term carries a specific coupling combination that cannot be reduced to the standard geometric factor as we previously calculated. More interestingly, the expression for $\mathit{Im}[K^{R}_{\hat{N}_{\alpha},\hat{N}_{\alpha^{\prime}}}]^{Inequiv}$ differs from Eq.\ref{eq:4GacKEq} both in structure and in Fermi distributions dependence: we consider it as the \textit{correction to the standard Kubo formula} due to inequivalent leads. This mismatch is not \textit{completely} unexpected: we are dealing with nonlocal correlators calculated for different electronic distribution of the leads. The lack of interaction offers the condition for decomposing the intricate correlations into simple product of one-particle Green's functions that are reducible merely to the local $G_{0,\alpha\alpha\mathbf{k}},G_{0,dd}$. However, the signature of the non-trivial configuration and electronics of the leads is proved by the specific couplings product and the asymmetric Fermi distributions combination at each term.\\
We study the \textit{dc-}limit of Eq.\ref{eq:4GacKInequiv}. We find agreement with the equivalent leads case in the term $\mathit{Im}[K^{R}_{\hat{N}_{\alpha},\hat{N}_{\alpha^{\prime}}}]^{Equiv}$. However, from the inequivalent component not much progress is gained. If $\alpha=\alpha^{\prime}$, $\mathcal{G}^{C,dc}(T)$ results in an undetermined ratio and if $\alpha\neq\alpha^{\prime}$ we have a finite expression $\mathcal{G}^{C,dc}(T)$ not reducible to any familiar expressions in the quantum transport literature.\\
In conclusion, in this section we show that in presence of different electronic nature of the leads, $\mathcal{G}^{C,ac}(\omega,T)$ inherits this complexity in terms of coupling and Fermi distribution arrangements not reducible to the standard results. Furthermore, the non-trivial expression does not obviously simplify under low-energy limit. The lack of interacting impurity ensures the final expression is determined only by leads and impurity noninteracting Green's functions characterised by local property only, leaving out any mixed and nonlocal contribution. We infer from this discussion that activating interactions in the model would prevent us from successful application of equation of motion technique. The final conductance expression would present both equivalent and inequivalent contributions featured by nonlocal correlators. This equation would be in significant discrepancy with the standard Meir-Wingreen and Landauer formulas we encounter previously. Thus, we state it is impossible to transform the interacting impurity model with inequivalent structured leads configuration into a known literature result.\\

\noindent{We} convey our discussion to completion by summarising our findings. We conceive the leads now with an energy dependence. Within the structured lead configuration, we can appreciate each electronic lead distribution - bringing to definition of equivalent and inequivalent leads configuration. As presented in this section, two different scenarios occur.\\ In fully interacting model under equivalent leads configuration, we have mathematical tools to derive the standard Meir-Wingreen under proportionate coupling and foremost the Landauer formula see Eq.\ref{eq:4LEquiv}. As significant outcomes, these equations are calculated without demanding the strict restrictions entailed in those approaches - see Secs.\ref{sec:MeirWingreen},\ref{sec:Landauer} respectively.\\
In noninteracting models under inequivalent leads configuration, we do find the \textit{corrected} \textit{ac-}Kubo formula Eq.\ref{eq:4GacKInequiv} as its version comprises the non-uniform conduction band configuration. However, no further observation can be make besides the locality of the final Green's functions appearing in the equation and the expression does not present any further simplified form. \\
Last, the very general case interacting impurity models with inequivalent leads configuration is beyond any desired reductions.

\section{Alternative formulation of the Meir-Wingreen current formula} \label{sec:AlternativeMW}
As we discuss in Sec.\ref{sec:MeirWingreen}, the term proportional to the lesser propagator in Eq.\ref{eq:MW} for non-PC model is the problematic aspect in the nonequilibrium Meir-Wingreen current formula \cite{Dias_noPC2017}. We aim to present alternative formulations of $G_{dd\sigma}^{<}$ such that we overcome the asymmetric explicit voltage bias dependence in the original MW formula. Ultimately, we seek for lesser propagator expression that makes the MW formula applicable to linear response conductance calculations, in non-proportionate coupling geometry. \\
In this section two different approaches are developed. The first strategy  makes use of Heisenberg equation of motion to factorise the nonequilibrium, interacting lesser impurity propagator into isolated, noninteracting lead propagators and higher-order impurity propagators. Although this attempt does no bring any actual simplification of the correlator and the final equation lacks of an explicit bias dependence, it is instructive to present the derivation. The second strategy takes inspiration from the Ng ansatz \cite{NgAnsatz1996}  and by means of perturbative calculations gives not only to an nonequilibrium distribution expression beyond the ansatz limitations, but also to a $G_{dd\sigma}^{<}$ equation - exact within the order of correction in the expansion - exhibiting the desired $V_{bias}$ dependence.\\
Hence, the section constitutes a reference for alternative manipulations on the $G^{<}_{dd\sigma}$ in order to make the MW formula tractable under linear response conductance.

\subsection{Alternative formulation for interacting $G_{dd}^{<}$}
Our first approach is to derive an alternative expression of the impurity lesser propagator. We seek for factorizing the nonequilibrium, interacting $G_{dd\sigma}^{<}(\omega)$ into uncoupled, noninteracting bath Green's functions $G^{0,<}_{bath}$ - where the electronic distribution dependence is clearly displayed, see Eq.\ref{A:eq:g_Def}. In the course of the derivation, we find that it is not possible to simplify the correlators such that an explicit voltage bias parameter appears in the terms and we give reasoning for this. \\
As the model Hamiltonian, we use the single-impurity two-lead Anderson model with one-dimensional chain representation. In particular, we adopt the geometry where the impurity occupies the $n=-1$ site of one-dimensional semi-infinite tight-binding chain that represents the bath. The resulting Hamiltonian reads
\begin{equation}
\begin{aligned}
\hat{H}^{SIAM} &= \sum_{\alpha\sigma} \sum_{n=0}^{+\infty} \Big( \epsilon_{\alpha n}c^{\dagger}_{\alpha n\sigma}c_{\alpha n\sigma} + t_{\alpha n}c^{\dagger}_{\alpha n\sigma}c_{\alpha n+1\sigma} + t^{\star}_{\alpha n}c^{\dagger}_{\alpha n+1\sigma}c_{\alpha n\sigma} \Big) +\epsilon_{d}\sum_{\sigma}\hat{n}_{d\sigma}+ U\hat{n}_{d\uparrow}\hat{n}_{d\downarrow} \\
& \hspace*{2cm} + \sum_{\alpha\sigma} \Big[ V_{\alpha}d^{\dagger}_{\sigma}c_{\alpha 0 \sigma} +  V^{\star}_{\alpha}c^{\dagger}_{\alpha 0 \sigma}d_{\sigma}\Big] ~,
\end{aligned}
\end{equation}
where we note the hybridization function is only defined between the impurity site at $n=-1$ and the first bath sites at $n=0$. Such a model could also represent the Wilson chain set-up in NRG, with properly chosen $\epsilon_{\alpha n}$ and $t_{\alpha n}$, since the Wilson chain is a discretized representation, see also Fig.\ref{F2:NRG}. In the derivation we now perform nonequilibrium techniques as presented in the appendix \ref{app:MW}.\\

\subsection*{Derivation of $G^{<}_{dd\sigma}$ from the single-impurity Anderson Model}
We start with applying the Heisenberg equations of motion in Eq.\ref{eq:EoM} to the impurity lesser propagator and we use Fourier transformation to bring it into the real-frequency domain to get 
\begin{equation}\label{eq:F<Def}
\begin{aligned}
&(\omega - \epsilon_{d})\sum_{\sigma} G^{<}_{dd\sigma}(\omega) =
V_{\alpha} \sum_{\sigma} G^{<}_{d,\alpha0\sigma}(\omega) + U\sum_{\sigma} \langle\langle [H^{U}_{imp},d_{\sigma}^{\dagger}];d_{\sigma} \rangle\rangle \\
&\Rightarrow \quad [G^{R}_{0,dd\sigma}(\omega)]^{-1} G^{<}_{dd\sigma}(\omega) = V_{\alpha}  G^{<}_{d,\alpha0\sigma}(\omega) +UF^{<}_{\sigma}(\omega) ~,
\end{aligned}
\end{equation}   
where the noninteracting, nonequilibrium isolated impurity Green's function $G^{R}_{0,dd\sigma}$ is given in Eq.\ref{A:eq:g_Def}, the mixed lesser Green's function is defined in Eq.\ref{A:eq:EoM-t-final} here in one-dimensional representation with respect to the $n=0$ site on the bath chain   $G^{<}_{d,\alpha0\sigma}(t,t^{\prime}) = i \langle c^{\dagger}_{\alpha0\sigma}(t^{\prime})d_{\sigma}(t)\rangle$ and its general form is derived in Eq.\ref{A:eq:G<Time}. In the second line of Eq.\ref{eq:F<Def}, we introduce the notation $F^{<}_{\sigma}(\omega)$ to define the \textit{higher-order impurity correlation}. \\
We now manipulate Eq.\ref{eq:F<Def} as follows. In the right hand side, the lesser mixed propagator is equal to $G^{<}_{d,\alpha0\sigma}(\omega)= G^{R}_{0,\alpha0\alpha0\sigma}(\omega) V^{\star}_{\alpha} G^{<}_{dd\sigma}(\omega)$ \cite{JauhoWingreeMeir1994}. We insert this expression in Eq.\ref{eq:F<Def} and collect equal terms on the left hand side to get:
\begin{equation}
	\begin{aligned}
	&\Big( (G_{0,dd\sigma}^{R}(\omega))^{-1}-\overbrace{G^{R}_{0,\alpha0\alpha0\sigma}(\omega) |V_{\alpha}|^2}^{\Delta^{R}_{\alpha0\alpha0}~\text{from Eq.\ref{eq:defDysonRA}}} \Big)G_{dd\sigma}^{<} (\omega)= UF^{<}_{\sigma}(\omega) ~,\\
	&\Rightarrow ~ (G_{0,dd\sigma}^{R}(\omega))^{-1} G_{dd\sigma}^{<} (\omega)= UF^{<}_{\sigma}(\omega) ~.
	\end{aligned}
\end{equation} 
Hence, we have calculated the \textbf{alternative local lesser nonequilibrium, interacting Green's function} $G_{dd\sigma}^{<}$ 
\begin{equation}\label{eq:Gdd<_F}
\boxed{ G_{dd\sigma}^{<} (\omega)= G_{0,dd\sigma}^{R}(\omega) U F^{<}_{\sigma}(\omega)} ~,
\end{equation}
where the upfront term on the right-hand side $G_{0,dd\sigma}^{R}$ is calculated from the nonequilibrium Dyson equation in Eq.\ref{eq:defDysonRA} for the noninteracting models using the hybridization function in Eq.\ref{eq:HybFunc}. \\
From these initial steps, we conclude that in order to solve the equation of motion for $G^{<}_{dd\sigma}$ in Eq.\ref{eq:Gdd<_F} we seek for an expression for the higher-order impurity correlation $F^{<}_{\sigma}$ in Eq.\ref{eq:F<Def}.

\subsection*{Derivation of higher-order impurity Green's function $F_{\sigma}^{<}$}
We repeat now the same operations for the Green's function $F_{\sigma}^{<}$ defined in Eq.\ref{eq:F<Def} and we obtain
\begin{equation}\label{eq:F<EoM_initial}
\underbrace{(\omega - \epsilon_{d})}_{[G^{R}_{0,dd\sigma}(\omega)]^{-1}} F_{\sigma}^{<}(\omega) =
-V_{\alpha}  G^{<}_{ddd,\alpha0\sigma}(\omega) - U \underbrace{ \langle\langle [H^{U}_{imp},d_{\sigma}^{\dagger}];[H^{U}_{imp},d_{\sigma}] \rangle\rangle}_{F^{\prime <}_{\sigma}(\omega)} ~,
\end{equation}
where on the right-hand side we have the higher order term $F_{\sigma}^{\prime <}$, the mixed lesser Green's function $G^{<}_{ddd,\alpha0\sigma}(t,t^{\prime})= i \langle d_{\sigma}^{\dagger}(t^{\prime})d_{\sigma}(t^{\prime})d_{\sigma}^{\dagger}(t^{\prime}) c_{\alpha 0 \sigma}(t)\rangle \equiv i \langle d_{\sigma}^{\dagger}(t^{\prime})c_{\alpha 0 \sigma}(t)\rangle$ and its expression is calculated again using equation of motion
\begin{equation}
G^{R}_{0,\alpha0\alpha0\sigma}(\omega) G^{<}_{dd,\alpha0\sigma}(\omega) = 
-t_{\alpha0} G^{<}_{dd,\alpha1\sigma}(\omega) -V^{\star}_{\alpha} F^{<}_{\sigma}(\omega) ~,
\end{equation}
where the appearance of $G^{<}_{dd,\alpha1\sigma}$ indicates that by calculating this equation we take the next site down the chain - compare with the initial equation of motion expression in Eq.\ref{eq:F<Def}. \\
We collect terms in Eq.\ref{eq:F<EoM_initial} and we insert in it the equation for $G^{<}_{dd,\alpha0\sigma}$ to obtain
\begin{equation}
F^{<}_{\sigma}(\omega) =\cfrac{- U F^{\prime<}_{\sigma}(\omega)}{[G^{0,R}_{0,dd\sigma}(\omega)]^{-1} - |V_{\alpha}|^{2} 
\underbrace{
\cfrac{1}{[G^{R}_{0,\alpha0\alpha0\sigma}(\omega)]^{-1} + t_{0\alpha} 
\cfrac{G^{<}_{dd,\alpha1\sigma}}{G^{<}_{dd,\alpha0\sigma} }(\omega)}}_{G^{0,R}_{0,\alpha0\alpha0\sigma}(\omega)} } ~,
\end{equation}
where $G^{0,R}_{0,\alpha0\alpha0\sigma}(\omega)$  is the uncoupled, noninteracting lead Green's function in the isolated system. By performing more iterations, we recognize  $G^{0,R}_{0,\alpha0\alpha0\sigma}(\omega)$ is written in continued fraction representation \cite{ContinuedFraction_Mori1965}: at each cycle, the equation of motion accounts for the next chain site, as we can easily check in the periodic trend shown in the ratio $G^{<}_{dd,\alpha x+1}/G^{<}_{dd,\alpha x}$. As we know already, the continued fraction structure allows to form a closed set of equation of motion. Using the definition of hybridization function for the single-impurity hybridization in Eq.\ref{eq:HybFunc} but now with respect to the $n=0$ chain site, the complete expression for the \textbf{lesser higher-order impurity Green's function} reads
\begin{equation}\label{eq:F<EoM_final}
\boxed{ F^{<}_{\sigma}(\omega) = \frac{- U F_{\sigma}^{\prime<}(\omega)}{[G^{0,R}_{0,dd\sigma}(\omega)]^{-1} - \Delta(\omega)}
\equiv - G^{R}_{0,dd\sigma}(\omega) U F_{\sigma}^{\prime<}(\omega) } ~,
\end{equation}
where in the equivalence we use the definition of the noninteracting, nonequilibrium Dyson equation.\\
In conclusion, we can now insert the expression of the higher order correlator $F^{<}_{\sigma}(\omega)$ in Eq.\ref{eq:F<EoM_final} into the equation the lesser impurity correlator in Eq.\ref{eq:Gdd<_F} to get
\begin{equation}
G_{dd\sigma}^{<}(\omega) = - \big(G^{R}_{0,dd\sigma}(\omega)\big)^{2} U^{2} F^{\prime<}_{\sigma}(\omega) ~.
\end{equation}
We comment on this obtained result. Although we deal with an interacting impurity model that usually requires approximation in order to be analytically solved, the calculated equation of motion forms a closed set that can be cast into continued fraction. However, we note the manifestation of higher order correlations in impurity operator at each equation of motion cycle calculated for the next chain site. We conclude that, even thought the continued fraction structure keeps under control the equations, they acquire a progressive increment of complexity at each iteration - hence, this derived expression does simplify the original $G^{<}_{dd\sigma}$. \\
According to our findings, we infer it is impossible to factorize the lesser interacting nonequilibrium Green's function into simple products of isolated, noninteracting correlators - whose definition does show an explicit $V_{bias}$ parameter dependence, as would be ideal for linear response conductance calculations. Hence, the attempt of extracting this element from the factorised expression is unsuccessful and this derived equation is not suitable yet for linear-response calculation. Further understanding on this conclusion are gained by calculating the corresponding lesser self-energy $\Sigma^{<}_{dd\sigma}$ - as we see in the next discussion.

\subsection*{Derivation of lesser self-energy $\Sigma^{<}_{dd\sigma}$}
From our previous findings, we verify the impracticality of decomposing the nonequilibrium, interacting $G^{<}_{dd\sigma}$ into noninteracting correlations. However, it can be decomposed into interacting retarded Green's function as it is showed in Eq.5.49 derived in reference \cite{komijani2013interaction}. We turn now our attention to find an expression for the corresponding lesser self-energy $\Sigma^{<}_{dd\sigma}$.\\
Recalling the Dyson-Keldysh equation in Eq.\ref{eq:defDysonKeldysh} and comparing it with the equation of motion for $G^{<}_{dd\sigma}$ in Eq.\ref{eq:Gdd<_F}, we identify
\begin{equation}
\boxed{ \Sigma^{<}_{dd\sigma}(\omega,T,V_{bias}) = - U^{2} \Bigg (\frac{G_{0,dd\sigma}^{R}(\omega)}{G_{dd\sigma}^{R}(\omega)}\Bigg )^{2} F_{\sigma}^{\prime<}(\omega) } ~,
\end{equation}
as the \textbf{alternative interacting, nonequilibrium lesser self-energy} corresponding to the alternative expression of $G^{<}_{dd\sigma}$ derived in Eq.\ref{eq:Gdd<_F}. This is an interesting result: it partially resembles the standard practice of writing the interacting part of the retarded self-energy as ratio of retarded correlations, namely
\begin{equation}\label{eq:4F/G}
\Sigma^{R}_{dd\sigma}(\omega,T)= \frac{F^{R}_{\sigma}(\omega,T)}{G^{R}_{dd\sigma}(\omega,T)} ~,
\end{equation}
where $F^{R}$ correlator can be derived from the Anderson model as in the original work \cite{Bulla_NRGSelfEnergyAIM_1998}. In this section, we generalize to the lesser correlator. Such a division acquires its appreciation in numerical implementation: only relative errors propagate in iterative calculations and the results undergo to an increment in accuracy. In particular, the conductance calculated by means of Eq.\ref{eq:4F/G} in interacting system under PC reproduces exact analytical results \cite{Mitchell_UniversalLowT2CK2012}. Despite of the similarity with Eq.\ref{eq:4F/G}, in our derived alternative self-energy, the numerically stable ratio of Green's functions $G_{0,dd\sigma}^{R}/G_{dd\sigma}^{R}$ is accompanied by the higher order lesser correlation $F^{\prime<}(\omega)$ defined in Eq.\ref{eq:F<EoM_initial}. As we previously discuss,  $F^{\prime<}(\omega)$ neither is subjected to further simplification into lower order composed by noninteracting correlations, nor is possible to extract any explicit voltage dependence. Any further iteration of equation of motion on $F^{<}$-type correlations bring to higher order terms solely.\\

\noindent{By} means of these expressions, our investigation towards an alternative lesser local $G^{<}_{dd\sigma}$ using equation of motion comes to an end. Thus, other avenues have to be pursed in order to utilise the MW formula for linear response conductance calculations - as we see in the next section.

\subsection{Corrected Fermi distribution in nonequilibrium: a different perspective of the Ng ansatz}
In the second strategy, we develop perturbative calculations out-of-equilibrium condition whose output offers an expression for the nonequilibrium Fermi distribution and an alternative expression for the nonequilibrium, interacting lesser Green's function - both expressions are correct up to order $\sim V_{bias}U^{2}$. We perform these calculations for the Anderson impurity model, but they illustrate the general methodology for transport calculations beyond the standard treatment. \\
In the background of this derivation there is a formula paradigm that is the fluctuation-dissipation theorem. In its form for Fermions in Eq.\ref{eq:defFDtheo}, it is a valid expression for equilibrium models. The model is then solved using the $f^{Eq}$ equilibrium Fermi distribution in Eq.\ref{eq:defFermieq} and an exact expression for $G_{0}$ in Eq.\ref{eq:defGeq}. The theorem defines the thermal equilibrium initial condition before the model enters in out-of-equilibrium regime by means of some applied external perturbation. This theorem represents a fertile expression for investigating system once the equilibrium is broken and in presence of interactions. Then, the theorem turns into an \textit{ansatz} since one has to identify the corresponding unknown expression for the nonequilibrium Fermi distribution and a solution for the interacting propagators \cite{Ryndyk}.\\
A well-known attempt to tackle this problem is the \textbf{Ng ansatz} of the nonequilibrium, interacting impurity self-energy \cite{NgAnsatz1996,Sergueev2002,Dong2002}. As developed in the original work by T-K. Ng, the model accounts for an interacting central region undergoing to an external voltage bias to maintain constant in time the unbalanced the electronic statistics in the leads. The Ng ansatz can be expressed as,
\begin{equation}\label{eq:4Ngansatz}
\Sigma^{Ng,<}_{dd \sigma} (\omega,T) = i \overline{f}^{Eq}(\omega) \big(-2 \mathit{Im}\Sigma^{R}_{dd \sigma} (\omega,T)\big) ~ \Leftrightarrow ~
 \overline{f}^{Eq}(\omega) = \frac{\Sigma^{Ng,<}_{dd \sigma} (\omega,T)}{-\big( \Sigma^{>}_{dd \sigma} (\omega,T) - \Sigma^{<}_{dd \sigma} (\omega,T) \big)} 
\end{equation}
and it relates the interacting, nonequilibrium lesser self-energy by means of the pseudo-equilibrium Fermi distribution $\overline{f}^{Eq}(\omega)$ defined in Eq.\ref{eq:defFermiPseudoEq} to the imaginary part of the retarded interacting self-energy. The second equivalence is obtained using the Keldysh identity in Eq.\ref{eq:identityKeldysh}. We note that $\overline{f}^{Eq}(\omega)$ is the  \textit{exact} nonequilibrium distribution of the noninteracting resonant level model given in Eq.\ref{eq:NI-RLM}.\\ 
The Ng ansatz is heuristic and somewhat uncontrolled. As a first issue, the ansatz attributes to the system the equilibrium electronic distributions for a noninteracting model, as those appear in the definition of $\overline{f}^{Eq}(\omega)$ in Eq.\ref{eq:defFermiPseudoEq}. This assumption is intrinsically wrong since the distributions are modified for interacting systems.
Furthermore, the interacting central region is treated as if uncoupled from the leads \cite{Ness_LandauerNonEqInt_2010} - such that its interactions are not contributing to the nonequilibrium statistics in the sample. A second issue, as we encountered already, there is no exact analytical equation for the many-body interacting nonequilibrium $G^{R}$. We infer, the Ng ansatz, already suffering from an approximation to the electronic statistics, also necessarily invokes approximated schemes to derive an expression for the nonequilibrium $\Sigma^{R}_{dd \sigma}$ prior to the evaluation  $\Sigma^{Ng,<}_{dd \sigma}$ in Eq.\ref{eq:4Ngansatz}. \\
Despite its shortcomings, it is easy to verify the Ng ansatz returns a correct description of the model under the following limiting cases:\\
$(i)$ in equilibrium at $V_{bias}= 0$ such that $\overline{f}^{Eq}(\omega)\equiv f^{Eq}(\omega)$ is the equilibrium Fermi distribution;\\
$(ii)$ in noninteracting models at $U=0$ such that $\mathit{Im}\Delta^{R}_{dd \sigma}(\omega)$ is fully analytical solvable and $\overline{f}^{Eq}(\omega)$ is the exact nonequilibrium distribution;\\ 
$(iii)$ in Fermi liquid system under linear response.\\
Considering the limited range of applicability of the Ng ansatz, unless significant approximations are taken, in the quantum transport literature we can find a number of different approaches that make use of the ansatz \cite{Ness_LandauerNonEqInt_2010,Aligia2014,ferretti2005first}. Here, we aim to follow a different route.\\
Our approach is motivated by points $(i),(ii)$ above: the Ng ansatz is exact in $V_{bias}=0,U=0$ limits. Here, we perform perturbation theory in $V_{bias}$ and $U$ to obtain rigorously the first order correction to the Ng ansatz. We achieve this for the AIM as a demonstration. In our derivation, we perform perturbative calculations on the full interacting, nonequilibrium $\Sigma^{</>}_{dd\sigma}$ in presence of both voltage bias and interactions.  The resulting expression is decomposed in its \textit{noninteracting, equilibrium} components $\Sigma^{</>}_{0,dd\sigma}$, see Eqs.\ref{eq:4Sigma0<all}-\ref{eq:4Sigma2<NEqT0} and Eqs.\ref{A:Ng4Sigma0>all}-\ref{A:Ng4Sigma2>NEqT0} the lesser and greater self-energies, respectively. Our strategy, opposed to other \cite{yeyati1993electron}, is then to arrange these elements in order of the voltage bias $V_{bias}$ such that we systematically identify the equilibrium and nonequilibrium components as obtained from the various manipulation. To this end, the general equation form we aim to derive reads as follows
\begin{equation}\label{eq:4SigmaLessCorrected}
\boxed{
\begin{aligned}
\Sigma^{<}_{dd \sigma} &= \Sigma^{Eq,<}_{dd \sigma} + \Sigma^{NEq,<}_{dd \sigma} +\mathcal{O}(V^{2}_{bias},U^{4}) ~,\\
&= \Sigma^{(n),<}_{0,dd \sigma} +U^2 \Sigma^{(n),<}_{0,dd \sigma} + V_{bias} \left( \Sigma^{\prime (n),<}_{0,dd \sigma} + U^2\Sigma^{\prime (n),<}_{0,dd \sigma} \right)+\mathcal{O}(V^{2}_{bias},U^{4})  
\end{aligned} } ~, 
\end{equation}
where $(n)$ indicates the order of perturbation expansion, the subscript $0$ means noninteracting self-energy, the prime symbol is a notation used to differentiate the self-energy derived in presence of finite $V_{bias}$ from the one without prime calculated at equilibrium in the perturbation expansion. We remark that the expressions corresponding to the self-energies $\Sigma^{(n),<}_{0,dd \sigma},\Sigma^{\prime (n),<}_{0,dd \sigma}$ in Eq.\ref{eq:4SigmaLessCorrected} are noninteracting and in equilibrium condition - as result of the various calculations. Later in the derivation in this section we calculate the explicit expressions and give the quantitative form of Eq.\ref{eq:4SigmaLessCorrected}. Moreover, we perform the calculation for general structured leads so we follow definition $(i)$ for the levelwidth function in Eq.\ref{eq:DoSinfWBL}. Hence, the first outcome from our methodology is the derivation of \textit{lesser and greater self-energy} expressions exact up to order $\mathcal{O}(V^{2}_{bias},U^{4})$ for the AIM.\\
Taking as reference the Ng ansatz equation structure in Eq.\ref{eq:4Ngansatz}, we use our results to write the \textbf{corrected Fermi distribution} up to order $\mathcal{O}(V^{2}_{bias},U^{4})$, namely
\begin{equation}\label{eq:4FermiCorrected}
\boxed{
\begin{aligned}
&f^{corr}(\omega,T)= \frac{\Sigma^{<}_{dd \sigma}(\omega,T)}{-\big(\Sigma^{>}_{dd \sigma}(\omega,T)-\Sigma^{<}_{dd \sigma}(\omega,T)\big)} \\
&\equiv \frac{\Sigma^{<}_{0,dd \sigma}(\omega,T)+V_{bias}\Sigma^{\prime,<}_{0,dd \sigma}(\omega,T)}{-\big[\Sigma^{>}_{0,dd \sigma}(\omega,T)+V_{bias} \Sigma^{\prime,>}_{0,dd \sigma}(\omega,T)-(\Sigma^{<}_{0,dd \sigma}(\omega,T)+V_{bias} \Sigma^{\prime,<}_{0,dd \sigma}(\omega,T))\big]} +\mathcal{O}(V^{2}_{bias},U^{4}) 
\end{aligned} } 
\end{equation}
and compare its final form in Eq.\ref{eq:4FermiCorrFinal} with the Ng ansatz, order by order. We obtain a faithful verification of the Ng ansatz accuracy: we prove the Ng ansatz in $U=0$ and/or $V_{bias}=0$ by means of our exact perturbative expressions and present also results beyond the orders $\mathcal{O}(V^{0}_{bias},U^{0}),\mathcal{O}(V^{1}_{bias},U^{0})$. The expression in Eq.\ref{eq:4FermiCorrFinal} is the main result we aim to derive in this section.\\
As third outcome, applying the Keldysh identity in Eq.\ref{eq:identityKeldysh} to our derived $\Sigma^{</>}_{dd\sigma}$, it is straightforward to calculate an exact expression for the interacting nonequilibrium $G^{<}_{dd\sigma}$ to be inserted into the Meir-Wingreen current formula. Hence, we obtain a solid  strategy to compute the MW formula without the aid of any approximation other than perturbation theory, to treat the problematic lesser term. Moreover, we use this current expression to derive linear response conductance - since the $G^{<}$ shows now an explicit voltage dependence as new feature appearing in our derived expressions.\\
In conclusion, we propose a clear methodology to derive out-of-equilibrium Fermi distribution to study nonequilibrium models and lesser propagators to be employed in the MW current formula.\\
we illustrate the derivation for the single-impurity Anderson model. However, we envisage our methodology is applicable to elaborate systems: the trade-off is just more involved perturbative calculations.\\
We continue now this section presenting the analytical derivation leading to our results for the lesser and greater impurity self-energies, the corrected Fermi distribution and the alternative lesser impurity Green's function.

\subsection*{Analytical derivation of the corrected Fermi distribution}
We start with the single-impurity Anderson model in presence of interaction. In time-dependent regime \cite{LangrethLinearRespNEq1976} we have
\begin{equation}
\begin{aligned}
&\hat{H} = \hat{H}^{AM} + \hat{H}^{\prime}(t) ~,\\
&\text{with}~\hat{H}^{AM}=\sum_{\alpha\mathbf{k}\sigma} \left( \epsilon_{\alpha\mathbf{k}} \hat{n}_{\alpha\mathbf{k}\sigma} +  ( V_{\alpha\mathbf{k}} c^{\dagger}_{\alpha\mathbf{k}\sigma}d_{\sigma} + V^{\star}_{\alpha\mathbf{k}}d^{\dagger}_{\sigma}c_{\alpha\mathbf{k}\sigma}) \right)  + \epsilon_{d}\sum_{\sigma}\hat{n}_{d\sigma}+B^{z}\hat{S}^{z}_{d\sigma} + U\hat{n}_{d\uparrow}\hat{n}_{d\downarrow} ~,
\end{aligned}
\end{equation}
where $\hat{H}^{\prime}(t)$ is time-dependent external field that drive the system $\hat{H}^{AM}$ in Eq.\ref{eq:AM} out of thermal equilibrium and it reduces to using the time-dependent parameters in \ref{A:HTimeDep}, $\hat{S}^{z}_{d\sigma} = 1/2 ( \hat{n}_{d\uparrow} - \hat{n}_{d\downarrow})$ is the \textit{z-}component of the impurity spin operator due to external magnetic filed $B^{z}$. Hence, $\hat{H}$ is defined in the \textit{interaction picture} where both operators and states are time-dependent evolving with respect to the thermal equilibrium $\hat{H}^{AM}$ Hamiltonian.\\
Thus, $\hat{H}$ results in out of equilibrium condition. Its corresponding propagators are the contour-ordered Green's functions defined on the Keldysh-Schwinger contour in Eq.\ref{eq:defGneq}. Within the Keldysh formalism,  we know two fundamental assumptions hold namely $(i)$ adiabatic switching-on of interactions in the central region and $(ii)$ absence of initial correlations. As already mentioned, the first assumption ensures the occupation numbers in the system are not changed whereas the second one implies the central region is initially fully uncoupled from the reservoirs. Furthermore, we take transient phenomena completed such that the system is characterised solely by its steady-state. Hence, this construction makes legitimate to use the Fourier transform from time to frequency in real domain. It is important to remark in generic nonequilibrium phenomena the system thermalization is \textit{not well-defined} and this causes an ambiguous concept of temperature away from equilibrium states. However, the concept of temperature is uncovered only after analytical continuation to the real-time axis. By means of the unified framework between equilibrium and nonequilibrium contours at strictly $T=0$, the resulting time-ordered propagators are in equilibrium condition $G(\omega,T)$ and acquire a temperature dependence \cite{Maciejko_NonEquilibrium,Stefanucci_EQ-NEqPT_2013}.\\

\noindent{The} first and most complicated step in the derivation is the nonequilibrium \textit{Kadanoff-Keldysh perturbation expansion} of the contoured-ordered impurity Green's function $G_{dd\sigma}(\tau,\tau^{\prime})$ up to order $\mathcal{O}(U^{4})$ \cite{Keldysh1965,LangrethLinearRespNEq1976,Werner_NEQ-MeanFieldWeakCoupling2013}. We remark that away from half-filling regime or in spin polarized phase due to $B^{z}\neq0$, the odd order diagrams are not necessarily vanishing and the treatment of the Hartree term with tadpole diagrams becomes non-trivial. However, due to the type of correlator we need to calculate i.e. $G^{</>}$, diagrams at odd orders are identically zero - as we discuss below. Hence, the expansion returns finite contribution at second order and this is what we apply now.\\
The propagator expansion is \textit{non-self consistent} meaning that it is performed only in terms of bare propagators, corresponding to one-particle irreducible diagrams from the fully contracted correlations. Hence, the \textit{dressed} in the sense of interacting propagator $G_{dd\sigma}(\tau,\tau^{\prime})$ is expanded in power of the $U$, that is to have
\begin{equation}
G_{dd\sigma}(\tau,\tau^{\prime}) = 
 -i \sum_{n=0}^{2} (-i)^{n} \int_{\mathcal{C}_{\tau}}  d\tau_{n} \langle 
T_{\mathcal{C}_{\tau}} \bigg[U(\tau_{n})\hat{n}_{d\uparrow}(\tau_{n})\hat{n}_{d\downarrow}(\tau_{n})d_{\sigma}(\tau) d^{\dagger}_{\sigma}(\tau^{\prime})\bigg]
\rangle_{con} ~,
\end{equation}
where $n=2$ is the maximum order calculated from the expansion that is the two-point particle Green's function and the subscript $\langle .. ~;~..\rangle_{con}$ indicates we retain only the connected diagrams after taking full contraction. By performing such an expansion, we end up with the integral version of the Keldysh-Dyson equation of the full interacting $G_{dd\sigma}(\tau,\tau^{\prime})$ in Eq.\ref{eq:defDysonKeldysh}. By means of the formal equivalence between nonequilibrium and equilibrium structure at $T=0$, we can use the Wick's theorem to contract the product of contour-ordered operators in the expansion into anti-symmetrized product of one-particle, noninteracting Green's functions. Hence, we decompose the full interacting $G_{dd\sigma}(\tau,\tau^{\prime})$ into noninteracting $G_{0,dd\sigma}(\tau,\tau^{\prime})$ - still defined on the Keldysh contour.\\ From the various contractions, we calculate that the first order term is Hartree type. Considering we are interested in lesser or greater components whose requirement is the different location of the complex time variables on the contour, the first order correction for lesser and greater function could have only  \textit{Fock type} terms - as standard result from the Kadanoff-Keldysh expansion \cite{Werner_NEQDynnamicalMFT2014}. However, since we have only the Hartree one, the first order correction in this derivation is then zero and so the expansion has non-vanishing only the second order correction. In case of finite $B^{z}$ field, the third order contribution would vanish for similar arguments - hence, regardless the applied magnetic field, the expansion has only even order contributions.\\
In order to project the complex time variables on the real-time axis, we apply the Langreth's continuation rules - see Appendix \ref{app:MW},\cite{LangrethTheorem} - to map the resulting noninteracting contour-ordered propagators to time-ordered ones and we obtain
\begin{equation}
G^{</>}_{dd\sigma}(t,t^{\prime}) = G^{</>}_{0,dd\sigma}(t,t^{\prime}) + \int_{\mathcal{C}_{t}}dt_{1} \int_{\mathcal{C}_{t}} dt_{2} G^{R}_{0,dd\sigma}(t,t_{1})\Sigma^{</>}_{dd\sigma}(t_{1},t_{2})G^{A}_{0,dd\sigma}(t_{2},t^{\prime}) ~,
\end{equation}
that is the integral Keldysh-Dyson equation now on real-time axis and the self-energy equation $\Sigma^{</>}_{dd\sigma} = \Sigma^{(0),</>}_{0,dd\sigma}+ \Sigma^{(2),</>}_{0,dd\sigma} + \mathcal{O}(U^{4})$ is given by the nonzero order correction of noninteracting self-energy. We remark that, in the spirit of Eq.\ref{eq:4SigmaLessCorrected}, each order of correction in $\Sigma^{</>}$ contains both the equilibrium and nonequilibrium parts. Moreover, from the mapping, the self-energy is now also function of temperature.\\
For the sake of conciseness, in the succeeding equations, we detail the derivation for the lesser component in the main text and we refer to the corresponding greater expressions as given in the Appendix \ref{app:ExpNg}: the $G^{>}$ correlators share the same structure with $G^{<}$, just with inverted time labels.\\
Recalling that the system is in steady state and so it fulfils the time translational invariance, we can perform the self-energy Fourier transform to real frequency domain, namely
\begin{equation}\label{eq:4SigmaExp}
\begin{aligned}
&\Sigma^{<}_{dd \sigma}(\omega,T) = \Sigma^{(0),<}_{0,dd \sigma}(\omega,T)+\Sigma^{(2),<}_{0,dd \sigma}(\omega,T) + \mathcal{O}(U^{4}) ~,\\
&= i \!\!\!\sum_{\alpha=L,R} \Gamma^{\alpha}(\omega)f^{\mu_{\alpha}}(\omega-\mu_{\alpha}) + U^{2} \sum_{\sigma \neq\overline{\sigma}} \int_{-\infty}^{\infty} \frac{d\omega_{1}}{2\pi} \int_{-\infty}^{\infty} \frac{d\omega_{2}}{2\pi} G^{<}_{0,dd\sigma}(\omega_{1})G^{<}_{0,dd\overline{\sigma}}(\omega_{2})G^{<}_{0,dd\overline{\sigma}}(\omega_{1}+\omega_{2}-\omega) ~,
\end{aligned}
\end{equation}
where the zero order is the trivial bare hybridization with general definition given by the noninteracting nonequilibrium self-energy as in Eq.\ref{A:eq:delta_Def} and the second order term is the direct Fourier transform of $\Sigma^{(2),<}_{0,dd\sigma}$ over its internal degree of freedom as it is indicated by the different spin labels. We note the temperature dependence in Eq.\ref{eq:4SigmaExp} is included in the Fermi distribution and in the self-energies.  Each term in Eq.\ref{eq:4SigmaExp} contains both the equilibrium and nonequilibrium component - the expansion is in terms of the interaction. In particular, the zero order term $\Sigma^{(0),<}_{0,dd \sigma}$ is an exact expression depending on the model hybridization and the Fermi distribution at constant $V_{bias}$ for a given temperature. The second order term $\Sigma^{(2),<}_{0,dd \sigma}$ shows an non-trivial structure due to the convolution of three noninteracting, nonequilibrium Green's functions. We avail of the knowledge on the Meir-Wingreen current formula in pseudo-equilibrium and we use the pseudo-equilibrium version of the lesser correlator in Eq.\ref{eq:G<PseudoEq} to unpack the second order term from the expansion
\begin{equation}
\begin{aligned}
\Sigma^{(2),<}_{0,dd \sigma}(\omega,T)=U^{2}\sum_{\sigma \neq\overline{\sigma}} \int_{-\infty}^{\infty} \frac{d\omega_{1}}{2\pi} \int_{-\infty}^{\infty} \frac{d\omega_{2}}{2\pi}
 &i 2\pi\mathcal{A}_{0,dd \sigma}(\omega_{1})i 2\pi\mathcal{A}_{0,dd \overline{\sigma}}(\omega_{2})(-i)2\pi\mathcal{A}_{0,dd \overline{\sigma}}(\omega_{1}+\omega_{2}-\omega) \times\\
&\times \left( \overline{f}^{Eq}(\omega_{1})\overline{f}^{Eq}(\omega_{2}) \overline{f}^{Eq}(\omega_{1}+\omega_{2}-\omega) \right) ~,
\end{aligned}
\end{equation}
where we recognise the pseudo-equilibrium Fermi distribution, $\overline{f}^{Eq}(\omega)$, and the noninteracting spectral function, $\mathcal{A}_{0,dd \sigma}$, as calculated for the impurity Green's function in Eq.\ref{eq:G_0dd} in a infinite wide-band-limit approximation, see Eq.\ref{eq:DoSinfWBL} and so we use for it the notation
\begin{equation}
\mathcal{A}_{0,dd\sigma}(\omega) \doteq \frac{2 \Gamma}{(\omega -\epsilon_{d\sigma})^{2} + \Gamma^{2}} ~,
\end{equation}
where we specify the spin label in the impurity energy dispersion, which incorporates a magnetic field via $\epsilon_{d\sigma}=\epsilon_{d} + \sigma/2 B^z$.\\
To summarise: staring from the initial full interacting nonequilibrium  $\Sigma^{<}_{dd \sigma}$ we calculate
\begin{equation}
\Sigma^{<}_{dd \sigma}(\omega,T,V_{bias}) =\Big( \Sigma^{(0),<}_{0,dd \sigma}(\omega,T,V_{bias}) \Big)^{Eq+NEq} + \Big(\Sigma^{(2),<}_{0,dd \sigma}(\omega,T,V_{bias}) \Big)^{Eq+NEq}  + \mathcal{O}(U^{4}) ~,
\end{equation}
as sum of nonequilibrium, noninteracting, time-ordered propagator exact up to order $\mathcal{O}(U^{4})$, see Eq.\ref{eq:4SigmaExp}. We observe both temperature and voltage bias dependence enter on an equal foot in each order of term. We need then to manipulate the equilibrium Fermi distribution in order to systematically separate the contribution in $T$ and $V_{bias}$ parameters. \\

\noindent{The} second step in the derivation focuses on extracting the parameter dependence from the equilibrium Fermi distribution $f^{\mu_{\alpha}}$. We achieve this by evaluating $f^{\mu_{\alpha}}$ at equally split voltage bias level between the two leads, namely
\begin{equation}
\sum_{\alpha} f^{\mu_{\alpha}}(\omega-\mu_{\alpha})=
\begin{cases}
f^{\mu_{L}}(\omega-\mu_{L})\bigg\rvert_{\mu_{L}=eV_{bias}/2}=f^{\mu_{L}}(\omega -eV_{bias}/2) \\
f^{\mu_{R}}(\omega-\mu_{R})\bigg\rvert_{\mu_{R}=-eV_{bias}/2}= f^{\mu_{R}}(\omega +eV_{bias}/2)
\end{cases}
\end{equation}
and then by calculating its linear expansion around the equilibrium condition $V_{bias} =0$
\begin{equation}\label{eq:4FermiLinExp}
\begin{aligned}
f^{\mu_{L}/\mu_{R}}\big( \omega \mp \frac{e V_{bias}}{2} \big) 
&= f^{Eq}(\omega) \pm \frac{eV_{bias}}{2} f^{\prime Eq}(\omega) +\mathcal{O}(V_{bias}^{2}) ~, \\
&=f^{Eq}(\omega) + \zeta_{\alpha} \frac{eV_{bias}}{2} \frac{1}{k_{B}T} \big(f^{Eq}(\omega)- (f^{Eq}(\omega))^{2} \big) +\mathcal{O}(V_{bias}^{2}) ~,\\
&\doteq f^{Eq}(\omega) + f^{NEq}(\omega) +\mathcal{O}(V_{bias}^{2}) ~,
\end{aligned}
\end{equation}
where in the first line we sort our findings in terms of equilibrium with no applied voltage bias and nonequilibrium with finite voltage bias contributions in similar spirit to Eq.\ref{eq:4SigmaLessCorrected} and in the second line we give the explicit result from the Taylor expansion with the prefactor $\zeta_{\alpha} = \pm1$ according to the expansion performed with respect to the left or right lead, respectively. \\
We now insert the distribution in Eq.\ref{eq:4FermiLinExp} into the perturbation expansion of $\Sigma^{<}_{dd\sigma}$ in Eq.\ref{eq:4SigmaExp} and we systematically get expressions with separate voltage $V_{bias}$ and interaction $U$ dependence.\\
For later reference, we define
\begin{equation}
\begin{aligned}
-\frac{\partial f^{Eq}(\omega)}{\partial \omega} &\equiv \frac{1}{k_{B}T}\Big( f^{Eq}(\omega) - (f^{Eq}(\omega))^{2} \Big)   ~,\\
&= \frac{1}{2k_{B}T \Big ( 1 + \cosh(\omega/k_{B}T) \Big)} = \frac{1}{4k_{B}T} \sech^{2}\Big( \frac{\omega}{2k_{B}T}\Big)
\doteq c(\omega, T) ~,
\end{aligned}
\end{equation}
as this function will ease the notation in the next results. Hence, the compact form of Eq.\ref{eq:4FermiLinExp} as we use in the following reads as
\begin{equation}
f^{\mu_{L}/\mu_{R}}\big( \omega \mp \frac{e V_{bias}}{2} \big) 
= f^{Eq}(\omega) + \zeta_{\alpha} \frac{eV_{bias}}{2}c(\omega, T) +\mathcal{O}(V_{bias}^{2}) ~,
\end{equation}
where the sign of the first order correction depends on which lead the expansion is calculated for.\\
We can also insert the linear expansion of the Fermi distribution up to order $\mathcal{O}(V_{bias}^{2})$ calculated in Eq.\ref{eq:4FermiLinExp} in the pseudo-equilibrium Fermi distribution $\overline{f}^{Eq}(\omega)$ defined in Eq.\ref{eq:defFermiPseudoEq} to derive an expansion for  it as follows:
\begin{equation}\label{eq:fEqNg}
	\begin{aligned}
&\overline{f}^{Eq}(\omega) = \frac{\Gamma^{L}(\omega)f^{\mu_{L}}(\omega-\mu_{L})+\Gamma^{R}(\omega)f^{\mu_{R}}(\omega-\mu_{R})}{\Gamma^{L}(\omega)+ \Gamma^{R}(\omega)} ~,\\
&= \frac{\Gamma^{L}(\omega)+ \Gamma^{R}(\omega)}{\Gamma^{L}(\omega)+ \Gamma^{R}(\omega)}f^{Eq}(\omega) +\\
&+ \frac{\Gamma^{L}(\omega)\left(\frac{+eV_{bias}}{2}\right)\frac{1}{k_{B}T}\left( f^{Eq}(\omega) - (f^{Eq}(\omega))^{2} \right)+ \Gamma^{R}(\omega)\left(\frac{-eV_{bias}}{2}\right)\frac{1}{k_{B}T}\left( f^{Eq}(\omega) - (f^{Eq}(\omega))^{2} \right)}{\Gamma^{L}(\omega) + \Gamma^{R}(\omega) } +\mathcal{O}(V_{bias}^{2}) ~,\\
&=f^{Eq}(\omega)  +\left(\frac{eV_{bias}}{2}\right)\frac{\sum_{\alpha}\zeta_{\alpha} \Gamma^{\alpha}(\omega)}{\sum_{\alpha}\Gamma^{\alpha}(\omega)}c(\omega, T)+\mathcal{O}(V_{bias}^{2}) ~,
\end{aligned}
\end{equation}
that is the pseudo-equilibrium function expansion in order of $V_{bias}$ at $U=0$.\\
 
\noindent{In} the last step of the derivation, we insert the linear expansion of the Fermi distribution in Eq.\ref{eq:4FermiLinExp} into the self-energy perturbative expression in Eq.\ref{eq:4SigmaExp}. We present results for both finite and zero temperature case.\\
In the \textit{lesser self-energy at zero order}, it is straightforward to insert the Fermi expansion and we get 
\begin{equation}\label{eq:4Sigma0<all}
\begin{aligned}
\Sigma^{(0),<}_{0,dd\sigma}(\omega,T, V_{bias}) &= \Big( \Sigma^{(0),<}_{0,dd \sigma}(\omega,T) \Big)^{Eq} + \Big( \Sigma^{(0),<}_{0,dd \sigma}(\omega,T, V_{bias}) \Big)^{NEq} ~,\\
&=\begin{cases}
2i \sum_{\alpha} \Gamma^{\alpha}(\omega) \big(f^{Eq}(\omega) +\zeta_{\alpha}\frac{eV_{bias}}{2} c(\omega,T)  \big) ~, \text{at finite}~T\\
2i \sum_{\alpha} \Gamma^{\alpha}(\omega)\big( \theta(\omega-\epsilon_{F}) +\zeta_{\alpha}\frac{eV_{bias}}{2} \delta(\omega-\epsilon_{F}) \big)~, \text{at}~T=0 
\end{cases} ~,\\
&\equiv \Sigma^{(0),<}_{0,dd \sigma} + V_{bias} \Sigma^{\prime(0),<}_{0,dd \sigma} +\mathcal{O}(V^{2}_{bias},U^{4}) ~,
\end{aligned}
\end{equation} 
where $\Sigma^{(0),<}_{0,dd \sigma},\Sigma^{\prime(0),<}_{0,dd \sigma}$ so defined, and the $T=0$ expression is calculated using standard operation of the Fermi-Dirac function at strictly zero temperature.\\
In the \textit{lesser self-energy at second order}, while inserting the Fermi distribution expansion in the product of pseudo-equilibrium distributions, we select again terms up to $\mathcal{O}(V_{bias}^{2})$ to be consistent with the order of correction in the expansion in Eq.\ref{eq:4FermiLinExp}. At finite temperature, the equilibrium term reads
\begin{equation}\label{eq:4Sigma2<EqT}
\begin{aligned}
\Big( \Sigma^{(2),<}_{0,dd \sigma}(\omega,T) \Big)^{Eq} 
&=2\pi i U^{2} \int_{-\infty}^{+\infty} d\omega_{1} \int_{-\infty}^{+\infty} d\omega_{2}
 \mathcal{A}_{0,dd \sigma}(\omega_{1}) \mathcal{A}_{0,dd \overline{\sigma}}(\omega_{2})\mathcal{A}_{0,dd \overline{\sigma}}(\omega_{1}+\omega_{2}-\omega) \times \\
& \quad \hspace*{4.0cm} \times \Big( f^{Eq}(\omega_{1})f^{Eq}(\omega_{2})(1-f^{Eq}(\omega_{1}+\omega_{2}-\omega)) \Big) ~,\\
&\equiv U^{2}\Sigma^{(2),<}_{0,dd \sigma} +\mathcal{O}(V^{2}_{bias},U^{4}) 
\end{aligned}
\end{equation}
and the nonequilibrium contribution
\begin{equation}\label{eq:4Sigma2<NEqT}
\begin{aligned}
\Big( \Sigma^{(2),<}_{0,dd \sigma}(\omega,T,V_{bias}) \Big)^{NEq} 
&=2\pi i U^{2} \int_{-\infty}^{+\infty} d\omega_{1} \int_{-\infty}^{+\infty} d\omega_{2}\mathcal{A}_{0,dd \sigma}(\omega_{1}) \mathcal{A}_{0,dd \overline{\sigma}}(\omega_{2})\mathcal{A}_{0,dd\overline{\sigma}}(\omega_{1}+\omega_{2}-\omega) \times \\
& \hspace*{-2cm}\times \sum_{\alpha}\zeta_{\alpha}\frac{e V_{bias}}{2}\bigg(
\Big( f^{Eq}(\omega_{1})c(\omega_{2},T)+f^{Eq}(\omega_{2})c(\omega_{1},T)\Big)\Big(1-f^{Eq}(\omega_{1}+\omega_{2}-\omega)\Big) + \\
& \hspace*{4cm} - f^{Eq}(\omega_{1})f^{Eq}(\omega_{2})c(\omega_{1}+\omega_{2}-\omega,T) \bigg) ~,\\
& \equiv V_{bias}U^{2} \Sigma^{\prime(2),<}_{0,dd \sigma} +\mathcal{O}(V^{2}_{bias},U^{4}) ~.
\end{aligned}
\end{equation}
At zero temperature, the equilibrium contribution reads
\begin{equation}\label{eq:4Sigma2<EqT0}
\begin{aligned}
\Big( \Sigma^{(2),<}_{0,dd \sigma}(\omega, T=0) \Big)^{Eq} 
&= 2\pi i U^2 \int_{\omega}^{\epsilon_{F}} d\omega_{2} \int_{\epsilon_{F}-\omega_{2}+\omega}^{\epsilon_{F}}d\omega_{1} 
\mathcal{A}_{0,dd \sigma}(\omega_{1})\mathcal{A}_{0,dd \overline{\sigma}}(\omega_{2})\mathcal{A}_{0,dd \overline{\sigma}}(\omega_{1}+\omega_{2}-\omega) ~,\\
&\equiv U^{2}\Sigma^{(2),<}_{0,dd \sigma} +\mathcal{O}(V^{2}_{bias},U^{4}) 
\end{aligned}
\end{equation}
and the nonequilibrium contribution
\begin{equation}\label{eq:4Sigma2<NEqT0}
\begin{aligned}
&\Big( \Sigma^{(2),<}_{0,dd \sigma}(\omega, T=0, V_{bias}) \Big)^{NEq} = \\
&=\sum_{\alpha}\zeta_{\alpha}\frac{e V_{bias}}{2} (2\pi)^{2} i U^2 
\bigg( \mathcal{A}_{0,dd\overline{\sigma}}(\epsilon_{F})
\int_{\omega}^{\epsilon_{F}} d\omega_{1}
\mathcal{A}_{0,dd \sigma}(\omega_{1})
\bigg( \mathcal{A}_{0,dd\overline{\sigma}}(\omega_{1}+\epsilon_{F}-\omega) 
-\mathcal{A}_{0,dd\overline{\sigma}} (-\omega_{1}+\epsilon_{F}+\omega)\bigg) +\\
& \hspace*{6cm} +\mathcal{A}_{0,dd\sigma}(\epsilon_{F})
\int_{\omega}^{\epsilon_{F}} d\omega_{2} 
\mathcal{A}_{0,dd \overline{\sigma}}(\omega_{2})\mathcal{A}_{0,dd \overline{\sigma}}(\epsilon_{F}+\omega_{2}-\omega) \bigg)\bigg) ~, \\
& \equiv V_{bias}U^{2} \Sigma^{\prime(2),<}_{0,dd \sigma} +\mathcal{O}(V^{2}_{bias},U^{4}) ~.
\end{aligned}
\end{equation}
The corresponding greater self-energy expressions are given in the Appendix \ref{app:ExpNg}.\\

\noindent{In} conclusion, we derive an exact expressions for the interacting impurity $\Sigma^{<}_{dd\sigma}$ up to order $\mathcal{O}(V_{bias}^{2},U^{4})$  in terms of noninteracting impurity correlators and equilibrium Fermi distribution. We put together now our findings to finally give a quantitative meaning to the Eq.\ref{eq:4SigmaLessCorrected}, namely 
\begin{equation}
\begin{aligned}
&\Sigma^{<}_{dd \sigma}(\omega,T,V_{bias}) = \\
&=\underbrace{ \Sigma^{(0),<}_{dd \sigma}(\omega,T)+U^{2}\Sigma^{(2),<}_{dd \sigma}(\omega,T)}_{\Sigma^{Eq,<}_{dd \sigma}(\omega,T)}
+\underbrace{V_{bias} \Big(\Sigma^{\prime(0),<}_{dd \sigma}(\omega,T) +U^{2}\Sigma^{\prime(2),<}_{dd \sigma}(\omega,T) \Big)}_{\Sigma^{NEq,<}_{dd \sigma}(\omega,T,V_{bias})} +\mathcal{O}(V_{bias}^{2},U^{4}) ~,
\end{aligned}
\end{equation}
where at finite temperature, the equilibrium elements are given in Eqs.\ref{eq:4Sigma0<all},\ref{eq:4Sigma2<EqT} and the nonequilibrium elements are given in Eqs.\ref{eq:4Sigma0<all},\ref{eq:4Sigma2<NEqT}. At zero temperature, in Eqs.\ref{eq:4Sigma0<all},\ref{eq:4Sigma2<EqT0} and in Eqs.\ref{eq:4Sigma0<all},\ref{eq:4Sigma2<NEqT0} respectively. We calculate the same expression also for the greater component, see again the Appendix \ref{app:ExpNg}. Furthermore, on the contrary of derivation using Heisenberg equations of motion previously presented, here we obtain an interacting self-energy with explicit $V_{bias}$ - and this aspect is inherited in every equation we calculate using it. We conclude we finally succeed in deriving a simplified correlator amenable of linear response calculations starting from a nonequilibrium noninteracting lesser correlator, but at the cost of a perturbation expansion.\\ 
By means of the derived lesser self-energies in Eqs.\ref{eq:4Sigma0<all}-\ref{eq:4Sigma2<NEqT0} and the greater self-energies in Eqs.\ref{A:Ng4Sigma0>all}-\ref{A:Ng4Sigma2>NEqT0}, we can write the corrected Fermi distribution in Eq.\ref{eq:4FermiCorrected} up to order $\mathcal{O}(V_{bias}^{2},U^{4})$, namely
\begin{equation} \label{eq:4FermiCorrFinal}\boxed{
\begin{aligned}
&f^{corr}(\omega,T) = \frac{\Sigma^{<}_{dd \sigma}(\omega,T,V_{bias})}{-\big(\Sigma^{>}_{dd \sigma}(\omega,T,V_{bias})-\Sigma^{<}_{dd \sigma}(\omega,T,V_{bias})\big)} :\\
&\text{numerator:}\quad  \Sigma^{(0),<}_{dd \sigma}(\omega,T)+ U^{2}\Sigma^{(2),<}_{dd \sigma}(\omega,T) +V_{bias} \big( \Sigma^{\prime(0),<}_{dd \sigma}(\omega,T)+U^{2}\Sigma^{\prime(2),<}_{dd \sigma}(\omega,T) \big) ~,\\
&\text{denominator:}\quad 
-\Big(\Sigma^{(0),>}_{dd \sigma}(\omega,T)+U^{2}\Sigma^{(2),>}_{dd \sigma}(\omega,T)+V_{bias} \big( \Sigma^{\prime(0),>}_{dd \sigma}(\omega,T)+U^{2}\Sigma^{\prime(2),>}_{dd \sigma}(\omega,T) \big) +\\
& \hspace*{3.5cm} -\big(\Sigma^{(0),<}_{dd \sigma}(\omega,T)+ U^{2}\Sigma^{(2),<}_{dd \sigma}(\omega,T) +V_{bias} \big( \Sigma^{\prime(0),<}_{dd \sigma}(\omega,T) + U^{2}\Sigma^{\prime(2),<}_{dd \sigma}(\omega,T) \big)  \Big) 
\end{aligned} }~.
\end{equation}
Although the linear expansion of the Fermi distribution in Eq.\ref{eq:4FermiLinExp} separates the parameters dependence in each term, the expression $f^{corr}(\omega)$ in Eq.\ref{eq:4FermiCorrFinal} is now given as ratio of self-energies and we cannot isolate terms according to their dependence in order of $V_{bias},U$. We perform a Taylor expansion of Eq.\ref{eq:4FermiCorrFinal} in two variables $V_{bias}$ and $U^{2}$ up to order $\mathcal{O}(V_{bias}^{3},(U^{2})^{3})$ such that we consistently keep the order of correction in our results. In the end, we retain terms up to order $V_{bias}U^2$, consistent with the order of perturbation theory used. We calculate three orders of terms, according to the following features:\\
$(i)$ the zero order term $\mathcal{O}((V_{bias})^{0},(U^{2})^{0})$. This is the limiting case of the trivially solvable noninteracting, equilibrium model;\\
$(ii)$ the first order terms: the order $\mathcal{O}((V_{bias})^{1},(U^{2})^{0})$ referring to the nonequilibrium, noninteracting model and the order $\mathcal{O}((V_{bias})^{0},(U^{2})^{1})$ referring to equilibrium, interacting model;\\
$(iii)$ the second order terms: the $\mathcal{O}((V_{bias})^{2},(U^{2})^{0})$ and $\mathcal{O}((V_{bias})^{0}(U^{2})^{2})$ contributions are excluded since they represent systems respectively beyond linear response regime voltage bias and quadratic in interaction. The remaining term of order $\mathcal{O}((V_{bias})^{1},(U^{2})^{1})$ represents the case of interacting and nonequilibrium system - which is the case of interest since it goes beyond the Ng ansatz prediction. \\
Hence, the relevant terms in $f^{corr}$ are the following
\begin{equation}
\begin{aligned}
f^{corr}(\omega,T)= &~~f^{(V_{bias})^{0},(U^{2})^{0}}(\omega,T) + f^{(V_{bias})^{1},(U^{2})^{0}}(\omega,T) +\\
&+ f^{(V_{bias})^{0},(U^{2})^{1}}(\omega,T)+  f^{(V_{bias})^{1},(U^{2})^{1}}(\omega,T) + \mathcal{O}((V_{bias})^{2},(U^{2})^{2})
\end{aligned}
\end{equation}
and here below we give each detailed expression, order per order:\\
$(i)$ the zero order correction $\mathcal{O}((V_{bias})^{0},(U^{2})^{0})$
\begin{equation}\label{eq:4fcorrV0U0}
f^{(V_{bias})^{0},(U^{2})^{0}}(\omega,T) = \frac{\Sigma^{(0),<}_{dd \sigma}(\omega,T)}{-\big[\Sigma^{(0),>}_{dd \sigma}(\omega,T)-\Sigma^{(0),<}_{dd \sigma}(\omega,T)\big]} ~;
\end{equation}
$(ii)$ the first orders $\mathcal{O}((V_{bias})^{1},(U^{2})^{0})$ 
\begin{equation}\label{eq:4fcorrV1U0}
 f^{(V_{bias})^{1},(U^{2})^{0}}(\omega,T) = V_{bias}
\frac{\Sigma^{\prime(0),<}_{dd \sigma}(\omega,T)\Sigma^{(0),>}_{dd \sigma}(\omega,T)-\Sigma^{\prime(0),>}_{dd \sigma}(\omega,T))\Sigma^{(0),<}_{dd \sigma}(\omega,T)}
{-\big[\Sigma^{(0),>}_{dd \sigma}(\omega,T)-\Sigma^{(0),<}(\omega,T)\big]^{2}} ~,
\end{equation}
and $\mathcal{O}((V_{bias})^{0},(U^{2})^{1})$
\begin{equation}\label{eq:4fcorrV0U1}
	\begin{aligned}
&f^{(V_{bias})^{0},(U^{2})^{1}}(\omega,T) =\\
&=U^{2}
\frac{\Sigma^{(2),<}_{dd \sigma}(\omega,T)\big[\Sigma^{(0),>}_{dd \sigma}(\omega,T)-\Sigma^{(0),<}_{dd \sigma}(\omega,T)\big] -\big[\Sigma^{(2),>}_{dd \sigma}(\omega,T)-\Sigma^{(2),<}_{dd \sigma}(\omega,T)\big]\Sigma^{(0),<}_{dd \sigma}(\omega,T) }
{-\big[\Sigma^{(0),>}_{dd \sigma}(\omega,T)-\Sigma^{(0),<}_{dd \sigma}(\omega,T)\big]^{2}} ~;
\end{aligned}
\end{equation}
$(iii)$ the second order correction $\mathcal{O}((V_{bias})^{1},(U^{2})^{1})$
\begin{equation}\label{eq:4fcorrV1U1}
\begin{aligned}
&f^{(V_{bias})^{1},(U^{2})^{1}}_{dd \sigma}(\omega,T)= \\
&=V_{bias}U^{2} \Bigg[
\frac{\Sigma^{\prime(2),<}_{dd \sigma}(\omega,T)}{-\big[\Sigma^{(0),>}_{dd \sigma}(\omega,T)-\Sigma^{(0),<}_{dd \sigma}(\omega,T)\big]} - \frac{\Sigma^{\prime(0),<}_{dd \sigma}(\omega,T)\big[\Sigma^{(2),>}_{dd \sigma}(\omega,T)-\Sigma^{(2),<}_{dd \sigma}(\omega,T)\big]}{-\big[\Sigma^{(0),>}_{dd \sigma}(\omega,T)-\Sigma^{(0),<}_{dd \sigma}(\omega,T)\big]^{2}} \\
& \hspace*{2cm} -\frac{\Sigma^{(2),<}_{dd \sigma}(\omega,T)\big[\Sigma^{\prime(0),>}_{dd \sigma}(\omega,T)-\Sigma^{\prime(0),<}_{dd \sigma}(\omega,T)\big]}{-\big[\Sigma^{(0),>}_{dd \sigma}(\omega,T)-\Sigma^{(0),<}_{dd \sigma}(\omega,T)\big]^{2}}\\
&\hspace*{2cm} -\frac{\big[\Sigma^{\prime(2),>}_{dd \sigma}(\omega,T)-\Sigma^{\prime(2),<}_{dd \sigma}(\omega,T)\big]\Sigma^{(0),<}_{dd \sigma}(\omega,T)}{-\big[\Sigma^{(0),>}_{dd \sigma}(\omega,T)-\Sigma^{(0),<}_{dd \sigma}(\omega,T)\big]^{2}}\\
&\hspace*{2cm} +\frac{2\big[\Sigma^{(2),>}_{dd \sigma}(\omega,T)-\Sigma^{(2),<}_{dd \sigma}(\omega,T)\big]\big[\Sigma^{\prime(0),>}_{dd \sigma}(\omega,T)-\Sigma^{\prime(0),<}_{dd \sigma}(\omega,T)\big]\Sigma^{(0),<}_{dd \sigma}(\omega,T)}{-\big[\Sigma^{(0),>}_{dd \sigma}(\omega,T)-\Sigma^{(0),<}_{dd \sigma}(\omega,T)\big]^{3}}\Bigg] ~. 
\end{aligned}
\end{equation}
We further study the analytical form of these equations.
We find these expressions are ratio of converging integrals. For finite temperature systems, we need to perform the integration over the all energy spectrum and we treat the equilibrium Fermi distribution through Sommerfeld expansion. For $T=0$ systems, the integrations end in converging results that are non-vanishing for $\omega=\epsilon_{F}$. However, strictly speaking, the boundary limits of integration in $T=0$ cases require $\epsilon_{F}$ should not be strictly equal to zero. We also note that denominators of the above expressions $\Sigma^{(0),>}_{dd \sigma}(\omega,T)-\Sigma^{(0),<}_{dd \sigma}(\omega,T) = \sum_{\alpha} \Gamma^{\alpha}(\omega)$. \\
We have now all the elements to compare our findings of the corrected Fermi distribution with the original Ng ansatz.

\subsection*{Comparison between Ng ansatz and perturbative result $f^{corr}(\omega,T)$}
To bring at completion our discussion, we compare order by order the Ng ansatz and the $f^{corr}$ as derived throughout the section. \\
In the Ng ansatz in Eq.\ref{eq:4Ngansatz}, the ratio $\Sigma^{<}/(-2i\mathit{Im}\Sigma^{R})$ is equal to the pseudo-equilibrium function. We can expand $\overline{f}^{Eq}$ in Eq.\ref{eq:fEqNg} to linear order in $V_{bias}$ to compare order by order in voltage bias and interaction with the Ng ansatz. The results are compared in Table \ref{table:fcorr}.\\
\!\!\noindent{As} outcome of our strategy using rigorous perturbation theory, linear expansions and \textit{without} invoking any further approximation, we derive the corrected Fermi distribution in Eq.\ref{eq:4FermiCorrFinal} up to order $\mathcal{O}((V_{bias})^{2},(U^{2})^{2})$. To facilitate the comparison order by order, we compile a similar table for $f^{corr}$, see Table \ref{table:fcorr}.\\
\begin{table}[H] 
	\centering
{\renewcommand{\arraystretch}{1.2} 
\begin{tabular}{ |c||c|c| } 
 \hline
 $f^{corr} (\omega,T)$ & $(V_{bias})^{0}$ & $(V_{bias})^{1}$ \\
 \hline\hline
 $(U^{2})^{0}$ & $f^{(V_{bias})^{0},(U^{2})^{0}}_{dd \sigma}(\omega,T)$ in Eq.\ref{eq:4fcorrV0U0} & $f^{(V_{bias})^{1},(U^{2})^{0}}_{dd \sigma}(\omega,T)$ in Eq.\ref{eq:4fcorrV1U0} \\ 
 $(U^{2})^{1}$& $f^{(V_{bias})^{0},(U^{2})^{1}}_{dd \sigma}(\omega,T)$ in Eq.\ref{eq:4fcorrV0U1} & $f^{(V_{bias})^{1},(U^{2})^{1}}_{dd \sigma}(\omega,T)$ in Eq.\ref{eq:4fcorrV1U1}\\
  \hline
\end{tabular} } ~.
\caption[Corrected Fermi distribution from perturbative results]{Corrected Fermi distribution as derived order by order in voltage bias and Coulomb potential from the previous perturbative results.}\label{table:fcorr}
\end{table}
\noindent{We} find consistency in our derived expressions since those include the Ng prediction for the cases of noninteracting model under both equilibrium and nonequilibrium condition, respectively given as
\begin{equation}
f^{(V_{bias})^{0},(U^{2})^{0}}(\omega,T) = \begin{cases}
 f^{Eq}(\omega,T)  \\
 \theta(\omega-\epsilon_{F}) 
\end{cases} ~, \\ 
\quad f^{(V_{bias})^{1},(U^{2})^{0}}(\omega,T) = \begin{cases}
e\frac{V_{bias}}{2} \frac{\sum_{\alpha}\zeta_{\alpha}\Gamma^{\alpha}(\omega)}{\sum_{\alpha}\Gamma^{\alpha}(\omega)}c(\omega,T) \\
e\frac{V_{bias}}{2} \frac{\sum_{\alpha}\zeta_{\alpha}\Gamma^{\alpha}(\omega)}{\sum_{\alpha}\Gamma^{\alpha}(\omega)} \delta(\omega-\epsilon_{F})
\end{cases}  ~,
\end{equation}
where the first line indicates the finite temperature value and the second one the limiting case at $T=0$. Hence, we confirm for noninteracting models the correctness of ansatz - though this is already a known result from the literature.\\
The key point of our strategy consists of separating the equilibrium from the out-of-equilibrium contribution in the self-energy in Eq.\ref{eq:4SigmaLessCorrected} by means of the nonequilibrium Kadanoff-Keldysh perturbation expansion in $U$ and the linear Taylor expansion in $V_{bias}$ - as we detailed in this section. From our calculations, we derive exact expressions up to order $\mathcal{O}(V_{bias}^{2},(U^{2})^{2})$ for the nonequilibrium lesser self-energies in Eqs.\ref{eq:4Sigma0<all}-\ref{eq:4Sigma2<NEqT0} and greater self-energy in Eqs.\ref{A:Ng4Sigma0>all}-\ref{A:Ng4Sigma2>NEqT0}. By going beyond the Ng ansatz prediction but staying within linear response $V_{bias}$ regime, the corresponding corrected distribution reads as
\begin{equation}
\begin{aligned}
f^{(V_{bias})^{1},(U^{2})^{1}}(\omega,T) &= ~
e\frac{V_{bias}}{2} \frac{\sum_{\alpha}\zeta_{\alpha}\Gamma^{\alpha}(\omega)}{\sum_{\alpha}\Gamma^{\alpha}(\omega)}c(\omega,T)
\frac{\pi U^{2}}{\sum_{\alpha}\Gamma^{\alpha}(\omega)} 
\int_{\infty}^{+\infty} d\omega_{1} \int_{\infty}^{+\infty} d\omega_{2} A(\omega_{1}, \omega_{2}, \omega) \times \\ 
& \hspace*{2cm} \times\Big(  f^{Eq}(\omega_{1}+\omega_{2}-\omega)(f^{Eq}(\omega_{1})+f^{Eq}(\omega_{2})-1)-f^{Eq}(\omega_{1})f^{Eq}(\omega_{2}) \Big) ~,\\
&\equiv e\frac{V_{bias}}{2} \frac{\sum_{\alpha}\zeta_{\alpha}\Gamma^{\alpha}(\omega)}{\sum_{\alpha}\Gamma^{\alpha}(\omega)}c(\omega,T) \times  U^{2}h(\omega) ~,
\end{aligned}
\end{equation}
where $A(\omega_{1}, \omega_{2}, \omega) \doteq \mathcal{A}_{0,dd \sigma}(\omega_{1}) \mathcal{A}_{0,dd \overline{\sigma}}(\omega_{2})\mathcal{A}_{0,dd \overline{\sigma}}(\omega_{1}+\omega_{2}-\omega)$ and $h(\omega)$ indicates the \textit{correction to the Ng ansatz due to interactions}. \\
Hence, we group the terms under linear response for the voltage regime at finite temperature and for general asymmetric barriers we have
\begin{equation} \label{eq:4fcorr}
	\boxed{
f^{corr}(\omega,T) = f^{Eq}(\omega,T) + e\frac{V_{bias}}{2}
\frac{\Gamma^{L}(\omega)-\Gamma^{R}(\omega)}{\Gamma^{L}(\omega)+\Gamma^{R}(\omega)} c(\omega,T) \Big( 1+ U^{2}h(\omega) \Big) + \mathcal{O}(V_{bias}^{2},U^{4}) }
\end{equation}
and for symmetric model with $\Gamma^{L}(\omega)=\Gamma^{R}(\omega)$,  Eq.\ref{eq:4fcorr} reduces to:
\begin{equation} \boxed{
f^{corr}(\omega,T) = f^{Eq}(\omega,T) +\mathcal{O}(V_{bias}^{2},U^{4}) }~.
\end{equation}
The Eq.\ref{eq:4fcorr} is the main result of this section: its version for symmetric model is a surprising consequence due to its simplicity. Although we note that this property also holds for $U=0$ from the pseudo-equilibrium function $\overline{f}^{Eq}(\omega)\rightarrow f^{Eq}(\omega)$ to linear order in $V_{bias}$ where $\Gamma^{L}(\omega) = \Gamma^{R}(\omega)$.\\ 
Our methodology brings a novelty in the sense that we obtain exact results beyond the Ng ansatz limitations, namely for models in presence of external voltage bias both with and without interactions. These self-energy equations can be inserted in Eq.\ref{eq:4FermiCorrected} to calculate the corrected Fermi distribution expression in Eq.\ref{eq:4FermiCorrFinal} that consistently describes the nonequilibrium interacting electronic distribution up to order $\mathcal{O}(V_{bias}^{2},(U^{2})^{2})$. As remarkable outcome, we have a finite contribution in the order $\mathcal{O}((V_{bias})^{1},(U^{2})^{1})$: this is the one we are generally interested for calculating differential conductance using Eq.\ref{eq:DiffCondLRac}. Furthermore, the structure of $f^{(V_{bias})^{1},(U^{2})^{1}}$ shows the same form as $f^{(V_{bias})^{1},(U^{2})^{0}}$ but now multiplied for an additional contribution $U^{2}h(\omega)$ that incorporates the interaction expansion contribution. This significant result for $f^{(V_{bias})^{1},(U^{2})^{1}}$ finds its completion in confirming the limiting case for symmetric leads. The calculation shows a non-trivial correction to the Ng ansatz arising in the linear response already at order $U^2$.\\
As mentioned at the beginning of this section, the demonstration is carried out for the single-impurity Anderson model and it can be extended to more elaborate systems - with the only remark of more involved non-self consistent perturbation expansion calculations.\\
We corroborate now our discussion by presenting the third significant result from our calculation: the lesser Green's function to derive an alternative Meir-Wingreen formula.

\subsection*{Meir-Wingreen current formula by perturbative expansion}
As we know from Sec.\ref{sec:MeirWingreen}, the difficult term is the Meir-Wingreen expression is the one proportional to the lesser Green's function: there is no exact analytical form for its nonequilibrium, interacting self-energy $\Sigma^{<}$ and it does not show an explicit $V_{bias}$ dependence such that the term is amenable for differential conductance calculations without explicit nonequilibrium calculations, even in linear response. We take advantage of the machinery we presented so far in this section and we derive an \textit{alternative lesser nonequilibrium, interacting lesser Green's function}.\\
We start with defining the retarded self-energy using the Kramers-Kroning relations, namely
\begin{equation}
\Sigma^{R}_{dd \sigma}(\omega) = \mathit{Re} \Sigma^{R}_{dd \sigma}(\omega) +i \mathit{Im} \Sigma^{R}_{dd \sigma}(\omega) 
=\frac{1}{\pi} \int_{-\infty}^{\infty} d\omega^{\prime} 
\mathcal{P} \Big[ \frac{1}{\omega-\omega^{\prime}}\Big] \mathit{Im} \Sigma^{R}_{dd \sigma} (\omega^{\prime}) + i \mathit{Im} \Sigma^{R}_{dd \sigma} (\omega^{\prime}) ~,
\end{equation}
where the symbol $\mathcal{P}[ ~]$ indicates the principal part \cite{Altland} and the imaginary part of $\Sigma^{R}_{dd \sigma}$ is calculated using the Keldysh identity in Eq.\ref{eq:identityKeldysh}, namely
\begin{equation}
2i \mathit{Im} \Sigma^{R}_{dd \sigma} (\omega) =  \Big(\Sigma^{>}_{dd \sigma} (\omega)-\Sigma^{<}_{dd \sigma} (\omega) \Big) ~.
\end{equation}
Hence, we can write the retarded nonequilibrium interacting self-energy using the lesser and greater components we derive in Eqs.\ref{eq:4Sigma0<all}-\ref{eq:4Sigma2<NEqT0} and in Eqs.\ref{A:Ng4Sigma0>all}-\ref{A:Ng4Sigma2>NEqT0}, respectively.\\
We use the Keldysh-Dyson equation in Eq.\ref{eq:defDysonKeldysh} in the form $G^{<} = \Sigma^{<}|G^{R}|^{2}$ because of $G^{A}=[G^{R}]^{\dagger}$. By means of these expression we derive 
\begin{equation} \label{eq:4G<Ng}
\boxed{
\begin{aligned}
&G^{<}_{dd\sigma}(\omega,T) =\\
&=  \Sigma^{<}_{dd\sigma}(\omega,T, V_{bias}) \Biggr\lvert \frac{1}{\frac{1}{G^{R}_{0,dd\sigma}(\omega,T)} - 
\Big( \frac{1}{\pi} \int d\omega^{\prime} 
\mathcal{P} \Big[ \frac{1}{\omega-\omega^{\prime}}\Big]+i \Big)\frac{-i}{2}\Big(\Sigma^{>}_{dd \sigma} (\omega,T, V_{bias})-\Sigma^{<}_{dd \sigma} (\omega,T, V_{bias}) \Big)} \Biggr\rvert ^{2} 
\end{aligned} } ~,
\end{equation}
where we use Eq.\ref{eq:defDysonRA} and the Kramers-Kroning relation presented above. The Eq.\ref{eq:4G<Ng} is an exact expression up to order $\mathcal{O}((V_{bias})^{2},(U^{2})^{2})$ using the equations derived in this section and it is fully determined in its parameters of temperature and external bias. This alternative version of the nonequilibrium, interacting lesser Green's function is now enclosing an explicit voltage bias dependence that is given by the perturbative expressions of $\Sigma^{</>}_{dd\sigma}$ in Eq.\ref{eq:4SigmaLessCorrected} as we derived in detail throughout this section. This significant result allows us to directly insert the lesser Green's function Eq.\ref{eq:4G<Ng} in the Meir-Wingreen current formula Eq.\ref{eq:MW}. This current expression presents now $V_{bias}$ dependence in all its terms and we calculate from it the differential conductance in Eqs.\ref{eq:DiffCondLRac},\ref{eq:DiffCondLRdc}.\\

\noindent{With} this last discussion we complete the analysis on the verification within and beyond the Ng ansatz assumptions - in the corrected Fermi distribution in Eq.\ref{eq:4FermiCorrFinal}. This derivation offers also a useful practice to derive equilibrium, noninteracting correlators from nonequilibrium, interacting ones. Furthermore, we obtain an alternative expression for the lesser Green's function to calculate linear response conductance using the Meir-Wingreen current formula.

\section{Improved Kubo formula for numerical renormalization group technique}\label{sec:ImprovedKubo}
In this section we demonstrate an alternative version of the Kubo formula for electric and heat conductance in Eqs.\ref{eq:defKuboLRel},\ref{eq:defKuboLRhea} respectively. Despite their simple analytical proof, the derived expressions find useful applications in their implementation in the numerical renormalization group technique we introduced in Sec.\ref{sec:RGtheo}. Hence, the expressions shown in this section represent not only an equivalent formulation on a mathematical ground of the standard Kubo formula, but its improved version when it is implemented in NRG algorithm. Other avenues have been attempted in the literature \cite{Orignac2002,GotzeWolfle1972}. Here the improvement is in the sense that derived formulae in this section overcome the problems usually met in the the standard Kubo formula as we discuss in Sec.\ref{sec:Kubo}.\\
We proceed with the analytical derivation of the new formulae and we follow in some detail with the NRG protocol to numerically quantify the advantage of this alternative formulation.

\subsection*{Analytical derivation of the improved Kubo formula}
The cornerstone of this derivation is to find an alternative and equivalent expression of the current-current correlator in Eq.\ref{eq:defCurrCurrCor} obeying the following properties:\\
$(i)$ the correlator has to depend only on baths degree of freedom - such that we avoid any information due to the coupling among those and the impurity, which might be complicated in case of elaborated systems;\\
$(ii)$ the expression returns finite values also in \textit{dc-}regime, stabilizing any numerical computation of such a term.\\
The requirement $(i)$ will be more clear in the next section when we present the NRG protocol; the requirement $(ii)$ relates to the frequency denominator in the regular Kubo formula in Eq.\ref{eq:defKuboLRel}, which is poorly controlled numerically as $\omega\rightarrow0$ \\
We consider the current-current correlator defined in Eq.\ref{eq:defCurrCurrCor}, here for its retarded version in the case of electrical and heat conductance, and using Eq.\ref{eq:defSuscept} we have
\begin{equation}
\begin{aligned}
K^{R}_{N_{\alpha},N_{\alpha^{\prime}}}(t, T) &= -i\theta(t-t^{\prime})
\langle [\dot{N}_{\alpha}(t),\dot{N}_{\alpha^{\prime}}(t^{\prime})]\rangle_{T} ~,\\
K^{R}_{H_{\alpha},H_{\alpha^{\prime}}}(t, T)& =  -i\theta(t-t^{\prime})
\langle [\dot{H}_{\alpha}(t),\dot{H}_{\alpha^{\prime}}(t^{\prime})]\rangle_{T}  ~,
\end{aligned}
\end{equation}
which are calculated at equilibrium. Taking the imaginary part of these equations in frequency domain, we aim to derive the linear response function for conductance in Eq.\ref{eq:defSusceptCondAC}. On those correlators we apply the equation of motion technique in Heisenberg picture as defined in Eq.\ref{eq:EoM} where we take as model Hamiltonian the standard single Anderson impurity system in Eq.\ref{eq:AM}. In this calculation, we introduce also the notation $\hat{X}_{\alpha}=\hat{N}_{\alpha},\hat{H}_{\alpha}$ for the number operator of the leads and the lead Hamiltonian, respectively.\\
Before we commence the derivation, we need to present a few properties on $\hat{X}_{\alpha}(t)$ and its time-derivative $\dot{X}_{\alpha}(t)$, namely:
\begin{equation}
\begin{aligned}
&(i)~ \hat{X}^{\dagger}_{\alpha}(t) =\hat{X}_{\alpha}(t): \quad \hat{X}_{\alpha}(t)~ \text{\textit{hermitian operator}}\quad\Rightarrow \langle \hat{X}_{\alpha}(t) \rangle \equiv\mathit{Re}\langle \hat{X}_{\alpha}(t) \rangle  ~,\\ 
&(ii)~ \dot{X}^{\dagger}_{\alpha}(t) =\dot{X}_{\alpha}(t): \quad \dot{X}_{\alpha}(t)~\text{\textit{hermitian operator}}\quad \Rightarrow \langle \dot{X}_{\alpha}(t) \rangle \equiv \mathit{Re}\langle \dot{X}_{\alpha}(t) \rangle ~,\\
&(iii)~ \Big([\dot{X}_{\alpha}(t),\dot{X}_{\alpha^{\prime}}(t^{\prime})]\Big)^{\dagger} = -[\dot{X}_{\alpha}(t),\dot{X}_{\alpha^{\prime}}(t^{\prime})]: \quad \dot{X}_{\alpha}(t)\dot{X}_{\alpha^{\prime}}(t^{\prime})~ \text{\textit{anti-hermitian operator}}\\
& \hspace*{8cm}\Rightarrow \langle [\dot{X}_{\alpha}(t),\dot{X}_{\alpha^{\prime}}(t^{\prime})]\rangle \equiv i\mathit{Im}
\langle [\dot{X}_{\alpha}(t),\dot{X}_{\alpha^{\prime}}(t^{\prime})]\rangle  ~.
\end{aligned}
\end{equation}
Moreover, the lead operators $\hat{X}_{\alpha}$ commute $[\hat{X}_{\alpha},\hat{X}_{\alpha^{\prime}} ] = 0$.\\

\noindent{We} keep in mind these information in our derivation and we perform the first cycle of equation of motion. In frequency domain, a good initial guess to start with is $\langle\langle \hat{X};\hat{X} \rangle\rangle$ so we have 
\begin{equation}
\begin{aligned}
\omega \langle\langle \hat{X}_{\alpha};\hat{X}_{\alpha^{\prime}} \rangle\rangle &= \langle \underbrace{[\hat{X}_{\alpha},\hat{X}_{\alpha^{\prime}} ]}_{=0}\rangle - 
\langle\langle \underbrace{[\hat{H},\hat{X}_{\alpha}]}_{\stackrel{Eq.\ref{A:eq:HeisenbergEoM}}{=}-i\dot{X}_{\alpha}};\hat{X}_{\alpha^{\prime}} \rangle\rangle ~,\\
&=+ i \langle\langle \dot{X}_{\alpha};\hat{X}_{\alpha^{\prime}} \rangle\rangle 
\end{aligned}
\end{equation}
and by taking the imaginary part of the correlators at both side we end up with
\begin{equation} \label{eq:4XX}
	\begin{aligned}
\mathit{Im}\left[\omega \langle\langle \hat{X}_{\alpha};\hat{X}_{\alpha^{\prime}} \rangle\rangle \right]	&= \mathit{Im}\left[+ i \langle\langle \dot{X}_{\alpha};\hat{X}_{\alpha^{\prime}} \rangle\rangle  \right] ~,\\
\Rightarrow\omega\mathit{Im}\langle\langle \hat{X}_{\alpha};\hat{X}_{\alpha^{\prime}} \rangle\rangle  &= \mathit{Re}\langle\langle \dot{X}_{\alpha};\hat{X}_{\alpha^{\prime}} \rangle\rangle ~.
\end{aligned}
\end{equation}
We continue with a second cycle of equation of motion on the expression just calculated and so we have
\begin{equation}
\begin{aligned}
\omega\langle\langle \dot{X}_{\alpha};\hat{X}_{\alpha^{\prime}} \rangle\rangle &=\langle[\dot{X}_{\alpha},\hat{X}_{\alpha^{\prime}} ]\rangle + \langle\langle\dot{X}_{\alpha}; \underbrace{[\hat{H},\hat{X}_{\alpha^{\prime}}]}_{\stackrel{Eq.\ref{A:eq:HeisenbergEoM}}{=}-i\dot{X}_{\alpha^{\prime}}} \rangle\rangle ~,\\
&=i\underbrace{\langle\left[\left[\hat{H},\hat{X}_{\alpha}\right];\hat{X}_{\alpha^{\prime}}\right]\rangle}_{const}-i \langle\langle\dot{X}_{\alpha};\dot{X}_{\alpha^{\prime}}\rangle\rangle ~,
\end{aligned} 
\end{equation}
where on the right hand side the first term is \textit{not} necessarily zero, but it results as Hermitian operator hence it is real and the corresponding $+i ~const$ is imaginary quantity.\\
We multiply times $+i$ the above expression on both sides, 
\begin{equation}
+i\omega\langle\langle\dot{X}_{\alpha};\hat{X}_{\alpha^{\prime}} \rangle\rangle = -const +\langle\langle\dot{X}_{\alpha};\dot{X}_{\alpha^{\prime}}\rangle\rangle ~,
\end{equation}
and then we take the imaginary part of it to get,
\begin{equation}\label{eq:4dotXX}
\mathit{Im}\left[\langle\langle\dot{X}_{\alpha};\dot{X}_{\alpha^{\prime}}\rangle\rangle\right]=\mathit{Im}\left[+i\omega\langle\langle\dot{X}_{\alpha};\hat{X}_{\alpha^{\prime}} \rangle\rangle \right] = \omega \mathit{Re}\langle\langle \dot{X}_{\alpha};\hat{X}_{\alpha^{\prime}} \rangle\rangle  ~. 
\end{equation}
Hence, by substituting Eq.\ref{eq:4dotXX} into Eq.\ref{eq:4XX},  we succeed in showing the following identity
\begin{equation}\boxed{
\begin{aligned}
&\langle\langle \dot X_{\alpha} ; \dot X_{\alpha^{\prime}} \rangle\rangle  \quad \Leftrightarrow \quad
 \langle\langle \dot{X}_{\alpha} ;  X_{\alpha^{\prime}} \rangle\rangle  \quad \Leftrightarrow \quad
 \langle\langle X_{\alpha} ; X_{\alpha^{\prime}} \rangle\rangle  \\
&\Longrightarrow ~ \mathit{Im}\langle\langle\dot{X}_{\alpha};\dot{X}_{\alpha^{\prime}}\rangle\rangle = +\omega \mathit{Re}\langle\langle \dot{X}_{\alpha};\hat{X}_{\alpha^{\prime}} \rangle\rangle = +\omega^{2}\mathit{Im}\langle\langle\hat{X}_{\alpha};\hat{X}_{\alpha^{\prime}} \rangle\rangle
 \end{aligned}} ~,
\end{equation}
and we can write the equivalent expressions for the retarded current-current correlators by means of the identity, namely
\begin{equation}
\begin{aligned}
&\mathit{Im}K^{R}_{N_{\alpha},N_{\alpha^{\prime}}}(\omega,T) =
\int_{-\infty}^{+\infty}dt \left(-\theta(t) \mathit{Im}
\langle [\dot{N}_{\alpha}(t),\dot{N}_{\alpha^{\prime}}(0)]\rangle_{T} \right) e^{i\omega t}   \equiv +\omega^{2}\mathit{Im}\langle\langle\hat{N}_{\alpha};\hat{N}_{\alpha^{\prime}} \rangle\rangle_{T}  ~,\\
&\mathit{Im}K^{R}_{H_{\alpha},H_{\alpha^{\prime}}}(\omega, T) =\int_{-\infty}^{+\infty}dt \left(-\theta(t)\mathit{Im}
\langle [\dot{H}_{\alpha}(t),\dot{H}_{\alpha^{\prime}}(0)]\rangle_{T}\right) e^{i\omega t}  \equiv +\omega^{2}\mathit{Im}\langle\langle\hat{H}_{\alpha};\hat{H}_{\alpha^{\prime}} \rangle\rangle_{T} ~,
\end{aligned} 
\end{equation}
where we take as usual $t^{\prime}=0$. We convert these results in Fourier space as we do in Eq.\ref{eq:defKuboLRel} and we insert our derived current-current correlator in the standard Kubo formula. At the end after the due simplifications, we derive the \textbf{improved Kubo formula for electrical conductance}, that is
\begin{equation}\label{eq:4defKuboLRel}\boxed{
\mathcal{G}^{C}(\omega,T) = -\mathcal{G}_{0} 2\pi\omega\mathit{Im}\langle\langle\hat{N}_{\alpha};\hat{N}_{\alpha^{\prime}} \rangle\rangle_{\omega,T} } ~,
\end{equation}
compare it with the form in the literature in Eq.\ref{eq:defKuboLRel} and the \textbf{improved Kubo formula for heat conductance}, that is
\begin{equation}\label{eq:4defKuboLRhea}\boxed{
\mathcal{G}^{E}(\omega,T) = -\mathcal{G}_{0} 2\pi\omega\mathit{Im}\langle\langle\hat{H}_{\alpha};\hat{H}_{\alpha^{\prime}} \rangle\rangle_{\omega,T}  } ~,
\end{equation}
compare it with the literature form in Eq.\ref{eq:defKuboLRhea}.\\
Eqs.\ref{eq:4defKuboLRel},\ref{eq:4defKuboLRhea} are calculated without invoking any approximations since they are exact reformulations and those represent an improvement over the standard formulation for the following reasons:
\begin{itemize}
\item the correlator $\langle\langle\hat{X}_{\alpha}(t);\hat{X}_{\alpha^{\prime}}(t^{\prime}) \rangle\rangle$\\
$(i)$ 
it depends only on the baths degree of freedom irrespective of the connections in the hybridization and of the details of the nanostructure itself. This aspect greatly simplify the numerical implementation: once the NRG code is designed for the correlator $\langle\langle \hat{X}_{\alpha} ; \hat{X}_{\alpha^{\prime}} \rangle\rangle$, the very same code can be used to compute the conductance for \textit{any} other model without any further modification on the numerics;\\
$(ii)$ the operator $\hat{X}_{\alpha}=\hat{N}_{\alpha}$ within the NRG paradigm, that represents the leads occupation is \textit{exactly} computed numerically through calculating the occupation of the respective Wilson chain - which we recall it is the representation used for the logarithmic discretized version of the original conduction band. As consequence, also the correlator for $\hat{X}_{\alpha}=\hat{H}_{leads}$ is exactly numerically  computed. In conclusion, numerical results obtained with $\langle\langle\hat{X}_{\alpha}(t);\hat{X}_{\alpha^{\prime}}(t^{\prime}) \rangle\rangle$ exhibits higher accuracy than the standard $\langle\langle\dot{X}_{\alpha}(t);\dot{X}_{\alpha^{\prime}}(t^{\prime}) \rangle\rangle$ for NRG implementations;
\item the \textit{dc-}limit of the expressions in Eqs.\ref{eq:4defKuboLRel},\ref{eq:4defKuboLRhea} acquires now great numerically stability since the frequency denominator is avoided in the $\omega\rightarrow 0$ limit.  Thus, this significant advantage makes possible numerical evaluation of these conductance formulae even in low-energy regimes - which is the relevant energy scale for studying Kondo effects.
\end{itemize}
In conclusion, the standard Eqs.\ref{eq:defKuboLRel},\ref{eq:defKuboLRhea} and the improved Kubo Eqs.\ref{eq:4defKuboLRel},\ref{eq:4defKuboLRhea} here derived are both exact formulae and completely equivalent to each other formally. They both employ correlators in Heisenberg representation at equilibrium conditions. Although both formulae are exact, any numerical method to evaluate the required correlators is necessarily approximate - see more comments at the end of Sec.\ref{sec:Kubo}. For the reasoning just presented, these alternative Kubo formulae tick all the requirements for better numerical performance.\\
We proceed with the next section describe the NRG algorithm for calculation of the new correlator $\langle\langle \hat{X}_{\alpha} ; \hat{X}_{\alpha^{\prime}} \rangle\rangle$.
 
\subsection{Improved Kubo formula in NRG}
In this section we devise the NRG calculation for the improved Kubo formulae as we analytically derived by means of equation of motion technique in Eqs.\ref{eq:4defKuboLRel},\ref{eq:4defKuboLRhea}. \\
As standard practice in the NRG technique \cite{Wilson1975,Bulla2008, Dias_ImpSolver2019}, we use the Lehmann representation to describe the single-particle Green's function of many-body systems \cite{Flensberg}. The Lehmann representation of generic spectral functions of the type $\mathcal{A}_{BC\sigma}(\omega,T) = \int \frac{dt}{2\pi} e^{i\omega t}\langle \hat{B}_{\sigma}(t)\hat{C}_{\sigma}\rangle_T$ - here we use the Fourier space version of Eq.\ref{eq:defSpectral} for the Green's function defined in Eq.\ref{eq:defG} - is given by
\begin{equation}\label{eq:4lehmann}
 \mathcal{A}_{BC\sigma}(\omega,T) = \sum_{a,b}\langle b| \hat{C}_{\sigma}|a\rangle \frac{e^{-\beta\epsilon_{a}}}{Z} \langle a| \hat{B}_{\sigma}|b\rangle \delta(\omega-\epsilon_{b}+\epsilon_{a}) ~,
\end{equation}
where $\{|a\rangle\}$ constitutes the \textit{complete} set of eigenstates of the Hamiltonian satisfying the Schr\"odinger equation and $Z=\sum_{a}exp(-\beta \epsilon_{a})$ is the partition function for the statistical canonical ensemble. In principle, the desired retarded correlator $\langle\langle\hat{X}_{\alpha};\hat{X}_{\alpha^{\prime}}\rangle\rangle_{\omega,T}$ can be obtained from Eq.\ref{eq:4lehmann} as 
\begin{equation}
\frac{-\mathit{Im}\langle\langle\hat{X}_{\alpha};\hat{X}_{\alpha^{\prime}} \rangle\rangle_{\omega,T}}{\pi}=
\mathcal{A}_{X_{\alpha}X_{\alpha^{\prime}}\sigma}(\omega,T) -\mathcal{A}_{X_{\alpha^{\prime}}X_{\alpha}\sigma}(-\omega,T)~.
\end{equation}
However, calculating such an object within NRG is somewhat unconventional because the domain of the lead operators $\hat{X}_{\alpha}=\hat{N}_{\alpha}$ is the Wilson chain, hence rather than local operators those are defined on the Wilson chain basis.\\
As we present in general fashion in Sec.\ref{sec:RGtheo}, the NRG protocol - see its schematic in Fig.\ref{F2:NRG} - requires that the logarithmic discretized version of the bare model is mapped to the one dimensional chain in Wilson chain representation. The chain $\hat{H}_{leads}^{WC}$ - with superscript Wilson chain (WC) - has model Hamiltonian of a tight-binding system with decaying hopping $t_{n}\sim D\Lambda^{-n/2}$ with $\Lambda>1$ the discretization parameter, $D$ bare conduction bandwidth, single-particle energy $\epsilon_{n}$ and chain operator $c^{\dagger}_{\alpha n \sigma}c_{\alpha n \sigma}$. We take $n=0,\dots,\infty$ sites departing from one side of the impurity. Thus, on the chain we define the lead number operator as 
\begin{equation}
\hat{N}_{\alpha}=\sum_{n=0}^{\infty} \hat{n}_{\alpha n}=\sum_{n=0}^{\infty} \sum_{\sigma}c^{\dagger}_{\alpha n \sigma}c_{\alpha n \sigma} ~,
\end{equation}
where $\alpha=L,R$ means \textit{each} continuum bath in the original model is logarithmically discretized and mapped to semi-infinite one dimensional chain representation that is the here presented Wilson chain.\\ 
As we know, the chosen mapping has a peculiar coupling with the impurity that is only with the $0^{th}$ orbital of the Wilson chain hybridises with the impurity. Hence, only the lead operator $c^{\dagger}_{\alpha 0 \sigma}c_{\alpha 0 \sigma}$ is included in $\hat{H}_{hyb}^{WC}$.\\  
As next step in the NRG protocol, starting from $\hat{H}_{imp}$ - that is invariant under the RG procedure - a sequence of Hamiltonians $\hat{H}_{n}$ is obtained iteratively by addition of a successive Wilson chain orbital, namely
\begin{equation}
\begin{aligned}
& \hat{H}_{0}=\hat{H}_{atomic} +\hat{H}_{hyb}^{WC} ~,\\ 
& \hat{H}_{n}= \hat{H}_{n-1}+ t_{n-1}\sum_{\alpha \sigma}\left( c^{\dagger}_{\alpha(n-1)\sigma}c_{\alpha n\sigma} + H.c. \right) ~.
\end{aligned}
\end{equation}
In practice $\hat{H}_{n}$ is rescaled at each step to ensure that its eigenvalues are of order $\mathcal{O}(1)$, but this detail is omitted here for the sake of clarity.\\
We remark that only the coupling between the impurity and the $0^{th}$ orbital of the Wilson chain corresponds to the \textit{original physical system} - the rest of the chain is an artefact required by the NRG procedure. The logarithmically discretized Hamiltonian $\hat{H}^{discr}=\hat{H}^{imp}+\hat{H}^{WC}_{leads}+\hat{H}^{WC}_{hyb}$ is equivalent to the original model but now defined with respect to a specific discrete version of the conduction band. Hence, $\hat{H}^{discr}$ is the exact discretized model Hamiltonian and we can recover $\hat{H}$ by taking limit $\hat{H}^{discr}=\lim_{n\rightarrow \infty} \hat{H}_{n}$ and $\Lambda\rightarrow 1$ limit.\\
Clearly the Fock space dimension of $\hat{H}_n$ grows exponentially with $n$, and the exact solution of even the discretized model is intractable. Instead, at each step, $\hat{H}_n$ is diagonalized and only the low-energy manifold of eigenstates is retained for the next step. This constitutes an RG transformation and the physics on successively lower energy scales is revealed on increasing $n$.\\
We continue now with the protocol focusing on the iterative scheme. It is usually initialized in some convenient basis e.g. $\{|\phi_a\rangle_0\}$ in absence of external perturbation spanning $\hat{H}_0$ by forming the corresponding Hamiltonian matrix $\mathbb{H}_{0}$, with elements $[\mathbb{H}_{0}]_{ab}={_0}\langle \phi_a | \hat{H}_0 | \phi_b \rangle_0$. We introduce the unitary matrix $\mathbb{U}_{0}$ constructed such that $\mathbb{U}_{0}^{\dagger}\mathbb{H}_{0}\mathbb{U}_{0} = \mathbb{D}_{0}$ is diagonal. The eigenstates $|\psi_{l}\rangle_0=\sum_a [\mathbb{U}_{0}]_{al}|\phi_a\rangle_0$ are used to construct $\mathbb{H}_{1}$, with matrix elements 
${_1}\langle \phi_{ji} | \hat{H}_1 | \phi_{j^{\prime}i{^\prime}} \rangle_1$ are defined in terms of 
basis states  $|\phi_{ji}\rangle_1=|\psi_j\rangle_0\otimes |\gamma_i\rangle_1$ spanning the Fock space of $\hat{H}_1$, where the first and second subscript indicates the old and new basis, respectively. We consider a general two-channel Fermionic model where $\{|\gamma_i\rangle_n\}$ are the 16 states defined on the $n^{th}$ Wilson shell corresponding to $c_{\alpha n \sigma}$. We now diagonalize the matrix $\mathbb{H}_{1}$ to find $\mathbb{U}_{1}$.\\
Truncation of the $M$-dimensional Fock space of $\hat{H}_n$ is accomplished in NRG by discarding high energy states at that iteration. The $M_{trunc}$ eigenstates of $\hat{H}_n$ are retained up to an energy cut-off $E_{cut}$. The matrix $\mathbb{U}_{n}$ with $dim(M \times M)$ is therefore reshaped to an $dim(M\times M_{trunc})$ matrix $\widetilde{\mathbb{U}}_{n}$. This implies that $\widetilde{\mathbb{U}}_{n}^{\dagger}\mathbb{H}_{n}\widetilde{\mathbb{U}}_{n} = \widetilde{\mathbb{D}}_{n}$ is a diagonal matrix of $dim(M_{ trunc}\times M_{ trunc})$ containing only the lowest eigenvalues of $\hat{H}_n$. The restricted set of $M_{trunc}$ retained eigenvectors at iteration $n$ are obtained as 
$|\psi_{l}\rangle_n=\sum_{ji} [\widetilde{\mathbb{U}}_{n}]_{(ji)l}|\phi_{ji}\rangle_n$. 
Only these retained states at iteration $n$ are used to construct the Hamiltonian matrix at iteration $n+1$. However, noted these operators are defined on the complete Anders-Schiller basis. The approximate Hamiltonian in the basis of $|\phi_{ji}\rangle_{n+1}=|\psi_j\rangle_n\otimes | \gamma_i\rangle_{n+1}$ is denoted  $\widetilde{\mathbb{H}}_{n+1}$. The truncation at each step means that the Fock space dimension is roughly independent of $n$.\\
Crucially, the structure of the Wilson chain with exponentially decaying hoppings implies that discarded states at a given iteration are unimportant for constructing the low-energy states at a later iteration. Furthermore, useful information can be extracted at each step since $\widetilde{\mathbb{H}}_{n}$ may be regarded as a renormalized version of the full Hamiltonian at an effective temperature $T_n \sim D\Lambda^{-n/2}$.\\

\noindent{Having} now this formalism established, we return to the calculation of the correlator $\langle\langle\hat{X}_{\alpha};\hat{X}_{\alpha^{\prime}}\rangle\rangle_{\omega,T}$.\\
On the contrary of $\langle\langle\dot{X}_{\alpha};\dot{X}_{\alpha^{\prime}}\rangle\rangle_{\omega,T}$, the new correlator is non-local as defined only on the Wilson chain basis. Thus, its basis does not include any information from the impurity site. Furthermore, at each iteration, the truncated set of eigenvectors is calculated from adding the contribution of every energy shell i.e. Wilson orbitals. Crucially, this gives access to all the energy scales at each iterative steps and it leads to higher precision results.\\
As we discuss at the beginning of this section, the exact result for the discretized model could in principle be obtained from Eq.\ref{eq:4lehmann} given the complete set of exact eigenstates in the discretized model Hamiltonian $\hat{H}^{discr}$. In NRG however, the iterative diagonalization and truncation procedure means that only the approximate - i.e. the \textit{renormalized} - eigenstates of $\widetilde{\mathbb{H}}_{n}$ at each step are known. The \textit{Anders-Schiller basis} (AS) \cite{AndersSchiller2005,AndersSchiller2006} comprises the \textit{discarded} states at each step and is a complete, albeit approximate, basis with which to compute spectral functions via the Lehmann sum in Eq.\ref{eq:4lehmann}. In the literature \cite{Weichselbaum2007,AndersPruschkeNRGGF2006}, the problem is reformulated in terms of the full density matrix established on the AS basis, wherein the spectrum $\mathcal{A}_{BC\sigma}(\omega,T)=\sum_n w_n \mathcal{A}_{BC\sigma,n}(\omega,T)$ consists of a weighted sum of contributions from each NRG iteration. This requires matrix representations of the operators $\hat{B}$ and $\hat{C}$ at iteration $n$ in the eigenbasis of $\widetilde{\mathbb{H}}_{n}$. For the correlator $\langle\langle\hat{X}_{\alpha};\hat{X}_{\alpha^{\prime}}\rangle\rangle_{\omega,T}$, we apply the usual NRG approximation to neglect the contributions to observables at iteration $n$ from the Wilson shells $m>n$. In particular, we have
\begin{equation}
\widetilde{\mathbb{N}}_{\alpha n} = \sum_{m=0}^n \widetilde{\mathbb{N}}^{m}_{\alpha n} ~,\quad \text{where}~ [\widetilde{\mathbb{N}}^{m}_{\alpha n}]_{ll^{\prime}} = {_n}\langle \psi_{l} |\hat{n}_{\alpha m} | \psi_{l^{\prime}} \rangle_n ~.
\end{equation}  
We examine the detailed contribution of this expression according to the number of iterations.\\
If $n=0$, we have $\widetilde{\mathbb{N}}_{\alpha 0} \equiv \widetilde{\mathbb{N}}^{0}_{\alpha 0}$ and it can be explicitly evaluated from the known exact eigenstates of $\hat{H}_0$. If $n>0$, we split $\widetilde{\mathbb{N}}_{\alpha n} =  \widetilde{\mathbb{N}}^{<}_{\alpha n} + \widetilde{\mathbb{N}}^{n}_{\alpha n}$ into two contributions. The on-shell term can be directly evaluated at each step, namely
\begin{equation}
[\widetilde{\mathbb{N}}^{n}_{\alpha n}]_{ll^{\prime}} =\sum_{ji} [\widetilde{\mathbb{U}}_{n}]^{\star}_{(ji)l} [\widetilde{\mathbb{U}}_{n}]_{(ji)l^{\prime}}\times{_n}\langle\gamma_i|\hat{n}_{\alpha n} |\gamma_i\rangle_n  ~,
\end{equation}
since the matrix elements involved are trivial while we have for $0\le m<n$
\begin{equation}
[\widetilde{\mathbb{N}}^{m}_{\alpha n}]_{ll^{\prime}} =\sum_{jj^{\prime}i} [\widetilde{\mathbb{U}}_{n}]^{\star}_{(ji)l} [\widetilde{\mathbb{U}}_{n}]_{(j^{\prime}i)l^{\prime}}\times [\widetilde{\mathbb{N}}^{m}_{\alpha n-1}]_{jj^{\prime}}   ~,
\end{equation}
where this last property implies the recursion relation
\begin{equation}
[\widetilde{\mathbb{N}}^{<}_{\alpha n}]_{ll^{\prime}} =\sum_{jj^{\prime}i}
 [\widetilde{\mathbb{U}}_{n}]^{\star}_{(ji)l} [\widetilde{\mathbb{U}}_{n}]_{(j^{\prime}i)l^{\prime}}\times [\widetilde{\mathbb{N}}_{\alpha n-1}]_{jj^{\prime}} ~.
\end{equation}
The required operator matrices $\widetilde{\mathbb{N}}_{\alpha n}$ can then all be obtained iteratively, starting from $\widetilde{\mathbb{N}}_{\alpha 0}$. Hence, we have obtained the lead operators are fully defined on the Wilson chain basis and any further iterations will involve only this type of elements - and each step carries also all the energy scales of the previous steps.\\
In conclusion, with this set of expressions, the correlator $\langle\langle\hat{X}_{\alpha};\hat{X}_{\alpha^{\prime}}\rangle\rangle_{\omega,T}$ can be obtained using standard NRG, and yields greatly improved accuracy for the conductance in the Eqs.\ref{eq:4defKuboLRel},\ref{eq:4defKuboLRhea}.\\

\noindent{This} brings to a completion our presentation on the technical implementation of the alternative version of the Kubo formula. This derivation based its root in straightforward analytical derivations and it acquires its max effectiveness in the NRG implementation. Here, the specific form of the current-current correlator not only makes accessible to trustful conductance values in the \textit{dc-}limit but also increases the numerical precision in all the energy scales since its expression is defined by exactly computable operators laying on the Wilson chain. 

\subsection{Numerical results for the Anderson impurity model}
\begin{figure}[H]
	\centering
	\includegraphics[width=0.75\linewidth]{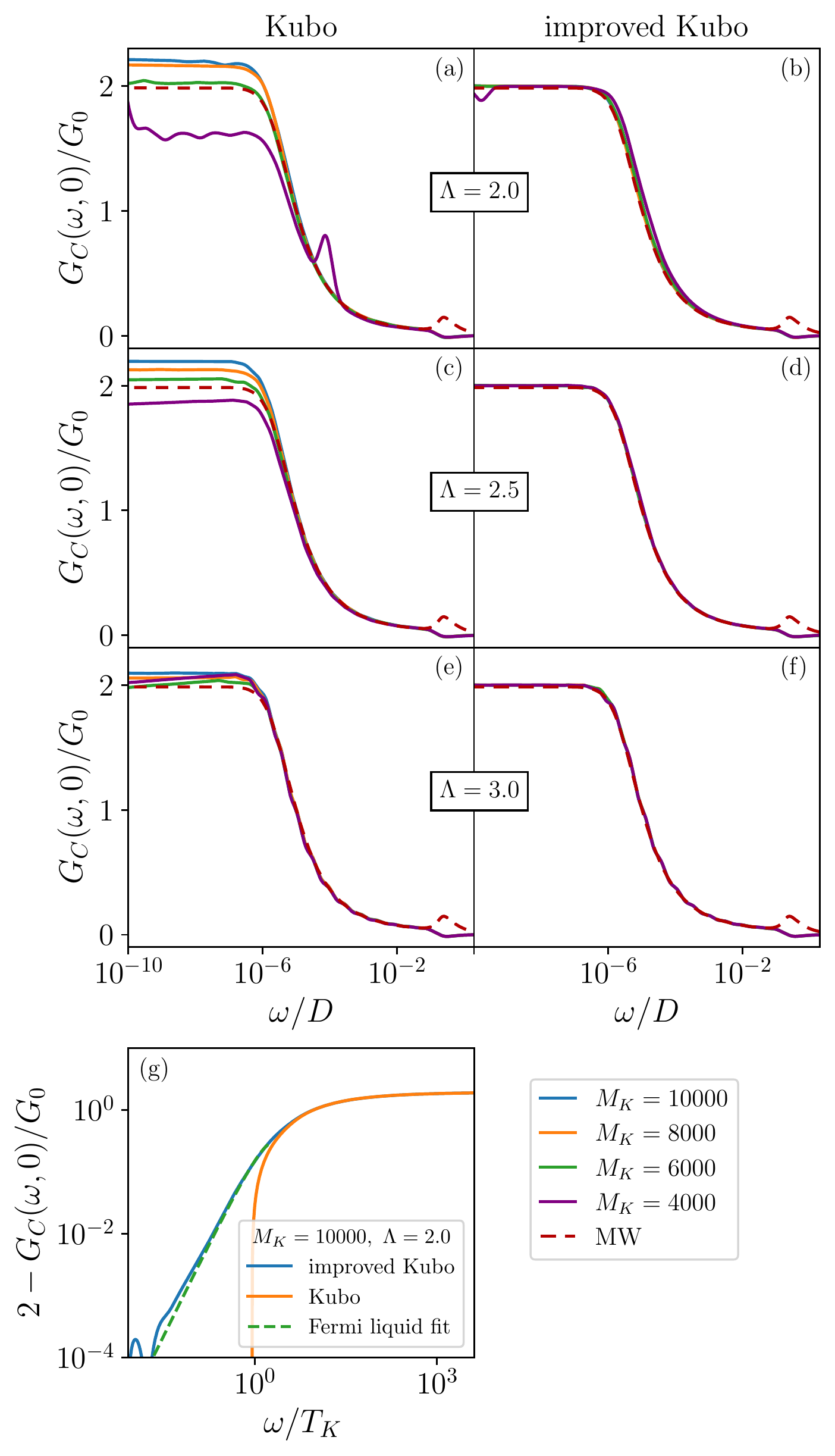}
	\caption[Kubo vs improved Kubo in Anderson impurity model]{ \textit{(a)-(f)} Comparison of NRG results for the Kubo (left) and \textit{improved} Kubo (right) formulae for the \textit{ac}-conductance $\mathcal{G}^C(\omega,0)$ at $T=0$ for the two-lead AIM, with different numbers $M_K$ of NRG kept states as given in the legend, and different NRG discretization parameters $\Lambda$. Results  benchmarked against the MW formula, evaluated within the equivalent single-channel AIM (red dashed line). \textit{(g)} Low-frequency behaviour, comparing Kubo and improved-Kubo with a Fermi liquid fit $2-\mathcal{G}^C(\omega,0)/\mathcal{G}_0= c_2(\omega/T_{K})^2+c_3|\omega/T_{ K}|^3$. AIM parameters: $U=0.4D$, $\epsilon_d=-\tfrac{1}{2}U$, $V_s=V_d=0.07D$.}\label{F4/KvsimpK}
\end{figure}

We aim for a systematic numerical investigation to substantiate the analytical results of improved Kubo formulae Eqs.\ref{eq:4defKuboLRel},\ref{eq:4defKuboLRhea} derived in this section. In Chapter \ref{ch:TDQD}, we give a similar numerical analysis for the triple quantum dots system.\\
\noindent{As} a proof-of-principle demonstration, we calculate the \textit{ac}-electrical conductance $\mathcal{G}^C(\omega,T)$ vs driving frequency $\omega$ at $T=0$ using FDM-NRG for the two-lead AIM Eq.~\ref{eq:AM} using the Kubo formula Eq.~\ref{eq:defKuboLRel}.\\
In Figs.~\ref{F4/KvsimpK}\textit{(a-f)} we compare the standard implementation of the Kubo formula \cite{Dias_ImpSolver2019} using the current-current correlator in Eq.~\ref{eq:defKuboLR}, see left panels, with the \textit{improved} Kubo formula Eq.~\ref{eq:4defKuboLRel}, see right panels, for different numbers $M_K$ of NRG kept states as specified in the legend. Plots \textit{(a-b)} are calculated for NRG discretization parameter $\Lambda=2$; plots \textit{(c,d)} for $\Lambda=2.5$; and plots \textit{(e,f)} for $\Lambda=3$. As a reference, we provide the Meir-Wingreen result in dashed line for the equivalent single-channel Anderson impurity model, see cartoon in Fig.\ref{F3:PC} and related discussion in Sec.\ref{sec:PC}. In this case, the linear response conductance is calculated within NRG using the PC version of the MW current formula in Eq.\ref{eq:MW->PC}. This computation should be considered the numerically-exact result for this system. \\
As we know from Sec.\ref{sec:RGtheo}, NRG technique recovers the exact solution only at $\Lambda=1$ where the continuum limit of the bath is restored. This is clearly not plausible due to the extremely expensive computational cost. Hence, by increasing the $\Lambda$ value, the discretization becomes more coarse. However, after each truncation step - see schematic in Fig.\ref{F2:NRG} - the number of kept states decrease and the computational time is sensibly reduced. Furthermore, once an appropriate $\Lambda$ value is found, regardless the number of kept state after the truncation step in each iteration, the results do not variate within a reasonable error bar. We say that we find \textit{convergence} with respect to the number kept states  $M_K$. In this case, we have the best trade-off between discretization value $\Lambda$ and retained states $M_K$ \cite{Weichselbaum_TradeMkLambda2011}. We note that the larger $\Lambda$ value, the faster the convergence is found since the fewer states are kept from the truncation step. \\
We start now analysing the numerical results. The Fig.~\ref{F4/KvsimpK}$(a)$ shows that the standard Kubo formula yields rather poor results, even at large $M_K=10000$ for $\Lambda=2$. The situation improves with increasing $\Lambda$ - see panels \textit{(c,e)} - due to the trade-off between $M_K$ and $\Lambda$. The best performance, with an error of $\sim 5\%$, is obtained with $\Lambda=3$ and $M_K=10000$ - see blue line, panel \textit{(e)} - although even here the results are still not fully converged with respect to $M_K$. By contrast, the improved Kubo results in panels \textit{(b,d,f)} are highly stable, and essentially fully converged even for remarkably low $M_K=4000$ for all $\Lambda$ considered. This shows that accurate results can be obtained within NRG using the improved Kubo formulation at relatively \textit{low} computational cost. To quantify this cost, we recall that NRG scales approximately as $N\times M_{K}^{3}$, here considering without the dynamics calculation. If we take as example Fig.\ref{F4/KvsimpK}\textit{(e,f)}, where we find that the standard Kubo needs $M_{K}=10000$ as opposed to the improved Kubo formulation requiring only $M_{K}=4000$ to correctly reach the numerically-exact MW result, the improved Kubo provides the correct solution approximately $\sim 15.5$ times faster than the standard Kubo.\\ 
\noindent{Note} that the Hubbard satellite feature at $\omega\sim U$ seen in the MW result is \textit{not} captured by either Kubo approach implemented via FDM-NRG, regardless the number of kept states and the discretization values. The MW and Kubo formulations should have the same resolution and the reasoning behind this discrepancy can be explain as follows. The MW formula is computed using dot Green's function solely whereas the both Kubo approaches are calculated via current-current correlator. The latter formulation is clearly much more sensible to the Wilson chain representation of the bath. As consequence, the lack of resolution in the Kubo conductance curve at high energies appears to be due to the logarithmic discretization of the bath Hamiltonian in the NRG protocol, see the first step in Fig.\ref{F2:NRG}. All other features are well-described; and the improved-Kubo approach confers a significant accuracy and efficiency gain.\\
Comparing Figs.~\ref{F4/KvsimpK}\textit{(a,c,e)} with Figs.~\ref{F4/KvsimpK}\textit{(b,d,f)}, it seems that the crossover from local moment to strong coupling behaviour found at $T_{K}/D\sim10^{-6}$ is equally captured from both the standard and the improved Kubo formulation. Hence, we may deduce that the Fermi liquid scale is reached by both formulations. The Fig.~\ref{F4/KvsimpK}\textit{(g)} in log-,log- scale shows the low-frequency behaviour, comparing Kubo and improved Kubo with a Fermi liquid fit in dashed green line. The fit indicates the asymptotic behaviour of the conductance is like $\mathcal{G}^C(\omega=0,T\rightarrow0)\sim\omega^2$ and the unitary limit is reached $\mathcal{G}^C(\omega,T=0) \rightarrow\mathcal{G}_0$. 
The improved Kubo formula in blue curve is found to capture the expected asymptotic form very accurately. Indeed, for $\Lambda=2$ and $M_K=10000$, the improved Kubo method satisfies the exact \textit{dc}-limit result $\mathcal{G}^C(0,0)/\mathcal{G}_0=2$ to within $0.01\%$. On the contrary, the standard Kubo formulation in orange curve for the same NRG parameters fails exactly at the crossover between trivial and Kondo phase at $\omega\simeq T_{K}$. In conclusion, only the improved Kubo formula correctly detects the Fermi liquid scale to high accuracy.\\
Although in this benchmark comparison, accurate results can be alternatively obtained at low computational cost from the MW formula using the equivalent single-channel model, we emphasize that for general non-PC systems this is not possible; and genuine two-channel NRG calculations are notoriously demanding.

\subsection{Failure of heat transport Kubo formula in NRG}
\begin{figure}[H]
	\centering
	\includegraphics[width=0.70\linewidth]{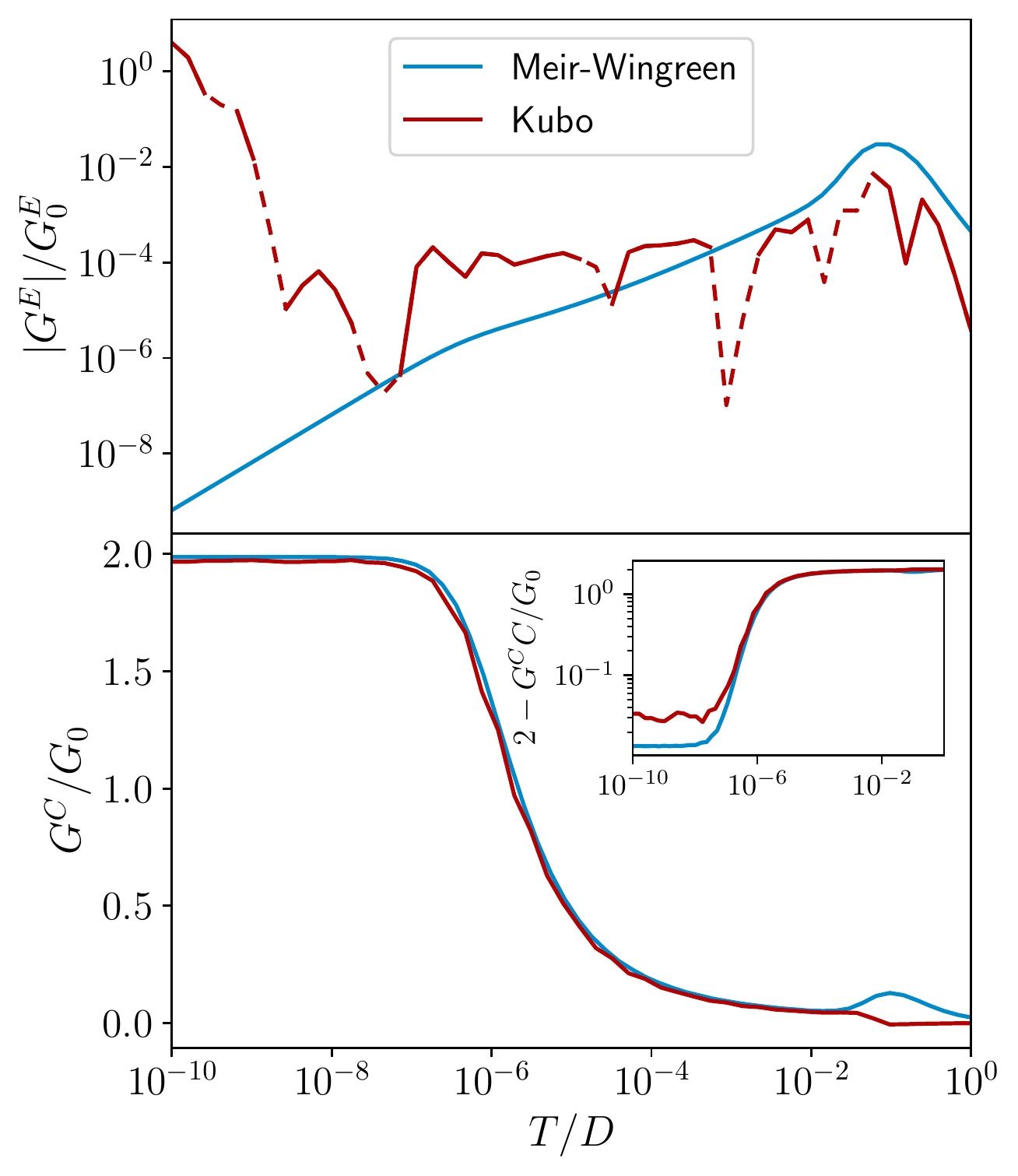}
	\caption[Heat vs electric transport using Kubo formula in NRG]{\textit{(top), log-scale} Heat conductance indicated as $|\mathcal{G}^{E}(T)|$ vs $T$ for the two-lead AIM in unit of quantised heat conductance indicated as $\mathcal{G}^{E}_0 =1/h$, calculated in NRG via the Kubo formula (Eq.~\ref{eq:defKuboLRhea}; red line), compared with the MW result (Eq.~\ref{eq:4MW_HeatPC}; blue line). Solid lines for $\mathcal{G}^{E}>0$, dashed for $\mathcal{G}^{E}<0$. Parameters used: $U=0.4D$, $\epsilon_d=-U_d/2$, $V_s=V_d=0.07D$, $\Lambda=3$,  $M_K=11000$ (two-channel Kubo) or $4000$ (single-channel MW). \textit{(bottom, main panel), linear-scale} Same as panel \textit{(top)} except for \textit{dc}-charge conductance $\mathcal{G}^C(T)$, obtained with NRG via the standard Kubo formula (Eq.~\ref{eq:defKuboLRel}; red line) or MW (Eq.~\ref{eq:4MW_HeatPC}; blue line). \textit{(bottom, inset), log-scale} The discrepancy between the two methods is much less than $\lll 10^{-1}$ order on the contrary to $\mathcal{G}^{E}(T)$ curves separated by $\sim 10^{8}$ orders.}\label{F4/heatkubo}
\end{figure}

In the previous section we presented a systematic numerical test on the better performance of the improved Kubo over the standard Kubo for electrical transport. In this section we shift to a short numerical study of the heat current. Although the Kubo formula for electrical transport as implemented in NRG can yield highly accurate results as we see in Fig.\ref{F4/KvsimpK}, the same is unfortunately \textit{not} true for heat transport. Perhaps surprisingly, we find that the standard Kubo formula for heat transport $\mathcal{G}^{E}(\omega,T)$ defined in Eq.~\ref{eq:defKuboLRhea}, does not yield even qualitatively correct results when implemented in NRG via Kubo formula. \\
Before presenting the numerical results, we pin down the discussion on the equations used in the NRG protocol. In FDM-NRG, the required current-current retarded correlation function $K^{R}_{Q,Q^{\prime}}(\omega)$ for linear response model in Eq.\ref{eq:defKuboLR} but now defined for energy current in Eq.\ref{eq:defEnCurr} takes the following form, namely
\begin{equation}\label{eq:4Kuboheat_corr}
	\begin{aligned}
(i) ~ K^{R}_{H_{leads}^s,H_{leads}^d}(\omega,T) &= 	
\int^{+\infty}_{-\infty} dt^{\prime} \Big(-i\theta(t^{\prime}) \langle [\dot{H}_{ leads}^s(t^{\prime}),\dot{H}_{leads}^d]\rangle_{T} \Big) e^{i \omega t^{\prime}} ~,\\
&\doteq \int^{+\infty}_{-\infty} dt^{\prime} K^{R}_{H_{leads}^s,H_{leads}^d}(t^{\prime})  e^{i \omega t^{\prime}}~, \\
&\hspace*{-4.7cm}(ii)~ K^{R}_{H_{leads}^s,H_{leads}^d}(t^{\prime})= \langle\langle \dot{H}_{ leads}^s(t^{\prime}),\dot{H}_{leads}^d \rangle\rangle =
		-V_sV_dt_0^2 \sum_{\sigma,\sigma^{\prime}}\langle\langle  \hat{O}_{s\sigma} ; \hat{O}_{d\sigma^{\prime}} \rangle\rangle ~,\\
&{\rm with}~~~\hat{O}_{\alpha\nu} = d^{\dagger}_{\alpha\nu}f_{\alpha 1 \nu} - f^{\dagger}_{\alpha 1 \nu} d_{\alpha \nu} \;,
	\end{aligned}
\end{equation}
where the current-current correlator in Eq.\ref{eq:defCurrCurrCor} is given $(i)$ in frequency domain and $(ii)$ in time domain. In the expression $(ii)$, $f_{\alpha 1 \nu}$ is the $n=1$ Wilson orbital defined with $c^{\dagger}_{n\sigma},c_{n\sigma}$ in Eq.~\ref{eq:2H_WC} - here we use a different notation to remark the difference among operators on Wilson chain and bare model. Although the $n=0$ Wilson orbital $f_{\alpha 0 \nu}$ may be interpreted as a discretized version of the \textit{physical} local lead orbital $c_{\alpha \sigma}$, the $n=1$ Wilson orbital $f_{\alpha 1 \nu}$ has no direct physical meaning in the bare model.\\
\noindent{In} Fig.\ref{F4/heatkubo} we plot $\mathcal{G}^{E}(\omega,T),\mathcal{G}^{C}(\omega,T)$ curves in standard Kubo formulation and numerically-exact MW approach. In order to highlight their level of agreement but to show a meaningful temperature range, we display \textit{(top)} $\mathcal{G}^{E}(\omega,T)$ in logarithm scale whereas \textit{(bottom, main panel)} $\mathcal{G}^{C}(\omega,T)$ in linear scale and \textit{(bottom, inset)} in logarithmic scale to validate the comparison.\\
We use Eq.~\ref{eq:4Kuboheat_corr} to calculate $\mathcal{G}^{E}(T)$ via Kubo formula for heat conductance Eq.~\ref{eq:defKuboLRhea} for the two-lead Anderson impurity model using NRG, and present numerical results in Fig.~\ref{F4/heatkubo}\textit{(top)} as the red line. In particular, with $\mathcal{G}^{E}_{0}=1/h$ in plot indicates the unit of quantised heat conductance. For comparison, we show the Meir-Wingreen result as the blue line for the effective single-channel system, see cartoon in Fig.\ref{F3:PC} and related discuss in Sec.\ref{sec:PC}. In this case, the linear response conductance is calculated within NRG using the PC version of the MW current formula in Eq.\ref{eq:MW->PC} but now for the heat current that is:
 \begin{equation}\label{eq:4MW_HeatPC}
 	\mathcal{G}^{E}(T)= \frac{1}{hT} \int d\omega~ \omega^2 \left(-\frac{\partial f^{Eq}(\omega)}{\partial \omega} \right)\widetilde{\mathbb{T}}_{PC}(\omega,T) ~,
 \end{equation}
where we use the generalized transfer matrix under PC in Eq.\ref{eq:GenTransfMatPC}.	The Eq.\ref{eq:4MW_HeatPC} should be regarded as the numerically-exact result. We see that $\mathcal{G}^{E}(T)$ from the Kubo formula fails to capture the correct physics, with apparently noisy data, erratic sign-changes, and diverging behaviour at low-temperature rather than vanishing as $\sim T$. At the lowest temperature scale, the discrepancy between the two curves covers several orders of magnitude. However, other physical quantities computed in the same NRG run, such as the electrical conductance $\mathcal{G}^{C}(T)$ obtained by the standard Kubo formula for electrical conductance in Eq.~\ref{eq:defKuboLRel}, correctly recover the expected results, see Fig.~\ref{F4/heatkubo}\textit{(bottom, main panel)}. As before, we observe the different resolution between the two methods at high-energy $\omega \sim U$ as consequence of their implementation in NRG with Wilson chain representation of the bath. In Fig.~\ref{F4/heatkubo}\textit{(bottom, inset)}, the logarithmic scale gives evidence for the almost similar behaviour of both curves at the lowest temperature scale.\\ 
The high-quality two-channel NRG calculations here are fully converged. We observed that Fig.~\ref{F4/heatkubo}\textit{(top)} is not improved by increasing the number of kept states $M_K$. We have also implemented an improved version of the heat Kubo formula derived in Eq.\ref{eq:4defKuboLRhea}, but the underlying failure of NRG demonstrated above is not resolved by this. \\
\noindent{We} note that the heat Kubo formula for the exactly-solvable $U=0$ resonant level model coupled to Wilson chains \textit{also} fails to capture the correct behaviour, including the low-$T$ asymptote $\mathcal{G}^{E}(T)\sim T$. This indicates that the problem lies with the representation of the lead Hamiltonian as a discretized Wilson chain, rather than with the iterative diagonalization procedure in NRG. We therefore believe the failure of NRG in this context to be because Wilson chains are not true thermal reservoirs \cite{esposito2015nature}, as argued in \cite{Rosch_WCReservoir2012} - arbitrary amounts of energy $\Delta E$ cannot be dissipated by Wilson chains without changing their temperature. We hypothesize that this issue may afflict thermal transport calculations using the Kubo formula for \emph{any} discretized system.\\
We emphasize that the above breakdown is not a problem with the heat Kubo formula itself, but rather a failure of NRG to capture the proper behaviour of the required current-current correlation function defined on the discretized Wilson chain, Eq.~\ref{eq:4Kuboheat_corr}. Note that for systems in the PC geometry, the MW formula Eq.~\ref{eq:4MW_HeatPC} can be used instead. This relates the heat conductance in linear response to nanostructure equilibrium retarded Green's functions, which are very accurately calculated within NRG \cite{Weichselbaum2007}. This approach within NRG yields the correct behaviour of $\mathcal{G}^{E}(T)$, as demonstrated in ~\cite{Costi_ThermoelectricQD2010}. The problem of how to obtain $\mathcal{G}^{E}(T)$ via NRG for non-PC systems therefore remains an open question.

\section{Alternative formulation of the Oguri formula}\label{sec:AlternativeOguri}
As discussed in Sec.\ref{sec:Oguri}, according to the argument developed by Oguri, beyond the Fermi liquid regime the Landauer-type formula cannot be derived as the limiting case of the Kubo formula for the electrical conductance for a general, non-proportionate coupling, interacting model. In case the Fermi liquid conditions are not satisfied, the perturbative correction on the current-current correlators is now much more complicated due to the intrinsically non zero, finite component of the interacting self-energy. \\
In this section, we take insight from the recursive relation applied on the isolated lead Green's function at the interface with the central region \cite{Oguri_QuasiParticlePRB2001} to elaborate an alternative derivation of the Oguri argument by enlarging the definition of central region. The advantage of applying the recursion on the redefined impurity region is that the central region then becomes merely a black-box of arbitrary complexity and all the relevant physical quantities required in the electrical conductance equation are formulated only in terms of lead scattering. We cast the transport process into scattering problem with $\mathrm{S}$-matrix defined on the enlarged impurity region. The scattering description of the problem allows a simpler analytical machinery to evaluate electrical conductance, and can be described in terms of $\mathrm{T}$-matrix. \\
Using the \textit{extended impurity model} framed by scattering theory, we can address two distinct systems. The first one, noninteracting impurity models at zero temperature and energy scales. The second one, interacting impurity model with renormalized energies at $T=0$ regime within Fermi liquid description. In both cases, we derive an alternative Oguri formula.\\ 
We proceed now with presenting the analytical calculation to redesign the impurity into an enlarged region such that alternative Oguri expressions can be obtained.

\subsection*{Analytical derivation of the alternative Oguri argument}
\begin{figure}[H]
\centering
\includegraphics[width=0.9\linewidth]{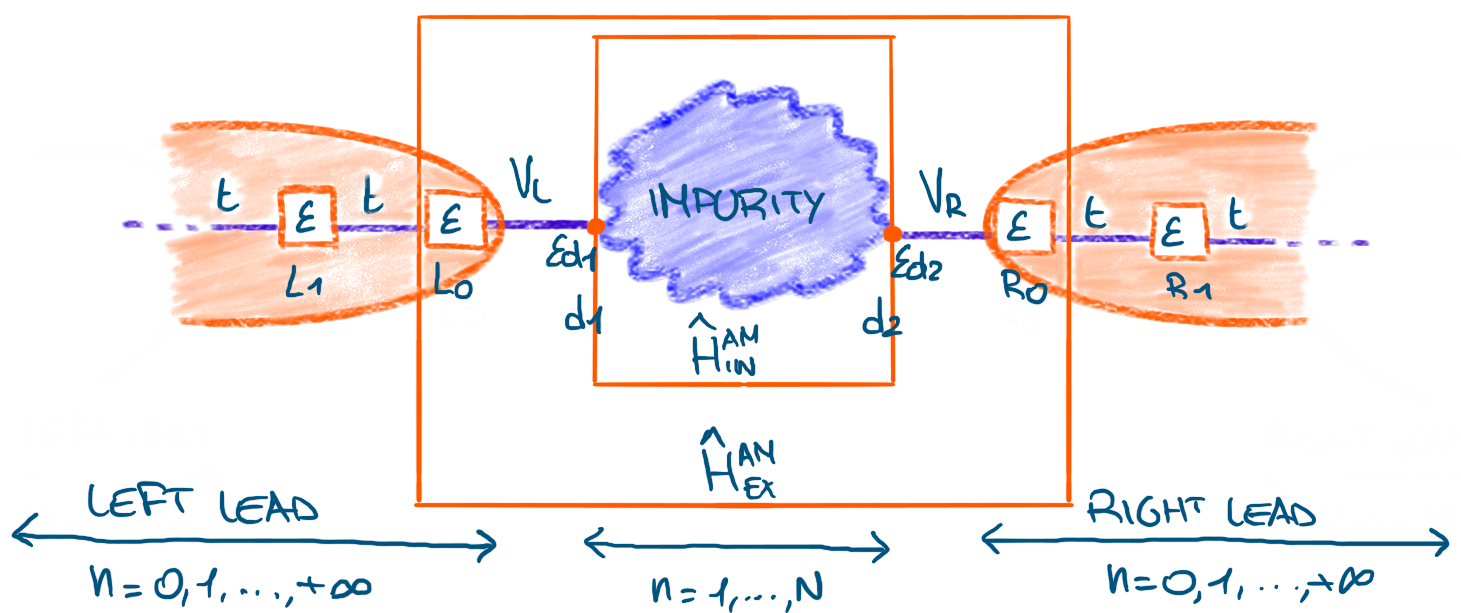}
\caption[Schematic of general Anderson impurity model with extended impurity region]{ The schematic shows a general Anderson impurity model with an arbitrary central region. Rather than connection between lattice sites as in  Fig.\ref{F3:Oguri}, here we tunnel electrons between orbitals. The leads are symmetrically labelled according to $\alpha n$ notation, $n=0,\dots,+\infty$ and the central region is independently labelled for $n=1,\dots,N$. In this schematic, we consider $N=2$. We indicate $\hat{H}^{AM}_{in}$ the initial and  $\hat{H}^{AM}_{ex}$ the extended impurity model.}\label{F4/AlternativeOguri}
\end{figure}

We consider a general Anderson impurity model at finite temperature regime. We treat the leads as structured and equivalent, see definitions in Eq.\ref{eq:4defEqIneq}, and the impurity as general interacting nanostructure. We generalised the one dimensional lattice representation in terms of sites used by Oguri in Fig.\ref{F3:Oguri} by considering the lead as noninteracting thermal equilibrium reservoir whose \textit{outer orbital} hybridizes with the \textit{frontier orbital} of the nanostructure, see Fig.\ref{F4/AlternativeOguri}. Thus, we transform the one dimensional real space representation as depicted in the original Oguri derivation into a general modelling with coupling occurring among external orbitals of the different system elements.\\
As presented in the introduction in the section, we split now our discussion into two parts. We start by focusing in initial noninteracting model at vanishing $\omega,T$, then we consider an initial interacting model where energy renormalization is required.
\noindent{We} start with the noninteracting impurity model and we apply again the real space representation adopted in Fig.\ref{F3:Oguri} but now for system of general dimension. As shown in the schematic, see Fig.\ref{F4/AlternativeOguri}, each label indicates a site in case of one dimensional lattice representation or an orbital in case of system with arbitrary dimension. \\
Without loss of generality, we consider the nanostructure characterised by two frontier orbitals $N=2$ hybridising with the two adjacent leads, such that the model Hamiltonian reads
\begin{equation}
\begin{aligned}
&\hat{H}^{AM}_{in} =\hat{H}_{leads}+\hat{H}_{imp-in}+\hat{H}_{hyb-in} ~,\\ 
&= \sum_{\sigma} \Bigg(  \sum_{n=0}^{+\infty}
\big( t_{Ln}c^{\dagger}_{Ln\sigma}c_{Ln+1\sigma} + t^{\star}_{Ln}c^{\dagger}_{Ln+1\sigma}c_{Ln\sigma} + \epsilon_{Ln}\hat{n}_{Ln\sigma} \big) +\\
&\hspace*{2cm} +  \sum_{n=0}^{+\infty}
\big( t_{Rn}c^{\dagger}_{Rn\sigma}c_{Rn+1\sigma} + t^{\star}_{Rn}c^{\dagger}_{Rn+1\sigma}c_{Rn\sigma} + \epsilon_{Rn}\hat{n}_{Rn\sigma} \big) +\\
&  + \sum_{n=1}^{2} \big(  \epsilon_{dn}\hat{n}_{dn\sigma} \big)
+t_{d}d^{\dagger}_{1\sigma}d_{2\sigma} + t^{\star}_{d}d^{\dagger}_{2\sigma}d_{1\sigma} 
+\big( V_{L} c^{\dagger}_{L 0\sigma}d_{1\sigma} +
V^{\star}_{L}d^{\dagger}_{1\sigma}c_{L 0\sigma} +
 V_{R} c^{\dagger}_{R 0\sigma}d_{2\sigma} +V^{\star}_{R}d^{\dagger}_{2\sigma}c_{R0\sigma} \big) \!\! \Bigg) ~,
\end{aligned}
\end{equation}
where the hybridization term is identified by the couplings $V_{\alpha}$ between the frontier orbitals on the central region and the outer orbital from the leads.\\
Considering the model specifics, it is straightforward to compute the Fisher-Lee formula in Eq.\ref{eq:FisherLee} using the generalized transfer matrix in Eq.\ref{eq:GenTransfMat} for the model Hamiltonian $\hat{H}^{AM}_{in}$, namely
\begin{equation}
\mathcal{G}^{dc}_{N=2}(\omega\to 0,T=0) = \mathcal{G}_{0} \sum_{\sigma} 4\Gamma^{L}(0)\Gamma^{R}(0)|G_{0,d1d2\sigma}^{R}(0,0)|^{2} ~,
\end{equation}
where the gamma functions are similarly defined as in Sec.\ref{sec:Oguri} as
\begin{equation}
\begin{aligned}
&\Gamma^{L}(0) = -|V_{L}|^{2} \mathit{Im}G^{0,R}_{0,L0L0\sigma}(0) = +\pi |V_{L}|^{2} \rho_{0,L0 \sigma} (0) ~,\\
&\Gamma^{R}(0) = -|V_{R}|^{2} \mathit{Im}G^{0,R}_{0,R0R0\sigma}(0) = +\pi |V_{R}|^{2} \rho_{0,R0 \sigma} (0) ~,
\end{aligned}
\end{equation}
and the noninteracting impurity correlator is identified as $G_{0,d1d2\sigma}^{R}(0)=\langle \langle d_{1\sigma};d^{\dagger}_{2\sigma} \rangle\rangle$ by means of the Green's function definition in Eq.\ref{eq:defG}.\\
We will make some remarks on equation $\mathcal{G}^{dc}(\omega \to 0,T=0)$ before considering the enlarged impurity. Its gamma functions are defined on the outer "zero" lead orbital that is the orbital physically coupled to the frontier orbital in the nanostructure. This is a good feature in case we want to compute the conductance using the numerical renormalization group technique we present in Sec.\ref{sec:RGtheo}. As we know, in the Wilson chain representation only the first site in the leads corresponds to a physical coupling with the impurity in the original model - see Fig.\ref{F2:NRG}. Despite this, the actual issue in the above conductance equation is the correlator of the central region. In case of interacting impurity belonging to systems outside the Fermi liquid regime, this element might bring to infeasible calculations. \\
In order to avoid any explicit dependence from the central region features, we proceed now by \textit{enlarging} it. That means to include the outer zero lead orbital on both sides into the definition of central region, see again Fig.\ref{F4/AlternativeOguri}. The resulting \textbf{extended impurity model Hamiltonian} reads
\begin{equation}
\begin{aligned}
&\hat{H}^{AM}_{ex} = \hat{H}_{leads-ex}+\hat{H}_{imp-ex}+\hat{H}_{hyb-ex} ~,\\
&= \sum_{\sigma} \Bigg(  \sum_{n=1}^{+\infty}
\big( t_{Ln}c^{\dagger}_{Ln\sigma}c_{Ln+1\sigma} + t^{\star}_{Ln}c^{\dagger}_{Ln+1\sigma}c_{Ln\sigma} + \epsilon_{Ln}\hat{n}_{Ln\sigma} \big) +\\ 
&\hspace*{2cm}+  \sum_{n=1}^{+\infty}
\big( t_{Rn}c^{\dagger}_{Rn\sigma}c_{Rn+1\sigma} + t^{\star}_{Rn}c^{\dagger}_{Rn+1\sigma}c_{Rn\sigma} + \epsilon_{Rn}\hat{n}_{Rn\sigma} \big) +\\
&  \hspace*{1cm}+ \hat{H}_{imp-ex} +\big( t_{L0} c^{\dagger}_{L 1\sigma}c_{L 0\sigma} +
t^{\star}_{L0}c^{\dagger}_{L 0\sigma}c_{L 1\sigma} +
 t_{R0} c^{\dagger}_{R1\sigma}c_{R0\sigma} +t^{\star}_{R0}c^{\dagger}_{R0\sigma}c_{R1\sigma} \big) \Bigg) ~,
\end{aligned}
\end{equation} 
where the $\hat{H}_{hyb-ex}$ represents the hybridization term now coupling zero and first orbitals on each lead and the \textit{extended impurity} $\hat{H}_{imp-ex}$ contains the initial impurity element $\hat{H}_{imp-in}$ in $\hat{H}_{in}^{AM}$ plus the two outer orbital on the leads, that is
\begin{equation}
\hat{H}_{imp-ex} = \sum_{\sigma} \left( \sum_{n=1}^{2} \big(  \epsilon_{dn}\hat{n}_{dn\sigma} \big) + \epsilon_{L0}\hat{n}_{L0\sigma}+ \epsilon_{R0}\hat{n}_{R0\sigma}
+t_{d}d^{\dagger}_{1\sigma}d_{2\sigma}  +V_{L}c^{\dagger}_{L0\sigma}d_{1\sigma}
+V_{R}c^{\dagger}_{R0\sigma}d_{2\sigma} + H.c. \right) ~.
\end{equation}
Under such a configuration, the hybridization between the extended impurity - spanning now from $L0$ orbital on the left lead to $R0$ orbital on the right one - and the \textit{new} coupled lead sites - respectively $L1$ and $R1$ - is now mediated by the hopping $t_{\alpha0}$ as we enter in the leads region.\\
We proceed again with computing the Fisher-Lee formula in Eq.\ref{eq:FisherLee} for the model Hamiltonian $\hat{H}^{AM}_{ex}$ and we obtain
\begin{equation}
\begin{aligned}
\mathcal{G}^{dc}_{N=2}(\omega\to 0,T=0) &= \mathcal{G}_{0} \sum_{\sigma} 4\overline{\Gamma}^{L}(0)\overline{\Gamma}^{R}(0)|G_{0,L0R0\sigma}^{R}(0,0)|^{2} ~,\\
& = \mathcal{G}_{0} \sum_{\sigma} 4 \pi |t_{L0}|^{2} \rho_{0,L1\sigma}(0)\pi |t_{R0}|^{2} \rho_{0,R1\sigma}(0) |G_{0,L0R0\sigma}^{R}(0,0)|^{2}~.
\end{aligned}
\end{equation}
This conductance expression for $\hat{H}^{AM}_{ex}$ is different from the one calculated for $\hat{H}^{AM}_{in}$. The $\hat{H}^{AM}_{ex}$  contains no details of the central region, with correlator now given by $G_{0,L0R0\sigma}^{R}(0,0) =\langle \langle c_{L0\sigma};c^{\dagger}_{R0\sigma}\rangle\rangle$. However, the gamma functions now take a different form,
\begin{equation}
\begin{aligned}
&\overline{\Gamma}^{L}(0) = -\sum_{\sigma}|t_{L0}|^{2} \mathit{Im}G^{0,R}_{0,L1L1\sigma}(0) = +\pi \sum_{\sigma}|t_{L0}|^{2} \rho_{0,L1\sigma} (0) ~,\\
&\overline{\Gamma}^{R}(0) = -\sum_{\sigma}|t_{R0}|^{2} \mathit{Im}G^{0,R}_{0,R1R1\sigma}(0) = +\pi \sum_{\sigma}|t_{R0}|^{2} \rho_{0,R1 \sigma} (0) ~,
\end{aligned} 
\end{equation}
as those are evaluated using the first orbital on the leads.\\
Hence, the enlarged impurity model successfully eliminates any impurity detail from $\mathcal{G}^{dc}$. However, the density of states is now defined for the first lead orbital. Hence, we aim now to cast the Fisher-Lee formula for the extended impurity model into $\mathrm{T}$-matrix form and to recover the outer orbital information by using only the leads property. In order to achieve that we need to find expressions for $G^{0,R}_{0,\alpha 0\sigma}$ and the $\rho_{0,\alpha1 \sigma}$.\\
The noninteracting correlator for the central region in the conductance in the model $\hat{H}^{AM}_{ex}$ can be evaluated using the $\mathrm{T}$-matrix equation in Eq.\ref{eq:TmatrixEq}, namely
\begin{equation}
G^{R}_{0,L0R0\sigma}(\omega,T) = \underbrace{G^{0,R}_{0,L0R0\sigma}(\omega,T)}_{0} + G^{0,R}_{0,L0L0\sigma}(\omega,T)\cdot \mathrm{T}_{LR}(\omega,T) \cdot G^{0,R}_{0,R0R0\sigma}(\omega,T) ~,
\end{equation}
where the first element indicates the atomic limit with isolated lead propagator that is identically equal to zero and the summation upon the $\mathrm{T}$-matrix is performed only on the noninteracting impurity degree of freedom - making it \textit{local} object, namely 
\begin{equation}
\mathrm{T}_{LR}(\omega,T) = V_{L}V_{R}^{\star}G_{0,d_{1}d_{2}\sigma}^{0,R}(\omega,T)~,
\end{equation}
where $G_{0,d_{1}d_{2}\sigma}^{0,R}(\omega,T)$ is the noninteracting, uncoupled  enlarged impurity Green's function.\\ 
Then, the isolated lead Green's functions at the interface - as appearing in the definition of the gamma function - are calculated using the identity $|G^{R}_{0}|^{2}= \mathit{Im}G^{R}_{0}/\mathit{Im}\Delta$ and so we obtain
\begin{equation}
\begin{aligned}
& |G^{0,R}_{0,L0L0\sigma}(0,0)|^{2} = \frac{\mathit{Im}G^{0,R}_{0,L0L0\sigma}(0,0)}{\mathit{Im}\Delta_{L1}(0)} \equiv \frac{\rho_{0,L0}(0)}{|t_{L0}|^{2}\rho_{0,L1}(0)}\\
&|G^{0,R}_{0,R0R0\sigma}(0,0)|^{2} = \frac{\mathit{Im}G^{0,R}_{0,R0R0\sigma}(0,0)}{\mathit{Im}\Delta_{R1}(0)} \equiv \frac{\rho_{0,R0}(0)}{|t_{R0}|^{2}\rho_{0,R1}(0)}
\end{aligned} ~,
\end{equation}
where we use Eq.\ref{eq:DoSinfWBL} to obtain the second equality. The usefulness of these expressions is that we can write the density of states for the first orbital on the leads in terms of the outer zero orbital only.\\
Finally, we insert these results into the conductance equation for the extended impurity model. After few steps, we obtain
\begin{equation}\label{eq:4GacFisherLeeEx} \boxed{
\mathcal{G}^{dc}_{N=2}(\omega=0,T=0)= \mathcal{G}_{0} \sum_{\sigma} (2\pi)^{2} \rho_{0,L0}(0)\rho_{0,R0}(0) \cdot |\mathrm{T}_{LR}(0,0)|^{2}} ~, 
\end{equation}
that is the conductance for a noninteracting impurity model at vanishing $\omega,T$ regime. The equation is fully defined in terms of the electronic properties of outer orbitals on the leads - i.e. $L0$, $R0$ - and the frontier orbitals on the nanostructure. Hence, the Eq.\ref{eq:4GacFisherLeeEx} is the \textbf{$\mathrm{T}$-matrix version of the Fisher-Lee formula} in Eq.\ref{eq:FisherLee} and it can be regarded as \textbf{alternative formulation of the Oguri argument} for transport for interacting models. That is we can replace the noninteracting Green's functions with their interacting, provided we can confine attention to $T=0$ and assume Fermi liquid property. \\
A last remark on our findings. Even if the gamma functions in $\hat{H}^{AM}_{in}$ and in  $\hat{H}^{AM}_{ex}$ are \textit{not} the same, the conductance equation calculated in the original and extended model in Eq.\ref{eq:4GacFisherLeeEx} \textit{are} the same. Hence, the operation of enlarging the impurity region allows to condense the physics of complicated systems in the electronic tunnelling occurring between the outer orbitals in the leads and the frontier orbitals in the nanostructure.  And this simplifies the calculation and makes suitable numerical computation of the conductance. It is also straightforward to generalise the Eq.\ref{eq:4GacFisherLeeEx} for an arbitrary multi-orbital nanostructure. \\

\noindent{Thus}, we can express the $T=0$ \textit{dc}-regime conductance from  Eq.\ref{eq:4GacFisherLeeEx} in terms of the $\omega=T=0$ scattering $\mathrm{T}$-matrix. To see this, we now turn interactions on the nanostructure. The strategy of enlarging the central region is applicable in this scenario, provided we stay at $T=0$ regime.\\
We consider the same model Hamiltonian $\hat{H}_{in}^{AM}$ introduced at the beginning of the section with the addition of $\sum_{n=1}^{2}U\hat{n}_{dn\uparrow}\hat{n}_{dn\downarrow}$ term. We now use the Oguri formula in Eq.\ref{eq:Oguri},
\begin{equation}
	\mathcal{G}^{dc}_{N=2}(\omega\rightarrow 0,T=0) = \mathcal{G}_{0} \sum_{\sigma} 4\Gamma^{L}(0)\Gamma^{R}(0)|G_{d1d2\sigma}^{R}(0,0)|^{2} ~,
\end{equation} 
where $\Gamma^{\alpha}(0)$ are defined as in the noninteracting impurity model and the impurity propagator $G_{d1d2\sigma}^{R}(\omega,T)$ is the full interacting one, evaluated at $\omega=T=0$. Hence, we apply again Fermi liquid description in the model.\\
We continue with extended impurity model as introduced for $\hat{H}_{in}^{AM}$ but now in the Fermi liquid picture for renormalized energy - hence, we can apply scattering theory. Following similar steps as we present before, we derive an analogue expression to Eq.\ref{eq:4GacFisherLeeEx}, namely
\begin{equation}\label{eq:4GdcOguriEx} 
	\boxed{
		\mathcal{G}^{dc}_{N=2}(\omega\rightarrow 0,T=0)= \mathcal{G}_{0} \sum_{\sigma} (2\pi)^{2} \rho_{0,L0}(0)\rho_{0,R0}(0) \cdot |\mathrm{T}_{LR}(0,0)|^{2}} ~,
\end{equation}
where $\mathrm{T}_{LR}(\omega,T)$ is now the fully \textit{renormalized} $\mathrm{T}$-matrix. \\
The Eq.\ref{eq:4GdcOguriEx} is the \textbf{alternative formulation of the Oguri argument} by mean of the $\mathrm{T}$-matrix version of the original Oguri formula given in Eq.\ref{eq:Oguri}. It is straightforward to generalize Eq.\ref{eq:4GdcOguriEx} for an arbitrary multi-orbital nanostructure. The advantage of this formulation on the original one it is the study of the transport problem on the extended impurity model using scattering theory. The conductance is fully determined by the local $\mathrm{T}$-matrix and the density of states of the outer lead orbitals directly coupled to the impurity. Hence,  the resulting $\mathcal{G}^{dc}(0,0)$ offers a convenient expression amenable to numerical computations.\\

\noindent{We} conclude this section by summarising our findings. The key idea is the redefinition on the impurity region such that its complicated properties does not affect the conductance calculation.\\ 
For a general noninteracting impurity model with no parameter restriction, we derive an alternative expression of the Fisher-Lee formula conductance in terms of $\mathrm{T}$-matrix, see Eq.\ref{eq:4GacFisherLeeEx}. This can also be considered an alternative Oguri formulation in presence of interactions. At finite $U$ term, but now exploiting the Fermi liquid properties of the system arising at $T=0$, we find an alternative expression of the Oguri formula, see Eq.\ref{eq:4GdcOguriEx}.\\
As last remark on this section, we comment that the idea redesigning the system with an extended impurity is a very useful strategy to reduce the system complexity and derive suitable formulation for numerical evaluation.

\chapter{Effective models: emergent proportionate coupling}
Except for the special cases we detailed in Sec.\ref{sec:PC}, most set-ups do not present a geometrical configuration fulfilling the proportionate coupling (PC) property. That implies the corresponding hybridization parameters do not satisfy Eq.\ref{eq:defPC}. The non-PC condition is quite common in generalized impurity systems, modelling, real nanostructures and quantum devices. It is therefore beneficial to find methods to manipulate the bare system such that the resulting effective model fulfils PC condition - with subsequent advantages in electrical conductance calculations. \\
This chapter aims to understand the low-temperature conductance properties of arbitrary systems under specific voltage and energy regimes, in terms of simple \textit{effective models}. The derived models are characterised by effective parameters and emergent properties arise. Among those, the effective models under appropriate energy constraints present an emergent PC propriety even though the bare model is not in PC. By studying the generic behaviour of such systems, we ultimately relate the conductance directly to the effective model parameters, avoiding the need to solve the effective models each time for every new system considered.\\
The general \textit{two-channel model Hamiltonian} we consider in this work is in real-space representation and it reads as:
\begin{equation}\label{eq:5H}
	\begin{aligned}
&\hat{H}= \hat{H}_{leads} + \hat{H}_{nano}(\epsilon_{d},U,t,J_{Hund},J_{RKKY},V_{gate}+B+\dots) + \hat{H}_{hyb} ~,\\
& \hat{H}_{leads}=\sum_{\alpha=s,d}\sum_{\sigma} \sum_{i=0}^{+\infty} \left( \epsilon_{\alpha i}c^{\dagger}_{\alpha i\sigma}c_{\alpha i\sigma} + t_{\alpha i}c^{\dagger}_{\alpha i\sigma}c_{\alpha i+1\sigma} + t^{\star}_{\alpha i}c^{\dagger}_{\alpha i+1\sigma}c_{\alpha i\sigma} \right) ~,\\
&\hat{H}_{hyb}= \sum_{\alpha=s,d}\sum_{\sigma}\sum_{m=1}^{M} \left(V_{\alpha m}d^{\dagger}_{m\sigma}c_{\alpha 0 \sigma} +  V^{\star}_{\alpha m}c^{\dagger}_{\alpha 0 \sigma}d_{m\sigma}\right) ~,
\end{aligned}
\end{equation}
where we use the one dimensional chain nanowire representation of the leads, such that the nanostructure orbital $m=1,\dots,M$ couples to the end of the nanowire $\alpha=s,d$ source and drain leads with tunnelling matrix elements $V_{\alpha m}$. The nanowire parameters $\left\{t_{\alpha i}\right\}$ and $\left\{ \epsilon_{\alpha i} \right\}$ allows us to encode any free leads density of states $\rho_{0,\alpha \sigma}(\omega)=-\mathit{Im}G^{0}_{bath,\alpha\sigma}(\omega)/\pi$ from Eq.\ref{eq:defDos} with $G^{0}_{bath,\alpha\sigma}(\omega) = \langle\langle c_{\alpha0\sigma};c^{\dagger}_{\alpha0\sigma} \rangle\rangle$ Green's function definition in Eq.\ref{eq:defG}.\\
We discuss further now each term appearing in the Hamiltonian Eq.\ref{eq:5H} which incorporates all the physical parameters of the system. Although we refer to chain site because of the real-space representation, each site has to be understood as an \textit{atomic orbital}. The first term $\hat{H}_{leads}$ indicates, as usual, the noninteracting reservoir at thermal equilibrium that couples with the interacting region, here we refer at it as the nanostructure. The $\hat{H}_{leads}$ here stands for $\alpha=s,d$ leads and it is the real-space representation of one dimensional semi-infinite tight-binding chain for $i$ sites, $\epsilon_{\alpha i}$ single-particle energy and $t_{\alpha i}$ hopping parameter. We remark that any noninteracting systems can be cast into chain form, therefore the specific mapping presented for $\hat{H}_{leads}$ brings no limitation on the leads modelling. The second term introduces $\hat{H}_{nano}$ as a generalized impurity Hamiltonian spanning between $m=1,\dots,M$ sites and whose parameter dependence includes, but is not limited to, the Hund coupling $J_{Hund}$ and RKKY interaction $J_{RKKY}$. In general, we also allow for a magnetic field $B$ acting on the nanostructure. Thus, we do not restrict our derivation to any specific class of nanostructure but we regard only those that do not satisfy the PC equality in Eq.\ref{eq:defPC} at the microscopic level. As previously seen in this thesis, because of the $0^{th}$-site on each lead is the only one directly coupled to the nanostructure - as it is defined in $\hat{H}_{hyb}$ - the hybridization term carries also the $m$-label indicating which site (orbital) on the nanostructure hybridises with the $0^{th}$-site (orbital) on $\alpha$-lead. Thus, this modelling enables direct control of the tunnelling among each nanostructure sites and the leads.\\
In the bare model defined in Eq.\ref{eq:5H}, we refer to  $V_{\alpha m}$ as \textit{bare hybridization} strength and the ratio:
\begin{equation}
	\nu_{m}= \dfrac{V_{sm}}{V_{dm}} ~,
\end{equation}
as \textit{intrinsic nanostructure asymmetry}. If $\nu_{m}\neq \nu ~~ \forall~ m$, the model does not hold the proportionate coupling property. Thus, for each $m^{th}$-degree of freedom on the nanostructure we have the pair $\lbrace V_{sm },V_{dm}\rbrace$ and we can write it in terms of the nanostructure asymmetry, namely:
\begin{equation}
	\lbrace V_{d1 },\nu_{1}V_{d1 } \rbrace~,~ \lbrace  V_{d2 },\nu_{2}V_{d2 } \rbrace~,~ \dots ~,~\lbrace V_{dM },\nu_{M}V_{dM } \rbrace ~,
\end{equation}
where each pair indicates the tunnelling per site through a nanostructure.\\
Both the physics of interest and the related mathematical operations we apply in the model Hamiltonian in Eq.\ref{eq:5H} take place at a specific energy window - which we discuss here in its general terms. First of all, in order to create the Kondo singlet, we need to impose the energy restriction $V_{\alpha m} \ll E_{C},U$ where $V_{\alpha m}$ is the physical hybridization required in electrical conductance calculations,  $E_{C}$ is the charging energy and $U$ is the on-site Coulomb repulsion. Considering $V_{\alpha m} \ll E_{C}$, the initial model can be regarded as an isolated nanostructure detached from leads i.e. $\hat{H}\equiv\hat{H}_{nano}$, whereas because of $V_{\alpha m} \ll U$ the nanostructure shows a high on-site Coulomb energy and we assume it is initially \textit{off-resonance} i.e. its eigenenergy $\epsilon_{d}$ is faraway from the Fermi level. As mentioned in Sec.\ref{sec:Kexp}, the electron scattering across the set-up is a single-electron transmission meaning that only unitary variation of charge $\Delta Q = \pm 1$ and spin $\Delta S = \pm 1/2$ occur. Then, we consider the electrical conductance as function of temperature and voltage. When we study $\mathcal{G}^{C}(T)$, we are generally interested in the region $k_{B}T \ll E_{C}$ where elastic and spin-flip cotunnelling transport occur until $k_{B}T \sim T_{K}$ where the system enters in strong coupling Kondo regime - see sketch in Fig.\ref{F3:QD_T} and related discussion. Moreover, within this temperature scale, the low-energy effective model we derive in this chapter correctly capture the physics.  When we analyse $\mathcal{G}^{C}(V_{bias},V_{gate})$ we regard the system as multi-level and by voltages tuning we explore different nanostructure occupancy regions in the stability diagram - see sketch in Fig.\ref{F3:QD_VV} and related discussion. Under appropriate choice of these parameters the electronic population on the ground-state is chosen to be even, odd or degenerate.  \\
Within these energy considerations, it is appropriate to apply perturbation theory - specifically the Brillouin-Wigner perturbation theory (BWPT) as main method to derive effective models in this chapter \cite{The1DHubbardModel_Korepin}. In order to define it for the various systems of interest, we need to introduce the general definition of basis state in \textit{charge number representation}.\\
We first consider the isolated nanostructure. The interacting nanostructure has $N$ electrons distributed in $m=1,\dots,M$ sites. Each $m^{th}$-site with $\sigma$ spin has occupation number $n_{m\sigma}=\lbrace0,1\rbrace$ because it is a Fermionic state. The \textit{charge basis states} $\ket{\phi}$ are spanning the whole Hilbert space $\mathcal{H}^{Q}$ and are defined as product of states in the occupation basis. According to the system characterization, the basis are labelled by the set of quantum numbers $Q$ charge number and $S^{z}$ \textit{z}-component of total spin operator - here, we include also splitting of degeneracy in energy levels labelled by $Q,S^{z}$ by adding an index $j$:
\begin{equation}
	\ket{\phi^{Q,S^{z}}_{l}} \doteq \ket{Q;\hat{S}^{z};j} ~,
\end{equation} 
for $l=0,1,2,3\dots$ eigenstates of the isolated $Q$-charge nanostructure where $l=0$ refers to the ground-state eigenstate. The quantum numbers satisfy the properties
\begin{equation}\label{eq:5QSzprop}
	\begin{aligned}
		&\text{charge} \quad Q= \sum_{m=1}^{M} \left(n_{m \uparrow} + n_{m \downarrow}\right) \quad,
		&\text{\textit{z}-spin} \quad \hat{S}^{z}= \sum_{m=1}^{M} \left(\dfrac{n_{m \uparrow} - n_{m \downarrow}}{2}\right) ~.
	\end{aligned}
\end{equation} 
The charge basis states satisfy the following properties:
\begin{enumerate}
	\item \textit{orthonormality} in $\mathcal{H}^{Q}$:
	\begin{equation}
	\braket{Q;S^z;j|Q^{\prime};S^{z^{\prime}};j^{\prime}}=\delta_{QQ^{\prime}}\delta_{S^{z}S^{z^{\prime}}}\delta_{jj^{\prime}}~;
  \end{equation}
    \item \textit{product state} in $\mathcal{H}^{Q}$:
	\begin{equation}
	\ket{Q;S^z;j}=(\ket{n_{1 \uparrow}}_{1 \uparrow} \bigotimes \ket{n_{1 \downarrow}}_{1 \downarrow}) 
	\times (\ket{n_{2 \uparrow}}_{2 \uparrow} \bigotimes \ket{n_{2 \downarrow}}_{2 \downarrow})
	\times \dots \times
	(\ket{n_{M \uparrow}}_{M \uparrow} \bigotimes \ket{n_{M \downarrow}}_{M \downarrow}) ~,
	\end{equation}
subject to the constraints on $\left\{ n_{m\sigma} \right\}$ satisfying Eq.\ref{eq:5QSzprop} for given $Q$ and $\hat{S}^{z}$. 
\end{enumerate}
From the basis states we construct the eigenstates $\ket{\Psi^{Q,S^{z}}_{l}}$ as a linear combination of the product state basis:
\begin{equation}\label{eq:5Q-BasisState-spin-Def}
	\begin{aligned}
	&\ket{\Psi^{Q,S^{z}}_{l}} =\sum_{comb}^{^{M}C_{Q}}c_{Q,S^{z},l}(n_{1},n_{2},n_{3}\dots n_{M})\ket{\phi^{Q}_{S^{z};l}}   ~, \\
	&\equiv \sum_{comb}^{^{M}C_{Q}}c_{Q,S^{z},l}(n_{1},n_{2},n_{3} \dots n_{M}) \times \\
	&\times\bigg( (\ket{n_{1 \uparrow}}_{1 \uparrow} \bigotimes \ket{n_{1 \downarrow}}_{1 \downarrow}) 
	\times (\ket{n_{2 \uparrow}}_{2 \uparrow} \bigotimes \ket{n_{2 \downarrow}}_{2 \downarrow})
	\times \dots \times
	(\ket{n_{M \uparrow}}_{M \uparrow} \bigotimes \ket{n_{M \downarrow}}_{M \downarrow}) \bigg)  ~,
	\end{aligned}
\end{equation}
where the multiplying constant $c_{Q,S^{z},l}(\dots)$ contains information on all the physics of the model and it is determined from the diagonal representation of the isolated $\hat{H}_{nano}$ in Eq.\ref{eq:5H}, whereas the multiplied string of states indicates the charge-number occupation of a specific charge state. This construction has to be evaluated for each combination of charges occupying a given state. The sum runs over all combinations $^{M}C_{Q}$ of $N$-particles distributed in $M$ sites consistent with the quantum numbers. The states $\ket{\Psi^{Q,S^{z}}_{l}}$ are eigenstates of $\hat{H}_{nano}$ and satisfy the Schr{\"o}dinger equation
\begin{equation}
 \hat{H}_{nano}\ket{\Psi^{Q,S^{z}}_{l}}=E^{Q,S^{z}}_{l}\ket{\Psi^{Q,S^{z}}_{l}} ~.
\end{equation}
In the BWPT, we project the whole system states on the ground-state. We can define it in charge-basis using Eq.\ref{eq:5Q-BasisState-spin-Def} and we have the ground-state $\ket{\Psi^{Q,S^{z}}_{0}}$  in term of the isolated nanostructure whose model Hamiltonian reduces to $\hat{H}\equiv\hat{H}_{nano}$. Due to single-electron tunnelling in $\hat{H}_{hyb}$, the matrix of $\hat{H}$ exhibits blocks structure with some empty sectors corresponding to forbidden tunnelling process. From the diagonalization of $\hat{H}_{nano}$, we determine $c_{Q,S^{z},l}(\dots)$ coefficients. Once this is completed, we can proceed with the actual projection operation as defined in the BWPT.\\
We introduce now the \textit{projector operator} $\widehat{P}_{0}$ on the charge basis belonging to the ground-state Hilbert subspace, namely:
\begin{equation}\label{eq:5defP}
	\widehat{P}_{0} = \sum_{S^{z}_{0}}\ket{\Psi^{Q_{0},S^{z}_{0}}_{0}}\bra{\Psi^{Q_{0},S^{z}_{0}}_{0}} \otimes \widehat{\mathbb{1}}_{leads} ~,
\end{equation}
where $Q_{0},S^{z}_{0}$ are the ground-state charge and spin projection, respectively. For example, if the ground-state has $Q_{0}=1,S^{z}_{0}=\tfrac{1}{2}$, we sum over the degenerate ground-state with $ S^{z}=\pm\tfrac{1}{2}$ and $Q=1$ charge sector.\\
By applying Eq.\ref{eq:5defP} on the model Hamiltonian Eq.\ref{eq:5H}, we project the whole system on the ground-state manifold.
Only those low-lying states are retained for the effective model. The specific form of the expression in Eq.\ref{eq:5defP} is determined by the type of ground-states - as we detail for each case later. In general, the operators $\widehat{P}_{x}$ satisfy the following properties:
\begin{equation}
	\begin{aligned}
		(i) &\quad \text{completeness} \quad \sum_{x} \widehat{P}_{x} = \mathbb{1} ~,\\
		(ii) &\quad \text{idempotence}  \quad \widehat{P}^{2}_{x} = \widehat{P}_{x} ~,\\
		(iii)&\quad \text{orthonormality} \quad \widehat{P}_{x}\widehat{P}_{y} = \delta_{xy}\widehat{P}_{x} ~.
	\end{aligned}
\end{equation}
The complementary projector operator is defined as $\widehat{Q}_{0} = \mathbb{1} - \widehat{P}_{0}$ and it belongs to the subspace of excited states $\mathcal{H}^{Q}-\mathcal{H}^{Q}_{0}$. The BWPT is a systematic way of incorporating the effect of virtual excitations on the ground-state manifold properties.\\
We have introduced all the elements required in the BWPT calculation and we proceed now with applying it to Eq.\ref{eq:5H}. As we discuss previously, the resulting effective Hamiltonian $\hat{H}_{eff}$ is derived from projecting out excited states.
The $\hat{H}_{eff}$  is a valid description of the system within energy window $k_{B}T \ll E_{C},U$. In general term, the projection operation satisfies the following equation:
\begin{equation}
\widehat{P}_{0} \hat{H} \widehat{P}_{0} \ket{\Psi^{Q_{0} ,S^{z}}_{0}} = E^{Q_{0},S^{z}_{0}}_{0} \widehat{P}_{0} \ket{\Psi^{Q_{0} ,S^{z}}_{0}} \quad\Leftrightarrow \quad \hat{H}_{eff}\ket{\Psi^{Q_{0} ,S^{z}}_{0}} = \epsilon_{0}\ket{\Psi^{Q_{0} ,S^{z}}_{0}} ~,
\end{equation} 
where $\epsilon_{0}$ indicates the ground-state eigenvalue after the projection. Hence, the general form of the effective model reads as:
\begin{equation}\label{eq:5Heff}
	\begin{aligned}
		&\hat{H}_{eff} \equiv 	\widehat{P}_{0} \hat{H} \widehat{P}_{0} \\
		&=\underbrace{\widehat{P}_{0}\hat{H}_{leads}\widehat{P}_{0}+\widehat{P}_{0}\hat{H}_{nano}\widehat{P}_{0}}_{0^{th}~order} + \underbrace{\widehat{P}_{0} \hat{H}_{hyb}\widehat{P}_{0}}_{1^{st}~order} + 
		\underbrace{\widehat{P}_{0}\hat{H}_{hyb}\widehat{Q}_{0} (E^{Q_{0},S^{z}_{0}}_{0} - \widehat{Q}_{0}\hat{H}_{nano}\widehat{Q}_{0})^{-1} \widehat{Q}_{0}\hat{H}_{hyb}\widehat{P}_{0} }_{2^{nd}~ order} +\mathcal{O}(\hat{H}_{hyb})^{4}
	\end{aligned} ~,
\end{equation}
where $\widehat{P}_{0}\hat{H}_{leads}\widehat{P}_{0}=\hat{H}_{leads}$ and $\widehat{P}_{0}\hat{H}_{nano}\widehat{P}_{0}=\epsilon_{0}$. The most important non-trivial information in $\hat{H}_{eff}$ comes from the second order correction in $\hat{H}_{hyb}$ in the Coulomb blockade regime, or as $\widehat{P}_{0} \hat{H}_{hyb}\widehat{P}_{0}$ in the mixed-valence regime, as we discuss further below.\\
The Eq.\ref{eq:5Heff} is obtained from degenerate perturbation theory and it can be understood as \textit{generalized Schrieffer-Wolff} transformation.\\

\noindent{As} we presented in this introduction to the chapter, the core of the BWPT method is the projection operation of the whole basis states to low-lying subspace that is the lowest spin multiplet of the ground-state manifold. We list below the types of ground-states according to their $Q$ charge number and $\hat{S}^{z}$ \textit{z}-component of the total nanostructure spin:
\begin{equation}
	\begin{aligned}
		(i)& \quad \ket{\Psi^{Q ,S^{z}}} = \ket{Q=\text{even}; S^{z}= \text{integer}} ~,\\ 
		(ii)& \quad \ket{\Psi^{Q ,S^{z}}} = \ket{Q=\text{odd}; S^{z}= \text{half-integer}} ~,\\
		(iii)& \quad \ket{\Psi^{Q ,S^{z}}} = \ket{Q,Q+1~\text{degenerate}; S^{z}~\text{degenerate multiplets both integer and half-integer}} ~.
	\end{aligned}
\end{equation}
We find that $(i),(ii)$ arise in systems at Coulomb blockade regime as discussed in Sec.\ref{sec:CBPC} whereas $(iii)$ happens for systems in mixed-valence regime as presented in Sec.\ref{sec:MVPC}.\\
In particular, in the case $(i)$ there is no extra degrees of freedom available for reaching virtual excited states and so there are no spin fluctuations. Hence, in this condition, we could perform perturbation calculation by introducing a trivial potential scattering. In Sec.\ref{sec:CBPC} we study this in the standard case of $S=0$, but others are possible. \\
In system satisfying $(ii)$ we have spin degeneracy. In this case, BWPT is successfully applied and the resulting effective model shows an emergent PC feature. As we discuss in Sec.\ref{sec:CBPC} for a general two-channel model, the actual PC condition appears only under the renormalisation flow of the coupling parameters at low-temperature regime.\\
The last ground-state configuration $(iii)$ allows both spin and charge degeneracy. Considering again a generic two-channel system in mixed-valence regime - the spin multiplet transition is determined by the parity of $Q$ as we discuss in Sec.\ref{sec:MVPC}, we apply BWPT and we obtain directly an effective low-energy model with an emergent PC property.\\
In this chapter we show the energy conditions, the analytical expressions, and the conductance from NRG results for effective models presenting an emergent PC condition starting from the above mentioned nanostructure ground-state manifold classified with respect to the lowest spin multiplet, namely in Coulomb blockade regime and in mixed-valence transition. We also present the effective model with numerical values of the effective parameters for serial two-impurity model in singlet-doublet MV regime and at the end, we briefly discuss the generalization of this machinery to high-spin ground-state multiplet.\\
Most of the results presented in this chapter are published in the paper \cite{transport}.

\section{Emergent proportionate coupling in Coulomb blockade regime} \label{sec:CBPC}
In most realistic systems - with the notable exception of single-dot devices - the orbital complexity and spatial structure of the nanostructure prevents a PC description, see examples of non-PC system presented in Sec.\ref{sec:PC}. However, in certain situations the low-energy effective model for the system may be in PC even though the bare model is not. This is particularly useful for the MW formulation of quantum transport involving only retarded single-particle nanostructure Green's functions. We explore these scenarios below for the case of systems in the Coulomb blockade (CB) regime.\\
In the stability diagram, see sketch in Fig.\ref{F3:QD_VV}, under appropriate voltages tuning, the CB region is located deep inside the so-called Coulomb diamond where the nanostructure has a well-defined number of electrons. In this regime, charge fluctuations are completely suppressed at low temperatures $k_{B} T \ll E_C$ - with $E_C$ the effective nanostructure charging energy. Incoherent transport, involving sequential tunnelling events between the nanostructure and leads, is therefore also suppressed \cite{NazarovCoTun1990}. This is the dominant transport mechanism for nanostructures with a net spin $S=0$. By contrast, Kondo-enhanced spin-flip scattering in spinful nanostructures can give a substantial boost to low-temperature coherent transport, with the unitarity limit of a perfect single-electron transistor attainable experimentally \cite{Kouwenhoven2000Exp}.\\
In both cases, the structure of the low-energy effective model can be obtained perturbatively to second order in $\hat{H}_{hyb}$, by projecting onto the ground manifold of $N$-electron states for the isolated nanostructure, and eliminating virtual excitations to nanostructure states with $N\pm 1$-electron states as described above. \\
We note that although the structure of the effective model for generalized quantum impurity systems can be obtained from the derived equation, see later Eq.~\ref{eq:5HeffCBPer}, the perturbative estimation of its coupling constants is accurate only in the limit of strong nanostructure interactions and weak hybridization. Recently machine learning techniques have been employed to determine the numerical value of the effective model parameters non-perturbatively \cite{rigo2020machine}. In the following we focus on the generic behaviour of the underlying effective models. Ultimately, the accuracy of predictions using these models depends on the accurate determination of the effective parameters.\\
We continue now with the generalities of the BWPT applied to a system under CB regime, then we present two prominent cases for nanostructure ground-states with $S=0$ and $S=\tfrac{1}{2}$ spin states.

\subsection*{Analytical method and physical observable}
In this part we present the general structure for calculating the effective model for systems in CB regime and we introduce the main observable we need to evaluate prior to electrical conductance computation using NRG.\\
We consider the general two-channel model Hamiltonian given in Eq.\ref{eq:5H} in CB regime as identified by the energy window $k_{B} T \ll E_C$. We take the general definition of spinful charge basis given in Eq.\ref{eq:5Q-BasisState-spin-Def} and the projector operator for a general $N$-electron nanostructure state according to the definition in Eq.\ref{eq:5defP}.
By applying $\widehat{P}_{0}$ on the bare Hamiltonian in Eq.\ref{eq:5H}, we project the whole set of states to the ground-state manifold characterised by the \textit{retained} $\ket{\Psi_{0}^{N, S^{z}}}$ states with $E^{N, S^{z}}_{0}$ energy, with $S^{z}=0$ for $S=0$ and $S^{z}=\pm\tfrac{1}{2}$ for $S=\tfrac{1}{2}$ ground-states.\\
According to the energy regime $k_{B} T \ll E_C$, the effective model is determined by the whole expression given by Eq.\ref{eq:5Heff}. The $0^{th}$-order is straightforward to obtain as in Eq.\ref{eq:5P0Leads} i.e. $\widehat{P}_{0}\hat{H}_{leads}\widehat{P}_{0}\equiv \hat{H}_{leads}$ because the projector is defined on the isolated nanostructured states only. A finite contribution arises also from  $\widehat{P}_{0}\hat{H}_{nano}\widehat{P}_{0} = \widehat{P}_{0}E^{N, S^{z}}_{0} \widehat{P}_{0}= \epsilon_{0}$ giving the nanostructure ground-state eigenvalue projected on subspace $\mathcal{H}^{N}_{0}$ - we comment on this result for specific $N$ parity. In the $1^{st}$-order, $\widehat{P}_{0} \hat{H}_{hyb}\widehat{P}_{0}$ is identically equal to zero since the hybridization implies a charge variation between nanostructure and leads. However, because in the CB regime at $k_{B} T \ll E_C$ the  model is nondegenerate in charge, we cannot tunnel between two different charge sectors. This results is indeed the opposite of what we have in the mixed-valence regime, see for comparison Eq.\ref{eq:5P0Hyb}. Hence, the involved  calculation is the $2^{nd}$-order contribution only.\\ 
In conclusion, the complete effective Hamiltonian in CB regime we want to calculate reads:
\begin{equation}\label{eq:5HeffCBPer}
	\begin{aligned}
	\hat{H}_{eff} \equiv \widehat{P}_{0} \hat{H} \widehat{P}_{0} &= \hat{H}_{leads} +\epsilon_{0}\widehat{P}_{0} +\hat{H}_{cotun} +\mathcal{O}(\hat{H}_{hyb})^{4} ~,\\
	&\text{with}\quad \hat{H}_{cotun}=\widehat{P}_{0}\hat{H}_{hyb}\widehat{Q}_{0} (E_{0}^{N,S^{z}} - \widehat{Q}_{0}\hat{H}_{nano}\widehat{Q}_{0})^{-1} \widehat{Q}_{0}\hat{H}_{hyb}\widehat{P}_{0} ~,
	\end{aligned}
\end{equation} 
where $\hat{H}_{cotun}$ indicates elastic cotunnelling higher-order spin scattering events, see sketch in  Fig.\ref{F3:QD_T} and its related discussion. The resolution of Eq.\ref{eq:5HeffCBPer} is given by BWPT degenerate perturbation theory that is in this case a generalization of the Schrieffer-Wolff transformation \cite{hubavc2010brillouin}.\\ 
In absence of other ground-states degeneracy, the effective model in local lead basis for $S=\tfrac{1}{2} $ takes the form of a \textit{generalized two-channel Kondo model}, namely
\begin{equation}
\hat{H}^{K} =  \hat{H}_{leads}+\epsilon_{0}\widehat{P}_{0}+ \sum_{\alpha\beta=s,d} \sum_{\sigma\sigma^{\prime}} \left( W_{\alpha\beta\sigma} c^{\dagger}_{\alpha\sigma}c_{\beta\sigma^{\prime}}\delta_{\sigma\sigma^{\prime}} +
J_{\alpha\beta\sigma\sigma^{\prime}} \Big(c^{\dagger}_{\alpha\sigma} \boldsymbol{\sigma}_{\sigma\sigma^{\prime}} c_{\beta\sigma^{\prime}}\Big) \cdot \hat{\mathbf{S}} \right) ~,
\end{equation}
which is in analogy to Eq.\ref{eq:K} where we have assumed $SU(2)$ spin symmetry that is $B=0$; here we adopt the notation $\alpha=s,d$ for source and drain \textit{leads}. The \textbf{effective model for system in CB regime} reads as
\begin{equation}\label{eq:5HeffCB}
	\boxed{
		\begin{aligned}
&\hat{H}_{eff} =\hat{H}_{leads} +\epsilon_{0}\widehat{P}_{0} + \sum_{\alpha\beta=s,d}\sum_{\sigma} W_{\alpha\beta} c^{\dagger}_{\alpha\sigma}c_{\beta\sigma} + \sum_{\alpha\beta=s,d} J_{\alpha\beta} \hat{\mathbf{s}}_{\alpha\beta} \cdot \hat{\mathbf{S}} ~, \\
&\text{with}\quad \hat{\mathbf{s}}_{\alpha\beta}=\sum_{\sigma\sigma^{\prime}} c^{\dagger}_{\alpha0\sigma}\boldsymbol{\sigma}_{\sigma\sigma^{\prime}}c_{\beta0\sigma^{\prime}}
	\end{aligned} } ~,
\end{equation} 
where $\hat{\mathbf{s}}_{\alpha\beta}$ is the spin-$1/2$ operator for the leads and $\hat{\mathbf{S}}$ is the operator for the overall nanostructure spin-$1/2$ ground-state degree of freedom - see more details in Eq.\ref{eq:K}.

\noindent{The} effective model parameters $J_{\alpha\beta}$ exchange coupling and $W_{\alpha\beta\sigma}$ potential scattering can be straightforwardly calculated from matrix elements of the isolated nanostructure states, as described in \cite{Mitchell2017KondoMolecule}. Hermiticity requires that $W_{\alpha\beta}=(W_{\beta\alpha})^{\star}$ and $J_{\alpha\beta}=(J_{\beta\alpha})^{\star}$. \\
The model in Eq.\ref{eq:5HeffCB} is still with no proportionate coupling property: both direct tunnelling lead-nanostructure and mediated tunnelling lead-nanostructure-lead are incorporated in the model Hamiltonian. Furthermore, the \textit{sd-}basis are the physical basis used to compute the conductance. Therefore, once the manipulation on the model Hamiltonian is completed, we need to transform back the Hamiltonian into the \textit{sd-}basis. Since the finite conductance is given by electrons flowing between leads, as we discussed in Sec.\ref{sec:Kexp}, the elements employed in conductance calculation are the inter-channel coupling components $J_{sd}$, $W_{sd}$. \\
Because the results of this chapter are published in the the paper \cite{transport}, we adopt here the notation used there which in places differs from that used in previously in this thesis.\\
We introduce the full Green's function connecting local orbitals in source and drain leads in the presence of $ W_{\alpha\beta}$, namely $G^{R}_{sd,\sigma}(\omega,T)= \langle\langle c_{s\sigma};c^{\dagger}_{d\sigma}\rangle\rangle$. This correlator can be expressed in terms of the free leads Green's functions of the isolated $\hat{H}_{leads}$ written as $G^{0,R}_{\alpha\beta,\sigma}(\omega)=\sum_{\mathbf{k}}\delta_{\alpha\beta}/(\omega+i\eta -\epsilon_{\mathbf{k}})\equiv \delta_{\alpha\beta}/(\omega+i\eta - \Delta(\omega))$ such that $-\mathit{Im}G^{0,R}_{bath,\alpha}(0)/\pi=\rho_{0,\alpha}$ is the free $\alpha$-lead density of states at the Fermi level. We introduce the definition \textit{effective hybridization} for the free $\alpha$-lead, namely
\begin{equation}
\Gamma_{eff}^{\alpha}= -\mathit{Im}\Delta(\omega=0)\equiv -\frac{1}{\pi \rho_{0,\alpha}} ~,
\end{equation}
where we use the definition for equivalent metallic leads with $\rho_{0,\alpha}=\rho_{0,\beta}\equiv\rho_{0}$ constant density of states in wide-flat conduction band limit defined in Eq.\ref{eq:DoSinfWBL}. In the context of nanostructure with multi-orbital $i$, the levelwidth expression previously presented in Eq.\ref{eq:DoSinfWBL} indicates the hybridization between the nanostructure and the $0^{th}$-orbital on the lead. On the contrary, $\Gamma_{eff}^{\alpha}$ refers to the coupling occurring between first and second orbital on $\alpha$-lead.\\
The dynamics of the effective model, discussed further below, are characterized by the spectrum of the scattering $\mathrm{T}$-matrix defined in Eq.\ref{eq:TmatSpe} for the case of equivalent metallic leads. For the effective model derived in Eq.\ref{eq:5HeffCB}, the $\mathrm{T}$-matrix expression can be expressed as
\begin{equation}
T_{\alpha\beta,\sigma}(\omega,T) = W_{\alpha\beta} + \langle \langle \hat{a}_{\alpha\sigma}^{\phantom{\dagger}} ; \hat{a}_{\beta\sigma}^{\dagger}\rangle\rangle \quad \text{where}\quad \hat{a}_{\alpha\uparrow} = \sum_{\gamma=s,d} \left (W_{\alpha\gamma} c_{\gamma\uparrow} + \tfrac{1}{2} J_{\alpha\gamma}(c_{\gamma\uparrow}\hat{S}^z + c_{\gamma\downarrow}\hat{S}^{-}) \right ) ~,
\end{equation}
and similarly for $\hat{a}_{\alpha\downarrow}$.\\
The retarded correlator $\langle \langle \hat{a}_{\alpha\sigma}^{\phantom{\dagger}} ; \hat{a}_{\beta\sigma}^{\dagger}\rangle\rangle $ can be computed directly in NRG. However, an improved version of the $\mathrm{T}$-matrix can be obtained in a similar fashion to the self-energy method for the retarded nanostructure Green's functions, see Eq.\ref{eq:4F/G}. We define a $2\times2$ matrix Dyson equation for the full impurity coupled lead Green's functions, $[\mathbb{G}_{\sigma}(\omega,T)]^{-1}=[\mathbb{G}_{\sigma}^{0}(\omega)]^{-1} - \mathbb{\Sigma}_{\sigma}(\omega,T)$, where the self-energy matrix $\mathbb{\Sigma}_{\sigma}(\omega,T)$ incorporates all the effects of the nanostructure. Rearranging the $\mathrm{T}$-matrix equation given Eq.\ref{eq:TmatrixEq} we have 
\begin{eqnarray}\label{eq:5SE_tm}
\mathbb{\Sigma}_{\sigma}(\omega,T) = [\mathbb{G}_{\sigma}(\omega,T)]^{-1}\mathbb{G}_{\sigma}^0(\omega) \mathbb{T}_{\sigma}(\omega,T) \;.
\end{eqnarray}
In practice we therefore compute both $\mathbb{G}_{\sigma}(\omega,T)$ and $\mathbb{T}_{\sigma}(\omega,T)$ in NRG to obtain $\mathbb{\Sigma}_{\sigma}(\omega,T)$ by Eq.\ref{eq:5SE_tm}. This gives an improved estimation of $\mathbb{G}_{\sigma}(\omega,T)$ from the Dyson equation, and hence an improved version of $\mathbb{T}_{\sigma}(\omega,T)$ through the $\mathrm{T}$-matrix equation. \\
In terms of quantum transport, the Kubo formula can be used to obtain both \textit{ac-} and \textit{dc-} linear response electrical conductance for the effective model, see Eqs.\ref{eq:defSusceptCondAC}, \ref{eq:defSusceptCondDC}, \ref{eq:defKuboLRel}. The retarded current-current correlator in Eq.\ref{eq:defCurrCurrCor} after Fourier transform in Eq.\ref{eq:defKuboLR} for the effective model in Eq.\ref{eq:5HeffCB} can be expressed in terms of the composite operators $\hat{\Omega}$ defined as follows
\begin{equation}\label{eq:5KuboKondo}
	\begin{aligned}
	&K^{R}_{N_{s},N_{d}}(\omega,T) = \langle\langle \hat{\Omega};\hat{\Omega} \rangle\rangle ~,\\
	&\text{where}\quad 	\hat{\Omega}=-i\left ( J_{sd}\hat{\mathbf{S}} \cdot \hat{\mathbf{s}}_{sd}+\sum_{\sigma}  W_{sd}^{\phantom{\dagger}}  c_{s\sigma}^{\dagger} c_{d\sigma}^{\phantom{\dagger}} 
	\right ) - {\rm H.c.} ~.
	\end{aligned}
\end{equation}
The form of the improved Kubo formula derived in Sec.\ref{sec:ImprovedKubo} is model-independent and so Eq.\ref{eq:4defKuboLRel} remains the same. The corresponding Kubo formula for heat conductance Eq.\ref{eq:4defKuboLRhea}  follows similarly.\\
To gain further insight into the expected behaviour and to make progress analytically, we consider various limiting cases below described by the effective model in Eq.\ref{eq:5HeffCB}.

\subsection{Spin $S=0$}
For isolated nanostructures with an even number of electrons $N$, the ground state is often (but not always) a unique spin singlet state with total $S=0$. Hence, in this part we consider the nanostructure ground-state manifold $\ket{\Psi^{Q=N,S^z}_{0}} = \ket{Q=\text{even}; S^z= \text{integer};j=0}$ belonging to $\mathcal{H}^{Q}_{0}$ Hilbert subspace only. The even $N$-electron charge sector defined on such manifold satisfies the equation:
\begin{equation}
	\hat{H} \ket{\Psi_{0}^{N,0}} = \hat{H} \ket{N;S^{z}=0;j=0}=E^{N,0}_{0}\ket{N;S^{z}=0;j=0} ~.
\end{equation}
Using Eq.\ref{eq:5defP}, the projector operator for the even charge sector reads as
\begin{eqnarray}
	\widehat{P}_{0} \doteq \ket{N;0;0}\bra{N;0;0} ~.
\end{eqnarray}
The application of this projector in Eq.~\ref{eq:5HeffCBPer} yields to an \textbf{effective model} in the form of a renormalized tunnel junction, namely
\begin{equation}\label{eq:5Heff_S0}
	\boxed{
	\hat{H}_{eff}=\hat{H}_{leads} +\epsilon_{0}\widehat{P}_{0}  + \underbrace{\sum_{\alpha,\beta}\sum_{\sigma}  W_{\alpha\beta} c_{\alpha\sigma}^{\dagger} c_{\beta\sigma}^{\phantom{\dagger}} }_{\hat{H}_{cotun}} }\;,
\end{equation}
where $\alpha,\beta \in s,d$ and $W_{\alpha\beta}=(W_{\beta\alpha})^{\star}$ are effective potential scattering parameters obtained after integrating out the nanostructure from the perturbation theory calculation. Note however that in the presence of a magnetic field on the nanostructure $B\neq0$, the excited states $|N\pm 1;S^{z} ;k\rangle$  and excitation energies $(E^{N,S^{z}}_{0}-E^{N\pm1,S^{z}}_{0})$ depend on spin $\sigma$, which in general endow $W_{\alpha\beta\sigma}$ with a spin dependence, which we retain here for generality. \\
Eq.~\ref{eq:5Heff_S0} is, in some sense, trivial in PC. Since the effective model is noninteracting, we may use the conductance in \textit{dc-}regime given by the alternative Oguri formulation in Eq.\ref{eq:4GdcOguriEx}, here in the present context written as,
\begin{equation}
	\mathcal{G}^{dc}(\omega=0,T=0) = \frac{e^2}{h}4 \Gamma_{eff}^{s}\Gamma_{eff}^{d}\sum_{\sigma} |G_{sd,\sigma}^{R}(\omega=0,T=0)|^{2} ~,
\end{equation}
where the required full Green's function can be obtained from standard equations of motion techniques as,
\begin{equation}
	G_{sd,\sigma}^{R}(\omega)=\frac{W_{sd\sigma}}{[\omega^+-W_{ss\sigma}-\Delta(\omega)][\omega^+-W_{dd\sigma}-\Delta(\omega)]-|W_{sd\sigma}|^2} ~,
\end{equation}
where $\omega^+ = \omega+i\eta$ for retarded correlators in the limit $\eta\rightarrow0^{+}$.\\
In the general case, one obtains the \textbf{electrical conductance} for $S=0$ system reads,
\begin{equation}\label{eq:5cond_qpc}
	\boxed{
	\mathcal{G}^{dc}(0,0)=\left ( \frac{e^2}{h}\right )\sum_{\sigma} \frac{4 |\widetilde{W}_{sd\sigma}|^2}{(1+|\widetilde{W}_{sd\sigma}|^2-\widetilde{W}_{ss\sigma}\widetilde{W}_{dd\sigma})^2+(\widetilde{W}_{ss\sigma}+\widetilde{W}_{dd\sigma})^2} }~,
\end{equation}
where we have defined the dimensionless quantities  $\widetilde{W}_{\alpha\beta\sigma}=\pi \rho_0 W_{\alpha\beta\sigma}$.\\
In the zero-field, particle-hole symmetric case where $W_{ss\sigma}=W_{dd\sigma}=0$ and $W_{sd\sigma}\equiv W_{sd}$, the standard result for the $T=0$ transmission of a tunnel-junction is recovered,
\begin{equation}\label{eq:5Gqpc}
	\boxed{
	\mathcal{G}^{dc}(0,0)=\left ( \frac{2e^2}{h}\right ) \frac{4 |\widetilde{W}_{sd}|^2}{(1+|\widetilde{W}_{sd}|^2)^2} }\;,
\end{equation}
and the maximum conductance $\mathcal{G}^{dc}(0,0)=2e^2/h$ is attained when $|\widetilde{W}_{sd}|^2=1$, which corresponds to intermediate values of $W_{sd}$.\\
The conductance is therefore seen to depend strongly on particle-hole asymmetry, generated and tuned in practice by application of gate voltages. In particular, the maximum conductance in case of finite $W_{ss}$ and/or $W_{dd}$ is always reduced $\mathcal{G}^{dc}(0,0) < 2e^2/h$ with respect to its particle-hole symmetric counterpart.

\subsection{Spin $S=1/2$}
In this section we discuss the case of spin degenerate ground-state within CB regime. The most interesting and subtle case arises when the isolated nanostructure has an odd number of electrons $N$ and hosts a net spin $S=\tfrac{1}{2}$ that means we consider the nanostructure ground-state manifold as $\ket{\Psi^{N,S^z}_{0}} = \ket{Q=\text{odd}; S^{z}= \text{half-integer};j=0}$ belonging to $\mathcal{H}^{Q}_{0}$ Hilbert subspace only. In this case, the odd $N$-electron charge sector satisfies the equation:
\begin{equation}
	\hat{H} \ket{\Psi^{N,S^z}_{0}} = \hat{H} \ket{N;S^{z}=\pm\tfrac{1}{2};j=0}=E^{N,S^z}_{0}\ket{N;S^{z}=\pm\tfrac{1}{2};j=0} ~.
\end{equation}
The projector operator for the odd charge sector ground-state manifold using Eq.\ref{eq:5defP} reads as
\begin{equation} 
		\widehat{P}_{0} \doteq \sum_{S^{z}=\pm\tfrac{1}{2}} \ket{N;S^{z} ;j=0} \bra{N; S^{z};j=0} \otimes \widehat{\mathbb{1}}_{leads} ~.
\end{equation}
By applying this projector to Eq.\ref{eq:5HeffCBPer} yields to the effective model in \textit{sd-}basis as given in Eq.\ref{eq:5HeffCB}. Later in this part we see manipulation of Eq.\ref{eq:5HeffCB} into different basis. In particular, the projection on $\hat{H}_{nano}=\epsilon_{0}\widehat{P}_{0}$ because we recall the number operator of the nanostructure is spin conserving, and $\sigma=\uparrow$ and $\sigma=\downarrow$ nanostructure states have the same energy for $B=0$.\\
In the absence of a magnetic field or spin-orbit coupling terms in the bare system, the effective model in Eq.\ref{eq:5HeffCB} therefore has full $SU(2)$ symmetry and $W_{\alpha\beta}\equiv W_{\alpha\beta}$ independent of spin. We consider this case explicitly below where the physics of $S=1/2$ ground-states is detailed according to the particle-hole symmetry and \textit{sd-}symmetry regimes. Extensions to higher spin multiplet ground-states is discussed at the end of the chapter.

\subsection*{No potential scattering: $W_{ss}=W_{sd}=W_{ds}=W_{dd}\equiv0$}
\begin{figure}[h]
	\centering
	\begin{subfigure}[b]{0.85\textwidth}
		\includegraphics[width=0.85\linewidth]{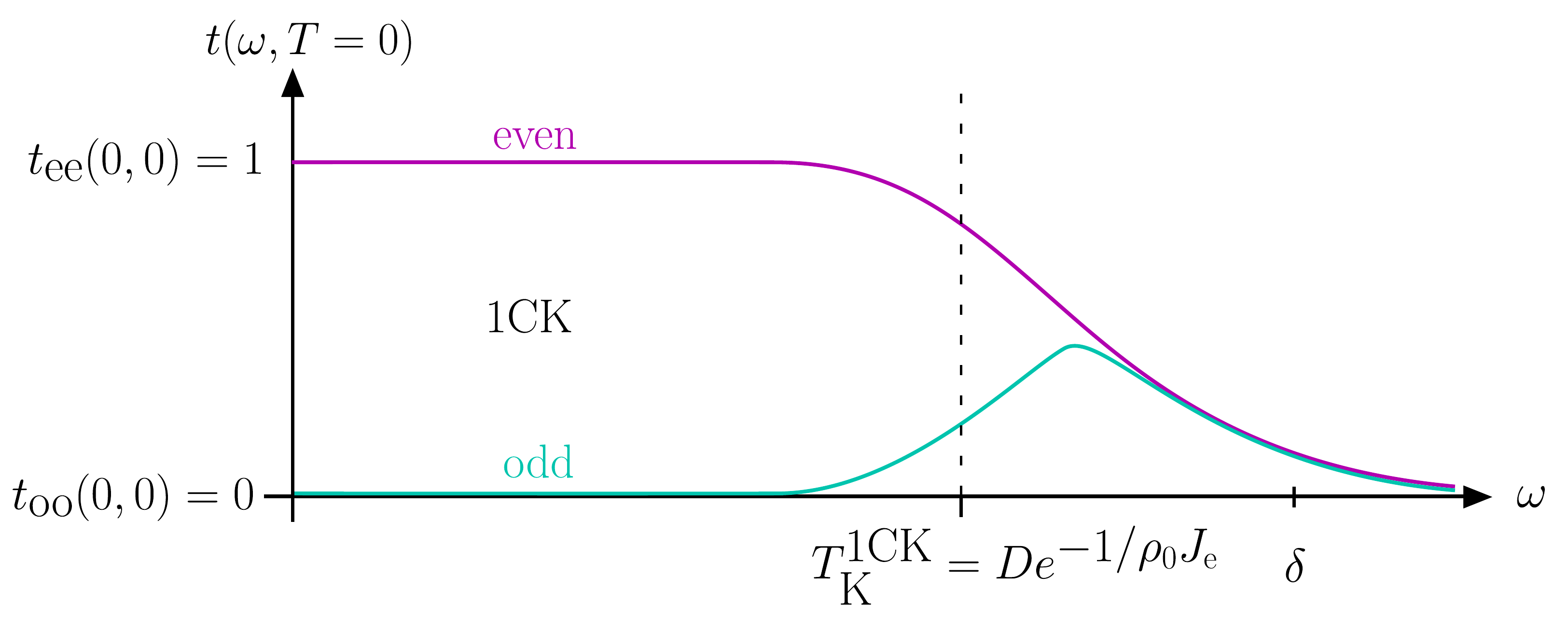}
		\caption{}
	\end{subfigure}
	\begin{subfigure}[b]{0.85\textwidth}
		\includegraphics[width=0.85\linewidth]{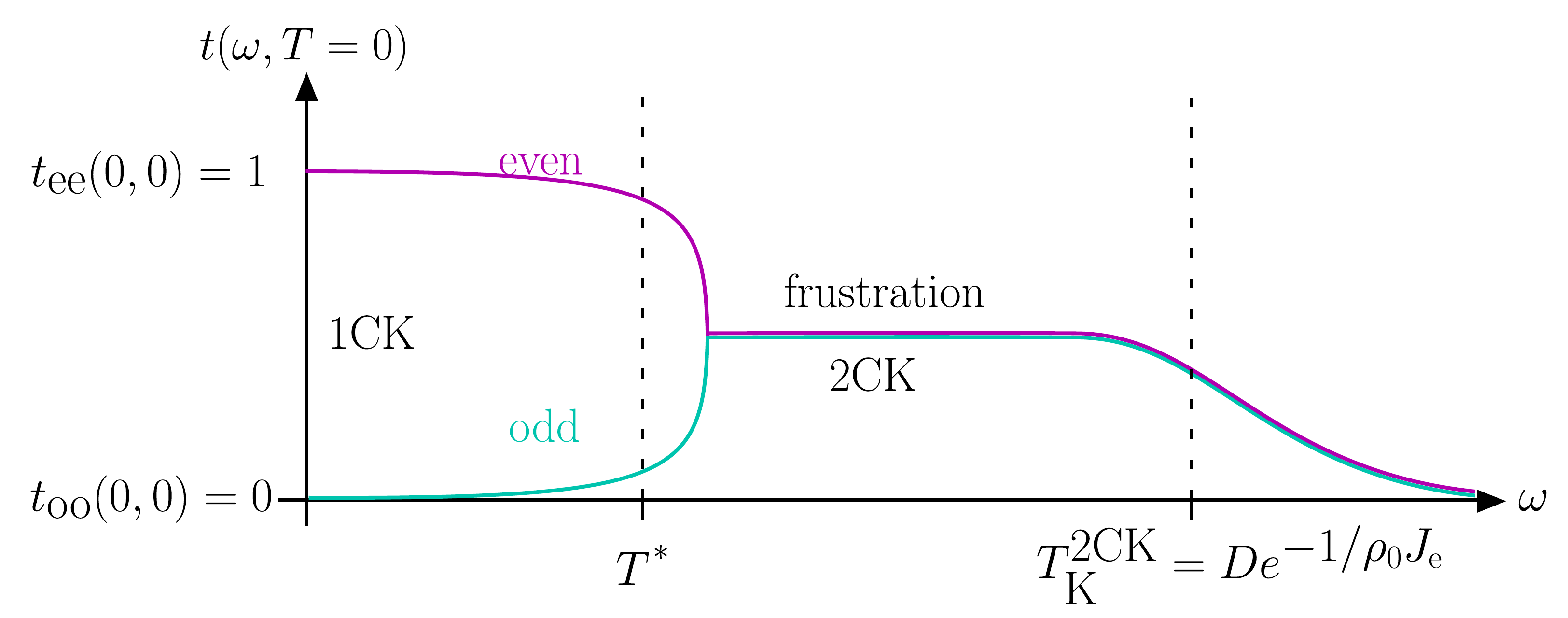}
		\caption{}
	\end{subfigure}
	\caption[Sketch of $\mathrm{T}$-matrix spectral function for $S=1/2$, $W=0$ system]{Sketch of $\mathrm{T}$-matrix spectral in even/odd basis at $T=0$, as function of energy scales $\omega$, for the particle-hole symmetric $S=1/2$ system. $(a)$ Strong coupling: $\delta =|J_{ee}-J_{oo}| > T^{e}_{K} \equiv T^{1CK}_{K}$ $(b)$ Weak coupling: $\delta = |J_{ee}-J_{oo}| < T^{e}_{K}\equiv T^{2CK}_{K}$, $T^{\star} \sim \delta^2/ T_{K}^{2CK}$.}\label{fig:5tm-W0}
\end{figure}

\begin{figure}[H]
	\centering
	\includegraphics[width=0.7\linewidth]{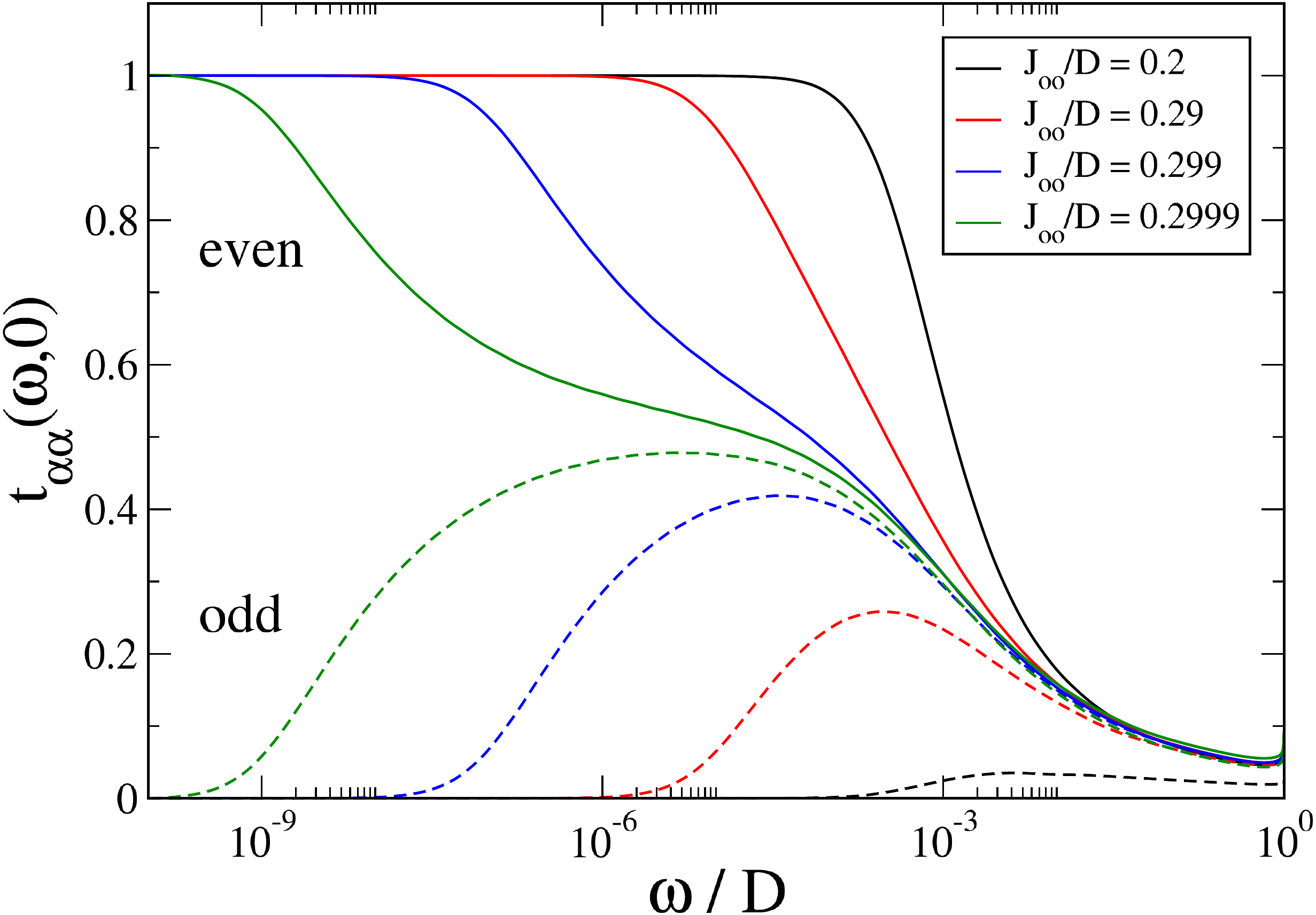}
	\caption[$\mathrm{T}$-matrix spectral function for $S=1/2$ and $W_{\alpha\beta}=0$]{NRG results for the $T=0$ $\mathrm{T}$-matrix spectrum  $t_{\alpha\alpha}(\omega,0)$ of the effective 2CK model, Eq.~\ref{eq:5H2ck_eo}. Even (odd) channel spectra shown as the solid (dashed) lines, for fixed $J_{ee}/D=0.3$ and $\mathbb{W}=\mathbb{0}$, with different $J_{oo}$. In all cases, $t_{ee}(0,0)=1$ and $t_{oo}(0,0)=0$, despite incipient frustration of Kondo screening for small $\delta=J_{ee}-J_{oo}$.}\label{fig:2ck_tm_W0}
\end{figure}

In particle-hole symmetric regime, there is no contribution from the potential scattering term $W_{\alpha\beta}$ in the system. We remark that, if $W_{sd}=W_{ds}=0$ it is not guaranteed breaking of particle-hole symmetry, but first we enforce vanishing contribution from \textit{all} the potential scattering terms, before relaxing this constraint in the next section.\\
Under such a condition, it proves useful to diagonalize the exchange coupling matrix by a \textit{canonical rotation} of the leads basis from $s,d$ to $e,o$ even/odd basis \cite{Mitchell2017KondoMolecule,Flensberg}. The transformation $\mathbb{U}^{\dagger}\mathbb{J}\mathbb{U}$ for the exchange scattering matrix $\mathbb{J}$ is defined by unitary matrix $\mathbb{U}$ with elements $U_{i\alpha}$ for $i=e,o$ and $\alpha=s,d$: 
\begin{equation}\label{eq:5CanT}
	\mathbb{U}=
	\begin{pmatrix}
		U_{es} & U_{ed} \\ U_{os} & U_{od} 
	\end{pmatrix} ~,
\end{equation}
and it satisfies the properties  $\mathbb{U}^{-1}=\mathbb{U}^{\dagger}$ such that $\mathbb{U}^{\dagger}\mathbb{J}\mathbb{U}$ is diagonal. By performing this transformation on the conduction electron operators we obtain
\begin{equation}\label{eq:Unitary-EO}
c^{\dagger}_{\alpha \sigma}=\sum_{i=e,o}c^{\dagger}_{i\sigma}U_{i \alpha} 
	\begin{cases}
		{c}^{\dagger}_{s\sigma} =
		{c}^{\dagger}_{e\sigma} U_{es} + {c}^{\dagger}_{o\sigma} U_{od} \\
		{c}^{\dagger}_{d\sigma} =
		{c}^{\dagger}_{e\sigma} U_{es} + {c}^{\dagger}_{o\sigma} U_{od}
	\end{cases} ~,
\end{equation}
and similarly for $c_{\alpha \sigma}$. Hence, we apply the transformation in Eq.\ref{eq:5CanT} to the exchange coupling Hamiltonian matrix form in Eq.\ref{eq:5HeffCB} and we derive
\begin{equation}\label{eq:H_J-EO}
	\begin{aligned}
\hat{H}^{K} &= \sum_{\sigma\sigma^{\prime}}\hat{\mathbf{S}} \cdot \boldsymbol{\sigma}_{\sigma\sigma^{\prime}} 	\begin{matrix} \begin{pmatrix} c^{\dagger}_{s\sigma} & c^{\dagger}_{d\sigma}\end{pmatrix} \\\mbox{} \end{matrix}
\begin{bmatrix}
	J_{ss} & J_{sd} \\  J_{ds} & J_{dd}
\end{bmatrix} 
\begin{pmatrix}  c_{s\sigma^{\prime}}  \\  c_{d\sigma^{\prime}} \end{pmatrix} ~,\\
&= \sum_{\sigma\sigma^{\prime}}
\hat{\mathbf{S}} \cdot \boldsymbol{\sigma}_{\sigma\sigma^{\prime}} 
\begin{matrix} \begin{pmatrix} c^{\dagger}_{e\sigma} & c^{\dagger}_{o\sigma}\end{pmatrix} \\\mbox{} \end{matrix}
\begin{bmatrix}
	J_{ee} & 0 \\  0 & J_{oo}
\end{bmatrix} 
\begin{pmatrix}  c_{e\sigma^{\prime}}  \\  c_{o\sigma^{\prime}} \end{pmatrix} =\sum_{i=e,o} J_{ii} \hat{\mathbf{S}}\cdot \hat{\mathbf{s}}_{ii} ~,
\end{aligned}
\end{equation}
where $J_{eo}=J_{oe}\equiv0$ by construction. As consequence, no direct spin tunnelling between the even and odd leads arises any more. We can further rewrite the diagonal elements in exchange coupling matrix as follows
\begin{subequations}\label{eq:5JeeJoo}
	\begin{align}
		&J_{ee} =\frac{1}{2} \left(J_{+}+ \sqrt{J^{2}_{-} +4|J_{sd}|^{2}} \right) \label{eq:Jee} ~,\\
		&J_{oo} = \frac{1}{2} \left(J_{+}- \sqrt{J^{2}_{-} +4|J_{sd}|^{2}} \right) \label{eq:Joo}  ~,
	\end{align} 
\end{subequations}
where $J_{\pm}=J_{ss}\pm J_{dd}$ and the potential scattering is zero. We also introduce the notation $\delta=J_{ee}-J_{oo}$.\\
We define the even/odd basis version of the $\mathrm{T}$-matrix spectral function in Eq.\ref{eq:TmatSpe} by applying rotation transformation in Eq.\ref{eq:5CanT} as follows
\begin{equation}
t_{\alpha\beta}(\omega,T)= \sum_{ij=e,o} U^{\star}_{i\alpha}t_{ij}(\omega,T)U_{j\beta} ~.
\end{equation}
The above expression for the inter-channel $\mathrm{T}$-matrix spectral function $t_{sd}(\omega,T)$ reads
\begin{equation}\label{eq:5tsdeo}
	\begin{aligned}
	t_{sd}(\omega,T) &\stackrel{(i)}{=}U^{\star}_{es}t_{ee}(\omega,T)U_{ed} + U^{\star}_{os}t_{oo}(\omega,T)U_{od} ~,\\
	&\stackrel{(ii)}{=}-\pi\rho_{0} \left( U^{\star}_{es}\mathit{Im}T_{ee}(\omega,T)U_{ed} + U^{\star}_{os}\mathit{Im}T_{oo}(\omega,T)U_{od} \right) \equiv -\pi\rho_{0}\mathit{Im}T_{sd}(\omega,T) ~,
	\end{aligned}  
\end{equation}
where in $(i)$ we use the fact that $J_{eo}=J_{oe}=W_{eo}=W_{oe}\equiv 0$ as result of the canonical transformation in Eq.\ref{eq:5CanT} such that the off-diagonal components $t_{eo}=t_{oe}\equiv 0$ and in $(ii)$ we give explicit expression for even/odd basis inter-channel $\mathrm{T}_{sd}(\omega,T)$ matrix. We note that Eq.\ref{eq:5tsdeo} can be further simplified under RG-flow of coupling parameters at $\omega=T=0$, as discussed further below.\\
Before we continue with the discussion, we want to comment on the physical implication of the canonical transformation. 
In the bare model, it is the $T_{sd}(\omega,T)$ component describing the spin scattering in the coupled lead-nanostructure-lead system to give a finite contribution to the  conductance. The application of canonical transformation makes $\hat{H}_{eff}$ diagonal such that the spin scattering is described only within even channel and nanostructure and only within odd channel and nanostructure i.e. $T_{ee}(\omega,T),T_{oo}(\omega,T)$ but no spin transfer occurs between even and odd channel through the nanostructure tunnelling i.e. $T_{eo}(\omega,T)=T_{oe}(\omega,T)=0$. The advantage of the diagonalised $\hat{H}_{eff}$ with no mixed terms is the independent RG-flow of coupling parameters for each even and odd channel separately. As consequence, once the appropriate energy scale is reached, the complete channel decoupling is possible by following the distinct channel evolution under RG - as we discuss more in detail below. It is important to remark that, although the even/odd basis is a convenient choice to study the renormalized system at low-energy scale, only the bare basis is the actual physical basis and only it can be used to define the conductance. In conclusion, once $\hat{H}_{eff}$ is found in even/odd basis and the RG-flow analysed, it is necessary to apply the inverse of the canonical transformation in order to restore the original basis. This is a straightforward operation since the transformation is unitary and therefore invertible.\\
We consider now two general scenarios happening in the particle-hole symmetric system.\\
The first is the special case arises when $J_{oo}=0$: it happens when the odd lead is decoupled on the level of $\hat{H}_{eff}$ and the \textbf{effective model} reduces precisely to a single-channel Kondo model with a single spin-$\tfrac{1}{2}$ impurity,
\begin{equation}\label{eq:5H1ck}
	\boxed{
	\hat{H}_{eff} = \hat{H}_{leads,e}  + J_{ee} \hat{\mathbf{S}} \cdot \hat{\mathbf{s}}_{ee} }\;.
\end{equation}
From Eq.~\ref{eq:Joo}, this occurs specifically when $|J_{sd}|^2=J_{ss}J_{dd}$. In fact, this condition is always automatically satisfied when the bare model is under PC condition and it \textit{exactly} holds for single-orbital model or single-impurity Anderson model. Indeed, this is physically natural since Eq.~\ref{eq:5H1ck} is the regular Schrieffer-Wolff transformation of the single-channel model using $\hat{H}_{hyb}$ as defined in Eq.\ref{eq:AM} but now for even channel obtained under PC, namely  $\sum_{i\sigma}(V_{i}c^{\dagger}_{e\sigma}d_{i\sigma} +H.c.)$ for given orbital $i$ of the nanostructure.\\
The conductance then follows from the MW formula under PC in Eq.\ref{eq:MW->PC} but now written in terms of the generalized transfer matrix under PC derived in Eq.\ref{eq:GenTransfMatPC}, namely
\begin{eqnarray}
	\widetilde{\mathbb{T}}_{PC}(\omega,T) \longrightarrow 4\frac{J_{ss}J_{dd}}{(J_{ss}+J_{dd})^2}\sum_{\sigma} t_{ee,\sigma}(\omega,T) \;,
\end{eqnarray}
where we use the $\mathrm{T}$-matrix spectral function in Eq.\ref{eq:TmatSpe} evaluated for the even channel.  We remark the prefactor $\tfrac{J_{ss}J_{dd}}{(J_{ss}+J_{dd})^2} \propto \tfrac{|V_{s}|^{2}|V_{d}|^{2}}{(|V_{s}|^{2}+|V_{d}|^{2})^2}$ is only valid within the Schrieffer-Wolff approximation for single-channel model, with equivalent lead in wide flat band limit.\\
In case of no potential scattering and no magnetic field as considered here, $t_{ee,\sigma}(\omega=0,T=0)=\sin^{2}(\pi/2)=1$ by FSR \cite{Hewson} and the resulting \textbf{electrical conductance} expression reads
\begin{equation}
	\boxed{
	\mathcal{G}^{dc}(\omega=0,T=0) = \frac{2e^{2}}{h}4\frac{J_{ss}J_{dd}}{(J_{ss}+J_{dd})^2}} ~,
\end{equation}
where $\mathcal{G}^{dc}$ is controlled by a pure geometric factor describing the nanostructure-lead coupling. Indeed for $\omega,T \ll T_K$, with $T_K\sim D \exp[-1/\rho_0 J_{ee}]$ the Kondo temperature for the effective model Eq.~\ref{eq:5H1ck}, we have $t_{ee,\sigma}(\omega,T)\simeq 1$ and so the conductance approximately saturates its low-temperature bound, $\mathcal{G}(T\ll T_K) \simeq \mathcal{G}^{dc}(0,0)$.\\
Of course, the above scenario is special, and for most realistic systems the exact PC condition $|J_{sd}|^2=J_{ss}J_{dd}$ is not expected to be satisfied. Indeed, in general multi-orbital systems we find $|J_{sd}|^2 \ll J_{ss}J_{dd}$. In such cases, $J_{oo}\ne 0$ and the odd channel remains formally coupled to the effective nanostructure spin-$\tfrac{1}{2}$. In the even/odd basis with $W_{\alpha\beta}=0$, Eq.~\ref{eq:5HeffCB} reduces to the \textbf{effective model},
\begin{equation}\label{eq:5H2ck_eo}
	\boxed{
	\hat{H}_{eff} = \hat{H}_{leads}  + J_{ee} \hat{\mathbf{S}} \cdot \hat{\mathbf{s}}_{ee} + J_{oo} \hat{\mathbf{S}} \cdot \hat{\mathbf{s}}_{oo} } \;.
\end{equation}
The physics of this channel-anisotropic 2CK model are rich but well-known, featuring a competition of Kondo screening between even and odd channels \cite{coleman1995simple2CK}. The model supports a non-Fermi liquid quantum critical point for $J_{ee}=J_{oo}$, but from Eq.~\ref{eq:5JeeJoo} we see that this only arises in these kinds of systems when both $J_{ss}=J_{dd}$ and $J_{sd}=0$. Although one may be able to tune to the \textit{sd-}symmetric condition $J_{ee}=J_{oo}$, to realize $J_{sd}=0$ requires the suppression of through-nanostructure exchange-cotunnelling processes. In principle, this could arise through gate-tunable many-body quantum interference effects \cite{Mitchell2017KondoMolecule} that conspire to produce an exact tunnelling node, but a real system exhibiting 2CK criticality driven by such effects has not yet been reported. On the other hand, for complex multi-orbital nanostructures, electron propagation across the entire structure embodied by the exchange cotunnelling $J_{sd}$ is typically small in magnitude (albeit finite) compared with the local terms $J_{ss}$ and $J_{dd}$. Therefore it may still be possible to access the quasi-frustrated quantum critical physics of the 2CK model at small $J_{sd}$ for intermediate temperatures or energies \cite{Mitchell2011_2COddChain,vzitko2006kondo}. In the following we simply regard $J_{ee}$ and $J_{oo}$ as free independent parameters of the effective model, however obtained, and consider the generic behaviour.\\
For concreteness, we now assume $J_{ss}$ is antiferromagnetic such that once the canonical transformation is applied, the inequality $J_{ee}>J_{oo}>0$ is fulfilled.\\ 
The perturbative scaling (poor man's scaling \cite{AndersonPoor1970}) starting from weak coupling, indicates that both $J_{ee}$ and $J_{oo}$ get renormalized upwards on reducing energy or temperature scale. The scaling invariant for the RG-flow of each is its respective Kondo temperature, $T_{K}^{\alpha} \sim D \exp[-1/\rho_0 J_{\alpha\alpha}]$. However, the low-energy physics is nonperturbative, and the 2CK strong coupling fixed point for a spin-$\tfrac{1}{2}$ impurity is unstable \cite{Nozieres_KondoRealMetal1980}. Since $J_{ee}> J_{oo}$ we have $T_{K}^e > T_{K}^o$ and the even channel ultimately flows under RG to strong coupling while the odd channel flows back to weak coupling. According to the couplings strength, two distinct regime appear, as we sketch in Fig.\ref{fig:5tm-W0}.\\
The first case happens at strong coupling regime $\delta =|J_{ee}-J_{oo}| > T^{e}_{K} \sim D e^{-1/\rho_{0}J_{ee}}$ where $T^{e}_{K}\equiv T^{1CK}_{K}$ single-channel Kondo temperature for the even channel. From high to low temperature range, the system undergoes to
\begin{enumerate}
	\item $T \gg T^{1CK}_{K}$: at high-temperature scale, the system is effectively two-channel model with both even, odd channel sectors are weakly coupled and increase progressively at equal rate. The nanostructure is nearly free local moment; 
	\item $T^{1CK}_{K} < T < \delta$: at intermediate temperature, channels start splitting with similar slope of increment/decrement for the even/odd channel respectively. The even coupling exceeds the odd one because of $J_{ee} > J_{oo}$;
	\item $T=T^{1CK}_{K}$: channels decoupling is completed and there is no more spin scattering between the odd channel sector and the nanostructure;
	\item $0<T < T^{1CK}_{K}$: the system undergoes to Kondo screening of the nanostructure spin due to the even electrons sector and it becomes effectively a single-channel model. At this energy scale we understand the system having an \textit{emergent} proportionate coupling condition;
	\item $T=0$: at strictly zero temperature, the system is exactly in single-channel characterised by $J_{oo}\equiv 0,~J_{ee}\neq0$ and $t_{ee}(0,0)=1$ is calculated exactly using the FRS.
\end{enumerate}
The second case happens at the opposite regime, namely at weak channel anisotropy $\delta= |J_{ee}-J_{oo}| < T^{e}_{K}\sim D e^{-1/\rho_{0}J_{ee}}$ where $T^{e}_{K}\equiv T^{2CK}_{K}$ two-channel Kondo temperature. Again, going from high to low temperature range, we find the system undergoing
\begin{enumerate}
	\item $T \gg T^{2CK}_{K}$: at high-temperature scale, the system is in two-channel configuration with both leads weakly coupled;
	\item $T^{\star}<T < T^{2CK}_{K}$: the system reaches a plateau where even and odd channels compete for screening the spin of the nanostructure and the couplings have similar strength i.e. $J_{ee} \simeq J_{oo}$. The system shows a frustrated state which lasts until the even channel overcomes the odd one with consequent channels decoupling and $T^{\star} \sim \delta^2/ T_{K}^{2CK}$ sets the crossover;
	\item $0<T  \leq T^{\star}$: the channel frustration is relived at $T^{\star}$
	and the model is effectively in single-channel i.e. $J_{ee}>J_{oo}\to0$. The Kondo singlet is formed from the screened nanostructure spin by the even electrons channel. At this energy scale the system presents an \textit{emergent} proportionate coupling property;
	\item $T=0$: as in the previous case, at strictly zero temperature the system is exactly in single-channel with $J_{oo}\equiv 0,~J_{ee}\neq0$ and the $t_{ee}(0,0)=1$ value is calculated exactly using the FRS.
\end{enumerate}
For a given $J_{ee}$, the system couplings flow in RG sense through both coupling regimes without undergoing to any phase transition. The continuous shifts from one regime to another of the RG-flow implies the crossover at $T^{1CK}_{K}$ for the strong coupling regime merges with the one at $T^{2CK}_{K}$ for the weak coupling. As result, we have $T^{1CK}_{K} = T^{2CK}_{K} \equiv T^{e}_{K} \sim D e^{-1/\rho_{0}J_{ee}}$. Hence, in order to incorporate both couplings regime, channel decoupling occurs at the minimum between $T^{\star}$ and $T^{e}_{K}$ temperature. The outcome is a single-channel model \cite{Anderson1961} but now only in even basis. We observe that unconstrained channel configuration and the PC found at $\omega=T=0$ regime are incipient properties of $\hat{H}_{eff}$ in even/odd basis: both the independent spin sectors and PC are not feature of the original bare model. The properties of the 2CK model $\mathrm{T}$-matrix were discussed in \cite{Fritz_Crossover2CK_2011}.\\
This physical picture is confirmed by numerically-exact NRG calculations for the effective 2CK model Eq.~\ref{eq:5H2ck_eo}, in Fig.~\ref{fig:2ck_tm_W0}. The RG-flow is vividly demonstrated in the $T=0$ spectrum of the $\mathrm{T}$-matrix for the even channel (solid lines) and odd channel (dashed lines), plotted for systems with fixed $J_{ee}$ but decreasing $\delta=J_{ee}-J_{oo}>0$ (and $W_{\alpha\beta}=0$). We highlight three features: $(i)$ in all cases, the $\mathrm{T}$-matrix spectrum of the even channel is always exactly pinned to 1 at zero frequency and zero temperature, $t_{ee}(0,0)=1$; $(ii)$ by contrast, the odd channel $\mathrm{T}$-matrix spectrum vanishes, $t_{oo}(0,0)=0$; $(iii)$ relief of incipient frustration is characterized by a new scale $T^{\rm FL}$, where the effective model can be described by Fermi liquid theory.\\
\noindent{Importantly}, we see an emergent decoupling of the odd lead at low energies $|\omega|\ll T^{\rm FL}$ in all cases, where $T^{FL}=\min(T_{K}^e,T^{\star})$ is a dynamically-generated Fermi liquid scale. Here $T^{\star} \sim \delta^2/ T_{K}^e$ is an energy scale embodying the frustration of Kondo screening from the two channels. For large $\delta > T_{K}^e $ we have $T^{\star} > T_{K}^e$ and the RG-flow of the odd channel is cut off on the scale of $T_{K}^e$ as the even channel flows to strong coupling. On the other hand, for small $\delta < T_{K}^e $ we have $T^{\star} < T_{K}^e $ and the system is described by the frustrated (non-Fermi liquid) 2CK fixed point over an intermediate temperature range $T^{\star} < T < T_{K}^e $. The even channel flows to strong coupling and the odd channel flows to weak coupling only for $T \ll T^{\star}$ in this case. \\
Without potential scattering, it is then guaranteed that $t_{ee}(0,0)=1$ and $t_{oo}(0,0)=0$, as dictated by phase-shift arguments: due to the decoupling at $\omega=T=0$, we may write $t_{\alpha\alpha}(0,0)=\sin^2(\delta_{\alpha})$, see discussion at Eq.\ref{eq:TmatSpe} for vanishing frequency energy and temperature. The Kondo effect in the even channel confers a phase shift $\delta_e=\pi/2$ whereas in the decoupled odd channel $\delta_o=0$. We note that although the non-Fermi liquid 2CK regime has been accessed experimentally in quantum dot devices, this is the result of fine-tuning and ingenious quantum engineering \cite{iftikhar2018tunable}. The more standard situation is for the odd channel to decouple at low temperatures, as described above.\\
Since such systems flow under RG to an effective single-channel description involving only the even lead combination, we have an \textit{emergent} PC condition. This greatly simplifies the calculation and interpretation of the low-temperature quantum transport. 
On the lowest temperature and energy scales $\omega,T \ll T^{\rm FL} $, the alternative formulation of the Oguri argument in \textit{dc-}regime derived in Eq.\ref{eq:4GdcOguriEx} yields the electrical conductance. We may now relate the inter-channel $\mathrm{T}$-matrix $T_{sd,\sigma}$ to the even channel $\mathrm{T}$-matrix $T_{ee,\sigma}$ using the rotation $\mathbb{U}$ in Eq.\ref{eq:5CanT} diagonalizing the exchange matrix as we shown in Eq.\ref{eq:5tsdeo}. Hence, we write the alternative Oguri formula now for the \textit{sd-}model with equivalent metallic leads, namely
\begin{equation}
	\begin{aligned}
	\mathcal{G}^{dc}(\omega=0,T=0)&= \frac{2e^2}{h} 4 |\pi \rho_{0} \mathit{Im}T_{sd,\sigma}(0,0)|^{2} ~,\\
	&=\frac{2e^2}{h} 4 |\pi \rho_{0} \left( U^{\star}_{es}\mathit{Im}T_{ee,\sigma}(\omega,T)U_{ed} + U^{\star}_{os}\mathit{Im}T_{oo,\sigma}(\omega,T)U_{od}\right)|^{2} ~,
	\end{aligned}
\end{equation}
here equivalently written in \textit{sd} and even/odd basis, where the latter expression is because $J_{eo}=W_{eo}=0$ so that $T_{eo,\sigma}(\omega,T)=T_{oe,\sigma}(\omega,T)=0$ from the canonical transformation in Eq.\ref{eq:5CanT}. As we mentioned previously, this equation further simplifies according to the RG-flow of coupling parameters: in the current case $W_{\alpha\beta}=0$, at $\omega=T=0$ we find $J_{oo}=0$  so that $T_{oo}(0,0)=0$ whereas $J_{ee}\neq0$ so that  $T_{ee}(0,0)\neq0$  and due to the emergent decoupling it contributes in
\begin{equation}
	\pi \rho_{0}\mathit{Im}T_{ee,\sigma}(0,0)=t_{ee,\sigma}(0,0) \equiv 1 ~,
\end{equation}
where we used the definition of the  $\mathrm{T}$-matrix spectral function in Eq.\ref{eq:TmatSpe} and the FSR in Eq.\ref{eq:FSR} \cite{Langreth1966,Georges2016}. We can indeed think of $\hat{H}_{eff}$ as now being an effective \textit{renormalized} model.
Therefore the $T=0$ \textbf{electrical conductance} follows as,
\begin{equation}\label{eq:5G0_emPC_W0}
	\boxed{
	\mathcal{G}^{dc}(\omega=0,T=0)=\left ( \frac{2e^2}{h}\right ) 4|U^{\star}_{es}U_{ed}|^{2}=\left ( \frac{2e^2}{h}\right ) \frac{4 |J_{sd}|^2}{4|J_{sd}|^2 + J_{-}^2} } \;.
\end{equation}
This result also applies accurately for all $T\ll T^{FL}$.
The low-temperature conductance can therefore be determined directly from the effective exchange couplings $J_{ss}$, $J_{dd}$, and $J_{sd}$.\\
Indeed, the factor $|J_{sd}|^2/(|J_{sd}|^2+J_{-}^2)$ appearing in Eq.~\ref{eq:5G0_emPC_W0} reduces as it must to $|V_{s}V_{d}|^2/(|V_{s}|^2+|V_{d}|^2)^2$ in the case of exact PC in the bare model, for which we have 
$|J_{sd}|^2=J_{ss}J_{dd}$ and $J_{ss}/J_{dd}=|V_{s}/V_{d}|^2$ by means of the Schrieffer-Wolff transformation \cite{Mitchell2017KondoMolecule}. In some sense, Eq.~\ref{eq:5G0_emPC_W0} contains the generalization of Eq.\ref{eq:GenTransfMatPC} but now for emergent PC condition. This is equivalent 
to the factor $ \Gamma_s \Gamma_d/(\Gamma_s + \Gamma_d)^2$ we used in the derivation of $\mathrm{T}$-matrix spectral function in Eq.\ref{eq:TmatSpe}. \\
As conclusion on this part: the even/odd basis transformation allows for independent channel sectors. Under RG-flow, the $\mathrm{T}$-matrix spectral function flows along different paths according to channel symmetry and takes specific values at $\omega=T=0$ in Fermi liquid regime, namely $t_{ee}(0,0)=1$ and $t_{oo}(0,0)=0$. We stress that, in case we had studied the RG-flow of coupling parameters in original basis, we would have dealt with a proper two-channel model with entangled contributions from lead-nanostructure,  lead-lead electronic flow and non-diagonal $\mathbb{J}_{\alpha\beta},\mathbb{W}_{\alpha\beta}$ matrices. Hence, under RG we would not have found the energy scale with incipient PC properties. On the contrary, the canonical transformation defined in Eq.\ref{eq:5CanT} combined with RG-flow of couplings in even/odd basis offers an emergent physics at low-energy scale.

\subsection*{Finite potential scattering $W_{\alpha\beta}\neq0$ in \textit{sd-}symmetry}
\begin{figure}[H]
	\centering
	\includegraphics[width=1.05\linewidth]{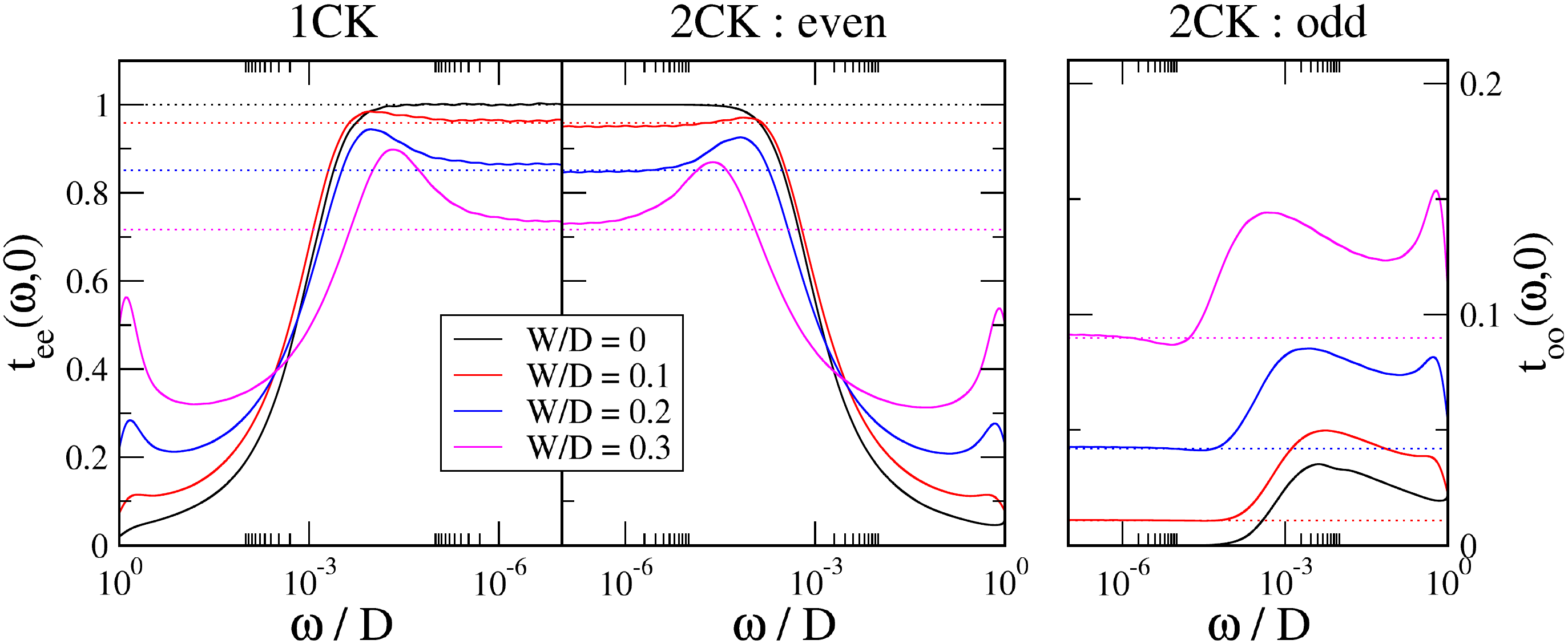}
	\caption[$\mathrm{T}$-matrix spectral function for \textit{sd-}symmetric $S=1/2$ and $W_{\alpha\beta}\neq0$]{NRG results for the $T=0$ $\mathrm{T}$-matrix spectrum of the effective 1CK model Eq.~\ref{eq:5H1ck} \textit{(left panel)}, and 2CK model Eq.~\ref{eq:5H2ck_eo_W} (even/odd channels in the centre/right panels). For the 2CK model, we take the \textit{sd-}symmetric case $J_{-}=W_{-}=0$, with fixed $J/D=0.25$ and $J_{sd}/D=0.05$, varying $W=3W_{sd}$. For the 1CK model we use the same parameters as for the even channel of the corresponding 2CK model ($J_{ee}/D=0.3$ and $W_{ee}=\tfrac{4}{3}W$). Dotted lines in the left/centre panels are given by Eq.~\ref{eq:5tm_ee_W_sd}, while Eq.~\ref{eq:5tm_oo_W_sd} is used in the right panel.}\label{fig:tm_sym}
\end{figure}

\begin{figure}[H]
	\centering
	\includegraphics[width=0.85\linewidth]{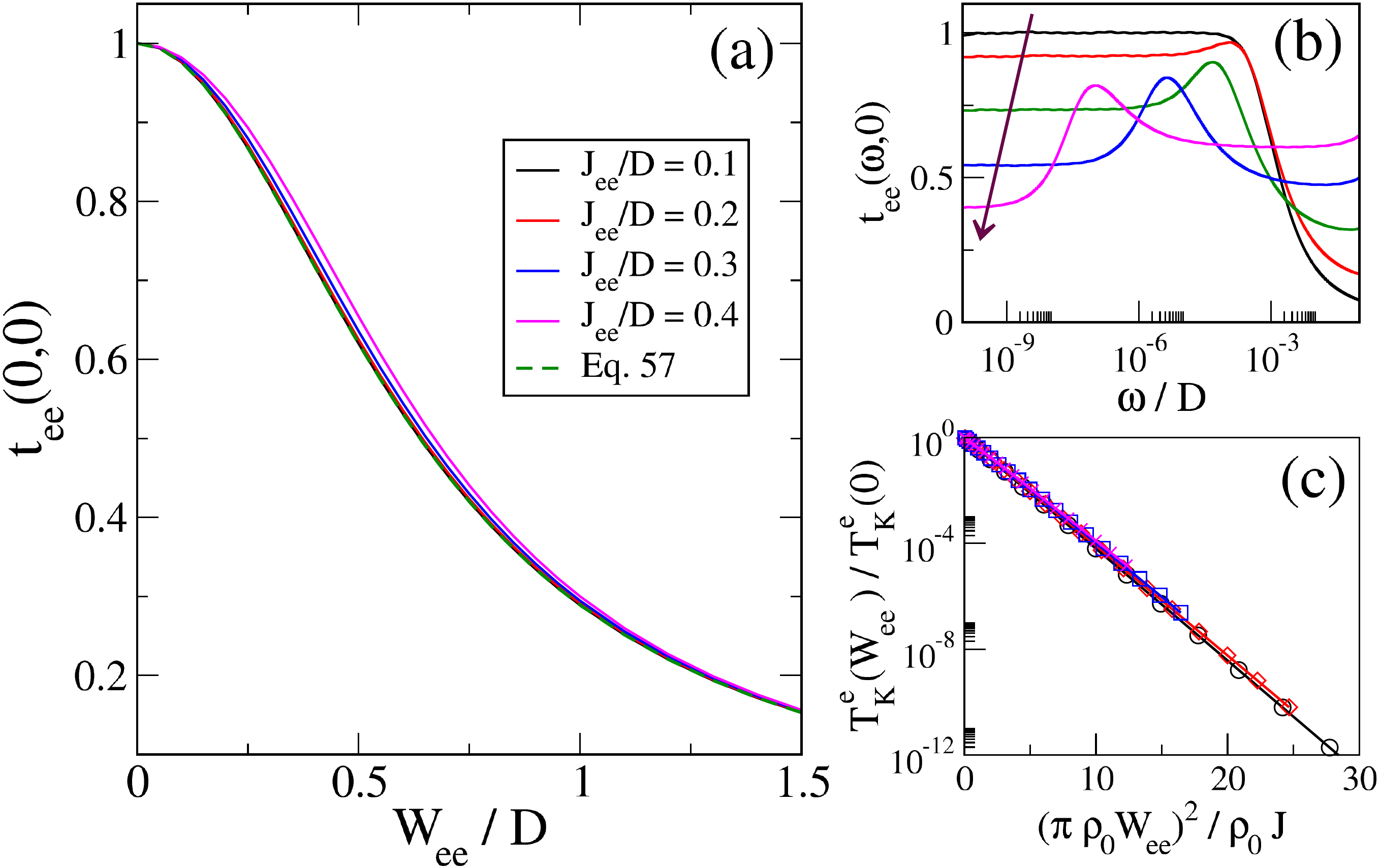}
	\caption[$\mathrm{T}$-matrix spectral function for single channel $S=1/2$ and $W_{\alpha\beta}\neq0$]{NRG results for the effective 1CK model, Eq.~\ref{eq:5HeffCB}. $(a)$ $T=0$ spectrum of the $\mathrm{T}$-matrix spectral function at the Fermi energy $t_{ee}(0,0)$, as a function of the local potential $W_{ee}$, for different $J_{ee}$. Dashed line is Eq.~\ref{eq:5tm_ee_W_sd}, showing only small deviations even at rather large $J_{ee}$. $(b)$ The full energy-dependence of the $\mathrm{T}$-matrix spectral function $t_{ee}(\omega,0)$ is shown for $J_{ee}/D=0.3$ with $W_{ee}/D=0, 0.2, 0.4, 0.6, 0.8$ increasing in the direction of the arrow. $(c)$ Dependence of the Kondo temperature $T_{K}^e$ on $W_{ee}$ for the same $J_{ee}$ used in $(a)$, confirming Eq.~\ref{eq:5tk_1ck}.}\label{fig:tm_1ck}
\end{figure}

\begin{figure}[H]
	\centering
	\includegraphics[width=0.85\linewidth]{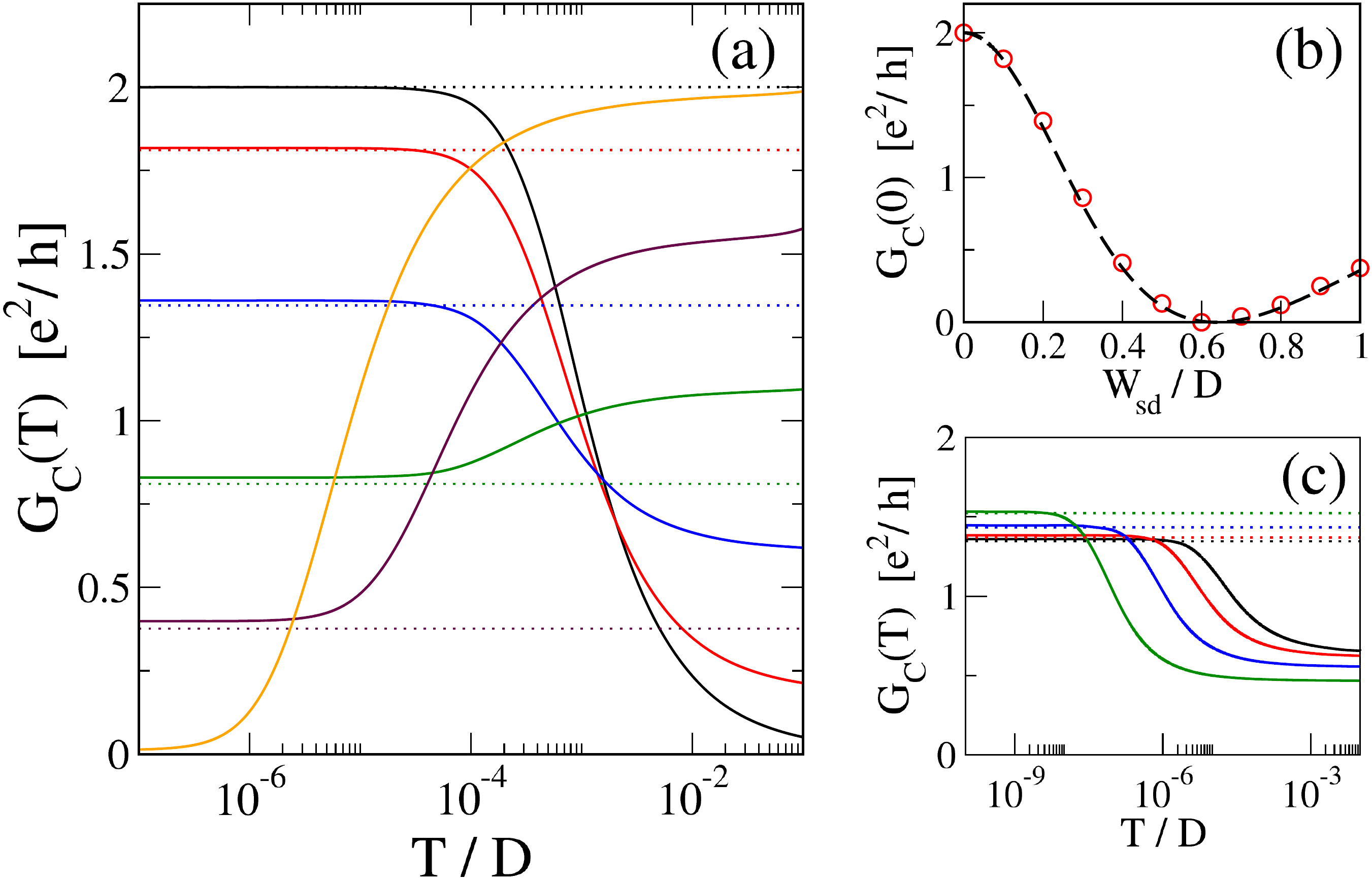}
	\caption[$\mathrm{T}$-matrix spectral function for generalized two-channel $S=1/2$ and $W_{\alpha\beta}\neq0$]{Conductance of the generalized 2CK model Eq.~\ref{eq:5HeffCB} at $sd$-symmetry, obtained by NRG using the improved Kubo method. $(a)$ $\mathcal{G}^{dc}(T)$ vs $T$ for fixed $J/D=0.25$, $J_{sd}/D=0.05$ and $W=0$ with different $W_{sd}/D=0, 0.1, 0.2, 0.3, 0.4, 0.6$ for the solid black, red, blue, green, violet and orange lines. Dotted lines are the low-$T$ predictions of Eq.~\ref{eq:5cond_2ck_sym}. $(b)$ $\mathcal{G}^{dc}(0)$ vs $W_{sd}$, for systems with the same parameters as in $(a)$, showing the conductance node in dimensionless quantity near $\widetilde{W}_{sd}=1$, compared with Eq.~\ref{eq:5cond_2ck_sym} (dashed line). \textit{(c)} $\mathcal{G}^{dc}(T)$ vs $T$ for fixed $J/D=0.15$, $J_{sd}/D=0.05$, $W/D=0.2$, varying $W_{sd}/D=0, 0.1, 0.2, 0.3$ for the black, red, blue and green solid lines (again compared with Eq.~\ref{eq:5cond_2ck_sym}, dotted lines).}\label{fig:cond_2ck_sym}
\end{figure}

The results of the previous section hold when $W_{\alpha\beta}=W_{\beta\alpha}=W_{\alpha\alpha}=W_{\beta\beta}=0$. Although an important simple limit, the resulting behaviour is not generic since even at particle-hole symmetry ($W_{ss}=W_{dd}=0$), the through-nanostructure cotunnelling term $W_{sd}$ is typically finite. In this section we consider finite potential scattering $W_{\alpha\beta}\ne 0$ in the \textit{sd}-symmetric limit - meaning $J_{ss}=J_{dd}\equiv J$ and $W_{ss}=W_{dd}\equiv W$ (or equivalently $J_{-}=W_{-}=0$). In this case, the full model has an overall mirror symmetry with respect to exchanging $s$ and $d$ leads.\\
At high temperatures $T \gg T_{K}$, renormalization due to the Kondo effect is weak, and the contribution to conductance is dominated by finite source-drain cotunnelling, $W_{sd}$. Conductance in this limit is therefore given approximately by Eq.~\ref{eq:5cond_qpc}, obtained for $J_{\alpha\beta}=0$. However, richer and more complex behaviour results at lower temperatures, where there is a subtle interplay between $\mathbb{W}$ and $\mathbb{J}$ matrices. We therefore focus on the low-$T$ physics in the following.\\
First, we note that in the \textit{sd}-symmetric limit, the even/odd channels are simply the symmetric/antisymmetric combinations of source and drain leads,
\begin{eqnarray}\label{eq:eo}
	c_{e\sigma} = \frac{1}{\sqrt{2}} \left(c_{s\sigma} + c_{d\sigma} \right) ~~~;~~~ c_{o\sigma} =\frac{1}{\sqrt{2}} \left( c_{s\sigma} - c_{d\sigma} \right) ~,
\end{eqnarray}
which yields,
\begin{subequations}
	\begin{align}
		J_{ee}=J+J_{sd} \quad &;\quad J_{oo}=J-J_{sd} ~,\\
		W_{ee}=W+W_{sd} \quad&;\quad W_{oo}=W-W_{sd}  ~,
	\end{align}
\end{subequations}
and $J_{eo}=J_{oe}=W_{eo}=W_{oe}=0$. Note that with \textit{sd-}symmetry, the lead transformation diagonalizing the exchange matrix $\mathbb{J}$ given in Eq.\ref{eq:5CanT} also diagonalizes the potential scattering matrix $\mathbb{W}$. We write the \textbf{effective model} as,
\begin{equation}\label{eq:5H2ck_eo_W}
	\boxed{
	\hat{H}_{eff} = \widetilde{H}_{leads}  + J_{ee} \hat{\mathbf{S}} \cdot \hat{\mathbf{s}}_{ee} + J_{oo} \hat{\mathbf{S}} \cdot \hat{\mathbf{s}}_{oo}} \;,
\end{equation}
where we have incorporated the local potentials $W_{ee}$ and $W_{oo}$ into the definition of the lead Hamiltonian, $\widetilde{H}_{leads} = \sum_{\alpha=e,o} [\hat{H}_{leads,\alpha} + W_{\alpha\alpha}^{\phantom{\dagger}} \sum_{\sigma} c_{\alpha\sigma}^{\dagger} c_{\alpha\sigma}^{\phantom{\dagger}}]$ where $\hat{H}_{leads,\alpha}$ is the free $\alpha$-lead Hamiltonian as defined in Eq.\ref{eq:5HeffCB}. \\
Analysis of Eq.~\ref{eq:5H2ck_eo_W} proceeds similarly to that of Eq.~\ref{eq:5H2ck_eo}, except now we account for the potential scattering by rediagonalizing the lead Hamiltonian $\widetilde{H}_{leads}$. This can be done implicitly using Green's function techniques, which allow us to relate the local lead Green's functions including the potential to those of the bare leads. We find  $[\widetilde{G}^0_{\alpha\alpha,\sigma}(\omega)]^{-1}=[G^0_{\alpha\alpha,\sigma}(\omega)]^{-1} - W_{\alpha\alpha}$ for $\alpha=e,o$. This implies that the renormalized Fermi level density of states is $\widetilde{\rho}_{\alpha}=-\tfrac{1}{\pi}{\rm Im} \widetilde{G}^0_{\alpha\alpha,\sigma}(0) = \rho_0/(1+\widetilde{W}_{\alpha\alpha}^2)$, where we again use the notation for dimensionless potential scattering $\widetilde{W}_{\alpha\beta}=\pi \rho_0 W_{\alpha\beta}$.\\
In fact, the quantities controlling the Kondo physics and RG-flow in such problems are the dimensionless parameters $j_{\alpha}=\widetilde{\rho}_{\alpha}J_{\alpha\alpha}$, rather than the bare exchange couplings $J_{\alpha\alpha}$. For $j_e>j_o$, the even lead flows to strong coupling, while the odd lead decouples; the opposite applies for $j_e<j_o$. Indeed, even with $J_{ee}>J_{oo}$ and $W_{ee}>W_{oo}$, one may be able to realize $j_e=j_o$, as required for 2CK criticality. It may therefore be possible to tune across the 2CK quantum phase transition in a given system by tuning gate voltages.\\
We now explore the effect of the local potentials $W_{ee}$ and $W_{oo}$ on the dynamics and conductance. For $W_{\alpha\beta}=0$ considered in the previous section, we argued that the emergent decoupling of the odd lead at low energies and temperatures $\omega,T\ll T^{FL}$ produced an effective PC condition, with $t_{ee,\sigma}(0,0)$ and hence $G_C(0)$ then being controlled by an effective \emph{single} channel Kondo (1CK) model, Eq.~\ref{eq:5H1ck}. With finite $W_{ee}$ and $W_{oo}$, the lead with the smaller $j_{\alpha}$ still decouples \textit{asymptotically}. However, $t_{ee,\sigma}(0,0)<1$ and $t_{oo,\sigma}(0,0)>0$ due to the potential scattering.\\
This physical picture is confirmed in Fig.~\ref{fig:tm_sym} where we plot the $T=0$ spectra of the even and odd $\mathrm{T}$-matrix for the 2CK model Eq.~\ref{eq:5H2ck_eo_W} in the centre and right panels, for fixed exchange couplings $J_{ee}$ and $J_{oo}$, but with different potential scattering strengths chosen for illustration to be $W=3W_{sd}$. Below an emergent scale $T^{FL} \sim T_{K}^e$ (here $\sim 10^{-3}D$), the $\mathrm{T}$-matrix of the even channel exhibits a Kondo resonance, while the incipient RG-flow of the odd channel is arrested.\\
\noindent{Since} the odd channel is decoupled at $\omega=T=0$, the  $\mathrm{T}$-matrix spectral function $t_{oo,\sigma}(0,0)$ is determined completely by $W_{oo}$. From the $\mathrm{T}$-matrix equation, Eq.~\ref{eq:TmatrixEq}, together with the bare and renormalized Green's functions 
$G^0_{oo,\sigma}(0)=-i\pi\rho_0$ and $\widetilde{G}^{0}_{oo,\sigma}(0,0)=-i\pi\widetilde{\rho}_o$, we immediately obtain,
\begin{equation}\label{eq:5tm_oo_W_sd}
	t_{oo,\sigma}(0,0)=\frac{\widetilde{W}_{oo}^2}{1+\widetilde{W}_{oo}^2} \;.
\end{equation}
This result is equivalent to a phase shift $\delta_o= \tan^{-1}[\widetilde{W}_{oo}]$ in the odd channel, since $t_{oo,\sigma}(0,0)=\sin^2(\delta_o)$ from Eq.\ref{eq:TmatSpe} evaluated at zero energy and zero frequency according to FSR. Thus, Eq.\ref{eq:5tm_oo_W_sd} is an \textit{exact} expression. The full complex odd-channel $\mathrm{T}$-matrix follows as
\begin{equation}
\pi\rho_0 T_{oo,\sigma}(0,0)=(\widetilde{W}^{-1}_{oo}-i)\times t_{oo,\sigma}(0,0) ~.
\end{equation}
For the systems considered in Fig.~\ref{fig:tm_sym}, we plot the corresponding values of $t_{oo,\sigma}(0,0)$ obtained from Eq.~\ref{eq:5tm_oo_W_sd} in the right panel as the dotted horizontal lines. The full odd-channel $\mathrm{T}$-matrix obtained by NRG is seen to saturate to these values for $|\omega|\ll T^{\rm FL}$, as anticipated.\\
More interestingly, comparison between the left and centre panels of Fig.~\ref{fig:tm_sym} shows that the low-energy dynamics of the \emph{even} channel can be accurately understood in terms of an effective single-channel Kondo model with the same $J_{ee}$ and $W_{ee}$. This is again a consequence of the emergent decoupling of the odd channel. Indeed, provided $\delta\equiv J_{ee}-J_{oo} \gg T_{K}^e$ - as is fairly standard condition for bare systems with at least one antiferromagnetic coupling - the correct Kondo scales are also well reproduced.\\
For the bare Andersonian nanostructure model, calculating the low-energy behaviour of the scattering $\mathrm{T}$-matrix is deeply nontrivial. Even for the single-impurity Anderson model, $t_{ee,\sigma}(0,0)=\sin^2(\delta_e)$ requires knowledge of the even-channel phase shift $\delta_e$. In this case, the Friedel sum rule dictates that $\delta_e=\tfrac{\pi}{2} n_{imp}$ in terms of the excess charge due to the impurity, $n_{imp}$. But the latter is renormalized by interactions and must itself be calculated using many-body techniques away from particle-hole symmetry. On the other hand, the above physical arguments and NRG results in Fig.~\ref{fig:tm_sym} establish that $t_{ee,\sigma}(0,0)$ can be obtained from an effective 1CK model, Eq.~\ref{eq:5HeffCB}. Treating the exchange coupling $J_{ee}$ as a \textit{perturbation} to the even-channel charge, we may then write $n_{imp} \simeq \tfrac{2}{\pi}\tan^{-1}[\widetilde{W}_{ee}]+1$, where the first term is the excess charge induced in a free bath due to a boundary potential $W_{ee}$, and we add $1$ for the singly-occupied impurity local moment in the Kondo model. Here, we use $\simeq$ symbol to indicate the independence of contribution in the phase-shift from the Kondo singlet formation and $\widetilde{W}_{ee}$ is taken into account. This approximation gives $\delta_e \simeq \tan^{-1}[\widetilde{W}_{ee}]+\tfrac{\pi}{2}$ where deviation from half-filling regime is given by the correction term $\tan^{-1}[\widetilde{W}_{ee}]$. The two contributions in $\delta_e$ can be either constructive or destructive, as we discuss in Fig.\ref{fig:cond_2ck_sym}. \\
By combining all the elements, the $\mathrm{T}$-matrix spectral function from the Friedel sum rule reads,
\begin{equation}\label{eq:5tm_ee_W_sd}
	t_{ee,\sigma}(0,0)\simeq \frac{1}{1+\widetilde{W}_{ee}^2} \;,
\end{equation}
and the full even-channel $\mathrm{T}$-matrix is therefore given by 
\begin{equation}
\pi\rho_0 T_{ee,\sigma}(0,0)=-(\widetilde{W}_{ee}+i)\times t_{ee,\sigma}(0,0) ~.
\end{equation}
\noindent{These} predictions are quantitatively substantiated in Fig.~\ref{fig:tm_1ck}, which shows NRG results for a pure 1CK model with exchange coupling $J_{ee}$ and potential scattering $W_{ee}$. The $\mathrm{T}$-matrix spectral function $t_{ee,\sigma}$ at $\omega=T=0$ is plotted in panel $(a)$ as a function of $W_{ee}$ for different $J_{ee}$. Even for rather large $J_{ee}/D=0.4$, the exact results agree accurately with Eq.~\ref{eq:5tm_ee_W_sd} (dashed line) over the entire range of $W_{ee}$.\\
\noindent{The} full energy dependence of the $\mathrm{T}$-matrix spectral function is shown in panel $(b)$ for fixed $J_{ee}$ with increasing $W_{ee}$. The asymptotic $\omega=0$ value reported in $(a)$ can be extracted for $|\omega|\ll T^e_{K}$, with the Kondo scale $T^e_{K}$ decreasing as the potential scattering strength $W_{ee}$ increases. Under RG-flow we have  $T^e_{K}(J_{ee},W_{ee}) = D \exp[-1/\widetilde{\rho}_e J_{ee}]$ and the Fermi level density of states incorporating $W_{ee}$ is $\widetilde{\rho}_e=\rho_0/(1+\widetilde{W}_{ee}^2)$ as before. Hence, for the 1CK model we may write,
\begin{equation}\label{eq:5tk_1ck}
	T^e_{K}(J_{ee},W_{ee}) \simeq T^e_{K}(J_{ee},0)\times \exp[-\widetilde{W}^2_{ee}/\rho_0 J_{ee}] \;\,
\end{equation}
and this is confirmed in Fig.~\ref{fig:tm_1ck}$(c)$ for the same systems as in panel $(a)$.\\
Eq.~\ref{eq:5tm_ee_W_sd} can therefore be used to find the low-energy behaviour of the even-channel $\mathrm{T}$-matrix spectral function in the generalized 2CK model, while Eq.~\ref{eq:5tk_1ck} provides an estimate of the regime of applicability of this result. Remarkably, the low-energy scattering behaviour can therefore be extracted purely from a knowledge of the effective model parameters $J_{ee}$ and $W_{ee}$, and full solution of the generalized 2CK model is not required. \\
\noindent{Finally}, we use this information on the dynamics to extract the generic low-temperature conductance properties. We use the alternative formulation of the Oguri argument in  Eq.~\ref{eq:4GdcOguriEx} but now in the even/odd basis of Eq.~\ref{eq:eo} to write 
\begin{equation}
	\mathcal{G}^{dc}(\omega=0,T=0)= \frac{2e^2}{h}4|\pi\rho_0 T_{ee,\sigma}(0,0)-\pi\rho_0 T_{oo,\sigma}(0,0)|^2
\end{equation}
and using the form of the even/odd $\mathrm{T}$-matrix spectral function derived in this section, we obtain the \textbf{electrical conductance},
\begin{equation}\label{eq:5cond_2ck_sym}
	\boxed{
	\mathcal{G}^{dc}(\omega=0,T=0)=\left ( \frac{2e^2}{h}\right ) \frac{4(1+\widetilde{W}_{ee}\widetilde{W}_{oo})^2}{(1+\widetilde{W}_{ee}^2)(1+\widetilde{W}_{oo}^2)}}\;.
\end{equation}
This is the main result of this section: it allows the Kondo-renormalized low-temperature conductance to be calculated purely from a knowledge of the effective model parameters. Note that the result holds approximately for all $T \ll T^{FL}$. \\
In the case of \textit{sd-}symmetry as considered here, only $W_{ee}=W+W_{sd}$ and $W_{oo}=W-W_{sd}$ are required. Furthermore,  Eq.~\ref{eq:5cond_2ck_sym} implies the existence of an exact quantum-interference effect conductance node, 
\begin{equation}\label{eq:5QInode}
	\mathcal{G}^{dc}(\omega=0,T=0)=0\qquad\text{when}\qquad W_{sd}^2=W^2+(1/\pi\rho_o)^2 ~.
\end{equation}
In principle, this condition can be satisfied on tuning gate voltages in multi-orbital nanostructures. Although conductance nodes due to single-particle quantum interference in effective noninteracting systems are well-known \cite{Nazarov}, their many-body counterparts are comparatively poorly understood - even though strong electron interactions are ubiquitous in e.g.~semiconductor quantum dot devices and single molecule transistors. We emphasize that in such systems, electron interactions are crucial to understand the low-temperature conductance properties. Indeed, the $T=0$ conductance node, Eq.~\ref{eq:5QInode}, is driven by the Kondo effect and a finite conductance is expected at higher temperatures, $T\gg T_{K}$, see Fig.~\ref{fig:cond_2ck_sym}$(a)$ the orange curve opposed to the black one. This \textit{Kondo blockade} has been proposed as a mechanism for efficient quantum interference effect transistors \cite{Mitchell2017KondoMolecule}. The above analysis puts such phenomena in the context of a more general framework.\\ 
\noindent{Another} way to present the quantum-interference effect in the system it is via phase-shift. At $T=0$, $\delta_e\simeq +\pi/2 +\pi/2 \simeq +\pi$: the correction due finite potential scattering interferes \textit{constructively} with the Kondo singlet formation. As result, the conductance reaches its unitary limit - see black curve in Fig.~\ref{fig:cond_2ck_sym}$(a)$ and the point from effective model where curve $\mathcal{G}^{dc}(T=0)/\mathcal{G}_0 =2$ in $(b)$. At $T\gg T_{K}$, $\delta_e\simeq -\pi/2 +\pi/2 \simeq 0$: now the conductance curve presents an inverted behaviour to the standard Kondo enhanced curve. The unitary limit is found again but in high-energy scale - see orange curve Fig.~\ref{fig:cond_2ck_sym}$(a)$. At $T=0$,  $\delta_e\simeq0$ indicates the potential scattering interferes \textit{destructively} and the Kondo effect is completely destroyed by the correction from $\widetilde{W}_{ee}$ term.  See point from effective model where the curve vanishes in Fig.~\ref{fig:cond_2ck_sym}$(b)$.  \\ 
\noindent{The} quantitative accuracy of Eq.~\ref{eq:5cond_2ck_sym} is established by NRG results in Fig.~\ref{fig:cond_2ck_sym}. The \textit{dc-}conductance $\mathcal{G}^{dc}(\omega=0,T)$ of the full 2CK model,  Eq.~\ref{eq:5HeffCB}, under \textit{sd-}symmetry is obtained within NRG using the improved Kubo approach, Eq.~\ref{eq:4defKuboLRel}.
In Fig.~\ref{fig:cond_2ck_sym}$(a)$ we plot $\mathcal{G}^{dc}$ as a function of $T$ for fixed $J$, $J_{sd}$ and $W=0$, with different $W_{sd}$. Although the high-$T$ conductance is rather accurately reproduced by \ref{eq:5cond_qpc}, Kondo correlations strongly affect the behaviour at low temperatures $T\ll T_{K}$, with $T_{K}$ the emergent Kondo temperature. The horizontal dotted lines are the result of Eq.~\ref{eq:5cond_2ck_sym}, and are seen to agree very well with the low-temperature conductance from NRG. In particular, the Kondo resonance at $W_{sd}=0$ (black line) is inverted to a Kondo blockade for $W_{sd} = 0.6D$ (orange line). $\mathcal{G}^{dc}(0)$ is plotted for the same parameters in Fig.~\ref{fig:cond_2ck_sym}$(b)$ as a function of $W_{sd}$, with NRG results (points) compared with Eq.~\ref{eq:5cond_2ck_sym} (dashed line), showing the conductance node predicted by Eq.~\ref{eq:5QInode}. Fig.~\ref{fig:cond_2ck_sym}$(c)$ confirms the more general form of Eq.~\ref{eq:5cond_2ck_sym} for finite $W$ and $W_{sd}$.

\subsection*{Finite potential scattering $W_{\alpha\beta}\neq0$ in broken \textit{sd-}symmetry}
Finally, we consider the most general case, in which both $\mathbb{J}$ and $\mathbb{W}$ are finite, and \textit{sd-}symmetry is broken  i.e. $J_{-}\ne 0$ and $W_{-}\ne 0$ for $J_{\pm}=J_{ss}\pm J_{dd}$ and $W_{\pm}=W_{ss}\pm W_{dd}$.\\ 
First, we note that for arbitrary $J_{\pm}$, $J_{sd}$ and $W_{\pm}$, $W_{sd}$ the lead basis rotation matrix $\mathbb{U}$ that diagonalizes the exchange matrix $\mathbb{J}$ will not in general diagonalize the potential scattering matrix $\mathbb{W}$. That is, although $\mathbb{U}^{\dagger} \mathbb{J}\mathbb{U}$ is constructed such that $J_{eo}=J_{oe}^{\star}=0$, $\mathbb{U}^{\dagger} \mathbb{W}\mathbb{U}$ still yields finite $W_{eo}=W^{\star}_{oe}\ne 0$. This means that the even/odd channels defined in this way are not distinct; and in particular the odd channel does not completely decouple at low energies, as required for the previous analysis. In the remaining of this part we develop additional basis transformations to bring the couplings into a basis form suitable for channel decoupling under RG.\\
The first step is therefore to define a new basis, in terms of which we can identify \emph{distinct} conduction electron channels, here labelled as $\textit{a}$ and $\textit{b}$. 
We introduce the transformation matrix $\mathbb{P}$ from \textit{sd} into \textit{ab} basis and we define for the potential scattering in Eq.\ref{eq:5Heff_S0} the equation $\mathbb{\mathcal{W}}=\mathbb{P}^{\dagger} \mathbb{W}\mathbb{P}$ to be diagonal, such that,
\begin{subequations}\label{eq:5WeeWoo}
	\begin{align}
		\mathcal{W}_{aa} &=\tfrac{1}{2} (W_{+}+\sqrt{W_{-}^2+4|W_{sd}|^2}) ~, \label{eq:Wee}\\
		\mathcal{W}_{bb} &=\tfrac{1}{2}(W_{+}-\sqrt{W_{-}^2+4|W_{sd}|^2}) ~, \label{eq:Woo}
	\end{align}
\end{subequations}
and where $\mathcal{W}_{ab}=\mathcal{W}^{\star}_{ba}=0$ by construction. The Eq.~\ref{eq:5WeeWoo} is as such the potential scattering analogue of Eq.~\ref{eq:5JeeJoo} for the exchange couplings. Incorporating these local potentials into the definition of the free lead Hamiltonian in the $a/b$ basis of $\mathcal{H}_{leads}$ with $\alpha=a,b$,
 we have renormalized Fermi level densities of states, namely  $\widetilde{\rho}_{\alpha}=\rho_0/[1+(\pi\rho_0\mathcal{W}_{\alpha\alpha})^2]$. \\
We remark that the same transformation $\mathbb{P}$ is applied to the exchange matrix $\mathbb{J}$, noting that $\mathbb{\mathcal{J}}=\mathbb{P}^{\dagger} \mathbb{J}\mathbb{P}$ is \textit{not} in general diagonal. We may now identify the relevant dimensionless quantities to analyse the RG-flow as $j_{\alpha\beta}=\sqrt{\widetilde{\rho}_{\alpha}\widetilde{\rho}_{\beta}} \mathcal{J}_{\alpha\beta}$ and the $2\times 2$ matrix of these couplings is denoted $\mathbb{j}$.\\
As second step, we introduce the transformation matrix $\mathbb{Q}$ from \textit{ab} into \textit{eo} basis. Instead of applying this transformation to $\mathbb{\mathcal{J}}$ that does not have a diagonal form generally, we perform the final rotation of the lead basis to diagonalize $\mathbb{j}$ matrix and the expression $\mathbb{j}^{\prime}=\mathbb{Q}^{\dagger} \mathbb{j}\mathbb{Q}$ yields the desired even/odd exchange couplings $j^{\prime}_{ee}$ and $j^{\prime}_{oo}$. It is in \emph{this} basis that we can say that the more strongly coupled even channel flows to strong coupling and the more weakly coupled odd channel decouples, analogous to the analysis of the previous section. This assumes that $j^{\prime}_{ee}>j^{\prime}_{oo}$, which is the conventional situation. However, according to strongest antiferromagnetic coupling in the bare model, it is also possible that $j^{\prime}_{oo}>j^{\prime}_{ee}$, in which case the roles of the even and odd channels in the following are reversed.\\
The same transformation  $\mathbb{Q}$ is applied to the potential scattering and it is given by $\mathbb{W}^{\prime}=\mathbb{Q}^{\dagger} \mathbb{\mathcal{W}}\mathbb{Q}$ where we define the dimensionless quantity $\widetilde{W}^{\prime}_{\alpha\beta}=\pi\rho_0 W^{\prime}_{\alpha\beta}$. As outcome, $\mathbb{W}^{\prime}$ is in the desired even/odd but it is \textit{not} diagonal, hence $\widetilde{W}^{\prime}_{ee}=\widetilde{W}^{\prime}_{eo}=\widetilde{W}^{\prime}_{oe}=\widetilde{W}^{\prime}_{oo}\neq0$.\\
The emergent decoupling of the odd channel in this basis at low energies and temperatures $\omega, T \ll T_{K}$ yields the nontrivial result $t^{\prime}_{eo}(0,0)=t^{\prime}_{oe}(0,0)=0$. This is a consequence of Kondo renormalization, and scattering between even and odd channels is finite at higher energies. In the fully asymmetric case, there is no lead rotation that yields $t^{\prime}_{eo}(\omega,T)=0$ at all energies and temperatures. The above prescription, although somewhat complicated, is needed in order to identify the specific lead combination where inter-channel scattering vanishes at low energies. We have confirmed this result explicitly with NRG. \\
In this basis, the analogue of Eqs.~\ref{eq:5tm_oo_W_sd} and \ref{eq:5tm_ee_W_sd} for the $\mathrm{T}$-matrix spectra are $t^{\prime}_{oo,\sigma}(0,0)=(\widetilde{W}^{\prime}_{oo})^2/(1+(\widetilde{W}^{\prime}_{oo})^2)$ and $t^{\prime}_{ee,\sigma}(0,0)=1/(1+(\widetilde{W}^{\prime}_{ee})^2)$, with the corresponding complex $\mathrm{T}$-matrices given by 
\begin{equation}
	\begin{aligned}
		&\pi\rho_0 T^{\prime}_{oo,\sigma}(0,0)=[(\widetilde{W}^{\prime})^{-1}_{oo}-i]\times t^{\prime}_{oo,\sigma}(0,0) ~,\\
		&\pi\rho_0 T^{\prime}_{ee,\sigma}(0,0)=-[\widetilde{W}^{\prime}_{ee}+i]\times t^{\prime}_{ee,\sigma}(0,0) ~.
	\end{aligned}
\end{equation}
The \textbf{low-temperature conductance} then follows from alternative version of the Oguri formulation in Eq.~\ref{eq:4GdcOguriEx},
\begin{equation}\label{eq:5cond_gen}
	\boxed{
	\mathcal{G}^C(0,0) = \left (\frac{2e^2}{h}\right)  | \Omega_e^{\phantom{^{\prime}}}  T^{\prime}_{ee,\sigma}(0,0) + \Omega_o^{\phantom{^{\prime}}}  T^{\prime}_{oo,\sigma}(0,0)|^2} \;,
\end{equation}
where the inter-channel $\mathrm{T}$-matrix $T_{ds,\sigma}$ in the physical basis of \textit{d} and \textit{s} leads is brought first into the basis of \textit{a} and \textit{b} channels using the transformation $\mathbb{P}$, and then into the even/odd basis via $\mathbb{Q}$. This yields the coefficients,
\begin{equation}
	\Omega_{\gamma} = 2\pi\rho_0\sum_{\alpha,\beta \in a,b} P_{d\alpha}Q_{\alpha \gamma} P^{\star}_{s\beta}Q^{\star}_{\beta \gamma} \;.
\end{equation}
The low-temperature conductance obtained from NRG is found to agree well with these predictions. \\
We emphasize that the benefit of the above framework is that the low-temperature conductance can be accurately estimated, purely from a knowledge of the effective 2CK model parameters $\mathbb{J}$ and $\mathbb{W}$. The 2CK model itself does not need to be solved. Furthermore, since any microscopic multi-orbital system hosting a net spin-$\tfrac{1}{2}$ coupled to source and drain leads can in principle be mapped to such a model, the problem of calculating low-temperature transport becomes one of determining effective model parameters. These can be estimated perturbatively using Eq.~\ref{eq:5HeffCBPer}, or nonperturbatively using more sophisticated techniques, such as model machine learning \cite{rigo2020machine}.

\subsection{Spin $S=1$}
When the nanostructure hosts an even number of electrons, a high-spin state with $S=1$ may result. This is especially prevalent in single molecule junctions with Hund's coupling. The effective model in this case is again Eq.~\ref{eq:5HeffCB}, but now with $\hat{\mathbf{S}}$ a spin-$1$ operator whereas $\hat{\mathbf{s}}_{\alpha\beta}$ is still spin-$\tfrac{1}{2}$ operator for the leads.\\
In the case with \textit{exact} PC on the level of the bare model, an effective single-channel spin-$1$ Kondo model results namely Eq.~\ref{eq:5H1ck} with $\hat{\mathbf{S}}$ a spin-$1$ operator. The low-energy physics is that of the underscreened Kondo effect - a singular Fermi liquid with a residual free spin-$\tfrac{1}{2}$ local moment surviving down to $T=0$ \cite{Coleman}. Conductance is again Kondo-enhanced in this case, but with logarithmic corrections to the low-temperature approach to the fixed point conductance \cite{mehta2005regular}. This situation can arise in parallel double quantum dots; although the exact PC condition on the hybridization matrix is challenging to realize in practice.\\
The more generic scenario is when the effective nanostructure spin-$1$ is \textit{delocalized} and the PC condition is not satisfied. In this case, both $J_{ee}$ and $J_{oo}$ remain finite in the even/odd lead basis. The nanostructure spin-$1$ is exactly screened on the lowest temperature scales - where we assume both exchange couplings are antiferromagnetic. However, since $J_{ee}>J_{oo}$ we have $T_{K}^e > T_{K}^o$ and a \textit{two-stage Kondo} effect results similar to what we discuss in Sec.\ref{sec:2CK}. In this case, the odd lead participates in screening of the nanostructure spin-$1$, and remains coupled down to $T=0$, in contrast to the spin-$\tfrac{1}{2}$ case discussed above. The nanostructure spin-$1$ is underscreened by the even lead to an effective spin-$\tfrac{1}{2}$ on the scale of $T_{K}^e$, and then fully screened down to $S=0$ by the odd lead on the scale of $T_{K}^o$.  Correspondingly, the conductance is Kondo enhanced for $T_{K}^o < T < T_{K}^e$ but suppressed on the lowest temperature scales $T\ll T_{K}^o$. These scenarios have been explored experimentally in \cite{paaske2006non,roch2009observation,parks2010mechanical,kurzmann2021kondo}.\\

\noindent{As} conclusion, we summarize our findings and method developed across this section. We start with a general non-proportionate coupling model with a net ground-state spin $S=0,1/2,1$ at energy scale $T \ll E_{C}$. The original system is characterized by spin degeneracy. Besides the $S=0$ system which satisfies the PC condition at the level of the effective model $\hat{H}_{eff}$, the other two cases require the application of basis transformation from the bare system basis into suitable even/odd basis which allow independent RG-flow of coupling in even and odd channel sector. Then, we study the resulting effective model under RG and we find which energy scale the channel decoupling occurs. At characteristic temperature depending on the ground-state spin, the model reduces to single-channel state: it acquires an incipient PC property and it has renormalised coupling in correspondence to the strongest channel. In order to evaluate the electrical conductance, we need to reverse engineer the canonical transformation to bring back the renormalized effective model in even/odd basis into original basis. We calculate the $\mathrm{T}$-matrix spectral function and study the RG-flow of couplings using it. The conductance is studied by means of the alternative Oguri formula; it is the crucial quantity in the analysis of emergent PC in Coulomb blockade systems.

\section{Emergent proportionate coupling in mixed-valence regime}\label{sec:MVPC}
This section studies systems with degeneracy \textit{both} in spin $\hat{S}^{z}$ and in charge $Q$.\\ 
We consider the generic electrical conductance properties of systems near a Coulomb peak, corresponding to the charge degeneracy point between $N$ and $N+1$ electrons on the nanostructure i.e. states separated by unitary charge variation. Typically, one can tune to the Coulomb peaks in experiment by applying suitable plunger gate voltages \cite{cronenwett1998tunable}. We expect enhanced cotunnelling conductance at such points due to strong nanostructure charge fluctuations arising in the mixed-valence regime (MV). In the stability diagram, see sketch in Fig.\ref{F3:QD_VV}, the MV regime is found at the crossover between two well-defined charge states where we find simultaneously $V_{bias}\to0$ and finite $V_{gate}$, here denoted as $V^{\star}_{gate}$. \\
By contrast with the Coulomb blockade situation of the previous Sec.\ref{sec:CBPC} where electrons visit intermediate virtual excited states and spin-flip processes occur, here electrons can tunnel from source to drain leads without leaving the nanostructure ground-state manifold; although electron interactions still typically renormalize the conductance. The fact that at charge degeneracy points the electrons do not tunnel to excited states - on the contrary to the Coulomb blockade system - stems into the different order of perturbation expansion in $\hat{H}_{hyb}$.  \\ 
Effective models near charge-degeneracy points can be obtained perturbatively to \emph{first} order in the hybridization $\hat{H}_{hyb}$ using BWPT. Here, $\hat{H}_{hyb}$ connects two adjacent electron states of the isolated nanostructure i.e. $N$ and $N+1$ electrons state. There are no tunnelling events to other charge states to first-order in $\hat{H}_{hyb}$.
The suitable energy scale where the derived effective model is a solid description of the system physics is identified as follows. The temperature has to be $(i)$ smaller than the energy difference of states having the same charge $Q$ but belonging to different eigenstates $l$ i.e. $E^{Q}_{l+1}-E^{Q}_{l}$ and $(ii)$ smaller than the energy difference of states belonging to the same eigenstates but in two different charge sectors i.e. $E^{Q+1}_{l}-E^{Q}_{l}$. To sum up, we assume $T \ll min(E^{Q}_{l+1}-E^{Q}_{l}, E^{Q+1}_{l}-E^{Q}_{l})\doteq \Delta E_{min}$, where this minimum energy is the lowest excited state not included in $\widehat{P}_{0}$ projector on the ground-state manifold defined in Eq.\ref{eq:5defP} - note that in these expressions we omit $S^{z}$ index for conciseness. Hence, $Q=N-1$ or $Q=N+2$ states are not energetically reachable from the ground-state energy levels in the chosen temperature range.\\
To place our discussion on a general ground, in the ground-state manifold characterised by $\left\{ \ket{\Psi^{Q,S^z}_{0}} \right\}$ state and $\left\{E^{Q,S^z}_{0}\right\}$ energy, we consider first a nondegenerate model where there is a finite energy gap $\Delta E_{0}$ between the ground-state energy levels, namely 
\begin{equation}
	\begin{aligned}
		& E^{N,S^z}_{0}(V_{gate}) = \epsilon ~~~;~~~ E^{N+1,S^{z}\pm1/2}_{0}(V_{gate}) = \epsilon+\epsilon^{\prime}~, \\
		& \Rightarrow \Delta E_{0} (V_{gate})\stackrel{non-deg}{=}  E^{N+1,S^{z}\pm1/2}_{0}- E^{N,S^z}_{0}=\epsilon^{\prime} ~,
	\end{aligned}
\end{equation}   
where $\epsilon^{\prime}$ is a gate voltage-dependent energy splitting to differentiate the ground-state levels in the two charge sectors. Next, we take a system in MV regime: the gate voltage is tuned so that all the ground-state energy levels are not degenerate, namely
\begin{equation}
		 \Delta E_{0}(V^{\star}_{gate}) \stackrel{MV}{=} 0 
		 \quad \Leftrightarrow\quad \epsilon^{\prime}=0 ~.
\end{equation}
This brief consideration on the energy difference between ground-state energy levels is relevant because although the derivation developed through this section stems into the MV regime, the method is also valid for nondegenerate systems in the close vicinity at the MV degenerate point - as we discuss more later.\\
We now continue with the general structure of the BWPT calculation for models at the valence transition.

\subsection*{Analytical method: generalities}
The model Hamiltonian of reference in this work is defined by the two-channel system in Eq.\ref{eq:5H}. We now present the general expressions used to calculate the effective model in MV regime. Then, we detail those according to the spinless and spinful electrons systems and on the latter for two different charge sector transitions.\\ 
At the crossover between $\ket{N}$ and $\ket{N+1}$ charge sectors, we define the projector operator for a general MV transition as follows
\begin{equation}\label{eq:5defPMV}
	\widehat{P}_{0} = \sum_{S^{z}\in G.S.}\ket{N;S^{z};j=0}\bra{N;S^{z};j=0} \otimes \widehat{\mathbb{1}}_{leads} +\sum_{S^{z}\in G.S.}\ket{N+1;S^{z};j=0}\bra{N+1;S^{z};j=0} \otimes \widehat{\mathbb{1}}_{leads} ~,
\end{equation}
where we use the definition in Eq.\ref{eq:5defP}, with $j=0$ is the ground-state in given charge sector and $\sum_{S^{z}\in G.S.}$ is the summation over ground-state $S^{z}$ components for a given $Q$. The Eq.\ref{eq:5defPMV} is the projector operator onto the manifold of the \textit{retained} states in the $\ket{N}$ and $\ket{N+1}$ charge sector of $\hat{H}_{nano}$. We keep only the ground-state manifold characterised by $\ket{\Psi^{N,S^{z}}_{0}}$ state, $E^{N,S^{z}}_{0}$ energy and  $\ket{\Psi^{N+1,S^{z\prime}}_{0}}$ state, $E^{N+1,S^{z\prime}}_{0}$ energy.\\
By applying the projector defined in the MV given in Eq.\ref{eq:5defPMV} to the bare model in Eq.\ref{eq:5H}, we calculate the effective model: according to the energy range $T\ll \Delta E_{min}$, it is given by the $0^{th}$ and $1^{st}$ orders of Eq.\ref{eq:5Heff}. In particular, the $0^{th}$-order is trivially obtained by:
\begin{equation}\label{eq:5P0Leads}
	\widehat{P}_{0}\hat{H}_{leads}\widehat{P}_{0} \equiv\hat{H}_{leads} ~,
\end{equation}
because the lead operators commute with the nanostructure states and
\begin{equation}\label{eq:5P0Imp}
	 \widehat{P}_{0}\hat{H}_{nano}\widehat{P}_{0}=  \epsilon^{\prime}\big(\ket{N+1}\bra{N}\ket{N}\bra{N+1} \big) \equiv \epsilon^{\prime} \ket{N+1}\bra{N+1} ~,
\end{equation} 
with $\epsilon^{\prime}=0$ precisely at the MV charge degeneracy point.\\
From Eqs.\ref{eq:5P0Leads},\ref{eq:5P0Imp}, we see that the actual form of $\hat{H}_{nano}$ does not enter in the projection calculation; only the relative splitting between states in the ground-state manifold. Furthermore, as we mentioned before, the perturbative scheme is not only applicable to MV regime: the whole calculation is valid also for nondegenerate systems near the MV point. When we present the various results for spinless and spinful models under MV regime, the generalization to nondegenerate model is given by simply adding the energy constant term $\epsilon^{\prime}$ to the effective MV expression.\\ 
The actual demanding calculation is the $1^{st}$-order of Eq.\ref{eq:5Heff}: $\widehat{P}_{0}\hat{H}_{hyb}\widehat{P}_{0}$. Recalling that we can only sandwich states that are separated by a unitary charge variation - hence tunnelling between states with same charge is zero - the projection for the hybridization term takes the general form:
\begin{equation}\label{eq:5P0Hyb}
	\begin{aligned}
\widehat{P}_{0}\hat{H}_{hyb}\widehat{P}_{0} =\sum_{S^{z},S^{z\prime}\in G.S.} \Big( \ket{N;S^{z}}\bra{N+1;S^{z\prime}}&\big(\bra{N,S^{z}}\hat{H}_{hyb}\ket{N+1;S^{z\prime}}\big) +\\
		&+\ket{N+1;S^{z\prime}}\bra{N;S^{z}}\big(\bra{N+1;S^{z\prime}}\hat{H}_{hyb}\ket{N;S^{z}}\big)  \Big) ~,
		\end{aligned}
\end{equation}
where all states are taken to be the $j=0$ ground-state for each $Q,S^{z}$ sector.\\
Hence, the complete effective Hamiltonian in the MV regime we aim to calculate reads:
\begin{equation}\label{eq:5HeffMVPer}
	\begin{aligned}
\hat{H}_{eff} = \hat{H}_{leads} &+ \epsilon^{\prime}\sum_{S^{z}\in G.S.}	\ket{N+1;S^{z}}\bra{N+1;S^{z}} +\\
&+\sum_{S^{z},S^{z\prime}\in G.S.}\Big(\bra{N;S^{z}}\hat{H}_{hyb}\ket{N+1;S^{z\prime}}\ket{N,S^{z}}\bra{N+1;S^{z\prime}}\big) +H.c. \Big)+\mathcal{O}(\hat{H}_{hyb})^{2}~.
\end{aligned}
\end{equation}
with $\epsilon^{\prime}=0$ precisely at the MV charge degeneracy point.\\
The remaining of this part is devoted to the calculation of Eq.\ref{eq:5P0Hyb}, according to the prominent systems in case of spinless and spinful models

\subsection{Spinless model}
When the nanostructure and the leads are effectively spinless - which may arise physically in the presence of a strong magnetic field as we have in the general two-channel model in Eq.\ref{eq:5H} - the effective model Hamiltonian in Eq.\ref{eq:5Heff} takes a simple form as we discuss in this section.\\
We use the same  model Hamiltonian in Eq.\ref{eq:5H} but we introduce some simplification for the case of spinless system in MV regime. We consider $\hat{H}_{hyb}$ to consist of tunnelling processes between the $0^{th}$-sites on the semi-infinite chain baths and the \textit{frontier orbitals} of the nanostructure. For a nanostructure of $m=1,\dots,M$ orbitals, we adopt the convention in which the lead $\alpha=s$ couples to $m=1$ and $\alpha=d$ couples to $m=M$.
The corresponding hybridization term reads as $\hat{H}_{hyb}=V_{s1} c^{\dagger}_{s0}d_{1} + V^{\star}_{s1}d^{\dagger}_{1}c_{s0} 
+V_{dM} c^{\dagger}_{d0}d_{M} + V^{\star}_{dM}d^{\dagger}_{M}c_{d0}$. Using the general definition of the charge-basis state given in Eq.\ref{eq:5Q-BasisState-spin-Def}, the \textbf{spinless charge basis states} now read:
\begin{equation}
	\ket{\phi^{Q}_{l}} =\ket{n_{1}}_{1} \bigotimes \ket{n_{2}}_{2} \bigotimes \ket{n_{3}}_{3} \bigotimes \dots \bigotimes \ket{n_{M}}_{M}~,
\end{equation}
with the unique combination of occupations $\left\{ n_{i}\right\}$ consistent with $Q=\sum_{j}n_{j}$ indexed by $j$, see Eq.\ref{eq:5QSzprop}. The eigenstates of $\hat{H}$ are then constructed as linear combination of these states:
\begin{equation}\label{eq:5Q-BasisState-spinless-Def}
\ket{\Psi^{Q}_{l}}=\sum_{j} c_{Q,l}(j)\ket{\phi^{Q}_{l}}~,\quad \text{with}~~c_{Q,l}(j)\equiv c_{Q}(n_{1},n_{2},\dots,n_{M})~,
\end{equation}
where summation $\sum_{j}$ involves all combinations of $\left\{  n_{i} \right\}$ consistent with the property $Q=\sum_{j}n_{j}$ in Eq.\ref{eq:5QSzprop} and $l=0$ is taken to be the ground-state in a given charge sector $Q$. \\
The two energy-degenerate ground-state with $N$ and $N+1$ electrons are $\ket{\Psi^{Q=N}_{0}} \equiv \ket{N},\ket{\Psi^{Q=N+1}_{0}} \equiv \ket{N+1}$. Those states satisfy the equations:
\begin{equation}
	\begin{aligned}
 &\hat{H}\ket{\Psi^{Q=N}_{0}}\equiv\hat{H}\ket{N}= E^{N}_{0}\ket{N} ~,\\
 &\hat{H}\ket{\Psi^{Q=N+1}_{0}}\equiv\hat{H}\ket{N+1}= E^{N+1}_{0}\ket{N+1} ~.
  \end{aligned}
\end{equation} 
The \textbf{spinless projector operator} $\widehat{P}_{0}$ under MV regime, see definition in Eq.\ref{eq:5defPMV}, now reads:
\begin{equation}\label{eq:5P0-spinless}
	\widehat{P}_{0} \doteq \ket{N}\bra{N} \otimes \widehat{\mathbb{1}}_{leads} +\ket{N+1}\bra{N+1} \otimes \widehat{\mathbb{1}}_{leads} ~.
\end{equation}
It is useful to introduce the operators in the reduced ground-state manifold of the nanostructure:
\begin{equation}\label{eq:5def-fnospin}
		f^{\dagger}=\ket{N+1} \bra{N} \quad;\quad f=\ket{N}\bra{N+1} ~,
\end{equation}
which are Fermionic operators satisfying the anti-commutation relations:
\begin{equation}
	\lbrace f,f\rbrace = 0 \quad , \quad \lbrace f^{\dagger},f^{\dagger}\rbrace = 0 \quad , \quad \lbrace f,f^{\dagger}\rbrace = \mathbb{1} ~,
\end{equation}
and those commute with the lead operators i.e. $[ f,c_{\alpha0} ] = 0$.  We note that the Fermionic operators in Eq.\ref{eq:5def-fnospin} are constructed over the nanostructure ground-state basis. Hence, those describe the physical process of unitary charge variation on a given $\ket{\Psi^{Q}_{0}}$ state and we can refer to them as \textbf{charge operators}. This interpretation is relevant when it comes to recognise the derived effective derived models under PC condition - as we see later in this chapter. We can use the $f^{\dagger},f$ operators in Eq.\ref{eq:5def-fnospin} to rewrite the spinless projector in Eq.\ref{eq:5P0-spinless} as
\begin{equation}
	\widehat{P}_{0} = (ff^{\dagger})\otimes \widehat{\mathbb{1}}_{leads} + (f^{\dagger}f)\otimes \widehat{\mathbb{1}}_{leads}.
\end{equation}
We observe that the physical process described by the $f^{\dagger},f$ operators is a single-transmission between the leads and the nanostructure ground-state of a spinless electron. Hence, $f^{\dagger},f$ operators represent only two possible transmission events separated by a unitary charge variation i.e. either empty or one electron state.\\
At this point, we have ready the analytical structure to implement degenerate perturbation calculations for a spinless model and we proceed with presenting its results.

\subsection*{Analytical result: the effective spinless Hamiltonian}
In this part we want to calculate the effective low-energy spinless Hamiltonian $\hat{H}_{eff}$ as expressed in Eq.\ref{eq:5HeffMVPer}. As discussed at the beginning of this section, the projection operation over $\hat{H}_{leads}$ and $\hat{H}_{nano}$ is directly given in Eq.\ref{eq:5P0Leads} and Eq.\ref{eq:5P0Imp}, respectively. Moreover, the expression for $\widehat{P}_{0}\hat{H}_{hyb}\widehat{P}_{0}$ requires the calculation of the matrix elements appearing in Eq.\ref{eq:5P0Hyb}, where the nanostructure operators act on the spinless charge basis states defined in Eq.\ref{eq:5Q-BasisState-spinless-Def}. The outcome of this projection calculation reads:
\begin{equation}
	\widehat{P}_{0}\hat{H}_{hyb}\widehat{P}_{0} =
	\left( c^{\dagger}_{s0} \widetilde{V}_{s} + c^{\dagger}_{d0} \widetilde{V}_{d} \right)\ket{N}\bra{N+1}
	+\ket{N+1}\bra{N} \left( \widetilde{V}^{\star}_{s}c_{s0}  + \widetilde{V}^{\star}_{d}c_{d0}  \right) ~,
\end{equation}
where the \textit{effective low-energy hybridization parameters} of the model $\widetilde{V}_{\alpha}$ now appear as consequence of the projection operation with expressions
\begin{equation}
\widetilde{V}_{s} \doteq V_{s1} \bra{N}d_{1}\ket{N+1} \quad,\quad \widetilde{V}_{d} \doteq V_{dM} \bra{N}d_{M}\ket{N+1} ~,
\end{equation}
where these expressions can be expressed in terms of matrix elements of the operators $d_{1},d_{M}$ in the product basis $\left\{ \phi_{l}^{Q}\right\}$. In particular,  $\widetilde{V}_{\alpha}~ \propto~ V_{\alpha m}c_{Q}(n_{1},n_{2},\dots,n_{M})n_{m}$,  $\widetilde{V}^{\star}_{\alpha} ~\propto~ V^{\star}_{\alpha m}  c_{Q}(n_{1},n_{2},\dots,n_{M})(1-n_{m})$ and $\widetilde{V}_{d},\widetilde{V}^{\star}_{d}$ include an extra $(-1)^Q$ factor due to anticommuting $M$ sites. 
These effective couplings are obtained from the physical hybridization strength $V_{\alpha m}$ defined in the bare model in Eq.\ref{eq:5H} multiplied by matrix element in the eigenbasis for the ground-states connected by $d_{1},d_{M}$, that is   $(i)$ the expansion constants $c_{Q}(\dots)$ defined in Eq.\ref{eq:5Q-BasisState-spinless-Def} and  $(ii)$ a combination of occupation number $n_{m}$ to show the anticommuting property of operators on the states.\\
Before proceeding with the derivation of $\hat{H}_{eff}$, we observe that the $\widetilde{V}_{\alpha}$ factors can be readily used to calculate the total levelwidth function defined in Eq.\ref{eq:DoSinfWBL} but now dressed by the effective hybridization. Assuming the standard scenario of equivalent metallic leads with constant density of states within the wide-flat band approximation, we have $\rho_{0,s}=\rho_{0,d}\equiv \rho_{0}$ and the \textbf{spinless effective levelwidth} function reads:
\begin{equation}\label{eq:5Gamma_Eff_spinless}
\Gamma^{\alpha}_{eff} =\pi |\widetilde{V}_{\alpha}|^{2} \rho_{0} ~.
\end{equation}
Ultimately, the electrical conductance is only governed by the effective coupling $\widetilde{V}_{\alpha}$ terms and those emerge as straightforward result from the applied perturbative scheme.\\
We return now to the derivation of the effective model: by inserting all these results in Eq.\ref{eq:5HeffMVPer}, we obtain the \textbf{effective low-energy spinless Hamiltonian in MV regime}:
\begin{equation}\label{eq:5Heff-spinless}
	\begin{aligned}
\hat{H}_{eff} & \stackrel{MV}{=} \hat{H}_{leads} + 
		\left( c^{\dagger}_{s0} \widetilde{V}_{s} + c^{\dagger}_{d0} \widetilde{V}_{d} \right)f 
		+f^{\dagger}  \left( \widetilde{V}^{\star}_{s}c_{s0}  + \widetilde{V}^{\star}_{d}c_{d0}  \right) ~,\\
		&\stackrel{non-deg}{=} \hat{H}_{leads} + 
		\left( c^{\dagger}_{s0} \widetilde{V}_{s} + c^{\dagger}_{d0} \widetilde{V}_{d} \right)f 
		+f^{\dagger}  \left( \widetilde{V}^{\star}_{s}c_{s0}  + \widetilde{V}^{\star}_{d}c_{d0}  \right) + \epsilon^{\prime}f^{\dagger}f ~,
	\end{aligned}
\end{equation} 
where we give both the charge degenerate MV regime and the nondegenerate case, using the $f$ operators definition in Eq.\ref{eq:5def-fnospin} and the effective hybridizations.\\
Before proceeding any further, we compare the derived model in Eq.\ref{eq:5Heff-spinless} with the original model presented in Eq.\ref{eq:5H}. The nanostructure region is initially a finite, interacting chain representing a multi-orbital configuration. As outcome of the projection, the whole nanostructure region is identified as an \textit{effective, noninteracting single-site} by the $f,f^{\dagger}$ operators. As consequence, the physical tunnelling matrices assume their effective version $\widetilde{V}_{\alpha}$ as we discuss above. And the spinless $f,f^{\dagger}$ operators correctly describe a single-transmission between the semi-infinite lead chains and the single-effective nanostructure.\\
We also note that Hamiltonian expression as written in Eq.\ref{eq:5Heff-spinless} reflects the fundamental requirement for proportionate coupling condition as discussed in Sec.\ref{sec:PC}. The effective nanostructure is coupled directly to both lead chains and the system satisfies the proportionality equation as given in Eq.\ref{eq:defPC} but now written for the effective levelwidth function $\Gamma^{\alpha}_{eff}$ defined in  Eq.\ref{eq:5Gamma_Eff_spinless}.\\
By means of these observations, the Hamiltonian in Eq.\ref{eq:5Heff-spinless} can be \textit{factorised} and we can combine both physical $s,d$ leads into a unified effective one. This is achieved by the canonical lead transformation $c_{o}=[\widetilde{V}_{s}c_{so}+\widetilde{V}_{d}c_{do}]/\widetilde{V}$ with $\widetilde{V}^{2}=\widetilde{V}^{2}_{s}+\widetilde{V}^{2}_{s}$.  
This \textit{single} lead couples with a \textit{unique} effective hybridization $\widetilde{V}$ that is the total \textit{spinless effective hybridization for the PC model in MV regime}. These manipulation brings the effective model in Eq.\ref{eq:5Heff-spinless} into the \textbf{effective Hamiltonian for a spinless PC model in MV regime}:
\begin{equation}\label{eq:5Heff-spinless-PC}
	\boxed{\hat{H}_{eff} \stackrel{MV}{=} \hat{H}_{leads} +
		c^{\dagger}_{0} \widetilde{V}f +  f^{\dagger}\widetilde{V}^{\star}c_{0}}~,
\end{equation} 
where this low-energy Hamiltonian is fully determined by the $0^{th}$-site of the combined lead chain and the $f^{\dagger},f$ operators on the effective nanostructure, mediated by an effective PC hybridization $\widetilde{V}$. The Eq.\ref{eq:5Heff-spinless-PC} is most important result of the spinless section: we have shown that performing BWPT calculation on a spinless interacting model under its MV condition, we obtain an effective Hamiltonian that has an emergent PC property. We conclude by saying that it is straightforward to generalize $\hat{H}_{eff}$ for the nondegenerate charge system by adding to Eq.\ref{eq:5Heff-spinless-PC} the term $\epsilon^{\prime}f^{\dagger}f\equiv(E^{N+1}_{0}-E^{N}_{0})\hat{n}_{f}$.

\subsection*{Spinless effective model as resonant-level model}
So far, we have developed an operative scheme to derive an effective model satisfying PC condition at low-energy scale while starting from a microscopic model that does not satisfy such a property. Now, we continue our discussion on the type of model the Eq.\ref{eq:5Heff-spinless-PC} can be mapped to.\\
Recalling the definition on the noninteracting resonant level model given in Eq.\ref{eq:NI-RLM}, we recognize the analogies between our derived effective model in Eq.\ref{eq:5Heff-spinless-PC} and the NI-RLM. If we release the spin index in the NI-RLM Hamiltonian in Eq.\ref{eq:NI-RLM}, the expression constitutes a two-level-charge system characterised by tunnelling processes with unitary charge variation between a unique bath chain and an nanostructure. Hence, the spinless NI-RLM is a proper \textit{two-level system}: the tunnelled electron can either be on one bath or on the nanostructure \\
Thus, the derived Hamiltonian for a spinless mixed-valence model in proportionate coupling in  Eq.\ref{eq:5Heff-spinless-PC} is recognised as \textbf{noninteracting resonant level model} as defined in Eq.\ref{eq:NI-RLM}, \cite{Schlottmann_SpinlessMVintoRLM_1980}. \\
As last observation, we notice that the mapped effective model shows specific features - for instance the PC property and the lack of Coulomb interaction - arising from the perturbative BWPT scheme. \\
In conclusion,  Eq.\ref{eq:5Heff-spinless-PC} shows PC property - opposed to the bare model. This incipient PC condition greatly facilitates quantum transport calculation in the low energy and temperature scales where the effective model $\hat{H}_{eff}$ provides a solid description. Furthermore, Eq.\ref{eq:5Heff-spinless-PC} - and its mapped system - is noninteracting despite of the strong electron correlations typically present on the physical bare nanostructure. We make use of these emergent properties in the next part where we calculate the  electrical conductance.

\subsection*{Analytical result: electrical conductance for spinless mixed-valence model}
Due to the incipient PC property appearing in the spinless effective model in Eq.\ref{eq:5Heff-spinless-PC} and to the mapping $\hat{H}_{eff} \equiv \hat{H}^{NI-RLM}$, the effective model is exactly solvable. In particular, we can write a single-particle description for the electrical conductance formula. \\
Considering $\hat{H}_{eff}$, the noninteracting retarded correlator of the effective $f$ single-level reads as $G^{R}_{0,ff}(\omega,T)=\lim_{\eta\rightarrow0^{+}}1/(\omega+i\eta -\epsilon^{\prime}-\Gamma^{\alpha}_{eff})$. Hence, we can apply the MW formula under PC in Eq.\ref{eq:MW->PC} to calculate the electrical conductance for the spinless $\hat{H}_{eff}$ by means of its effective parameters. The current expression then reads:
\begin{equation}
I=\frac{e}{h} \int d\omega \big( f^{\mu_{L}} (\omega)-  f^{\mu_{R}}(\omega \big) Tr [\widetilde{\mathbb{T}}_{PC}(\omega,T)] ~,
\end{equation}
where we use $\widetilde{\mathbb{T}}_{PC}(\omega,T)$ generalised transfer matrix under PC in Eq.\ref{eq:GenTransfMatPC} for standard equivalent metallic leads in wide-flat band approximation - the explicit expression for the noninteracting RLM reads as:
\begin{equation}
	\begin{aligned}
\widetilde{\mathbb{T}}_{PC}(\omega,T) =&4\frac{\Gamma^{s}_{eff}\Gamma_{eff}^{d}}{(\Gamma_{eff}^{s}+\Gamma_{eff}^{d})} \frac{\Gamma^{s}_{eff}+\Gamma_{eff}^{d}}{\Gamma_{eff}^{s}+\Gamma_{eff}^{d}}(-\mathit{Im}G^{R}_{0,ff}(\omega,T)) ~,\\
&=4\frac{\Gamma_{eff}^{s}\Gamma_{eff}^{d}}{(\Gamma_{eff}^{s}+\Gamma_{eff}^{d})^{2}}(\Gamma_{eff}^{s}+\Gamma_{eff}^{d})\left(- \lim_{\eta\rightarrow0^{+}} \frac{-(\eta+\Gamma_{eff}^{s}+\Gamma_{eff}^{d})}{(\omega-\epsilon^{\prime})^{2}+(\eta+\Gamma_{eff}^{s}+\Gamma_{eff}^{d})^{2}}   \right) ~, \\
&=4 \frac{\Gamma_{eff}^{s}\Gamma_{eff}^{d}}{(\omega-\epsilon^{\prime})^{2}+(\Gamma_{eff}^{s}+\Gamma_{eff}^{d})^{2}} ~,
	\end{aligned}
\end{equation}
where we use the effective gamma function definition in Eq.\ref{eq:5Gamma_Eff_spinless} and $\epsilon^{\prime}=E^{N+1}_{0}-E^{N+1}_{0}$ to account for nondegenerate charge states. To conclude, the \textbf{MW formula under PC $\hat{H}_{eff}$ spinless model} is equal to:
\begin{equation}\label{eq:5Gcspinless}
\boxed{
I=\frac{e}{h} \int d\omega \big( f^{\mu_{L}} (\omega)-  f^{\mu_{R}}(\omega \big) Tr \left(4 \frac{\Gamma_{eff}^{s}\Gamma_{eff}^{d}}{(\omega-\epsilon^{\prime})^{2}+(\Gamma_{eff}^{s}+\Gamma_{eff}^{d})^{2}}\right)} ~,
\end{equation}
whose spectral function - the term inside round parenthesis - is described by a Lorentzian function centred at $\omega=\epsilon^{\prime}$ with full-width at half-max equal to $\Gamma_{eff}^{s}+\Gamma_{eff}^{d}$. Moreover, the conductance is generally decaying as $\Gamma_{eff}^{\alpha} ~ \propto ~ |\widetilde{V}_{\alpha}|^{2}$. We now discuss the possible scenarios occurring for Eq.\ref{eq:5Gcspinless}.\\
At strictly zero energy and temperature scales i.e. $\omega=T=0$ and in full MV regime i.e. $\epsilon^{\prime}\equiv 0$, we can use the MW current formula in Eq.\ref{eq:5Gcspinless} under linear response to calculate the differential conductance:
\begin{equation}
\mathcal{G}^{dc}(0,0)=\frac{e^{2}}{h} \left( 4\frac{|\widetilde{V}_{s}|^{2}|\widetilde{V}_{d}|^{2}}{(|\widetilde{V}_{s}|^{2}+|\widetilde{V}_{d}|^{2})^{2}}\right) \equiv \mathcal{G}_{0}\widetilde{\Gamma}~,
\end{equation}
where we use the definition in Eq.\ref{eq:5Gamma_Eff_spinless}. From this expression, the conductance is fully determined by the effective hybridization parameters in the definition of the geometric factor $\widetilde{\Gamma}$ arising from BWPT - which ultimately are related to the bare hybridization of the bare model, see Eq.\ref{eq:defGeomFactor}. Furthermore, we obtain the \textit{maximum} conductance value in the spinless system. The value is enhanced to $e^2 /h$ in case of symmetric effective hybridizations $\widetilde{V}_{s} \equiv \widetilde{V}_{d}$ which does not necessarily imply the symmetry in the bare hybridization $V_{s1} \equiv V_{dM}$. By means of the current expression, we can also identify the MV region extending around the crossover point between $N$ and $N+1$ charge sector for a radius equal to $\Gamma_{eff}^{s}+\Gamma_{eff}^{d} ~ \propto ~ |\widetilde{V}_{s}|^{2}+|\widetilde{V}_{d}|^{2}$. \\
At $\omega$, $T\simeq0$ and $\epsilon^{\prime}\simeq0$, the small energy and temperature regimes allow to study the system in the vicinity of the charge degenerate crossover with radius again equal to $\Gamma_{eff}^{s}+\Gamma_{eff}^{d}$ - hence, we regard the system as in \textit{pseudo}-MV regime. The corresponding spectral function has the same half-width at half-max of the previous case but now it is centred at $\epsilon^{\prime}$.\\ 
In the general situation for finite $\omega,T$ and nondegenerate charge system $\epsilon^{\prime}\neq0$, the system is outside the MV area identified by the radius $\Gamma_{eff}^{s}+\Gamma_{eff}^{d}$. The conductance is derived from the MW formula in Eq.\ref{eq:5Gcspinless} where we need to solve the integration over the whole energy spectrum. The electrical conductance is now function of: 
\begin{equation}
\mathcal{G}^{C}(\omega,T)= \mathit{f} \Big\lbrace   \epsilon^{\prime}, \widetilde{V}_{s},\widetilde{V}_{d}  \Big\rbrace ~,
\end{equation}
which follows directly from Eq.\ref{eq:5Gcspinless}.

\subsection{Spin model: singlet-doublet transition}
A more generic and interesting situation arises in spinful systems where states of $\hat{H}_{nano}$ form a spin multiplet structure. Here we focus on the most common case where the ground state of the nanostructure with an \textit{even} number of electrons $N$ is a degenerate spin-singlet state, while with odd $N+1$ electrons we have a degenerate spin-doublet state. Both singlet and doublet must be taken into account in the effective model near the charge degeneracy point between $N$ and $N+1$ electrons.\\
In the derivation of the effective model for singlet-doublet valence transition, we use the general two-channel model defined in Eq.\ref{eq:5H} with \textbf{spin charge basis state} introduced in Eq.\ref{eq:5Q-BasisState-spin-Def}. As it will be clear from the derivation, we insert in the bare hybridization also the spin label i.e. $V_{\alpha m \sigma}$ such that we can define the coupling for each spinful channel of the system. The bare model presents $3$-fold degeneracy including only $\ket{\Psi^{N,S^z=0}_{0}},\ket{\Psi^{N+1,S^z=\pm1/2}_{0}}$ - as consistent with the energy regime $T \ll \Delta E_{min}$ considered here and with the order of perturbation expansion $\mathcal{O}(\hat{H}_{hyb})^{2}$. The energy-degenerate charge sectors $N$ and $N+1$ defined on the nanostructure ground-state manifold satisfy the equations:
\begin{equation}
	\begin{aligned}
		&\hat{H}\ket{\Psi^{Q=N,S^z=0}_{0}}\equiv\hat{H}\ket{N;S^z=0}= E^{N,S^z=0}_{0}\ket{N;S^{z}=0} ~,\\
		&\hat{H}\ket{\Psi^{Q=N+1,S^z=\pm1/2}_{0}}\equiv\hat{H}\ket{N+1;S^z=\pm1/2}= E^{N+1,S^z=\pm1/2}_{0}\ket{N+1;S^z=\pm1/2} ~.
	\end{aligned}
\end{equation} 
and extensions to higher spin multiplet ground-states is discussed at the end of the chapter although, in practice, the $T\to 0$ transport of our interest is typically controlled by the \textit{lowest} spin multiplet in $N$ and $N+1$ electron sectors, and we now focus only on these situations.\\
The \textbf{spin projector operator} $\widehat{P}_{0}$ in the MV regime, see definition in Eq.\ref{eq:5defP}, now reads:
\begin{equation}\label{eq:5P0-spin}
		\widehat{P}_{0} \doteq \ket{N;0} \bra{N;0} \otimes \widehat{\mathbb{1}}_{leads} +\sum_{S^z=\pm1/2}\ket{N+1;S^z} \bra{N+1;S^z} \otimes \widehat{\mathbb{1}}_{leads} ~.
\end{equation}
In analogy to the spinless case, it is helpful to introduce the spinful operators:
\begin{equation}\label{eq:5def-fspin}
	f^{\dagger}_{\sigma}=\ket{N+1; \sigma}\bra{N;0} \quad;\quad f_{\sigma}=\ket{N;0}\bra{N+1; \sigma} ~,
\end{equation}
with $\sigma=\uparrow$ for $S^z=+1/2$ and $\sigma=\downarrow$ for $S^z=-1/2$. These \textbf{charge operators} are again Fermionic operators satisfying the anti-commutation relations $\lbrace f_{\sigma},f_{\sigma^{\prime}}\rbrace = 0~,~\lbrace f^{\dagger}_{\sigma},f^{\dagger}_{\sigma^{\prime}}\rbrace = 0~,~\lbrace f_{\sigma},f^{\dagger}_{\sigma^{\prime}}\rbrace = \mathbb{1}$ and the commutation relation with lead operators i.e. $\lbrace f_{\sigma},c_{0 \sigma} \rbrace = 0$. We also observe that the $f^{\dagger}_{\sigma},f_{\sigma}$ conserve the spin. With this respect, these operators are understood as like the $\hat{S}^{z}$-spin operator. 
This demonstrates the fact that  $\hat{H}_{hyb}$ returns non-vanishing matrix elements only if it is calculated between two different charge states separated by $\Delta Q = \pm1$ but with spin conservation during the whole transmission process. Analogously to the spinless case, we use $f^{\dagger}_{\sigma},f_{\sigma}$ operators to rewrite the spin projector in Eq.\ref{eq:5P0-spin} as:
\begin{equation}\label{eq:5P0-with-f-spin}
	\widehat{P}_{0} = \sum_{\sigma \in \lbrace \uparrow,\downarrow \rbrace}
	(f_{\sigma}f^{\dagger}_{\sigma})\otimes \widehat{\mathbb{1}}_{leads} + (f^{\dagger}_{\sigma}f_{\sigma})\otimes \widehat{\mathbb{1}}_{leads} ~.
\end{equation} 
We continue in the next part to implement the BWPT calculation on the spinful MV model.

\subsection*{Analytical result: the effective spin Hamiltonian}
We aim now to calculate the effective low-energy spin Hamiltonian $\hat{H}_{eff}$ as expressed in Eq.\ref{eq:5HeffMVPer}. The projection on $\hat{H}_{leads}$ is trivially given by Eq.\ref{eq:5P0Leads} and on $\hat{H}_{nano}$ by Eq.\ref{eq:5P0Imp} - just now introducing also the spin label. We focus on $\widehat{P}_{0}\hat{H}_{hyb}\widehat{P}_{0}$ where we need to compute the matrix elements appearing in Eq.\ref{eq:5P0Hyb}, where the nanostructure operators act on the spin charge basis states defined in Eq.\ref{eq:5Q-BasisState-spin-Def}. The outcome of the projection calculation reads as:
\begin{equation}
	\widehat{P}_{0}\hat{H}_{hyb}\widehat{P}_{0} = \!\!\!
	\sum_{\sigma \in \lbrace \uparrow, \downarrow \rbrace} \sum_{\alpha=s,d} 
	\left( c^{\dagger}_{\alpha0\sigma} \widetilde{V}_{\alpha\sigma}\ket{N;0}\bra{N+1;\sigma}
	+\ket{N+1;\sigma}\bra{N;0}  \widetilde{V}^{\star}_{\alpha\sigma}c_{\alpha0\sigma} \right) ~,
\end{equation}
where the \textit{effective low-energy hybridization parameters} of the model $\widetilde{V}_{\alpha\sigma}$ per $\alpha$ channel with $\sigma$ spin are emergent parameters of arising from the perturbation calculation. These effective elements are defined by the physical hybridization strength $V_{\alpha m \sigma}$ from the bare spinful model - here accounting all the degree of freedom - multiplied by matrix elements, namely
\begin{equation}\label{eq:5Hyb-Eff-Bare_spin}
\widetilde{V}_{s \sigma} \doteq V_{s1 \sigma} \bra{N;0}d_{1\sigma}\ket{N+1;\sigma} \quad,\quad \widetilde{V}_{d \sigma} \doteq V_{dM \sigma} \bra{N;0}d_{M\sigma}\ket{N+1;\sigma} ~,
\end{equation}
where these matrix elements can again be expressed in terms of the original product state basis for the actual evaluation. In particular, $\widetilde{V}_{\alpha\sigma}~ \propto~ V_{\alpha m\sigma} c_{Q}(n_{1},n_{2},\dots,n_{M})n_{m}(\Pi_{y=1}^{m-1} (-1)^{n_{y\uparrow}+n_{y\downarrow}})$ and  $\widetilde{V}^{\star}_{\alpha\sigma} ~\propto~ V^{\star}_{\alpha m\sigma} c_{Q}(n_{1},n_{2},\dots,n_{M})(1-n_{m})(\Pi_{y=1}^{m-1} (-1)^{n_{y\uparrow}+n_{y\downarrow}})$ where the matrix elements are expressed by $(i)$ the expansion constants $c_{Q}(\dots)$ given in Eq.\ref{eq:5Q-BasisState-spin-Def} and $(ii)$ a string of occupation numbers $n_{m \sigma}$ according to the anticommuting property of the Fermionic spinful operators on the charge state - here according to the spin indexing order in Eq.\ref{eq:5Q-BasisState-spin-Def} where each $m^{th}$-site is filled first by a spin up electron and then by a spin down electron.\\
As for the spinless case, it is straightforward to implement the \textbf{spinful effective levelwidth} function using the effective hybridization $\widetilde{V}_{\alpha \sigma}$ in Eq.\ref{eq:5Hyb-Eff-Bare_spin}, namely:
\begin{equation}\label{eq:Gamma_Eff_spin}
 \Gamma^{\alpha}_{eff,\sigma}= \pi |\widetilde{V}_{\alpha \sigma}|^{2}\rho_{0} ~,
\end{equation}
and the electrical conductance is controlled only by these effective parameters arising from the BWPT calculation.\\
We continue now with the derivation of the effective model: by inserting all the above results in Eq.\ref{eq:5HeffMVPer}, we calculate the \textbf{effective low-energy spin Hamiltonian in MV regime}:
\begin{equation}\label{eq:5Heff-spin}
	\begin{aligned}
	\hat{H}_{eff} & \stackrel{MV}{=} \hat{H}_{leads}  
	+ \sum_{\sigma \in \lbrace \uparrow, \downarrow \rbrace} \sum_{\alpha=s,d}
	\left( c^{\dagger}_{\alpha0\sigma} \widetilde{V}_{\alpha \sigma} f_{\sigma} +
	f^{\dagger}_{\sigma} \widetilde{V}^{\star}_{\alpha \sigma}c_{\alpha0\sigma} \right) +B_{eff} \hat{S}^{z}_{f} ~, \\
	&\stackrel{non-deg}{=} \hat{H}_{leads}+ \sum_{\sigma \in \lbrace \uparrow, \downarrow \rbrace} \sum_{\alpha=s,d}
	\left( c^{\dagger}_{\alpha0\sigma} \widetilde{V}_{\alpha \sigma} f_{\sigma} +
	f^{\dagger}_{\sigma} \widetilde{V}^{\star}_{\alpha \sigma}c_{\alpha0\sigma} \right) +B_{eff} \hat{S}^{z}_{f}+ \epsilon^{\prime} \hat{n}_{f}   ~,
	\end{aligned}
\end{equation}
where we use Eqs.\ref{eq:5def-fspin},\ref{eq:5Hyb-Eff-Bare_spin}, the $z$-component of the total nanostructure spin $\hat{S}^z_f=\tfrac{1}{2}(\hat{n}_{f\uparrow}-\hat{n}_{f\downarrow})$ for $\hat{n}_{f_{\sigma}}= f_{\sigma}^{\dagger}f_{\sigma}^{\phantom{\dagger}}$ and $B_{ eff}=(E^{N+1;\uparrow}_{0}-E^{N+1;\downarrow}_{0})$. In the second line for nondegenerate charge system we use $\hat{n}_f=\hat{n}_{f\uparrow}+\hat{n}_{f\downarrow}$ and $\epsilon^{\prime}=\tfrac{1}{2}(E^{N+1;\uparrow}_{0}-E^{N+1;\downarrow}_{0})-E^{N;0}_{0}$. We remark that the Eq.\ref{eq:5Heff-spin} ascribes a \textit{noninteracting} model. Before we proceed with the derivation, in analogy to the discussion of the spinless model,  the interacting multi-orbital nanostructure is condensed to a unique effective nanostructure with $f^{\dagger}_{\sigma},f_{\sigma}$ operators. The effective hybridizations $\widetilde{V}_{\alpha \sigma}$ in Eq.\ref{eq:5Hyb-Eff-Bare_spin} couple this unique site, that is the lowest energy single-orbital, with the $0^{th}$-site of both chain leads. \\
By following the same steps we did for the spinless system, we would attempt to directly factorise the terms in Eq.\ref{eq:5Heff-spin}. However, we cannot apply this yet because the spin states set is now \textit{enlarged} with respect to the bare system states. We need to remove the extra spurious spin states under appropriate energy constraints - as we discuss now.\\
According to the energy regime $T\ll\Delta E _{min}$ as we assume for the MV modelling, states $Q=N-1$ or $Q=N+2$ are neither energetically favourable, nor are appearing within $\mathcal{O}(\hat{H}_{hyb})^{2}$ order of perturbation expansion from the ground-states to first order. We denote then the three states $\ket{N;0}$, $\ket{N+1;\pm\tfrac{1}{2}}$ described by the bare model in Eq.\ref{eq:5H} as \textit{admissible states} of system. However, we have four effective states of the spinful \textit{f}-orbital, namely:
\begin{equation}
\ket{0}_{f} \quad , \quad \ket{\uparrow}_{f}~,~\ket{\downarrow}_{f}\quad , \quad \ket{\uparrow\downarrow}_{f} ~.
\end{equation} 
We must eliminate the spurious state $\ket{\uparrow\downarrow}_{f}$.\\
The $\hat{H}_{eff}$ in Eq.\ref{eq:5Heff-spin} has a $4$-fold  degeneracy and that is \textit{not} consistent with single-electron transmission within energy scale $T\ll\Delta E _{min}$, which gives $3$-fold  degeneracy MV states.
In order to eliminate the extra unwanted states that lead to energetically unfavoured tunnelling events, we introduce a \textit{fictitious on-site Coulomb potential} as a constraint, $U_{c}$.
We add this to the derived effective model in Eq.\ref{eq:5Heff-spin} that is:
\begin{equation}\label{eq:5Heff-spin-int}
	\hat{H}^{\prime}_{eff} = \hat{H}_{eff} + U_{c} \hat{n}_{f_{\uparrow}} \hat{n}_{f_{\downarrow}} ~.
\end{equation}
The Eq.\ref{eq:5Heff-spin-int} describes the effective single Fermionic degree of freedom with operators $f^{\dagger}_{\sigma},f_{\sigma}$ defining the three accessible states $\lbrace \ket{0}, \ket{\uparrow}, \ket{\downarrow} \rbrace$ under the \textit{energy constraint}  $U_{c}\rightarrow +\infty$ to ensure $\braket{\hat{n}_{f_{\uparrow}} \hat{n}_{f_{\downarrow}}} = 0$. Thus, the effective \textit{interacting} model in Eq.\ref{eq:5Heff-spin-int} returns the correct $3$-fold  degeneracy consistent with the bare model spin degeneracy. \\
We note that after implementing the Coulomb potential constraint, $\hat{H}^{\prime}_{eff}$ is subjected to an additional energy scale. Next to $T \ll \Delta E_{min}$ required for applying the perturbative scheme up to order $\mathcal{O}(\hat{H}_{hyb})^{2}$, we now have also $k_{B}T \ll U_{c}$ to project out spurious states in Eq.\ref{eq:5Heff-spin} such as the doubly occupied one. This condition is always satisfied for $U_{c}\to\infty$ in the effective model in Eq.\ref{eq:5Heff-spin-int}, or implemented in practice. It is indeed known that the Coulomb interaction and the energy level spacing between states at same $Q$ charge sector control the conductance curve for a given temperature \cite{MeirWingreenLee_StronglyInteractingElec_PRL1991}. \\ 
Having restored the correct ground-state degeneracy in the effective model by means of introducing the fictitious constraint $U_c$ and taking its limit to infinity, we proceed with the factorization of $\hat{H}^{\prime}_{eff}$ in Eq.\ref{eq:5Heff-spin-int} to formulate an effective single-channel model as previously,
\begin{equation}
	c_{0\sigma}=\frac{1}{\widetilde{V}_{\sigma}}\left( \widetilde{V}_{s \sigma}c_{s0\sigma} +\widetilde{V}_{d \sigma}c_{d0\sigma}\right) ~,
\end{equation}
with
\begin{equation}
\widetilde{V}_{\sigma}^{2}=\widetilde{V}_{s \sigma}^{2}+\widetilde{V}_{d \sigma}^{2} ~.
\end{equation}
As before, we define $\Gamma_{eff,\sigma}=\pi |\widetilde{V}_{\sigma}|^{2}\rho_{0}$ that is the \textit{spinful effective hybridization for the PC model in the singlet-doublet MV transition}.\\
At the end of these manipulation on Eq.\ref{eq:5Heff-spin-int} we obtain the \textbf{effective Hamiltonian for a spinful PC model in singlet-doublet transition}:
\begin{equation}\label{eq:5Heff-spin-PC}
	\boxed{\hat{H}^{\prime}_{eff} = \hat{H}_{leads} +
		\sum_{\sigma \in \lbrace \uparrow,\downarrow \rbrace} \left( c^{\dagger}_{0 \sigma} \widetilde{V}_{\sigma}f_{\sigma} +  f^{\dagger}_{\sigma}\widetilde{V}^{\star}_{\sigma}c_{0\sigma} \right)
		+ U_{c} \hat{n}_{f_{\uparrow}} \hat{n}_{f_{\downarrow}} } ~,
\end{equation}
with the constraint $U_{c}\rightarrow + \infty $. The low-energy Hamiltonian in Eq.\ref{eq:5Heff-spin-PC} is fully determined by the $0^{th}$-site of the combined lead chain and the $f^{\dagger}_{\sigma},f_{\sigma}$ operators on the effective nanostructure, mediated by the effective PC hybridization $ \widetilde{V}_{\sigma}$. This is the most important result of this part: under the appropriate energy condition, the effective model calculated from BWPT presents an emergent PC property. It is straightforward to  generalize $\hat{H}^{\prime}_{eff}$ model to nondegenerate charge states by adding a term $+\epsilon^{\prime}\sum_{\sigma}f^{\dagger}_{\sigma}f_{\sigma}$ with $\epsilon^{\prime}=\tfrac{1}{2}(E^{N+1;\uparrow}_{0}+E^{N+1;\downarrow}_{0})-E^{N;0}_{0}$.

\subsection*{Spinful effective model as Anderson impurity model}
In the previous part we calculated the effective model satisfying proportionate coupling condition starting from a microscopic model that does not satisfy such a property. To complete the analysis, it remains to identify a mapping for Eq.\ref{eq:5Heff-spin-PC}.\\
According to the discussion in Sec.\ref{sec:QImpModel}, the infinitely strong interacting effective model $\hat{H}^{\prime}_{eff}$ in Eq.\ref{eq:5Heff-spin-PC} is equivalent to spinful \textbf{infinite-U Anderson impurity model} in the MV transition between singlet and doublet charge sector. By means of the large $U$ limit, tunnelling to any multi-occupied states are inhibited and the effective model can be manipulated into a form that satisfies PC condition. This property is of great advantage for conductance calculation, as we discuss in the next part.

\subsection*{Analytical and NRG results: electrical conductance for spinful singlet-doublet mixed-valence model}
An important consequence of the emergent PC in $\hat{H}^{\prime}_{eff}$ is that the quantum transport at low energy $T\ll \Delta E_{min}$ can be obtained using the MW formula under PC - as we calculate below. \\
In the derived effective model, the interacting retarded correlator of the effective single Fermionic degree of freedom is, in analogy to Eq.\ref{eq:G_dd}, given by $G^{R}_{ff\sigma}(\omega,T) =\lim_{\eta \rightarrow 0^{+}} 1/(\omega+i\eta -\epsilon^{\prime}-\Gamma_{eff,\sigma}-\Sigma_{ff\sigma}(\omega,T))$ where $\Sigma_{ff\sigma}(\omega,T)$ is the $f$-level interaction self-energy according to the standard equivalent single metallic lead model under wide-flat conduction band limit. Unlike for the previous spinless case, here the dynamics are nontrivial due to the constraint $U_{c}\to\infty$. However, $G^{R}_{ff\sigma}(\omega,T)$ can be calculated numerically using NRG and hence, the quantum transport properties can be obtained. This requires \textit{only} the solution of the effective model, rather than the full bare model. Reference results for the effective model can therefore be calculated and reused for any microscopic system.\\
We start the derivation of MW current formula under PC for the singlet to doublet valence transition by writing its expression in Eq.\ref{eq:MW->PC} as
\begin{equation}
	I=\frac{2e}{h} \int d\omega \big( f^{\mu_{L}} (\omega)-  f^{\mu_{R}}(\omega \big) Tr [\widetilde{\mathbb{T}}_{PC}(\omega,T)] ~,
\end{equation} 
for $\widetilde{\mathbb{T}}_{PC}(\omega,T)$ the generalised transfer matrix in PC, see Eq.\ref{eq:GenTransfMatPC}, now for the model Hamiltonian in Eq.\ref{eq:5Heff-spin-PC} is given by the expression
\begin{equation}
	\widetilde{\mathbb{T}}_{PC}(\omega,T) =4\sum_{\sigma} \frac{|\widetilde{V}_{s \sigma}|^{2}|\widetilde{V}_{d \sigma}|^{2}}{(|\widetilde{V}_{s \sigma}|^{2}+|\widetilde{V}_{d \sigma}|^{2})^{2}} t_{ff,\sigma}(\omega,T) ~,
\end{equation}
where the $\mathrm{T}$-matrix spectral function defined in Eq.\ref{eq:TmatSpe} now for the effective single Fermionic degree of freedom reads as 
\begin{equation}
	\begin{aligned}
&t_{ff,\sigma}(\omega,T) = \Gamma_{eff} \left(-\mathit{Im}G^{R}_{ff\sigma}(\omega) \right) ~,\\
& t_{ff,\sigma}(\omega=0,T=0) \stackrel{FL}{=} \frac{ \Gamma_{eff,\sigma}^{2}}{(\epsilon^{\star})^{2}+\Gamma^{2}_{eff,\sigma}}	~,
\end{aligned}
\end{equation}
where in the second line we use $G^{R}_{ff\sigma}(\omega,T)$ correlator for Fermi liquid system characterised by interaction-renormalised single Fermionic level energy  $\epsilon^{\star}=\epsilon^{\prime}+{\mathit{Re}}\Sigma_{ff}(\omega\to0,T\to0)$, and vanishing  $\mathit{Im}\Sigma_{ff}(\omega\to0,T\to0)\to0$ - see discussion for Eq.\ref{eq:G_ddFL}.
We remark that the Fermi liquid results are exact \textit{only} at $\omega=T=0$ where $t_{ff,\sigma}(0,0)$ spectra is described by a Lorentzian curve centred at $\epsilon^{\star}$ and having half-width at half-max as $\Gamma_{eff,\sigma}$. Hence, any deviation from zero temperature and energy scales presents $t_{ff,\sigma}(\omega,T)$ spectra with deformed Lorentzian shape due to non-negligible $\mathit{Im}\Sigma_{ff}(\omega,T)$. \\

\noindent{In} the absence of a magnetic field in the bare model, we have $E^{N+1;\uparrow}_{0}=E^{N+1;\downarrow}_{0}$ such that the effective field vanishes, $B_{eff}=0$, and the effective tunnelling amplitudes become independent of the spin label $\sigma$ hence $\widetilde{V}_{\alpha \sigma}\equiv \widetilde{V}_{\alpha}$ and $\Gamma_{eff,\sigma}\equiv\Gamma_{eff}$.\\
The $\mathrm{T}$-matrix spectral function is identified by the following expression
\begin{equation}
 t_{ff}(\omega=T=0) = \lim_{\omega\rightarrow 0} \frac{ (\Gamma_{eff})^{2}}{(\omega-\epsilon^{\star})^{2}+(\Gamma_{eff})^{2}}  ~,
\end{equation}
We notice that $t_{ff}(\omega=0,T=0)$ is the core quantity of interest for the electrical conduction calculation and we can evaluate it easily once we know both the effective hybridization $\Gamma_{eff}$ - as we derive already in Eq.\ref{eq:5Hyb-Eff-Bare_spin} - and the renormalised $\epsilon^{\star}$ value. Once the effective model is fully derived, $\epsilon^{\star}$ still remains an unknown quantity we are left to calculate and about it we discuss more below. However, for the infinite-$U$ metallic AIM, $\Sigma_{ff}(0,0)$ and hence $\epsilon^{\star}$ are completely determined by $\Gamma_{eff}$ and $\epsilon^{\prime}$.  Therefore, the conductance in the MV spinful model is function of:
\begin{equation}
	\mathcal{G}^{C}(\omega, T \to0)= \mathit{f} \Big\lbrace\epsilon^{\prime};\Gamma_{eff}\Big\rbrace ~.
\end{equation}  
In conclusion, we can use the \textbf{MW current formula} for $\hat{H}_{eff}$ spinful model to calculate the \textbf{conductance} at $T=0$ and vanishing $\omega$, namely
\begin{equation} \label{eq:5Gcspin_FL}
	\boxed{ \begin{aligned}
&\mathcal{G}^{dc}(0,0)=\frac{2e^{2}}{h} \left(4\frac{|\widetilde{V}_{s}|^{2}|\widetilde{V}_{d }|^{2}}{(|\widetilde{V}_{s }|^{2}+|\widetilde{V}_{d}|^{2})^{2}} t_{ff}(\omega=T=0)\right)  ~,\\
&t_{ff}(\omega=T=0) = \frac{1	}{1+(\epsilon^{\star})^{2}/(\Gamma^{s}_{eff}+\Gamma^{d}_{eff})^{2} } 
\end{aligned} } ~.
\end{equation}
In particular, we can use the Friedel sum rule result, see Eq.\ref{eq:FSR}, to write the spectra $t_{ff}(\omega=T=0)$ in term of the excess charge of the single Fermionic degree of freedom, namely
\begin{equation}
t_{ff}(\omega=T=0) =\sin^{2}\left( \frac{\pi}{2}\hat{n}_{f}(T=0)\right)  ~.
\end{equation}
The $\hat{n}_{f}$ quantity undergoes to renormalization under RG-flow and it presents a highly non-trivial behaviour, hence the above equation is not ideal for practical implementations without a full solution of $\hat{H}_{eff}$. For this reason, it is more convenient to compute using NRG the spectra $t_{ff}(\omega=T=0)$ defined by the effective parameters as derived from the BWPT. We comment that a numerical demonstration of the accuracy of Eq.\ref{eq:5Gcspin_FL} in a complex impurity model - the benzene molecule - is given in Chapter \ref{ch:benz}. What we find is the analytical prediction is perfectly verified by NRG data. \\ 
In case we need to compute the under linear response at $T=0$ but finite $\omega$, the $\mathrm{T}$-matrix spectral function expression to use is as follows,
\begin{equation}\label{eq:5Gcspin_OmFinite}
	t_{ff}(\omega,T\to0) = \sum_{\alpha} \frac{(\Gamma^{\alpha}_{eff})^{2}}{(\omega-\epsilon^{\star})^{2}+( \Gamma^{\alpha}_{eff})^{2}} ~.
\end{equation}
\begin{figure}[H]
	\centering
	\includegraphics[width=0.85\linewidth]{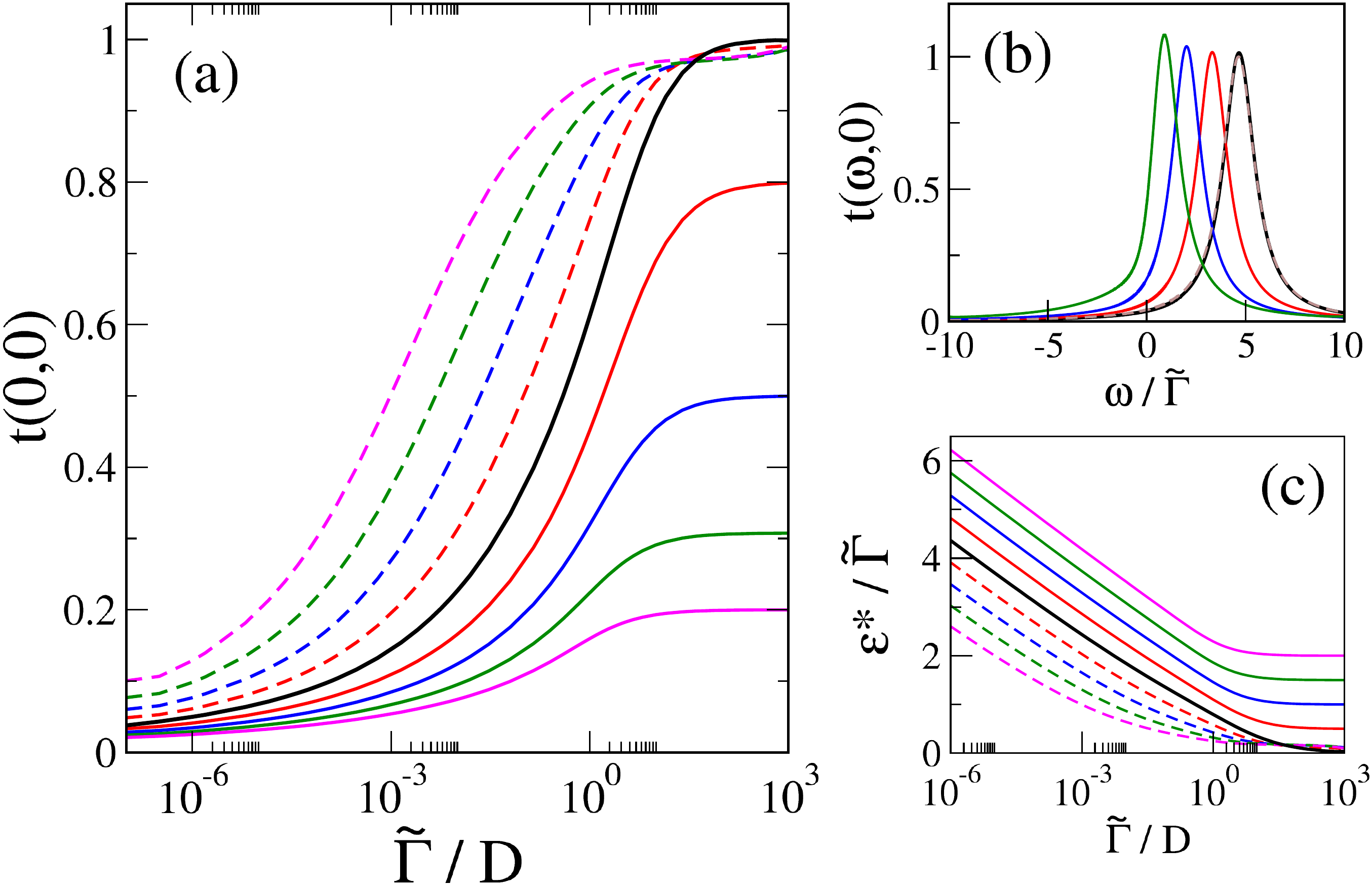}
	\caption[$\mathrm{T}$-matrix spectral function for valence singlet-doublet multiplet ]{$\mathrm{T}$-matrix for the effective model Eq.\ref{eq:5Heff-spin-PC} at $T=0$, describing the dynamics in the vicinity of a singlet-doublet charge degeneracy point, calculated from NRG. $(a)$ $t(0,0)$ as a function of $\tilde{\Gamma}$ for $\epsilon^{\prime}=0$ at charge degeneracy (black line) and $|\epsilon^{\prime}|/\tilde{\Gamma}=0.5, 1, 1.5, 2$ (red, blue, green and magenta lines) with $\epsilon^{\prime}>0$ shown as solid lines and $\epsilon^{\prime}<0$ as dashed lines. $(b)$ $t(\omega,0)$ vs $\omega/\tilde{\Gamma}$ at $\epsilon^{\prime}=0$ charge degeneracy point for $\tilde{\Gamma}/D=10^{-1},10^{-3},10^{-5},10^{-7}$ for the green, blue, red and black lines. $t(\omega,0)$ in Eq.~\ref{eq:5Gcspin_OmFinite} is shown as the dashed line for comparison to black line. $(c)$ Behaviour of $\epsilon^{\star}/\tilde{\Gamma}$ as a function on $\tilde{\Gamma}$ for the same parameters as in panel $(a)$, showing small $\tilde{\Gamma}/D\ll 1$ asymptotes following Eq.~\ref{eq:5MV_sd_eps}.}\label{fig:tmatrix_MV_sd}
\end{figure}
\noindent{In} Fig.~\ref{fig:tmatrix_MV_sd}, we now consider the spectrum of $\mathrm{T}$-matrix $t_{ff}(\omega,T\to0)$ for the effective model in Eq.\ref{eq:5Heff-spin-PC} with $B_{eff}=0$ and $U_{c}\to\infty$ obtained by full NRG calculation. For given effective parameters $\epsilon^{\prime}$ and $\Gamma_{eff}$, the low-$T$ transport can then be obtained via Eqs.\ref{eq:5Gcspin_FL},\ref{eq:5Gcspin_OmFinite}.\\ 
We start with the charge degenerate system for $\epsilon^{\prime}=0$ in the valence transition between singlet and doublet spin multiplet, as we show in panel $(b)$ of Fig.~\ref{fig:tmatrix_MV_sd} for different effective $\Gamma_{eff}$ - in the plot published in the paper \cite{transport} this is indicated as $\tilde{\Gamma}$ so we adopt this notation for the sake of clarity from now on. According to the ratio $\tilde{\Gamma}/D$, we distinguish two opposite regimes. At $\tilde{\Gamma}/D \ll 1$, $t_{ff}(\omega,T\to0)$ spectrum is accurately approximated by that of an equivalent NI-RLM with $U_{c}=0$ under Fermi liquid description. The results of $t_{ff}(\omega,T\to0)$ in Eq.\ref{eq:5Gcspin_OmFinite} are shown for comparison as dashed lined in Fig.~\ref{fig:tmatrix_MV_sd}$(b)$. On the contrary, at $\tilde{\Gamma}/D \gg 1$, discrepancies from the Lorentzian form of $t_{ff}(\omega,T\to0)$ in Eq.\ref{eq:5Gcspin_OmFinite} becomes more pronounced at finite frequency, due mainly to $\mathit{Im}\Sigma_{ff}(\omega,T\to0) \neq 0$ leading to an enhanced peak height and asymmetric shape.\\    
As we mentioned previously, once the effective parameters are derived, the only remaining quantity we need to calculate in order to determine the electrical conductance is the interaction-renormalised energy $\epsilon^{\star}$. Hence, it is useful to find its asymptotic behaviour as function of the effective $\tilde{\Gamma}$ as we show in Fig.~\ref{fig:tmatrix_MV_sd}$(c)$. In the physically-relevant regime $\tilde{\Gamma}/D\ll 1$, the behaviour is found to be simply,
\begin{equation}\label{eq:5MV_sd_eps}
	\epsilon^{\star}/\tilde{\Gamma} \simeq a\epsilon^{\prime}/\tilde{\Gamma} - b\log(\tilde{\Gamma}/D) +c \;,
\end{equation}
where numerically we find $a\simeq 0.93$, $b\simeq 0.29$ and $c\simeq 0.35$.
By contrast, at large $\tilde{\Gamma}/D\gg 1$, we have two scenarios according to the sign of $\epsilon^{\prime}$. If $\epsilon^{\prime}>0$, the self-energy correction can be neglected such that $\epsilon^{\star} \to \epsilon^{\prime}$ and the Kondo effect is suppressed. If  $\epsilon^{\prime}<0$, the Kondo effect renormalises the effective energy giving $\epsilon^{\star}\to 0$. This is indeed confirmed directly by the saturation values and the low/high $\tilde{\Gamma}$ regime behaviour of $\epsilon^{\star}$, which yields $t_{ff}(\omega=T=0)$ exactly from Eq.~\ref{eq:5Gcspin_FL}, as presented in Fig.~\ref{fig:tmatrix_MV_sd}$(c)$.\\
The quantity $t_{ff}(\omega=T=0)$ is required to compute the zero energy and temperature scales conductance and reference NRG results for it as a function of $\tilde{\Gamma}$ are presented in Fig.~\ref{fig:tmatrix_MV_sd}$(a)$ for different $\epsilon/\tilde{\Gamma}$. 
At $\tilde{\Gamma}/D\ll 1$ for any $\epsilon^{\prime}$, the spectra $t_{ff}(\omega=T=0)\simeq0$. At $\tilde{\Gamma}/D\gg 1$, for $\epsilon^{\prime} \leq 0$, $t_{ff}(\omega=T=0)\simeq1$ whereas at $\epsilon^{\prime}>0$, for increasing $\epsilon^{\prime}$ value we have decreasing $t_{ff}(\omega=T=0)$. At intermediate $\tilde{\Gamma}$ regime, the spectra of $t_{ff}(\omega=T=0)$ representing states away from the valence singlet-doublet multiplet transition, spread around the charge degeneracy curve - in black line - where maximum conductance is measured. In panel $(b)$ we verify an important observation: there is a one-to-one correspondence between the effective hybridizations and the spectra $t_{ff}(\omega=T=0)$. This is precisely the crucial advantage of computing the conductance by means of effective quantities rather than solving the microscopic model per se.  \\
Equipped with these results, the low-temperature conductance of a system in the vicinity of a singlet-doublet charge-degeneracy point can be straightforwardly predicted, once the effective model parameters $\epsilon^{\prime}$ and $\widetilde{V}_{\alpha}$ have been determined.

\subsection{Numerical study for serial configuration of two-impurity with single-orbital model: example of singlet-double mixed-valence case}
So far, in the modelling of systems under MV regime, we always implicitly consider a single-site on each lead coupled to the $m^{th}$ degree of freedom on the nanostructure via $V_{\alpha m \sigma}$ hybridization. Prior to any form of diagonalization, this bare hybridization matrix is composed by finite entries in each matrix element. We remark that the $m$-label stands for either an impurity site or an orbital in more complex structures.\\ 
To complete our examination of emergent proportionate coupling MV systems, we consider now a nanostructure with two-orbital and we couple it with leads in non-PC set-up. Under this condition, a \textit{unique} correspondence between effective and bare couplings can be established. In this configuration, we intend to have one unique $V_{s\sigma}$ coupling between the source lead and nanostructure characterised by a single-orbital and similarly for $V_{d\sigma}$ on the drain lead side. \\
In this section, starting from a single-orbital system whose microscopic model does not show PC property, we aim to numerically compute its effective model showing emergent new properties and using those to compute the electrical conductance. We pin down the discussion considering a \textit{serial configuration of two-impurity} which is by construction lacking of  PC property, see more discussion on PC vs no-PC in Sec.\ref{sec:PC}. We chose this impurity system to be in MV regime. In the following, we give the derivation with explicit numerical values. Extension to multiple-impurity set-up follows as direct generalization. \\ 
We start with the two-impurity model Hamiltonian, namely:
\begin{equation}
	\begin{aligned}
		\hat{H}= \hat{H}_{leads} &+
		\sum_{\sigma, \sigma^{\prime} \in \lbrace \uparrow,\downarrow \rbrace} \sum_{\beta=\lbrace 1,2\rbrace} \big(
		\epsilon_{d_{\beta}} d^{\dagger}_{\beta\sigma} d_{\beta\sigma}+ U_{\beta} \hat{n}_{d\beta \sigma} \hat{n}_{d\beta \sigma\prime} + t(d^{\dagger}_{1\sigma}d_{2\sigma}+d^{\dagger}_{2\sigma}d_{1\sigma})
		+V_{gate} \hat{n}_{d_{\beta} \sigma} \big) +\\
		&+ \sum_{\sigma \in \lbrace \uparrow,\downarrow \rbrace}
		\Big(   V_{s \sigma}  c^{\dagger}_{ s\sigma}d_{1\sigma}  + 
		V_{d\sigma} c^{\dagger}_{ d\sigma}d_{2\sigma} + H.c. \Big) ~,
	\end{aligned}
\end{equation}
where we explicitly take $\epsilon_{d 1} \neq \epsilon_{d 2}$ to avoid further charge degeneracy in the model and particle-hole symmetry regime meaning that $\epsilon_{d_{\beta} } = - U_{\beta}/2$. The \textit{intrinsic two-impurity asymmetry factor} is given only by $\nu= V_{d\sigma}/V_{s\sigma}$. \\
We follow now the protocol detailed in Sec.\ref{sec:MVPC} to derive the effective model with an emergent proportionate coupling property from the above two-impurity single-orbital model Hamiltonian in MV regime. At $V_{gate}=0$, the lowest state is a collective two-impurity spin singlet state. We tune $V_{gate}$ to the MV transition point between this singlet state and a state with a single electron on the two-impurity structure giving a doublet state.\\
The two-impurity structure has basis states in $N=2$ electrons sector,
\begin{equation}
	\ket{\uparrow,\uparrow} \quad,\quad \ket{\uparrow,\downarrow}\quad,\quad \ket{\downarrow,\uparrow}\quad,\quad \ket{\downarrow,\downarrow} ~,
\end{equation}
and in $N=1$ electrons sector,
\begin{equation}
	\ket{\uparrow,-} \quad,\quad \ket{\downarrow,-}\quad,\quad \ket{-,\uparrow}\quad,\quad \ket{-,\downarrow} ~.
\end{equation}
This many-body basis describes states in the isolated nanostructure, which is then used to construct the matrix Hamiltonian of $\hat{H}_{nano}$.
Upon diagonalization, we obtain the normalized eigenvectors numerically.
Furthermore, we can identify the lowest spin multiplet and at which $V_{gate}$ value it crossed another spin multiplet state - this is the MV point. \\
In our numerical analysis we use as bare parameters $\epsilon_{d1}=-10~eV,~U_{1}=20~eV$,  $\epsilon_{d2}=-6~eV,~U_{2}=12~eV$, hopping $t=1.5~eV$. We find the two lowest energy-degenerate two-impurity ground-state are the charge $N=1,~N+1=2$ sectors and their crossover occurs at $V^{\star}_{gate}=6.06~eV$. To sum up, the energy degenerate eigenbasis states defining the two-impurity ground-state read as:
\begin{equation}
	\begin{aligned}
		&\ket{\Psi^{N=1,S^z=\sigma}_{0}} = -0.95\ket{\sigma,-} +0.32\ket{-,\sigma} ~, \\
		&\ket{\Psi^{N=2,S^z=0}_{0}}  = -0.70 \big(\ket{\uparrow,\downarrow} -\ket{\downarrow,\uparrow} \big) + 0.13 \big(\ket{\uparrow\downarrow,-} -\ket{-,\uparrow\downarrow}\big) ~,
	\end{aligned} 
\end{equation}
for $\sigma=\uparrow,\downarrow$ .\\
By means of these degenerate ground-state, in analogy to Eqs.\ref{eq:5defP},\ref{eq:5defPMV}, we define the projector operator of the two-impurity model, namely:
\begin{equation}
\widehat{P}_{0} = \sum_{\sigma}\left(\ket{\Psi^{N=1,S^z=\sigma}_{0}}\bra{\Psi^{N=1,S^z=\sigma}_{0}}+\ket{\Psi^{N=2,S^z=0}_{0}} \bra{\Psi^{N=2,S^z=0}_{0}}  \right) ~,
\end{equation}
and from it we can define $f^{\dagger}_{\sigma},f_{\sigma}$ Fermionic operators analogously to Eq.\ref{eq:5def-fspin}. \\
Once we complete the projection operation up to $\mathcal{O}(\hat{H}_{hyb})^{2}$, the resulting effective model is the same as Eq.\ref{eq:5Heff-spin}, just with different spin label for $f^{\dagger}_{\sigma},f_{\sigma}$ and leads operators, 
\begin{equation}
	\hat{H}_{eff} = \hat{H}_{leads}  
	+ \sum_{\sigma,\sigma^{\prime} \in \lbrace \uparrow, \downarrow \rbrace} \sum_{\alpha=s,d}
	\left( c^{\dagger}_{\alpha0\sigma} \widetilde{V}_{\alpha} f_{\sigma^{\prime}} +
	f^{\dagger}_{\sigma^{\prime}} \widetilde{V}^{\star}_{\alpha}c_{\alpha0\sigma} \right) ~, 
\end{equation}
with effective hybridizations
\begin{equation}
	\begin{aligned}
		\widetilde{V}_{s}  = -0.3475 V_{s \sigma}
		\quad;\quad \widetilde{V}_{d} = 0.7066 V_{d \sigma} ~,
	\end{aligned}
\end{equation}
where these numerical pre-factors result form matrix element tunnelling amplitude from the projection calculation - see for comparison Eq.\ref{eq:5Hyb-Eff-Bare_spin}. Due to the single-orbital coupling, there is a unique \textit{effective 2-impurity asymmetry factor}, namely: $\tilde{\nu}=\widetilde{V}_{d}/\widetilde{V}_{s}$. After introducing  fictitious on-site Coulomb potential as we previously, see Eq.\ref{eq:5Heff-spin-int}, to render the effective model only $3$-fold spin degenerate, the final effective model is equivalent to Eq.\ref{eq:5Heff-spin-PC} - here again with opposite spin label for $f_{\sigma}$ and lead operators. At this point, we can identify the direct proportionality among couplings, namely:
\begin{equation}
	\widetilde{V}_{s}\doteq \widetilde{V} \quad;\quad\widetilde{V}_{d}=\tilde{\nu}\widetilde{V}_{s} \doteq \tilde{\nu}\widetilde{V}_{\sigma} ~,
\end{equation}
where the single-lead effective model with $\widetilde{V}^{\prime}= \widetilde{V}\sqrt{1+\tilde{\nu}^{2}}$ and $c_{0\sigma}=\widetilde{V}/\widetilde{V}^{\prime}\left(c_{s0\sigma} +\tilde{\nu}c_{d0\sigma}\right)$ 
such that the effective asymmetric factor $\tilde{\nu}$ can be written as ratio of the numerical values resulting from the exact diagonalization the isolated nanostructure multiplied for the intrinsic asymmetric factor from the bare model. Hence, due to the emergent proportionality, the ratio of effective couplings corresponds to the ratio of the bare ones. \\
In conclusion, the effective model of spinful serial two-impurity model in MV transition regime reads
\begin{equation}
	\hat{H}^{\prime}_{eff} = \hat{H}_{leads} +
	\sum_{\sigma \in \lbrace \uparrow,\downarrow \rbrace} \left( c^{\dagger}_{0 \sigma} \widetilde{V}^{\prime}f_{\sigma} +  f^{\dagger}_{\sigma}\widetilde{V}^{\star\prime}c_{0\sigma} \right)
	+ U_{c} \hat{n}_{f_{\uparrow}} \hat{n}_{f_{\downarrow}} ~,
\end{equation}
that is analogous to Eq.\ref{eq:5Heff-spin-PC} with $\widetilde{V}^{\prime}=\sqrt{0.121V^{2}_{s}+0.5V^{2}_{d}}$ and $U_{c}\to+\infty$.\\
We have completed the derivation of the effective model giving a numerical sense to the various effective parameters. Now, we make use of the incipient PC property of $\hat{H}^{\prime}_{eff}$ to derive the electrical conductance using the MW formula under PC as we show in Eqs.\ref{eq:5Gcspin_FL},\ref{eq:5Gcspin_OmFinite}. With this effective model and the results of Fig.\ref{fig:tmatrix_MV_sd}, the conductance can now be readily calculated.\\
In conclusion, in this section we show from the exact diagonalization of the isolated $\hat{H}_{nano}$ with double-orbital, we derived the essential quantity not only to define the effective model with emergent PC property but also to evaluate the electrical conductance.

\subsection{Spin model: doublet-triplet transition}
The charge degeneracy point between spin-doublet and spin-triplet multiplet states also yields nontrivial strongly-correlated electron behaviour - yet it can again be captured in a simple effective model in PC. The doublet-triplet transition scenario is more common in the context of single molecule junctions, see Chapter \ref{ch:benz}, where Hund's Rule or other complex multi-orbital interactions can yield a high spin $S=1$ ground-state in the even electron sector of the isolated nanostructure. This is opposed to an even-electron $S=0$ singlet, as we considered in the singlet to doublet valence transition.\\
Here we shall take $N$ to be odd, with the $N$-electron nanostructure hosting a net $S=\tfrac{1}{2}$ doublet state; while in the (even) $N+1$ electron sector, the nanostructure ground state is an $S=1$ triplet. Focusing again on the low-$T$ transport, we construct an effective model in which only the lowest energy spin multiplet of each sector is retained. In this case, by means of the BWPT, the derived effective nanostructure Hamiltonian comprises five states: two for the $N$ electron doublet, and three for the $N+1$ electron triplet. We refer to those spin states described by the bare model as \textit{admissible}. For simplicity we consider below the $SU(2)$ spin-symmetric case arising when no external magnetic field acts i.e. $B=0$ in the bare model in Eq.\ref{eq:5H}, such that states $\ket{N;S^{z}=\pm1/2}$ for the ground-state spin doublet are degenerate and eigenstate $E_{0}^{N;S^{z}=\pm1/2} \equiv E_0^N$ is independent of $S^{z}$. \\
We construct an effective model using Eq.~\ref{eq:5HeffMVPer}, casting the result in terms of one \textit{effective Fermionic site} $f_{\sigma}$, and one \textit{effective spin}-$\tfrac{1}{2}$ \textit{degree of freedom} $\hat{\mathbf{S}}_g$, together with carefully chosen constraints - which are implemented through additional hard-core terms in the Hamiltonian. This allows us to capture the various spin-multiplet transitions from doublet to triplet by adding a single electron. By applying the projector operator in Eq.\ref{eq:5defPMV} up to first order in $\hat{H}_{hyb}$, the resulting \textbf{effective model at doublet to triplet valence transition} reads 
\begin{equation}\label{eq:5Heff_mv_d-t}
	\boxed{
	\begin{aligned}
		\hat{H}_{eff} &\stackrel{MV}{=}\hat{H}_{leads} - J_c \hat{\mathbf{S}}_f\cdot \hat{\mathbf{S}}_g 	+ \sum_{\substack{\alpha=s,d\\ \sigma=\uparrow,\downarrow}} \left ( \widetilde{V}_{\alpha\sigma}^{\phantom{\dagger}} f^{\dagger}_{\sigma}c^{\phantom{\dagger}}_{\alpha\sigma} + \rm{H.c.} \right )  + \delta_c \hat{n}_f ~,\\
		&\stackrel{no-deg}{=} \hat{H}_{leads}  - J_c \hat{\mathbf{S}}_f\cdot \hat{\mathbf{S}}_g 
		+ \sum_{\substack{\alpha=s,d\\ \sigma=\uparrow,\downarrow}} \left ( \widetilde{V}_{\alpha\sigma}^{\phantom{\dagger}} f^{\dagger}_{\sigma}c^{\phantom{\dagger}}_{\alpha\sigma} + \rm{H.c.} \right ) + \delta_c \hat{n}_f +\epsilon^{\prime} (\hat{n}_f-1)^2~, 
	\end{aligned} }
\end{equation}
and we refer to it as an $fg$-system whose elements are defined as follows. The $\hat{\mathbf{S}}_g$ is a spin-$\tfrac{1}{2}$ operator for the effective local moment, $\hat{\mathbf{S}}_f=\tfrac{1}{2}\sum_{\nu,\nu'} f_{\nu}^{\dagger} \boldsymbol{\sigma}_{\nu\nu'}f_{\nu'}^{\phantom{\dagger}}$ is the spin density, $\hat{n}_f=\sum_{\sigma} f_{\sigma}^{\dagger}f_{\sigma}^{\phantom{\dagger}}$ is as before the number operator for the effective $f$ level, and
\begin{equation}
	\begin{aligned}
f^{\dagger}_{\uparrow} &= |N\text{+1;}S^z\text{=1}\rangle\langle N\text{;}S^z\text{=}\tfrac{1}{2} | + \tfrac{1}{\sqrt{2}} |N\text{+1;}S^z\text{=0}\rangle\langle N\text{;}S^z\text{=-}\tfrac{1}{2} | ~,\\
f^{\dagger}_{\downarrow} & = |N\text{+1;}S^z\text{=-1}\rangle\langle N\text{;}S^z\text{=-}\tfrac{1}{2} | + \tfrac{1}{\sqrt{2}} |N\text{+1;}S^z\text{=0}\rangle\langle N\text{;}S^z\text{=}\tfrac{1}{2} | ~,
\end{aligned}
\end{equation}
with $f_{\sigma}=(f_{\sigma}^{\dagger})^{\dagger}$ as usual. Here, the $\tfrac{1}{\sqrt{2}}$ factor is Clebsch-Gordon coefficient implied by the Wigner-Eckart theorem. The $\widetilde{V}_{\alpha\sigma}$ is effective tunnelling amplitude between allowed valence transitions and it is equal to the matrix elements derived from diagonal representation of the isolated nanostructure, namely
\begin{equation}
	\widetilde{V}_{\alpha\sigma}=
	\begin{cases}
		V_{\alpha\sigma}\langle N+1;\hat{S}^z=1 | d_{\alpha\uparrow}^{\dagger} | N;\hat{S}^z=\tfrac{1}{2}\rangle ~, \\
		V_{\alpha\sigma}\langle N+1;\hat{S}^z=-1 | d_{\alpha\downarrow}^{\dagger} | N;\hat{S}^z=-\tfrac{1}{2}\rangle , 
	\end{cases}
\end{equation}
here we recall $d_{\alpha\sigma}^{\dagger},d_{\alpha\sigma}$ is nanostructure operators in the bare model Eq.\ref{eq:5H} where we use the standard $d_{s\sigma}=d_{1\sigma}$ and $d_{d\sigma}=d_{M\sigma}$ and by $SU(2)$ spin symmetry and  Wigner-Eckart theorem it is straightforward to obtain the other transitions, that is
\begin{equation}
	\widetilde{V}_{\alpha\sigma}=
	\begin{cases}
		\sqrt{2}V_{\alpha\sigma}\langle N+1;\hat{S}^z=0 | d_{\alpha\uparrow}^{\dagger} | N;\hat{S}^z=-\tfrac{1}{2}\rangle ~, \\
		\sqrt{2}V_{\alpha\sigma}\langle N+1;\hat{S}^z=0 | d_{\alpha\downarrow}^{\dagger} | N;\hat{S}^z=\tfrac{1}{2}\rangle ~.
	\end{cases}
\end{equation} 
In the second line of Eq.\ref{eq:5Heff_mv_d-t} we present the expression for the effective model away from the doublet triplet valence transition regime. Those deviations from the precise charge-degeneracy point are captured by the parameter $\epsilon^{\prime}=E^{N+1}_{0}-E^{N}_{0}$.\\
We note the $fg$-system in Eq.~\ref{eq:5Heff_mv_d-t} includes eight effective states: we wish to retain only the five admissible states that represent the $N$ electron doublet and the $N+1$ electron triplet of the bare model, namely: 
\begin{equation}
\begin{aligned}
	|N;S^z=\tfrac{1}{2}\rangle &\equiv |-\rangle_f\otimes|\uparrow\rangle_g \;,\\ |N;S^z=-\tfrac{1}{2}\rangle &\equiv |-\rangle_f\otimes|\downarrow\rangle_g \;,\\ 
	|N+1;S^z=1\rangle &\equiv |\uparrow\rangle_f\otimes|\uparrow\rangle_g \;,\\ 
	|N+1;S^z=0\rangle &\equiv \tfrac{1}{\sqrt{2}}( |\uparrow\rangle_f\otimes|\downarrow\rangle_g + |\downarrow\rangle_f\otimes|\uparrow\rangle_g) \;,\\ 
	|N+1;S^z=-1\rangle &\equiv |\downarrow\rangle_f\otimes|\downarrow\rangle_g \;.
\end{aligned}
\end{equation} 
We identify as spurious states $|\uparrow\downarrow\rangle_f\otimes|\uparrow\rangle_g$, $|\uparrow\downarrow\rangle_f\otimes|\downarrow\rangle_g$ and $\tfrac{1}{\sqrt{2}}( |\uparrow\rangle_f\otimes|\downarrow\rangle_g - |\downarrow\rangle_f\otimes|\uparrow\rangle_g)$ and we can eliminate them by the energy constraints implemented in Eq.~\ref{eq:5Heff_mv_d-t} by setting $\delta_c=\tfrac{1}{4}J_c$ and sending then $J_c\to \infty$.\\
By means of the energy constraint, the effective model in Eq.~\ref{eq:5Heff_mv_d-t} is therefore a mixed-valence single impurity Anderson model, side-coupled to an additional spin-$\tfrac{1}{2}$ local moment by a ferromagnetic interaction. Because the AM is by construction fulfilling the PC property, $\hat{H}_{eff}$ presents such an emergent property that facilitates the evaluation of the electrical conductance - as we see in the next part. This can be seen from the fact that in Eq.\ref{eq:5Heff_mv_d-t} is an effective one-channel model, with $c_{0\sigma}=[\widetilde{V}_{s\sigma}+\widetilde{V}_{d\sigma}]/\widetilde{V}_{\sigma}$ and $\widetilde{V}_{\sigma}^{2}=\widetilde{V}_{s \sigma}^{2}+\widetilde{V}_{d \sigma}^{2}$, describing the first site of a combined lead.

\subsection*{NRG results: electrical conductance for spinful doublet-triplet mixed-valence model}
\begin{figure}[H]
	\centering
	\includegraphics[width=0.7\linewidth]{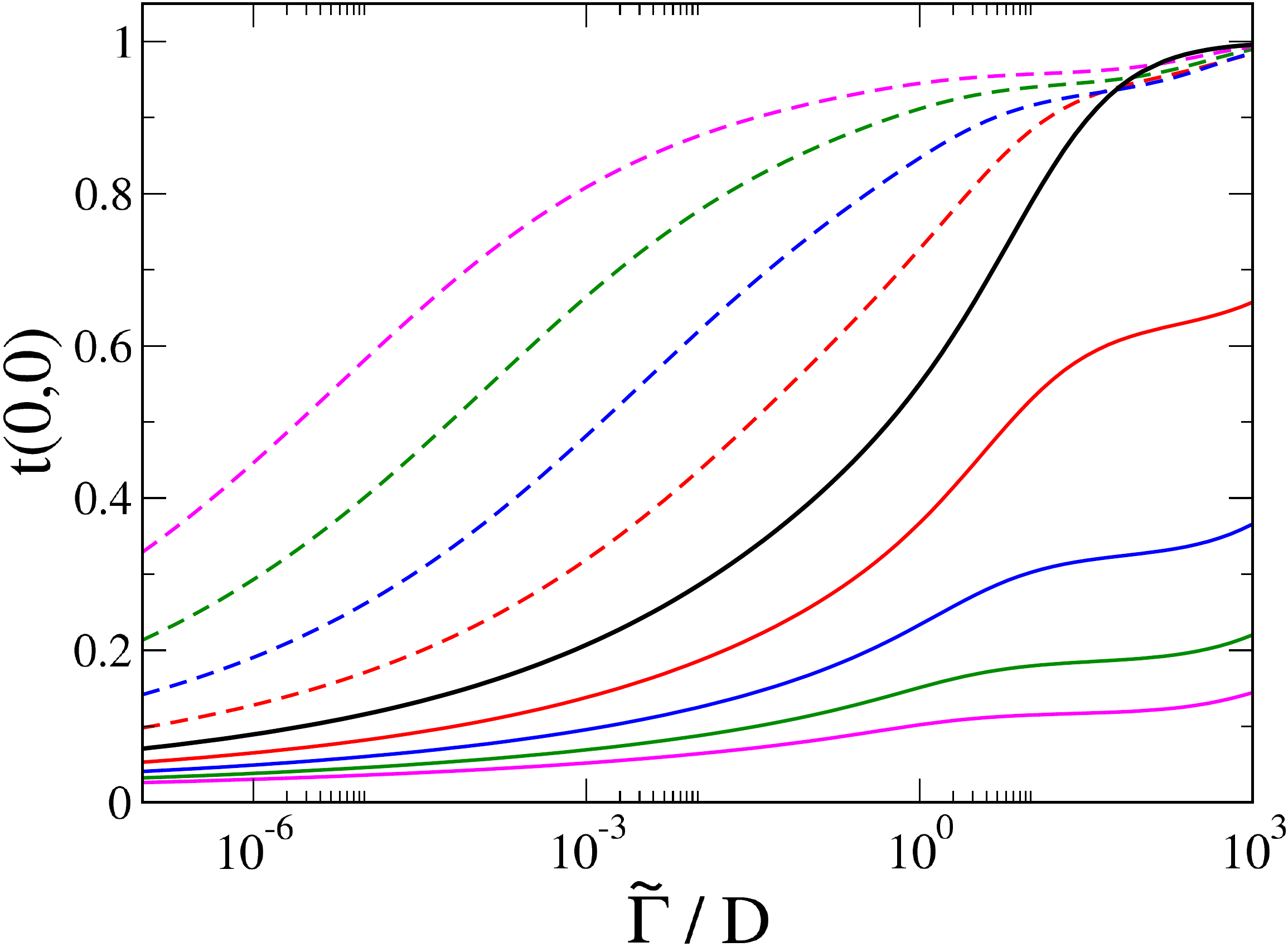}
	\caption[$\mathrm{T}$-matrix spectral function for valence doublet-triplet  multiplet]{$\mathrm{T}$-matrix $t_{ff}(\omega=0,T=0)$ at $T=0$ as a function of the hybridization $\tilde{\Gamma}$ near a doublet-triplet charge degeneracy point, described by the effective model Eq.~\ref{eq:5Heff_mv_d-t}, obtained by NRG. Plotted for $\epsilon^{\prime}=0$ (black line) charge degenerate and $|\epsilon^{\prime}|/\tilde{\Gamma}=0.5, 1, 1.5, 2$ (red, blue, green and magenta lines) with $\epsilon^{\prime}>0$ shown as solid lines and $\epsilon^{\prime}<0$ as dashed lines.} \label{fig:tmatrix_MV_dt}
\end{figure}

Turning now to the low temperature and low energy scales quantum transport, the PC form of the $\hat{H}_{eff}$ in Eq.~\ref{eq:5Heff_mv_d-t} - once we implement the energy constraints to discard the additional unwanted states - implies that we may use the MW formula in PC version, see Eq.\ref{eq:MW->PC}. Therefore, the electrical conductance can be successfully evaluated from the current equations in Eqs.\ref{eq:5Gcspin_OmFinite},\ref{eq:5Gcspin_FL}. \\
As before, we consider the conductance at $T\to 0$. Moreover, we adopt the notation of effective hybridization labelled $\tilde{\Gamma}=\pi |\widetilde{V}_{\sigma}|^{2}\rho_{0}$ as in Fig.~\ref{fig:tmatrix_MV_dt} published in the paper \cite{transport}, so we use it from now on for the sake of clarity.\\
The $T\to 0$ electrical conductance of the physical system depends on the dynamics of the effective model Eq.~\ref{eq:5Heff_mv_d-t} as $T\to 0$ and $\omega\to 0$, as encoded in $\mathrm{T}$-matrix spectral function $t_{ff}(\omega=0,T=0)$. In Fig.~\ref{fig:tmatrix_MV_dt} we plot $t_{ff}(\omega=0,T=0)$ as a function of the effective $\tilde{\Gamma}$ for different $\epsilon^{\prime}/\tilde{\Gamma}$, obtained from NRG. This is the analogous plot for the doublet-triplet transition as Fig.~\ref{fig:tmatrix_MV_sd}$(a)$ for the singlet-doublet transition. Although we find qualitatively similar behaviour to that of the singlet-doublet case according to $\epsilon^{\prime}$ parity, the details are different due to the different valence transition. Indeed, here the full spectrum $t_{ff}(\omega,T=0)$ shows much more pronounced asymmetries than the simple Lorentzian function in the spectra of Eq.~\ref{eq:5Gcspin_OmFinite}, even at very small $\tilde{\Gamma}$. However, in the physically-relevant regime $\tilde{\Gamma}\ll 1$ we find $t_{ff}(\omega=0,T=0)$ for the effective model is still accurately approximated by the spectra defined in Eq.~\ref{eq:5Gcspin_FL} with asymptotic behaviour given by Eq.~\ref{eq:5MV_sd_eps} but now with modified $a\simeq 1.15$, $b\simeq 0.18$ and $c\simeq 0.7$ parameters numerically computed. \\

\noindent{With} these NRG results, the low-temperature conductances for real systems near a doublet-triplet charge degeneracy point can be obtained simply from the effective model parameters $\epsilon^{\prime}$ and effective hybridization defining the tunnelling amplitudes $\widetilde{V}_{\alpha\sigma}$. Comparison of Fig.~\ref{fig:tmatrix_MV_sd}$(a)$ and Fig.~\ref{fig:tmatrix_MV_dt} shows that for a given effective $\tilde{\Gamma}$ and $\epsilon^{\prime}$, the electrical conductance is larger in the vicinity of the doublet-triplet charge degeneracy point that than near the singlet-doublet charge degeneracy point, which we attribute to enhanced spin fluctuations in the former.\\
Finally, we remark that the above analysis of the effective model Eq.~\ref{eq:5Heff_mv_d-t}, obtained perturbatively using BWPT up to $\mathcal{O}(\hat{H}_{hyb})^{2}$ leading order in the nanostructure-lead hybridization, accounts for the dominant contributions to conductance. However, higher-order corrections will in general result in an effective coupling to the neglected odd conduction electron channel, taking the effective model out of PC condition. Although irrelevant for the singlet-doublet case considered in the previous section - whose many-body ground-state is always a nondegenerate singlet due to the Kondo effect - in the doublet-triplet case the involvement of the odd channel is expected to induce a second-stage Kondo screening, quenching the $\ln(2)$ residual entropy found for Eq.~\ref{eq:5Heff_mv_d-t}. Since such terms arise only at higher order in perturbation theory, the associated Kondo scale for the second-stage screening is expected to be very small, and hence may be neglected at experimentally-relevant temperatures. We therefore argue that the effective model Eq.~\ref{eq:5Heff_mv_d-t}, and results of this section, remain valid close to the charge degeneracy point where there is a clear separation of scales.

\section{Generalized spinful model}
As common feature of this chapter, the BWPT requires the projection of the whole system states onto the lowest spin-multiplet ground-state manifold of the nanostructure. Considering only single-electron tunnelling events and appropriate energy and temperature scales, real quantum impurity systems may present \textit{higher} spin multiplet ground-states. In those situations, the BWPT is again a valid method to successfully derive effective models exhibiting an emergent PC condition - confirming the general applicability of degenerate perturbation theory. \\
In this section, we briefly mention the generalization of the main equations used in BWPT for general spin models. The resulting $\hat{H}_{eff}$ takes analogue form to the one derived for the lowest spin multiplet case - hence, we find again emergent PC property that was lacking in the microscopic bare model. At the end, we give some short quantitative analysis for the general spin model under CB and MV regime.\\
We consider $N$ electron spin-$S$ multiplet ground-state and $N+1$ electron spin-$S+\tfrac{1}{2}$ multiplet ground-state. The \textbf{generalized charge spin multiplet ground-state} $\ket{\Psi_{0}^{Q}}$, based on the definition of charge state in Eq.\ref{eq:5Q-BasisState-spin-Def}, is defined as:
\begin{equation}
	\ket{Q=N;S^{z}} =
\begin{cases}
		 \ket{N; \left\{-S, -S + 1, \dots, S + 1, S\right\} }  ~,\\
\ket{N+1; \left\{  -S-\dfrac{1}{2}, -S+\dfrac{1}{2}, \dots, S-\dfrac{1}{2}, S+\dfrac{1}{2} \right\} } ~,
	\end{cases}
\end{equation}
with corresponding \textbf{generalized projector operator} defined as:
\begin{equation}
	\begin{aligned}
		\widehat{P}_{0} &= \left( \ket{N;-S}\bra{N;-S}+\ket{N;-S+1}\bra{N;-S+1}+\dots+
		\ket{N;S}\bra{N;S}\right)\otimes\widehat{\mathbb{1}}_{leads} ~,\\
		\widehat{P}_{0} &= \Big( \ket{N+1;-S-\tfrac{1}{2}}\bra{N+1;-S-\tfrac{1}{2}}+\ket{N+1;-S+\tfrac{1}{2}}\bra{N+1;-S+\tfrac{1}{2}}+\dots+ \\
		&\hspace*{1cm }+\ket{N+1;+S+\tfrac{1}{2}}\bra{N+1;+S+\tfrac{1}{2}}\Big)\otimes\widehat{\mathbb{1}}_{leads}~,
	\end{aligned}
\end{equation}
as straightforward generalization of Eq.\ref{eq:5defP}. Using these generalized formulas we can perform the projection operation to any spin multiplet ground-state manifold.\\
In the case of the MV regime in Sec.\ref{sec:MVPC}, we presented the two most common scenarios for singlet-doublet and doublet-triplet valence transition. However, within the energy window $T \ll \Delta E_{min}$, other valence transitions are compatible with unitary charge and spin variation.\\
In the context of singlet-doublet crossover,  for example $N$ charge sector state comprises next to $\ket{N;S^{z}=0}$ also $\ket{N;S^{z}=-1,0,+1}$ that is consistent with single-electron tunnelling events with $\ket{N+1;S^{z}=\pm\tfrac{1}{2}}$ state. Hence, this is the complete spin charge basis set we use to construct the generalized projector and derive $\hat{H}_{eff}$ which takes a similar form to Eq.\ref{eq:5Heff-spin-PC}. In this case, because the effective model is a generalization of the Anderson impurity model now evaluated for multiple spin sectors, $\hat{H}_{eff}$ is mapped to so-called \textit{ionic models} \cite{Hewson}. \\
This analysis can be extended even further taking now the general charge degeneracy point between an $N$ electron spin-$S$ multiplet ground-state, and an $N+1$ electron spin-$S+\tfrac{1}{2}$ multiplet ground-state. The corresponding effective model is of the same form as Eq.~\ref{eq:5Heff_mv_d-t}, but with $\mathbf{\hat{S}}_g$ now a spin-$S$ operator, and with 
$ \widetilde{V}_{\alpha\sigma}=V_{\alpha\sigma}\bra{ N+1;S^z=S+\tfrac{1}{2}} d_{\alpha\uparrow}^{\dagger} \ket{ N;S^z=S} $. However, $\epsilon^{\prime}=E^{N+1}_{0}-E^N_{0}$ as before. The admissible $4S+3$ physical states of the spin $S$ and $S+\tfrac{1}{2}$ multiplets are retained in the ground-state manifold of the effective model, meaning that $4S+1$ spurious states must be eliminated in Eq.~\ref{eq:5Heff_mv_d-t} by implementing the energy constraints $\delta_c$ and $J_c$. This implies setting $\delta_c=\tfrac{1}{2}S J_c$ and sending $J_c\to \infty$. The resulting effective model correctly captures only the admissible states of the bare model and shows an emergent PC property. The low temperature  quantum transport can be obtained in the same way as we study in Sec.\ref{sec:MVPC} from a knowledge of $t_{ff}(\omega=0,T=0)$, which can be calculated for the effective model using numerical methods e.g.~NRG.\\
With this we conclude our brief discussion of general spinful model. The take-home message is that the BWPT is a solid method to derive effective models presenting emerging PC property which were not present in the bare system - if appropriate energy constraints are implemented to make $\hat{H}_{eff}$ in agreement with the admissible spin states of the physical system.

\chapter{Application and comparison of quantum transport techniques to the triple quantum dot system}\label{ch:TDQD}
The goal of this chapter is to demonstrate the results of Chapter $3$ on the theory of mesoscopic quantum transport and chapter $4$ on Improved calculations for quantum transport in interacting multi-orbital nanostructures by application to specific multi-orbital strongly-correlated systems. \\
This chapter, and chapter $5$ on quantum transport from effective PC models, feature material contained in the publication \cite{transport}. We adopt the notation used there which differs from the one employed in the thesis for: electrical conductance $G_{C}(\omega,T)$, heat conductance $K_{Q}(T)$ and we introduce a few definitions useful for the discussion.\\
We introduce for a PC set-up the definition of single linear combination of nanostructure operators, $\bar{\bar{d}}_{\sigma}^{\dagger}$, to which the even/strongest channel couples, given by $\bar{\bar{d}}_{\sigma}^{\dagger}=\tfrac{1}{V}\sum_i V_i d_{i\sigma}^{\dagger}$ with $V^2=\sum_i |V_i|^2$. The hybridization Hamiltonian then takes the same form as that of a single-channel, single-impurity model, $H_{ hyb} = V\sum_{\sigma} (\bar{\bar{d}}_{\sigma}^{\dagger} c_{e\sigma}^{\phantom{\dagger}} + \rm{H.c.} )$ where we define the real-space lead orbitals localized at the nanostructure position, given by $c_{\alpha\sigma}=\sum_{k}U^{\alpha}_k c_{\alpha k \sigma}^{\phantom{\dagger}}$ for $\sum_k | U^{\alpha}_k|^2=1$. \\
We introduce the Lorenz ratio $L=\kappa / TG_C$ which plays an important role in understanding the nature of thermoelectric transport in nanoelectronic devices. In particular, quantum transport in the Fermi liquid regime typically satisfies the Wiedemann-Franz (WF) law, \cite{franz1853ueber,Costi_ThermoelectricQD2010} in which the Lorenz ratio as $T\to 0$ takes the special value $L_0=\pi^2 k_{ B}^2/3e^2$. This is expected when the carriers of both charge and heat are the same fermionic quasiparticles (or simply the bare electrons themselves). Although violations of the WF law are often associated with non-Fermi liquid physics in bulk systems, \cite{mahajan2013non} in the context of mesoscopic quantum devices, violations are known in Fermi liquid systems \cite{vavilov2005failure,kubala2008violation,bergfield2009thermoelectric}. This highlights the richer thermoelectric capabilities of complex nanostructures.\\
Having completed this short preamble to adjust this chapter content with the existing concepts and formulas of the thesis, we can start with the work.\\

\noindent{Our} chosen test system is the two-lead triangular triple quantum dot (TQD). The outline of the chapter runs as follows: we start with introducing the TQD model Hamiltonian and a numerical comparison between standard and improved Kubo formula using NRG for the TQD system. Then, we continue with the conductance results for proportionate coupling and non-PC TQD configurations.

\section{Model: triple quantum dot}\label{sec:tqd}
\begin{figure}[H]
	\centering
	\includegraphics[width=0.75\linewidth]{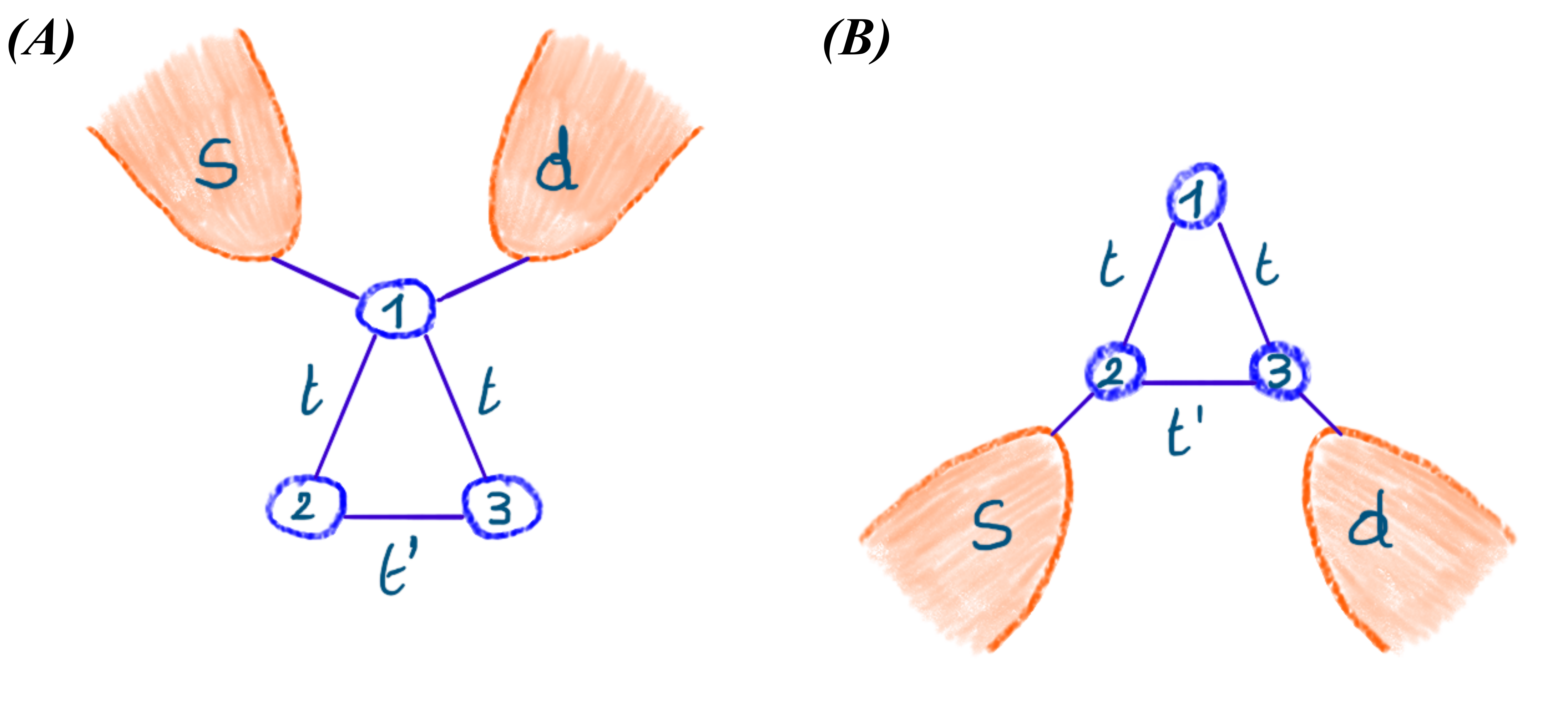}
	\caption[Illustration two-lead triple quantum dot model]{Illustration of the triple quantum dot devices studied. \textit{(a)} PC coupling geometry; \textit{(b)} non-PC geometry.  } \label{fig:tqd}
\end{figure}

Here we consider the TQD model $H=H_{ leads}+H_{ TQD}+H_{ hyb}$, with $H_{ leads}= \sum_{\alpha}H_{ leads}^{\alpha}=\sum_{\alpha,k,\sigma} \epsilon_{k}^{\phantom{\dagger}} c_{\alpha k \sigma}^{\dagger} c_{\alpha k \sigma}^{\phantom{\dagger}}$ and,
\begin{equation}\label{eq:tqd}
		H_{ TQD} = \sum_{j=1,2,3} \left ( \epsilon_j \hat{n}_j + U_j \hat{n}_{j\uparrow}\hat{n}_{j\downarrow} \right )  +\sum_{i\ne j} \left (U_{ij}'\hat{n}_i\hat{n}_j   + \sum_{\sigma} t_{ij}^{\phantom{\dagger}} d^{\dagger}_{i\sigma}d^{\phantom{\dagger}}_{j\sigma} \right) \;,
\end{equation}
where $\hat{n}_j=\sum_{\sigma}\hat{n}_{j\sigma}$, and the inter-dot tunnelling matrix elements satisfy $t_{ij}=t_{ji}^*$. We take for simplicity equivalent dots with $\epsilon_j\equiv \epsilon$, $U_j\equiv U$, $U'_{ij}\equiv U'$, and consider the mirror symmetric case  $t_{12}=t_{13}\equiv t$ but $t_{23}\equiv t'$. We consider two geometries for the TQD-lead hybridization,
\begin{subequations}
	\begin{align}
		H_{ hyb}^{I}  &= \sum_{\alpha=s,d} \sum_{\sigma} \left ( V_{\alpha}^{\phantom{\dagger}}d^{\dagger}_{1\sigma}c_{\alpha\sigma}^{\phantom{\dagger}} +{\rm H.c.} \right ) \;,\label{eq:tqd_hyb_1}\\
		H_{ hyb}^{II} &= \sum_{\sigma} \left (V_s^{\phantom{\dagger}} d^{\dagger}_{2\sigma}c_{s\sigma}^{\phantom{\dagger}}+V_d^{\phantom{\dagger}} d^{\dagger}_{3\sigma}c_{d\sigma}^{\phantom{\dagger}} +{\rm H.c.} \right ) \;. \label{eq:tqd_hyb_2}
	\end{align}
\end{subequations}
For simplicity we also take $V_s=V_d$.
Importantly, note that $H_{ hyb}^{I}$ in Eq.~\ref{eq:tqd_hyb_1} (illustrated in Fig.~\ref{fig:tqd}\textit{a}) satisfies the PC geometry condition while $H_{ hyb}^{II}$ in Eq.~\ref{eq:tqd_hyb_2} (Fig.~\ref{fig:tqd}\textit{b}) is non-PC and therefore irreducibly 2-channel.\\
\noindent{The} behaviour of TQD systems is notoriously rich, with aspects of their complex physics having been uncovered in both experiments (see e.g.~Refs.~\cite{vidan2004triple,schroer2007electrostatically,rogge2008two,granger2010three,gaudreau2012coherent,seo2013charge}) and theory (see e.g.~Refs.~\cite{mitchell2009quantum,mitchell2010two,mitchell2013local,bonvca2008numerical,vernek2009kondo,numata2009kondo,oguri2011kondo,saraga2003spin,wrzesniewski2018dark,koga2013field,wang2013enhancement,ramos2017spin}). Our purpose here is to examine the quantum transport properties in a systematic and consistent fashion, applying and comparing the techniques discussed above. We solve the underlying quantum impurity problems using NRG. \\
\noindent{Before} proceeding with the core of our results, we present in the next section a numerical comparison between Kubo formula and the improved version derive in the thesis on the TQD system.

\subsection{Kubo vs improved Kubo formulation in NRG for triple quantum dot system}
As a proof-of-principle demonstration, we calculate the \textit{ac}-electrical conductance $G_C(\omega,T)$ vs driving frequency $\omega$ at $T=0$ with NRG for the two-lead AIM Eq.~\ref{eq:AM} using the Kubo formula Eq.~\ref{eq:defKuboLRel}. In Sec.\ref{sec:ImprovedKubo}, we present a systematic numerical investigation to corroborate the improved Kubo formula analytical derivation with NRG results on the Anderson impurity model. In this section we compute a similar comparison between standard and improved Kubo formulation for the TQD system in PC configuration, see Fig.\ref{fig:tqd}\textit{(a)}.\\
\begin{figure}[H]
	\centering
	\includegraphics[width=0.75\linewidth]{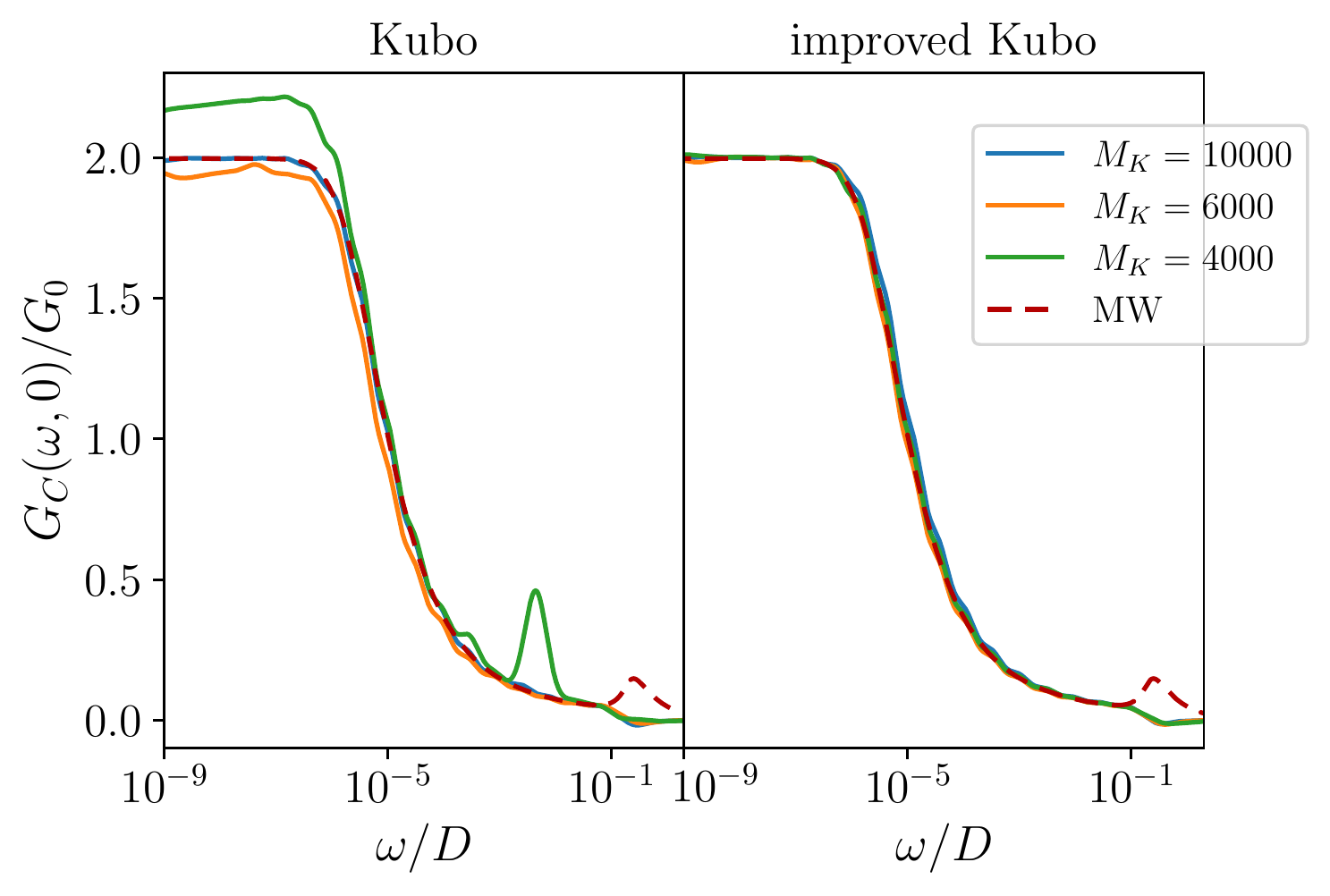}
	\caption[Kubo vs improved Kubo in triple quantum dot model]{Comparison of NRG results for Kubo \textit{(left)} and improved Kubo \textit{(right)} formulae for the \textit{ac}-conductance $G_{C}(\omega,T=0)$ for TQD in PC configuration for different number $M_K$ NRG kept states and fixed $\Lambda=3$. Parameters used: $t=5\times10^{-4}$, $t^{\prime}=0.01$, $U=0.4D$, $\epsilon_{d}=-U/2$, $V_{s}=V_{d}=0.07D$. The MW result shown as dashed red line for comparison is calculated on the effective single-channel, NRG and physical parameters are $M_K =4000$, $V=0.01D$.}\label{F7/TQD_KubovsImpKubo}
\end{figure}
\noindent{In} Fig.~\ref{F7/TQD_KubovsImpKubo} we compare the standard implementation of the Kubo formula using the current-current correlator in Eq.~\ref{eq:defKuboLR}, see left panel, with the \textit{improved} Kubo formula Eq.~\ref{eq:4defKuboLRel}, see right panel, for different numbers $M_K$ of NRG kept states as specified in the legend. Both plots are calculated for NRG discretization parameter $\Lambda=3$. As a reference, we provide the Meir-Wingreen result in dashed red line for the equivalent single-channel Anderson impurity model, see cartoon in Fig.\ref{F3:PC} and related discuss as in Sec.\ref{sec:PC}. In this case, the linear response conductance is calculated within NRG using the PC version of the MW current formula in Eq.\ref{eq:MW->PC}. This computation should be considered the numerically-exact result for this system.\\
The Fig.~\ref{F7/TQD_KubovsImpKubo} shows that the standard Kubo formula is able to capture the correct behaviour of $G_{C}(\omega,T=0)$, but only for very large number of kept states, $M_{k}\sim 10000$ in blue curve - for which the two channel NRG calculations become very computationally costly. By contrast, highly accurate results are obtained by the improved Kubo method even at unexpectedly low $M_K \sim 4000$ in green curve.\\
As we note already in the Anderson impurity model comparison plotted in Fig.\ref{F4/KvsimpK}, despite the different $M_{K}$ states in the FDM-NRG implementation, the Kubo formulation is not able to capture the Hubbard satellite features at $|\omega|\sim U$ as opposed to the MW formulation. This seems to suggest that the current-current correlator is more responsive to the logarithmic discretization of the bath Hamiltonian than the dot Green's function employed in the MW formula. And this causes lack of resolution at high energy scale.

\section{Triple quantum dot system in proportionate coupling configuration}\label{sec:TQDpc}
We first consider the PC case, Eq.~\ref{eq:tqd_hyb_1} (Fig.~\ref{fig:tqd}\textit{a}) with $U'=0$ and present conductance calculations under various conditions. \\

\begin{figure}[H] 
	\begin{subfigure}[b]{0.25\linewidth}
		\centering
		\includegraphics[width=0.75\linewidth]{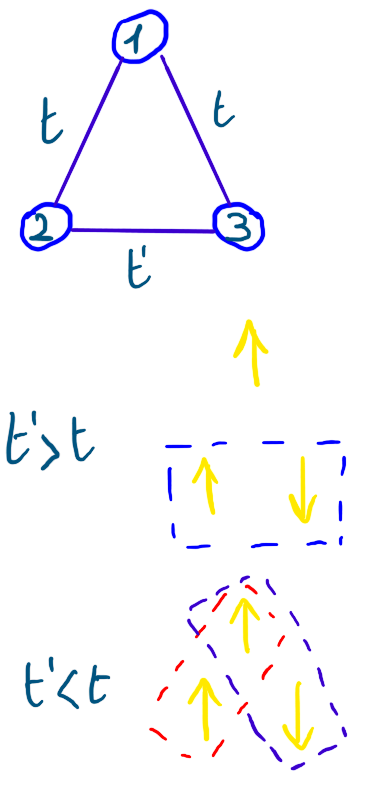} 
	\end{subfigure}
	\begin{subfigure}[b]{0.75\linewidth}
		\centering
		\includegraphics[width=1.05\linewidth]{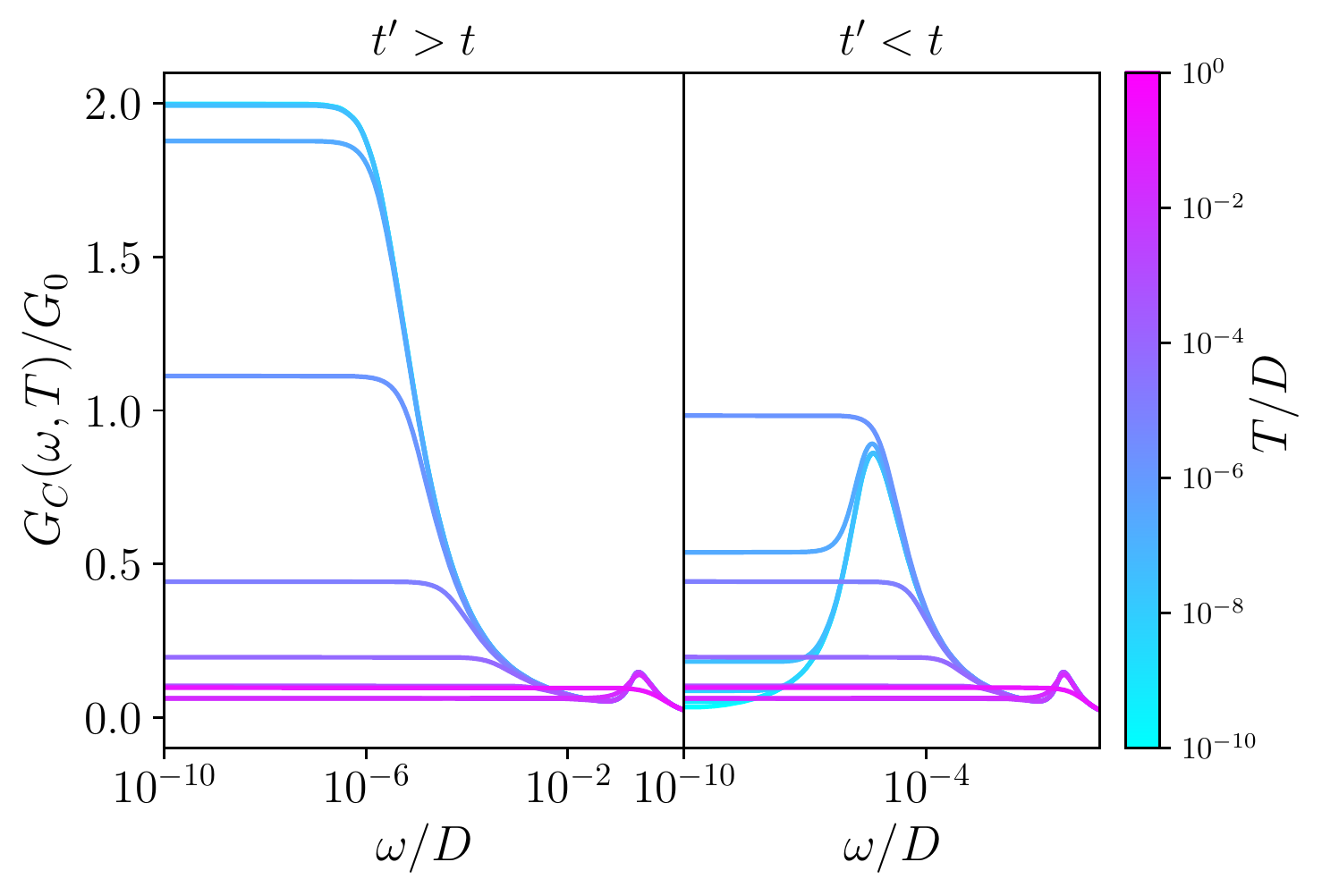} 
	\end{subfigure} 
\caption[ac-conductance for PC triple quantum dot system]{Sketch of spin state configurations according to the phase and \textit{ac}-conductance $G_C(\omega,T)$ as a function of driving frequency $\omega$ at different temperatures $T$ as indicated in the colour scale, for the TQD model in the PC geometry (see Fig.~\ref{fig:tqd}\textit{(a)} and Eq.~\ref{eq:tqd_hyb_1}). Results shown for the Kondo screened phase $t'>t$ \textit{(left)} and local moment phase $t'<t$ \textit{(right)}. NRG results are obtained using the \textit{ac}-generalized MW formula, Eq.~\ref{eq:Kubo_PC}, within the effective single-channel model. Model parameters: $U=0.4D$, $\epsilon=-U/2$, $U'=0$, $V_s=V_d=0.071D$, $t=5\times10^{-4}D$ and $t'=10^{-2}D$ \textit{(left)} or $t'=0$ \textit{(right)}. NRG parameters: $\Lambda=2.5$, $M_K=3000$.}\label{fig:Gac}
\end{figure}

\noindent{The} behaviour of the linear \textit{dc} electrical conductance $G_C(T)$ for this system was reported in Ref.~\cite{mitchell2009quantum}. In Fig.~\ref{fig:Gac} we consider instead the \textit{dynamical} conductance $G_C(\omega,T)$, plotted vs \textit{ac} driving frequency $\omega$ at different temperatures $T$ and we also include a schematic of the spin state configurations according to the phase.\\

\noindent{The} TQD in this geometry supports a quantum phase transition driven by geometric frustration in the triangular arrangement of dots \cite{mitchell2009quantum}. The transition, tuned by the ratio $t/t'$,  embodies the competition between an antiferromagnetic-Kondo Fermi liquid with a screened spin-singlet ground state, and a ferromagnetic-Kondo singular Fermi liquid with a degenerate local moment ground state. We show how this plays out in the \textit{ac} conductance in Fig.~\ref{fig:Gac}, which we calculate via 
\begin{equation}\label{eq:Kubo_PC}
	G_C(\omega,T) \overset{\text{PC}}{\longrightarrow} \left(\frac{e^2}{h}\right) \frac{4\Gamma^d\Gamma^s}{\Gamma^d+\Gamma^s}\sum_{\sigma} \int_{-\infty}^{\infty}d\omega' ~{\rm Im}~ \overline{\overline{G}}_{\sigma}(\omega',T)
	\times \left [\frac{f^{Eq}(\omega'+\omega)- f^{Eq}(\omega'-\omega) }{2\omega}\right ] ~,
\end{equation}
that is the \textit{ac} Kubo formula for structured, equivalent leads in Eq.\ref{eq:4GacKEq} under PC. As such, the conductance in Eq.~\ref{eq:Kubo_PC} is controlled by the equilibrium retarded Green's function for dot 1 (interacting, lead-coupled), since in this geometry $\overline{\overline{G}}_{\sigma}(\omega,T)=G_{11,\sigma}(\omega,T)\equiv  \langle\langle d_{1\sigma}^{\phantom{\dagger}} ; d_{1\sigma}^{\dagger}\rangle\rangle$. This is in turn obtained from the Dyson equation Eq.~\ref{eq:defDysonRA}, with the TQD self-energy matrix calculated in NRG via Eq.\ref{eq:4F/G}.\\
\noindent{For} $t'>t$ \textit{(left panel)} we see a Kondo resonance, with strongly enhanced \textit{ac} conductance for $|\omega|, T \ll T_{ K}$. As the temperature is increased, the resonance is lost. To understand this dynamical transport behaviour in the TQD, the screening mechanism \cite{mitchell2009quantum} must be understood. At temperatures $T\lesssim U$, the dots become essentially singly-occupied (assuming $t, t'\ll U$ as here). On the scale $J\sim t^2/U\ll U$, effective antiferromagnetic exchange interactions then bind the three spins of the TQD into a combined doublet ground state,
\begin{eqnarray}\label{eq:plusstate}
	|+;S_{ TQD}^z=\tfrac{1}{2}\rangle  = \tfrac{1}{\sqrt{2}}[|\uparrow\uparrow\downarrow\rangle - |\uparrow\downarrow\uparrow\rangle ] \;,
\end{eqnarray}
where the basis states $|\sigma_1,\sigma_2,\sigma_3\rangle$ are labelled by the spins on the three dots, and the other component of the doublet with $S_{ TQD}^z=-\tfrac{1}{2}$ is obtained by replacing $\uparrow \leftrightarrow \downarrow$ - see sketch of the spin state configuration in Fig.~\ref{fig:Gac}. This doublet then couples to the leads with an effective exchange interaction $J_{eff}^+\sim +V^2/U$, which generates spin-flip scattering of conduction electrons, and results in the Kondo effect below the Kondo temperature \cite{Hewson} $T^+_{ K} \sim D \exp(-1/\rho_0 J^+_{eff})$. For the parameters used, we have $T^+_{ K}\approx 10^{-5}D$, with $D$ the conduction electron bandwidth.\\
\noindent{For} $T\ll T^+_{ K}$ the conduction electron scattering rate is on the order of $T^+_{ K}$. Therefore, when a small \textit{dc} bias voltage is applied, the enhanced electronic scattering boosts the net current flowing from source to drain \cite{PustilnikGlazman_review2004,Pustilnik_2CTransport2004}. The conductance in this case can reach its maximum value, $G_C=2e^2/h$. In the \textit{ac} case, conductance is still strongly enhanced, provided that the electronic scattering time is much shorter than the period of the voltage bias oscillations, or $|\omega| \ll T^+_{ K}$. As the frequency increases, the current does not have time to build up fully before the bias voltage changes sign. At large frequencies, the TQD sees an `averaged' bias and the conductance is low. The resonance condition is therefore $|\omega|\sim T^+_{ K}$ when $T\ll T^+_{ K}$. This is seen in the low-$T$ limit (turquoise line) in the left panel of Fig.~\ref{fig:Gac}, which is found to have asymptotic behaviour in the Kondo-screened phase following,
\begin{subequations}\label{eq:Gtpgtrt}
	\begin{align}
		G_C(\omega,0)/G_0 &= 2 - \bar{a}(\omega/T^+_{ K})^2 ~:~|\omega|\ll T^+_{ K}~,\\
		&=\bar{b}/\ln^2(|\omega|/T^+_{ K}) ~:~|\omega|\gg T^+_{ K}~,
	\end{align}
\end{subequations}
with $\bar{a},\bar{b}$ constants of order 1. \\
\noindent{As} the temperature is increased, the Kondo effect is destroyed, the resonance condition is lost, and the low-frequency \textit{ac} conductance decreases. For $T\gg T^+_{ K}$, the conductance is low, being the result of incoherent sequential tunnelling only (it is not boosted by spin-flip scattering from the Kondo effect). This is seen in the left panel of Fig.~\ref{fig:Gac} by the sequence of lines from turquoise to purple on increasing temperature.\\ 
\noindent{Upon} tuning $t'$ from $>t$ to $<t$, the isolated TQD ground state changes, and this results in a quantum phase transition in the lead-coupled system \cite{mitchell2009quantum}.   For $t'<t$ (right panel, Fig.~\ref{fig:Gac}), the collective TQD doublet ground state (in the singly-occupied limit) is,
\begin{eqnarray}
	|-;S_{ TQD}^z=\tfrac{1}{2}\rangle  = \tfrac{1}{\sqrt{6}}[2|\downarrow\uparrow\uparrow\rangle - |\uparrow\uparrow\downarrow\rangle-|\uparrow\downarrow\uparrow\rangle ] \;,\label{eq:minusstate}
\end{eqnarray}
see the sketch of spin state configuration in Fig.~\ref{fig:Gac}. This doublet state again forms on the scale of $J\sim t^2/U$, but its effective coupling to the leads is now  \emph{ferromagnetic}, $J^-_{ eff} <0$. This results in the \textit{suppression} of spin-flip scattering on reducing the temperature \cite{Hewson} below $T \sim J$, and hence a suppression of the conductance. In the \textit{dc} limit, we have  $G_C(0,0)=0$, embodying the emergent decoupling of the TQD from the leads. However, singular Fermi liquid corrections to the local moment fixed point at finite energy \cite{koller2005singular} lead to logarithmic corrections to the low-frequency \textit{ac} conductance,
\begin{eqnarray}
	G_C(\omega,0)/G_0 =\bar{\alpha}_1/\ln^2(\bar{\alpha}_2|\omega|/J) \quad :~|\omega|\ll J ~.
\end{eqnarray}
Since the underscreened TQD doublet has a finite energy $J$, it has a finite lifetime due to the energy-time uncertainty principle. Even at $T=0$, scattering from this state can therefore be probed at finite frequencies. \\
\noindent{At} energies $\gg J$, the collective TQD doublet $|-;S_{ TQD}^z\rangle$ has not developed, and lead conduction electron scattering is dominated by the direct coupling between the leads and dot 1. In this regime the physics is controlled by an effective single-impurity Kondo temperature $T^1_{ K} \simeq T_{ K}^+$. If $J\gg T_{ K}^1$, then scattering is incoherent and weak at energies $\gg J$. Correspondingly, the \textit{ac} conductance for $|\omega| \gg J$ is small. However, if $J\ll T_{ K}^1$ then Kondo-enhanced scattering with a rate $\sim T_{ K}^1$ controls conductance at frequencies $|\omega|\gg J$, following Eq.~\ref{eq:Gtpgtrt}. In particular, we expect in this regime Kondo resonant \textit{ac} conductance for $|\omega| \sim T^1_{ K} \gg J$, but suppressed conductance for $|\omega|\ll J$.\\ 
\noindent{This} is precisely what is observed in the right panel of Fig.~\ref{fig:Gac} at low-$T$ (turquoise line), where we have chosen $J\sim T_{ K}^1$. This gives rise to a non-monotonic behaviour as the temperature is increased, with the Kondo effect suppressed at $T\ll J$ by ferromagnetic correlations, activated around $T\sim T_{ K}^1$ due to incipient screening of dot 1, and then again suppressed thermally for $T\gg T_{ K}^1$.\\
\noindent{The} dynamical \textit{ac-}conductance $G_C(\omega,T)$ therefore contains much richer information on the scattering and screening processes than the steady-state \textit{dc-}conductance $G_C(T)$.\\

\begin{figure}[H]
	\centering
	\includegraphics[width=0.6\linewidth]{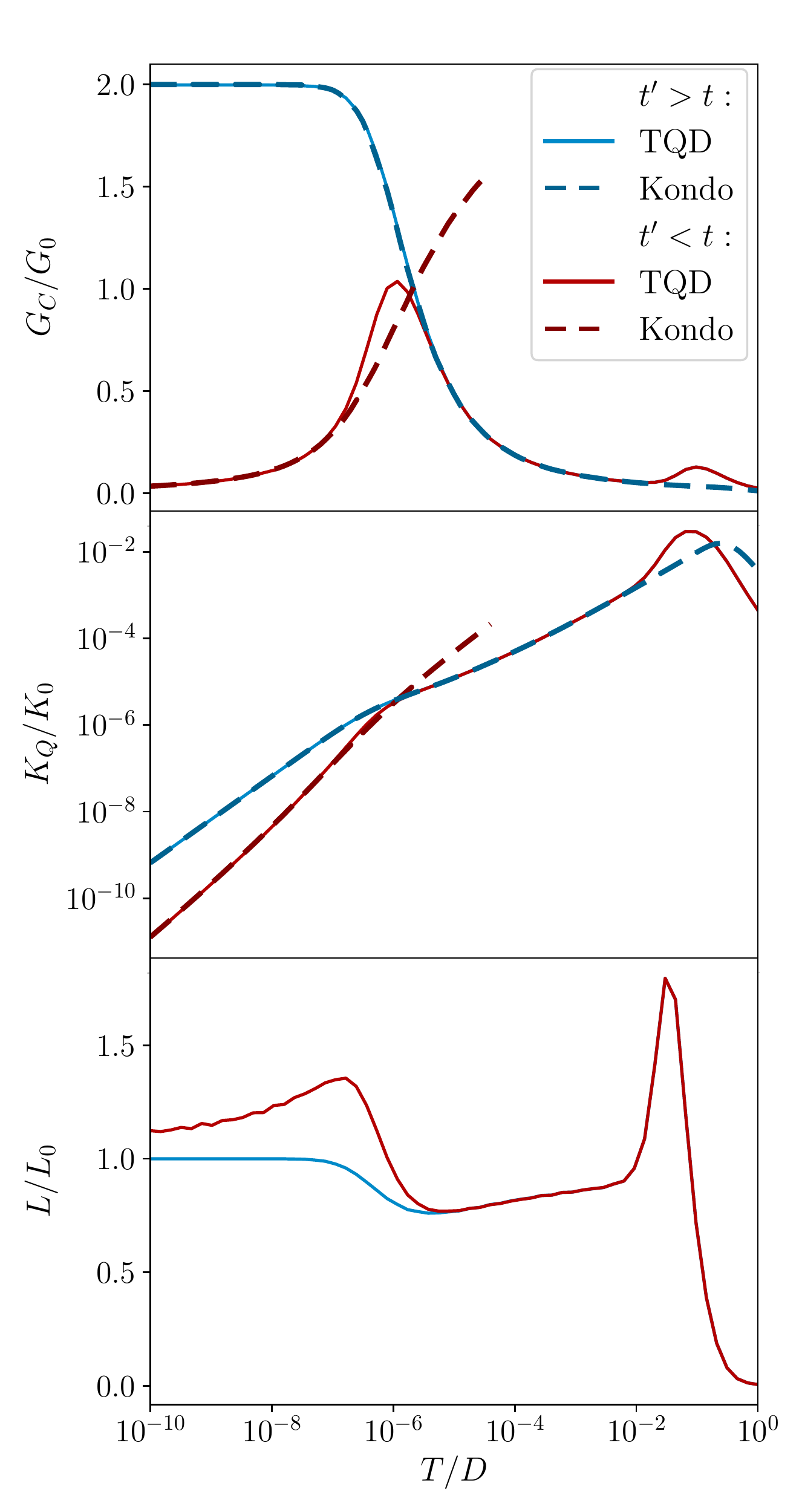}
	\caption[Electric, heat conductance and Lorenz ratio for PC triple quantum dot system]{\textit{dc}-electrical conductance $G_C(T)$ \textit{(top)}, heat conductance $K_Q(T)$ \textit{(middle)}, and Lorenz ratio $L(T)$ \textit{(bottom)} as a function of temperature, for the same TQD system as in Fig.~\ref{fig:Gac}. NRG results obtained using the MW formulae Eq.~\ref{eq:MW_cond} (solid lines) for both the Kondo screened phase ($t'>t$, blue) and the unscreened local moment phase ($t'<t$, red). Dashed lines show the result of the effective single-impurity, single-channel Kondo model, with effective exchange coupling $J_{ eff}$ obtained from model machine learning methods \cite{rigo2020machine}.}\label{fig:MWthermoelectric}
\end{figure}

\noindent{In} Fig.~\ref{fig:MWthermoelectric} we investigate the temperature-dependence of \textit{dc} thermoelectric properties in the same system. The solid lines in the top panel for the \textit{dc} electrical conductance $G_C(T)$ are obtained by taking the  $\omega\to 0$ limit of $G_C(\omega,T)$, as shown in Fig.~\ref{fig:Gac}. In practice, the \textit{dc} limit is realized for $|\omega| \ll \max(T_{ K},T)$ and the NRG data are well-converged in this regime. This is formally and numerically equivalent to using the linear response conductance using the PC version of the MW current formula in Eq.\ref{eq:MW->PC}, that is
\begin{equation}\label{eq:MW_cond}
G_C(T) ~\overset{\text{PC}}{\longrightarrow}~ \frac{e^2}{h} \int d\omega~ \left(-\frac{\partial f^{Eq}(\omega)}{\omega}  \right)\widetilde{\mathbb{T}}_{ PC}(\omega,T) ~.
\end{equation}
We again show the representative behaviour in each phase (blue line for $t'>t$, red line for $t'<t$). The middle panel shows the corresponding curves for the heat conductance $K_Q(T)$, obtained from the MW formula Eq.~\ref{eq:4MW_HeatPC}, with transmission function under PC in Eq.~\ref{eq:GenTransfMatPC} using $\overline{\overline{G}}_{\sigma}(\omega,T)$, and $\mathrm{T}$-matrix $t_{ee,\sigma}(\omega,T)=-\Gamma~{\rm Im}G_{11,\sigma}(\omega,T)$. The lower panel shows the Lorenz ratio $L=\kappa/T G_C$ in units of the limiting WF value \cite{franz1853ueber}, $L_0=\pi^2 k_{ B}^2/3e^2$.\\
\noindent{We} consider first the behaviour in the Kondo-screened phase for $t'>t$ (blue lines). At low temperatures $T\ll T_{ K}^+$, the physics of the TQD is typical of a Fermi liquid \cite{Costi_ThermoelectricQD2010}, with quadratic corrections to the strong coupling fixed point. Here we find,
\begin{subequations}\label{eq:tpgtrtdcFL}
	\begin{align}
		G_C(T)/G_0 &= 2 - a(T/T^+_{ K})^2 ~,\\
		K_Q(T)/T K_0 &= \frac{2\pi^2}{3}-b(T/T^+_{ K})^2 ~,\\
		L(T)/L_0 &=1-c(T/T^+_{ K})^2 ~,
	\end{align}
\end{subequations}
where $a,b,c$ are constants of order 1 depending on the specific definition of $T_{ K}^+$ used. Fitting to the NRG data yields $b/a \approx 9$, while $c=3b/2\pi^2-a/2$. At the lowest temperatures, the TQD behaves like a renormalized non-interacting system with $G_c \sim e^2/h$ and $K_Q\sim k_{ B}T/h$ while $L \to L_0$. Note that the geometric factor $4\Gamma^s\Gamma^d/(\Gamma^s+\Gamma^d)^2$ in Eq.~\ref{eq:GenTransfMatPC} describing the relative coupling strength of the TQD to source and drain leads drops out here for $V_s=V_d$. In general for $V_s\ne V_d$, both $G_C$ and $K_Q$ are multiplied by this overall constant (independent of temperature), such that the Lorenz ratio $L$ is always unaffected by the coupling geometry in PC. This is not necessarily the case for non-PC systems.\\
\noindent{At} higher temperatures $T\gg T_{ K}^+$ in the Kondo phase, spin-flip scattering of conduction electrons from the TQD show up in thermoelectric properties as,
\begin{subequations}\label{eq:tpgtrtdcLM}
	\begin{align}
		G_C(T)/G_0 &= \frac{\alpha_1}{\ln^2(\alpha_2 T/T_{ K}^+)} ~,\\
		K_Q(T)/T K_0 &= \frac{\gamma_1}{\ln^2(\gamma_2 T/T_{ K}^+)} ~,\\
		L(T)/L_0 &=1-\frac{\xi}{\ln(\gamma_2 T/T_{ K}^+)} ~,
		\label{eq:L1}
	\end{align}
\end{subequations}
where we find $\gamma_1/\alpha_1=\pi^2/3$ and $\gamma_2/\alpha_2\approx 1.7$. It then follows that $\xi=\ln(\gamma_2/\alpha_2)>0$.\\
\noindent{The} numerical results in Fig.~\ref{fig:MWthermoelectric} show the full crossover behaviour. Note also that although $G_C(\omega,0)$ and $G_C(0,T)\equiv G_C(T)$ appear similar, the details of their functional forms are different. The Lorenz ratio $L$ is found to approach the WF value $L_0$ from below for both $T\ll T_{ K}^+$ and $T\gg T_{ K}^+$ in the Kondo phase, although in the latter case the limit is saturated logarithmically slowly. At intermediate temperatures, especially $T\sim T_{ K}^+$, we see strong violations of the WF law, driven by electronic scattering from interactions \cite{wang2013enhancement}. This phenomenology is also known experimentally in single-electron transistors \cite{kubala2008violation}. Such violations are in general expected for interacting nanostructures. Note however that at intermediate temperatures, $L\ne L_0$ does not imply non-Fermi liquid physics: the WF law strictly applies \cite{franz1853ueber} only in the limit $T \to 0$, where most conventional systems exhibit Fermi liquid behaviour and satisfy the WF law. \\
\noindent{Turning} now to the ferromagnetic local moment phase realized in the TQD for $t'<t$ (red lines in Fig.~\ref{fig:MWthermoelectric}), singular Fermi liquid corrections appear in the thermoelectric properties at low temperatures $T\ll J$, namely
\begin{subequations}\label{eq:tplesst}
	\begin{align}
		G_C(T)/G_0 &= \frac{\alpha'_1}{\ln^2(\alpha'_2 T/J)} ~,\\
		K_Q(T)/T K_0 &= \frac{\gamma'_1}{\ln^2(\gamma'_2 T/J)} ~,\\
		L(T)/L_0 &=1-\frac{\xi}{\ln(\gamma'_2 T/J)} ~,
		\label{eq:L2}
	\end{align}
\end{subequations}
where the ratios $\gamma'_1/\alpha'_1=\pi^2/3$ and $\gamma'_2/\alpha'_2\approx 1.7$ as well as $\xi=\ln(\gamma'_2/\alpha'_2)$ are the same as for Eq.~\ref{eq:tpgtrtdcLM}. In this local moment phase, the electrical and heat conductances are suppressed relative to the Kondo phase; in particular for $T\to 0$ we have $G_C\to 0$, and $K_Q/T \to 0$ corresponding to a transmission node. However, their ratio remains finite and $L \to L_0$ as $T\to 0$. The fact that the effective ferromagnetic Kondo effect in the TQD asymptotically satisfies the WF law is a signature of its low-energy singular Fermi liquid correlations. By contrast, conventional transmission nodes typically yield strongly enhanced Lorenz ratios \cite{bergfield2009thermoelectric}.
\noindent{For} $T\gg J$ the incipient Kondo screening of dot 1 results in behaviour described by Eq.~\ref{eq:tpgtrtdcLM}. The crossover between Eqs.~\ref{eq:L1} and \ref{eq:L2} upon lowering the  temperature from $T\gg J \sim T_{ K}^+ $ to $T\ll J$ is therefore predicted to give a sign-change in $L(T)-L_0$, and this fingerprint of the ferromagnetic Kondo effect is observed in the lower panel of Fig.~\ref{fig:MWthermoelectric}.\\
\noindent{The} above results support the perturbative prediction \cite{mitchell2009quantum} that the physics of the half-filled TQD in the PC coupling geometry is always described by an effective single-channel, single spin-$\tfrac{1}{2}$ Kondo model (Eq.~\ref{eq:5H1ck}) at the lowest temperatures.
This is to be expected from RG theory, since the Kondo model is the \textit{minimal} effective model, comprising only the most RG-relevant terms \cite{Hewson}. The resulting universality implies that the low-energy behaviour of thermodynamical observables in the TQD is the same as that of the Kondo model when rescaled. \\
\noindent{Since} the observables of interest here are the thermoelectric transport coefficients, we expect that $G_C(T)$ and $K_Q(T)$ calculated via Eqs.~\ref{eq:GenTransfMatPC} using $\overline{\overline{G}}_{\sigma}(\omega,T)$, \ref{eq:TmatSpe} for the effective Kondo model will precisely match those of the full TQD at low temperatures. This is because the electrical and heat conductances are controlled by the spectrum of the scattering $\mathrm{T}$-matrix $t_{ee,\sigma}(\omega,T)$, which should agree in bare and effective models at low temperatures and energies -- even though the single-channel Kondo model itself has no such capacity for quantum transport measurements. This illustrates a broader point that low-energy effective models with greatly reduced complexity can still be used to obtain the correct transport properties at low temperatures -- provided the effective model parameters have been accurately determined.

\subsection{Triple quantum dot system with finite magnetic field}
We consider now transport through the TQD in the presence of a finite local magnetic field $B$. In order to do this, we add a term to the TQD Hamiltonian $H_{ field}=-B\sum_j \hat{S}^z_j$. The $\mathrm{T}$-matrix controlling quantum transport in this coupling geometry becomes spin-dependent $t_{ee,\uparrow}(\omega,T)\ne t_{ee,\downarrow}(\omega,T)$, and therefore the current for up and down spin electrons is unequal. Within the MW formalism, the total (spin-summed) linear conductances follows from Eqs.~\ref{eq:MW->PC}, \ref{eq:GenTransfMatPC}, \ref{eq:TmatSpe} using $\overline{\overline{G}}_{\sigma}(\omega,T)$.\\ 
\begin{figure}[H]
	\centering
	\includegraphics[width=0.75\linewidth]{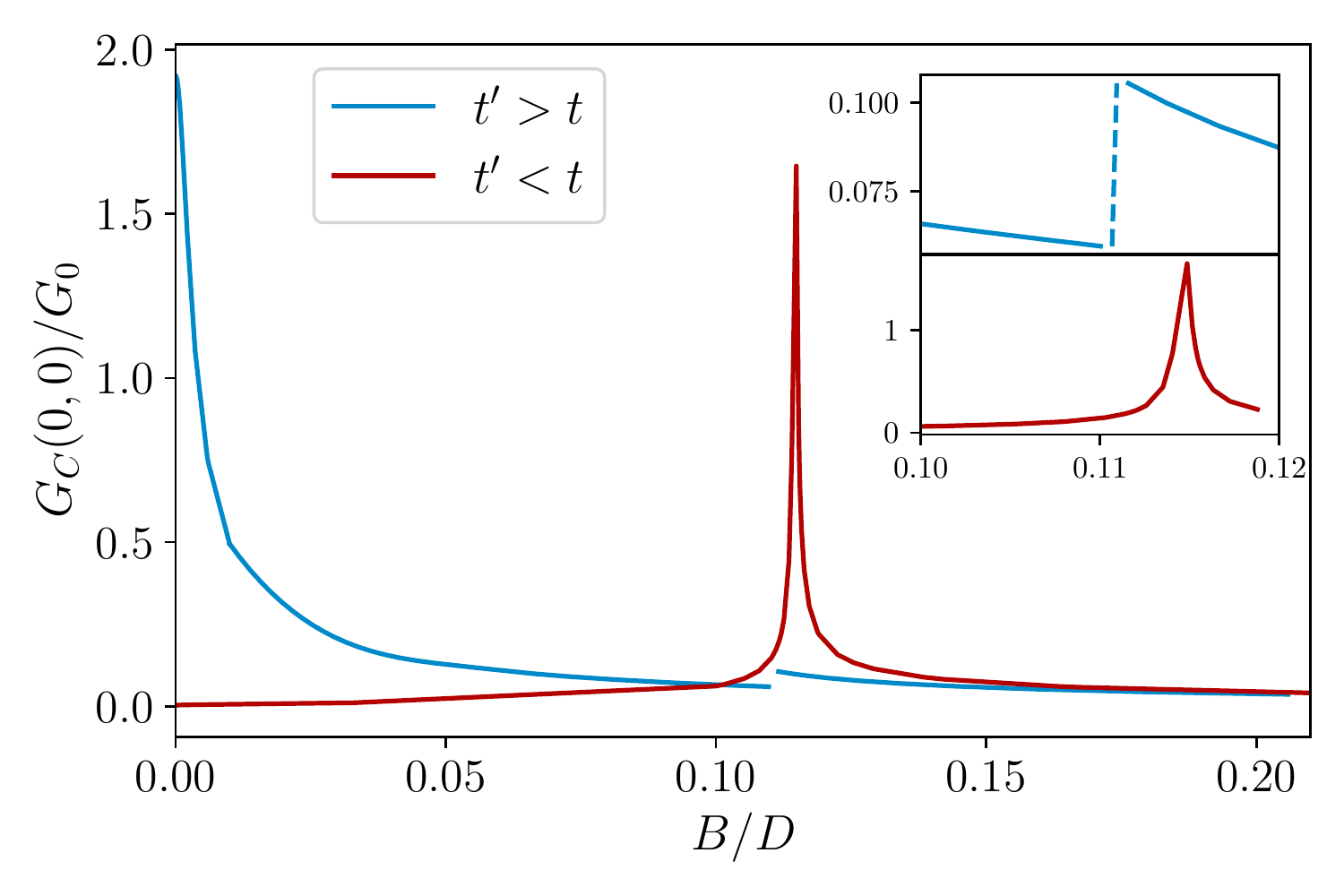}
	\caption[ac-conductance for triple quantum dot system with magnetic field]{\textit{dc}-electrical conductance $G_C(0,0)$ at $T=0$ for the TQD model in PC geometry, as a function of applied magnetic field $B$. Blue line for Kondo screened phase $t'>t$, red line for local moment phase $t'<t$. Inset shows magnified view around the field-induced ground-state transition in the isolated TQD, which results in a level-crossing QPT in the lead-coupled system for $t'>t$ but re-entrant Kondo for $t'<t$. Model parameters: $U=0.4D$, $U'=0$, $\epsilon=-U/2$, $V_s=V_d=0.1D$ and $(t,t')=(0.094,0.04)$ and $(0.08,0.1)$ for red and blue lines, respectively. NRG parameters: $\Lambda=2.5$, $M_K=2000$. }\label{fig:bfield}
\end{figure}
\noindent{Focusing} on the $T=0$ \textit{dc} linear electrical conductance $G_C(0)$ as a function of magnetic field strength $B$ in Fig.~\ref{fig:bfield}, we again compare the behaviour in the Kondo phase ($t'>t$, blue line) and the local moment phase ($t'<t$, red line). For $B=0$, we see from Fig.~\ref{fig:MWthermoelectric} that $G_C(0)=0$ for $t'<t$ but $G_C(0)\simeq 2e^2/h$ for $t'>t$ (deviations from the ideal unitarity limit for a single-electron transistor arise if particle-hole symmetry is broken). For $t'>t$, the degenerate TQD doublet ground state $|+;S^z_{ TQD}=\pm\tfrac{1}{2}\rangle$ at $B=0$ gets split at finite $B$. The Kondo effect is still operative, providing the Zeeman splitting $\sim |B|$ is much less than the Kondo binding energy $T_{ K}^+$. However, the Kondo effect is completely destroyed \cite{Hewson} and the low-$T$ conductance collapses \cite{PustilnikGlazman_review2004} for $B \gg T_{ K}^+$, since then the unique TQD ground state $|+;+\tfrac{1}{2}\rangle$ is singled out and spin-flip excitations involving $|+;-\tfrac{1}{2}\rangle$ become suppressed. The blue line in Fig.~\ref{fig:bfield} shows the expected \cite{koga2013field} rapid decrease in the conductance as $B$ increases. \\
\noindent{As} $B$ increases further to $B\sim J$, the ground state of the isolated TQD with $t'>t$ changes from $|+;+\tfrac{1}{2}\rangle$ to the extremal weight state of the spin-quartet $|\uparrow\uparrow\uparrow\rangle$. Exactly at the transition $B=B^*$, the $|+;+\tfrac{1}{2}\rangle$ and $|\uparrow\uparrow\uparrow\rangle$ TQD states are precisely degenerate. However, they cannot be interconverted by tunnelling to the leads because they have different \emph{parity} under the mirror symmetry operation \cite{mitchell2009quantum} that swaps the labels on dots 2 and 3 i.e. $|+;S^z_{ TQD}\rangle$ is symmetric while $|\uparrow\uparrow\uparrow\rangle$ is antisymmetric - see for completeness the sketch of spin state configuration in Fig.~\ref{fig:Gac}. This means that $\langle \uparrow\uparrow\uparrow | \hat{S}^+_1 | +;+\tfrac{1}{2}\rangle=0$, as can be seen directly from the form of Eq.~\ref{eq:plusstate} (although the symmetry argument holds to all orders and does not require the half-filled limit approximation). As a consequence, there can be no field-induced Kondo effect \cite{PustilnikKondoMagnetic2001,DeFranceschi-Kouwenhoven_KondoSpin1-2000Exp,PustilnikGlazman_realQD2001,golovach2003kondo,costi2000KondoBfield,Dias_2QD-SIAM_2008} in the TQD system for $t'>t$, and we see a first-order (level-crossing) quantum phase transition in the full lead-coupled system, with a change in many-body ground states at the critical field strength $B^*$. This shows up in the $T=0$ conductance as a small discontinuity -- see inset, Fig.~\ref{fig:bfield} (blue line). \\
\noindent{The} $t'<t$ phase is more interesting because, although no Kondo effect occurs at $B=0$, we now have the possibility of a field-induced Kondo effect at $B=B^*$. This is because the ground state of the isolated TQD at finite $B$ crosses over from $| -;S^z_{ TQD}=+\tfrac{1}{2}\rangle$ to $|\uparrow\uparrow\uparrow\rangle$ as a function of $B$ for  $t'<t$, and both states have the same parity i.e. they are both antisymmetric under $2\leftrightarrow 3$ exchange - see for completeness the sketch of spin state configuration in Fig.~\ref{fig:Gac}. At $B^*$, the two states are degenerate and can therefore be interconverted by coupling to the leads. The matrix element $\langle \uparrow\uparrow\uparrow | \hat{S}^+_1 | -;+\tfrac{1}{2}\rangle \ne 0$, as can be seen from Eq.~\ref{eq:minusstate}, and therefore an effective Kondo model of the form Eq.~\ref{eq:5H1ck} arises at $B=B^*$, where the two components of the effective impurity spin-$\tfrac{1}{2}$ are played here by the degenerate TQD states $| -;+\tfrac{1}{2}\rangle$ and $|\uparrow\uparrow\uparrow\rangle$, similar to the finite-field singlet-triplet Kondo effect in quantum dots \cite{DeFranceschi-Kouwenhoven_KondoSpin1-2000Exp,PustilnikGlazman_realQD2001}. This novel prediction for a field-induced Kondo effect involving the crossing of spin doublet and spin quartet states in the TQD is borne out by the NRG results shown as the red lines in Fig.~\ref{fig:bfield}, where a dramatic conductance enhancement arises for $t'<t$ at $B\sim J$. There is no quantum phase transition as a function of $B$ for $t'<t$. As last comment, we note that the conductance curve reaches only asymptotically the unitary limit. This is because of numerical limitation we could not fully resolved around $B=B^{\star}$ but we verified the corresponding spectral function is normalized for the whole $B$ field interval.\\
\noindent{The} TQD therefore hosts Kondo-boosted conductance at $B=0$ for $t'>t$, but at $B\sim J$ for $t'<t$. This provides sensitive control over the current through the TQD, with a high on-off ratio, by tuning magnetic fields. We conclude this section with Table \ref{table:TQD_B} summarizing the physics of the TQD in presence of a magnetic field.
\begin{table}[H]
	\centering
	\begin{tabular}{ |c||c|c|c| } 
		\hline
		 \textcolor{white}{b} & \textit{process} & $t^{\prime}>t$ & $t^{\prime}<t$ \\
 		\hline\hline
		$B \gg T^{+}_{K}$ & Kondo suppression &\textcolor{white}{b}& \textcolor{white}{b}  \\  
		\hline
		$B \sim J $& isolated TQD: inversion ground-state & $|+;S^z_{TQD}  \rangle\rightarrow|\uparrow\uparrow\uparrow\rangle$ & $|-;S^z_{TQD} \rangle\rightarrow|\uparrow\uparrow\uparrow\rangle$ \\
		\hline
		$B = B^{\star} $ & coupled TQD system & level-crossing transition & field-induce Kondo effect \\
		\hline\hline
	\end{tabular}  ~.
	\caption[Summary for TQD in PC configuration with finite magnetic field]{Summary of processes occurring at TQD system in PC configuration under finite magnetic field.}\label{table:TQD_B}
\end{table}

\section{Triple quantum dot system in non-proportionate coupling configuration}\label{sec:TQDnopc}
We now turn to the more complex case in which leads are attached to dots 2 and 3 (see Fig.~\ref{fig:tqd}\textit{b} and Eq.~\ref{eq:tqd_hyb_2}). Since the coupling geometry does not satisfy the PC condition, the MW formulation involving only retarded TQD Green's functions used in the previous section cannot be applied here. Instead, we obtain the linear electrical conductance by NRG from the Kubo formula Eq.~\ref{eq:defKuboLRel}, using the `improved' version of the current-current correlator, Eq.~\ref{eq:4defKuboLRel}. In this section, we present conductance calculations under various conditions\\

\begin{figure}[H]
	\centering
	\includegraphics[width=0.75\linewidth]{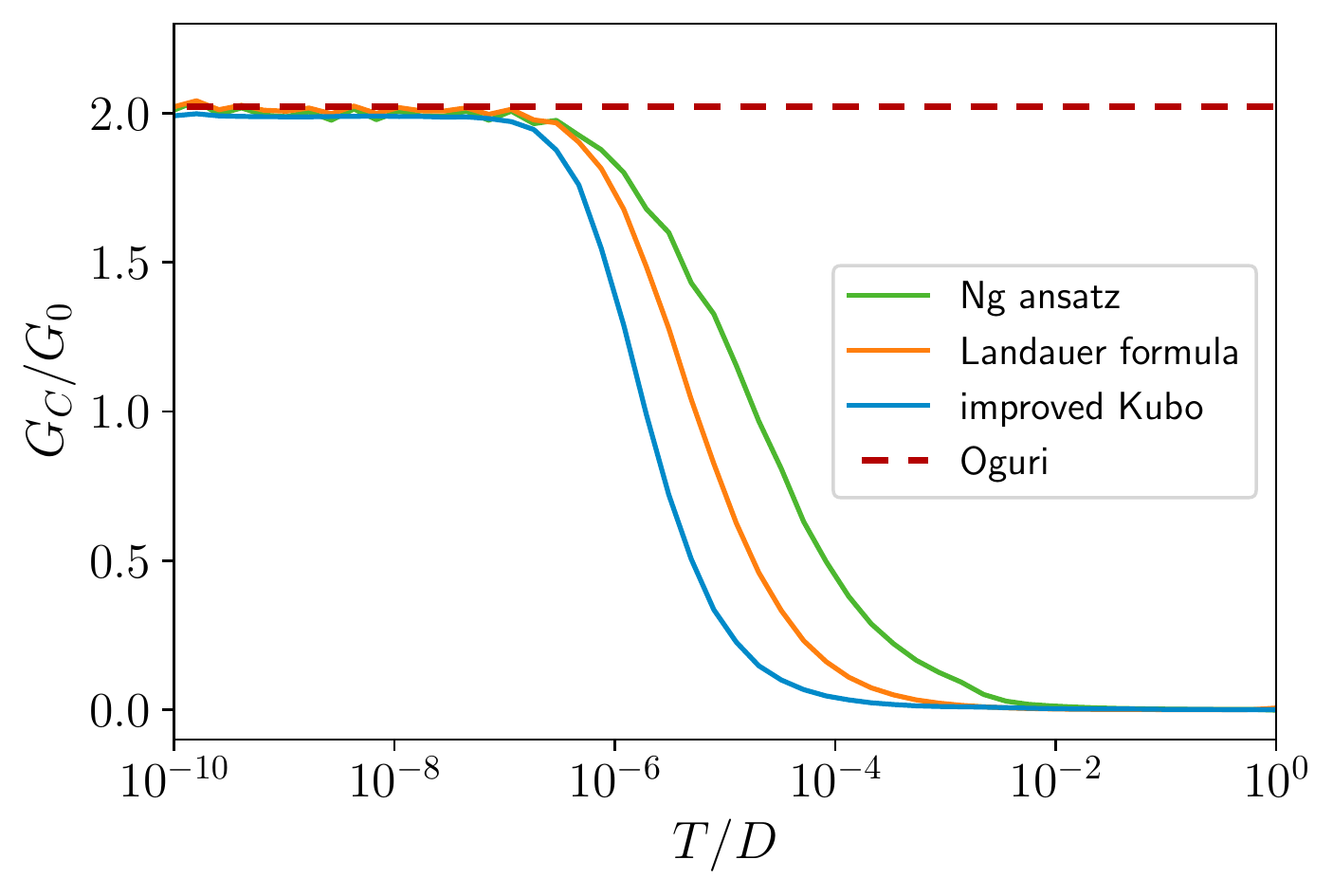}
	\caption[Techniques comparison for triple quantum dot system in non-proportionate coupling]{Serial \textit{dc}-electrical conductance $G_C(T)$ through the TQD in non-PC coupling geometry (see Fig.~\ref{fig:tqd}\textit{(b)} and Eq.~\ref{eq:tqd_hyb_2}), comparing the numerically-exact improved Kubo formula (Eqs.~\ref{eq:defKuboLRel} and its \textit{dc}-limit, Eq.\ref{eq:4defKuboLRel}, blue line) with common approximations based on Green's functions: Ng ansatz (Eqs.~\ref{eq:MW_cond}, \ref{eq:ng_T}, green line), Landauer (Eq.~\ref{eq:MW_cond} but using generalised transfer matrix without PC condition in Eq.\ref{eq:GenTransfMat} with interacting Green's functions, orange line), and Oguri renormalized Landauer (Eq.~\ref{eq:Oguri}, red dashed). Model parameters: $U = 0.4D$, $U'=0$,  $\epsilon=-U/2$, $V_s = V_d =0.12D$, $t=0.02D$, $t'=0.005D$. NRG parameters: $\Lambda=3$, $M_K = 10000$.}\label{fig:nonPC}
\end{figure}

\begin{figure}[H]
	\centering
	\includegraphics[width=0.6\linewidth]{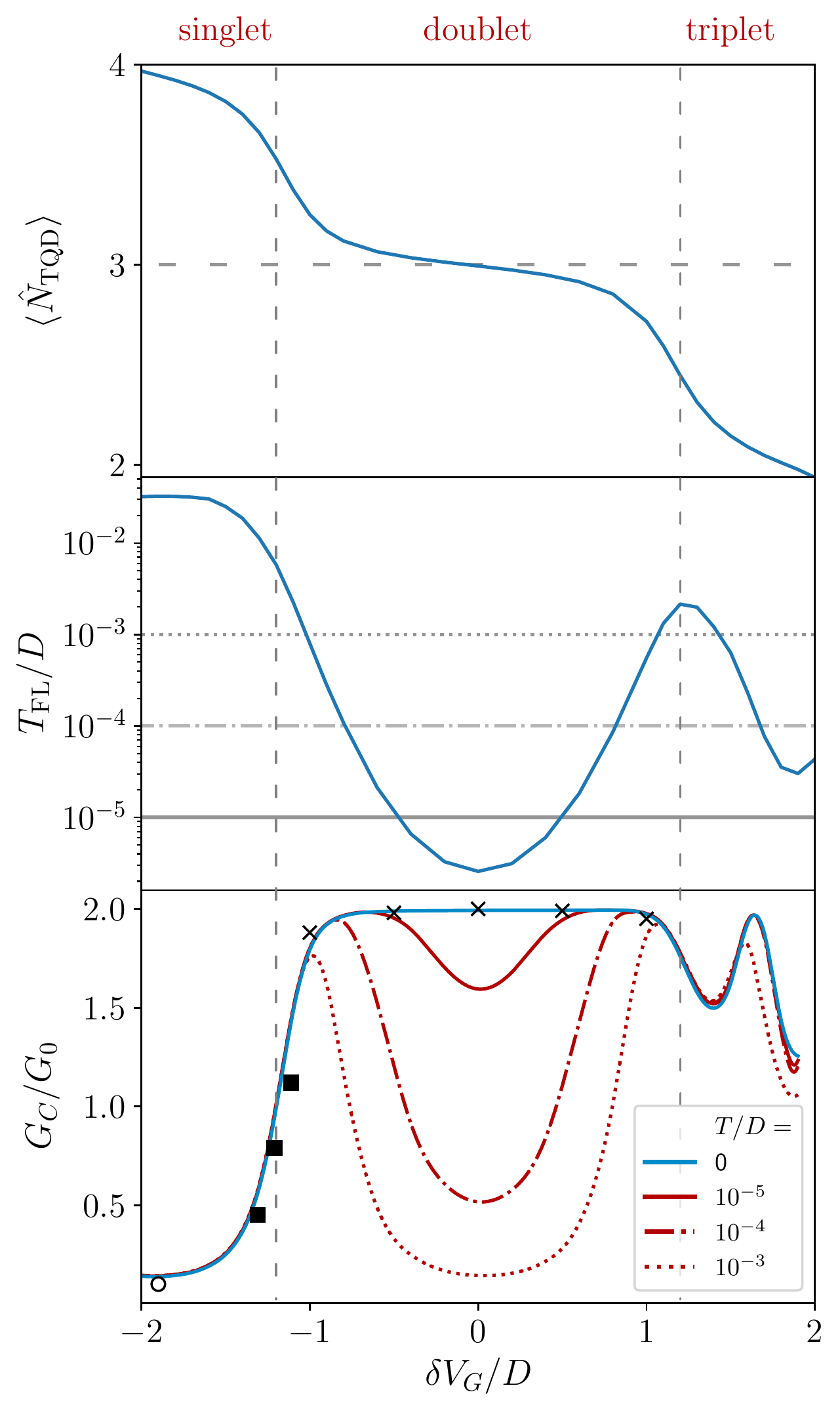}
	\caption[Triple quantum dot system in non-proportionate coupling under gate-controlled]{Gate-controlled TQD in the non-PC coupling geometry. \textit{Top panel:} TQD occupancy $\langle \hat{N}_{ TQD}\rangle$ in the full lead-coupled, interacting system, showing Coulomb blockade staircase. \textit{Middle panel:} Low-energy Fermi liquid scale $T_{ FL}$, defined as the temperature below which TQD degrees of freedom are quenched (in practice via the TQD entropy $S_{ TQD}(T_{ FL})=0.1$). \textit{Bottom panel:} \textit{dc} series electrical conductance, obtained from NRG using the improved Kubo formula,
		Eqs.~\ref{eq:defKuboLRel} and its \textit{dc}-limit, Eq.\ref{eq:4defKuboLRel}	(lines) at different temperatures given in the legend. Symbols show the $T=0$ effective model predictions: $S=1/2$ CB regime (cross points via Eq.~\ref{eq:5cond_2ck_sym}), $S=0$ CB regime (circle point via Eq.~\ref{eq:5Gqpc}), singlet-doublet MV crossover (square points via Eq.~\ref{eq:5Heff-spin-PC}). Vertical dashed lines show MV transitions in the isolated TQD for reference. Model parameters: $U=3D$, $U'=D$, $t=0.3D$, $t'=0.1D$ and $V_s=V_d=0.3D$. Gate voltage defined such that $\delta V_G=0$ is at half-filling. NRG parameters: $\Lambda=3$, $M_K=6000$.} \label{fig:gateTQD}
\end{figure}

\noindent{In} Fig.~\ref{fig:nonPC} we consider $G_C(T)$ for the TQD at half-filling with typical parameters, comparing the result of the improved Kubo formula (blue line) with approximations involving only retarded equilibrium single-particle Green's functions. The \textit{dc} conductance shows a strong Kondo enhancement at low-temperatures $T\ll T_{ K}\sim 10^{-5}D$, signifying the involvement of collective TQD spin states in series transport.\\
\noindent{The} red dashed line is the Oguri result Eq.~\ref{eq:Oguri}, for the asymptotic $T\to 0$ \textit{dc} conductance $G_C(0)$. The agreement with the result from the Kubo formula for $T\ll T_{ K}$ is as expected for a system with a Fermi liquid ground state. The green line is obtained using the Ng Ansatz, in which the approximate transmission function, namely
\begin{equation}\label{eq:ng_T}
\widetilde{\mathbb{T}}_{ Ng}(\omega,T)  = 4\sum_{\sigma} {\rm Tr}[\mathbb{G}_{\sigma}(\omega,T)\mathbb{\Gamma}^{d}\mathbb{\Lambda}_{\sigma}(\omega,T)\mathbb{G}_{\sigma}(\omega,T)^* \mathbb{\Gamma}^{s}],~\mathbb{\Lambda}_{\sigma} = \mathbb{{\rm I}}-[\mathbb{\Gamma}^s+ \mathbb{\Gamma}^d + 2\delta \mathbb{{\rm I}}]^{-1} [{\rm Im}\mathbb{\Sigma}_{\sigma}] ~,
\end{equation}
for $\delta\to 0^+$ is employed in Eq.~\ref{eq:MW_cond} \cite{Sergueev2002,zhang2002spin,ferretti2005first}. It reduces to the Oguri result, as expected, for $T\ll T_{ K}$, and captures the qualitative features of the true result, but significant deviations are observed for $T\sim T_{ K}$. Ironically, simply using the interacting retarded Green's functions in the non-interacting Landauer formula, Eq.~\ref{eq:LB_2probe} (orange line) does a better job.\\
\noindent{For} linear response electrical transport in non-PC geometries, we conclude that the improved Kubo formula is the method-of-choice. Within NRG, the accuracy of the method was validated in Fig.~\ref{F4/KvsimpK}; it is in fact also simpler and less computationally expensive than the standard approximate alternatives.\\
\noindent{In} the lower panel of Fig.~\ref{fig:gateTQD} we consider the series \textit{dc} electrical conductance through the TQD as a function of gate voltage $\delta V_{gate}=\epsilon-\epsilon_0$, defined such that $\delta V_{gate} =0$ corresponds to half-filling. The conductance $G_C$ is calculated using NRG via the improved Kubo formula at different temperatures, corresponding to the different gate voltage trace lines. We use this to test the analytic predictions of Secs.~\ref{sec:MVPC} and \ref{sec:CBPC}, which are shown as the points in Fig.~\ref{fig:gateTQD}.\\
\noindent{Before} discussing transport, we comment on the underlying Kondo physics of this system \cite{bonvca2008numerical,bonvca2007fermi}. The number of electrons on the \textit{isolated} TQD is conserved and takes integer values. We define $\langle \hat{N}_{ TQD}\rangle$ in terms of the TQD number operator $\hat{N}_{ TQD}=\sum_{j,\sigma} d_{j\sigma}^{\dagger}d_{j\sigma}^{\phantom{\dagger}}$. With finite interdot repulsion $U'>0$, the half-filled $N_{ TQD}=3$ electron sector is well-separated in gate voltage from the $N_{ TQD}=2,4$ sectors. The transitions between these filling sectors are marked as the vertical grey dashed lines in Fig.~\ref{fig:gateTQD}, and correspond to the charge-degeneracy points in the isolated TQD (for the parameters used, they arise at $\delta V_{gate}/D \approx \pm 1.2$). Analysis of the TQD shows that for $N_{ TQD}=3$ we have a spin-doublet ground state (effective impurity $S=\tfrac{1}{2}$), while for $N_{ TQD}=2$ we have an effective $S=1$ spin-triplet ground state, and for $N_{ TQD}=4$ the ground state is an effective $S=0$ spin-singlet. This results in a rich range of gate-tunable many-body physics on attaching the leads. In the top panel of Fig.~\ref{fig:gateTQD} we see how the sharp Coulomb blockade (CB) steps are now smoothed into a continuous evolution of $\langle \hat{N}_{ TQD}\rangle$. However, one can still identify clear CB regimes and mixed-valence (MV) crossovers. In particular, we have \textit{defined} $\langle \hat{N}_{ TQD}\rangle=3$ at $\delta V_{gate}=0$, while, for the parameters used, the centre of the triplet and singlet CB regimes with $\langle \hat{N}_{ TQD}\rangle=2,4$ occur at $\delta V_{gate}/D \approx \pm 2$.\\
\noindent{Deep} in the spin-doublet CB regime, one can derive an effective $S=\tfrac{1}{2}$ two-channel Kondo model of the type given in Eq.~\ref{eq:5HeffCB}. Note that in standard situations, the cross terms $J_{sd}$ and $W_{sd}$ are finite; these are the processes in the effective model responsible for mediating a source-drain current. We consider explicitly the $sd$ symmetric case with $V_s=V_d$, and so $J_{ss}=J_{dd}\equiv J$ and $W_{ss}=W_{dd}\equiv W$. Deviations from $\langle \hat{N}_{ TQD}\rangle=3$ at finite $\delta V_{gate}$ lead to finite potential scattering terms $W_{ss}$ and $W_{dd}$.\\ 
\noindent{From} the results of Sec.~\ref{sec:CBPC}, we expect an effective spin-$\tfrac{1}{2}$ single-channel Kondo effect on the lowest energy scales, forming with the even lead combination. The resulting Kondo temperature $T_{ K}$ in this regime is shown in the middle panel of Fig.~\ref{fig:gateTQD}, and exhibits the typical behaviour of Eq.~\ref{eq:5tk_1ck}. At $T=0$ we therefore expect Kondo-boosted conductance. This is shown in the lower panel of Fig.~\ref{fig:gateTQD} as the blue line, where the conductance is seen to reach up to $G_C(0) \simeq 2e^2/h$ at the particle-hole symmetric point $\delta V_{gate}=0$. In fact, the conductance remains close to this maximum value throughout most of the doublet regime, despite significant deviations in $\langle \hat{N}_{ TQD}\rangle$ away from half-filling. The $T=0$ conductance only begins to attenuate when close to the mixed-valence points. \\
\noindent{This} behaviour can be understood from the effective low-energy model. 
With the chosen bare TQD parameters $U=3D$, $U'=D$ and $V_s=V_d=U/10$, the perturbative estimation of the effective model parameters via Eq.~\ref{eq:5HeffCBPer} is quite accurate. The low-$T$ conductance can then be estimated from Eq.~\ref{eq:5cond_2ck_sym}. For strong interactions, we find that the effective potential scattering parameters $W_{ee}$ and $W_{oo}$ remain small until close to the MV points (where the applicability of Eq.~\ref{eq:5HeffCBPer} and hence Eq.~\ref{eq:5cond_2ck_sym} breaks down anyway). The numerical values of $G_C(0)$ predicted in the doublet CB regime are shown in the lower panel of Fig.~\ref{fig:gateTQD} as the cross points, and agree rather well with the full NRG calculations. \\
\noindent{The} red lines in Fig.~\ref{fig:gateTQD} show NRG results for the conductance at higher temperatures. Since the conductance remains Kondo-enhanced for $T\ll T_{ K}$ but is strongly reduced as the Kondo effect is destroyed for $T\gg T_{ K}$, we expect a non-trivial behaviour of $G_C(T)$ at a fixed $T$ on changing the gate voltage $\delta V_{gate}$, because $T_{ K}$ itself depends on $\delta V_{gate}$. We plot the conductance at $T/D=10^{-5}, 10^{-4}, 10^{-3}$ in the lower panel (these same temperatures are marked as the horizontal lines in the middle panel). As expected, we see reduced conductance as the ratio $T/T_{ K}$ increases.\\
\noindent{Deep} in the CB spin-singlet regime around $\delta V_{gate}/D=-2$, the low-energy effective model is described by Eq.~\ref{eq:5Heff_S0}, with conductance controlled simply by the effective cotunnelling amplitude $W_{sd}$ via Eq.~\ref{eq:5Gqpc}. There is no low-energy scale in this regime; the effective model is valid for all $T\ll U$ (we denote the scale where the spin singlet ground state forms as $T_{\rm K}$, but this should be understood as a Fermi liquid scale rather than a Kondo scale \textit{per se}). The perturbative estimation of $W_{sd}$ and hence $G_C(0)$ is shown as the circle point in the lower panel of Fig.~\ref{fig:gateTQD} and agrees well with NRG calculations in this regime. As anticipated, the conductance is small since electrons must tunnel through the interacting TQD unaided by the Kondo effect. \\
\noindent{In} the CB spin-triplet regime, both even and odd conduction electron channels are involved in screening the collective TQD spin states (see Sec.~\ref{sec:CBPC}). The delicate interplay between the effective couplings $J_{ee}$ and $J_{oo}$ (and hence the Kondo scales $T_{ K}^e$ and $T_{ K}^o$) produces more complex behaviour in the conductance. The Fermi liquid scale $T_{ FL}$  corresponds here to the temperature below which the TQD spin is exactly screened, $T_{ FL} \equiv \min(T_{ K}^e,T_{ K}^o) = T_{ K}^o$. From NRG we see a Kondo peak reaching up to the maximum $G_C=2e^2/h$ at around $\delta V_{gate} \approx 1.7$ (that is, away from the center of the CB window at $\delta V_{gate} \simeq 2$ where $\langle \hat{N}_{ TQD}\rangle=2$). This can be viewed as a many-body quantum interference effect resulting from different conductance pathways around the TQD ring \cite{Mitchell2017KondoMolecule}.  The insensitivity of the conductance in this regime to the ratio $T/T_{ K}$ implies a two-stage screening with $T_{ K}^e\gg T_{ K}^o$. The first-stage partial screening arises at relatively high temperatures $T_{ K}^e >T$ and is chiefly responsible for the enhanced conductance.\\
\noindent{We} now consider the MV crossovers between the above regimes, where the TQD charge is strongly fluctuating and the CB approximation breaks down. For $\delta V_{gate}/D \approx -1.2$ the isolated TQD has  degenerate ground states with $N_{ TQD}=3$ and $4$ electrons, corresponding to a singlet-doublet MV transition. This is smoothed into a crossover on attaching the leads and the TQD occupancy varies smoothly as a function of gate; the many-body ground states in the Kondo-screened doublet CB regime and the CB singlet regime are continuously connected. In the vicinity of the singlet-doublet MV point we use the effective model Eq.~\ref{eq:5Heff-spin-PC}, calculating the effective parameters $\tilde{\Gamma}$ and $\epsilon$ from the isolated TQD eigenstates and energies. Since this effective model is in PC, the transmission function within the MW formalism then follows from $\mathrm{T}$-matrix as plotted in Fig.~\ref{fig:tmatrix_MV_sd}. The low-$T$ conductance obtained using this approximation are shown as the solid square points in Fig.~\ref{fig:gateTQD}, and capture the behaviour of the exact $T=0$ result (blue line).\\
\noindent{At} $\delta V_{gate}/D=+1.2$ the system is at the MV point between the $N_{ TQD}=3$ doublet and the $N_{ TQD}=2$ triplet. An effective model describing the coupling of these 5 TQD states to the leads to first order in the hybridization was derived in Eq.~\ref{eq:5Heff_mv_d-t}. However, diagonalization of the isolated TQD shows multiple low-lying excited states in this regime. The analysis of Sec.~\ref{sec:MVPC} requires that the retained 5-fold-degenerate ground state TQD manifold be well-separated from excited states, and hence breaks down in this case. In addition, the coupling to the odd conduction channel is not captured by first-order perturbation theory, but 2-stage Kondo screening involving both even and odd channels is observed in NRG calculations for the full lead-coupled system. This becomes important when $T \lesssim T_{ K}^o$. Predictions of the  conductance using Eq.~\ref{eq:5Heff_mv_d-t} in this regime (not shown) considerably underestimate the exact results. This demonstrates that care must be taken when using results of the effective models to check that the conditions used to derive them are satisfied for a given system. Of course, in such cases the effective models could be generalized to include more TQD states and/or go beyond first-order perturbation theory.

\chapter{Transport in molecular junction: benzene transistor}\label{ch:benz}
In this chapter we investigate the interplay of symmetry and Kondo effect in a benzene single-electron transistor using a combination of NRG numerical computation as well as first- and second- (generalised Schrieffer-Wolff transformation) order perturbation expansion in the hybridization. Depending on the connectivity of the leads and the applied gate voltages, we uncover spin-$1/2$ and spin-$1$ Kondo effects including a two-stage Kondo effect, a first order quantum phase transition, and zero conductive state at an emergent $SU(4)$ symmetry point. The interplay between emergent many-body effects and molecular symmetry is discussed in the context of quantum-boosted device functionality based on the conductance characteristics.

\section{Introduction: single-molecule electronics}
Individual molecules when used as active electronic components show great potential for novel, low cost device functionalities beyond semiconductor devices \cite{Evers_SingMolTran2020,VanDerZant_BreakJunc2019}. Quantum effects become inevitable at such small length scales producing exotic, emergent phenomena that have no classical analogues~\cite{Aradhya_SingMolJunc2013}. Recent progress in nano-fabrication techniques has helped us realise quantum boosted device functionalities. \\
Currently there exists a vast range of molecules that may be exploited and controlled to produce robust functionalities at the nanoscale. The charge states of these molecules can be tuned with the application of an external gate voltage
in the lead coupled system - this is referred to as a gate-controlled set-up. Once the single-molecule is attached to source/drain electrodes, the lead-molecule-lead coupled system with a back gate is a proper three-terminal device. Electronic population in the leads is regulated via the potential difference $V_{bias}$ while the charge occupancy on the molecule is fine tuned via the back gate voltage $V_{gate}$. The coupled system recovers then the \textit{transistor} device functionality. At low temperatures, a small $V_{gate}$ variation causes the electrical conductance to change from blockade to maximum intensified value $\mathcal{G}_{0}$. Hence, the \textbf{molecule junction} behaves either as switch or as an amplifier according to temperature and voltage scale. We note that the conductance fluctuation values are observed both at given temperature by changing charge regime and within the same charge sector by lowering progressively the temperature scale. \\
The most common among the molecule class employed in these coupled systems are the \textit{organic molecular} junctions~\cite{Evers_SingMolTran2020}. In particular, molecules containing benzene rings or similar phenyl groups have been studied extensively over the years~\cite{Reed_CondMolJunc1997,Danilov_SingMolJunc2008}. \\
Single-molecule junctions are understood as  more elaborate quantum impurity models from those introduced in Sec.\ref{sec:QImpModel}. This is because of the multi-orbital structure and degeneracies due to internal symmetries in single molecules disclose a richer physics than the one of single or few orbital systems. For instance, we consider the ground-state of a benzene molecule junction whose state could be a nondegenerate singlet, a two-fold spin degenerate doublet or a three-fold spin degenerate triplet. Those represent the ground-state of a neutral, a singly charged and a doubly charged molecular state respectively, as a function of external gate voltage. Based on the charge state, the \textbf{benzene molecule} can thus behave like a magnetic impurity with a residual spin $S$ which at low enough temperatures, gets entangled with the conduction electrons of the leads realizing the many-body Kondo effect~\cite{Scott_KondoMol2010}. In addition, symmetries result in pseudospin degrees of freedom. In conclusion, the benzene single-molecule exhibits a vast range of many-body quantum effects and it is a promising candidate for future nanoelectronic devices. Despite its complexity, the electronic structure is correctly described by an effective two channel impurity model at low energy and temperature scales - whose theory is the same as we discussed in detail in Secs.\ref{sec:QImpModel},\ref{sec:2CK}.\\
The outline of this chapter starts with presenting the lead-benzene-lead modelling and its symmetry properties, then we derive the effective model at mixed-valence regime showing emergent proportionate coupling condition that allows transport formulation under MW current expression in PC. The effective parameters and NRG data are used to calculate the benzene junction observables meaning the molecule entropy and electrical conductance to fully characterise the coupled system functionality. At the end, we conclude with a brief discussion. \\
The results of this chapter are based on the paper \cite{benzene}.

\section{Benzene single-molecule junction: model and symmetries}
As discussed in the introduction to this chapter, in this work we study the benzene molecule junction. Here, we first present the bare modelling and properties of both isolated benzene single-molecule and benzene junction. Then, we address the discussion to the molecular symmetry. This is because at low-energy and temperature scales, thermodynamics and quantum electrical transport properties are largely influenced by the lead-molecule-lead system symmetry. Furthermore, the leads geometry intrinsically determines not only the $SU(2)$ Kondo effect characteristics but also gives access to higher $SU(4)$ symmetry Kondo states. Hence, we conclude the section by introducing the concept of a pseudospin degree of freedom and its role in tunnelling events.\\

\begin{figure}[H] 
	\centering
 \hspace*{-0.6cm}	\includegraphics[width=1.1\linewidth]{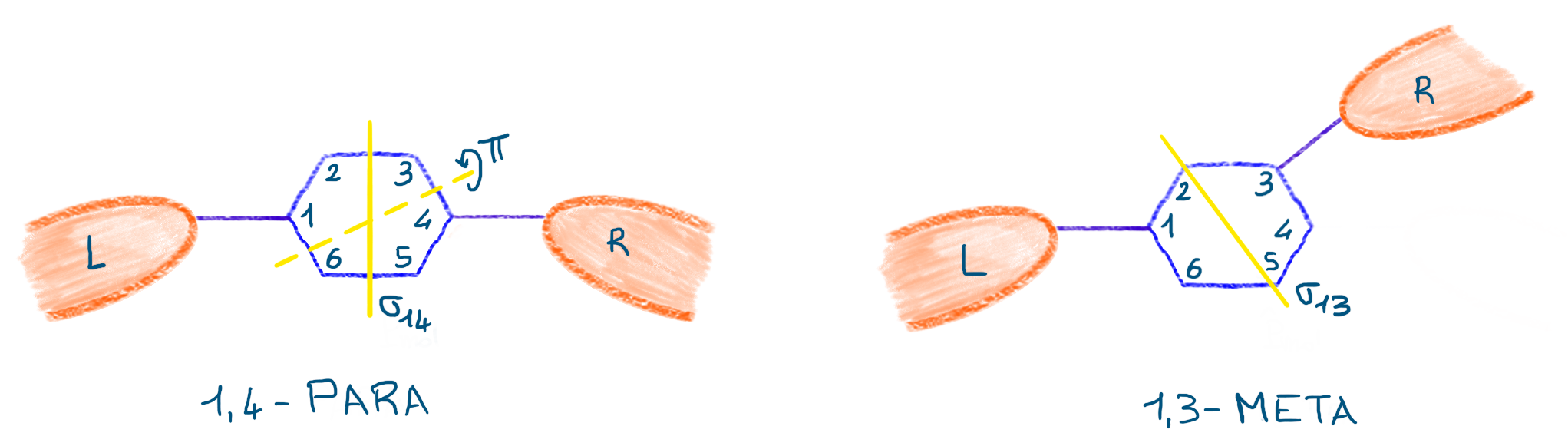}
	\caption[Schematic of the benzene junction \textit{para} and \textit{meta} configuration]{Schematic of the benzene junction \textit{para} \textit{(left)} and \textit{meta} \textit{(right)} configuration. In the $1,4$-configuration we find mirror symmetry along both $\sigma_{14}$ plane (solid line) and $\pi$ axis (dashed line); in the $1,3$-configuration we find mirror symmetry only along $\sigma_{13}$ plane (solid line).}\label{F6:schematic}
\end{figure}

\noindent{To} uncover the role of symmetry on the many-body physics we simulate the \textit{benzene single-molecule junction} in two geometries, namely the $1,4$- or \textit{para}- configuration and the $1,3$- or \textit{meta}- configuration as shown in the schematic in Fig.\ref{F6:schematic}. In the $1,4$ case, the right/drain lead is are situated at $180^{\circ}$ with respect to the left/source one. We find mirror symmetry along $\sigma_{14}$ plane (solid line) and rotation along $\pi$ axis (dashed line, into the page through molecule centre). The model is invariant under left$\leftrightarrow$right exchange. In the $1,3$ case, the right/drain lead is are situated at $120^{\circ}$ with respect to the left/source one. We find invariance only under left$\leftrightarrow$right exchange in the mirror symmetry along $\sigma_{13}$ plane (solid line), but no rotation symmetry about the axis into the page. This difference in the symmetries turns out to be important, as we shall see.\\ 
\noindent{The} molecular junction set-up is described by the interacting, non-proportionate coupling model Hamiltonian, namely:
\begin{equation}\label{eq:6Hbare}
	\begin{aligned}
		&\hat{H} = \hat{H}_{leads} + \hat{H}_{mol} + \hat{H}_{hyb} ~,\\
	&\hat{H}_{leads}= \sum_{\alpha=L,R}\sum_{\sigma,\mathbf{k}} \epsilon_{\mathbf{k}} c^{\dagger}_{\alpha  \mathbf{k}\sigma}c_{\alpha  \mathbf{k}\sigma} ~,\\
& \hat{H}_{mol}=-t\sum_{ij,\sigma}(d_{i\sigma}^\dagger d_{j\sigma}+ H.c.)+\sum_{i,\sigma} (\epsilon-eV_{gate}) d_{i\sigma}^\dagger d_{i\sigma}+U\sum_{i}\hat{n}_{i\uparrow}\hat{n}_{i\downarrow}+\sum_{i\ne j}V_{ij}\hat{n}_{i}\hat{n}_{j} ~, \\
& \hat{H}_{hyb}= 	\sum_{\alpha,\sigma}  V_{\alpha}\big(c^{\dagger}_{\alpha \sigma}d_{i_{\alpha}\sigma}  +c^{\dagger}_{\alpha \sigma}d_{i_{\alpha}\sigma} \big)  ~.
	\end{aligned}
\end{equation}
We discuss now each term appearing in Eq.\ref{eq:6Hbare}. The $\hat{H}_{leads}$ indicates the noninteracting, thermal equilibrium conduction electron leads for $c^{\dagger}_{\alpha \mathbf{k}\sigma}(c_{\alpha  \mathbf{k}\sigma})$ creation (annihilation) Fermionic operator acting on $\alpha=L,R$ lead with $\sigma=\uparrow,\downarrow$ spin and momentum $\mathbf{k}$. The isolated molecule $\hat{H}_{mol}$ is represented by the \textit{Pariser-Parr-Pople model} (PPP model) \cite{pariser1953,pople1953}, where $d^{\dagger}_{i\sigma}(d_{i\sigma})$ operator creates (annihilates) an electron with spin $\sigma$ in site $i$ of the isolated molecule which represents an extended $\pi-$conjugate system of electrons. These operators identify \textit{localised} orbitals corresponding to the carbon $P_{z}$ orbitals in the benzene molecule. In the PPP model we have the parameters $U$ representing the local on-site Hubbard interactions and $V_{ij}$ being the non-local Coulomb interaction experienced by the electrons for $\hat{n}_i=\hat{n}_{i\uparrow}+\hat{n}_{i\downarrow}$ with $\hat{n}_{i\sigma}=d^{\dagger}_{i\sigma}d_{i\sigma}$. According to the standard \textit{Ohno parametrization} \cite{ohno}, the $\pi-$conjugate systems are typically described by $U\sim 11~eV$ and $V_{ij}=U/\sqrt{1+|\vec{r}_{ij}|U^{2}/207.3}$, where $\vec{r}_{ij}$ is the distance between $i$-th and $j$-th sites. In the present work, we choose the parameters $U=11.26~eV$ and nearest-neighbour hopping $t=2.4~eV$. Lastly, $\hat{H}_{mol}$ includes also the application of an external voltage $V_{gate}$ that is macroscopically tuned. In the third Hamiltonian term $\hat{H}_{hyb}$, we define  $ c_{\alpha\sigma} = V^{-1}_{\alpha} \sum_{\mathbf{k}} V_{\alpha\mathbf{k}} c_{\alpha \mathbf{k}\sigma}$ the annihilation operator for the localised orbital in $\alpha$-lead at the junction and $i_\alpha$ represents the frontier orbital of the molecule that couples with $\alpha$-lead via a tunnel coupling energy parameter $V_\alpha$ where $V^{2}_{\alpha}=\sum_{\mathbf{k}}V^{2}_{\alpha\mathbf{k}}$. In this modelling, we consider the case where the fully coupled system preserve invariance under \textit{mirror symmetry} i.e. left$\leftrightarrow$right exchange, thus $V_{L}=V_{R}\equiv V$. In this work we choose $V=1~eV$. \\ 
We remark that the PPP model does not incorporate any supplementary interacting effects such as Hund's coupling, RKKY interaction etc. In case vibrational models would be included, $(i)$ additional secondary peaks appear symmetrically at the sides of the main Kondo resonance with periodicity given by the phonon frequency \cite{paaske2005vibrational} and $(ii)$ the Fermi liquid description of the system is affected with renormalization of the hybridization \cite{da2009phonon}.\\
The only parameter left unspecified in the PPP model in $\hat{H}_{mol}$ is the single-particle value $\epsilon$. We select it such that the benzene molecule is in neutral thermal average state: it means its ground-state is populated by $N=6$ electrons with one electron occupying each site. This is the charge system condition we set as half-filling regime and we can think of $d_{i\sigma},d^{\dagger}_{i\sigma}$ operators describing six localised orbitals on each site of the hexagonal benzene ring. As consequence, in this modelling, the benzene molecule hosts maximum $N=12$ electrons. Considering the usual single-electron transmission process and the parameters of choice, by tuning the voltage gate in the system we can visit $N=6,7,8$ charge sectors with six, seven and eight localised orbitals on the sites respectively. These values are confirmed by the NRG calculation, as we discuss next.\\
\begin{figure}[H]
	\centering
	\includegraphics[width=0.85\linewidth]{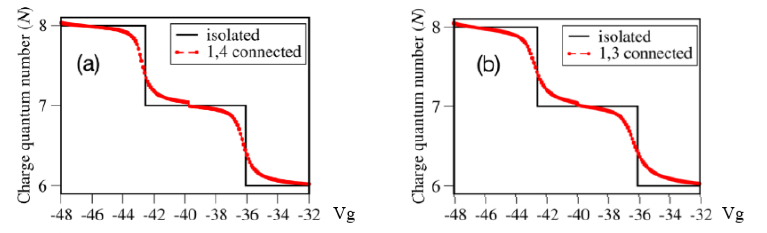}
	\caption[Occupation staircase of the benzene molecule using NRG data for both isolated molecule and coupled molecule-leads system.]{Benzene molecule ground-state charge occupancy staircase $N$ both in the isolated case (black solid line) and in the coupled molecule-leads (red dashed curve with symbols) for $1,4$ and $1,3$ configurations respectively in $(a),(b)$ as function of gate voltage $V_{g}$. In the benzene junction, the hybridization introduces valence fluctuations which round the occupation line at the crossover between integer charge sectors.}\label{F6:MolCharge}
\end{figure}
\noindent{The} coupled system as modelled in Eq.\ref{eq:6Hbare} is a gate-controlled device meaning that by tuning $V_{gate}$ we explore the so-called Coulomb electronic occupation staircase - see a sketch in Fig.\ref{F3:QD_VV}. In Fig.\ref{F6:MolCharge}, we show NRG data from the thermal average total ground-state occupation giving the charge sector numbers. This confirms the values $N=6,7,8$ electrons and the corresponding ground-state spin values which are found within specific $V_{gate}$ intervals. We present results for the molecule ground-state charge occupancy $N$ both in the isolated and in the coupled molecule-leads case. \\
We discuss first the charge distribution in the isolated molecule - see black solid line in Fig.\ref{F6:MolCharge}. We should note that these results on ground-state occupancy and spin state can be found also from diagonalization of the isolated system, namely $\hat{H}=\hat{H}_{leads}+\hat{H}_{mol}$. However, this is usually not very accurate and so we proceed with the NRG results only. In the plot, the segmented line depicts sharp discontinuities at $V_{gate}\sim-43~ eV$ in correspondence of the mixed-valence doublet-triplet transition from $N=7$ to $N=8$ charge sectors and at $V_{gate}\sim-36~ eV$ at mixed-valence singlet-doublet transition from $N=6$ to $N=7$ charge sectors. The integer charge regions are identified at $-36~eV< V_{gate}< -32~eV$ for $N=6$ characterised by total molecule spin $S=0$, at $-43~eV< V_{gate}< -36~eV$ for $N=7$ corresponding to $S=1/2$ and at $-48~eV< V_{gate}< -43~eV$ for $N=8$ for $S=1$. We remark that the voltage gate scale is computed with respect to the chosen model parameters - thus, the half-filling condition happens at $V_{gate}\sim -30 ~eV$. Applying an off-set, we can reset $V_{gate} = 0$ in the middle of $N=6$ sector corresponding to the natural configuration of neutral benzene and then rescale accordingly the values. We now focus on the molecule charge occupancy in the lead connected benzene single-molecule junction - see red curve in Fig.\ref{F6:MolCharge}. In NRG, the charge quantum number is computed as $N=\hat{N}_{tot}-\hat{N}^{0}_{tot}$ given by the difference of the total charge in the full coupled system minus the total charge of the detached leads. The coupling between leads and molecule introduces valence fluctuations where deviations from integer charge and spin states occur and molecule charge is no longer conserved. As consequence, in correspondence to the crossover between adjacent even/odd charge sectors, the molecule occupancy curve shows a smooth behaviour.\\
Several tunnelling regimes, from Coulomb blockade to enhanced conductive states, develop in the benzene junction as we explore various charge sectors in Fig.\ref{F6:MolCharge} accessible through appropriate $V_{gate}$ tuning. The quantitative analysis of the conductance is given with NRG data in the last section.\\

\noindent{We} now shift the discussion on the benzene molecular symmetries. As standard scenario in multi orbital systems, we usually encounter nondegenerate orbitals where the system satisfies $SU(2)$ spin symmetry. However, at low energy and temperature scales, additional orbital symmetries may arise such that the system explores higher symmetry states.\\
This is indeed what happens in the $N=7$ charge sector which we consider first. By decreasing energy and temperature, extra symmetries appear to supplement the physical spin state with new orbital degeneracies. Those degeneracies contribute as \textbf{pseudospin degrees of freedom} and we count those in addition to the physical spin degrees of freedom in the effective model. In both $1,3$ and $1,4$ configurations, each spin $\uparrow$ and spin $\downarrow$ component of the spin doublet is two-fold degenerate with respect to the pseudospin degree of freedom. Hence, the whole benzene ground-state doublet is \textit{four-fold} degenerate in spin-pseudospin degrees. Although we could classify also the $N=6,8$ charge sectors with one of the several point group symmetry of the benzene molecule, because each of those spin state has only one ground-state multiplet, we introduce the pseudospin degree solely for odd $N=7$ charge sector. \\
In order to shed light on the extra orbital degeneracies arising in the two doublet spin states, we analyse first the isolated benzene molecule. We find three conserved quantities, namely: $(i)$ the charge quantum number on the molecule $N$, $(ii)$ the \textit{z}-component of the total spin on the ground-state molecule $\hat{S}^{z}$ and $(iii)$ parity $p$ through the \textit{parity operator} $\widehat{P}_{mol}$ of the isolated molecule corresponding to the $\sigma_{14}$ mirror symmetry in the $1,4$ configuration or to the $\sigma_{13}$ mirror symmetry in the $1,3$ configuration, see Fig.\ref{F6:schematic}. In some sense, we can regard $\widehat{P}_{mol}$ as \textit{spacial} parity symmetry operator. Thus, we can use those as good quantum numbers to construct molecular state as $\ket{\Psi^{Q=N;S^z}_{l}}\doteq\ket{N;S^{z};p;j}$, where $l$ indexes eigenstates with the same quantum numbers and $l=0$ refers to the ground-state eigenstate, $j$ degeneracy in energy levels - in analogy to the state definition we use in Sec.\ref{sec:MVPC} and fully defined by Eq.\ref{eq:5Q-BasisState-spin-Def}. There are total of $4^6$ eigenstates $\ket{\Psi^{Q=N;S^z}_{l}}$ where this power indicates four Fermionic degree of freedom spanned over six sites of the benzene ring and among all these states we select the ground-state. If we apply the operator $\widehat{P}_{mol}$ to the doublet multiplet states we find:
\begin{equation}
	\begin{aligned}
\widehat{P}_{mol} \ket{7;\pm\tfrac{1}{2};+}= +\ket{7;\pm\tfrac{1}{2};+} \quad,\quad
\widehat{P}_{mol} \ket{7;\pm\tfrac{1}{2};-}= -\ket{7;\pm\tfrac{1}{2};+-} ~,
	\end{aligned}
\end{equation}
where even (odd) molecular parity is represented by the parity $+(-)$ and corresponds to symmetric (antisymmetric) eigenstates. We can therefore associate the molecule parity $p=\pm1$ with a pseudospin degree of freedom.
Hence, the correspondence between parity and pseudspin degrees is used both in the derivation of the effective model and in the discussion of the thermodynamic properties of the system. To sum up: at any $T\ll U$, both parities - the two pseudospin degrees - per doublet component are contributing to the electronic structure of the isolated molecule with a net total spin $S=\tfrac{1}{2}$ on the ground-state.\\
In the isolated molecule there are several symmetries, some of which we show in the schematic in Fig.\ref{F6:schematic}. In the lead-coupled system, not all those symmetries remain invariant. However, the mirror symmetries in both $1,3$ and $1,4$ configurations - indicated by solid lines in Fig.\ref{F6:schematic} - are \textit{still} conserved, as is the axial $\pi$ rotation in $1,4$. Hence, it is consistent to classify the pair of degenerate doubles according to the molecule parity $\widehat{P}_{mol}$ under simultaneous exchange of left and right leads. We remark that without the invariance in the leads exchange the parity is \textit{not} conserved in the coupled system. To quantify this operation we introduce the \textit{total parity operator} defined as $\widehat{P} = \widehat{P}_{mol}\otimes\widehat{P}_{LR}$, where $\widehat{P}_{LR}$ swaps left and right leads, and has even (odd) lead parity eigenvalues $+1(-1)$. The operator  $\widehat{P}$ is \textit{always} conserved the fully coupled system in both $1,3$ and $1,4$ configurations because it commutes with the Hamiltonian in Eq.\ref{eq:6Hbare} i.e. $[\widehat{P},\hat{H}]=0$. On the contrary, the invariance of the molecule parity depends on the coupling geometry of the leads. We can write the operator $\widehat{P}_{mol}$ in term of $\widehat{P}_{ij}$ for each pair of $i,j$ exchanged benzene orbitals under mirror symmetry. In the $1,4$ case on the right of Fig.\ref{F6:schematic}, we have  $\widehat{P}_{mol}=\widehat{P}_{14}\widehat{P}_{23}\widehat{P}_{56}$ and this parity is \textit{conserved} in the coupled system under mirror symmetry $\sigma_{14}$. This means that different parity spin-doublet states cannot mix. Hence, we can distinguish the doublet components according to the parity value so we have $\ket{\Psi^{N=7,S^z=\pm1/2}_{0}}=\ket{N=7;S^z=\pm\tfrac{1}{2};p=\pm1;0}$.
In the $1,3$ case on the left of Fig.\ref{F6:schematic}, we have instead  $\widehat{P}_{mol}=\widehat{P}_{13}\widehat{P}_{46}$ but this molecular parity is \textit{not} conserved in the coupled system under mirror symmetry $\sigma_{13}$. Hence, spin-pseudospin degrees are intrinsically entangled and become mixed on coupling to leads. The true many-body ground-state of the lead-coupled molecule therefore comprises of a mixture of $\ket{N=7;S^z=\pm\tfrac{1}{2};p=\pm1;0}$ states.\\
At $T=T^{\star} \ll \delta E$ with $\delta E$ the energy splitting between the lowest two $N=7$ doublets in the lead-coupled system, the system undergoes to a \textbf{pseudospin freezing} process meaning that the higher energy pseudospin degree is frozen out. Hence, at any $T\leq T^{\star}$, there is only one active doublet component in tunnelling events. Hence, we indicate the residual parity doublet state after the pseudospin freezing process as:
\begin{equation}\label{eq:6def_BasisStateParity}
	\ket{\Psi^{N=7,S^z}_{0}}=\ket{N=7;S^{z};p^{\star};0} ~,
\end{equation}
where $p^{\star}$ is the parity of the overall molecule ground-state.\\
Considering that the physics of our interest occurs at $T_{K} \ll T^{\star}$, the effective model, thermodynamics and quantum transport properties are studied with respect to the parity of the lowest energy state - calculated from NRG data according to each set-up configuration. \\
We conclude the discussion on the molecular symmetries with two final observations. The first one regards the ground-state molecule parity showing different values according to gate voltage. As we see in Fig.\ref{F6:MolCharge}, by varying $V_{gate}$, we access both to deep inside the Coulomb diamond area characterised by fixed charge and spin and to boundary where valence and spin fluctuations arise. 
The second note is about how the perturbation order in $\hat{H}_{hyb}$ affects the parity. Up to order $\mathcal{O}(\hat{H}_{hyb})^{2}$, we find the molecular parity $p=\pm1$ is conserved in the full lead-coupled system for $1,4$ configuration but \textit{not} conserved for $1,3$. We consider only terms up to  $\mathcal{O}(\hat{H}_{hyb})^{2}$ which involve a subset of the total symmetries.

\section{Effective model for benzene junction in mixed-valence regime}
The bare modelling in Eq.\ref{eq:6Hbare} is evidently lacking of an equal tunnelling among each frontier orbital on the benzene molecule with source and drain leads. The microscopic model in non-PC requires alternative quantum transport formulation from those currently encompassed in literature. As discussed in Sec.\ref{sec:MVPC}, the derivation of effective model describing the low energy and temperature scale system implies a new set of properties and parameter values. Among those, we find the effective hybridizations hold an emergent proportionate coupling property. Thus, in virtue of the effective parameters, we can disclose the low temperature dynamics of general system without PC set-up. \\
In this section we use the analytical theory derived in Sec.\ref{sec:MVPC} to calculate the effective model of the benzene junction at the mixed-valence regime (MV) between $6\leftrightarrow7$ singlet to doublet state at $V_{gate} = -36~eV$ and between $7\leftrightarrow8$ doublet to triplet at  $V_{gate} = -43~eV$ crossovers, see again the charge occupancy staircase in Fig.\ref{F6:MolCharge}. Within the MV transition, we consider as previously the single-electron transmission occurring between degenerate molecule ground-state spin $S$ and spin $S+1/2$ multiplet. We discard other spin states such as spin $S-1/2$ multiplet and spin $S$ multiplet with doubly occupied states, by restricting to transitions only within the ground-state manifold. That is, we assume $T \ll \Delta E_{min}$, with $\Delta E_{min}$ the energy gap to the first excited state above the ground-state.\\
As we discuss in the previous section, the Kondo temperature typically happens at lower scale than the pseusospin freezing temperature  $T^{\star}$. The operations used to derive the effective model i.e. diagonalization of the isolated $\hat{H}_{mol}$ matrix and degenerate perturbation expansion in order of $\mathcal{O}(\hat{H}_{hyb})^{2}$, do not return quantitative information on the preferred parity state. The operator $\widehat{P}_{mol}$ is computed via NRG. Hence, the multiplet active in tunnelling events with leads in the $N=7$ charge regime is always the residual parity doublet state given in Eq.\ref{eq:6def_BasisStateParity}. In the current derivation, we leave $p^\star$ unspecified and we refer to  the next section for its specifics according to the set-up geometry, temperature scale and $V_{gate}$ value.\\
The derivation of effective model at the MV transition in the benzene junction remains in its essence the same as the methodology applied in Sec.\ref{sec:MVPC}. The $\hat{H}_{eff}$ is derived perturbatively from projecting the whole system to the ground-state molecule manifold using first order expansion in $\hat{H}_{hyb}$ in the BWPT. The resulting effective parameters have an incipient PC property. In order to use the effective parameters in quantum transport calculation, we need to relate the new set of parameters from  $\hat{H}_{eff}$ to the bare parameters in the physical basis, in this case $\alpha=L,R$. The $\hat{H}_{eff}$ is mapped to a minimal Fermionic effective model and only within specific energy scales both models are equivalent. Once the mapping is completed, we can use parameters from Fermionic effective model and relate those to the bare ones to compute quantum transport analysis.\\ 
In the doublet-triplet MV case, we can think of this Fermionic effective model as a general two orbital system. It comprises an Anderson-type site (orbital) both coupled to the effective lead and coupled to a local-moment spin $S=1/2$ via antiferromagnetic exchange coupling $J$. We note the latter spin does not hybridise with the effective lead, which corresponds to the symmetric leads combination. In the singlet-doublet- MV case, the Fermionic effective model consists only of the Anderson-type site coupled to the single effective lead. We also assume the antisymmetric or weakest lead combination is decoupled from the system. In the section we elaborate more about this derivation, however with this summary we highlight the differences and similarities to the derivation in Sec.\ref{sec:MVPC}.  \\
In the next sections we proceed with the analytical derivation of the effective model for singlet-doublet and doublet-triplet mixed-valence crossovers, giving also the corresponding MW current formula for the derived effective models, now showing an incipient PC property. 

\subsection{$\hat{H}_{eff}$ in doublet-triplet mixed-valence regime}
At the crossover between $N=7$ to $N=8$ charge sector, the bare model in Eq.\ref{eq:6Hbare} spans over the Hilbert space $\mathcal{H}^{5}$ formed by two spin components of doublet multiplet $S=1/2$ and three components of triplet spin state $S=1$, namely:
\begin{equation}\label{eq:6admState}
	\ket{\Psi^{7,S^z}_{0}}=\sum_{{S}^{z}=\pm1/2}\ket{7,S^{z};p^\star;0} \quad,\quad \ket{\Psi^{8,S^z}_{0}}=\sum_{{S}^{z}=\pm1,0} \ket{8,S^{z};0} ~,
\end{equation}
where in the $N=7$ regime we use the parity doublet state in Eq.\ref{eq:6def_BasisStateParity} after pseudospin freezing. We refer to the spin states in Eq.\ref{eq:6admState} as \textit{admissible spin states}.\\
We define the projector operator for the doublet-triplet MV transition as follows
\begin{equation}\label{eq:6defPMV78}
	\widehat{P}_{0} = \sum_{{S}^{z}=\pm1/2}\left(\ket{7;S^{z};p^\star;0}\bra{7;S^{z};p^\star;0} \otimes \widehat{\mathbb{1}}_{leads}\right) +\sum_{{S}^{z}=\pm1,0}\left(\ket{8;S^{z};0}\bra{8;S^{z};0} \otimes \widehat{\mathbb{1}}_{leads} \right) ~,
\end{equation}
where we use the definition given in Eq.\ref{eq:5defP}. From now on we suppress the $p^\star$ label. The Eq.\ref{eq:6defPMV78} is the projector operator onto the manifold of the admissible states in $\ket{\Psi^{7}_{0}},\ket{\Psi^{8}_{0}}$ of $\hat{H}_{mol}$. By applying $\widehat{P}_{0}$ to the bare model in Eq.\ref{eq:6Hbare}, we calculate the effective model. Within the energy window $T \ll \Delta E_{min}$ as defined at the beginning of this section, the effective benzene molecule model is determined by zero and first order of the perturbation expansion in $\hat{H}_{hyb}$ written in Eq.\ref{eq:5Heff}. That is the $0^{th}$-order as projection on $\hat{H}_{leads}$ in Eq.\ref{eq:5P0Leads} and on $\hat{H}_{mol}$ in Eq.\ref{eq:5P0Imp}. Whereas the $1^{st}$-order projection on $\hat{H}_{hyb}$ is given by unspecified expression in Eq.\ref{eq:5P0Hyb}. In conclusion, the effective $\hat{H}_{eff}$ model at $7\leftrightarrow8$ MV transition is determined by deriving the expression in Eq.\ref{eq:5HeffMVPer}. Up to this point there is no difference from what we calculated before. The prominent calculation we need to solve here in order to calculate $\hat{H}_{eff}$ in the benzene junction is $\widehat{P}_{0}\hat{H}_{hyb}\widehat{P}_{0}$. In this case, that is given by:
\begin{equation}
	\begin{aligned}\label{eq:6P0Hyb78}
		\widehat{P}_{0}\hat{H}_{hyb}&\widehat{P}_{0} =\sum_{\alpha=L,R} \Big( \ket{8;0}\underbrace{\braket{8;0 |V_{\alpha} d^{\dagger}_{i_{\alpha}\uparrow}|7;-1/2}}_{\widetilde{V}^{\prime}_{\alpha}}\bra{7;-1/2} c_{\alpha\uparrow} + \ket{8;0}\underbrace{\braket{8;0 |V_{\alpha} d^{\dagger}_{i_{\alpha} \downarrow}|7;1/2}}_{\widetilde{V}^{\prime}_{\alpha}}\bra{7;1/2} c_{\alpha\downarrow} +\\
		&  +\ket{8;1}\underbrace{\braket{8;1 |V_{\alpha} d^{\dagger}_{i_{\alpha}\uparrow}|7;1/2}}_{\widetilde{V}_{\alpha}}\bra{7;1/2} c_{\alpha\uparrow} + \ket{8;-1}\underbrace{\braket{8;-1 |V_{\alpha} d^{\dagger}_{i_{\alpha} \downarrow}|7;-1/2}}_{\widetilde{V}_{\alpha}}\bra{7;-1/2} c_{\alpha\downarrow} +h.c. \Big) ~,
	\end{aligned}
\end{equation}
where we have eased the notation by omitting the parity labels. In Eq.\ref{eq:6P0Hyb78}, the elements $\widetilde{V}^{\prime}_{\alpha}, \widetilde{V}_{\alpha}$ are the \textit{effective hybridization} evaluated on the set of admissible spin states in Eq.\ref{eq:6admState}.\\ 
We note the calculation of the effective hybridization reduces to the find the allowed transitions between the five physical ground-states, meaning that we regard only transitions among states with a \textit{unitary charge and spin variation} i.e. $\Delta N = \pm 1$, $\Delta S = \pm 1/2$. A general framework to discuss the selection of states is within the Wigner-Eckart theorem, as we proceed with in the next section.

\subsection*{Selection rules: the Wigner-Eckart theorem}
We briefly present now the generalities on the Wigner-Eckart theorem \cite{Cohen2019quantum}.\\
For a well-defined basis $\ket{ S_{1}, S_{1}^{z}},\ket{ S_{2}, S_{2}^{z}}$, we introduce the spherical tensor operator $\mathcal{T}^{k}_{q}$ with $(2k+1)$-rank and $q=-k,-k+1, \dots ,k$-components \cite{Bouzas_WE2016}. Then, the \textbf{Wigner-Eckart theorem} calculates the transition amplitude from the initial state $\ket{ S_{1}, S_{1}^{z}}$ added to $\ket{k,q}$ state by projecting it onto the final state $\ket{ S_{2}, S_{2}^{z}}$:
\begin{equation} \label{eq:6WEdef}
	\braket{Q; S_{2}, S_{2}^{z} |\mathcal{T}^{k}_{q}|Q; S_{1}, S_{1}^{z}} = \braket{S_{1}, k ; S_{1}^{z}, q |S_{2}, S_{2}^{z}} \braket{Q, S_{1} ||\mathcal{T}^{k} || Q; S_{2}} ~.
\end{equation} 
This amplitude is determined by a multiplication of two terms on the right-hand side of Eq.\ref{eq:6WEdef}. The first one is the \textit{Clebsh-Gordan coefficient} (CGC): that is the expansion coefficient of the total spin basis in tensor product basis. The second one, indicated by the standard notation $\braket{~. ||\dots||.~}$, is the so-called \textit{reduced matrix element} since it is independent from all $S^{z}$ quantum numbers. We conclude that, once the reduced matrix is evaluated, we can derive the allowed process for which the left-hand side of Eq.\ref{eq:6WEdef} is non-vanishing by calculating the non-vanishing CGCs. In general, the CGC is non-zero if the summation $S_{1}+S_{2}+k $ is an even number. From this, we derive the \textit{selection rules}:
\begin{equation}
	\begin{aligned}
		&\Delta S^{z} = S_{2}^{z} - S_{1}^{z} = q ~,\\
		& |\Delta S| = |S_{2} - S_{1}| \leq k ~.
	\end{aligned}
\end{equation}
The transitions at MV regime are based on the spin representation class $k=1/2$. However, we can use a \textit{generalization of the Wigner-Eckart theorem} where any transitions for which the reduced matrix written as $\braket{Q+1; k+1/2=1 || d^{\dagger} ||Q; k=1/2}$ correctly describe the transitions, then, its effective hybridization ratio is independent from the coupled lead and it is equal to $1/\sqrt{2}$ - as result of the CGC. We infer that, in general, without calculating any matrix element of complicated object, once the model spin representation is defined with its corresponding selection rules, any non-zero transition is fully determined by these allowed transitions i.e. non-vanishing CGCs. \\

\noindent{We} turn the discussion of the Wigner-Eckart theorem in the context of the mixed-valence benzene model. According to the spin-$1/2$ basis representation, the smallest spin representation is given by $k=1/2$ such that we identify $q=\pm 1/2$-components of the irreducible tensor of $2$-rank $\mathcal{T}^{1/2}_{\pm1/2}$. Then, we associate the tensor to the corresponding operators: 
\begin{align}
	\mathcal{T}^{1/2}_{1/2} \Rightarrow d^{\dagger}_{\uparrow} \quad \text{and} \quad \mathcal{T}^{1/2}_{-1/2} \Rightarrow d^{\dagger}_{\downarrow} ~,
\end{align}
and $\mathcal{T}^{k}$ in Eq.\ref{eq:6WEdef} is denoted as $d^{\dagger}$ without spin label.\\
Clearly, by means of the spin representation, we have obtained the selection rules for the doublet-triplet transition: 
\begin{align}
	\Delta S_{z} = \pm 1/2  \quad , \quad |\Delta S| = 1/2 ~,
\end{align}
and we have verified $S_{1}+S_{2}+k = 1/2 + 1 + 1/2 = 2$ is even number. From the selection rule we need to specify also the total spin $\hat{S}$, thus we introduce it as additional quantum number in the definition of the basis state i.e. $\ket{\Psi^{N}_{0}}=\ket{N;S;S^{z}}$. We can now write down the allowed transitions:
\begin{equation}\label{eq:6admTrans} 
\begin{aligned}
	(i)& \quad \braket{8;1,0 |d^{\dagger}_{\downarrow}|7;1/2,1/2} = \braket{1/2,1/2;1/2,-1/2 |1,0} \braket{8,1 || d^{\dagger}||7,1/2} = 1/\sqrt{2} \braket{ || d^{\dagger}||} ~,\\
	(ii)& \quad \braket{8;1,0 |d^{\dagger}_{\uparrow}|7;1/2,-1/2} = \braket{1/2,1/2;-1/2,1/2 |1,0}\braket{8,1 || d^{\dagger}||7,1/2} = 1/\sqrt{2} \braket{ || d^{\dagger}||} ~,\\
	(iii)& \quad\braket{8;1,+1 |d^{\dagger}_{\uparrow}|7;1/2,1/2} = \braket{1/2,1/2;1/2,1/2 |1,1}\braket{8,1 || d^{\dagger}||7,1/2} = 1 \braket{ || d^{\dagger}||} ~,\\
	(iv)&\quad \braket{8;1,-1 |d^{\dagger}_{\downarrow}|7;1/2,-1/2} = 
	\braket{1/2,1/2;-1/2,-1/2 |1,-1}\braket{8,1 || d^{\dagger}||7,1/2} = 1 \braket{ || d^{\dagger}||} ~,
\end{aligned}
\end{equation}
where in the second equality we insert the CGC values and the usual short-hand notation $\braket{ || d^{\dagger}||}$ for the reduced matrix. The allowed transitions in Eq.\ref{eq:6admTrans} equate transmission amplitude from $\ket{7;1/2,S^{z}}$ to $\ket{8;1,S^{z}}$ states to the non-zero CGC of the process times the reduced matrix - that is the same for any non-vanishing transition regardless the $S^{z}$ quantum number. \\
We apply the expressions in Eq.\ref{eq:6admTrans} to evaluate the relation between the effective hybridizations in Eq.\ref{eq:6P0Hyb78} by replacing their ratio with the corresponding transitions derived from the Wigner-Eckart theorem. As outcome, we obtain:
\begin{equation}\label{eq:6VeffRatio}
	\begin{aligned}
		 \frac{\widetilde{V}^{\prime}_{\alpha}}{\widetilde{V}_{\alpha}} \stackrel{(a)}{=} \frac{V_{\alpha}\braket{8;1,0 |d^{\dagger}_{i_{\alpha}\uparrow}|7;1/2,-1/2}}{V_{\alpha}\braket{8;1,1 | d^{\dagger}_{i_{\alpha}\uparrow}|7;1/2,1/2}} = &\frac{1}{\sqrt{2}} \quad,\quad
		\frac{\widetilde{V}^{\prime}_{\alpha}}{\widetilde{V}_{\alpha}} \stackrel{(b)}{=} \frac{V_{\alpha}\braket{8;1,0 |d^{\dagger}_{i_{\alpha}\downarrow}|7;1/2,1/2}}{V_{\alpha}\braket{8;1,-1 | d^{\dagger}_{i_{\alpha}\downarrow}|7;1/2,-1/2}} = \frac{1}{\sqrt{2}} ~,\\
		& \Rightarrow \widetilde{V}^{\prime}_{\alpha} = \frac{\widetilde{V}_{\alpha}}{\sqrt{2}} ~,
	\end{aligned}
\end{equation}
where ratio $(a)$ is defined by the transitions $(ii)/(iii)$ and ratio $(b)$ by $(i)/(iv)$ from expressions in Eq.\ref{eq:6admTrans}.
We also give the short-hand notation for the effective hybridization, namely $\widetilde{V}_{\alpha}=V_{\alpha}c_{CG}\braket{8;1||d^{\dagger}||7;1/2}$ where the factor $c_{CG}$ indicates the Clebsh-Gordan coefficient in the transitions: 
\begin{equation}\label{eq:6CGdef78}
	\begin{aligned}
	&\ket{8,S^{z}=0} \leftrightarrow \ket{7,S^{z}=\pm1/2} & \Rightarrow &~c_{CG}=\braket{1/2,1/2,\pm1/2, \mp1/2|1,0} =1/\sqrt{2} ~,\\
	&\ket{8,S^{z}=\pm1} \leftrightarrow \ket{7,S^{z}=\pm1/2} &\Rightarrow &~ c_{CG}=\braket{1/2,1/2,\pm1/2, \pm1/2|1,\pm1} =1~.
	\end{aligned}	
\end{equation}
We conclude that for any transitions - characterised by unitary charge and spin variation - occurring among the degenerate doublet and triplet ground-states defined in Eq.\ref{eq:6admState} in the mixed-valence regime, the effective hybridization $\widetilde{V}^{\prime}_{\alpha},\widetilde{V}_{\alpha}$ - resulting from the projection operation in Eq.\ref{eq:6P0Hyb78} - are not \textit{independent} but in fact are related by 
the non-vanishing CGC from the selection rules. It would be then logical to manipulate the hybridization term in order to rearrange it in the form satisfying the PC property - see more comments on this in Sec.\ref{sec:PC}. However, the model is still in the same original $\alpha=L,R$ basis of the bare model in Eq.\ref{eq:6Hbare}. Under such basis, the conduction electron of each lead are not independent hence we cannot use this basis to define a decoupled system where then PC condition is satisfied. It is necessary to transform the system basis into an independent set - as we calculate in the section. 

\subsection*{Rotation to even-odd basis: effective low-energy model and emergent PC} 
Utilizing the Wigner-Eckart theorem result, we can now perform a lead basis transformation to decouple one lead combination in $\hat{H}_{eff}$. This is possible because the model satisfies PC condition.
Hence, we apply a canonical rotation of the leads basis from $L,R$ to $e$ even/ $o$ odd basis as we perform in Sec.\ref{sec:CBPC}, \cite{Flensberg}. The conduction electrons in the new basis set reads as:
\begin{equation}
	\begin{aligned}
		 c_{e\sigma} = \frac{\widetilde{V}_{L}c_{L\sigma}+\widetilde{V}_{R}c_{R\sigma}}{\sqrt{|\widetilde{V}_{L}|^{2}+|\widetilde{V}_{R}|^{2}}} \quad,\quad
		 c_{o\sigma} = \frac{-\widetilde{V}_{R}c_{L\sigma}+\widetilde{V}_{L}c_{R\sigma}}{\sqrt{|\widetilde{V}_{L}|^{2}+|\widetilde{V}_{R}|^{2}}} ~.
	\end{aligned}
\end{equation}
A model Hamiltonian whose lead operators are written in even/odd basis allows for independent manipulation of conduction electrons in each lead. As consequence, the odd lead (antisymmetric combination) is decoupled from $\hat{H}_{eff}$. In virtue of Eq.\ref{eq:6VeffRatio}, the even channel (symmetric combination) is defined by the conduction electron operators $c_{e\sigma} =\widetilde{V}^{-1}\sum_{\alpha} \widetilde{V}_{\alpha} c_{\alpha\sigma}$. We plug the operators $c^{\dagger}_{e\sigma},c_{e\sigma}$ and the hybridization ratio in Eq.\ref{eq:6VeffRatio} into the projection $\widehat{P}_{0}\hat{H}_{hyb}\widehat{P}_{0}$ in Eq.\ref{eq:6P0Hyb78} and the final form of the \textbf{low-energy effective model} reads:
\begin{equation}\label{eq:6Heff_even} 
	\boxed{
	\begin{aligned}
\hat{H}_{eff}
		\stackrel{MV}{=} \hat{H}_{leads}  +
		\widetilde{V} \bigg[ &\bigg( \frac{1}{\sqrt{2}}\ket{8;0}\bra{7;-\frac{1}{2}} +    \ket{8;1}\bra{7;\frac{1}{2}}\bigg)c_{e\uparrow} +\\
		&+\bigg( \frac{1}{\sqrt{2}}\ket{8;0}\bra{7;\frac{1}{2}} + \ket{8;-1}\bra{7;-\frac{1}{2}}\bigg)c_{e\downarrow} +h.c.\bigg] +\mathcal{O}(\hat{H}_{hyb})^{2} 
	\end{aligned} }~,
\end{equation}
where the benzene molecule couples to the even lead combination through the effective hybridization $\widetilde{V}$ and the odd combination is detached. Each transition is multiplied upfront by the corresponding CGC. The model consists of only non-vanishing transition appearing at the mixed-valence crossover between doublet and triplet degenerate ground-states. We observe that the hybridization matrix element $\widetilde{V}=\sqrt{\widetilde{V}_{L}^{2}+\widetilde{V}_{R}^{2}}$ appears now without any leads index, then the effective model in Eq.\ref{eq:6Heff_even} is an equivalent single-channel model. As it is written, Eq.\ref{eq:6Heff_even} is indeed under PC condition. This is an \textit{emergent PC property} belonging to $\hat{H}_{eff}$ only as result from first order perturbation in $\hat{H}_{hyb}$.\\ 
As we mentioned at the beginning of this section, this effective model is not written solely in terms of Fermionic operators but contains also many-body projectors. It is necessary to map it to a general Fermionic system such that its effective parameters can be related to the physical system. This is indeed the scope of the next calculation.

\subsection*{Mapping the effective Hamiltonian: equivalence between $\hat{H}_{eff}$ and minimal Fermionic $f,g$-model}  
In the $\hat{H}_{eff}$ derived in Eq.\ref{eq:6Heff_even}, in similar sprit to what we presented in Sec.\ref{sec:MVPC}, one may think of the charge distribution on the benzene molecule composed by one Anderson-type site coupled directly to: $(i)$ the even lead combination through the effective hybridization $\widetilde{V}$ and $(ii)$ the spin-$1/2$ local moment attached through 
exchange coupling $J$. There is no direct tunnelling between the single spin and the even lead. Hence, we may turn the effective model into an effective two Fermionic orbital system, denoted as $f$ and $g$ orbitals, where we indicate with $f$-orbital as Anderson-type site and with $g$-orbital as local moment site. Hence, the \textbf{$f,g$-effective model} now reads:
\begin{equation}\label{eq:6Heff_fg}
	\begin{aligned}
		\widetilde{H}_{eff} &= \widetilde{H}_{leads} + \widetilde{H}_{mol} + \widetilde{H}_{hyb} ~,\\
		&\widetilde{H}_{leads} = \sum_{\mathbf{k} \sigma} \epsilon_{\mathbf{k}} c^{\dagger}_{e\mathbf{k}\sigma}c_{e\mathbf{k}\sigma} ~,\\
		& \widetilde{H}_{mol} = \epsilon \sum_{\sigma} f^{\dagger}_{\sigma} f_{\sigma} - J \hat{\mathbf{S}}_{g} \cdot \hat{\mathbf{s}}_{f} ~,\\
		& \widetilde{H}_{hyb} = \widetilde{V} \sum_{\sigma} \big(c_{e\sigma}f_{\sigma} + f^{\dagger}_{\sigma}c_{e\sigma}  \big)  \quad \text{with} \quad c^{\dagger}_{e\sigma}= \frac{1}{\widetilde{V}} \sum_{\mathbf{k}} \widetilde{V}_{\mathbf{k}}c_{e\mathbf{k}\sigma} ~.
	\end{aligned}
\end{equation}
We observe that in the modelling in Eq.\ref{eq:6Heff_fg}, the odd lead combination is still detached and does not feature explicitly. Hence, the incipient PC property of the effective model $\hat{H}_{eff}$ is still valid \textit{also} in $f,g$-model $\widetilde{H}_{eff}$. We remark that there is no Schrieffer-Wolff transformation involved in Eq.\ref{eq:6Heff_fg}: this model Hamiltonian serves to describe a system where spin-doublet and spin-triplet ground-state multiplets are degenerate at the mixed-valence crossover.\\
In order to recognise correctly the Hilbert space dimension of $\widetilde{H}_{eff}$, we need to consider each spin degree of freedom in $f$ and $g$ orbital. It is straightforward to see that the $g$-orbital can accommodate no more than singly occupied spin state $\ket{\sigma}$, and the corresponding operator $\hat{\mathbf{S}}_{g}$ recovers the role of local moment spin with total spin $S_{g}=1/2$ \cite{Tsay_SingletGS_1971}. On the contrary, the $f$-orbital includes four spin degrees of freedom, namely both empty $\ket{0}$, singly $\ket{\sigma}$ and doubly $\ket{\uparrow\downarrow}$ occupied states. The larger number of states contained in the $f$-orbital can be further supplied by the direct coupling with the electronic reservoir in the even lead - that indeed is like the source lead.
By construction $\widetilde{H}_{eff}$ belongs to an enlarged manifold relative to the original one, that is the Hilbert space $\mathcal{H}^{8}$ spanned by eight states - one local moment $S=1/2$ spin and one Fermionic site. We can also see the space dimension from the summation of one triplet, one singlet and two doublets multiplets in the $f$ and $g$ orbitals respectively given as $(S+1/2)\times1+(S-1/2)\times1+S\times2=8$, for $S=1/2$. However, our original model in Eq.\ref{eq:6Hbare} had but five states, not eight.\\
In order to make the $f,g$-effective model $\widetilde{H}_{eff}$ compatible with $\hat{H}_{eff}$ and as consequence to the bare model in Eq.\ref{eq:6Hbare}, in the  initial step we need to relate the basis spin state $\ket{\Psi^{N}_{0}}=\ket{N;S,S^{z}}$ in the original model with the basis spin states in the $f,g$-model. We denote those according to the \textit{z}-component of the spin in the state $f$,$g$ orbitals, namely $\ket{\Psi_{i}}=\ket{\phi_{f};\phi_{g}}$ with $i=1,\dots,8$ states. We formalise the eight spin states in $\widetilde{H}_{eff}$ as follows. First, the five admissible spin states in the original model given in Eq.\ref{eq:6admState} now in the $f,g$-model read as:
\begin{equation}
	\begin{aligned}
		\ket{\Psi^{N}_{0}} &= \ket{N;S,S^{z}} &\Rightarrow& \ket{\phi_{f};\phi_{g}} ~, \\
		\ket{\Psi^{8}_{0}} &= \ket{8;1,0} &\Rightarrow&  \frac{1}{\sqrt{2}}\big(\ket{\uparrow;\downarrow} + \ket{\downarrow;\uparrow} \big) \doteq \ket{\Psi_{1}} ~,\\
		\ket{\Psi^{8}_{0}}&= \ket{8;1,+1} &\Rightarrow&  \ket{\uparrow; \uparrow} \doteq \ket{\Psi_{2}} ~,\\
		\ket{\Psi^{8}_{0}} &= \ket{8;1,-1} &\Rightarrow&  \ket{\downarrow;\downarrow} \doteq \ket{\Psi_{3}} ~,\\
		\ket{\Psi^{7}_{0}} &= \ket{7;\frac{1}{2}, \frac{1}{2}} &\Rightarrow&  \ket{-,\uparrow} \doteq \ket{\Psi_{4}} ~,\\
		\ket{\Psi^{7}_{0}} &= \ket{7;\frac{1}{2}, -\frac{1}{2}} &\Rightarrow&  \ket{-,\downarrow} \doteq \ket{\Psi_{5}} ~,
	\end{aligned}
\end{equation}
where $\ket{\Psi_{1}},\ket{\Psi_{2}},\ket{\Psi_{3}}$ states are the $N=8$ spin multiplet $S+1/2$ of the triplet and $\ket{\Psi_{4}},\ket{\Psi_{5}}$ states are the $N=7$ spin multiplet $S$ of the doublet with $S=1/2$. Then, we define the extra three spin states appearing only in $\widetilde{H}_{eff}$, namely: 
\begin{equation}
	\begin{aligned}
		 \frac{1}{\sqrt{2}}\big(\ket{\uparrow;\downarrow} - \ket{\downarrow;\uparrow} \big) &\doteq \ket{\Psi_{6}} ~,\\
		 \ket{\uparrow\downarrow;\uparrow} &\doteq \ket{\Psi_{7}} ~,\\
		 \ket{\uparrow\downarrow;\downarrow} &\doteq \ket{\Psi_{8}}~,
	\end{aligned}
\end{equation}
where $\ket{\Psi_{6}}$ state corresponds to the $S-1/2$ spin singlet and $\ket{\Psi_{7}},\ket{\Psi_{8}}$ states are the spin multiplet $S$ with doubly occupied state coupled $f$ orbital. These states do not have any counterpart in the original model. Hence, in order to make the mapping consistent with the physical spin state in the bare model, we need to find the condition on $\epsilon,J$ in Eq.\ref{eq:6Heff_fg} such that the $f,g$-model retains only the physical states $\ket{\Psi_{1}}, \dots, \ket{\Psi_{5}}$ and eliminates the additional $\ket{\Psi_{6}},\ket{\Psi_{7}},\ket{\Psi_{8}}$ states. \\
In practice, this operation starts with studying the spin state algebra of the $f,g$-model by applying the $f^{\dagger}_{\sigma},f_{\sigma}$ operators on the $\ket{\Psi_{i}}=\ket{\phi_{f};\phi_{g}}$ basis. Below, we list the results for the creation operator per spin:
\begin{equation}
	\begin{aligned}
		&f^{\dagger}_{\uparrow} \ket{\Psi_{1}} = \frac{1}{\sqrt{2}} \ket{\Psi_{7}} ~, && f^{\dagger}_{\downarrow} \ket{\Psi_{1}} = -\frac{1}{\sqrt{2}} \ket{\Psi_{8}} ~,\\
		&f^{\dagger}_{\uparrow} \ket{\Psi_{2}}=0~, && f^{\dagger}_{\downarrow} \ket{\Psi_{2}} = -\ket{\Psi_{7}}~,\\
		&f^{\dagger}_{\uparrow} \ket{\Psi_{3}} =\ket{\Psi_{8}}~, && f^{\dagger}_{\downarrow} \ket{\Psi_{3}} = 0 ~,\\
		&f^{\dagger}_{\uparrow} \ket{\Psi_{4}} = \ket{\Psi_{2}}~, && f^{\dagger}_{\downarrow} \ket{\Psi_{4}} =  \frac{1}{\sqrt{2}} \big(\ket{\Psi_{1}} - \ket{\Psi_{6}} \big) ~,\\
		&f^{\dagger}_{\uparrow} \ket{\Psi_{5}}= \frac{1}{\sqrt{2}} \big(\ket{\Psi_{1}} + \ket{\Psi_{7}} \big) ~, && f^{\dagger}_{\downarrow}\ket{\Psi_{5}}= \ket{\Psi_{3}}~,\\
		&f^{\dagger}_{\uparrow} \ket{\Psi_{6}}= -\frac{1}{\sqrt{2}} \ket{\Psi_{7}} ~,&& f^{\dagger}_{\downarrow}\ket{\Psi_{6}} = -\frac{1}{\sqrt{2}} \ket{\Psi_{8}} ~,\\
		&f^{\dagger}_{\uparrow} \ket{\Psi_{7}} =0~, && f^{\dagger}_{\downarrow} \ket{\Psi_{7}} =0~,\\
		&f^{\dagger}_{\uparrow} \ket{\Psi_{8}}=0 ~,&& f^{\dagger}_{\downarrow}  \ket{\Psi_{8}}=0~.
	\end{aligned}
\end{equation}
In order to find the effective couplings affecting \textit{only} the spurious states, we apply $\widetilde{H}_{eff}$ to the triplet and doublet states:
\begin{equation}
	\begin{aligned}
		&\widetilde{H}_{eff} \left(\ket{\Psi_{1}}+\ket{\Psi_{2}}+\ket{\Psi_{3}} \right) =\left(\epsilon - \frac{J}{4}\right)\left(\ket{\Psi_{1}}+\ket{\Psi_{2}}+\ket{\Psi_{3}} \right)  ~,\\
		&\widetilde{H}_{eff}\left(\ket{\Psi_{4}}+\ket{\Psi_{5}} \right) = 0 ~.
		\end{aligned}
\end{equation}
Then, due to the degeneracy of triplet and doublet states at MV crossover, we  impose equality of their eigenvalues, namely $\epsilon -J/4\equiv 0$. From this we infer the  \textit{$f,g$-model constraint}, here defined in general for $S=1/2$:
\begin{equation}\label{eq:6constraint}
	\boxed{
	\epsilon= JS^{2}= \frac{J}{4} \quad \text{for} \quad S=\frac{1}{2}} ~.
\end{equation}
We comment on the logic behind the expression in Eq.\ref{eq:6constraint} before applying to $\widetilde{H}_{eff}$. The derived constraint is pinned on the mixed-valence transition between $S=1/2$ (symmetric) and $S+1/2$ multiplet states. These are high-spin states but they are coupled ferromagnetically with the local moment $g$-orbital such that they are energetically favoured. On the other hand, the unwanted transition between $S=1/2$ (antisymmetric) and $S-1/2$ multiplet states occur among low-spin states that are in energetically at higher levels. Due to the energy scale assumed in the bare model $T \ll \Delta E_{min}$, these states live outside this energy window. Thus, they do not participate in the tunnelling and we can discard them in the BWPT derivation. However, in the $f,g$-model in Eq.\ref{eq:6Heff_fg}, these spurious states appear again and we need to impose specific energy constraint to the model. Generalization of this argument to any spin $S$ multiplet state at MV transition is briefly discuss at the end of this section.\\
We return now to our derivation and we implement the parameters in Eq.\ref{eq:6constraint} in the  $f,g$ effective model. As consequence, we correctly find that:
\begin{equation}
	\widetilde{H}_{eff} \big(\ket{\Psi_{1}} \dots \ket{\Psi_{5}} \big)=0 ~,
\end{equation}
meaning that the admissible states under the constraint in $\epsilon,J$ have zero energy. On the contrary, we apply the constraints on the additional states and we find a non-zero eigenvalues, namely:
\begin{equation}
	\begin{aligned}
		 \widetilde{H}_{eff}\ket{\Psi_{6}} &= J \ket{\Psi_{6}} ~,\\
     	\widetilde{H}_{eff}\ket{\Psi_{7}} &= J/2\ket{\Psi_{7}}~, \\
		\widetilde{H}_{eff}\ket{\Psi_{8}} &= J/2\ket{\Psi_{8}}~.
	\end{aligned}
\end{equation}
Those contributions can be eliminated by simply taking the limit of the exchange coupling $J\rightarrow+\infty$, such that $\ket{\Psi_{6}},\ket{\Psi_{7}} ,\ket{\Psi_{8}}$ states are placed at infinite energy out of the energy bandwidth for electrons tunnelling from the molecule ground-state.\\
In conclusion, if and only if we impose the constraint on $\epsilon=J/4$ in Eq.\ref{eq:6constraint} \textit{and} we take the limit to infinity on $J$, the effective model $\hat{H}_{eff}$ in Eq.\ref{eq:6Heff_even} is equivalent to the Fermionic $f,g$-model $\widetilde{H}_{eff}$ in Eq.\ref{eq:6Heff_fg}. Then, the Hilbert space of both systems is the identically the same, encompassing only the physical states of bare model in Eq.\ref{eq:6Hbare}.\\
Finally, we can complete our discussion on the mapped model by identifying the mapping between the $f,g$-spin states and the original ones, as we proceed in the final section.	

\subsection*{Mapping spin states: equivalence between $\hat{H}$ and $f,g$- model}
As mentioned earlier, the last step to complete the mapping involves relating the $f_{\sigma}$-site operator of the $f,g$-model with the original molecule operators. The reason for this it is that the conductance is always calculated in terms of the physical operators. The mapping of the effective model to the $f,g$ two orbital system is indeed a clever strategy to derive a minimal description of the model consisting of Fermionic physical states only.\\
We start with the expression for $f^{\dagger}_{\sigma},f_{\sigma}$ operators in term of the spin basis of the  $f,g$-model, namely $f^{\dagger}_{\sigma} = f^{\dagger}_{\sigma} \sum_{i=1}^{8} \ket{\Psi_{i}}\bra{\Psi_{i}}$ and we obtain: 
\begin{equation}
	\begin{aligned}
	 f^{\dagger}_{\uparrow} &= \frac{1}{\sqrt{2}}\ket{\Psi_{1}}\bra{\Psi_{5}} + \ket{\Psi_{2}}\bra{\Psi_{4}} \equiv \frac{1}{\sqrt{2}} \ket{8;1,0}\bra{7;\frac{1}{2},-\frac{1}{2}} + \ket{8;1,1}\bra{7;\frac{1}{2},\frac{1}{2}} ~,\\
	 f^{\dagger}_{\downarrow} &= \ket{\Psi_{3}}\bra{\Psi_{5}}+ \frac{1}{\sqrt{2}}\ket{\Psi_{1}}\bra{\Psi_{4}} \equiv \ket{8;1,-1}\bra{7;-\frac{1}{2},\frac{1}{2}} + \frac{1}{\sqrt{2}} \ket{8;1,0}\bra{7;\frac{1}{2},\frac{1}{2}} ~,
	\end{aligned}
\end{equation}
in both expressions, the first equality is obtained by imposing the constraint in Eq.\ref{eq:6constraint} and the second one relates the $f,g$-spin basis to the original many-body state in Eq.\ref{eq:6admState}. By means of these two equations we obtain the correspondence between spin states both in the bare and in the $f,g$-effective model. It is straightforward to verify that $f^{\dagger}_{\sigma},f_{\sigma}$ operators are Fermionic since they obey to the anti-commutation relations $\lbrace f^{\dagger}_{\sigma},f_{\sigma^{\prime}}\rbrace = \delta_{\sigma,\sigma^{\prime}}$ within the reduced $5$ dimension state-space.\\
We proceed now by writing the admitted transitions in the model using the above expressions for $f^{\dagger}_{\sigma},f_{\sigma}$ operators. Crucially, in analogy to Eq.\ref{eq:6VeffRatio}, we find
\begin{equation}
	\begin{aligned}
		&\braket{8;1,0 |f^{\dagger}_{\uparrow}|7;1/2,-1/2}= \braket{8;1,0 |f^{\dagger}_{\downarrow}|7;1/2,1/2}= \frac{1}{\sqrt{2}}~,\\
		&\braket{8;1,1 |f^{\dagger}_{\uparrow}|7;1/2,1/2}= \braket{8;1,-1 |f^{\dagger}_{\downarrow}|7;1/2,-1/2}=1 ~,
	\end{aligned}
\end{equation}
where these Fermionic states represent matrix elements of tunnelling events that are in the \textit{same} proportion as we obtain from Wigner-Eckart theorem in Eq.\ref{eq:6admTrans}. With these calculations, we complete the verification of the algebra of $\ket{\Psi_{i}}$ spin basis, their application on $\hat{H}_{eff}$ and consequent requirements on parameters to turn the effective $f,g$-model consistent with bare model. We also have defined the correspondence between physical and $f,g$ spin states.\\
With this association, we can regard the low-energy effective model derived from the BWPT projection in Eq.\ref{eq:6Heff_even} equivalent to the  minimal $f,g$-effective model with an emergent PC property given by the following expression:
\begin{equation} \label{eq:6Heffmin_78}
	\boxed{
		\widetilde{H}_{eff} = \sum_{\mathbf{k} \sigma} \epsilon_{\mathbf{k}} c^{\dagger}_{e\mathbf{k}\sigma}c_{e\mathbf{k}\sigma} +
		J \bigg( \frac{1}{4}\sum_{\sigma} f^{\dagger}_{\sigma} f_{\sigma} - \hat{\mathbf{S}}_{g} \cdot \hat{\mathbf{s}}_{f}\bigg) +
		\widetilde{V} \sum_{\sigma} \big(c_{e\sigma}f_{\sigma} + f^{\dagger}_{\sigma}c_{e\sigma}  \big) ~, J\rightarrow + \infty} ~.
\end{equation}
The equivalence between the perturbatively derived $\hat{H}_{eff}$ and $\widetilde{H}_{eff}$ in Eq.\ref{eq:6Heffmin_78} is set under proper choice of the single-particle energy of the $f$-site including the limit to infinity of the exchange coupling. This is the conclusive part of the derivation of the effective model for the benzene junction in doublet-triplet MV regime. In the next section we can make use of the property of Eq.\ref{eq:6Heffmin_78} to calculate the transport.

\subsection*{Analytical results: electrical conductance for doublet-triplet MV benzene junction}
The important outcome of the emergent PC property in $\hat{H}_{eff}$ in Eq.\ref{eq:6Heff_even} - and in its mapped $f,g$-model $\widetilde{H}_{eff}$ in Eq.\ref{eq:6Heffmin_78} - is that we can study quantum transport events at low energy and temperature scale using the MW formula under PC given in Eq.\ref{eq:MW->PC}. \\
In particular, at zero energy and temperature scale and under linear response, we derive an analytical expression for the \textit{dc}-regime MW electrical conductance, namely:
\begin{equation}\label{eq:6Gc_MV78} 
	\boxed{
	\mathcal{G}^{dc}(0,0)=	\frac{2e^{2}}{h}\left( 4\frac{|\widetilde{V}_{L}|^{2}|\widetilde{V}_{R}|^{2}}{(|\widetilde{V}_{L}|^{2}+|\widetilde{V}_{R}|^{2})^{2}} t_{ff}(0,0) \right) } ~,
\end{equation}
where $\widetilde{V}_{\alpha}$ is given in Eq.\ref{eq:6P0Hyb78} as $\widetilde{V}_{\alpha}=V_{\alpha}c_{CG}\braket{8;1||d^{\dagger}||7:1/2}$ for $c_{CG}$ coefficients defined in Eq.\ref{eq:6CGdef78} and the $\mathrm{T}$-matrix spectral function $t_{ff}(0,0)$ is given for the $f,g$ model in Fig.\ref{fig:tmatrix_MV_dt}.\\  
The MW conductance can directly be computed to characterise the transport in the benzene junction at crossover between doublet and triplet spin states. The required ingredients are only the bare hybridization and the matrix element of allowed transitions in the isolated benzene molecule. The emergent PC property makes possible the study of tunnelling events within MW formulation.

\subsection{$\hat{H}_{eff}$ in singlet-doublet mixed-valence regime}
At the crossover between $N=6$ to $N=7$ charge sector, the bare model in Eq.\ref{eq:6Hbare} spans over the Hilbert space $\mathcal{H}^{3}$ formed by two spin components of doublet multiplet $S=1/2$ in $N=7$ sector and one component of the singlet $S=0$ in $N=6$ sector, namely:
\begin{equation}\label{eq:6admState67}
	\ket{\Psi^{6, S^z}_{0}}= \ket{6,S^{z}=0;0} \quad,\quad \ket{\Psi^{7, S^z}_{0}}=\sum_{{S}^{z}=\pm1/2} \ket{7,S^{z};p^\star;0}   ~,
\end{equation}
where in the $N=7$ regime we use the lowest parity doublet state in Eq.\ref{eq:6def_BasisStateParity} after pseudospin freezing. We refer to these three states in Eq.\ref{eq:6admState67} as \textit{admissible spin states} as we did before.\\
We define the projector operator for the singlet-doublet MV transition as:
\begin{equation}\label{eq:6defPMV67}
	\widehat{P}_{0} = \ket{6;S^{z}=0;0}\bra{6;S^{z}=0;0} \otimes \widehat{\mathbb{1}}_{leads} + \sum_{{S}^{z}=\pm1/2}\left( \ket{7;S^{z};p^\star;0}\bra{7;S^{z};p^\star;0} \otimes \widehat{\mathbb{1}}_{leads} \right) ~,
\end{equation}
where we use the definition in Eq.\ref{eq:5defP}. We now abbreviate these states as $\ket{N;S^{z}}$. In this MV regime, the derivation follows analogously to Sec.\ref{sec:MVPC}. Hence, we directly give the final expression of the low-energy effective model:
\begin{equation}\label{eq:6Heff_67} \boxed{
		\hat{H}_{eff} \stackrel{MV}{=}\hat{H}_{leads}
		+ \sum_{i\neq j  \in \uparrow,\downarrow}\sum_{\alpha=L,R} \left( c^{\dagger}_{\alpha i} \widetilde{V}_{\alpha} f_{j} +h.c. \right) + \mathcal{O}(\hat{H}_{hyb})^{2}  }~,
\end{equation}
where we identify the Fermionic operators $f^{\dagger}_{i},f_{i}$ with the states $f^{\dagger}_{i} = \ket{7, S^{z}= \pm1/2}\bra{6,S^{z}=0}$, $f_{i}= \ket{6, S^{z}=0}\bra{7, S^{z}= \pm1/2}$ and spin $i\neq j$ in order to have singlet-doublet transition.  We note in Eq.\ref{eq:6Heff_67}, the molecule is represented by an effective single-site with creation (annihilation) operator $f^{\dagger}_{j}$ ($f_{j})$. Hence, $\hat{H}_{eff} $ model Hamiltonian is equivalent to a single-impurity Anderson model and by construction the set-up fulfils proportionate coupling condition.\\
In the singlet-doublet mixed-valence regime, there is only one allowed transition according single-electron transmission. We identify it with the effective hybridization reads now as:
\begin{equation}\label{eq:6Veff_67}
	\boxed{
	\widetilde{V}_{\alpha} =V_{\alpha}\braket{6;0||d^{\dagger}_{i_{\alpha}}||7;1/2} = 
	V_{\alpha} c_{CG} \braket{6;0||d^{\dagger}||7;1/2}  }~,
\end{equation} 
where in the first equality $i_{\alpha}=1$ for $\alpha=L$, $i_{\alpha}=3,4$ for the benzene junction in either $1,3$ or $1,4$ configurations. In the second equality, using the  Wigner-Eckart theorem in Eq.\ref{eq:6WEdef}, the Clebsh-Gordan coefficient are defined as $c_{CG} = \braket{1/2,1/2,\pm1/2, \pm1/2|0,0} =1$ corresponding to non-vanishing transition and $\braket{||d^{\dagger}||}$ reduced matrix element transition. 
From Eq.\ref{eq:6Veff_67} it is straightforward to write the effective Gamma function for the singlet-doublet crossover as per Eq.\ref{eq:LevelWidth}.\\
Considering the initial physical spin states, we can again think of mapping of $\hat{H}_{eff}$ to the minimal Fermionic model - such that we can use the effective parameters to calculate the electrical conductance in the physical system. In this case, we have one Anderson-type  orbital $f$. 
However, the Hilbert space of the effective model is now enlarged to $\mathcal{H}^{4}$, since $f$-states with $\ket{0},\ket{\uparrow},\ket{\downarrow},\ket{\uparrow\downarrow}$ are possible in Eq.\ref{eq:6Heff_67}. 
We need to enforce a constraint on the charge occupancy $\braket{\hat{n}_{f_{\uparrow}}\hat{n}_{f_{\downarrow}}}=0$ to eliminate the spurious state $\ket{\uparrow\downarrow}$. In order to do so, we introduce an on-site Coulomb potential term $ U \hat{n}_{f_{\uparrow}}\hat{n}_{f_{\downarrow}}$ in the effective model in Eq.\ref{eq:6Veff_67}. By taking its limit to infinity, we ensure any double occupied $f$-state is placed at such high energy level that no tunnel event from the ground-state can occur. Hence, we inhibit any high-energy state and the mapped $f$-model is equivalent effective model, namely:
\begin{equation}\label{eq:6Heffmin_67}
	\boxed{
	\hat{H}_{eff} \Leftrightarrow \widetilde{H}_{eff} = \hat{H}_{eff} + U \hat{n}_{f_{\uparrow}}\hat{n}_{f_{\downarrow}} \quad \text{with} \quad U \rightarrow +\infty} ~,
\end{equation}
such that the minimal model is understood as infinite-interacting Anderson model. By means of the energy constraint, the mapped model in Eq.\ref{eq:6Heffmin_67} satisfies PC condition. Hence, we can proceed with using its effective parameters to evaluate electrical conductance. Although the simple steps we take to derive the effective model for singlet-doublet MV regime benzene junction, these analytical results give perfect agreement with the NRG computation of the conductance, as we discuss in the next section.

\subsection{NRG results: electrical conductance for singlet-doublet MV benzene junction}
\begin{figure}[H]
	\centering
	\includegraphics[width=0.85\linewidth]{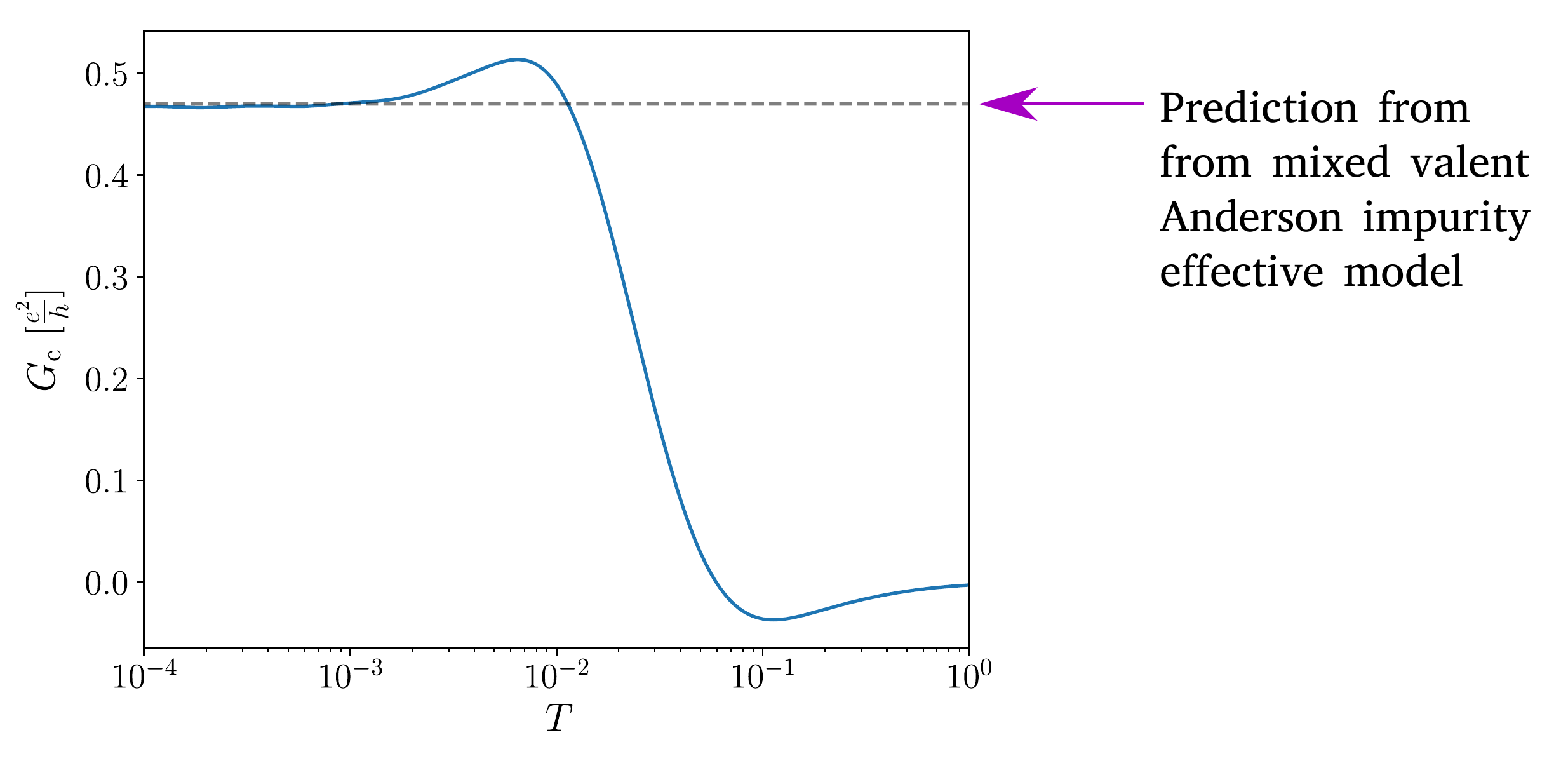}
	\caption[Numerical verification of the conductance calculated from effective model in mixed-valence regime of $1,4$- benzene junction.]{Numerical verification of analytical current in Eq.\ref{eq:6Gc_MV67} for $1,4$-configuration benzene junction in singlet-doublet MV regime. The electrical conductance $\mathcal{G}^{C}(T)$ as function of temperature $T$ is calculated: (blue curve) $\mathcal{G}^{C}(T)$ in NRG using improved Kubo formula, (dashed line) $\mathcal{G}^{C}(T=0)$ analytically using the current in Eq.\ref{eq:6Gc_MV67} from the effective model in Eq.\ref{eq:6Veff_67}. The numerical data exactly overlaps asymptotically with the analytical prediction at $T \lesssim 10^{-3}$.}\label{F6:analytic}
\end{figure}

From the previous section, in Eq.\ref{eq:6Heffmin_67} we derive an effective model with emergent PC property. This crucial aspect allows to apply the MW form under PC in Eq.\ref{eq:MW->PC} to calculate electrical conductance. In particular, as we did for the doublet-triplet crossover, we can use Eqs.\ref{eq:5Gcspin_OmFinite},\ref{eq:5Gcspin_FL} for our case of emergent PC in the benzene molecule. Here, we give a numerical verification of the accuracy of the analytical derivation developed in this chapter and in Sec.\ref{sec:MVPC} for the benzene junction in the $1,4$-configuration.\\
We consider the regime $T=0$ such that the system is described as Fermi liquid. The MW  conductance using Eq.\ref{eq:5Gcspin_FL} for the single-molecule benzene junction reads as:
\begin{equation}\label{eq:6Gc_MV67} 
	\boxed{
		\begin{aligned}
&\mathcal{G}^{dc}(0,0)=\frac{2e^{2}}{h} \left(
4\frac{\Gamma^{L}_{eff}\Gamma^{R}_{eff}}{(\Gamma^{L}_{eff}+\Gamma^{R}_{eff})^{2}} t_{ff}(0,0) \right) ~,\\
&t_{ff}(0,0) = \frac{1}{1+(\epsilon^{\star})^{2}/(\Gamma^{L}_{eff}+\Gamma^{R}_{eff})^{2} } \quad;~ \Gamma^{\alpha}_{eff}=\pi |\widetilde{V}_{\alpha}|^{2}\rho_{0}~,~\epsilon^{\star}=\mathit{Re}\Sigma_{ff}(0,0)
\end{aligned} } ~,
\end{equation}
where we use the effective hybridization in Eq.\ref{eq:6Veff_67} to compute also the effective Gamma function and the $\mathrm{T}$-matrix spectral function is given by $\mathit{Im}G^{R}_{ff\sigma}(0,0)$ exactly analytically calculated from Eq.\ref{eq:6Heffmin_67} including interaction-renormalized level $\epsilon^{\star}$ given in terms of the Fermi liquid self-energy $\Sigma_{ff}(0,0)$. The value of $t_{ff}(0,0)$ as function of $\Gamma^{L}_{eff}+\Gamma^{R}_{eff}$ obtained by NRG are given in Fig.\ref{fig:tmatrix_MV_sd}.  The conductance derived using the current expression in Eq.\ref{eq:6Gc_MV67} is given in dashed line in Fig.\ref{F6:analytic}. In the plot, we compute also the electrical conductance using improved Kubo formula in Eq.\ref{eq:4defKuboLRel} via NRG. The numerical values (blue curve) are evaluated for $T\leq D$, where $D=1$ is the conduction bandwidth. As remarkable result, the NRG data are asymptotically flattening at the analytical value at the temperature scale $T \lesssim 10^{-3}$, see Fig.\ref{F6:analytic}. This demonstrates the validity of the analytical derivation of effective model. The new set of parameters in $\widetilde{H}_{eff}$ presents a solid description of the low energy and low temperature physics of the bare system whose microscopic model does not satisfy PC condition. \\
The precision of the effective model in quantum transport calculation is of general validity. At first order in $\hat{H}_{hyb}$ perturbation, we find accurate description at $T=0$ scale for the mixed-valence regime. By extending the second order in perturbation, we incorporate also the description of Coulomb blockade regime for  namely $N=6,7,8$. Those effective models \cite{benzene} are used in the next section to compute the electrical conductance. 

\section{Thermodynamics and transport}
\begin{figure}[H] 
	\begin{subfigure}[b]{0.5\linewidth}
		\centering
		\includegraphics[width=0.95\linewidth]{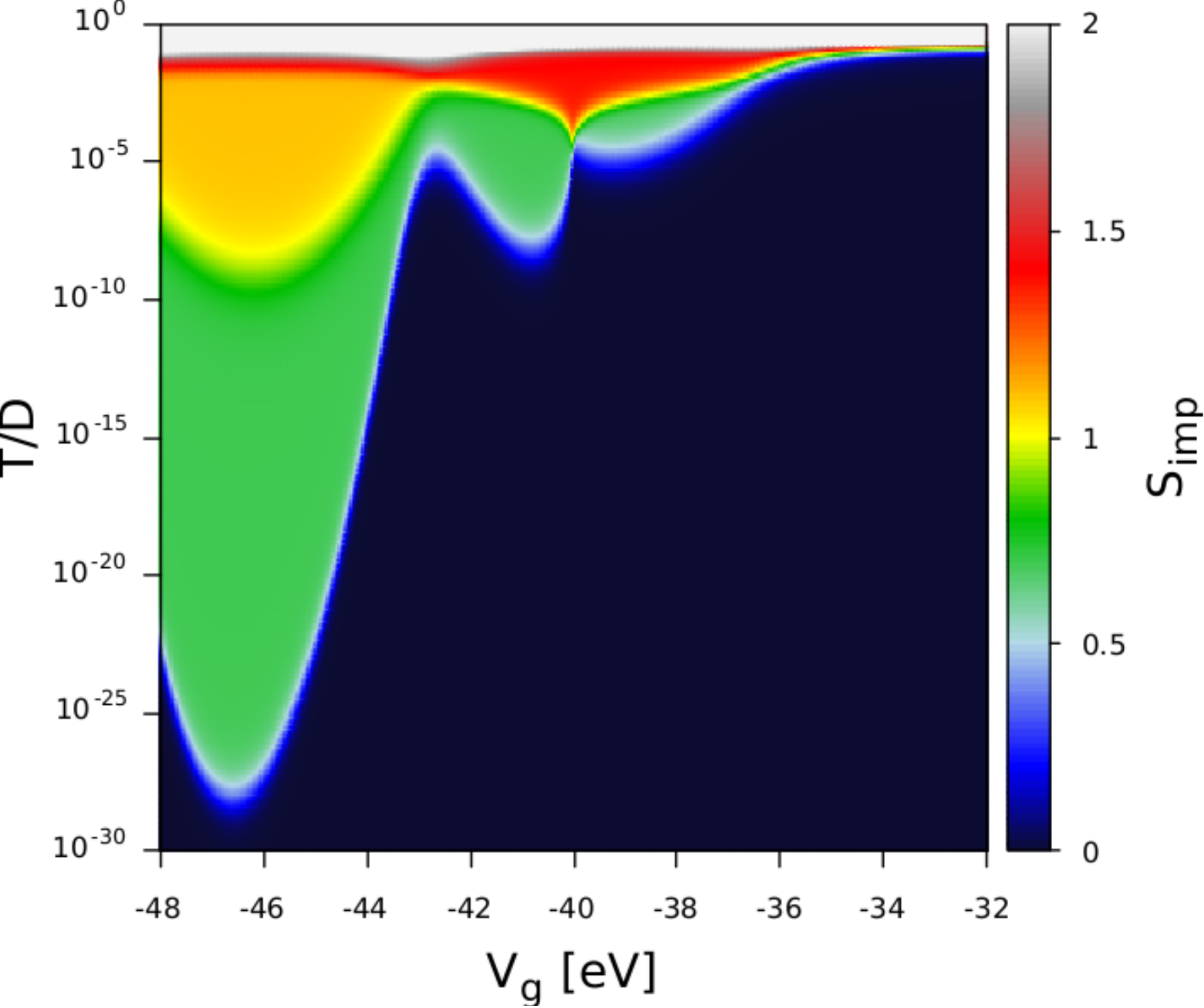} 
		\caption{Entropy in $1,3$ configuration.} 
		\label{F6:S_VgT_13} 
		\vspace{4ex}
	\end{subfigure}
	\begin{subfigure}[b]{0.5\linewidth}
		\centering
		\includegraphics[width=0.95\linewidth]{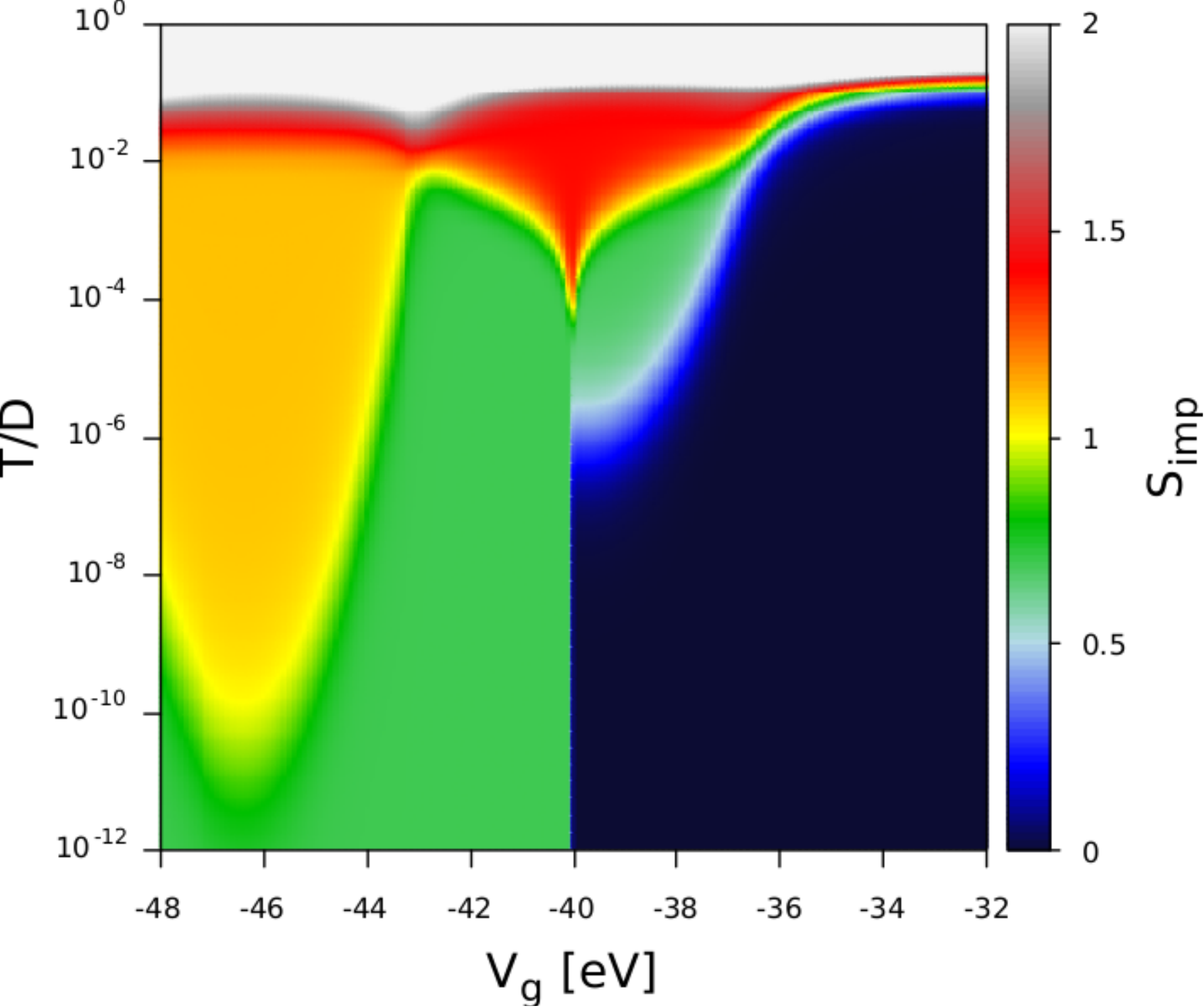} 
		\caption{Entropy in $1,4$ configuration.} 
		\label{F6:S_VgT_14} 
		\vspace{4ex}
	\end{subfigure} 
	\begin{subfigure}[b]{0.5\linewidth}
		\centering
		\includegraphics[width=0.95\linewidth]{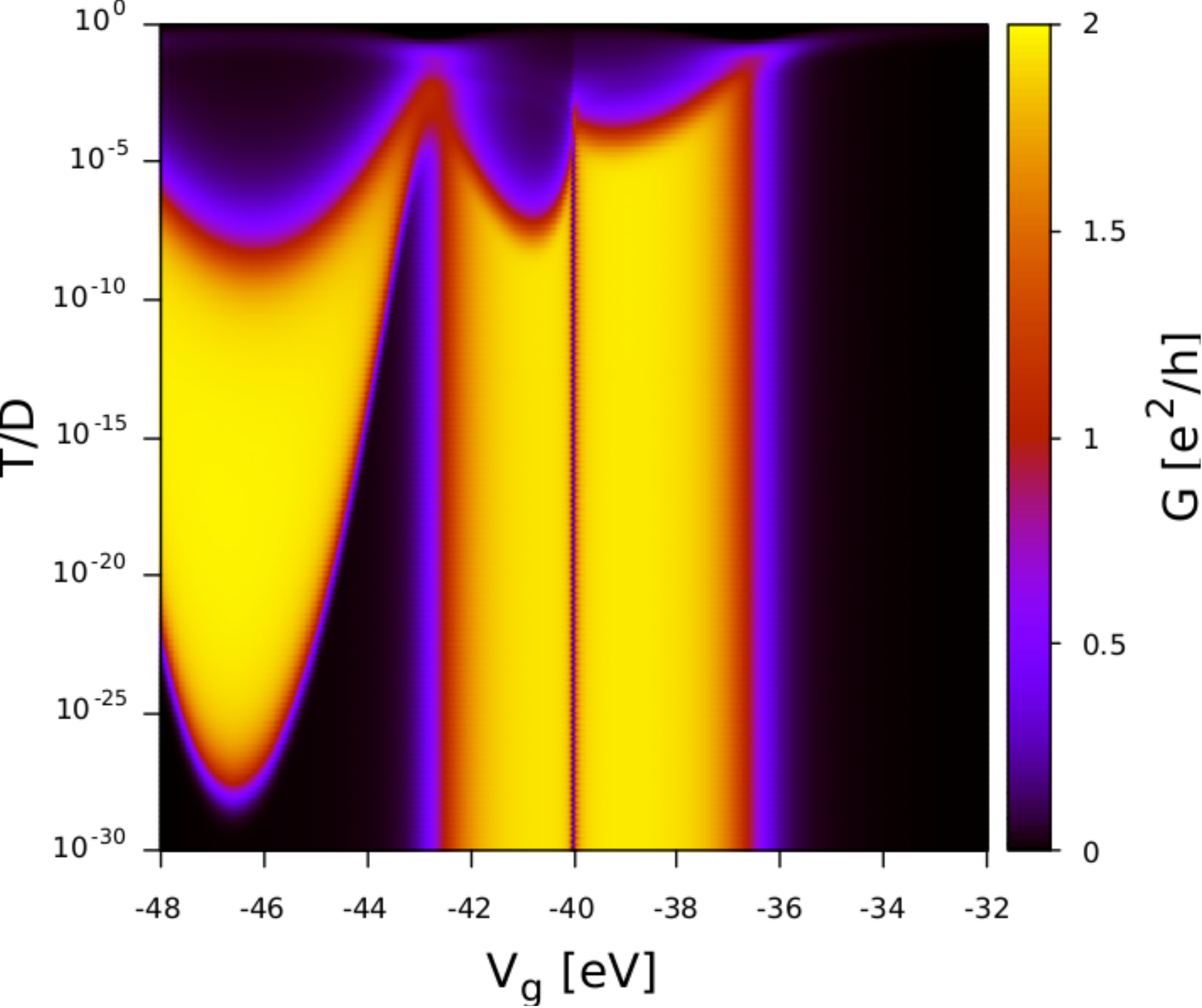} 
		\caption{Conductance in $1,3$ configuration.} 
		\label{F6:G_VgT_13} 
	\end{subfigure}
	\begin{subfigure}[b]{0.5\linewidth}
		\centering
		\includegraphics[width=0.95\linewidth]{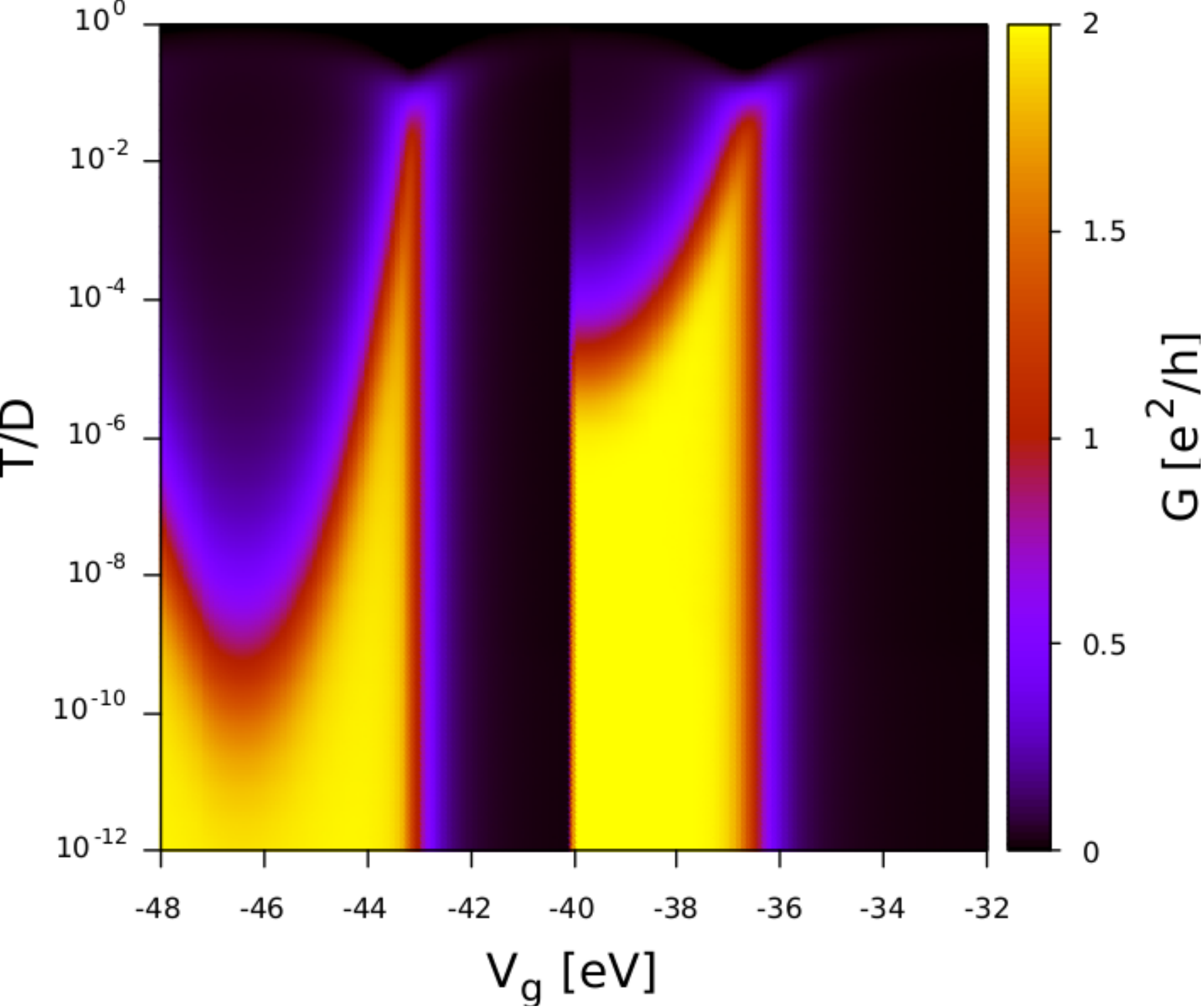} 
		\caption{Conductance in $1,4$ configuration.} 
		\label{F6:G_VgT_14} 
	\end{subfigure} 
	\caption[NRG data for molecule entropy and electrical conductance in $1,3$ and $1,4$ configuration]{NRG data for thermodynamics and quantum transport in the $1,3$ and $1,4$ configurations. \textit{(top)} Benzene entropy $S_{mol}$ - indicated as $S_{imp}$ in the plot - and \textit{(bottom)} electrical conductance $\mathcal{G}^{C}(T)$ in unit of $\mathcal{G}_{0}=e^{2}/h$  in the benzene junction as function of temperature $T/D$ and voltage gate $V_{gate}/D$, given $D$ conduction bandwidth. In Figs.\ref{F6:S_VgT_13},\ref{F6:S_VgT_14}, the different colours in the left bar depict the number of active degrees of freedom involved in transport across the junction. The different phases are: (dark blue) strong coupling phase $S_{mol}=\ln(1)=0$, (green) Spin-$1/2$ doublet state in local moment phase $S_{mol}=\ln(2)$, (yellow) Spin-$1$ triplet state, (red) phase with spin and pseudospin degrees contribution $S_{mol}=\ln(4)$ and (grey) high temperature phases involving several thermally accessible states. In Figs.\ref{F6:G_VgT_13},\ref{F6:G_VgT_14}, the different colours in the left bar indicates the ratio $\mathcal{G}^{C}(T)/\mathcal{G}_{0}$ varying from (yellow) unitary limit $\mathcal{G}^{C}(T)/\mathcal{G}_{0}=2$ here considering also the spin degeneracy to (black) Coulomb blockade $\mathcal{G}^{C}(T)/\mathcal{G}_{0}=0$. See text for a complete discussion.} \label{F6:SmolGc} 
\end{figure}

In this section we discuss the NRG data for molecular entropy $S_{mol}$ and \textit{dc}-regime electrical conductance $\mathcal{G}^{C,dc}(T)$ for both $1,3$ and $1,4$ configuration, see plots in Fig.\ref{F6:SmolGc}. The effective models in Coulomb blockade regimes are calculated from perturbation expansion up to order $\mathcal{O}(\hat{H}_{hyb})^{2}$.\\
For each set-up, the results are presented for successively lower $V_{gate}$ values corresponding to increasing charge occupancy of the $N=6$ ground-state molecule - compare with the occupation staircase in Fig.\ref{F6:MolCharge}. Main properties are identified for any temperature scale spanning from high-temperature scales $T,U$ to low energy scales $\delta E , T^{\star} , T_{K}\to 0$. \\
The observables are computed from full two-channel NRG calculations: the entropy of the molecule is obtained using the usual prescription $S_{mol}=S_{tot}-S_{leads}^{0}$ with $S_{tot}$ the total entropy at the molecule-leads coupled system and $S_{leads}^{0}$ the entropy of the free leads. 
The electrical conductance under linear response regime is calculated from the improved Kubo formula derived in Eq.\ref{eq:4defKuboLRel}. The effective exchange coupling parameters are derived in second order expansion in $\hat{H}_{hyb}$ - these appear in the paper \cite{benzene}. \\
We remark that the difference in the temperature scale range in the entropy plots Figs.\ref{F6:S_VgT_13},\ref{F6:S_VgT_14} is because in the $1,3$ case we find in the spin triplet ground-state a two-stage perferct screened Kondo effect that is absent in the $1,4$ configuration. Hence, in order to appreciate the double spin screening occurring in the $1,3$ coupled system, we extend further the temperature scale. 

\subsection{\textit{meta}-connected benzene transistor}
In this section we focus on the \textit{meta}- or $1,3$- connected benzene transistor, see right schematic of Fig.\ref{F6:schematic}, with the right/drain lead connected on the third site of the benzene molecule. We focus on the different charge occupation values $N=6,7,8$ of the Coulomb staircase as seen from the right plot in Fig.\ref{F6:MolCharge} in terms of molecular entropy $S_{mol}$ in Fig.~\ref{F6:S_VgT_13} and electrical conductance $\mathcal{G}^{C}(T)$ in Fig.~\ref{F6:G_VgT_13} as a function of temperature and gate voltage.

\subsection*{Spin singlet ground-state $S=0$ in $N=6$ charge sector}
At $-36~eV<V_{gate}\lesssim 32~eV$, the isolated benzene junction is populated by an even number of electrons such that the isolated molecule hosts a spin singlet ground-state with a net spin $S=0$. In this work, as we show in Fig.\ref{F6:MolCharge}, the lowest even charge sector corresponds to $N=6$. 
The lack of a free molecule spin degrees of freedom prevents any Kondo singlet formation, or Kondo-boosted conductance. \\
Hence, there is no finite molecular entropy in the $N=6$ sector for all $T\ll U$, see the extended dark blue region within $-36~eV<V_{gate}\lesssim 32~eV$ in Fig.\ref{F6:S_VgT_13} - and analogously in Fig.\ref{F6:S_VgT_14}. The conductance shows a Coulomb blockade valley resulting in a zero conductive state for the entire temperature range. This is represented by the dark column down to $T=0$ located between $-36~eV<V_{gate}\lesssim 32~eV$ in Fig.\ref{F6:G_VgT_13} - and similarly in Fig.\ref{F6:G_VgT_14}. Although, at very high temperature scale relevant high-energy states contribute to $S_{mol}\neq0$, at $T\gtrsim \delta E$  no many-body quantum effects develops hence no electronic tunnelling. \\
Around $V_{gate}\sim-36~eV$, the crossover from $6 \leftrightarrow7$ charge sectors occurs. 
The strong fluctuations in both charge and spin contribute to a spike in the conductance value up.

\subsection*{Spin doublet ground-state $S=\tfrac{1}{2}$ in $N=7$ charge sector: $SU(4)$ Kondo effect}
\begin{figure}[H]
	\centering
	\includegraphics[width=0.6\linewidth]{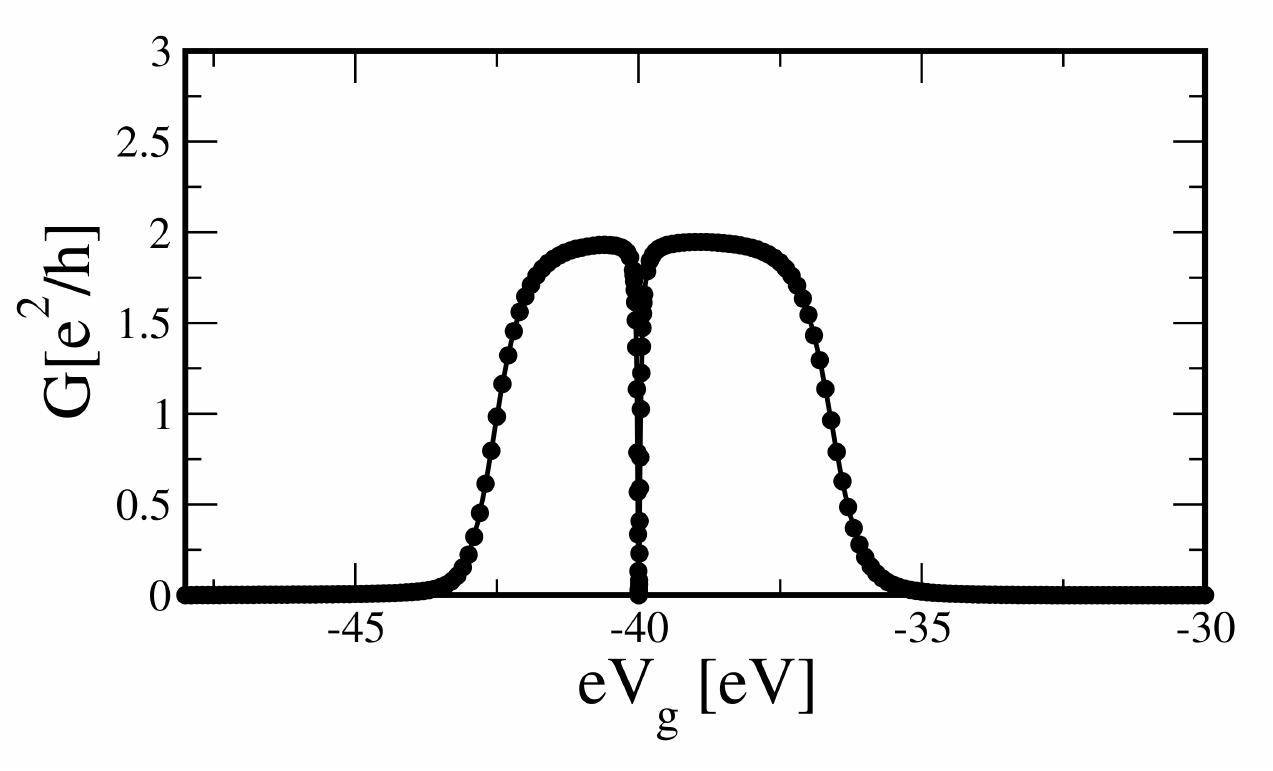}
	\caption[Electrical conductance $\mathcal{G}^{C}(T=0)$ in $1,3$ configuration]{Electrical conductance $\mathcal{G}^{C}(T=0)$ as a function of $V_{gate}$ encompassing $N=6,\,7,\,8$ sector in the $1,3$-connected molecule-lead coupled benzene junction. In the $N=7$ charge sector, under RG-flow the benzene junction reaches $SU(4)$ state that manifests as a conductance node at $V_{gate}\sim -40~eV$. The conductance further recovers the Kondo enhanced unitarity limit at any other $V_{gate}$ within $N=7$ sector. Furthermore, the conductance eventually decays to zero in the $N=8$ regime due to the underlying two stage Kondo effect of the $S=1$ spin triplet state.}\label{F6:G_T0_Vg_13}
\end{figure}

As we tune $V_{gate}$ to lower values, the molecule accommodates an additional electron and enters into the $N=7$ electron region. As shown in Fig.~\ref{F6:MolCharge}, this regime prevails in the gate voltage range $-43~eV< V_{gate}\lesssim 36~eV$, when the many-body ground state of the isolated molecule can be represented by an effective doublet spin $S=1/2$ state which is delocalized over the whole molecule. Several interesting observations for the benzene single-molecule coupled to leads can be deduced in this regime that are intricately associated with the underlying symmetry of the single-molecule junction. \\
We start with analysing the molecule entropy in Fig.\ref{F6:S_VgT_13} in the $N=7$ sector. \\
In the whole voltage range $-43~eV< V_{gate}\lesssim -36~eV$ at high temperature $T\sim \delta E$, both spin and pseudospin degrees are effectively free.
We have two pseudospin states per doublet spin $\uparrow,\downarrow$ to contribute in $S_{mol}=\ln(2+2)$, shown as red region. By lowering the temperature at $T=T^{\star}$, the molecule enters deeper into the $N=7$ regime. At this temperature scale, the pseudospin parity degree of freedom is frozen out such that only the doublet $p^{\star}$ is thermally accessible 
and $S_{mol}=\ln(2)$, indicated in green region. Below $T^{\star}$, the system is described as a generalized two-channel spin-$1/2$ Kondo model. By lowering the temperature even further to Kondo temperature $T_{K}$, under RG-flow, the $\ket{\Psi^{7}_{0}}=\ket{7;\pm1/2;p^{\star}}$ doublet state is screened by the conduction electrons from the strongest channel - usually that is the even lead combination - while the weakest, i.e. odd, channel decouples. This process persists down to $T\to 0$. Hence, at $0 \leq T \leq T_{K}$ the coupled system enters in strong coupling phase indicating the formation of the Kondo singlet as $S_{mol}=\ln(1)=0$, that is the large dark blue region arsing below the green one.\\
We observe that the pseudospin freezing develops for any $V_{gate}$ in the $N=7$ sector \textit{except for} $V_{gate}=-40~ eV$ where at $T =T_{K}$, the molecule entropy value jumps from $S_{mol}=\ln4 \rightarrow \ln1=0$ without passing through an intermediate local moment phase and strong coupling phase lasts till $T=0$. Thus, spin and pseudospin degrees of freedom remain entangled down to the lowest temperature and the four-fold degenerate molecular spin-$1/2$ states undergoes full Kondo screening below $T_{K}$. We know that the standard spin-$1/2$ $SU(2)$ state gives a crossover $S_{mol}=\ln2 \rightarrow \ln1=0$. However, at $V_{gate}=-40~ eV$, we see a single-stage crossover $S_{mol}=\ln4  \rightarrow \ln1=0$, that is the signature of 
$SU(4)$ Kondo effect.
Hence, we can rephrase that the entangled spin-pseudospin complex undergoes complete $SU(4)$ \textbf{Kondo effect} at characteristic $T_{K}^{SU(4)}$ \cite{Simon_SU4_PRB2003}, with screening involving all four conduction electron flavours $L,R,\uparrow,\downarrow$. We further investigate this special point at $V_{gate}=-40~ eV$ and in its vicinity.\\
From NRG data, we compute the average of $\widehat{P}_{mol}$ in the lead-coupled system. We find the values continuously change from $\sim -1$ at $V_{gate}=-41~eV$ to $\sim +1$ at $V_{gate}=-38~eV$. In particular, at $V_{gate}=-40~eV$ we have $\langle \widehat{P}_{mol} \rangle=0$. This proves an equal mixture of $p=\pm1$ parities takes place at this specific $V_{gate}$ value. Furthermore, no abrupt change in the doublet ground-state happens at the special $SU(4)$ point.\\
\begin{figure}[H]
	\centering
	\includegraphics[width=0.65\linewidth]{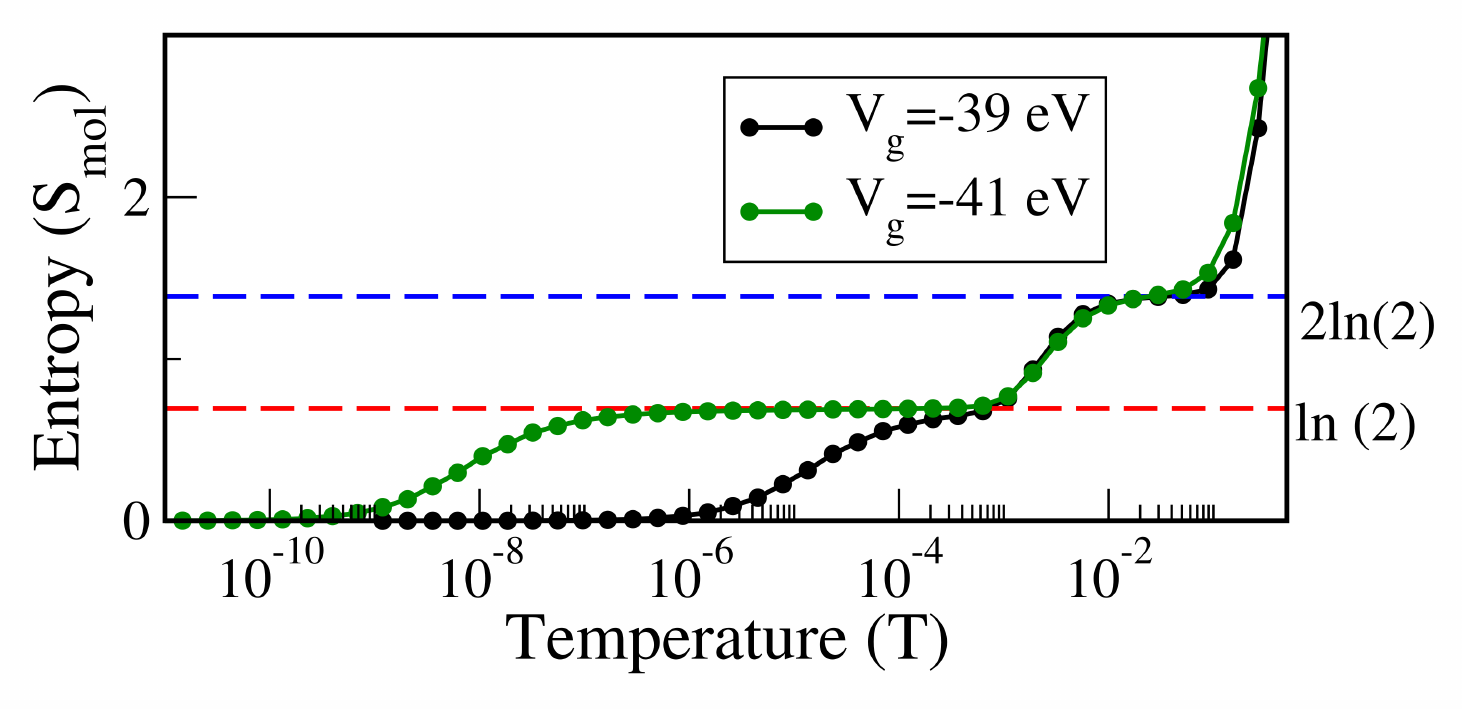}
	\caption[Molecule entropy in $1,3$ configuration at $N=7$ sector around $V_{gate}=-40~eV$]{Molecular entropy in the $N=7$ regime, as a function of temperature $T$ for two different gate voltages $V_{gate}=-39~eV$ (black curve) and $V_{gate}=-41~eV$ (green curve) on either side of $V_{gate}\sim -40~eV$ value where the $SU(4)$ Kondo effect occurs in the $1,3$-connected benzene junction. The doublet spin-$1/2$ state with lowest energy parity state is Kondo quenched before and after the $SU(4)$ point at any temperature scale  $T\leq T_{K}$. }\label{F6:S_40-G_13}
\end{figure}
\noindent{To} get insight about the molecule entropy at the edge $V_{gate}$ values around $SU(4)$ point, we plot $S_{mol}$ as function of temperature at $V_{gate}=-39~eV$ (black curve) and $V_{gate}=-41~eV$ (green curve) in Fig.\ref{F6:S_40-G_13} as such $\sim 1eV$ either side of the $SU(4)$ point. The entropy curves go through two distinct plateaux at $S_{mol}=\ln(2+2)$ and $S_{mol}=\ln(2)$, marked with blue and red dashed lines respectively. The $\ln(4)$ value indicates both spin and pseudospin degrees are manifested at high temperatures - here the two curves overlap. The $\ln(2)$ value is determined by the residual pseudospin degree bringing the coupled system to the local moment phase - again the two curves for a short temperature range are overlapped. The crossover between the two states occurs at pseudospin freezing scale $T^{\star}$, here around $10^{-3}eV$. At $T\leq T_{K}$, the spin-$1/2$ state surviving after pseudospin freezing is Kondo screened \textit{both} cases.
A further observation is that $T_{K}(V_{gate}=-41~eV)\ll T_{K}(V_{gate}=-39~eV)$, as indicated by longer plateau on the local moment phase in the green curve. However, at $V_{gate}=-40~eV$ no change in the molecule ground-state manifold happens: the $SU(4)$ state is \textit{not} accompanied by quantum phase transition.\\
To sum up: the additional orbital degeneracies encoded in the pseudospin degrees $(i)$ remain entangled to the spin degrees for any $T$ at $V_{gate}=-40~eV$ $(ii)$ bring $1,3$-connected system to higher symmetry state $SU(4)$.\\ 
\noindent{By} means of thermodynamics analysis we study now the electrical conductance in Fig.\ref{F6:G_VgT_13}. \\
At $T^{\star}\ll T$, the temperature is too large to give Kondo-enhanced transport. Hence, the conductance is small almost negligible - that is the black top region in the plot. At the pseudospin freezing temperature $T^{\star}$, the conductance slightly increases as indicated by the violet sharp region. However, it is only at $T \leq T_{K}$ that the Kondo singlet is formed by quenching the spin-$1/2$ from the preferred parity doublet state $\ket{\Psi^{7}_{0}}$ with the conduction electrons tunnelling from the strongest (even) channel. In this strong coupling phase, the conductance is Kondo enhanced reaching the unitary limit $\mathcal{G}^{C}(T \ll T_{K}; -43~eV<V_{gate}<-40~eV~\wedge~-40~eV<V_{gate}<-36~eV)=\mathcal{G}_{0}$, showed bright yellow region in the figure. However, at any temperature for $V_{gate}=-40~eV$ we have a \textbf{conductive node}, that is the black line splitting in two the yellow conductive region.\\
The appearance of the node is more evident in the conductance curve at zero temperature and zero $V_{bias}$, see plot in Fig.\ref{F6:G_T0_Vg_13}. From the Coulomb blockade valley in the $N=6$ sector, by lowering $V_{gate}$ the tunnelling becomes finite as soon as the valence transition with the $N=7$ sector occurs. Within the doublet multiplet ground-state, the conductance reaches the unitary limit except for $V_{gate}=-40~eV$. The conductance can be analytically calculated  \cite{PustilnikGlazman_review2004} using the phase shifts in the formula  $\mathcal{G}^{C}(T=0) \sim \sin^{2}(\delta_{e}-\delta_{o})$ - which we  use in every discussion for $T=0$ conductance results. At $V_{gate}=-40~eV$ we have an equal contribution from both leads i.e. $\delta_{e}=\delta_{o}=\pi/4$ such that $\mathcal{G}^{C}(T=0)\sim 0$. Anywhere else, the strongest channel has full Kondo effect $\delta_{e} \sim \pi/2$ whereas $\delta_{o}\sim0$ for the decoupled lead, hence $\mathcal{G}^{C}(T=0) \sim 1$. At $SU(4)$ point, both channels remain coupled equally down to $T=0$.\\
\noindent{In} analogy to valence transition between $6 \leftrightarrow7$ sector, around $V_{gate}\sim-43~eV$ the crossover from $7 \leftrightarrow8$ charge sectors occurs and is accompanied by by unitary limit $\mathcal{G}_{0}$ value of conductance curve.

\subsection*{Spin triplet ground-state $S=1$ in $N=8$ charge sector: two-stage perfect screened Kondo effect}
As the gate voltage is tuned further, in the $1,3$-\textit{meta} configuration the molecule accommodates another electron and moves into the $N=8$ electron regime. The molecular ground-state crosses from doublet over to $S=1$ spin triplet state. This is observed in the gate voltage range $-48~eV \leq V_{gate}\leq -43~eV$, see Fig.\ref{F6:MolCharge}. Although the pseudospin degrees are not included any more in the effective modelling as we discussed (there is a single lowest triplet state of the isolated molecule), the triplet ground-state allows us to explore another type of Kondo effect.\\
In order to understand the physics of the triplet spin state in $N=8$ Coulomb blockade benzene junction, we avail of second order perturbation expansion results. From the effective model, the calculated exchange couplings are both antiferromagnetic meaning that both $J_{e},J_{o}$ are RG relevant parameters. Here, we assume again the even channel is the strongest i.e.  $J_{e}>J_{o}>0$. The couplings indicate that deep in the $N=8$ sector the benzene junction behaves as a two-channel spin-$1$ Kondo model, that undergoing a \textbf{two-stage perfect screened Kondo effect} - see full discussion of the two-channel model in Sec.\ref{sec:2CK}. Once $T=0$ limit is reached, no residual spin degrees are left in the final state.\\
We now discuss the thermodynamics and the conductance as showed in Figs.\ref{F6:S_VgT_13},\ref{F6:G_VgT_13} respectively.\\
At $T\sim \delta E$, the molecule entropy is determined by the three spin degrees in the triplet state and $S_{mol}=\ln(3)$, this is the yellow region in the entropy plot.\\
By lowering the temperature, at $T\leq T_{K}^{e}$ the first Kondo singlet is formed from the screening of spin-$1/2$ of the molecule with the conduction electron from the even channel. The molecule entropy counts only two spin degrees left such that  $S_{mol}=\ln(2)$ - see green region in the plot. These spin degrees of freedom are enough to allow finite transmission between leads and molecule due to the Kondo effect. Thus, the conductance increases up to $\mathcal{G}^{C}(T_{K}^{o}<T\leq T_{K}^{e})=\mathcal{G}_{0}$ as signature of enhanced tunnelling due to the Kondo singlet formation, given by light yellow region. However, at $T\ll T_{K}^{o}$ the second Kondo singlet is formed from the previously unquenched molecule spin-$1/2$ with the conduction electron from the odd channel. The molecule entropy counts only one spin degree left and it drops to $S_{mol}=\ln(1)=0$, that is the dark blue region in the plot. At this stage, there are no residual molecule spin degree available for screening process and the conductance vanishes $\mathcal{G}^{C}(T\ll T_{K}^{o})=0$, as we see from the dark region in the bottom.\\
The vanishing of the conductance from the unitary limit due to the two stage Kondo effect \cite{Coleman_conductanceS1_PRL2005} is more evident Fig.\ref{F6:G_T0_Vg_13}. In particular, we clearly see the lack of any conductive state for any $T\ll T_{K}^{o}$ and is persistent down to $T=0$. In this case, the couplings are of the same nature indicating that $\delta_{e}=\delta_{o}=\pi/2$. Hence, we have  $\mathcal{G}^{C}(T=0) \sim \sin^{2}(\delta_{e}-\delta_{o})\sim 0$ in the whole zero temperature $N=8$ charge regime.

\subsection{\textit{para}-connected benzene transistor}
In this section we focus on the \textit{para}- or $1,4$- connected benzene transistor, see the left schematic in Fig.\ref{F6:schematic}, with the right/drain lead connected on the fourth site of the benzene molecule. We focus on the different charge occupation values $N=6,7,8$ of the Coulomb staircase as seen from the left plot in Fig.\ref{F6:MolCharge} in terms of molecular entropy $S_{mol}$ in Fig.~\ref{F6:S_VgT_14} and \textit{dc}-regime electrical conductance $\mathcal{G}^{C}(T)$ in Fig.~\ref{F6:G_VgT_14} as a function of temperature $T$ and gate voltage $V_{gate}$. \\
We note that the $N=6$ regime behaves similar to the $1,3$-geometry and the same physics follows as before. Moreover, the similar conductive spikes in the conductance plots appear in correspondence of the valence transitions as in the  $1,3$-geometry, compare Figs.~\ref{F6:G_VgT_13},\ref{F6:G_VgT_14}. Hence, for the sake of the more interesting $N=7$ and $N=8$ regimes, we omit those comments in this section.

\subsection*{Spin doublet ground-state $S=\tfrac{1}{2}$ in $N=7$ charge sector: Kondo effect with level crossing quantum phase transition}
\begin{figure}[H]
	\centering
	\includegraphics[width=0.6\linewidth]{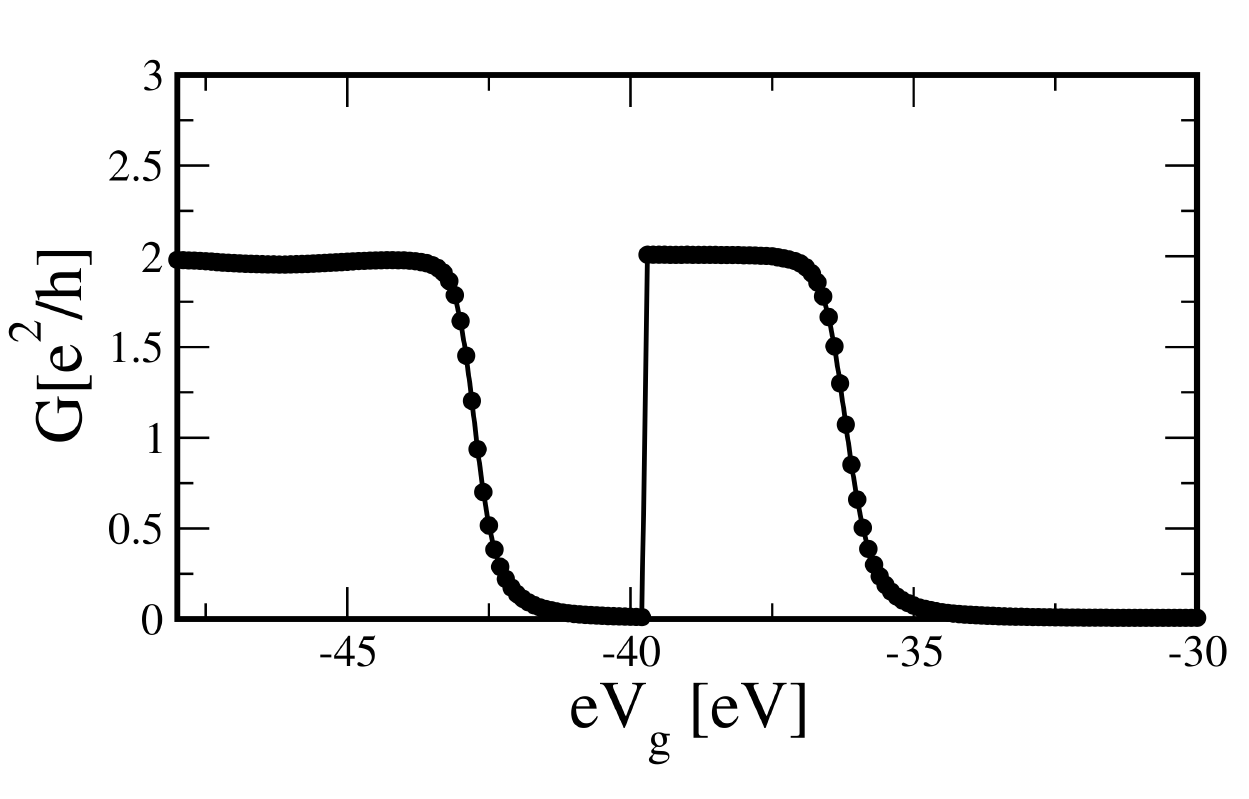}
	\caption[Electrical conductance $\mathcal{G}^{C}(T=0)$ in $1,4$ configuration]{Electrical conductance $\mathcal{G}^{C}(T=0)$ as a function of $V_{gate}$ encompassing $N=6,\,7,\,8$ sector in the $1,4$-connected molecule-lead coupled benzene junction. In the $N=7$ charge sector, the benzene junction behaves as a switch where the Kondo enhanced conductance prior to the level crossing phase transition switches to zero at $V_{gate}\sim -40~eV$. As $V_{gate}$ is further tuned to more negative values, the molecule enters the $N=8$ regime and the conductance eventually recovers the unitarity limit due to the underlying partial Kondo screening of the $S=1$ spin triplet state.}\label{F6:G_T0_Vg_14}
\end{figure}

Similar to the \textit{meta}-connected configuration, decreasing $V_{gate}$ smoothly increases the molecule charge when leads are connected. However, we can still identify the Coulomb blockade regimes and mixed-valence crossovers. In the voltage range $-43~eV< V_{gate}\lesssim 36~eV$, the ground-state of the isolated molecule is essentially identified by a doublet spin $S=1/2$ state. However, the different lead-molecule-lead configuration exhibits new physics from the $1,3$-connected model. This is further evidence of the major role of symmetries in the low-energy behaviour of the system.\\
We start with analysing the molecule entropy in Fig.\ref{F6:S_VgT_14} in the $N=7$ sector.\\
In analogy to the $1,3$ configuration, at high temperature $T\sim \delta E$ for the entire voltage range $-43~eV< V_{gate}\lesssim -36~eV$, the doublet ground-state comprises both spin and pseudospin degrees of freedom. 
This is the red region in the plot characterised by $S_{mol}=\ln(2+2)$. The pseudospin freezing out of the weakest parity doublet occurs only at lower temperature $T=T^{\star}$. Only the lowest pseudospin $p^{\star}$ is accessible,
such that the entropy is reduced to $S_{mol}=\ln(2)$, as seen in the green region. At $T\leq T^{\star}$, the system is described as a generalized two-channel spin-$1/2$ Kondo model. By lowering even further the temperature, the thermodynamics before and after $V_{gate}\sim - 40~eV$ is evidently of different nature. At $-40~eV< V_{gate}\lesssim -36~eV$ in the temperature scale $T_{K}\ll T^{\star}$, under RG-flow of coupling parameters we find that the lowest energy doublet with parity $p^{\star}$ is screened by the conduction electrons from the strongest channel, while the weakest channel has to detach from the molecule. The Kondo singlet forms $T \ll T_{K}$ and the signature of this process is the entropy value $S_{mol}=\ln(1)=0$  which persists down to $T=0$ -
that is the dark blue region developing from the green area to the bottom of the diagram in Fig.\ref{F6:S_VgT_14}. Opposed to this strong coupling phase, at $-43~eV< V_{gate}< -40~eV$ we find a persistence of the local moment phase for any $T\leq T^{\star}$, as indicated by the large green region sharply confined on the right at $V_{gate}= -40~eV$ and developing till $T=0$. \\
We note that $SU(2)$ symmetry holds in the system for the whole voltage range $-43~eV< V_{gate}\lesssim -36~eV$. However, the sudden jump in thermodynamics values at $V_{gate}\sim - 40~eV$ suggests a first order level crossing transition from Kondo screened to local moment phase should take place in the coupled system. This observation is indeed corroborated by the eigenspectrum analysis from NRG. As consequence of the pseudospin freezing process happening at $T\leq T^{\star}$, the lowest ground-state doublet is $\ket{\Psi^{7}_{0}}=\ket{7;\pm1/2;-}$ \textit{before} $V_{gate}\sim - 40~eV$. However, at $V_{gate}\sim - 40~eV$ we find an \textit{inversion} of the ground-state parity eigenvalue and the previously frozen doublet $\ket{\Psi^{7}_{0}}=\ket{7;\pm1/2;+}$ becomes the ground-state doublet. In conclusion at $V_{gate}\sim - 40~eV$ the system undergoes to a switch of the ground-state doublet state: an unequivocal signature of \textbf{level crossing quantum phase transition}. Now, we further investigate the special point at $V_{gate}\sim - 40~eV$ and in its boundary.\\
As we analyse for the $1,3$ system, we compute the average of $\widehat{P}_{mol}$ using NRG data. Opposed to the previous result, we never find a gate value where both parity doublet multiplets equate. Moreover, $\langle \widehat{P}_{mol} \rangle$ runs from negative values with limit at $\sim -1$ to positive ones with limit at $\sim +1$ with a discontinuity at $V_{gate}\sim - 40~eV$. This discontinuity evidently confirms the level crossing transition already seen in the switching of ground-state energy at the special $V_{gate}$ value. \\
\begin{figure}[H]
	\centering
	\includegraphics[width=0.65\linewidth]{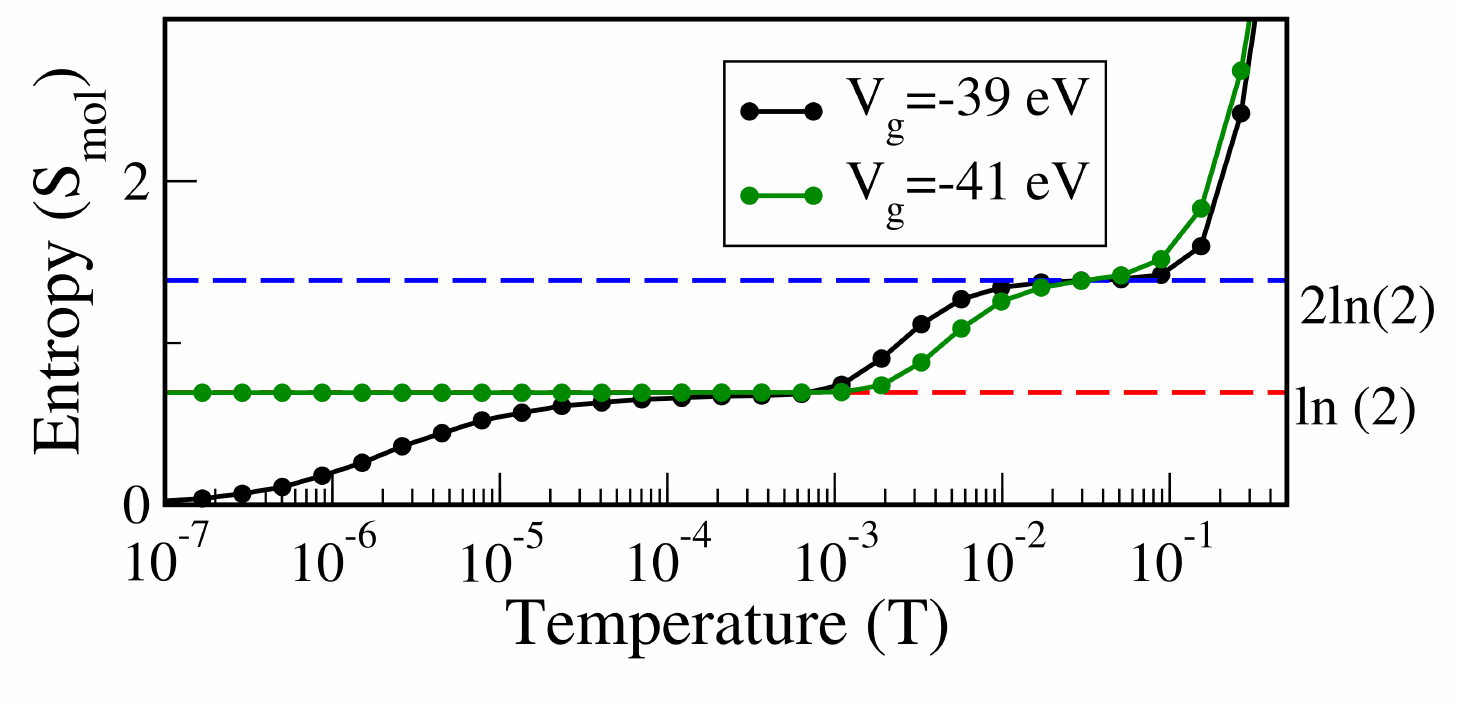}
	\caption[Molecule entropy in $1,4$ configuration at $N=7$ sector around $V_{gate}=-40~eV$]{Molecular entropy in the $N=7$ regime, as a function of temperature $T$ for two different gate voltages $V_{gate}=-39~eV$ (black curve) and $V_{gate}=-41~eV$ (green curve) on either side of $V_{gate}\sim -40~eV$ value where the level crossing quantum phase transition occurs in the $1,4$-connected benzene junction. At $T\ll T_{K}$, for $V_{gate}=-39~eV$ the doublet spin-$1/2$ state with parity $p=+1$ is Kondo quenched whereas for $V_{gate}=-41~eV$ the doublet spin-$1/2$ state with parity $p=-1$ remains unscreened down to $T=0$.}\label{F6:S_40-G_14}
\end{figure}
\noindent{To} get a complete intuition about the pseudospin freezing process with related system phases, we plot $S_{mol}$ as function of temperature in the phases either side of $V_{gate}\sim - 40~eV$, namely at $V_{gate}= -39~eV$ and $V_{gate}=-41~eV$ in black and green curve respectively in Fig.\ref{F6:S_40-G_14}. We distinguish again two plateaux at $S_{mol}=\ln(2+2)$ and $S_{mol}=\ln(2)$, marked with blue and red dashed lines respectively. At high temperature, the entropy value is determined by entangled spin and pseudospin degrees both before and after the special $V_{gate}$ point - at this scale the two curves approximately overlap. At $V_{gate}= -39~eV$ once pseudospin freezing temperature $T=T^{\star}$ is reached, the molecule ground-state is represented solely by $\ket{\Psi^{7}_{0}}=\ket{7;\pm1/2;+}$ doublet. This spin doublet state is associated to $S_{mol}=\ln(2)$ value, that is the local moment phase as indicated by intermediate plateau in the black curve. This residual spin-$1/2$ degree is Kondo quenched at lower temperature $T_{K}\ll T^{\star}$, reaching $S_{mol}=\ln(1)$ value as seen in the black tail approaching zero entropy value as $T\to0$ at the left corner of the plot. On the contrary, at $V_{gate}= -41~eV$ below the pseudospin freezing temperature $T=T^{\star}$, 
the molecule ground-state is now represented solely by $\ket{\Psi^{7}_{0}}=\ket{7;\pm1/2;-}$ state. This is now the prevailing spin doublet state to which corresponds $S_{mol}=\ln(2)$ entropy value. The system preserves this local moment phase till $T=0$ with residual molecule spin-$1/2$ degrees of freedom, as showed by the long plateau in green curve.\\
To sum up: the additional orbital degeneracies encoded in the pseudospin degrees $(i)$ are separately distinct from the spin degrees at $T\ll T^{\star}$ $(ii)$ bring $1,4$-connected system at $V_{gate}=-40~eV$ to an inversion of parity in the molecule ground-state leading to a level crossing phase transition.\\ 
\noindent{Considering} the interplay between parity and Kondo singlet formation, we complete our understanding by extracting information on the couplings nature. We find that at $V_{gate}= -39~eV$, the exchange coupling with parity $p=+1$ is antiferromagnetic in the even channel and ferromagnetic in the odd  channel whereas the exchange coupling with parity $p=-1$ is ferromagnetic in both channels. At $V_{gate}= -41~eV$, the coupling nature is the same with the exception for the exchange coupling with parity $p=+1$ turns antiferromagnetic also in the odd channel. Across the transition, the sign of the coupling of the ground-state doublet therefore changes.\\
\noindent{We} study now the electrical conductance in Fig.\ref{F6:G_VgT_14}. \\
As before, at $T^{\star}\ll T\sim \delta E$, temperature is beyond the regime for observing any many-body quantum effect. Hence, the conductance is almost negligible, that is the black top region in the plot. We analyse separately the transport for $V_{gate} > -40~eV$ and $V_{gate}< -40~eV$. In the range $-40~eV< V_{gate}\lesssim -36~eV$, at temperature $T^{\star}$ the freezing out the doublet state with parity $p=-1$ occurs and the conductance slightly increases as indicated by the violet sharp region. However, it is only at $T \ll T_{K}$ that under RG-flow the Kondo singlet is formed from the residual spin-$1/2$ doublet with $p=+1$ screened by the conduction electrons from the even channel (coupling $J^{+}_{e}>0$) while the odd channel (coupling $J^{+}_{o}<0$) separates at $T\to0$. The system reaches strong coupling phase which determines a Kondo enhanced conductance $\mathcal{G}^{C}(T \ll T_{K}; -40~eV<V_{gate}<-36~eV)=\mathcal{G}_{0}$, showed by the large bright yellow region in the plot. At $V_{gate}= -40~eV$ the conductance abruptly changes, confirming the level crossing occurring at this point. At $-43~eV< V_{gate}<-40~eV$, the system undergoes initially at $T=T^{\star}$ to pseudospin freezing of spin doublet with $p=-1$ followed then by a ground-state inversion that turns back the lowest energy doublet with $p=-1$ parity. The resulting two-channel spin-$1/2$ is a ferromagnetic Kondo model. Under RG-flow, residual spin-$1/2$ doublet with $p=-1$ remains unquenched till $T=0$ since both even and odd channels with couplings $J^{-}_{e},J^{-}_{o}<0$.
Within this voltage range, the coupled system remains in a zero conductive state at any temperature, as indicated by the black column occupying half of the $N=7$ regime. Thus, we conclude the benzene junction at $V_{gate}= -40~eV$ behaves like a \textbf{switch}: the conductance changes from unitary limit $\mathcal{G}_{0}$ to zero just by tuning the gate voltage value in the vicinity of $V_{gate}=-40~eV$. \\ 
\noindent{The} switching behaviour is more evident in the conductance curve at $T=0$, see plot in Fig.\ref{F6:G_T0_Vg_14}. Only at the crossover from $N=6$ to $N=7$ sector there is finite electron tunnelling to lift the system from the Colomb blockade regime. Within the doublet multiplet ground-state, the conductance takes maximum value at $V_{gate}> -40~eV$ then vanishes at $V_{gate}=-40~eV$ in discontinuous fashion. Ultimately, the conductance increases again at the valence transition between $7\leftrightarrow 8$ sectors. We confirm this result also analytically, namely before $V_{gate}= -40~eV$, the antiferromagnetic $J^{+}_{e}>0$ contributes to finite $\delta_{e} \sim \pi/2$ whereas the ferromagnetic $J^{+}_{o}<0$ has vanishing phase shift $\delta_{o} =0$  so we obtain maximum conductance $\mathcal{G}^{C}(T=0) \sim \sin^{2}(\pi/2-0)\sim 1$. After $V_{gate}= -40~eV$, both couplings are ferromagnetic giving  $\delta_{e} =\delta_{o} \sim 0$ and so we have zero conductance.

\subsection*{Spin triplet ground-state $S=1$ in $N=8$ charge sector: effective single-channel underscreened Kondo model}
As the gate voltage is tuned further down, in the $1,4$-\textit{meta} configuration the molecule accommodates another electron to give $N=8$ and the molecular ground-state crosses from  doublet $S=1/2$ to  $S=1$ spin triplet regime. This is observed in the gate voltage range $-48~eV \leq V_{gate}\leq -43~eV$, see Fig.\ref{F6:MolCharge}. As in the \textit{meta}-configuration, the triplet ground-state offers enough spin degrees for realising Kondo screened phase. However, the different couplings nature determines half screened state from $1,3$ case - as we discuss now.\\
From second order perturbation expansion results in $\hat{H}_{hyb}$, we find the exchange coupling in the even channel is antiferromagnetic $J_{e}>0$ whereas in the odd channel is ferromagnetic $J_{o}<0$. As consequence, $J_{e}$ is RG relevant parameter while $J_{o}$ is RG irrelevant. Under RG-flow, deep in the $N=8$ sector the benzene junction behaves as an effective single-channel spin-$1$ Kondo model undergoing to \textbf{partial Kondo screening} - see more comments in Sec.\ref{sec:2CK}. In the limit $T\to 0$, there are residual unquenched spin degrees of freedom.\\
We now discuss the thermodynamics and the conductance as showed in Figs.\ref{F6:S_VgT_14},\ref{F6:G_VgT_14} respectively.\\
As before, at $T\sim \delta E$, the molecule entropy is determined by the three spin degrees in the triplet state as $S_{mol}=\ln(3)$, this is the yellow region in the entropy plot. As seen previously, in this regime we have weak electron tunnelling - low conductance means no Kondo effect, see  black area in the conductance plot.\\
By lowering the temperature, at $T\ll T_{K}^{e}$ the Kondo singlet is formed from the screening of spin-$1/2$ of the molecule with the conduction electron tunnelling from the even channel with positive $J_{e}$. The molecule entropy is reduced by a unit as $S_{mol}=\ln(2)$ as represented by the green region of the diagram. Thus, the conductance increases up to $\mathcal{G}^{C}(T\ll T_{K}^{e})=\mathcal{G}_{0}$ as signature of enhanced tunnelling due to the Kondo singlet formation, given by light yellow region in the plot. By decreasing even further the temperature down to $T\to 0$, under RG-flow the Kondo singlet state persists while the odd channel with ferromagnetic $J_{o}$ decouples from the molecule. Although there is still a spin-$1/2$ degree of freedom on the molecule now, there is no second stage Kondo because the odd channel couples ferromagnetically and decouples under RG-flow. This model can be regarded as \textbf{single-channel underscreened Kondo model}.
The system reaches a partial Kondo screened state with a local moment ground-state accompanied by unitary limit of conductance  $\mathcal{G}^{C}(T\ll T_{K}^{e})=\mathcal{G}_{0}$ which persists down to $T=0$.\\
The enhancement in conductance due to the partial Kondo screening \cite{Coleman_conductanceS1_PRL2005,PustilnikGlazman_realQD2001} is evident also from Fig.\ref{F6:G_T0_Vg_14}. Subsequently to the gradual increment in the valence transition, the conductance remains at maximum value for the whole $N=8$ charge regime from $T_{K}^{e}$ till $T=0$.  In this case, the couplings have opposite nature namely $\delta_{e}\sim\pi/2$ and $\delta_{o}\sim0$. Hence, we have  $\mathcal{G}^{C}(T=0) \sim \sin^{2}(\delta_{e}-\delta_{o})\sim 1$. 

\section{Conclusion}
Nanoscale junctions involving real molecules require a description encompassing strong interactions, molecular geometry and its orbital structure on the same footing. The molecular geometry embodies certain structural symmetries that determine the transport characteristics of the contributing orbitals in contact with the lead electronic wave functions. This interplay between the molecular symmetry and Kondo type strong interactions, between both molecular electrons and conduction electrons of the leads may direct to a plethora of rich characteristics in the device.\\ 
In this chapter, we have shown that the simple single-benzene molecule coupled to source and drain electrode can host Kondo enhanced conductance due to additional pseudospin degrees. According to the system configuration, we have found the following cases:
\begin{enumerate}
	\item $N=7$ charge sector: $S=1/2$ Kondo effect leading to $(i)$ a conductance node that is tunable via the application of a back gate voltage and is determined by the underlying higher $SU(4)$ symmetry of the full lead-coupled transistor, $(ii)$ first order quantum phase transition representing a molecular current switch from blockade to maximally enhanced conductance;
	\item $N=8$ charge sector: $(i)$ two-stage $S=1$ perfectly screened Kondo effect, $(ii)$ partially screened $S=1$ Kondo model leading to an effective single-channel underscreened Kondo model.
\end{enumerate}
In particular, we demonstrate the interplay between molecular symmetry and Kondo effect driven strong interactions via NRG calculation on the full benzene ring to determine the conductance characteristics, and comprehend these using effective models.\\ 
These theoretical investigation demonstrate the high-tuned conductive states of the lead-benzene-lead junction via external voltages. The coupled system embraces the transistor functionality where small $V_{gate}$ variation leads to different tunnelling regime in the system. Hence, the single-molecule benzene coupled to leads is a promising candidate for future engineering of nanoelectronic devices.\\
As remarkable results, the effective model both in first and in second order perturbation expansion in $\hat{H}_{hyb}$ correctly predicts the $\omega=T=0$ conductance behaviour in the mixed-valence and in the Coulomb blockade charge sectors, respectively. Furthermore, from second order results we calculate the exchange couplings. The advantage of the effective model is to reduce the electronic complexity from the bare model and to create suitable configuration under PC that is absent in the microscopic model. In this regards, even elaborate systems can be studied through minimal element of a generalized Fermion class of effective two-lead under PC models.

\chapter{Two-Channel Charge-Kondo physics in graphene quantum dots}\label{ch:graphene}
Nanoelectronic quantum dot devices exploiting the charge-Kondo paradigm have been established as versatile and accurate analogue quantum simulators of fundamental quantum impurity models. In particular, hybrid metal-semiconductor dots connected to two metallic leads realize the two-channel Kondo (2CK) model, in which Kondo screening of the dot charge pseudospin is frustrated. Here, we consider theoretically a two-channel charge-Kondo device made instead from graphene components, realizing a pseudogapped version of the 2CK model. We solve the model using Wilson's Numerical Renormalization Group method, and uncover a rich phase diagram as a function of dot-lead coupling strength, channel asymmetry, and potential scattering. The complex physics of this system is explored through its thermodynamic properties, scattering $\mathrm{T}$-matrix, and experimentally measurable conductance. We find that the strong coupling pseudogap Kondo phase persists in the channel-asymmetric two-channel context, while in the channel-symmetric case frustration results in a novel quantum phase transition.  Remarkably, despite the vanishing density of states in the graphene leads at low energies, we find a \emph{finite} linear conductance at zero temperature at the frustrated critical point, which is of non-Fermi liquid type. Our results suggest that the graphene charge-Kondo platform offers a unique possibility to access multichannel pseudogap Kondo physics.\\
The results of this chapter are published in the paper \cite{minarelli2022two} - here in this version we highlight more physical explanations to adhere this material with  concepts and techniques we presented in previous chapters. As an outline, we set the work with respect to the existing literature and define our objectives. Then, we present the model and methods employed prior to the main section, the results, where we address our findings. We conclude suggesting future extension of this work.

\section{Introduction}
The Kondo effect \cite{Kondo1964} was originally discussed in the context of local magnetic impurities such as Fe, embedded in non-magnetic metallic hosts like Au. By progressively decreasing the temperature $T$, experimental measurements revealed an unexpected resistivity minimum, attributed to enhanced electronic scattering from the impurity local moments \cite{Hewson}. The low-energy physics of such systems is explained by the deceptively simple Kondo model, which features a single spin-$\tfrac{1}{2}$ local moment exchange coupled to a featureless bath of metallic, non-interacting conduction electrons. The `Kondo effect' refers to the universal physics of this model, appearing at $T \sim T_{K}$ with $T_{K}$ an emergent low-energy scale called the Kondo temperature, in which the impurity spin is dynamically screened by a surrounding many-body entanglement cloud of conduction electrons \cite{Wilson1975}.\\
Since then, variants of the basic Kondo effect that arise when magnetic impurities are embedded in unconventional host materials have been studied. Examples include ferromagnets \cite{martinek2003kondo} and superconductors \cite{franke2011competition}, as well as topological materials such as topological insulators \cite{mitchell2013kondo} or Dirac/Weyl semimetals \cite{mitchell2015kondo}. However, local moments in graphene have attracted the most attention \cite{chen2011tunable,fritz2013physics,Dias_Graphene2017}. In neutral graphene, the Dirac point is at the Fermi level \cite{neto2009electronic} and so a spin-$\tfrac{1}{2}$ impurity couples to a bath of conduction electrons with a density of states (DoS) featuring a low-energy pseudogap $\rho(\omega)\sim |\omega|^r$ with $r=1$. This has dramatic consequences for the resulting Kondo physics \cite{fritz2013physics} due to the depletion of low-energy degrees of freedom in graphene that can participate in screening the impurity spin.\\
Deeper insights into strongly correlated electron physics and Kondo have been gained from tunable circuit realizations of fundamental models in nanoelectronics devices, made possible by remarkable recent advances in nano-fabrication and characterization techniques \cite{heinrich2021quantum,barthelemy2013quantum}. This provides a route to probing and manipulating quantum matter at the nanoscale in a way that would be impossible in bulk systems. In particular, semiconductor quantum dots (QDs) have been shown to behave like artificial atoms \cite{Kastner_Artificialatoms1994}, with the extreme quantum confinement producing a discrete orbital structure and strong electron-electron interactions. Coupling such quantum dots to metallic electrodes at quantum point contacts (QPCs) gives rise to the Kondo effect at low temperatures \cite{Goldhaber-Gordon-Kondo1998Exp,cronenwett1998tunable,Kouwenhoven2000Exp}, with a single local moment trapped on the dot facilitating spin-flip scattering of lead conduction electrons. In such devices, entanglement spreads across the QD and the leads in a macroscopic \textit{Kondo cloud} \cite{mitchell2011real,yoo2018detecting}, producing the famous Kondo resonance in electrical conductance \cite{PustilnikGlazman_review2004}. Quantum transport properties of QDs can be tuned \textit{in situ} by applying gate voltages to control the QPC transmissions and dot potential. A  bias voltage can drive the system out of equilibrium.\\
Quantum dot devices also allow more complex quantum impurity models to be realized experimentally by controlling the microscopic interactions. As such, they constitute a versatile platform to study a rich range of physics \cite{barthelemy2013quantum}, including quantum phase transitions (QPTs) \cite{vojta2006impurity,mitchell2009quantum}, emergent symmetries \cite{keller2014emergent,mitchell2021so}, and non-Fermi liquid (NFL) physics \cite{GoldhaberGordon-2CK-2007Exp,keller2015universal,Mitchell_UniversalLowT2CK2012,mitchell2012two}. The two channel Kondo (2CK) model \cite{Nozieres_KondoRealMetal1980} is a famous example which embodies the frustration of Kondo screening of a single impurity by two distinct channels of metallic conduction electrons, and displays all these features \cite{Affleck_CriticalOverscreened_1991}. The standard 2CK model Hamiltonian reads,
	\begin{equation}\label{eq:9H2ck}
		\hat{H}^{2CK} = \hat{H}_{leads} + J_1 \hat{\mathbf{S}}\cdot\hat{\mathbf{s}}_1  + J_2 \hat{\mathbf{S}}\cdot\hat{\mathbf{s}}_2 ~,
	\end{equation}
where $\hat{H}_{leads} =\sum_{\alpha\sigma\mathbf{k}} \epsilon_{\mathbf{k}}^{\phantom{\dagger
}} c_{\alpha\sigma \mathbf{k}}^{\dagger}c_{\alpha\sigma \mathbf{k}}^{\phantom{\dagger
}}$ describes two leads $\alpha=1,2$, each with spin $\sigma=\uparrow,\downarrow$ electrons with  momentum $\mathbf{k}$. The Eq.\ref{eq:9H2ck} is indeed the two-channel version of the Kondo model we introduce in Eq.\ref{eq:K}. In the original formulation of 2CK model, the dispersion $\epsilon_{\mathbf{k}}$ is taken to be linear at low energies such that the electronic DoS of the leads at the impurity position is flat. The metallic flat band approximation is typically employed for the free conduction electrons, $\rho(\omega)\sim \sum_{\mathbf{k}} \delta(\omega-\epsilon_{\mathbf{k}}) \equiv \overline{\rho} \theta(|\omega|-D)$, describing a flat density of states $ \overline{\rho}$ inside a band of half-width $D$. In Eq.~\ref{eq:9H2ck},  $\hat{\mathbf{S}}$ is a spin-$\tfrac{1}{2}$ operator for the impurity and $\hat{\mathbf{s}}_{\alpha}$ is a spin-$\tfrac{1}{2}$ operator for the spin density in lead $\alpha$ at the impurity position, such that $\hat{H}^{2CK}$ possesses $SU(2)$ spin symmetry. The metallic 2CK model supports a QPT, with an non-Fermi liquid critical point at $J_1=J_2$ \cite{Affleck_CriticalOverscreened_1991}. Signatures of the critical point in this model have been observed experimentally in semiconductor quantum dot devices \cite{GoldhaberGordon-2CK-2007Exp,keller2015universal}.\\
More recently, a new nanoelectronics paradigm has emerged, based on charge-Kondo quantum dots \cite{2CK-RenormalisationFlow_IftikharPierre2015Exp,mitchell2016universality,iftikhar2018tunable,han2021fractional,pouse2021exotic}. In the standard set-up, a large QD is coupled at QPCs to source and drain leads. These devices realize anisotropic multichannel Kondo models through Matveev's mapping \cite{matveev1995coulomb,furusaki1995theory} of the macroscopic charge states of the large QD to an effective pseudospin that is flipped by electronic tunnelling at the QPCs. This approach has led to unprecedented control over the frustrated 2CK state, and has uncovered the full renormalization group (RG) flow diagram through transport measurements \cite{2CK-RenormalisationFlow_IftikharPierre2015Exp}. \\
Motivated by these developments, in this chapter we consider combining the two-channel charge-Kondo set-up of Ref.~\cite{2CK-RenormalisationFlow_IftikharPierre2015Exp} with the pseudogap Kondo physics of graphene in Ref.~\cite{fritz2013physics}, to realize a novel two-channel pseudogap charge-Kondo effect. We envisage a charge-Kondo device made from graphene components, see Fig.~\ref{Fig:setup}, such that the dot charge pseudospin is coupled to two channels of conduction electrons, each with the characteristic linear pseudogap DoS of graphene. This work is a theoretical exploration of such a system and its phase diagram. We characterize the phases and phase transitions through thermodynamic quantities, and focus on experimentally-relevant physical observables such as the conductance. However, we do not claim to address the practical complexities that will inevitably arise in the experimental realization of a graphene charge-Kondo device.\\  
We note that the generic properties of fully spin- and channel-symmetric two-channel pseudogap Kondo models were discussed in Ref.~\cite{schneider2011two}, although the $r=1$ linear pseudogap case relevant to graphene was not analysed in detail and a device realization was not proposed. Furthermore, our charge-Kondo implementation leads to crucial differences in the model and transport measurement set-up which have not previously been considered. These differences and our new results are highlighted in the following.\\
Before proceeding with the modelling and the results, we give a short overview to the electronic structure we study in this work.

\subsection*{Graphene density of states}
In the general scenario, the energy spectrum in the Kondo model - see discussion in Sec.\ref{sec:QImpModel} - can be analysed by the power-law density of states defined for a conduction bandwidth $D$ as:
	\begin{equation}\label{Eq:DoSPse}
		\rho(\omega)=\rho \biggr\lvert \frac{\omega}{D} \biggr\rvert^{r} \quad \text{at}~ |\omega|<D~.
	\end{equation}
According to the sign of the exponent $r$, the system undergoes to different behaviour. At $-1<r <0$, the DoS diverges at the Fermi level $\epsilon_{F}$ leading to a critical or Van-Hove singularity and the system is characterised by power-law Kondo model; at $r \geq 0$, there are two special cases i.e. $r=0$ metallic and $r=1$ pseudogap spectrum. The former case describes the standard set-up composed by metallic leads with continuum energy spectrum and finite $\rho(\omega)$ at the Fermi level. It is customary to approximate the conduction band structure into wide-flat band limit where $\rho(\omega) \equiv \overline{\rho}$ is constant within the conduction bandwidth $D$ and the Kondo temperature is defined as $T_{K} = \sqrt{DJ}e^{(-1/\overline{\rho}J)}$. The latter case, the DoS vanishes at the Fermi level. In this special case, the system has very few electronic states available for scattering events in the vicinity of $\epsilon_F$ - that is the relevant energy scale for Kondo effect - and the formation of the Kondo singlet entails a modification of $\rho(\omega)$. \\
An example of material with zero $\rho(0)$ where we set the Fermi level at $\epsilon_{F}=0$,  is graphene \cite{neto2009electronic}. It is a two dimensional lattice of carbon atoms with two atoms $A,B$ per unit cell defining two sublattices on the hexagonal Brillouin zone. Hopping between nearest $\braket{..~;~..}$ and next-nearest $\braket{\braket{..~;~..}}$ neighbours is described in the tight-biding Hamiltonian, namely:
	\begin{equation}
		\hat{H}= -t \sum_{\braket{i;j}\sigma}(a^{\dagger}_{i\sigma}b_{j\sigma} +h.c.) - t^{\prime}\sum_{\braket{\braket{i;j}}\sigma}(a^{\dagger}_{i\sigma}a_{j\sigma}+ b^{\dagger}_{i\sigma}b_{j\sigma} +h.c)~,
	\end{equation}
with $a_{i\sigma}$ ($b_{j\sigma}$) Fermionic operator acting on electron in $i$ ($j$) site of sublattice $A$ ($B$). Once the Hamiltonian is diagonalised, we obtain the valence $\epsilon_{-}$ and the conduction $\epsilon_{+}$ dispersion bands, that is:
	\begin{equation}\label{Eq:EnDispGr}
		\begin{aligned}
			&\epsilon_{\pm}(\mathbf{k})= \pm t \sqrt{3+f_{\mathbf{k}}} - t^{\prime}f_{\mathbf{k}} ~, \\ 
			&\text{with}\quad f_{\mathbf{k}}= 2\cos{(\sqrt{3}k_{y}a_{1})}+ 4\cos{\left(\frac{\sqrt{3}}{2}k_{y}a_{1}\right)}\cos{\left(\frac{3}{2}k_{x}a_{1}\right)} ~,
		\end{aligned}
	\end{equation}
where $a_{1}=1.21${\AA} is the bond length \cite{Kivelson_LocDoSGraphitePRB2005}. At the Dirac points identified by the wavevectors $\mathbf{K},\mathbf{K}^{\prime}$, the two bands touch the Brillouin zone and the energy dispersion in Eq.~\ref{Eq:EnDispGr} becomes linear in the form:
	\begin{equation}\label{Eq:EnDispGrL}
		\epsilon_{\pm}(\mathbf{q}) \approx \pm v_{F}|\mathbf{q}| + \mathcal{O}(q/K)^{2}~,
	\end{equation}
with $\mathbf{q}=\mathbf{k}-\mathbf{K} (or ~\mathbf{K}^{\prime})$ and Fermi velocity $v_{F}=3ta_{1}/2$.\\
In the special case of charge-neutral graphene, all the valence states are filled and all the conduction ones are empty. This symmetric band structure sets the Fermi level at the Dirac points and the DoS vanishes linearly at the Fermi level as:
	\begin{equation}\label{Eq:DoSGr}
		\rho(\omega)=\frac{2}{\sqrt{3}\pi t^{2}} |\omega|^{2}~.
	\end{equation}
Because of the similar behaviour in the electronic structure at Fermi level for charge-neutral graphene layer and for pseudogap Kondo model, we can study a pseudogap Kondo graphene system whose DoS in the low-energy spectrum given by Eq.~\ref{Eq:DoSGr} is now replaced by Eq.~\ref{Eq:DoSPse} for $r=1$ and for higher energy levels the DoS is derived from the energy bands dispersion in Eq.~\ref{Eq:EnDispGr}. In practice we choose $t^{\prime}=0$, for which case the pristine graphene host possesses particle-hole symmetry that is $\rho(\omega)=\rho(-\omega)$.

\section{Model, methods and observables}
\begin{figure}[H]
	\centering
	\includegraphics[width=0.75\linewidth]{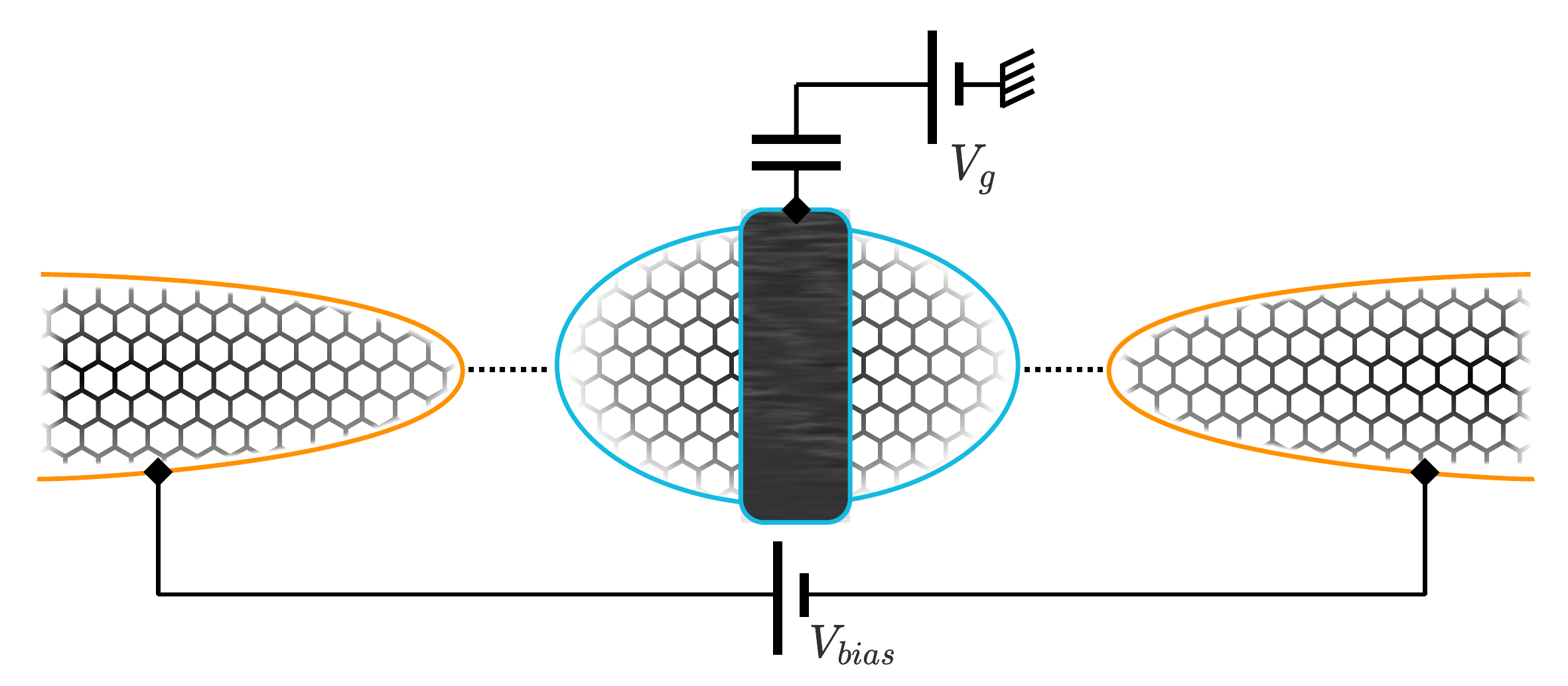}
	\caption[Schematic two-channel graphene charge-Kondo model]{Schematic of the two-channel graphene charge-Kondo quantum dot system. A net current flows from source to drain graphene leads through the large graphene dot in response to a bias voltage. A gate voltage $V_g$ controls the dot filling. The black bar denotes the decoherer.\label{Fig:setup}}
\end{figure}

\begin{figure}[H]
	\centering
	\includegraphics[width=0.75\linewidth]{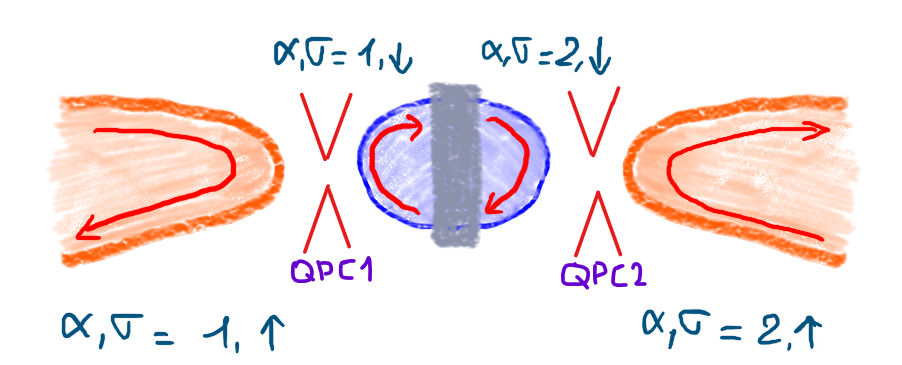}
	\caption[Schematic of the spin flux in two-channel charge Kondo device]{Schematic of the spin flux in two-channel charge Kondo device with model Hamiltonian derived in Eq.\ref{Eq:9HK} where $\alpha=1,2$ indicates the conduction electron reservoir are distinct per quantum point contact (QPC) and $\sigma=\uparrow,\downarrow$ the real spin flavour. \label{Fig:schemeSpinFLux}}
\end{figure}

We consider a two-channel charge-Kondo (CK) device in which both dot and leads are made from graphene, as illustrated in Fig.~\ref{Fig:setup}. The large dot is tunnel-coupled to leads $\alpha=1,2$ at QPCs with transmission $\tau_{\alpha}$ which can be controlled \textit{in situ} by gate voltages. A plunger gate voltage $V_g$ controls the dot potential and hence the dot filling. A decoherer is interjected between the leads via an Ohmic contact on the dot (black bar in Fig.~\ref{Fig:setup}) which gives rise to a long dwell time and an effective continuum dot level spectrum - this was achieved in the experiments of Ref.~\cite{2CK-RenormalisationFlow_IftikharPierre2015Exp} using a metallic component. This results in two effectively independent electronic reservoirs around each of the two QPCs; these form the two independent channels in the 2CK model. However, tunnelling events onto and off the dot are correlated by the large dot charging energy, $E_C$. The whole device is operated in a strong magnetic field so that the electrons are effectively spinless - that is, the Zeeman splitting is the largest energy scale in the problem.\\
\noindent{Before} proceeding with the implementation of charge-Kondo model composed of graphene, we would like to give a qualitative explanation of the mathematical operations we apply below. Both in the standard set-up with metallic elements and in the version we propose here with graphene material, the charge-Kondo paradigm comprises a large central nanostructure coupled to two leads, source and drain, through quantum point contacts. The nanostructure in the original work is semiconductor quantum dot - but the following features are also applicable to graphene dot. The dot presents electronic levels with negligible level spacing $\delta E$ such that it is considered a bath itself. Finite charge transfer is measured both in the lead-dot junction and in the coupled lead-dot-lead system such that leads are probes for dynamical quantities. Upon appropriate selection of gate voltage and temperature scale, only two macroscopic degenerate charge ground-states on the dot are energetically favourable for single-electron tunnelling through QPCs. These identify the charge pseudospin states and any higher energy states are discarded from the spin flipping processes. Furthermore, because of finite charging energy $E_{c}$ on the dot, spin-flip events are confined to charge pseudospin-$1/2$ states and the charge-Kondo model is linked via exact mapping to the standard spin Kondo model. This significant aspect allows to determine the coupling parameters exchange coupling $J$, potential scattering $W$ from the macroscopic tunnelling matrices in the charge basis model and to map those into the spin basis model - rather than deriving the couplings from perturbative method as the Schrieffer-Wolff transformation - we explain it in Sec.\ref{sec:QImpModel}. The CK model is in fact a gate-controlled device where experimentally controlled voltages give direct access to the essential low-energy physics of the system.\\
We continue now with the model Hamiltonian for the device illustrated in Fig.~\ref{Fig:setup}. It is given by,
\begin{equation}\label{Eq:9Heff}
	\begin{aligned}
		\hat{H} = \sum_{\alpha=1,2} \sum_{\mathbf{k}} \sum_{\sigma=L,D} \epsilon_{\mathbf{k}} c^{\dagger}_{\alpha\sigma\mathbf{k}}c_{\alpha\sigma\mathbf{k}}
		+\sum_{\alpha=1,2} \sum_{\mathbf{k}\mathbf{k}^{\prime}} \Big( &J_{\alpha} (\hat{Q}^{+} c^{\dagger}_{\alpha D\mathbf{k}}c_{\alpha L\mathbf{k}^{\prime}}+ c^{\dagger}_{\alpha L\mathbf{k}^{\prime}}c_{\alpha D\mathbf{k}} \hat{Q}^{-}) +\\
		&+ W_{\alpha}(c^{\dagger}_{\alpha L\mathbf{k}}c_{\alpha L \mathbf{k}^{\prime}}+c^{\dagger}_{\alpha D\mathbf{k}}c_{\alpha D \mathbf{k}^{\prime}}) \Big) + E_{C}(\hat{N}_{D}-N_{g})^{2} ~,
	\end{aligned}
	\end{equation}
we know discuss one by one the terms appearing in Eq.\ref{Eq:9Heff}. The first term is $\hat{H}_{leads}$ and it describes the distinct conduction electron reservoirs around each QPC labelled by $\alpha=1,2$, and with $\sigma=L,D$ corresponding to lead or dot electrons - rather than physical spin $\uparrow$,$\downarrow$. For graphene components, we use the energy dispersion $\epsilon_{\mathbf{k}}$ around the Fermi level as in Eq.~\ref{Eq:EnDispGrL} or at higher energy as in Eq.~\ref{Eq:EnDispGr}. The second term comprises the exchange coupling $J_{\alpha}$. The terms proportional to it describe electronic tunnelling at the QPCs between leads and dot. The tunnelling matrix elements $J_{\alpha}$ are related to the bare (unrenormalized) QPC transmissions via \cite{lebanon2003coulomb} $\tau_{\alpha}(\omega)=4\pi^2\rho_{\alpha L}(\omega)\rho_{\alpha D}(\omega)J_{\alpha}^2/[1+\pi^2\rho_{\alpha L}(\omega)\rho_{\alpha D}(\omega)J_{\alpha}^2]^2$, which are in general energy-dependent for structured leads. States of the isolated dot with a macroscopic number of electrons $N_D$ are denoted $|N_D\rangle$, with corresponding (total charge) dot number operator $\hat{N}_D=\sum_{\alpha\mathbf{k}}c_{\alpha D\mathbf{k}}^{\dagger}c_{\alpha D\mathbf{k}}^{\phantom{\dagger}}\equiv \sum_{N_D} N_D\ket{N_D}\bra{ N_D}$. Recalling that the dot is \textit{macroscopic}, it can host a Brillouin zone and the corresponding Block wavefunctions with creation $c_{\alpha D\mathbf{k}}^{\dagger}$ (annihilation $c_{\alpha D\mathbf{k}}$) operator. Tunnelling at the QPCs changes the dot charge, which we describe \cite{matveev1995coulomb} using the charge raising and lowering pseudocharge operators $\hat{Q}^{\pm}= \sum_{N_{D}} \ket{N_{D}\pm1}\bra{N_{D}}$. The ladder operator definition ensures only unitary charge variation occurs in the system. The third term proportional to $W_{\alpha}$ identifies the charge potential scattering at QPCs. The last term is the $\hat{H}_{charge}$: the dot has a finite charging energy that depends on the filling via the term proportional to $E_{C}$. The filling can be adjusted by tuning $N_g$ in Eq.~\ref{Eq:9Heff}, which is controlled in experiment by the gate voltage  $V_{gate}=V_{gate}^0-2E_{C} N_g/e$. We define $\delta V_{gate}=-2E_{C} (N_g-N_D^0-\tfrac{1}{2})/e$ such that the macroscopic dot charge states $\ket{N_D^0}$ and $\ket{N_D^0+1}$ are degenerate at $\delta V_{gate}=0$.\\
The physics described by the model Hamiltonian in Eq.~\ref{Eq:9Heff} is robust to variation in $V_{gate}$ provided that any $\delta V_{gate}$ is the smallest energy scale in the system, namely $\delta V_{gate} \ll E_{C}$. In the special case where $V_{gate} \rightarrow V_{gate}^{0}$, $N_{g}$ takes only half-integer values: the system is set at the charge degeneracy crossover between $\ket{N_{D}^0},\ket{N_{D}^0+1}$ dot ground-states. At the energy window $\delta E \ll k_{B}T \ll E_{C}$, only these two degenerate charge states are energetically accessible for tunnelling through QPCs and any higher charge states at $N<N_{D}^0,~N>N_{D}^0+1$ are neglected. Within these conditions, the pseudocharge ladder dot operators are converted into pseudospin-$1/2$ operators, namely:
\begin{equation}
	\begin{aligned}
		&\hat{Q}^{+}\rightarrow\hat{S}^{+}_{D}=  \ket{N_{D}^0+1}\bra{N_{D}^0} \quad,\quad \hat{Q}^{-}\rightarrow\hat{S}^{-}_{D}=  \ket{N_{D}^0}\bra{N_{D}^0+1} ~,\\
		&\hat{S}^{z}_{D}=\tfrac{1}{2}( \ket{N_{D}^0+1}\bra{N_{D}^0+1}-\ket{N_{D}^0}\bra{N_{D}^0}) ~,
	\end{aligned}
\end{equation}
where we also introduce the pseudospin operator $z$-component. Thus, the $\hat{S}^{+}_{D}$ flips the dot charge pseudospin from down spin to up spin while $\hat{S}^{-}_{D}$ flips it back and the second term in Eq.~\ref{Eq:9Heff} regulates the spin-$1/2$ flipping between dot and leads via QPCs - as it does the Kondo exchange term, see Eq.\ref{eq:K}. Finally, we perform a trivial relabelling $\sigma=\{ L,D\}\to \{\uparrow,\downarrow\}$ the physical spin-$1/2$ states such that the electronic operators become  $c_{\alpha L \mathbf{k}} \to c_{\alpha \uparrow \mathbf{k}}$ and $c_{\alpha D \mathbf{k}} \to c_{\alpha \downarrow \mathbf{k}}$. With this, we map the model in Eq.~\ref{Eq:9Heff} into the \textbf{effective pseudogap two-channel charge-Kondo model} (2CCK) studied in this paper:
	\begin{equation}\label{Eq:9HK}
		\boxed{
		\begin{aligned} 
		\hat{H}^{2CCK} =  
		\sum_{\alpha=1,2} \sum_{\mathbf{k}} \sum_{\sigma=\uparrow,\downarrow} \epsilon_{\mathbf{k}} c^{\dagger}_{\alpha\sigma\mathbf{k}}c_{\alpha\sigma\mathbf{k}} +
		\sum_{\alpha=1,2}\sum_{\mathbf{k}\mathbf{k}^{\prime}} \Big(
		&J_{\alpha} (\underbrace{\hat{S}_{D}^{+}c^{\dagger}_{\alpha \downarrow \mathbf{k}}c_{\alpha  \uparrow \mathbf{k}^{\prime}} }_{\hat{S}_{D}^{+}\hat{s}_{\alpha}^{-}} + \underbrace{c^{\dagger}_{\alpha \uparrow \mathbf{k}^{\prime}}c_{\alpha \downarrow \mathbf{k}} \hat{S}_{D}^{-})}_{\hat{s}_{\alpha}^{+}\hat{S}_{D}^{-}}  +\\ 
		&+W_{\alpha} \sum_{\sigma=\uparrow,\downarrow}c^{\dagger}_{\alpha\sigma\mathbf{k}}c_{\alpha\sigma\mathbf{k}^{\prime}} \Big) +e\delta V_{gate}\hat{S}^{z}_{D} 
		\end{aligned}  } ~,
	\end{equation}
where this mapping is exact and it breaks down only at $k_{B}T \approx E_{C}$ when multiple pairs of charge states are available for spin-flip process. A schematic of the spin flux in the 2CCK device described by $\hat{H}^{2CCK}$ Hamiltonian is given in Fig.\ref{Fig:schemeSpinFLux}. The model Hamiltonian in Eq.\ref{Eq:9HK} is a variant of the famous 2CK model, Eq.~\ref{eq:9H2ck} - but with a few important differences. Firstly, the DoS of the conduction electrons described by $\hat{H}_{leads}$ first term is not metallic, but has a low-energy $r=1$ pseudogap. Hence, it is the same as Eq.\ref{Eq:9Heff} since $\hat{H}$ there and $\hat{H}^{2CCK}$ describe the very same model. Secondly, tunnelling at the QPCs give an effective \textit{anisotropic} exchange coupling between the dot charge pseudospin and the conduction electrons. The $SU(2)$ symmetry of Eq.~\ref{eq:9H2ck} is broken in Eq.~\ref{Eq:9HK} since the $z$-component of the coupling is missing. However, we find that this effective spin anisotropy is RG irrelevant in the two-channel pseudogap Kondo problem (just as for the single-channel \cite{Hewson,kogan2018poor} and two-channel \cite{Affleck_CriticalOverscreened_1991} metallic case, as well as the single-channel pseudogap case \cite{fritz2004phase}). Only the spin-flip terms are important for Kondo, and these are captured by the effective model. It should also be emphasized that the effective exchange couplings $J_{\alpha}$ originate from the QPC tunnellings; there is no underlying Anderson model, so the $J_{\alpha}$ need not be perturbatively small. In fact, since they are related to the QPC transmissions, they can become large simply by opening the QPCs \cite{2CK-RenormalisationFlow_IftikharPierre2015Exp}. This is important because Kondo physics is only realized in the pseudogap model at relatively large bare coupling strengths. We remark also that the $s_{\alpha}^{\pm}$, $S_{D}^{\pm}$ spin-$1/2$ operators are acting on conduction and dot electrons, respectively. Thirdly, we have an additional potential scattering term $W_{\alpha}$ that is spin state invariant and identifies the particle-hole symmetry regime in the coupled system. This is traditionally omitted in Eq.~\ref{eq:9H2ck} because potential scattering is RG irrelevant in the metallic Kondo problem \cite{Hewson}. However, we must keep it because potential scattering is known to be important in the single-channel pseudogap Kondo model \cite{fritz2004phase}. Indeed, we find that it is strongly RG relevant in our two-channel pseudogap variant, Eq.~\ref{Eq:9HK}. Finally, the gate voltage $\delta V_g$ appears as an effective impurity magnetic field. In order to reach the charge degeneracy crossover at the edge of the Coulomb blockade valley as we know we need $\delta V_{gate} \equiv 0$ - thus, this last term does not appear in the energy regime of our interest.\\
\noindent{Another} important difference in terms of the experimental realization is the nature of the transport measurement - meaning the spin current relevant to Eq.\ref{Eq:9HK} compared to the charge current in Eq.\ref{Eq:9Heff}, as we discuss more now. As illustrated in Fig.~\ref{Fig:setup}, a series current of spinless electrons is measured between the physical source and drain leads through the dot, in response to a bias voltage. But in the mapped spin model, this is an unconventional measurement: we effectively apply a bias between leads $\alpha=1,2$ but only to the $\sigma=\uparrow$ conduction electrons. Even though there is no charge current possible between leads in the original 2CK model Eq.~\ref{eq:9H2ck}, the charge-Kondo set-up Eq.~\ref{Eq:9HK} allows an effective \textit{spin} current to be measured - see schematic in Fig.\ref{Fig:schemeSpinFLux}. \\
The $ac$-linear response electrical conductance through the device is defined,
	\begin{equation}\label{Eq:9Gc}
		\mathcal{G}^{C}(\omega,T) = \frac{\langle \hat{I}_{2\uparrow}\rangle}{V_{bias}}\biggr\rvert_{V_{bias}\rightarrow 0} ~,
	\end{equation}
due to an oscillating bias described by $\hat{H}_{bias}=-eV_{bias}\cos(\omega t)\hat{N}_{1\uparrow}$ with $ac$ frequency $\omega$ - see more discussion in Sec.\ref{sec:Kubo}. Here $\hat{I}_{\alpha \uparrow}=-e\tfrac{d}{dt}\hat{N}_{\alpha\uparrow}$ is the current operator for lead $\alpha$ and $\sigma=\uparrow$ while $\hat{N}_{\alpha \uparrow}=\sum_{\mathbf{k}}c_{\alpha \uparrow \mathbf{k}}^{\dagger}c_{\alpha \uparrow \mathbf{k}}^{\phantom{\dagger}}$. We obtain the ac linear conductance from the Kubo formula \cite{IzumidaSakai1997}, 
	\begin{equation}\label{Eq:9Kubo}
		\begin{aligned}
			\mathcal{G}^{C}(\omega,T) 
			= \frac{-\mathit{Im} \langle\langle\hat{I}_{1\uparrow};\hat{I}_{2\uparrow}\rangle\rangle_{\omega,T}}{\omega} 
			\equiv -2\pi \mathcal{G}_{0} \omega {\mathit{Im}}\langle\langle\hat{N}_{1\uparrow};\hat{N}_{2\uparrow} \rangle\rangle_{\omega,T} ~,
		\end{aligned}
	\end{equation}
where $\langle\langle\cdot~;~\cdot \rangle\rangle $
denotes a retarded real-frequency correlation function evaluated at equilibrium, and 
$\mathcal{G}_{0}=e^{2}/h$ is the conductance quantum per channel or spin and $\hbar= 1$. The second equality in Eq.~\ref{Eq:9Kubo} follows from equations of motion and is found to greatly improve the accuracy of numerical calculations \cite{transport}, as we derive in Eq.\ref{eq:4defKuboLRel} and see the complete discussion of the improved version of the Kubo formula in Sec.\ref{sec:ImprovedKubo}. Note that the system is not in proportionate coupling, and so correlated electron transport coefficients cannot be expressed in terms of a Landauer-type formula \cite{MeirWingreen_1992} involving the dot spectral function.\\
In addition to the conductance, we explore the phase diagram and RG fixed points (FPs) of the model using physical thermodynamic observables. We define the dot contribution to a thermodynamic quantity $\Omega$ at temperature $T$ as $\Omega_D(T)=\Omega(T)-\Omega^0(T)$, where $\Omega(T)$ is calculated for the full lead-dot-lead system, while $\Omega^0(T)$ is calculated only for the free conduction electrons without the dot pseudospin. For the \textbf{entropy} $S_D(T)$ we use $S^{(0)}=-\partial F^{(0)}/\partial T$, with $F^{(0)}=-k_{B}T\ln Z^{(0)}$ the free energy and $Z$ partition function to have:
\begin{equation}\label{Eq:9S}
	S_{D}(T)=  S_{tot}(T)-S^{0}_{tot}(T) ~.
\end{equation}
Recently this entropy has been extracted experimentally in similar quantum dot devices by exploiting a Maxwell relation connecting the entropy change for a process to measurable changes in the dot charge \cite{child2021entropy,han2021fractional}. For the \textbf{magnetic susceptibility} $k_{B} T \chi_D(T)$, at zero field $\delta V_g=0$, we evaluate:
\begin{equation}\label{Eq:9chi}
	k_{B}T\chi_{D}^{(0)}= \braket{(\hat{S}^{z}_{tot})^{2}}^{(0)} -(\braket{\hat{S}^{z}_{tot}}^{(0)})^{2} ~,
\end{equation}
with $\hat{S}^{z}_{tot}$ the $z$-projection of the total spin of the system. The role of particle-hole asymmetry will be assessed through the conduction electron \textbf{excess charge}: 
\begin{equation}\label{Eq:9exChar}
	N_{\alpha}=\langle \hat{N}_{\alpha}\rangle - \langle \hat{N}_{\alpha}\rangle^0 ~,
\end{equation}	
where we have defined $\hat{N}_{\alpha}=\sum_{\sigma}\hat{N}_{\alpha\sigma}$. \\
Dynamics of the system are characterized by the channel-resolved $\mathrm{T}$-matrix, which describes how conduction electrons are scattered from the dot pseudospin. The $\mathrm{T}$-matrix equation, see definition in Eq.\ref{eq:TmatrixEq}, reads,
	\begin{equation}\label{Eq:9TmatEq}
		\mathbb{G}_{\alpha\beta}(\omega,T) - \delta_{\alpha\beta}\mathbb{G}^{0}_{\alpha\beta}(\omega) = \mathbb{G}^{0}_{\alpha\alpha}(\omega) \mathbb{T}_{\alpha\beta}(\omega,T)  \mathbb{G}^{0}_{\beta\beta}(\omega) ~,
	\end{equation}
where $\mathbb{G}_{\alpha\beta}(\omega,T)$ and $\mathbb{G}_{\alpha\beta}^0(\omega)$ are, respectively, the full and free retarded electronic Green's functions at the dot position. Due to decoherence between the QPCs resulting in separately conserved charge in each channel in Eq.~\ref{Eq:9HK}, we have  $\mathbb{T}_{\alpha\beta} = \mathbb{T}_{\beta\alpha} =0$ and the $\mathrm{T}$-matrix equation becomes channel-diagonal. Furthermore,  $-\tfrac{1}{\pi}{\rm Im}\mathbb{G}^0_{\alpha\alpha}(\omega)=\rho(\omega)$ is the free graphene DoS. We introduce now the $\mathrm{T}$-spectral function for channel $\alpha$:
\begin{equation}\label{Eq:9Tmatspec}
	t_{\alpha} (\omega,T) = - \frac{\mathit{Im}\mathbb{T}_{\alpha\alpha}(\omega,T)}{\pi}~,
\end{equation}
and we observe that	$t_{\alpha} (\omega)$ is sensitive to the scattering rate in each channel. The definition in Eq.\ref{Eq:9Tmatspec} returns nonzero results for vanishing density of states, compare it with the expression in Eq.\ref{eq:TmatSpe}, as we have a pseudogap spectrum in this model.\\
Before we continue with the results, we discuss the NRG applied to the two-channel charge-Kondo model in Eq.\ref{Eq:9HK} - compare with Sec.\ref{sec:RGtheo}.

\subsection*{Numerical Renormalization Group}
The two-channel pseudogap charge-Kondo model, Eq.~\ref{Eq:9HK}, is solved using Wilson's Numerical Renormalization Group (NRG) technique \cite{Wilson1975,Bulla2008,Weichselbaum2007}, which provides numerically-exact access to the physical quantities discussed in the previous section.\\
Here we go briefly through the NRG for this model, see also the schematic in Fig.\ref{F2:NRG}. The first step is the logarithmic discretization of the conduction electron DoS, and subsequent mapping of $\hat{H}_{leads}$ to Wilson chains \cite{Wilson1975,Bulla2008},
	\begin{equation}\label{Eq:9HBathWC}
		\hat{H}_{leads} \to \hat{H}^{WC}_{ leads}=\sum_{\alpha,\sigma}~\sum_{n=0}^{\infty}t_{n}\left( f^{\dagger}_{\alpha \sigma n}f_{\alpha \sigma n+1}^{\phantom{\dagger}} + f^{\dagger}_{\alpha \sigma n+1}f_{\alpha \sigma n} ^{\phantom{\dagger}} \right) ~.
	\end{equation} 
The dot then couples to the end of the Wilson chains, at site $n=0$. 
The logarithmic discretization is parametrized by $\Lambda$, with the continuum description being recovered as $\Lambda \to 1$ (in this work we use a standard choice of $\Lambda=2.5$). The key feature of the Wilson chain is the behaviour of the hopping parameters $t_n$. For the metallic flat band, $t_n\sim \Lambda^{-n/2}$ at large $n$ \cite{Wilson1975}. This exponential energy-scale separation down the chain justifies a numerical scheme based on iterative diagonalization and truncation: starting from the dot, successive sites of the Wilson chain are coupled into the system and this intermediate Hamiltonian is diagonalized. Only the lowest $M_K$ eigenstates at iteration $n$ are used to construct the Hamiltonian at iteration $n+1$. High-energy states discarded at a given iteration do not affect the retained low-energy states at later iterations because of the ever-decreasing couplings $t_n$. This constitutes an RG procedure since the physics of the system at successively lower energy scales is revealed as more Wilson orbitals are added. The computational complexity is \textit{constant} as new Wilson orbitals are added (rather than exponentially growing) because the same number $M_K$ of states is kept at each step. Importantly, it was shown in Ref.~\cite{bulla1997anderson} that although the detailed structure of the Wilson chain coefficients are modified in the pseudogap DoS case, the energy scale separation down the chain is maintained, and hence the NRG can still be used in this case. We use the exact graphene DoS in this work rather than a pure pseudogap, and keep $M_K=6000$ states at each iteration. Dynamical quantities are calculated using the full-density-matrix NRG approach \cite{Weichselbaum2007,peters2006numerical}, established on the complete Anders-Schiller basis \cite{AndersSchiller2005}.

\section{Results}
In this section, initially we present in Sec.\ref{Sec:Overview} a summary of our results by means of the phase-diagram of the system couplings and the fixed point  analysis on the $3D$ version of RG-flow diagram, then we detail our findings with respect to the thermodynamics in Sec.~\ref{Sec:Thermo} and dynamical quantities in Sec.~\ref{Sec:TmatGc} according to the system degree of freedom.\\
In the following we confine our attention to the charge-degeneracy point $\delta V_g=0$ and  we also assume $W>0$ such that the Kondo effect can happen. We also introduce the \textit{channel-asymmetry parameter} $\Delta=J_2/J_1 \equiv W_2/W_1$ and discuss the physics in the space of $J\equiv J_1$, $W\equiv W_1$ and $\Delta$. Note that $\Delta=0$ corresponds to the situation in which channel $\alpha=2$ is decoupled on the level of the bare model, while for $0 < \Delta < 1$ both channels are coupled to the dot but channel $1$ couples more strongly.  $\Delta=1$ describes the frustrated two-channel situation. We confine our attention to regime $0\le \Delta \le 1$ but it should be noted that $\Delta>1$ simply corresponds to stronger coupling for channel $2$, and results follow from the duality $1 \leftrightarrow 2$ and $\Delta \leftrightarrow 1/\Delta$. A last observation: the reason behind the same numerical value $\Delta$ for the couplings ratio is because we want to deform in equal proportion each dimension of the phase-diagram. In principle, the parameters are all independent but in practice these are \textit{not} and any variation has to be controlled at equal footing in any axis. \\
Before we start presenting our findings, we give below a list of abbreviations we use in the next parts:  FP i.e. fixed-point, LM i.e. local moment, (F)ALM i.e. (frustrated) asymmetric local moment, (F)ASC i.e. (frustrated) asymmetric strong coupling, (F)SSC i.e. (frustrated) symmetric strong coupling and (F)ACR i.e. (frustrated) asymmetric critical.

\subsection{Overview: phase-diagram and fixed-point analysis}\label{Sec:Overview}
\begin{figure}[H]
	\centering
	\includegraphics[width=0.85\linewidth]{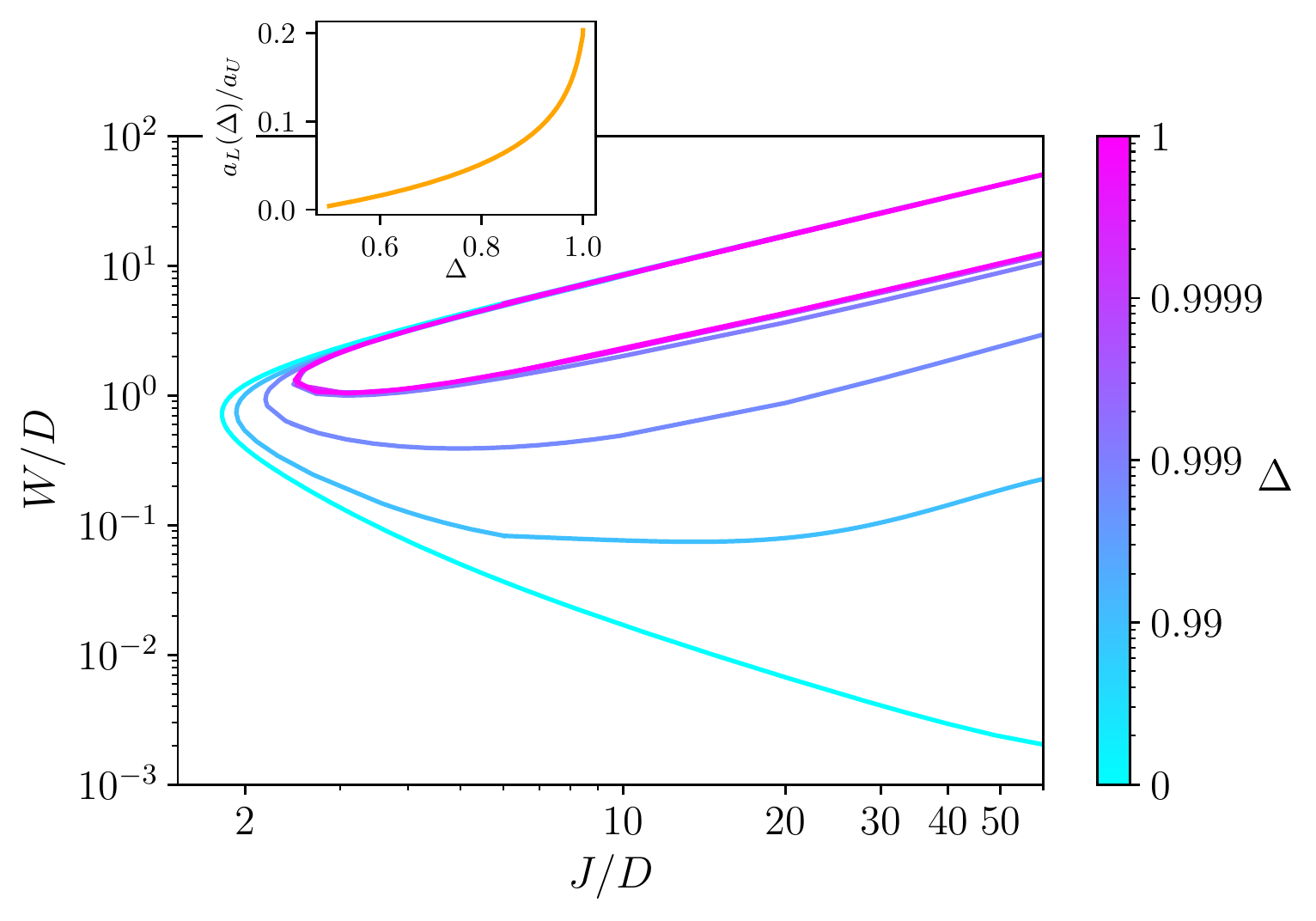}
	\caption[Phase-diagram for $0\leq\Delta\leq1$ from NRG.]{Full NRG phase diagram for $0\leq\Delta\leq1$. The enclosed region in each case is the Kondo-screened ASC phase (frustrated FASC for $\Delta=1$); the exterior region is the unscreened LM phase. Inset shows asymptotic behaviour of phase boundaries, see text.\label{Fig:PD}}
\end{figure}

\begin{figure}[H]
	\centering
	\includegraphics[width=0.9\linewidth]{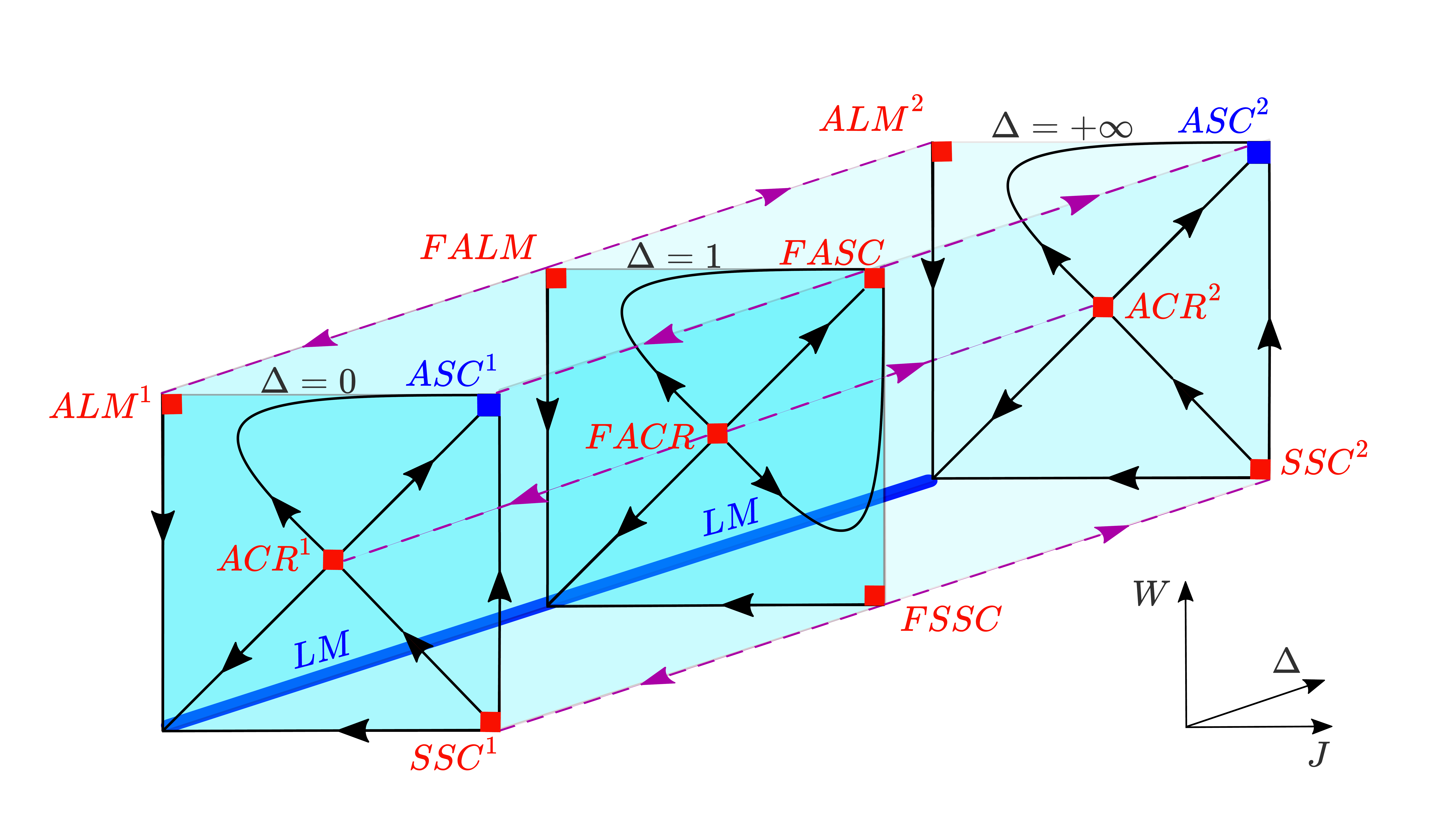}
	\caption[Schematic of $3D$ RG-flow diagram for $0\leq\Delta\leq+\infty$]{$3D$ RG-flow diagram for the pseudogap 2CCK model Eq.~\ref{Eq:9HK}, in the space of effective  exchange coupling $J$, potential scattering $W$ and channel asymmetry $\Delta$. Stable (unstable) FPs are denoted as blue (red) squares. Couplings flow from $\Delta=1$ to $\Delta=0,+\infty$ are indicated with magenta lines. $\Delta=0$ ($+\infty$) is the pure single-channel for $\alpha=1$ ($\alpha=2$) strongest channel and $\Delta=1$ is frustrated two-channel model. For an explanation of the FPs, see text.\label{Fig:3DRG}}
\end{figure}

In Fig.\ref{Fig:PD} we show the \textit{quantitative} phase-diagram in the $(J,W)$ plane for different $\Delta$, with the exact phase boundaries obtained with NRG. The phase in each region depends on: $(i)$ particle-hole symmetry where symmetric (asymmetric) FP indicates zero (finite) $W$, $(ii)$ channel asymmetry parameter $\Delta$ varying from asymmetric channel configuration at $0\leq \Delta<1$ to symmetric at $\Delta=1$ with corresponding frustrated FPs. The region outside the curves is the local moment phase characterised by the dot spin uncoupled from both leads whereas the region inside each curve corresponds to asymmetric strong coupling phase where the Kondo effect develops as the dot spin is quenched by the surrounding conduction electrons. On the curve, the value pairs $(J_{CR},W_{CR})$ identify the critical phase boundary. \\
\noindent{From} the non-perturbative NRG results displayed in Fig.\ref{Fig:PD}, we deduce the $3D$ version of the RG-flow  diagram in the space of $(J,W,\Delta)$, see schematic in Fig.\ref{Fig:3DRG}, and it gives a good overview of the physics of Eq.~\ref{Eq:9HK}. We discuss first the phase-diagram and then go through the fixed-point analysis.\\

\noindent{We} start discussing the most significant aspect in Fig.\ref{Fig:PD}: in the phase-diagram we find a finite ASC region at any $\Delta$ and this area does \textit{not} disappear even at the \textit{asymptotes}, as it follows from our analysis. The upper branch asymptote equation is $W^{U}_{CR}(J,\Delta)=a_{U}(\Delta)J_{CR}$ fitting and for all $\Delta$ we find  $a_{U}(\Delta) \equiv a_{U} \approx 1.0$ such that $W^{U}_{CR}~\propto~ J_{CR}$ has linear behaviour. In the lower branch asymptote at $\Delta=0$, considering the effective exchange coupling expression we would expect the curve to decay as $W^{L}_{CR}~ \propto~ J_{CR}^{-1}$. However, our numerical analysis breaks down at $10^2 \lesssim J$.
This numerical limitation prevents from resolving completely the asymptotic behaviour at $\Delta=0$ and we obtain from the numerical $W^{L}_{CR} ~\propto~ J_{CR}^{-1.35}$. On the contrary, at $0<\Delta\leq 1$ we find no numerical restrictions and the lower branch asymptote equation is $W^{L}_{CR}(J,\Delta)=a_{L}(\Delta)J_{CR}$. We find $a_{L}$ increases with increasing $\Delta$, as shown in the inset in Fig.~\ref{Fig:PD}. However, $a_{L}(\Delta)/a_{U}< 1$ for all $\Delta$ and the ratio reaches its maximum $\approx 0.2$ as $\Delta \to 1$. This asymptotically proves the upper and lower phase boundaries never cross, and the ASC phase persists out to infinite $J$ and $W$. \\
We analyse now the system at $\Delta=0$, that is the turquoise line in the phase-diagram in Fig.\ref{Fig:PD} and the front plane in the $3D$ RG-flow in Fig.\ref{Fig:3DRG}. The basic physics of this single-channel Kondo model are well-known from previous studies of the $r=1$ pseudogap Anderson and Kondo models \cite{bulla1997anderson,gonzalez1998renormalization,logan2000local,fritz2004phase,vojta2004upper,vojta2010gate,fritz2013physics,Dias_Graphene2017}, although note that our graphene charge-Kondo set-up gives a spin-anisotropic model, and we use the full graphene DoS rather than a pure pseudogap.\\
At particle-hole symmetry regime, the system shows no spin screening except for arbitrary small but finite $W$ in the limit $J \rightarrow +\infty$ corresponding to symmetric strong coupling FP (SSC). In this sense, the $\Delta=0$ curve does not have a $W^{min}$. Once particle-hole symmetry is broken, we find on the curve a $J^{min}$: for Eq.~\ref{Eq:9HK}, $J_{CR}^{min}\simeq 1.81D$. For $J<J_{CR}^{min}$ no Kondo state is possible at any $W$ value. The $J^{min}$ value denotes both the minimum exchange coupling required to activate the Kondo singlet formation and a critical value at the transition between LM and ASC regions. 
Inside the ASC phase, we distinguish two phases: $(i)$ $W<W^{L}_{CR}(J^{min}_{CR})$ implies the exchange coupling is not large enough to compensate the rather small $W$, $(ii)$ $W>W^{U}_{CR}(J^{min}_{CR})$ implies $W$ is very large and suppresses the effective spin-exchange. Under RG scaling the effective exchange coupling between $f_{1\sigma0}$ and $f_{1\sigma1}$ lead orbitals
becomes very small at large $W$ in this region. Hence, in both cases the system enters in LM phase and for $J\geq J_{CR}^{min}$. The ASC phase is stable for $J\geq J_{CR}^{min}$ and within $W_{CR}^{L}(J)<W<W_{CR}^{U}(J)$. This behaviour found from the analysis of the full NRG phase boundary deserves more comments. That is recognised as \textit{re-entrant} behaviour back into the LM phase at large $W$: only suitable intermediate couplings bring the system inside the ASC phase. We are not aware of a discussion of this in the literature, but the physical arguments we describe for re-entrant LM behaviour in the single and two-channel spin-anisotropic pseudogap Kondo model are also valid for the regular spin-\textit{isotropic} single-channel pseudogap Kondo case. NRG data for the full phase boundary for the single channel spin-isotropic Kondo model confirm qualitatively the same curve we found in Fig.\ref{Fig:PD}. This re-entrant LM behaviour is physically intuitive since $J$ and $W$ work antagonistically: at very large $W$ the lead orbital $f_{1\sigma 0}$ becomes depopulated, and hence the exchange coupling to that site $J$ gets switched off. Perturbative arguments suggest that the residual coupling to the $f_{1\sigma 1}$ bath orbital is then $J_{eff}\sim t_0^2 J/W^2$, which is consistent with $W^{U}_{CR} ~\propto~ J_{CR}$. This is indeed confirmed by NRG calculations. \\
\noindent{The} main focus of this work is the situation when the coupling to the second channel is switched on, hence we have $\Delta>0$, and we find several differences form the pure single-channel model.\\
The system at $0<\Delta <1$ presents finite threshold values of the couplings to realize ASC physics - see curves from light blue to light violet in Fig.\ref{Fig:PD} and the intermediate volume connecting $\Delta=1$ towards $\Delta=0$ plane through magenta lines in the $3D$ RG-flow in Fig.\ref{Fig:3DRG}. Those are $J^{min}_{CR}(\Delta)>0$ - which increases slightly from $\simeq 1.81D$ at $\Delta=0$ up to $\simeq 2.47D$ as $\Delta \to 1$ - and $W^{min}_{CR}(\Delta)>0$ - which reaches its maximum value $\simeq D$ as $\Delta \to 1$. For $W^{min}_{CR}(J_{min}(\Delta))>W^{min}$ and $J^{min}_{CR}(\Delta)>J^{min}$ we observe again the re-entrant LM phase with stable ASC phase within $W^{L}_{CR}(J^{min}_{CR}(\Delta))<W<W^U_{CR}(J^{min}_{CR}(\Delta))$. \\
\noindent{Finally}, see magenta curve in Fig.\ref{Fig:PD} and the middle plane $\Delta=1$ in Fig.\ref{Fig:3DRG}, an analogue phase-digram topology to $0<\Delta <1$ persists also in the system at $\Delta=1$ channel frustrated regime.\\
\noindent{The} different FPs we identify from the full NRG phase-diagram can be schematically represented in the RG-flow diagram, see Fig.\ref{Fig:3DRG}. The FPs are classified according to particle-hole symmetry (asymmetry) and channel asymmetry parameter $\Delta$. We can further distinguish the FPs based on the type of Fermi liquid system. As it will be clear from the $\mathrm{T}$-matrix spectral function analysis in Sec.~\ref{Sec:TmatGc}, both LM and ASC phases are regular Fermi liquids but the ACR PF is of non-Fermi liquid type. In the RG-flow the system undergoes simultaneous perturbations in exchange coupling $J$, potential scattering $W$ and asymmetry parameter $\Delta$. The different FPs at $(J,W,\Delta)$ - which are understood as the values taken \textit{under RG} at the FPs themselves - and their trajectories are presented in the $3D$ version of RG-flow, see Fig.\ref{Fig:3DRG}. The set of FPs located on the $\Delta=0$ plane mirrors the ones at $\Delta=+\infty$ for the mentioned duality $1\leftrightarrow2$. The $\Delta=1$ plane is also characterised by additional channel frustration. We explain now the various FPs and their interconnection under RG.\\
In the $\Delta=0$ plane, that correspond to the standard single-channel Kondo model, we find the stable LM FP at $(0,0,0)$ and the stable ASC FP at $(\infty,\infty,0)$. The unstable ACR FP at $(J_{CR},W_{CR},0)$ represents the crossover between LM and ASC FPs where the flow moves to. This is a $1^{st}$-order level crossing transition meaning that the degrees of freedom at the CR phase are the addition of those at LM and ASC phases \cite{fritz2004phase,vojta2004upper}. At the unstable SSC FP located in $(\infty,0,0)$, the system can flow arbitrarily close to it before ultimately crossing over to either LM or ASC FPs. The latter ASC FP is accessible from SSC due to the lack of finite $W^{min}$. At $(0,\infty,0)$ we find the unstable asymmetric LM FP (ALM): the existence of finite $J^{min}$ prevents from direct flowing from ALM towards ASC and the ALM FP is not directly reachable under RG. However, the eigenspectrum levels in ALM phase coincide with the levels in the LM FP, hence 
from ALM the couplings flow towards LM FP.  \\
In the $\Delta=1$ plane,  symmetry dictates a channel degeneracy down to $T=0$ and therefore \textit{no channel freezing} - meaning decoupling of the less strongly coupled channel that is $\alpha=2$ in $0\leq\Delta<1$ case. In this configuration, we have only one stable FP that is the LM FP at $(0,0,1)$, in which both channels flow symmetrically to weak coupling and to particle-hole symmetry. The frustrated ASC (FASC) at $(\infty,\infty,1)$ acquires instability because of the channel degeneracy: it is a doubled version of ASC FP in $\Delta=0$ with the Kondo effect and conduction electron hole form in either channel $\alpha=1$ or $2$. The unstable frustrated ACR (FACR) at $(J_{CR},W_{CR},1)$ is \textit{tricritical} because it remains at the crossover between LM, FASC and ASC phases. It indicates again a $1^{st}$-order quantum phase transition. Due to finite $W^{min}$, from the unstable frustrated SSC FP (FSSC) at $(\infty,0,1)$ there is no crossover to FASC phase but from FSSC the coupling moves only towards LM FP. Due to finite $J^{min}$, from the unstable frustrated ALM FP (FALM) at $(0,\infty,1)$ there is again no crossover to FASC phase under RG and couplings flow only towards LM FP. \\
We note that the frustrated FPs are delicate because they sit precisely on the separatrix between RG-flow to states with dominant channel $1$ for $\Delta<1$, and flow to states with dominant channel $2$ for $\Delta>1$. Any finite perturbation in  $|\Delta-1|$ - as we discuss more in Sec.\ref{Sec:Thermo} - relieves the channel frustration and leads ultimately to channel freezing on the lowest energy scales. \\
Perturbing FPs located on the $\Delta=1$ plane such that $\Delta<1$ leads to RG-flow continuously connected towards the FPs on $\Delta=0$ plane, see magenta lines pointing from the middle plane towards the front plane in Fig.\ref{Fig:3DRG}. For all $0<\Delta<1$ - identified by the intermediate planes $J=0$, $W=0$, $J=\infty$ and $W=\infty$ - the stable FPs are all in the $\Delta=0$ plane. Analogously, starting from $\Delta=1$ plane, we find flow arrows towards the back plane $\Delta=+\infty$ due to perturbation in $\Delta>1$. Because of relevant perturbation in $\Delta$, under RG, the couplings flow starting from FPs corresponding to asymmetric channel configurations move away from $\Delta=1$ plane until ending on single-channel FPs in either $\Delta=0$ or $\Delta=\infty$ planes. In this sense, the single-channel  model is understood as limiting case of the asymmetric two-channel model with the weakest channel fully decoupled under RG-flow. Any asymmetric two-channel model represents intermediate instances of the single-channel limit because any channel asymmetry 
leads asymptotically to the decoupling of the less strongly coupled channel. This can be regarded as \textit{channel freezing} at low temperatures opposed to \textit{free channel} degree of freedom $\alpha=1,2$ found at at high temperatures. As consequence of this insight, with respect to $\Delta=1$ plane, symmetric RG-flow trajectories move either towards $\Delta=0$ if $\Delta<1$ and so $\alpha=2$ channel is decoupled or towards $\Delta=\infty$ if $\Delta>1$ and so $\alpha=1$ channel is decoupled, see Fig.~\ref{Fig:3DRG}. Hence, each frustrated FPs find its asymmetric counterpart either on $\Delta=0$ or $\Delta=\infty$ plane according to the channel strength. This smooth connection between the two planes fails for the LM FP: $\Delta$ is defined as ratio of couplings hence on $J=W=0$ axis it is undetermined. We find more appropriate to identify the $J=W=0$ axis as LM FP line in the \textit{limit} $J,W\rightarrow0$: this axis represents the LM phase at any $\Delta$.\\
To conclude, giving the importance of the $\Delta$ parameter in attributing the FPs properties, we classify the system degree of freedom as follows: at $0 \leq \Delta <1$ frozen channel degrees - since all the couplings will inevitably flow under RG towards the $\Delta=0$ FPs and at $\Delta=1$ frustrated channel degrees of freedom. We use this classification to present the thermodynamics in the next section.\\
In Table~\ref{tabFP}, we summarize the FPs discussed above in relation to Figs.~\ref{Fig:PD},\ref{Fig:3DRG} and classify them according to their physical properties. These properties are extracted from the limiting behaviour of the full thermodynamic and dynamic observables presented in Sec.\ref{Sec:Thermo} and Sec.\ref{Sec:TmatGc}, respectively. In particular, the excess charge for the strongest channel $\alpha=1$ can be directly computed using NRG via the formula in Eq.\ref{Eq:9exChar}. These are discussed below.
\begin{center}
\begin{table} [H]
	\begin{tabular}{ |c||c|c|c|c|c|c| } 
		\hline
		asymmetry & fixed-point	& $S_{D}(T=0)$	& $T\chi_{D}(T=0)$ & $N_1$ & $t_{1}(\omega,T\rightarrow0)$ &$\mathcal{G}^{C}(\omega,T\rightarrow0)$\\
		\hline\hline
			$\forall \Delta$	& LM line & $\ln2$ & $1/4$ & $0$ & $|\omega|$ & $\omega^{2}$\\
		\hline
		$0 \leq \Delta<1$	& ASC	& $0$	& $0$ & $-1$  & $|\omega|$& $ \omega^{2}$\\
		$0 \leq \Delta <1$	& ACR$^{\star}$ & $\ln3$	& $1/6$	& $-1/3$ & $1/\omega \ln^{2}(\lambda_{CR}/\omega)$ & $0$\\
		\hline
		$\Delta=1$		&  FASC$^{\star}$	& $\ln2$	& $0$ & $-1/2$ & $|\omega|$ & $\omega^{2}$\\
		$\Delta=1$		&  FACR$^{\star}$	& $\ln4$	& $1/8$ & $-1/4$ & $1/\omega \ln^{2}(\lambda_{CR}/\omega)$ & $const$\\
		\hline
	\end{tabular} ~.
\caption[Classification fixed-point in two-channel charge-Kondo graphene model]{Classification of FPs according to their physical observables, with superscript $\star$ denoting unstable FPs.\label{tabFP}}
\end{table}
\end{center}

\subsection{Thermodynamics} \label{Sec:Thermo}
The temperature dependence of the dot contribution to entropy $S_D(T)$ and magnetic susceptibility $T\chi_D(T)$ are obtained from NRG \cite{Bulla2008} and presented in Fig.~\ref{Fig:STChi} for different channel asymmetries $\Delta$. We focus on the behaviour near the critical points by fixing $W$ and tuning $J$ across the transition. From this, information on the fixed points is deduced.

\subsubsection{Frozen channel degree of freedom: $0\leq \Delta<1$}
\begin{figure}[H]
	\centering
	\hspace*{-1.7cm}\includegraphics[width=1.15\linewidth]{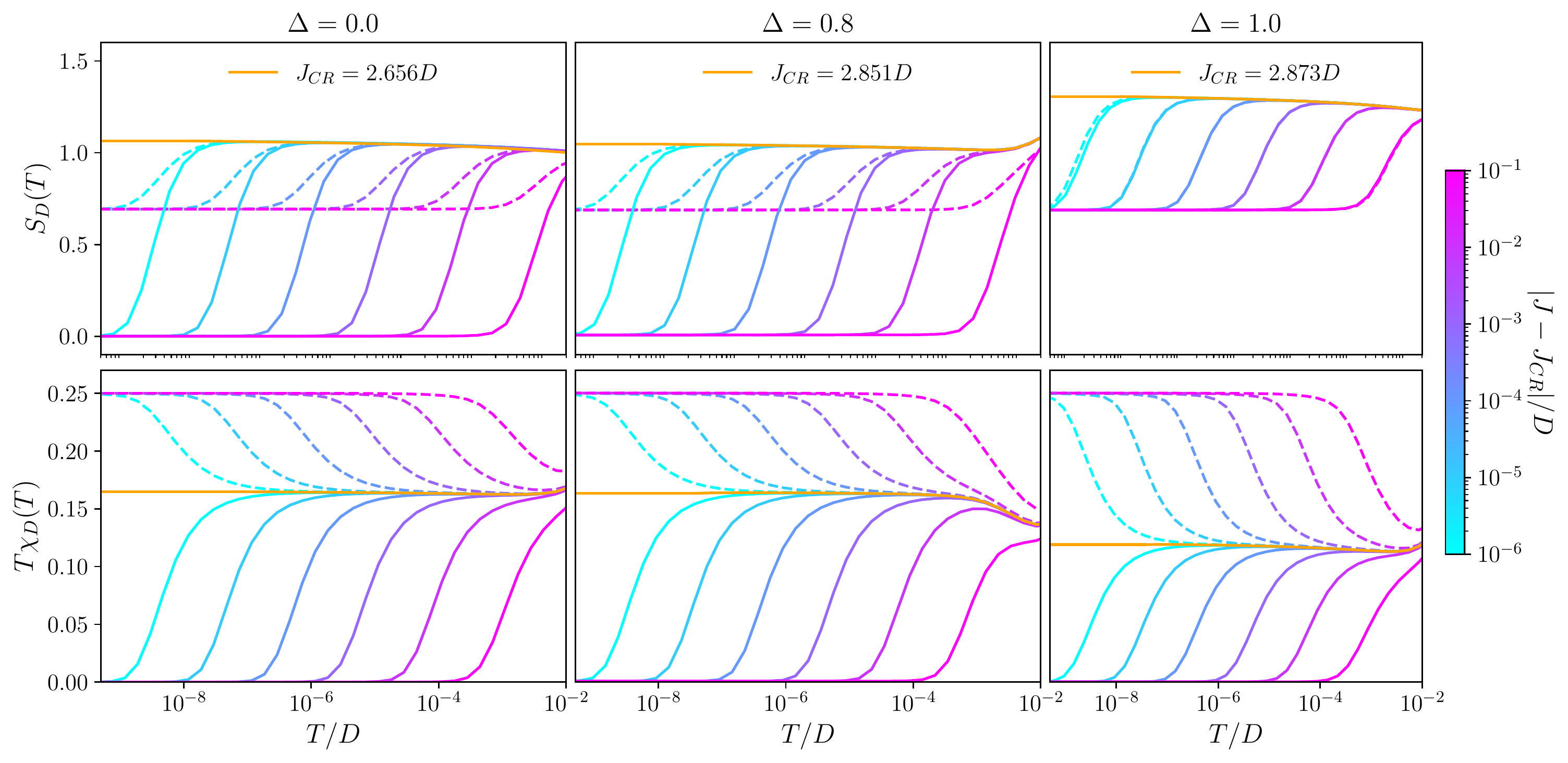}
	\caption[Thermodynamic quantities for the graphene-dot in two-channel charge-Kondo model obtained by NRG]{Dot contribution to thermodynamic quantities for the graphene 2CCK model obtained by NRG. \textit{Top row:} entropy $S_D(T)$; \emph{bottom row:} magnetic susceptibility $T\chi_D(T)$. Left, middle, and right columns correspond to $\Delta=0$ (pure 1CCK), $\Delta=0.8$ (asymmetric 2CCK), and $\Delta=1$ (symmetric 2CCK) respectively. Shown for fixed $W=2D$, varying $J$ across the QPT according to the colour scale, with solid lines for $J>J_{CR}$ in the ASC (FASC) phase, and dashed lines for $J<J_{CR}$ in the LM phase. Orange lines show the behaviour at the ACR (FACR) critical point. \label{Fig:STChi}}
\end{figure}

We first consider the regime $0\leq \Delta<1$ in the left and middle columns of Fig.~\ref{Fig:STChi}. Solid lines show the behaviour in the ASC phase for $J>J_{CR}$, dashed lines for the LM phase with $J<J_{CR}$, and orange line at the critical point $J=J_{CR}$. \\
In the LM phase, $S_D=\ln(2)$ and $T\chi_D=\tfrac{1}{4}$ at $T=0$ in all cases, characteristic of the asymptotically-free spin-$\tfrac{1}{2}$ dot pseudospin - compare also with first row in Table~\ref{tabFP}. The excess conduction electron charge - not shown in the plot - is zero in both channels, suggesting an emergent particle-hole symmetry. This is confirmed by analysis of the NRG many-particle level spectrum (finite size spectrum) at the LM FP, which is \textit{identical} to that of the free leads. The dot remains unscreened in LM because of the depleted conduction electron DoS at low energies in graphene \cite{fritz2013physics}. The FP Hamiltonian in the LM phase is therefore given by,
\begin{eqnarray}\label{eq:9HLM}
\boxed{	\hat{H}_{LM}=\hat{H}^{2CCK} } \quad {\rm with} \quad J_1=J_2=W_1=W_2=\delta V_g=0 ~.
\end{eqnarray}
In the ASC phase at $T=0$, the spin degree of freedom is completely quenched by the conduction electrons. As consequence, the system has quenched dot entropy $S_D=\ln1=0$ and $T\chi_D=0$ because the dot spin is in a singlet state.
These thermodynamical values apply in all ASC cases, see Table~\ref{tabFP}, characteristic of Kondo singlet formation. However, the conduction electron excess charge is $N_1=-1$ and $N_2=0$ for $W>0$, implying hole formation in the more strongly coupled channel $\alpha=1$ with $W_1 \to \infty$ under RG, while the less strongly coupled channel $\alpha=2$ recovers an effective low-energy particle-hole symmetry, $W_2 \to 0$ under RG. This suggests the screening mechanism in the generic two-channel case which we explain now with the aid of the Wilson chain representation given in Eq.~\ref{Eq:9HBathWC} and the dimensionless coupling $j_{\alpha}=J_{eff}\rho_{\alpha n}$ defined on $\hat{H}_{leads}^{WC}$.\\ 
The two-channel model in the NRG scheme corresponds to the graphene-dot coupled to two Wilson chains. In general, the potential scattering applies only to the $0^{th}$-site - corresponding to $f_{\alpha\sigma0}$ Wilson orbital - in which it creates a hole at $T_{hole}$ temperature. This is because its effect corresponds to an unoccupied orbital in the physical system. Once we introduce channel asymmetry, only the channel with the largest $W$ forms the hole on the $0^{th}$-site. If we take $W_1 > W_2$ as it occurs in the $\Delta<1$ case, the hole develops only on the $\alpha=1$ channel at $f_{1\sigma0}$ Wilson orbital  with pseudogap DoS $\rho_{10}(\omega)=\rho |\omega/D|$ at $n=0$ site as given in Eq.~\ref{Eq:DoSPse} with $r=1$. Because of $f_{1\sigma 0}$ Wilson orbital becomes depopulated, thereby it generates an effective coupling between the dot pseudospin and the Wilson $f_{1\sigma 1}$ orbital, $J_{eff}\sim t_0^2 J_1/W_1^2$. Then, the $0^{th}$-site does not contribute to the Kondo singlet because it has no electronic levels available to screen the dot spin and the $0^{th}$-site is like removed. The remaining shortened chain modifies the pseudogap DoS at $1^{st}$-site into a power-law $\rho_{11}(\omega)=\rho |\omega/D|^{-1}$. The $1^{st}$-site now has enough electronic states to quench the dot spin and the Kondo singlet is formed at temperature $T_K$. Despite of $J_{eff}$ coupling on $\alpha=1$ channel being smaller than $J_{2}$ under RG-flow, the $\rho_{11}(\omega)$ is strongly enhanced such that $j_{1}=J_{eff}\rho_{11}(\omega)$ is larger than $j_{2}=J_{2}\rho_{20}(\omega)$ at the Fermi level due to hole-singlet mechanism. As consequence, the weakest $\alpha=2$ channel decouples from the graphene-dot under RG such that the resulting system is effectively single-channel model. This is explained because no hole forms in the weakly coupled channel, and so $j_2=\rho_{20} J_2$ remains small due to the depleted bare DoS in channel $2$, and flows under RG to weak coupling. This argument shows that the Kondo singlet must form in the \textit{same} channel in which the hole forms and this is confirmed by analysis of the NRG level spectrum. Furthermore, hole and Kondo singlet formation on the same chain side is a \textit{simultaneous} process hence it manifests at the same energy scale $T_{hole} \equiv T_K$. Throughout the ASC phase $T_K\sim D$ since $J_{CR}^{min}>0$. Thus, this Kondo scale for the screening process is strongly enhanced because of the diverging effective DoS \cite{mitchell2013quantum}. \\
By means of this \textit{two-channel screening} or \textit{hole-singlet} mechanism, the $0< \Delta<1$ model represents the system with intermediate hole-singlet stage and partial decoupling of the weakest channel: once these processes are completed the system reaches the limiting case of the $\Delta=0$ single-channel model.\\
In general we therefore have two distinct ASC phases and two distinct ASC fixed points, depending on whether $\Delta<1$ or $\Delta>1$. For $\Delta<1$, the hole-singlet complex forms in channel $\alpha=1$ and channel $\alpha=2$ decouples (FP denoted ASC$^1$), while for $\Delta>1$ (ASC$^2$) it is the other way around.  
The ASC$^{\alpha}$ FP Hamiltonian obtained when channel $\alpha$ is more strongly coupled reads,
\begin{eqnarray}\label{Eq:9HASC}
	\boxed{
	\hat{H}^{\alpha}_{ASC} = \hat{H}_{leads} + J_{\alpha}\left (\hat{S}_D^+ f_{\alpha \downarrow 1}^{\dagger}f_{\alpha \uparrow 1}^{\phantom{\dagger}} + \hat{S}_D^- f_{\alpha \uparrow 1}^{\dagger}f_{\alpha \downarrow 1}^{\phantom{\dagger}}\right ) + W_{\alpha}\sum_{\sigma}f_{\alpha \sigma 0}^{\dagger}f_{\alpha \sigma 0}^{\phantom{\dagger}} } ~{\rm with}~ J_{\alpha},W_{\alpha}\to \infty~.
\end{eqnarray}
We now consider the situation in the close vicinity of the QPT, by fixing $W$ and tuning $J$. At the critical point itself - indicated in the orange line for $J=J_{CR}$ - we find $S_D=\ln(3)$ and $T\chi_D=\tfrac{1}{6}$ at $T=0$, compare with the third row in Table~\ref{tabFP}. This suggests a level-crossing transition in which the critical FP ACR comprises uncoupled sectors corresponding to LM and ASC. This gives an overall dot ground state degeneracy of $2+1=3$ states, that is $2$ for LM and $1$ for ASC, consistent with the $\ln(3)$ entropy, and a magnetic susceptibility $(\tfrac{1}{4}+\tfrac{1}{4}+0)/3=\tfrac{1}{6}$ corresponding to the average of $(S^z)^2$ for these three degenerate states. This is further supported by the conduction electron excess charge $N_1=-\tfrac{1}{3}$, since a single hole appears in channel $1$ for only one of the three degenerate ground states and $N_2=0$ for the decoupled free channel $2$.\\
The first-order transition is also consistent with the \textit{linear} emergent crossover scale $T^{\star} \sim |J-J_{CR}|$, describing the flow from ACR to either LM or ASC due to a small detuning perturbation. This scale is evident in Fig.~\ref{Fig:STChi} by the sequence of lines for different  $(J-J_{CR})$. Indeed, one can cross the QPT by fixing $J$ and tuning $W$ through $W_{CR}$, which also gives a linear scale $T^{\star}$. We also checked this behaviour along the entire critical phase boundary lines $(J_{CR},W_{CR})$ in Figs.~\ref{Fig:3DRG},\ref{Fig:PD} for different $\Delta$. We find,
\begin{eqnarray}\label{eq:9tstar}
	T^{\star} = b|J-J_{CR}| + c|W-W_{CR}| \;,
\end{eqnarray}
where $b\equiv b(J_{CR},W_{CR},\Delta)$ and $c\equiv c(J_{CR},W_{CR},\Delta)$. This implies a \textit{universal scaling} in terms of a single reduced parameter $T/T^{\star}$, independent of the combination of bare perturbations that act.
The FP Hamiltonian describing the critical point is,
	\begin{equation}\label{Eq:9HACR} \boxed{
		\hat{H}^{\alpha}_{ACR} = \left(\frac{1+\hat{\tau}^{z}}{2}\right)\hat{H}_{LM} + \left(\frac{1-\hat{\tau}^{z}}{2}\right)\hat{H}^{\alpha}_{ASC} }~.
	\end{equation}
where the $\alpha$ label denotes the more strongly coupled lead with which the dot forms the Kondo effect in ASC, and $\hat{\tau}^z$ is a Pauli-$z$ operator. In Eq.~\ref{Eq:9HACR}, $\tau^z=+1$ gives the doubly-degenerate LM ground state ($\hat{H}_{LM}$ given in Eq.~\ref{eq:9HLM}) while $\tau^z=-1$ gives the ASC ground state ($\hat{H}_{ASC}$ given in Eq.~\ref{Eq:9HASC}). At the ACR FP, the three many-body ground states are degenerate and uncoupled - $\hat{\tau}^z$ has no dynamics. Since ACR is unstable, we also consider the leading RG \textit{relevant} perturbations to the FP Hamiltonian, $\delta \hat{H}^{\alpha}_{ACR} \sim + T^{\star}\hat{\tau}^z$, which has the effect of biasing towards either the LM or ASC ground states on the scale of $T^{\star}$. In such a way, the ACR ground-state degeneracy is lifted.\\
We briefly comment about the SSC and ALM FPs Hamiltonian: both present instability due to relevant correction in the RG-flow of the same type of the ACR phase, namely $\delta\hat{H}_{SSC},\delta\hat{H}_{ALM}\sim+T^{\star}\hat{\tau}^{z}$.\\
We conclude the analysis for system presenting frozen channel degree with two comments. As first: the qualitative behaviour of the thermodynamics shown in Fig.~\ref{Fig:STChi} for $\Delta=0$ and $\Delta=0.8$ is similar, but it should be noted that both channels are involved for $\Delta \ne 0$ at finite temperatures. However, the less strongly coupled channel decouples asymptotically because finite $0<\Delta<1$ flows to $\Delta=0$ under RG on reducing the temperature or energy scale - see Fig.~\ref{Fig:3DRG}. As second, we report that from the eigenspectrum analysis for both FPs Hamiltonian $\hat{H}^{\alpha}_{ASC}$ and free Wilson chain, we observe a matching in the energy levels. This confirms an even number of sites, i.e. $0^{th}$ and $1^{st}$, have to be decoupled from the chain after the hole-singlet formation. This fact is verified also from the constant phase-shift computed in NRG.

\subsubsection{Frustrated channel degree of freedom: $\Delta=1$}
We turn now to the frustrated case $\Delta=1$, with pristine channel symmetry, see Fig.~\ref{Fig:STChi}, right column. Although the $T=0$ entropy is $S_D=\ln(2)$ everywhere except on the phase boundary in the top right panel of Fig.~\ref{Fig:STChi}, the origin of the ground state degeneracy is different in the two phases separated by it. \\
In the LM phase realized for $J<J_{CR}$, no reduction of spin degree of freedom occurs in the symmetric two-channel model and the fixed-point Hamiltonian is again given by Eq.~\ref{eq:9HLM}. Thus, we again have a free dot pseudospin decoupled from two symmetric baths of free conduction electrons; the $\ln(2)$ entropy here derives from the free dot pseudospin-$\tfrac{1}{2}$ degree of freedom. This is confirmed by the magnetic susceptibility in this phase, which reaches $T\chi_D=\tfrac{1}{4}$, see dashed lines in the bottom right panel of Fig.~\ref{Fig:STChi}.\\ 
The other phase realized for $J>J_{CR}$ is described by the FASC FP: due to the channel symmetry, the ASC state can form in either channel $\alpha=1,2$.  The $\ln(2)$ entropy in this case derives from the free channel degree of freedom \cite{schneider2011two}, which embodies the choice of forming the hole-singlet complex of ASC with either of the two channels. Hence, this is a \textit{residual} entropy arising from the frustrated choice of which channel to form the Kondo effect in and it opposes to the results in the channel asymmetric case.
The finite entropy value is similarly reflected in the $T=0$ value of $T\chi_D=0$ in the FASC phase, see solid lines in the bottom right panel of Fig.~\ref{Fig:STChi} and fourth row in Table~\ref{tabFP}, since the dot pseudospin is Kondo screened in both of the degenerate ground states. Furthermore, we find that the average conduction electron excess charge in FASC is $N_{\alpha}=\tfrac{1}{2}$ for both channels - that is, a single hole forms, with equal probability to be in either channel $1$ or $2$.\\
A Kondo strong coupling state involving both channels simultaneously is not stable. To see this, we consider two holes forming symmetrically in the $f_{1\sigma 0}$ and $f_{2\sigma 0}$ Wilson $n=0$ orbitals with $W_{1}=W_{2}~,~J_{1}=J_{2}$. We have then a channel symmetric version of the hole-singlet mechanism we describe in ASC phase for $0\leq\Delta<1$. However, the dot entropy is not quenched in this case, since the ground state of the complex is a spin-doublet. At the end of the Kondo singlet formation, this effective doublet state couples to the Wilson $n=2$ orbitals $f_{1\sigma 2}$ and $f_{2\sigma 2}$. But since the DoS of these sites is again $\rho_{\alpha 2}(\omega)=\rho |\omega/D|$ the effective local moment cannot be screened and the system flows to the LM FP. The dot pseudospin can only be screened by an \textit{asymmetric} ASC state. Channel symmetry is restored by having two such \textit{degenerate} states, one in each channel. To sum up: the resulting Kondo singlet derives from the superposition of the singlet formation only on $\alpha=1$ channel and only $\alpha=2$ channel due to channel degeneracy. The final system is in a \textit{frustrated channel} state. In each of the two overlapping state, there are residual degrees of freedom from the unscreened channel i.e. (pseudospin) free channel degrees from the leads which are added to the spin-$1/2$ degrees from the screened channel.\\
The FASC FP Hamiltonian comprises a combination of $\hat{H}_{ASC}^1$ and $\hat{H}_{ASC}^2$ from Eq.~\ref{Eq:9HASC}, controlled by an emergent channel degree of freedom $\hat{\alpha}$,
	\begin{equation}\label{Eq:9HFASC} \boxed{
		\hat{H}_{FASC} = \left(\frac{1+\hat{\alpha}^{z}}{2}\right)\hat{H}^{1}_{ASC}+\left(\frac{1-\hat{\alpha}^{z}}{2}\right)\hat{H}^{2}_{ASC} }~.
	\end{equation}
Here $\hat{\alpha}^z$ is a Pauli-$z$ operator that selects ASC$^1$ when $\alpha^z=+1$ and ASC$^2$ when $\alpha^z=-1$ - since no mixed channel terms can be defined through $\hat{\alpha}^z$. Restricting to the symmetric $\Delta=1$ plane, FASC is stable. However, there is an instability with respect to breaking channel degeneracy - not shown in the plot, since then either ASC$^1$ or ASC$^2$ will be selected on the lowest energy scales. A finite perturbation $|\Delta-1|$ generates a flow from FASC to ASC$^1$ or ASC$^2$; from NRG we find that this QPT is also first-order. The low-energy scale determining the emergent crossover, namely
\begin{equation}
T_{\Delta} \sim |\Delta-\Delta_{CR}|\equiv |\Delta-1|~,
	\end{equation}
This can be captured in the effective model by including the leading RG \textit{relevant} perturbation to the FASC FP, $\delta \hat{H}_{FASC} \sim+ T_{\Delta} \hat{\alpha}^z$.	\\
Finally, we consider the quantum critical point in the $\Delta=1$ plane between LM and FASC. Here we find a level crossing first-order transition, with entropy $S_D=\ln(4)$ and magnetic susceptibility $T\chi_D=\tfrac{1}{8}$ at the FACR FP \cite{schneider2011two} and see last row in Table~\ref{tabFP}, which derives from the composition of uncoupled LM and FASC sectors. We have two spin-$\tfrac{1}{2}$ states from the LM degenerate with two spin-singlet states with a (pseudospin) free channel degree of freedom in FASC. The excess conduction electron charge is therefore $N_{\alpha}=-\tfrac{1}{4}$ per channel. We describe the FACR FP with the Hamiltonian,
	\begin{equation}\label{Eq:9HFACR} \boxed{
		\hat{H}_{FACR} = \left(\frac{1+\hat{\tau}^{z}}{2}\right)\hat{H}_{LM}+\left(\frac{1-\hat{\tau}^{z}}{2}\right)\hat{H}_{FASC} }~,
	\end{equation}
where $\hat{H}_{LM}$ is given in Eq.~\ref{eq:9HLM} and $\hat{H}_{FASC}$ in Eq.~\ref{Eq:9HFASC}, and we have introduced the operator $\hat{\tau}^z$ to distinguish the sectors, similar to Eq.~\ref{Eq:9HACR}. As with ACR, the FP is destabilized by RG \textit{relevant} detuning perturbations that favour either LM of FASC ground-state, which collectively generate the scale $T^{\star}$ given in Eq.~\ref{eq:9tstar}. This leads to an FP correction $\delta \hat{H}_{FACR}^{\star}\sim +T^{\star} \hat{\tau}^z$. This is shown by the sequence of lines in the right column of Fig.~\ref{Fig:STChi}. However, FACR is also destabilized by relieving the channel frustration through the perturbation $|\Delta-1|$ which generates the scale $T_{\Delta}$, since FACR contains an FASC sector with this instability. Therefore FACR has a second RG \textit{relevant} correction $\delta \hat{H}_{FACR}^{\Delta}\sim + T_{\Delta} \hat{\alpha}^z$. FACR is in this sense \textit{tricritical} since it sits between LM, FASC and ASC.

\subsection{Dynamics and transport}\label{Sec:TmatGc}
\begin{figure}[H]
	\centering
	\includegraphics[width=0.95\linewidth]{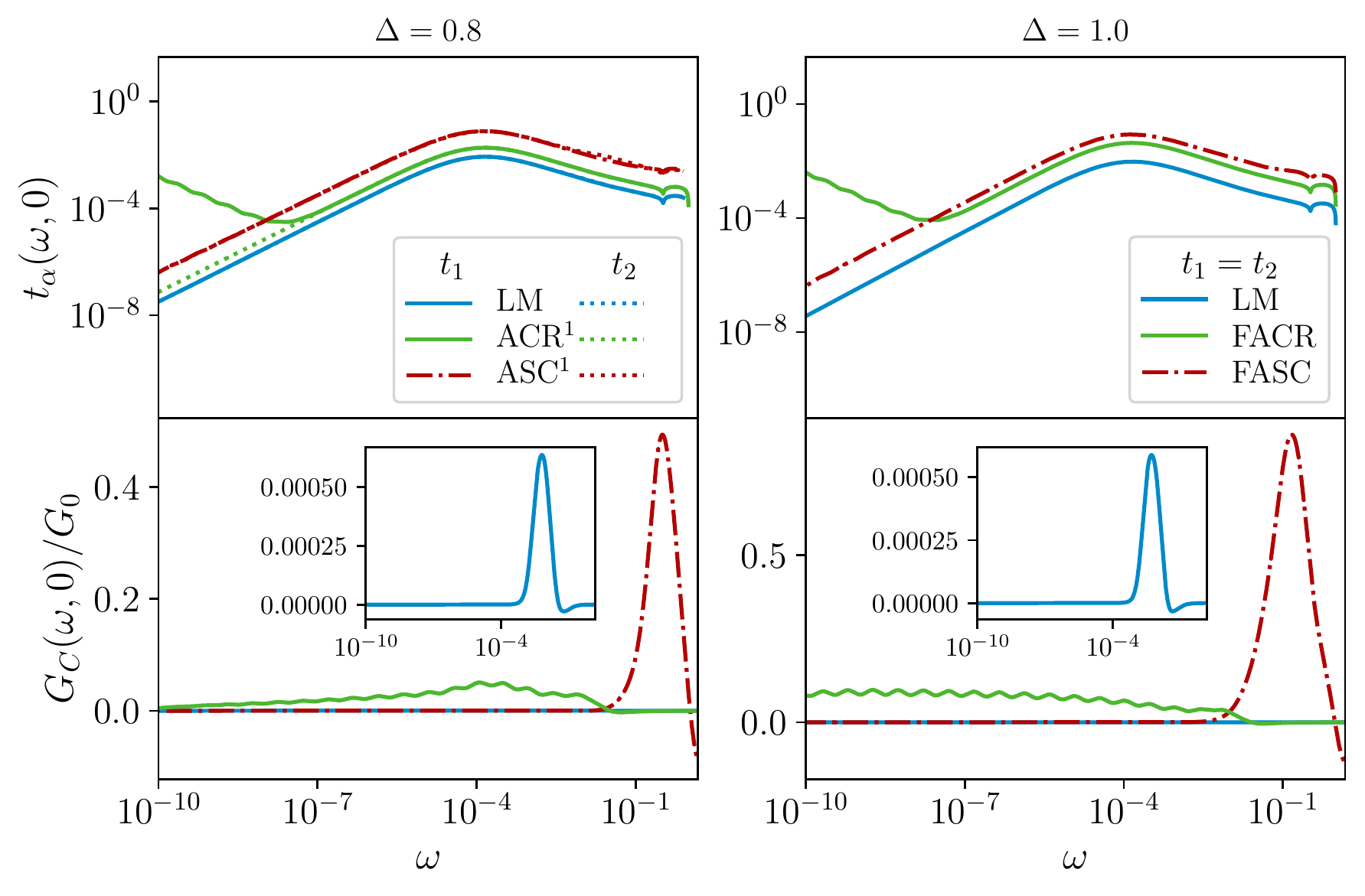}
	\caption[NRG results for dynamics and transport in the graphene two-channel charge-Kondo model at $T=0$]{NRG results for dynamics and transport in the graphene 2CCK model at $T=0$. The legends refers to both top and bottom panels. \textit{Top row:} Channel-resolved spectrum of the $\mathrm{T}$-matrix $t_{\alpha}(\omega,0)$. \textit{Bottom row:} Linear response ac electrical conductance $G_C(\omega,0)$. \textit{Left}: channel asymmetry $\Delta=0.8$; \textit{Right} frustrated case $\Delta=1$. Model parameters: $J/D=10$ with $W/D=1$ (LM); $12$ (ASC); $12$ (FASC); $\simeq 8.605$ (ACR); $\simeq 8.572$ (FACR). \label{Fig:TGc}}
\end{figure}

The FPs analysis, discussed in Sec.\ref{Sec:Overview}, includes the FPs coordinates on the $3D$ RG-flow in Figure~\ref{Fig:3DRG} by means of the parameters $(J,W,\Delta)$ and how those points are connected from $\Delta=1$ to $\Delta=0$ plane and to $\Delta=\infty$ as well under channel symmetry breaking perturbations. In the fixed-point Hamiltonian expressions, see Sec.\ref{Sec:Thermo}, we find also the FPs stability property with related relevant perturbation parameters and crossover energy scales. We now aim to complete the FPs characterization by analysing dynamical properties. We can also differentiate between Fermi liquid and non-Fermi liquid states by examining the $\mathrm{T}$-matrix spectral function given in Eq.~\ref{Eq:9Tmatspec} via NRG in the $\omega\to 0$ limit. \\
In this final section of the results, we now discuss the low-temperature behaviour of the scattering  $\mathrm{T}$-matrix and linear response $ac$-electrical conductance in the graphene 2CCK device - see Fig.~\ref{Fig:TGc}. We first consider the $T=0$ spectrum of the $\mathrm{T}$-matrix as a function of energy in the top row of Fig.~\ref{Fig:TGc}, for the channel asymmetric case $\Delta=0.8$ on the left and frustrated case $\Delta=1$ on the right. In all cases we identify an emergent low-energy scale $\lambda$ (which is $\approx 10^{-4}D$ for the parameters chosen) which characterizes the RG-flow through a crossover behaviour in the pseudogap dynamics \cite{fritz2004phase,schneider2011two}. \\
\noindent{Deep} in the LM phase, see blue solid and dotted lines in Fig.~\ref{Fig:TGc}, the bare potential scattering $W$ modifies the bare conduction electron pseudogap DoS of graphene $\rho_{\alpha0}(\omega)\sim |\omega|$, to give an effective DoS $\rho_{\alpha1}(\omega)\sim 1/|\omega|$ - up to logarithmic corrections - that corresponds to $\pi/2$ phase shift. This produces leading behaviour in the $\mathrm{T}$-matrix $t_{\alpha}(\omega,0) \sim 1/|\omega|$, as seen in Fig.~\ref{Fig:TGc} for $|\omega| \gg \lambda$. However, under RG $W \to 0$ in the LM phase; this flow is controlled by the scale $\lambda$. Therefore, on the scale of $\lambda$ the effective DoS returns to $\rho_{\alpha1}(\omega)\sim |\omega|$ - indicating now zero phase shift - and hence $t_{\alpha}(\omega,0) \sim |\omega|$ for $|\omega|\ll \lambda$. Since the emergent particle-hole symmetry in the LM phase occurs in both channels for any $\Delta$, compare with Table~\ref{tabFP}, we see the same behaviour for $t_1(\omega,0)$ and $t_2(\omega,0)$ for both $\Delta=0.8$ and $1$. \\
In the ASC phase for $\Delta=0.8$, the weakly coupled channel $\alpha=2$ shows the same behaviour as LM since it decouples from the dot and gains particle-hole symmetry. For the strongly coupled channel $\alpha=1$, we have the hole-singlet mechanism in which both the effective $W,J \to \infty$. Counter-intuitively we again see similar dynamical behaviour as for LM. This is because, for $|\omega| \gg \lambda$ we have a developing conduction electron hole which gives $t_{1}(\omega,0) \sim 1/|\omega|$, while for $|\omega|\ll \lambda$ the Kondo singlet forming with the $n=1$ Wilson orbital effectively removes a second site from the bath. The remaining conduction electrons experience a $\pi$ phase shift from the modified boundary, and the effective DoS is back to $\rho_{\alpha2}(\omega)\sim |\omega|$. Therefore in ASC we also have $t_1(\omega,0) \sim |\omega|$ for $|\omega|\ll \lambda$. That we have identical behaviour for $\Delta=1$ in both channels confirms that FASC is indeed a superposition of ASC$^1$ and ASC$^2$ as argued above.\\
On the lowest energy scales we have $t_{\alpha}(\omega,0)\sim|\omega|$ in both channels, at any $\Delta$, and in either phase. Given the bare DoS $\rho_{\alpha0}(\omega)\sim |\omega|$, this confirms that both phases are \textit{regular Fermi liquids} (FL) with well-defined, long-lived quasiparticles \cite{Hewson,fritz2004phase,vojta2004upper}.\\
More interesting is the behaviour at the critical point (F)ACR, since here we have both spin and charge fluctuations associated with the degenerate LM and (F)ASC ground states. A new dynamical scale is generated, $\lambda_{CR}\sim \lambda^2/D$ ($\approx 10^{-8}D$ for the chosen parameters) which characterizes the low-energy RG-flow \cite{fritz2004phase,schneider2011two}. We find from NRG that in the channel-asymmetric graphene 2CCK model e.g.~at $\Delta=0.8$ as shown in the plot, the $\mathrm{T}$-matrix of the more strongly-coupled channel $\alpha=1$ \textit{diverges} at low energies. Specifically, $t_1(\omega ,0)\sim 1/[|\omega/\lambda_{CR}|\times\ln^{2}(|\omega/\lambda_{CR}|)]$ as $|\omega|\to 0$ see solid green lines in the top panels of Fig.~\ref{Fig:TGc} \cite{vojta2001kondo,VojtaMultiChPseudoPRL2001}, indicating that ACR is an \textit{non-Fermi liquid} (NFL) FP. We remark that the nature of this non-Fermi liquid FP is \textit{not} ascribable to the standard two-channel Kondo critical fixed-point found in system with Majorana states and fractional residual impurity entropy at $T=0$. The dynamical crossover, and hence the minimum in $t_1(\omega,0)$ occurs on the scale of $|\omega|~\propto~ \lambda_{CR}$. However, the weakly coupled channel $\alpha=2$ has FL correlations $t_2(\omega ,0)\sim |\omega|$ as $|\omega|\to 0$ in dotted green lines, confirming that it decouples from the critical complex formed from the dot and channel $1$. In the frustrated case $\Delta=1$, both channels behave identically - and both exhibit the same NFL critical divergence at low energies. This again suggests that FACR comprises two copies of ACR, one in each channel. The enhanced conduction electron scattering at the critical point has implications for the conductance, as now shown.\\
In the bottom row of Fig.~\ref{Fig:TGc} we plot the $T=0$ dynamical $ac$-conductance as a function of $ac$-driving frequency $\omega$, for the same set of systems. The $dc$-conductance is obtained in the $\omega\to 0$ limit, which we consider first. In the charge-Kondo system, series transport proceeds by the following mechanism: an electron tunnels from the source lead onto the dot (say at QPC $\alpha=1$), thus flipping the dot charge pseudospin from $\Downarrow$ to $\Uparrow$. A second electron then tunnels from the dot to the drain lead (at QPC $\alpha=2$), thus flipping the dot charge pseudospin back to $\Downarrow$ and resetting the device ready for transport of another electron. A bias voltage between source and drain produces a net current flow. The amplitude for such a process depends on the conduction electron density of states $\rho(\omega)$ and the tunnelling rate at the QPCs. For graphene we have $\rho(\omega)\sim |\omega|$ at low energies, suggesting that the low-temperature $dc$ conductance should vanish, since there are not enough low-energy electrons in the graphene leads to tunnel through the nanostructure. On the other hand, the tunnelling rate gets renormalized by the interactions - the energy-dependent scattering at the QPCs is characterized by the $\mathrm{T}$-matrices discussed above. Indeed, strong renormalization of the bare QPC transmission at low temperatures due to Kondo physics was measured experimentally in the metallic leads version of the present system in Ref.~\cite{2CK-RenormalisationFlow_IftikharPierre2015Exp}. \\
The measured $dc$-conductance of the graphene 2CCK system involves a subtle interplay between the conduction electron DoS and interaction-renormalized scattering rates. We expect the $T=0$ $dc$ conductance to vanish in all channel-asymmetric systems because the less strongly coupled channel always decouples on the lowest energy scales. Both leads must remain coupled to ensure a finite series current. This is indeed seen in the $\omega \to 0$ limit of each of the curves in the bottom left panel of Fig.~\ref{Fig:TGc} for $\Delta=0.8$. \\
However, in the frustrated channel-symmetric case $\Delta=1$, both channels remain coupled down to $T=0$. Although the scattering rates and bare DoS both vanish as $\sim |\omega|$ in the LM and FASC phases implying a suppression of $dc$ conductance, at the critical point FACR the electronic scattering diverges as $|\omega|\to 0$. We find from NRG that these effects conspire to give a \textit{finite} linear $dc$ conductance in this case - see green line in the bottom right panel of Fig.~\ref{Fig:TGc}.\\
For an $ac$-bias the conductance is measured as a function of the driving frequency $\omega$. Conductance resonances are expected when the $ac$ frequency matches the QPC tunnelling rate. At high energies the pseudospin flip rate in the 2CCK model is given by the bare $J$ or effective $J_{eff}$. We therefore expect to see a peak in the $ac$ conductance when $|\omega|\sim J,J_{eff}$; this is observed from NRG results in the LM, ASC and FASC phases in Fig.~\ref{Fig:TGc}. However, at low energies $|\omega|\ll \lambda$, the pseudospin flip rate is renormalized and we find $\mathcal{G}^C(\omega,0)\sim \omega^2$ in these cases, independent of $\Delta$. At the critical point ACR for $0< \Delta <1$, both charge and spin fluctuations give an enhanced $ac$ conductance around $|\omega|\sim \lambda$. However, channel $\alpha=2$ decouples for $|\omega|\ll \lambda$ and so the conductance also decays at low frequencies. We find from NRG a slow attenuation $\mathcal{G}^C(\omega,0) \sim -1/\ln|\omega|$ in this regime. However, in the channel-symmetric case $\Delta=1$ at the critical point FACR, $\mathcal{G}^C(\omega,0) \sim \textit{const}.$ for $|\omega|\ll \lambda_{CR}$. The finite dynamical conductance here persists down to the $dc$-limit. This is the smoking gun signature of the NFL frustrated critical point in the graphene 2CCK system.

\section{Discussion}\label{Sec:Discus}
In this chapter we proposed a charge-Kondo quantum dot device made from graphene components that realizes a linear-pseudogap two-channel Kondo model. This exotic system has a complex phase diagram in the space of dot-lead coupling strength, potential scattering, and channel asymmetry. We analysed the thermodynamic and dynamical properties of the model using NRG, and used this to gain a detailed understanding of the renormalization group flow and fixed points. We classify the fixed points and construct the corresponding fixed-point Hamiltonians, together with their leading corrections. In particular, we uncover a channel-frustrated Kondo phase, with a non-Fermi liquid quantum critical point at the first-order quantum phase transition. Despite the depleted electronic density of the neutral graphene leads at low energies, critical fluctuations give rise to diverging scattering rates at the critical point, which produce a finite conductance even as $T\to 0$.\\
The model supports other interesting but as yet unexplored regimes. We have confined attention to the dot charge degeneracy point $\delta V_g=0$; but finite $\delta V_g$ appears in the effective model like a magnetic field on the dot pseudospin. One could also investigate the effect of doping/gating the graphene so that the Fermi level is not at the Dirac point. This will give rise to a quantum phase transition between metallic 2CK and pseudogap 2CK, with the competition controlled by the ratio of the doping to the Kondo temperature. Other physical quantities could also be investigated, such as thermoelectric transport on including a temperature gradient between leads.\\
The pseudogap 2CCK model we have studied theoretically here is likely a simplified description of any real graphene charge-Kondo nanoelectronics device. There may be complexities and subtleties in an experimental realization which are not included in our model or analysis. For example, we have assumed that the conduction electrons on both leads and dot have the same DoS. In particular, gate voltage tuning of the dot to achieve charge-degeneracy, and the addition of the decoherer, may affect the dot electronic DoS. However, we do not expect our basic results to be qualitatively modified by this because the Kondo exchange interaction derives from tunnelling at the QPC and hence involves the DoS of both lead and dot - see expression for $\tau(\omega)$. Therefore even if only the lead DoS is pseudogapped at low energies, an effective pseudogap Kondo model should still result. To make quantitative connection to experiment, such effects would have to be taken into account, as well as the possible involvement of more than just 2 dot charge states - that is, relaxing the condition $T\ll E_C$. We believe the predicted conductance signature of the frustrated quantum critical point should however still be observable in experiment.

\chapter{Discussion and conclusion} 
\subsection*{Main results}
In this conclusion chapter we aim to summarise our most important contributions to the field of mesoscopic transport in interacting nanostructures obtained in Chapters $4-8$.\\
Chapter $4$, \textit{Improved calculations for quantum transport in interacting multi-orbital nanostructures}, contains all the analytical derivations of the new sets of formulae found in this work.\\
In Sec.\ref{sec:AlternativeLandauer}, we introduce the concept of structured leads, both for equivalent and for inequivalent leads, see definition in Eq.\ref{eq:4defEqIneq} - such that we go beyond the usual wide flat band limit approximation in Eq.\ref{eq:DoSinfWBL} and the standard choice of metallic leads. We find there is no Landauer-type of formula for structured, inequivalent leads whereas we find an expression for generic interacting models under proportionate coupling (PC) with equivalent leads in Eq.\ref{eq:4LEquiv} and for noninteracting models with inequivalent leads in Eq.\ref{eq:4GacKInequiv}. The summary of our results is given in Table \ref{table:strucEqIneq}. 
In Sec.\ref{sec:AlternativeMW}, we follow two distinct approaches to derive an alternative Meir-Wingreen current formula. In the first case, we attempt to calculate an expression for $G^{<}_{dd\sigma}(\omega)$ containing an explicit $V_{bias}$ dependence such that the correlator can be readily employed in linear response transport. This calculation fails, since the final equation for $G^{<}_{dd\sigma}(\omega)$ in Eq.\ref{eq:Gdd<_F} not only does not present explicitly the voltage parameter but it is also calculated for higher order impurity correlator $F^{<}_{\sigma}(\omega)$ in Eq.\ref{eq:F<EoM_final} which cannot be further simplified. In the second case, taking insight from the Ng ansatz, we derive a corrected Fermi distribution out of equilibrium condition $f^{corr}(\omega)$ under perturbation expansion up to order $\mathcal{O}(V_{bias},U^{2})$. The final expression of $f^{corr}(\omega)$ in Eq.\ref{eq:4fcorr} contains corrections to the standard Ng ansatz to order in $V_{bias}$ and $U$, compared with Table \ref{table:fcorr}, and it reduces to the equilibrium distribution $f^{Eq}(\omega)$ in the case of symmetrically-coupled equivalent leads. This result can be used to derive the $G^{<}_{dd}(\omega)$ correlator in Eq.\ref{eq:4G<Ng}. By means of this expression, each term in the Meir-Wingreen formula has an explicit $V_{bias}$ dependence and we can evaluate the conductance under linear response.
In Sec.\ref{sec:ImprovedKubo}, we calculate an improved Kubo formula for electric and heat conductance, see Eqs.\ref{eq:4defKuboLRel},\ref{eq:4defKuboLRhea} respectively. We find that the implementation in NRG of the derived analytical expression for  $\mathcal{G}^{C}(\omega,T)$ shows better numerical precision. We systematically demonstrate this for the AIM, see plot in Fig.\ref{F4/KvsimpK}. The improved electrical Kubo formula shows finite numerically-stable values in the \textit{dc}-regime and identifies correctly the crossover between LM and Fermi liquid regime even for surprisingly inexpensive NRG calculations (modest number of kept states $M_{K}$ for any given discretization $\Lambda$) - unlike the Kubo standard formula. We perform also a similar investigation for the heat Kubo formula $\mathcal{G}^{E}(\omega,T)$ in NRG for both standard and improved versions. Regardless the choice of $M_{K},\Lambda$ parameters, $\mathcal{G}^{E}(\omega,T)$ in  NRG implementation fails completely to capture the expected physics, see plot in Fig.\ref{F4/heatkubo}. Our understanding of the issue is that in the NRG protocol, the Wilson chain does not represent a true thermal reservoir. We conclude that it is yet unsolved how to obtain the calculation of $\mathcal{G}^{E}(\omega,T)$ in NRG for genuinely non-PC impurity models. 
In Sec.\ref{sec:AlternativeOguri}, within the framework of extended impurity model, see schematic in Fig.\ref{F4/AlternativeOguri}, starting from noninteracting model we derive an alternative Fisher-Lee formula using $\mathrm{T}$-matrix in Eq.\ref{eq:4GacFisherLeeEx}. Then, turning on interactions in the model but confining the study at zero energy and temperature scale, we derive an alternative Oguri formula using renormalized $\mathrm{T}$-matrix in Eq.\ref{eq:4GdcOguriEx}. In general, formulations obtained from the calculation of the $\mathrm{T}$-matrix, instead of diagrammatic expansion, facilitate derivations and present a tangible quantity to be measured in experiments. \\
In Chapter $5$, \textit{Effective models: emergent proportionate coupling}, we study the RG-flow of $\mathrm{T}$-matrix spectrum via NRG using the effective parameters derived from the effective low energy model showing an emergent PC condition.\\
We analyse two prominent cases.
In Sec.\ref{sec:CBPC}, we consider a non-PC two-channel Kondo model in Coulomb blockade regime for both even and odd charge ground-states. The most important result occurs in the case of the odd charge ground-state with half-integer spin $S=1/2$ state in \textit{sd}-symmetry regime. We succeed in analytically deriving an expression for the electrical conductance $\mathcal{G}^{dc}(0,0)$ in terms only of renormalised potential scatterings, see Eq.\ref{eq:5cond_2ck_sym}. We numerically verify it in comparison to the fully temperature and energy dependent improved Kubo formula $\mathcal{G}^{ac}(\omega,T)$ using NRG, see plots in Fig,\ref{fig:cond_2ck_sym}\textit{(a),(b)}.
In Sec.\ref{sec:MVPC}, we consider a non-PC two-channel Kondo model in mixed-valence regime for singlet-doublet and doublet-triplet crossovers. The main output from this analysis is the fact that we can map effective models to minimal one- or two-site system satisfying an emergent PC property, see Eq.\ref{eq:5Heff_mv_d-t}. It is straightforward to generalise this modelling to higher ground-states spin. Our findings are supported by NRG results of the $\mathrm{T}$-matrix spectrum.\\
In Chapter $6$, \textit{Application and comparison of quantum transport techniques}, we give a comprehensive numerical test of the analytical formulations derived in this thesis. The system of choice is the TQD, as it can be tuned to a regime and configuration of difficult - if not inaccessible - implementation using literature methods.\\
For the TQD in PC configuration, we show via NRG the improved Kubo formula, $\mathcal{G}^{C}(\omega,T)$ compared to the standard formulation, performs higher accuracy results in the \textit{dc}-regime despite being computationally cheaper with respect to regular Kubo, see Fig.\ref{F7/TQD_KubovsImpKubo}.
Furthermore, in Sec.\ref{sec:TQDpc} for the TQD in PC, using Meir-Wingreen formula we discover that there is a revival of the Kondo effect in a strong magnetic field, see Fig.\ref{fig:bfield}, which was unknown - to the best of our knowledge.
In Sec.\ref{sec:TQDnopc} for the TQD in non-PC, we compare in  Fig.\ref{fig:nonPC} the various transport methods studied in this thesis. We find that all formulations agree with energy and temperature scale with the Oguri formula, that is low temperature \textit{dc}-regime. However, only the improved Kubo formula for electrical conductance $\mathcal{G}^{C}(\omega,T)$ manages to display the behaviour in line of the expected Kondo scale whereas the Landauer-B{\"u}ttiker formula and Ng ansatz fail to do so. Interestingly, Landauer-B{\"u}ttiker formula is closer to the real scale than the Ng ansatz - although we have not checked if Landauer formula consistently outperforms the Ng prediction for other models.\\
In Chapter $7$, \textit{Transport in molecular junction: benzene transistor}, we investigate the practicality of the benzene molecule in a break junction as an electronic switch with a back gate voltage actuator. \\
We derive the effective model $\hat{H}_{eff}$ at the mixed-valence transition between singlet-doublet and doublet-triplet points. The  $\hat{H}_{eff}$ presents an effective PC configuration that allows to be readily evaluated using the Meir-Wingreen formula under PC, see corresponding conductance formulae in Eqs.\ref{eq:6Gc_MV67},\ref{eq:6Gc_MV78} for singlet-doublet and doublet-triplet crossovers. In particular, the analytical prediction at zero energy and zero temperature scale in the singlet-doublet MV regime is in great agreement with the numerical calculation using improved Kubo formula in NRG, see Fig.\ref{F6:analytic}. Then, we compute in NRG the molecular entropy and the electrical conductance using improved Kubo formula in both configurations, see Fig.\ref{F6:SmolGc}. Our major findings are in the $N=7$ charge sector. In the \textit{meta} case, we discover the appearance of the $SU(4)$ Kondo effect and consequent conductance node. This result persists till $T=0$, see Fig.\ref{F6:G_T0_Vg_13}. In the \textit{para} case, we observe a level crossing quantum phase transition corresponding to a sudden conductance switch from maximally enhanced to identically vanishing. Again, this feature is seen till $T=0$, see Fig.\ref{F6:G_T0_Vg_14}. This latter property is ideal for the application in electronic circuits where the molecule plays  the role of a transistor.\\
In Chapter $8$, \textit{Two-channel charge-Kondo physics in graphene quantum dots}, we propose a device to study Kondo physics in graphene by means of a two-channel charge-Kondo paradigm. \\We compute the entire phase diagram numerically with NRG. The most prominent observation from our studies is the re-entrant LM behaviour with persistence of the strong coupling phase even asymptotically, see Fig.\ref{Fig:PD}. We give the fixed-point Hamiltonians for each phase and the leading perturbation order for all fixed-point, as classified in the $3D$ RG-flow diagram in Fig.\ref{Fig:3DRG}. Surprisingly, in the symmetric two-channel set-up, we find finite electrical conductance down to $\omega\to0$ at the frustrated asymmetric critical fixed-point, see Fig.\ref{Fig:TGc}. This conductance behaviour is a signature of exotic non-Fermi liquid physics that constitutes a new frustrated asymmetric critical fixed-point. 

\subsection*{Thinking forward}
In this work we give our contribution to the following question: how to provide well-defined and numerically controlled quantum transport solutions for elaborate systems without diminishing the rich physics of the bare model. \\
Our general strategy is to decompose complicated interacting systems under nonequilibrium condition into simpler exact analytical expressions which limiting cases at $\omega\to0$ and $T\to0$ can be benchmarked numerically. In this thesis we employ NRG as the numerical method to validate our analytical predictions. However, because the class of quantum transport problems we solved in this work is under linear response theory, we envisage other numerical methods based on the linear response principle are amenable for testing the alternative and the improved quantum transport formulae derived in this thesis. Furthermore, these new formulae are defined through physical quantities which can be probed in realistic experimental devices.\\
Another crucial aspect of our strategy is to allow the maximum flexibility of the components in the coupled system. About the leads, we designed formulae considering: $(i)$ structured object with energy and spin dependence, $(ii)$ arbitrary coupling  configuration. About the impurity, we formulate it in terms of  arbitrary $(i)$ type of local interactions, $(ii)$ orbital complexity, $(iii)$ number of sites. The flexibility of our formulations opens to the applicability of our results not only to strongly correlated many-body systems but also to organic single-molecules with functionality as single-molecule transistors (e.g. \textit{p-Phenylene} \cite{xin2019concepts}) and nanoparticles network (e.g. to mimic programmable logic gates \cite{bose2015evolution} for near term neuromorphic computers).

\subsection*{What can be done next}
An important future direction of this work is the extension of these new transport formulations to multi-channel models. With this respect, we suppose two main issues would arise. \\
The first is the analytical complexity to extend our alternative and improved formulae to an arbitrary number of channels. It may result that some formulations do not have a counterpart in systems with more than two channels.\\
The second is related to the numerical implementation with NRG of multi-channel systems as the complexity grows exponentially with the number of channels. To tackle this problem, we propose to carry future investigation in multi-channel systems by means of impurity solvers with more favourable scaling.

\begin{singlespace}
\clearpage

\begin{appendices}
\noappendicestocpagenum

\chapter{The Meir-Wingreen current formula}\label{app:MW}  
In this appendix we derive the \textit{electrical current formula in the time-independent case}, as originally introduced by Meir and Wingreen \cite{MeirWingreen_1992}
\begin{equation}
\begin{aligned}
\hat{I} =  \frac{ie}{2h}\sum_{\sigma}\bigintsss d\omega Tr \bigg[ 
& \Big( f^{\mu_{L}} (\omega-\mu_{L}) \Gamma^{L} -  f^{\mu_{R}}(\omega-\mu_{R}) \Gamma^{L} \Big) \Big( G_{dd\sigma}^{R}(\omega) -G_{dd\sigma}^{A}(\omega) \Big)\\
&+ \Big( \Gamma^{L} - \Gamma^{L} \Big) G_{dd\sigma}^{<}(\omega) \bigg] ~,
\end{aligned}
\end{equation}
that is used to study transport in strongly interacting mesoscopic nanostructure. We refer for the \textit{nonequilibrium theory} through the \textit{Keldysh formalism} in the Sec.\ref{sec:Gtheo}. Further technical readings can be found in the references \cite{Jauho_QuantumKineticsTransport,Ryndyk,Spicka_NGF,Maciejko_NonEquilibrium}. The physics of this formula is discussed in the corresponding Sec.\ref{sec:MeirWingreen}. \\

\noindent{We} start with the general impurity model Hamilonian in Eq.\ref{eq:defImpModelGen}, here we write down its explicit form as used in the original derivation with a two-channel model, namely
\begin{equation}\label{A:eq:H}
\hat{H} = \sum_{\alpha\mathbf{k}\sigma} \epsilon_{\alpha\mathbf{k}} c^{\dagger}_{\alpha\mathbf{k}\sigma} c_{\alpha\mathbf{k}\sigma} 
+ \hat{H}^{U}_{imp} \lbrace d^{\dagger}_{n \sigma} ; d_{n \sigma}\rbrace
+ \sum_{n\mathbf{k}\sigma} \Big( V_{\alpha\mathbf{k}} c^{\dagger}_{\alpha\mathbf{k}\sigma} d_{n \sigma} + V^{\star}_{\alpha\mathbf{k}} d_{n \sigma}^{\dagger} c_{\alpha\mathbf{k} \sigma}  \Big ) ~,
\end{equation}
where $c^{\dagger}_{\alpha\mathbf{k}\sigma} c_{\alpha\mathbf{k}\sigma}$ is set of spinful electron operators acting on $\alpha \in L,R$ leads and $d_{n \sigma}^{\dagger},d_{n \sigma}$ is set of spinful electron operators acting in the central region for $n$ number of degree of freedom. This impurity or central region contains the one-body Coulomb potential to represent the interaction part. The hybridization between lead and central region is described by coupling strength $V_{\alpha\mathbf{k}}$.  \\ 
The Hamiltonian model as defined in Eq.\ref{A:eq:H} is used to derive a \textit{steady state current} - that is the key input to calculate the time-independent conductance. We will develop the analytical derivation of the Meir-Wingreen formula as calculated in the literature.

\section*{Steady state current definition}
As initial step, we seek for an expression for the steady state current. In the system current is conserved i.e. $\hat{I}_{L} (t) = -\hat{I}_{R} (t) \Rightarrow \hat{I}(t)$. Hence, without loss of generality, we define the \textit{current with respect to left lead} as
\begin{equation}
\hat{I}_{L} (t) \equiv -e \langle\dot{N}_{L} (t) \rangle ~,
\end{equation}
that is calculated from the expectation value of time-derivative of the number operator $\hat{N}_{L}$. We evaluate the current using the Heisenberg equation of motion i.e. $\dot{\Psi}(t) = i [H,\Psi(t)]$ such that we obatin 
\begin{equation}\label{A:eq:HeisenbergEoM}
-i\langle \dot{N_{L}} (t) \rangle= \langle [ \hat{H}(t),\hat{N}_{L}(t)] \rangle ~,
\end{equation}
where $\hat{H}$ indicates the full Hamiltonian expression given in eq.\ref{A:eq:H}.\\
It can be shown the equal time Green's function $G(t,t)$ is equivalent to calculate the Green's function evaluated in static expectation value: $G(t,t) \equiv -i \langle \Psi \Psi^{\dagger} \rangle$ and this equality is also consistent with the fact we are interested in steady state transport. At this stage, we note the steady state condition indicates also time-translational invariance in the system. Its propagator is a the function of time difference only $G(t-t^{\prime})$ and we can apply Fourier transformation to bring the final result into energy domain. Hence, in this derivation, we refer to time-translational, steady-state Green's functions. \\
Taking into account these considerations, we compute the current in its static expectation value 
\begin{equation}
\hat{I} = -e(-i) \langle \dot{N}_{L} \rangle= e\langle [ \hat{H},\hat{N}_{L}] \rangle ~,
\end{equation}
where we note the number operator $\hat{N}$ commutes in general with $\hat{H}_{leads}, \hat{H}_{imp}$. At the end of the calculation, we can restore the time parameters by transforming the operators into interacting representation. After applying the equation of motion technique (EoM) see Eq.\ref{eq:EoM}, the result is a combination of anti-Hermitian operators. By means of the Keldysh-contour, see Fig.\ref{F2:KeldyshCont}, we recognize the calculated Green's function expression is in lesser type form. In particular, we find the so-called \textit{mixed lesser Green's function} namely
\begin{equation}
G_{n,\alpha \mathbf{k}\sigma}^{<} (t -t^{\prime}) =i \langle c^{\dagger}_{\alpha\mathbf{k}\sigma}(t^{\prime}) d_{n\sigma}(t)\rangle  ~,
\end{equation}
since both lead and nanostructure operators are included hence the subscripts in the Green's function are group accordingly. We also note, that as it is written, $G_{n,\alpha \mathbf{k}\sigma}^{<} (t -t^{\prime})$ is a time-ordered Green's function.\\
In conclusion, we obtain the \textbf{steady state current} equation with respect to left lead 
\begin{equation}\label{A:eq:J}
\hat{I}_{L} (t)  = 2e \mathit{Re} \Big[ \sum_{n\alpha\mathbf{k}\sigma} V_{\alpha\mathbf{k}} G_{n,\alpha\mathbf{k}\sigma}^{<} (t,t^{\prime}) \Big]
\end{equation}
where the time-ordered operator current is given by the real part of the hybridization term $V$ multiplied for the mixed lesser $G^{<}$ This expression in Eq.\ref{A:eq:J} requires the calculation of the equilibrium form $G^{<}$, once transient phenomena are completed. However, the problem is set in the out-of-equilibrium regime. We will study in the next section the connection between equilibrium and nonequilibrium quantities in order to calculate the time-translational invariant lesser Green's function $G^{<} (t -t^{\prime})$.

\section*{Mixed $G^{<}$ in nonequilibrium form}
In the previous section, we have derived the expression for the steady state current in time domain, see Eq.\ref{A:eq:J}. We seek now an expression for the lesser Green's function $G_{n,\alpha \mathbf{k}\sigma}^{<}$ in nonequilibrium, steady state. \\ 
Taking the standard results from nonequilibrium perturbation theory, at $T=0$ it can be shown the formal structure of equilibrium theory is the same as the nonequilibrium one. Due to this crucial correspondence at zero temperature, we can adopt and apply the familiar equilibrium calculation techniques to the corresponding nonequilibrium expressions. Thus, we make use of the equation of motion method to calculate Green's functions and of the perturbation expansion with its diagrammatic interpretation. The conversion to nonequilibrium theory is simply provided by substituting the real-time axis with the semi-closed path Keldysh-contour and by changing the time-ordered operators in real-time variables with the contour-ordered ones in complex-time variables.\\
Having these concepts in mind, we start evaluating $G_{n,\alpha \mathbf{k}\sigma}^{<}(t-t^{\prime})$ that is the equilibrium mixed lesser Green's functions defined on real-time contour. As we will see later, the Fourier transformed expression of  $G^{<}$ can be inserted in Eq.\ref{A:eq:J} and the final current expression is depending only on energy arguments. Hence, the lesser Green's function from EoM calculation results in
\begin{equation}\label{A:eq:EoM-t-initial}
\sum_{n\alpha\mathbf{k}\sigma}\Big( -i \frac{d}{d{t^{\prime}}} - \epsilon_{\alpha\mathbf{k}} \Big) G_{n,\alpha\mathbf{k}\sigma}^{<} (t-t^{\prime}) =\sum_{nm\alpha\mathbf{k}\sigma} V^{\star} _{\alpha\mathbf{k}} G_{nm\sigma}^{<}(t-t^{\prime})~,
\end{equation}
where we made use of noninteracting leads to get a closed set of EoM and the Green's function on the right hand side is identified as the equilibrium lesser Green's function of the central-region
\begin{equation}\label{A:eq:GnmDef}
\sum_{nm\sigma}G_{nm\sigma}^{<} (t-t^{\prime})\equiv i \langle d^{\dagger}_{n\sigma}(t^{\prime});d_{m\sigma}(t) \rangle ~.
\end{equation}
As we will see in later steps, an important aspect of the Meir Wingreen approach is that only the local Green's function $G_{nm}$ will appear in the final formula.\\
In order to identify and then to invert the term multiplying upfront $G_{n,\alpha \mathbf{k}\sigma}^{<}$ in Eq.\ref{A:eq:EoM-t-initial}, we need to bring the expression from time to frequency domain by applying the Fourier transform and we get
\begin{equation}
\sum_{n\alpha\mathbf{k}\sigma}G_{n,\alpha\mathbf{k}\sigma}^{<} (\omega) = \sum_{nm\alpha\mathbf{k}\sigma} G_{0,\alpha \mathbf{k}\alpha\mathbf{k}\sigma}^{R} (\omega) V^{\star} _{\alpha\mathbf{k}} G_{nm\sigma}^{<}(\omega) ~,
\end{equation}
where we can uniquely identify $ G_{0,\alpha \mathbf{k}\alpha\mathbf{k}\sigma}^{R}$ as the exact retarded Green's function of the uncoupled leads - hence, when the system is at $t \rightarrow -\infty$. \\
We can transform back this expression to real time domain using the convolution theorem \cite{WeberArfken} and finally we calculate the time-ordered \textbf{mixed lesser equilibrium Green's function}
\begin{equation}\label{A:eq:EoM-t-final}
\sum_{n\alpha\mathbf{k}\sigma}G_{n,\alpha\mathbf{k}\sigma}^{<} (t-t^{\prime}) = \sum_{nm\alpha\mathbf{k}\sigma}  V^{\star} _{\alpha\mathbf{k}} G_{nm\sigma}^{<} (t-t^{\prime})  G_{0,\alpha \mathbf{k}\alpha\mathbf{k}\sigma}^{R}  (t-t^{\prime}) ~, 
\end{equation}
where 
\begin{equation}
G_{0,\alpha \mathbf{k}\alpha\mathbf{k}\sigma}^{R}  (t-t^{\prime}) \equiv -i \theta (t-t^{\prime}) \langle \lbrace c_{\alpha\mathbf{k}\sigma}(t), c^{\dagger}_{\alpha\mathbf{k}\sigma}(t^{\prime})  \rbrace \rangle = ( -id_{t}  - \epsilon_{\alpha\mathbf{k}} ) 
\end{equation}
is the \textbf{exact time-independent retarded Green's function of isolated leads}. The expression given in Eq.\ref{A:eq:EoM-t-final} is the explicit version of Eq.\ref{A:eq:EoM-t-initial}, since now all the elements are identified and $G_{n,\alpha \mathbf{k}\sigma}^{<}$ takes a proper analytical form.\\
We conclude now by showing the correspondence at zero temperature of Eq.\ref{A:eq:EoM-t-final} with its nonequilibrium counterpart. As we know already, the mixed lesser nonequilibrium Green's function in complex-time arguments $G_{n,\alpha \mathbf{k}\sigma}^{<}(\tau,\tau^{\prime})$ has the same formal structure as $G_{n,\alpha\mathbf{k}\sigma}^{<} (t-t^{\prime})$ derived in Eq.\ref{A:eq:EoM-t-final}. However, due to the replacement of the real-time variables, the real-time axis is substituted by the Keldysh-contour $\mathcal{C}_{\tau}$ where the contour-ordered Green's functions are integrated on. Hence, the \textbf{mixed lesser nonequilibrium Green's function} reads
\begin{equation}\label{A:eq:G<contour}
\sum_{n\alpha\mathbf{k}\sigma}G_{n,\alpha\mathbf{k}\sigma}^{<} (\tau,\tau^{\prime}) = \sum_{nm\alpha\mathbf{k}\sigma}  V^{\star} _{\alpha\mathbf{k}} \int_{\mathcal{C}_{\tau}}  d\tau_{1} G_{nm\sigma}^{<} (\tau,\tau_{1})  G_{0,\alpha\mathbf{k}\alpha\mathbf{k}\sigma}^{R}  (\tau_{1},\tau^{\prime}) ~,
\end{equation}
where performing this integral on general contour is subject of the next section. \\

\noindent{For} the sake of conciseness in the equations, we will omit the upfront summation over parameters and consider it as implicitly performed. 

\section*{Analytic continuation from the Keldysh-contour to real-time axis}
As next step, we directly evaluate of the nonequilibrium $G^{<}(\tau,\tau^{\prime})$ Eq.\ref{A:eq:G<contour} and we attempt it through two different approaches.\\
The first one consists in deforming the original Keldysh-contour $\mathcal{C}_{\tau}$ in Eq.\ref{A:eq:G<contour} into two sub-contours such that the integration is split over these new branches. By considering the new time-label order, we are able to recognise the type of Green's functions. We manipulate the integral on complex-time variables in Eq.\ref{A:eq:G<contour} using the Keldysh identities in Eq.\ref{eq:identityKeldysh} to turn it into into the form of \textbf{mixed lesser equilibrium Green's function}
\begin{equation}\label{A:eq:G<Time}
G_{n,\alpha \mathbf{k}\sigma}^{<} (t-t^{\prime}) =  V^{\star} _{\alpha\mathbf{k}} \int_{-\infty}^{\infty}  dt_{1} \Big( G_{nm\sigma}^{R} (t-t_{1})  G_{0,\alpha \mathbf{k}\alpha \mathbf{k}\sigma}^{<} (t_{1}-t^{\prime}) + G_{nm\sigma}^{<} (t-t_{1})  G_{0,\alpha \mathbf{k}\alpha \mathbf{k}\sigma}^{A} (t_{1}-t^{\prime}) \Big) ~,
\end{equation}
where the mixed $G^{<}$ is written again as time-ordered Green's function. This result is expected: the process of deformation the complex-time contour and subsequent extrapolation of real-time variables, is a projection of complex variables onto real-time axis. We note that, on a practical level, manipulating contour-ordered Green's functions might bring to cumbersome calculation - suggesting different methods should be explored.\\
The second approach is developed in the \textit{Langreth theorem} \cite{LangrethTheorem}: the Keldysh-contour modification is completely avoided by introducing a specific matrix representation of the nonequilibrium $G$ that encapsulates the internal structure of the Keldysh-contour and absorbs the  integration over $\mathcal{C}_{\tau}$. The outcome is a successive matrix multiplication rules that can be directly evaluated as integrand in real-time integration. In this procedure, from the complex-time contour we have developed an \textit{analytical continuation} to the real-time axis. To the purpose of this derivation, we employ the multiplicative rules explained in the Langreth theorem for Fermionic models only. For completeness, we mention that the Langreth theorem is also applicable to Bosonic and mix of Bosonic-Fermionic systems \cite{Langreth_MasterEq-TimeDepAM-PRB1992}.\\
We proceed now with a quantitative discussion of the Langreth theorem for analytical continuation. We start with writing the contour-ordered Green's functions as matrix with $dim(2 \times 2)$ and its matrix elements are identified according to which branch the time labels are located on. We take as example the lesser Green's function $G^{<} (\tau,\tau^{\prime})$ in matrix form becomes
\begin{equation}
\mathbb{G}^{<} =
\begin{pmatrix}
G^{R} & G^{<} \\
0 & G^{A}
\end{pmatrix} ~.
\end{equation}
According to type of Green's function we aim to calculate, we write its factors in the corresponding matrix form and we perform their matrix multiplication. At the end, we select the matrix element corresponding to the desired type of Green's function. For instance, if we want to calculate the correlators product $G^{<}G_{0}^{R}$ in Eq.\ref{A:eq:G<contour} into the corresponding matrix  multiplication rule we proceed as follows
\begin{equation}
\begin{aligned}
\mathbb{G}_{n,\alpha k}^{<} = \mathbb{G}^{<}_{nm\sigma} \cdot\mathbb{G}^{<}_{0,\alpha \mathbf{k}\alpha\mathbf{k}\sigma} &= 
\begin{bmatrix}
G^{R} & G^{<} \\
0 & G^{A}
\end{bmatrix}_{nm\sigma} \cdot
\begin{bmatrix}
G_{0}^{R} & G_{0}^{<} \\
0 & G_{0}^{A}
\end{bmatrix}_{\alpha\mathbf{k}\alpha \mathbf{k}\sigma} \\
&=\begin{bmatrix}
G_{nm\sigma}^{R} G_{0,\alpha \mathbf{k}\alpha \mathbf{k}\sigma}^{R} &  G_{nm\sigma}^{R}G_{0,\alpha \mathbf{k}\alpha \mathbf{k}\sigma}^{<}+G_{nm\sigma}^{<}G_{0,\alpha \mathbf{k}\alpha \mathbf{k}\sigma}^{A} \\
0 & G_{nm\sigma}^{A}G_{0,\alpha \mathbf{k}\alpha \mathbf{k}\sigma}^{A}
\end{bmatrix} \\ 
&\equiv
\begin{bmatrix}
G^{R} & G^{<} \\
0 & G^{A}
\end{bmatrix}_{n,\alpha \mathbf{k}\sigma} ~,
\end{aligned}
\end{equation}
where in the second line, the top right element represents the \textit{lesser function analytical continuation} on the real-time axis of the product $G^{<}(\tau,\tau^{\prime})G^{R}_{0}(\tau,\tau^{\prime})$ on $\mathcal{C}_{\tau}$ contour as given in eq.\ref{A:eq:G<contour}. As we expected, it is the same result of Eq.\ref{A:eq:G<Time} obtained by direct integration on the Keldysh contour: the matrix representation incorporates the contour calculation on the modified branches. Hence, from the matrix multiplication $\mathbb{G}^{<} \cdot \mathbb{G}_{0}^{<}$ we take the top right quantity as the lesser element of the matrix and we plug it in directly as integrand of corresponding real-time integration: 
\begin{equation}
\begin{aligned}
G^{<} (\tau,\tau^{\prime}) = \int_{\mathcal{C}_{\tau}} G^{<} G_{0}^{R} \quad \Leftrightarrow \quad
G^{<} (t,t^{\prime}) = \int_{\mathcal{C}_{t}} \big( G^{R}G_{0}^{<} + G^{<}G_{0}^{A} \big) ~.
\end{aligned}
\end{equation}
The matrix representation of contour-ordered Green's function as derived by Langreth is clearly a more practical tool to derive equations in nonequilibrium condition. For completeness, we summarize the Langreth rules between Keldysh and real-time contour for the case of two terms multiplication 
\begin{equation}
C(\tau,\tau^{\prime}) = \int_{\mathcal{C}_{\tau}} AB \Rightarrow
\begin{cases}
C^{< , >} (t,t^{\prime})= \int_{\mathcal{C}_{t}} \big ( A^{R}B^{< , >} + A^{< , >}B^{A} \big) ~,  \\
C^{R , A} (t,t^{\prime})= \int_{\mathcal{C}_{t}} A^{R,A}B^{R,A} ~,
\end{cases}
\end{equation}
and multiplication for more terms follows similarly.\\
In conclusion, we have transformed the mixed lesser complex-time variable $G^{<}(\tau,\tau^{\prime})$ Green's function derived in Eq.\ref{A:eq:G<contour} into the mixed lesser real-time variable $G^{<}(t-t^{\prime})$ as computed in Eq.\ref{A:eq:G<Time}. This discussion completes the correspondence between equilibrium and nonequilibrium forms. As next step we are ready to insert these analytical expressions into the steady state current in Eq.\ref{A:eq:J}. 

\section*{Levelwidth function}
We introduce the expression for the mixed lesser equilibrium $G^{<}(t-t^{\prime})$ derived in Eq.\ref{A:eq:G<Time} into the current in Eq.\ref{A:eq:J}
\begin{equation}
\hat{I}_{L} (t)  = 2e \mathit{Re} \sum_{nm\alpha\mathbf{k}\sigma} 
\Big[ V_{\alpha\mathbf{k}} V^{\star}_{\alpha\mathbf{k}} \int_{-\infty}^{\infty}  dt_{1} \Big( G_{nm\sigma}^{R} (t-t_{1})  G_{0,\alpha \mathbf{k}\alpha \mathbf{k}\sigma}^{<} (t_{1}-t^{\prime}) + G_{nm\sigma}^{<} (t-t_{1})  G_{0,\alpha \mathbf{k}\alpha \mathbf{k}\sigma}^{A} (t_{1}-t^{\prime}) \Big) \Big] ~\
\end{equation}
where we have restored the overall parameter summation for the clarity in the derivation.\\
We are interested in deriving the \textit{steady state time-independent current}, where correlators depend on time differences only. In the retarded and advanced Green's functions, using the $\mp i \theta(\pm t \mp t^{\prime})$ factor in their definition in Eq.\ref{eq:defGeq}, we select the non-vanishing time difference. However, in case of lesser Green's functions, there is no time-ordering symbol in their definition see in Eqs.\ref{eq:defGeq},\ref{eq:defGneq}. Hence, we can twist as we need its real-time argument. \\
We start manipulating the integrand in the current equation by introducing the exact time-independent Green's functions for isolated leads with noninteracting contacts, namely
\begin{equation}\label{A:eq:g_Def}
\begin{aligned}
& G_{0,\alpha \mathbf{k}\alpha \mathbf{k}\sigma}^{<} (t - t^{\prime}) = i f^{\mu_{\alpha}} (\omega- \mu_{\alpha}) e^{-i \omega (t - t^{\prime})} \\
& G_{0,\alpha \mathbf{k}\alpha \mathbf{k}\sigma}^{R/A}(t - t^{\prime}) = \mp \theta(\pm t \mp  t^{\prime}) e^{-i \omega (t - t^{\prime})} ~,
\end{aligned}
\end{equation}
where $f^{\mu_{\alpha}}$ is the Fermi-Dirac distribution of the initial $\alpha-$lead population uncoupled from the central region given in Eq.\ref{eq:defFermiNeq}. In this context, we conclude $f^{\mu_{\alpha}}$ is always indicating an equilibrium electronic distribution that is set at distant past, when neither interactions nor tunnelling events occur. \\
By using these definition for noninteracting in Eq.\ref{A:eq:g_Def} and interacting $G^{<},G^{R}$ propagators, we transform the integration variable in $\int d(t-t_{1})$. Then, we shuffle terms in the integrand such that we can assemble the definition of a Fourier transform to energy domain namely
\begin{equation}
\int_{-\infty}^{+\infty } d(t-t_{1}) G(t-t_{1}) e^{-i \omega(t-t_{1})} \equiv G(\omega) ~,
\end{equation}
since in steady state time-independent current the energy is tunable quantity. The corresponding energy domain expressions after Fourier transform read
\begin{equation}\label{A:eq:g_Def}
\begin{aligned}
& G_{0,\alpha \mathbf{k}\alpha \mathbf{k}\sigma}^{<} (\omega) = \int d(t - t^{\prime}) G_{0,\alpha \mathbf{k}\alpha \mathbf{k}\sigma}^{<} (t - t^{\prime}) e^{i \omega (t - t^{\prime})} =     i f^{\mu_{\alpha}} (\omega- \mu_{\alpha}) 2\pi \mathcal{A}_{0,\alpha\mathbf{k}\sigma} (\omega)  ~,\\
& G_{0,\alpha \mathbf{k}\alpha \mathbf{k}\sigma}^{>} (\omega) = -i( 1-f^{\mu_{\alpha}} (\omega- \mu_{\alpha})) 2\pi\mathcal{A}_{0,\alpha\mathbf{k}\sigma} (\omega) ~, \\
& G_{0,\alpha \mathbf{k}\alpha \mathbf{k}\sigma}^{R/A}(\omega) = \int d(t - t^{\prime}) G_{0,\alpha \mathbf{k}\alpha \mathbf{k}\sigma}^{R/A}  e^{i \omega (t - t^{\prime})} =  \frac{1}{\omega \pm i \eta - \epsilon_{\mathbf{k}}} ~,
\end{aligned}
\end{equation}
where the equation for $G_{0}^{>}$ is just obtained by using the general identity $G^{<}(\omega) = e^{-\beta\omega}G^{>}(\omega)$ and the $G^{</>}$ equations are the noninteracting version of the fluctuation-dissipation theorem in Eq.\ref{eq:defFDtheo}. For completeness, we also write the noninteracting self-energies 
\begin{equation}\label{A:eq:delta_Def}
\begin{aligned}
&\Delta_{\alpha \mathbf{k}\alpha \mathbf{k}\sigma}^{<}(\omega) =  V_{\alpha\mathbf{k}} V^{\star} _{\alpha\mathbf{k}} G_{0,\alpha \mathbf{k}\alpha \mathbf{k}\sigma}^{<}(\omega) = i f^{\mu_{\alpha}} (\omega- \mu_{\alpha}) 2\Gamma^{\alpha}(\omega)  ~,\\
&\Delta_{\alpha \mathbf{k}\alpha \mathbf{k}\sigma}^{>}(\omega) =  V_{\alpha\mathbf{k}} V^{\star} _{\alpha\mathbf{k}} G_{0,\alpha \mathbf{k}\alpha \mathbf{k}\sigma}^{>}(\omega) = -i(1- f^{\mu_{\alpha}} (\omega- \mu_{\alpha})) 2\Gamma^{\alpha}(\omega)  ~,\\
&\Delta_{\alpha \mathbf{k}\alpha \mathbf{k}\sigma}^{R/A}(\omega) =  V_{\alpha\mathbf{k}} V^{\star} _{\alpha\mathbf{k}} G_{0,\alpha \mathbf{k}\alpha \mathbf{k}\sigma}^{R/A}(\omega) =  \mathit{Re}\Delta_{\alpha \mathbf{k}\alpha \mathbf{k}\sigma}^{R} (\omega) \mp \frac{i}{2} \mathit{Im}\Delta_{\alpha \mathbf{k}\alpha \mathbf{k}\sigma}^{R} (\omega)  ~,
\end{aligned}
\end{equation}
where in the lesser and greater equations we derived the second equality using the Fluctuation-Dissipation theorem in Eq.\ref{eq:defFDtheo} and $ -\mathit{Im}\Delta_{\alpha \mathbf{k}\alpha \mathbf{k}\sigma}^{R} (\omega) \equiv \mathrm{\Gamma}^{\alpha}(\omega)$ as we defined in Eq.\ref{eq:DoSinfWBL} for general \textit{structured} leads such that the Keldysh identity in Eq.\ref{eq:identityKeldysh} is fulfilled.\\
The final current expression is obviously only function of energy, we convert the discrete summation over momentum $\mathbf{k}$ into a integral of the density of states $\rho_{0,\alpha\sigma}(\epsilon_{\mathbf{k}})$ with respect to $\alpha$-lead
\begin{equation}
\sum_{\mathbf{k}} \longrightarrow \int \frac{d\epsilon_{\mathbf{k}}}{2\pi} \rho_{0,\alpha\sigma}(\epsilon_{\mathbf{k}}) ~.
\end{equation}
As final step, we introduce the time-independent \textbf{levelwidth function}, that is the matrix form in $n,m$ indices of the gamma function seen in Eq.\ref{eq:DoSinfWBL} 
\begin{equation}\label{A:eq:LevelWidth}
[ \mathbb{\Gamma}^{\alpha}(\epsilon_{\mathbf{k}})]_{nm} = \pi  \mathbf{V}_{\alpha\mathbf{k}n} \mathbf{V}^{\star} _{\alpha\mathbf{k}m}\rho_{0,\alpha\sigma}(\epsilon_{\mathbf{k}})
\end{equation}
where the expression satisfies $ \forall~ n,m$ central-region degrees with $\rho_{0,\alpha\sigma}$ is the noninteracting density of states function with respect to tunnelling in $\alpha$-bath and $\mathbb{V}_{\alpha\mathbf{k}}$ is the matrix hybridization strength between $\alpha$-bath and the central interacting region. The gamma expression in Eq.\ref{A:eq:LevelWidth} is fully dependent on the leads properties and their coupling with the central region. We note also the relation between the single-particle energy $\epsilon_{\mathbf{k}}$ and momentum dependence in the hybridization is given by $V_{\alpha\mathbf{k}n}=V_{\alpha n}(\epsilon_{\mathbf{k}})= V_{\alpha n} u_{0 k}$ with $u_{0 k}$ expansion in momentum basis.\\
Collecting all these results, we derive the \textbf{steady state time-independent current with respect to the left lead}
\begin{equation}\label{A:eq:J_L}
\hat{I}_{L} = i e \sum_{nm\sigma}\int \frac{d\epsilon_{\mathbf{k}}}{2 \pi} \Big[ Tr \Big( [ \Gamma^{L}(\epsilon_{\mathbf{k}})]_{nm} f^{\mu_{L}} (\epsilon_{\mathbf{k}}) (G^{R}_{nm\sigma}(\epsilon_{\mathbf{k}}) -G^{A}_{nm\sigma}(\epsilon_{\mathbf{k}})) \Big) + Tr \Big( [ \Gamma^{L}(\epsilon_{\mathbf{k}})]_{nm} G^{<}_{nm\sigma}(\epsilon_{\mathbf{k}}) \Big) \Big] ~,
\end{equation}
that is the exact expression for interacting mesoscopic transport and it is given with respect to its matrix elements $n,m$. In Eq.\ref{A:eq:J_L} some crucial properties of the current emerge: $(i)$ it is a function of energy, $(ii)$ it is determined by local correlations i.e. impurity Green's functions defined in Eq.\ref{A:eq:GnmDef} and $(iii)$ it contains indication of the initial thermal equilibrium population in the leads through the Fermi distribution $f^{\mu_{\alpha}}$.\\
We are only left with deriving the full form of the Meir-Wingreen equation by adding also the contribution to the  electronic flow from the right lead.

\section*{Time-independent steady state current}
Recalling the current conservation property satisfied by our model in steady-state, by performing the same steps presented in this Appendix on the current now flowing from the right lead, we obtain an identical result to Eq.\ref{A:eq:J_L}. We can then symmetrize the total current composed by these two current equations $\hat{I}_{L},\hat{I}_{R}$ calculated separately:
\begin{equation}
\hat{I} =\frac{\hat{I}_{L} + \hat{I}_{L} }{2} =\frac{\hat{I}_{L} - \hat{I}_{R} }{2} 
\end{equation}
and we combine the current flow contribution from the left and right leads. The outcome  yields to the \textit{dc-}current, time-independent in steady state that is the \textbf{Meir-Wingreen formula} 
\begin{equation}
\begin{aligned} 
\hat{I} =  \frac{i e}{2 } \sum_{nm\sigma}\int \frac{d\epsilon_{\mathbf{k}}}{2 \pi}  Tr & \bigg[ 
\Big( f^{\mu_{L}} (\epsilon_{\mathbf{k}}) [ \Gamma^{L}(\epsilon_{\mathbf{k}})]_{nm} -  f^{\mu_{R}}(\epsilon_{\mathbf{k}}) [ \Gamma^{R}(\epsilon_{\mathbf{k}})]_{nm} \Big) \Big( G_{nm\sigma}^{R}(\epsilon_{\mathbf{k}}) -G_{nm\sigma}^{A}(\epsilon_{\mathbf{k}}) \Big) +\\
& \hspace*{2cm} + \Big([ \Gamma^{L}(\epsilon_{\mathbf{k}})]_{nm} -  [ \Gamma^{R}(\epsilon_{\mathbf{k}})]_{nm} \Big) G_{nm\sigma}^{<}(\epsilon_{\mathbf{k}}) \bigg] 
\end{aligned}
\end{equation}
as it has been derived in \cite{MeirWingreen_1992}, bring to a completion our derivation. We note the gamma function in the final expression are still evaluated over the matrix components in Eq.\ref{A:eq:LevelWidth} and all the propagators are locally calculated for the central region.

\subsection*{Meir-Wingreen formula under equilibrium}
It is interesting to discuss explicitly the Meir-Wingreen current formula under equilibrium condition.\\
The Fermi distributions of leads are the same i.e. $f^{\mu_{L}}=f^{\mu_{R}} \equiv f^{Eq}$, provided $\mu_{L} = \mu_{R} \equiv \mu$, hence it reduces to the equilibrium expression in Eq.\ref{eq:defFermieq}. It follows, the Meir-Wingreen formula under equilibrium condition 
\begin{equation}
\hat{I} =  \frac{i e}{2 } \sum_{nm\sigma}\int \frac{d\epsilon_{\mathbf{k}}}{2 \pi} 
Tr \bigg[ f^{Eq} (\epsilon_{\mathbf{k}}) [\Gamma^{Eq}(\epsilon_{\mathbf{k}})]_{nm} \Big( G_{nm\sigma}^{R}(\epsilon_{\mathbf{k}}) -G_{nm\sigma}^{A}(\epsilon_{\mathbf{k}}) \Big)  + [\Gamma^{Eq}(\epsilon_{\mathbf{k}})]_{nm} G_{nm\sigma}^{<}(\epsilon_{\mathbf{k}})\bigg] ~,
\end{equation}
where $G^{<}_{nm\sigma}(\epsilon_{\mathbf{k}}) = if^{Eq} (\epsilon_{\mathbf{k}}) 2\pi\mathcal{A}_{dd\sigma} (\epsilon_{\mathbf{k}})$ with the spectral function calculated for local impurity Green's function as defined in Eq.\ref{eq:defSpectral}.\\
Thus, we conclude at equilibrium condition the Meir-Wingreen formula is identically zero, as lack of unequal electronic distribution in the leads $\mu_{L} \neq \mu_{R}$ prevents any net current flow through the system.

\section*{Meir-Wingreen current formula in time-dependent models}
The nonequilibrium procedure to derive steady-state current for time-independent electrical conductance comprises also the study of time-dependent mesoscopic transport. We present now the main results for the \textit{time-dependent current} flowing from noninteracting leads to a central interacting region - as completion of the Meir-Wingreen current formula derivation.\\
In general, the wavefunction coherence phase can be modified by either external magnetic field or temperature variation. In case an external time-dependence perturbation is applied, the wave function acquires an extra phase factor $exp(-i \int dt^{\prime} \epsilon(t^{\prime}))$. However, the electronic distribution for a given state does not vary applying such a perturbation. Under these conditions, the derivation of the time-dependent current is not conceptually different from the steady-state current we developed so far in this appendix - since the time-dependent feature just appears in the explicit time-parameter definition in the model. Nevertheless, some extensions are required \cite{JauhoWingreeMeir1994}, as we aim to briefly present in this section. \\
The external time-dependent driving force is physically the time-dependent voltage bias. Due to the perturbation, the time-dependent model Hamitonian reads the same as presented in Eq.\ref{A:eq:H} but now with an explicit time-dependence in its single-particle energy states and tunnelling terms i.e.
\begin{equation}\label{A:HTimeDep}
\begin{aligned}
& \epsilon_{\alpha\mathbf{k}} \longrightarrow \epsilon_{\alpha \mathbf{k}}(t) =   \epsilon_{\alpha \mathbf{k}} + \delta_{\alpha}(t) ~,\\
& c^{\dagger}_{\alpha\mathbf{k}\sigma} c_{\alpha\mathbf{k}\sigma} \longrightarrow c^{\dagger}_{\alpha\mathbf{k}\sigma}(t) c_{\alpha\mathbf{k}\sigma}(t) ~,\\
& H^{U}_{imp} \lbrace d^{\dagger}_{n \sigma} ; d_{n \sigma}\rbrace \longrightarrow 
H^{U}_{imp} \lbrace d^{\dagger}_{n \sigma} (t) ; d_{n \sigma} (t)\rbrace ~,\\
&  V_{\alpha\mathbf{k}} \longrightarrow  V_{\alpha\mathbf{k}}(t) ~.
\end{aligned}
\end{equation}
We note the single-particle energy of the uncoupled lead is split into static and time-dependent components, such that under the limit $\delta_{\alpha}(t)\rightarrow0$, the energy level coincides with its time-independent one. The corresponding time-dependent nonequilibrium leads Green's function of the isolated model are now defined as
\begin{equation}
\begin{aligned}
& G_{0,\alpha \mathbf{k}\alpha \mathbf{k}\sigma}^{<} (t, t^{\prime}) = i f^{\mu_{\alpha}} (\omega- \mu_{\alpha}) exp\bigg[-i \int_{t^{\prime}}^{t} dt_{1}\omega (t_{1})\bigg] ~,\\
& G_{0,\alpha \mathbf{k}\alpha \mathbf{k}\sigma}^{R/A}(t ,t^{\prime}) = \mp \theta(\pm t \mp  t^{\prime}) exp\bigg[-i \int_{t^{\prime}}^{t} dt_{1}\omega (t_{1})\bigg] ~.
\end{aligned}
\end{equation}
By comparing those expressions to their time-independent counterpart and Fourier transformed ones in Eq.\ref{A:eq:g_Def}, we note now that $(i)$ the Green's functions depend on each time label separately, hence time-translation invariance is broken, $(ii)$ the Fourier transform to energy domain is not applicable any longer and the final electrical conductance formula is in time domain, $(iii)$ the electronic distribution for a given $\alpha,k$ state is not modified, hence we can still use the equilibrium Fermi distribution also in time-dependent transport phenomena.\\
We introduce now the \textbf{time-dependent levelwidth function} valid for $ \forall~ n,m$
\begin{equation}
[\mathbb{\Gamma}^{\alpha}(\epsilon_{\mathbf{k}}, t_{1}, t)]_{nm} =\pi  \mathbf{V}_{\alpha\mathbf{k}n} \mathbb{V}^{\star}_{\alpha\mathbf{k}m}\rho_{0,\alpha\sigma}(\epsilon_{\mathbf{k}}) exp\bigg[i \int_{t_{1}}^{t} dt_{2}\delta_{\alpha} (\epsilon_{\mathbf{k}}, t_{2})\bigg] ~,
\end{equation}
that incorporates now the time-dependent part of the lead energy level, compare it with its steady-state version in Eq.\ref{A:eq:LevelWidth}.\\
The \textbf{general time-dependent Meir-Wingreen current formula} with respect to the left lead reads 
\begin{equation}
\begin{aligned}
\hat{I}_{L} (t) = - 2e \sum_{nm\sigma} \int_{-\infty}^{t} dt_{1} \int \frac{d\epsilon_{\mathbf{k}}}{2\pi}
\mathit{Im} Tr \Big( & e^{-i \epsilon_{\mathbf{k}}(t_{1}-t)} [\Gamma^{L}(\epsilon_{\mathbf{k}}, t_{1},t)]_{nm}
G^{<}_{nm\sigma}(t,t_{1}) + \\
& \hspace*{2cm}+ f^{\mu_{L}}(\epsilon_{\mathbf{k}}- \mu_{L})G^{R}_{nm\sigma}(t,t_{1})] \Big) ~,
\end{aligned}
\end{equation}
as presented in later work by Meir and Wingreen \cite{MeirWingreenJauho_TimeDepCurrent_PRB1993}. An analogous current $\hat{I}_{R} (t)$ expression with respect to the right lead can be derived. However, due to the time-dependence in the transport process, $\hat{I}_{L} (t),\hat{I}_{R} (t)$ do not have the same form.  As observed in the steady state case, the current expression is fully determined by the local Green's function of the central region. It is straightforward to reduce the time-dependent result to the time-independent case by switching off the time varying component in the single-particle energy $\delta_{\alpha}(t) \longrightarrow 0$.\\
We conclude this discussion by introducing the\textit{ generalised proportionate coupling condition}, as straightforward time-dependent extension of Eq.\ref{eq:defPC}, namely
\begin{equation} 
[\mathbb{\Gamma}^{L}(\epsilon_{\mathbf{k}},t_{1},t)]_{nm}= \gamma [\mathbb{\Gamma}^{R}(\epsilon_{\mathbf{k}},t_{1},t)]_{nm} ~,
\end{equation} 
that is valid for $ \forall~ n,m$ central-region degrees and it holds only if $\delta_{L}(t)=\delta_{R}(t)\equiv\delta(t)$.\\
By means of this extension in time domain, it can be derived a \textit{time averaged current} whose form is recognisable as of Landauer-type \cite{JauhoWingreeMeir1994}. Hence, even in time-dependent interacting transport, by introducing the appropriate constraints, the Meir-Wingreen formula reduces to the Landauer formula in Eq.\ref{eq:LB_2probe}.\\

\noindent{We} this last discussion we complete an exhaustive derivation on all the aspects involved in the Meir-Wingreen approach. We refer to Sec.\ref{sec:MeirWingreen} for the physical implication of this formulation in mesoscopic transport.

\chapter{Non Equilibrium perturbation expansion of interacting $\Sigma^{</>}$}\label{app:ExpNg} 
In the section \ref{sec:AlternativeMW} on alternative formulation of the Meir-Wingreen current equation, we propose two different approaches. In this appendix we focus on the perturbative one, where we take as prototype the Ng ansatz in Eq.\ref{eq:4Ngansatz} and we develop perturbative calculations to derive a corrected Fermi distribution exact up to order of $\mathcal{O}(V^{2}_{bias},U^{4})$ for interacting models out-of-equilibrium condition - see Eq.\ref{eq:4FermiCorrFinal}. The key idea behind this strategy is to systematically separate the equilibrium from the nonequilibrium contributions such that an explicit voltage bias dependence is found in the expression, see the equation structure in Eq.\ref{eq:4SigmaLessCorrected}.\\
In the main text in Sec.\ref{sec:AlternativeMW}, we describe the analytical steps required to derive expression for the $\Sigma^{<}_{dd\sigma}$. In this appendix,  we present the greater components which share the same formal structure with the lesser ones - just with inverted time variables if we consider their initial location on the Keldysh contour. Further literature references are found in \cite{Spicka_NGF,Ryndyk,LangrethLinearRespNEq1976,Werner_NEQ-MeanFieldWeakCoupling2013}\\
We adopt the same notation as in the main text and we report it here for convenience. The noninteracting spectral function is calculated for the impurity Green's function in Eq.\ref{eq:G_0dd} in a infinite wide-band-limit approximation, see  Eq.\ref{eq:DoSinfWBL} so we use the notation
\begin{equation}
\mathcal{A}_{0,dd\sigma}(\omega) \doteq \frac{2 \Gamma}{(\omega -\epsilon_{d\sigma})^{2} + \Gamma^{2}} ~.
\end{equation}
As derived in the main text, the linear expansion of the Fermi distribution reads
\begin{equation}
f^{\mu_{L}/\mu_{R}}\big( \omega \mp \frac{e V_{bias}}{2} \big) 
= f^{Eq}(\omega) + \zeta_{\alpha} \frac{eV_{bias}}{2}c(\omega, T) +\mathcal{O}(V_{bias}^{2}) ~,
\end{equation}
where we used the notation $\zeta_{\alpha}=\pm1$ to indicate the expansion calculated with respect to left or right lead respectively and we define the function 
\begin{equation}
\frac{1}{k_{B}T}\Big( f^{Eq}(\omega) - (f^{Eq}(\omega))^{2} \Big) = \frac{1}{2k_{B}T \Big ( 1 + \cosh(\omega/k_{B}T) \Big)} = \frac{1}{4k_{B}T} \sech^{2}\Big( \frac{\omega}{2k_{B}T}\Big) \doteq c(\omega, T) ~.
\end{equation} 
We go through now to the list of expression in parallel to those in the main text for the lesser components.\\

\noindent{The} analogue of the lesser component in Eq.\ref{eq:4Sigma0<all}, the \textit{greater self-energy at zero order} reads
\begin{equation} \label{A:Ng4Sigma0>all}
\begin{aligned}
\Sigma^{(0),>}_{0,dd\sigma}(\omega, T, V_{bias}) &= \Big( \Sigma^{(0),>}_{0,dd \sigma}(\omega,T) \Big)^{Eq} + \Big( \Sigma^{(0),>}_{0,dd \sigma}(\omega,T, V_{bias}) \Big)^{NEq} ~,\\
&=\begin{cases}
-2i \sum_{\alpha} \Gamma^{\alpha}(\omega) \big((1-f^{Eq}(\omega)) -\zeta_{\alpha}\frac{eV_{bias}}{2} c(\omega,T)  \big) ~,~ \text{at finite}~T\\
-2i \sum_{\alpha} \Gamma^{\alpha}(\omega)\big(\theta(\epsilon_{F}-\omega) -\zeta_{\alpha}\frac{eV_{bias}}{2} \delta(\omega-\epsilon_{F}) \big)~,~ \text{at}~T=0 
\end{cases} ~,\\
&\equiv \Sigma^{(0),>}_{0,dd \sigma} + V_{bias} \Sigma^{\prime(0),>}_{0,dd \sigma} +\mathcal{O}(V^{2}_{bias},U^{4}) ~.
\end{aligned}
\end{equation} 
The analogous of the lesser component in Eq.\ref{eq:4Sigma2<EqT}, the \textit{greater self-energy at second order} at finite temperature, the equilibrium term reads
\begin{equation}\label{A:Ng4Sigma2>EqT}
\begin{aligned}
\Big( \Sigma^{(2),>}_{0,dd \sigma}(\omega,T) \Big)^{Eq} 
&=-2 \pi i U^{2} \int_{-\infty}^{+\infty} d\omega_{1} \int_{-\infty}^{+\infty} d\omega_{2}
 \mathcal{A}_{0,dd \sigma}(\omega_{1}) \mathcal{A}_{0,dd \overline{\sigma}}(\omega_{2})\mathcal{A}_{0,dd \overline{\sigma}}(\omega_{1}+\omega_{2}-\omega) \times \\
& \hspace*{3.5cm}\times \Big((1- f^{Eq}(\omega_{1}))(1-f^{Eq}(\omega_{2}))f^{Eq}(\omega_{1}+\omega_{2}-\omega) \Big)~,\\
&\equiv U^{2}\Sigma^{(2),>}_{0,dd \sigma} +\mathcal{O}(V^{2}_{bias},U^{4})  
\end{aligned}
\end{equation}
and the analogous of the lesser component in Eq.\ref{eq:4Sigma2<NEqT}, the nonequilibrium contribution
\begin{equation}\label{A:Ng4Sigma2>NEqT}
\begin{aligned}
&\Big( \Sigma^{(2),>}_{0,dd \sigma}(\omega,T,V_{bias}) \Big)^{NEq} = \\
&=-2 \pi i U^{2} \int_{-\infty}^{+\infty} d\omega_{1}\int_{-\infty}^{+\infty} d\omega_{2} \mathcal{A}_{0,dd \sigma}(\omega_{1}) \mathcal{A}_{0,dd \overline{\sigma}}(\omega_{2})\mathcal{A}_{0,dd\overline{\sigma}}(\omega_{1}+\omega_{2}-\omega) \times \\
& \hspace*{3cm}\times \sum_{\alpha}\zeta_{\alpha}\frac{e V_{bias}}{2} \bigg(
(1- f^{Eq}(\omega_{1}))(1- f^{Eq}(\omega_{2}))c(\omega_{1}+\omega_{2}-\omega,T) + \\
&\hspace*{4cm}- f^{Eq}(\omega_{1}+\omega_{2}-\omega)
\Big( (1- f^{Eq}(\omega_{1}))c(\omega_{2},T)+(1- f^{Eq}(\omega_{2}))c(\omega_{1},T)\Big) \bigg) ~,\\
& \equiv V_{bias}U^{2} \Sigma^{\prime(2),>}_{0,dd \sigma} +\mathcal{O}(V^{2}_{bias},U^{4}) ~.
\end{aligned}
\end{equation}
The analogous of the lesser component in Eq.\ref{eq:4Sigma2<EqT0}, at zero temperature the equilibrium contribution reads
\begin{equation} \label{A:Ng4Sigma2>EqT0}
\begin{aligned}
\Big( \Sigma^{(2),>}_{0,dd \sigma}(\omega,T=0) \Big)^{Eq} 
&= -2 \pi i U^2 \int_{\epsilon_{F}}^{\omega+\epsilon_{F}} d\omega_{2} \int_{\epsilon_{F}}^{\omega-\omega_{2}+\epsilon_{F}}d\omega_{1}
\mathcal{A}_{0,dd \sigma}(\omega_{1})\mathcal{A}_{0,dd \overline{\sigma}}(\omega_{2})\mathcal{A}_{0,dd \overline{\sigma}}(\omega_{1}+\omega_{2}-\omega) ~,\\
&\equiv U^{2}\Sigma^{(2),>}_{0,dd \sigma} +\mathcal{O}(V^{2}_{bias},U^{4}) 
\end{aligned}
\end{equation}
and the analogous of the lesser component in Eq.\ref{eq:4Sigma2<NEqT0}, the nonequilibrium contribution
\begin{equation} \label{A:Ng4Sigma2>NEqT0}
\begin{aligned}
&\Big( \Sigma^{(2),>}_{0,dd \sigma}(\omega,T=0,V_{bias}) \Big)^{NEq} =\\
&=\sum_{\alpha}\zeta_{\alpha}\frac{e V_{bias}}{2} (2\pi)^{2} i U^2 
\bigg( \mathcal{A}_{0,dd\overline{\sigma}}(\epsilon_{F})
\int_{\epsilon_{F}}^{\omega} d\omega_{1} 
\mathcal{A}_{0,dd \sigma}(\omega_{1})
\bigg( \mathcal{A}_{0,dd\overline{\sigma}}(\omega_{1}+\epsilon_{F}-\omega) 
-\mathcal{A}_{0,dd\overline{\sigma}} (-\omega_{1}+\epsilon_{F}+\omega)\bigg) +\\
& \hspace*{4cm} +\mathcal{A}_{0,dd\sigma}(\epsilon_{F})
\int_{\epsilon_{F}}^{\omega} d\omega_{2} 
\mathcal{A}_{0,dd \overline{\sigma}}(\omega_{2})\mathcal{A}_{0,dd \overline{\sigma}}(\epsilon_{F}+\omega_{2}-\omega) \bigg)~, \\
& \equiv V_{bias}U^{2} \Sigma^{\prime(2),<}_{0,dd \sigma} +\mathcal{O}(V^{2}_{bias},U^{4}) ~.
\end{aligned}
\end{equation}\\

\noindent{We} collect these findings to give a quantitative version of the Eq.\ref{eq:4SigmaLessCorrected}, now for the greater component
\begin{equation}
\begin{aligned}
&\Sigma^{>}_{dd \sigma}(\omega,T,V_{bias}) = \\
&=\underbrace{ \Sigma^{(0),>}_{dd \sigma}(\omega,T)+U^{2}\Sigma^{(2),>}_{dd \sigma}(\omega,T)}_{\Sigma^{Eq,>}_{dd \sigma}(\omega,T)}
+\underbrace{V_{bias} \Big(\Sigma^{\prime(0),>}_{dd \sigma}(\omega,T) +U^{2}\Sigma^{\prime(2),>}_{dd \sigma}(\omega,T) \Big)}_{\Sigma^{NEq,>}_{dd \sigma}(\omega,T,V_{bias})} +\mathcal{O}(V_{bias}^{2},U^{4}) ~,
\end{aligned}
\end{equation}
where at finite temperature, the equilibrium elements are given in Eqs.\ref{A:Ng4Sigma0>all},\ref{A:Ng4Sigma2>EqT} and the nonequilibrium elements are given in Eqs.\ref{A:Ng4Sigma0>all},\ref{A:Ng4Sigma2>NEqT}. At zero temperature, in Eqs.\ref{A:Ng4Sigma0>all},\ref{A:Ng4Sigma2>EqT0} and in Eqs.\ref{A:Ng4Sigma0>all},\ref{A:Ng4Sigma2>NEqT0} respectively.\\
These equations for the greater components, together with the lesser components in the main text, we compute the corrected Fermi distribution that its final form is equal to Eq.\ref{eq:4FermiCorrFinal}.\\
The expressions in Eqs.\ref{A:Ng4Sigma0>all}-\ref{A:Ng4Sigma2>NEqT0} are also employed in the derivation of nonequilibrium interacting $G^{<}_{dd\sigma}$, resulting in Eq.\ref{eq:4G<Ng}. This equation has direct application into the Meir-Wingreen formula and so we have access to the linear response conductance, see discussion in the main text.

\end{appendices}

\bibliographystyle{unsrt}  
\bibliography{biblio}

\end{singlespace}

\end{document}